\documentclass[acmlarge]{acmart}

\setcopyright{none}

\usepackage{booktabs}   
\usepackage{subcaption} 
\usepackage{amsmath}

\usepackage{amssymb}
\usepackage{tikz-cd}
\usepackage{ifthen}
\usepackage{xspace}
\usepackage{xparse}
\usepackage{xstring}
\usepackage{mathpartir}
\usepackage{cleveref}
\usepackage{enumitem}
\usepackage{stmaryrd}
\usepackage{multicol}
\usepackage{tikz-cd}
\usepackage{mathtools}
\usepackage{stackengine}
\usepackage{scalerel}
\usepackage{wasysym}
\usepackage{fontawesome}
\usepackage{xcolor}
\usepackage{tcolorbox}
\usepackage{accents}
\usepackage{ulem}
\usepackage{multirow}
\usepackage{array,longtable}
\usepackage{turnstile}

\usetikzlibrary{matrix}
\usetikzlibrary{calc}
\usetikzlibrary{positioning}
\usetikzlibrary{decorations.pathreplacing}


\DeclareFontFamily{OT1}{pzc}{}
\DeclareFontShape{OT1}{pzc}{m}{it}{<-> s * [1.10] pzcmi7t}{}
\DeclareMathAlphabet{\mathpzc}{OT1}{pzc}{m}{it}

\crefname{figure}{fig.}{fig.}
\Crefname{figure}{Fig.}{Fig.}
\Crefname{section}{Sec.}{Sec.}

\definecolor{light-gray}{gray}{0.85}

\allowdisplaybreaks

\DeclareUnicodeCharacter{21D2}{$\Rightarrow$}
\DeclareUnicodeCharacter{2113}{$\mathpzc{l}$}
\DeclareUnicodeCharacter{2248}{$\approx$}
\DeclareUnicodeCharacter{2200}{$\forall$}
\DeclareUnicodeCharacter{2218}{$\circ$}
\DeclareUnicodeCharacter{2294}{$\sqcup$}
\DeclareUnicodeCharacter{393}{$\Gamma$}
\DeclareUnicodeCharacter{3A8}{$\Psi$}
\DeclareUnicodeCharacter{3B1}{$\alpha$}
\DeclareUnicodeCharacter{3B2}{$\beta$}
\DeclareUnicodeCharacter{3B7}{$\eta$}
\DeclareUnicodeCharacter{3BB}{$\lambda$}
\DeclareUnicodeCharacter{3BC}{$\mu$}
\DeclareUnicodeCharacter{2080}{$_0$}
\DeclareUnicodeCharacter{2081}{$_1$}
\DeclareUnicodeCharacter{2082}{$_2$}
\DeclareUnicodeCharacter{2261}{$\equiv$}
\DeclareUnicodeCharacter{2E1}{$^l$}
\DeclareUnicodeCharacter{2B3}{$^r$}
\DeclareUnicodeCharacter{2032}{$'$}
\DeclareUnicodeCharacter{21D4}{$\Leftrightarrow$}
\DeclareUnicodeCharacter{224A}{$\approxeq$}
\DeclareUnicodeCharacter{27E8}{$\langle$}
\DeclareUnicodeCharacter{27E9}{$\rangle$}
\DeclareUnicodeCharacter{27F6}{$\longrightarrow$}
\DeclareUnicodeCharacter{220E}{$\blacksquare$}
\DeclareUnicodeCharacter{27FA}{$\Leftrightarrow$}
\DeclareUnicodeCharacter{22A4}{$\top$}
\DeclareUnicodeCharacter{22A5}{$\bot$}
\DeclareUnicodeCharacter{03A3}{$\Sigma$}
\DeclareUnicodeCharacter{03C0}{$\pi$}
\DeclareUnicodeCharacter{25A1}{$\square$}
\DeclareUnicodeCharacter{2237}{$::$}
\DeclareUnicodeCharacter{2308}{$\lceil$}
\DeclareUnicodeCharacter{2309}{$\rceil$}
\DeclareUnicodeCharacter{22A2}{$\vdash$}

\newcolumntype{L}{>{$}l<{$}}
\newcolumntype{R}{>{$}r<{$}}
\newcommand{\codelevel}{\textsc{c}\xspace}
\newcommand{\proglevel}{\textsc{d}\xspace}

\newcommand{\mltt}{\textsc{MLTT}\xspace}

\theoremstyle{remark}
\newtheorem*{remark}{Remark}

\makeatletter
\def\namedlabel#1#2{\begingroup
   \def\@currentlabel{#2}%
   \label{#1}\endgroup
}
\makeatother

\makeatletter
\newcommand{\customlabel}[2]{%
   \protected@write \@auxout {}{\string \newlabel {#1}{{#2}{\thepage}{#2}{#1}{}} }%
   \hypertarget{#1}{}
}
\makeatother

\makeatletter
\DeclareFontFamily{U}{MnSymbolA}{}
\DeclareFontShape{U}{MnSymbolA}{m}{n}{
    <-6>  MnSymbolA5
   <6-7>  MnSymbolA6
   <7-8>  MnSymbolA7
   <8-9>  MnSymbolA8
   <9-10> MnSymbolA9
  <10-12> MnSymbolA10
  <12->   MnSymbolA12}{}
\DeclareFontShape{U}{MnSymbolA}{b}{n}{
    <-6>  MnSymbolA-Bold5
   <6-7>  MnSymbolA-Bold6
   <7-8>  MnSymbolA-Bold7
   <8-9>  MnSymbolA-Bold8
   <9-10> MnSymbolA-Bold9
  <10-12> MnSymbolA-Bold10
  <12->   MnSymbolA-Bold12}{}
\DeclareSymbolFont{MnSyA}{U}{MnSymbolA}{m}{n}
\SetSymbolFont{MnSyA}{bold}{U}{MnSymbolA}{b}{n}

\DeclareRobustCommand{\overleftharpoon}{\mathpalette{\overarrow@\leftharpoonfill@}}
\DeclareRobustCommand{\overrightharpoon}{\mathpalette{\overarrow@\rightharpoonfill@}}
\def\leftharpoonfill@{\arrowfill@\leftharpoondown\mn@relbar\mn@relbar}
\def\rightharpoonfill@{\arrowfill@\mn@relbar\mn@relbar\rightharpoonup}

\DeclareMathSymbol{\leftharpoondown}{\mathrel}{MnSyA}{'112}
\DeclareMathSymbol{\rightharpoonup}{\mathrel}{MnSyA}{'100}
\DeclareMathSymbol{\mn@relbar}{\mathrel}{MnSyA}{'320}
\makeatother

\DeclareFontFamily{U}{mathx}{}
\DeclareFontShape{U}{mathx}{m}{n}{<-> mathx10}{}
\DeclareSymbolFont{mathx}{U}{mathx}{m}{n}
\DeclareMathAccent{\widecheck}{0}{mathx}{"71}

\newcommand{\vect}[1]{\overrightarrow{#1}}

\newcommand{\vGamma}{\vect \Gamma}

\newcommand{\STo}{\Rightarrow}
\newcommand{\D}{\ensuremath{\mathrm{D}}}
\newcommand{\Pf}{\ensuremath{\mathcal{D}}}

\newcommand{\Nat}{\ensuremath{\texttt{Nat}}}

\newcommand{\ze}{\textsf{zero}}
\newcommand{\tsucc}{\textsf{succ}}
\newcommand{\su}[1]{\tsucc~#1}

\newcommand{\Nf}{\textsf{Nf}}
\newcommand{\Ne}{\textsf{Ne}}
\newcommand{\Botp}{\textsf{Bot}}
\newcommand{\Topp}{\textsf{Top}}

\newcommand{\Env}{\textsf{Env}}

\newcommand{\Ctx}{\textsf{Ctx}}

\newcommand{\Exp}{\textsf{Exp}}

\newcommand{\Subst}{\textsf{Subst}}

\newcommand{\N}{\mathbb{N}}
\newcommand{\intp}[2]{\ensuremath{\llbracket #1 \rrbracket(#2)}}
\newcommand{\intpbr}[1]{\ensuremath{\llbracket #1 \rrbracket}}

\newcommand{\Rnf}{\textsf{R}^{\Nf}}
\newcommand{\Rne}{\textsf{R}^{\Ne}}
\newcommand{\Rty}{\textsf{R}^{\textsf{Ty}}}

\newcommand{\zed}{\texttt{ze}}
\newcommand{\sud}[1]{\texttt{su}(#1)}

\newcommand{\telimn}{\textsf{elim}_{\Nat}}
\newcommand\elimn[4]{\ensuremath{\telimn~(#1)~#2~(#3)~#4}}

\newcommand\recd[5]{\mathsf{rec}(#1, #2, #3, #4, #5)}

\newcommand{\sep}{\;|\;}

\newcommand{\nbe}{\textsf{nbe}}

\newcommand\id{\ensuremath{\textsf{id}}}

\newcommand{\Se}{\texttt{Ty}}

\newcommand{\Ty}[1]{\ensuremath{\Se_{#1}}}
\newcommand{\tEl}{\texttt{El}}

\newcommand\Uc{\mathcal{U}}
\newcommand\El{\textbf{El}}

\newcommand\tand{\text{ and }}
\newcommand\byIH{\tag{by IH}}

\newcommand{\at}[1]{\scalebox{0.75}{\color{gray}$\,@~#1$}}

\newcommand{\vertrule}[1][1ex]{\rule{.45pt}{#1}}

\newcommand{\tttstileA}{\ensuremath{\mathrel{\raisebox{0.01pt}{\vertrule[1.6ex]}\hspace{.15em}{\vertrule[1.6ex]}{\hspace{.1em}\vDash}}}}
\newcommand{\sttstileA}{\ensuremath{\mathrel{\raisebox{0.01pt}{\vertrule[1.6ex]}{\hspace{.1em}\vDash}}}}

\DeclareDocumentCommand{\genjudge}{ m m m } {
  \ensuremath{#1 #2 #3}
}

\DeclareDocumentCommand{\singlejudge}{ o m m } {
  \IfNoValueTF {#1}
  {\genjudge{\Gamma}{#2}{#3}}
  {\genjudge{#1}{#2}{#3}}
}

\DeclareDocumentCommand{\dualjudge}{o o m m m } {
  \IfNoValueTF {#1}
  {\IfNoValueTF {#2}
    {\genjudge{\Psi #3 \Gamma}{#4}{#5}}
    {\genjudge{\Psi #3 #2}{#4}{#5}}}
  {\IfNoValueTF {#2}
    {\genjudge{#1 #3 \Gamma}{#4}{#5}}
    {\genjudge{#1 #3 #2}{#4}{#5}}}
}

\DeclareDocumentCommand{\triplejudge}{o o o m m } {
  \IfNoValueTF {#1}
  {\IfNoValueTF {#2}
    {\IfNoValueTF {#3}
      {\genjudge{L \sep \Psi; \Gamma}{#4}{#5}}
      {\genjudge{#3 \sep \Psi; \Gamma}{#4}{#5}}}
    {\IfNoValueTF {#3}
      {\genjudge{L \sep \Psi; #2}{#4}{#5}}
      {\genjudge{#3 \sep \Psi; #2}{#4}{#5}}}}
  {\IfNoValueTF {#2}
    {\IfNoValueTF {#3}
      {\genjudge{L \sep #1; \Gamma}{#4}{#5}}
      {\genjudge{#3 \sep #1; \Gamma}{#4}{#5}}}
    {\IfNoValueTF {#3}
      {\genjudge{L \sep #1; #2}{#4}{#5}}
      {\genjudge{#3 \sep #1; #2}{#4}{#5}}}}  
}

\DeclareDocumentCommand{\dcjudge}{o o m m } {
  \dualjudge[#1][#2]{;}{#3}{#4}
}

\DeclareDocumentCommand{\dSemjudgebf}{ o o m } {
  \dcjudge[#1][#2]{\sttstileA}{#3}
}

\DeclareDocumentCommand{\judge}{ o m } {
  \singlejudge[#1]{\vdash}{#2}
}

\DeclareDocumentCommand{\gequiv}{ o m m } {
  \IfNoValueTF {#1}
  {\judge[L]{#2 \approx #3}}
  {\judge[#1]{#2 \approx #3}}
}

\DeclareDocumentCommand{\cont}{ o m } {
  \square (\judge[#1]{#2})
}

\DeclareDocumentCommand{\rjudge}{ o m } {
  \IfNoValueTF {#1}
  {\singlejudge[\vGamma]{\vdash_r}{#2}}
  {\singlejudge[#1]{\vdash_r}{#2}}
}

\DeclareDocumentCommand{\semjudge}{ o m } {
  \singlejudge[#1]{\vDash}{#2}
}

\DeclareDocumentCommand{\semweak}{ o m m } {
  \singlejudge[#1]{\vDash_w}{#2 : #3}
}

\newcommand{\semctx}[1]{\ensuremath{\vDash #1}}

\DeclareDocumentCommand{\Semjudge}{ o m } {
  \singlejudge[#1]{\Vdash}{#2}
}

\DeclareDocumentCommand{\mjudge}{ o m } {
  \IfNoValueTF {#1}
  {\singlejudge[\vGamma]{\vdash}{#2}}
  {\singlejudge[#1]{\vdash}{#2}}
}

\DeclareDocumentCommand{\djudge}{ o o m } {
  \dualjudge[#1][#2]{;}{\vdash}{#3}
}

\DeclareDocumentCommand{\ljudge}{ o m m } {
  \IfNoValueTF {#1}
  {\genjudge{\Psi}{\vdash_{#2}}{#3}}
  {\genjudge{#1}{\vdash_{#2}}{#3}}
}

\DeclareDocumentCommand{\ljudgel}{ o m m } {
  \IfNoValueTF {#1}
  {\genjudge{\Gamma}{\vdash_{#2}}{#3}}
  {\genjudge{#1}{\vdash_{#2}}{#3}}
}

\DeclareDocumentCommand{\DTyp}{o m m} {
  \ensuremath{(\ljudgel[#1]{#2}{\at{#3}})}
}

\DeclareDocumentCommand{\DTrm}{o m m m} {
  \ensuremath{(\ljudgel[#1]{#2}{#3\at{#4}})}
}

\DeclareDocumentCommand{\CTyp}{o m} {
  \ensuremath{\square \DTyp[#1]{\codelevel}{#2}}
}

\DeclareDocumentCommand{\CTrm}{o m m} {
  \ensuremath{\square \DTrm[#1]{\codelevel}{#2}{#3}}
}

\DeclareDocumentCommand{\TPI}{m o m m m}{
  \ensuremath{(#1 : \DTyp[#2]{\proglevel}{#3}) \STo^{#4} #5}
}

\DeclareDocumentCommand{\lpgenjudge}{ o o m m } {
  \IfNoValueTF {#1}
  {\IfNoValueTF {#2}
    {\genjudge{L \sep \Psi}{#3}{#4}}
    {\genjudge{#2 \sep \Psi}{#3}{#4}}}
  {\IfNoValueTF {#2}
    {\genjudge{L \sep #1}{#3}{#4}}
    {\genjudge{#2 \sep #1}{#3}{#4}}}
}

\DeclareDocumentCommand{\pjudge}{ o o m } {
  \lpgenjudge[#1][#2]{\vdash}{#3}
}

\DeclareDocumentCommand{\ptyping}{o o m m } {
  \pjudge[#1][#2]{#3 : #4}
}

\DeclareDocumentCommand{\ptyequiv}{o o m m m } {
  \pjudge[#1][#2]{#3 \approx #4 : #5}
}

\DeclareDocumentCommand{\lpjudge}{ o o m m } {
  \pjudge[#1][#2]{_{#3}#4}
}

\DeclareDocumentCommand{\lpequiv}{ o o m m m } {
  \lpjudge[#1][#2]{#3}{#4 \approx #5}
}

\DeclareDocumentCommand{\lpcobj}{ o o m m m } {
  \lpjudge[#1][#2]{#3}{#4 : #5}
}

\DeclareDocumentCommand{\tjudge}{ o o o m } {
  \triplejudge[#1][#2][#3]{\vdash}{#4}
}

\DeclareDocumentCommand{\ltjudge}{o o o m m} {
  \tjudge[#1][#2][#3]{_{#4} #5}
}

\DeclareDocumentCommand{\lttypwf}{o o o m m m} {
  \ltjudge[#1][#2][#3]{#4}{#5 \at{#6}}
}

\DeclareDocumentCommand{\lttyping}{o o o m m m m} {
  \ltjudge[#1][#2][#3]{#4}{#5 : #6 \at{#7}}
}

\DeclareDocumentCommand{\lttypingd}{o o o m m m m} {
  \ltjudge[#1][#2][#3]{#4}{#5 : \Ty{#6}\;(\at{#7})}
}

\DeclareDocumentCommand{\ltsubst}{o o o m m m} {
  \ltjudge[#1][#2][#3]{#4}{#5 : #6}
}

\DeclareDocumentCommand{\ltsubstv}{o o o m m m} {
  \ltjudge[#1][#2][#3]{#4}{^{\#} #5 : #6}
}

\DeclareDocumentCommand{\ltsubeq}{o o o m m m m} {
  \ltjudge[#1][#2][#3]{#4}{#5 \approx #6 : #7}
}

\DeclareDocumentCommand{\lttypeq}{o o o m m m m} {
  \ltjudge[#1][#2][#3]{#4}{#5 \approx #6 \at{#7}}
}

\DeclareDocumentCommand{\lttyequiv}{o o o m m m m m} {
  \ltjudge[#1][#2][#3]{#4}{#5 \approx #6 : #7 \at{#8}}
}

\DeclareDocumentCommand{\lttyequivv}{o o o m m m m m} {
  \ltjudge[#1][#2][#3]{#4}{^{\#} #5 \approx #6 : #7 \at{#8}}
}

\DeclareDocumentCommand{\lttyequivd}{o o o m m m m m} {
  \ltjudge[#1][#2][#3]{#4}{#5 \approx #6 : \Ty{#7}\;(\at{#8})}
}

\DeclareDocumentCommand{\lsemjudge}{ o m m } {
  \IfNoValueTF {#1}
  {\genjudge{\Psi}{\sttstileA_{#2}}{#3}}
  {\genjudge{#1}{\sttstileA_{#2}}{#3}}
}

\DeclareDocumentCommand{\lsemvjudge}{ o m m } {
  \IfNoValueTF {#1}
  {\genjudge{\Psi}{\Vdash_{#2}}{#3}}
  {\genjudge{#1}{\Vdash_{#2}}{#3}}
}

\DeclareDocumentCommand{\lmjudge}{ o o m m } {
  \djudge[#1][#2]{_{#3} #4}
}

\DeclareDocumentCommand{\dpjudge}{ o o m } {
  \lpgenjudge[#1][#2]{\sttstileA}{#3}
}

\DeclareDocumentCommand{\dtjudge}{ o o o m } {
  \triplejudge[#1][#2][#3]{\tttstileA}{#4}
}

\DeclareDocumentCommand{\ldpjudge}{ o o m m m } {
  \dpjudge[#1][#2]{_{#3}^{#4} #5}
}

\DeclareDocumentCommand{\ldjudge}{ o o o m m m } {
  \dtjudge[#1][#2][#3]{_{#4}^{#5} #6}
}

\DeclareDocumentCommand{\dpSjudge}{ o o m } {
  \lpgenjudge[#1][#2]{\sttstileA}{#3}
}

\DeclareDocumentCommand{\dSjudge}{ o o o m } {
  \triplejudge[#1][#2][#3]{\sttstileA}{#4}
}

\DeclareDocumentCommand{\dsjudge}{ o o o m } {
  \triplejudge[#1][#2][#3]{\Vdash}{#4}
}

\DeclareDocumentCommand{\ldpSjudge}{ o o m m m } {
  \dpSjudge[#1][#2]{_{#3}^{#4} #5}
}

\DeclareDocumentCommand{\ldSjudge}{ o o o m m m } {
  \dSjudge[#1][#2][#3]{_{#4}^{#5} #6}
}

\DeclareDocumentCommand{\dsemjudge}{ o o m } {
  \dualjudge[#1][#2]{;}{\sttstileA}{#3}
}

\DeclareDocumentCommand{\dSemjudge}{ o o m } {
  \dualjudge[#1][#2]{;}{\Vdash}{#3}
}

\DeclareDocumentCommand{\wmjudge}{ o m } {
  \IfNoValueTF {#1}
  {\vGamma \vdash_{\! w} #2}
  {#1 \vdash_{\! w} #2}
}

\DeclareDocumentCommand{\msemjudge}{ o m } {
  \IfNoValueTF {#1}
  {\vGamma \vDash #2}
  {#1 \vDash #2}
}

\DeclareDocumentCommand{\mSemjudge}{ o m } {
  \IfNoValueTF {#1}
  {\vGamma \Vdash #2}
  {#1 \Vdash #2}
}

\DeclareDocumentCommand{\typing}{ o m m } {
  \judge[#1]{#2 : #3}
}

\DeclareDocumentCommand{\wkjudge}{ o m } {
  \IfNoValueTF {#1}
  {\Gamma \vdash_{\! w} #2}
  {#1 \vdash_{\! w} #2}
}

\DeclareDocumentCommand{\weakening}{ o m m } {
  \wkjudge[#1]{#2 : #3}
}

\DeclareDocumentCommand{\kweak}{ o m m } {
  \IfNoValueTF{#1}
  {\Gamma \vdash_{\! \mathsf{k}} #2 : #3}
  {#1 \vdash_{\! \mathsf{k}} #2 : #3}
}

\DeclareDocumentCommand{\rtyping}{ o m m } {
  \rjudge[#1]{#2 : #3}
}

\DeclareDocumentCommand{\semtyp}{ o m m } {
  \semjudge[#1]{#2 : #3}
}

\DeclareDocumentCommand{\semvtyp}{ o m m } {
  \IfNoValueTF {#1}
  {\semtyp[\Psi]{ #2}{#3}}
  {\semtyp[#1]{ #2}{#3}}
}

\DeclareDocumentCommand{\Semtyp}{ o m m } {
  \Semjudge[#1]{#2 : #3}
}

\DeclareDocumentCommand{\tyequiv}{ o m m m } {
  \judge[#1]{#2 \approx #3 : #4}
}

\DeclareDocumentCommand{\semtyeq}{ o m m m } {
  \semjudge[#1]{#2 \approx #3 : #4}
}

\DeclareDocumentCommand{\semvtyeq}{ o m m m } {
  \IfNoValueTF {#1}
  {\semjudge[\Psi]{ #2 \approx #3 : #4}}
  {\semjudge[#1]{ #2 \approx #3 : #4}}
}

\DeclareDocumentCommand{\lSemtypPrime}{ o o m m m } {
  \dSemjudgebf[#1][#2]{_{#3} #4 : #5}
}

\DeclareDocumentCommand{\mtyping}{ o m m } {
  \mjudge[#1]{#2 : #3}
}

\DeclareDocumentCommand{\lmtyping}{ o m m m } {
  \mjudge[#1]{_{#2} #3 : #4}
}

\DeclareDocumentCommand{\dtyping}{ o o m m } {
  \djudge[#1][#2]{#3 : #4}
}

\DeclareDocumentCommand{\ltyping}{ o o m m m } {
  \djudge[#1][#2]{_{#3} #4 : #5}
}

\DeclareDocumentCommand{\wmtyping}{ o m m } {
  \wmjudge[#1]{#2 : #3}
}

\DeclareDocumentCommand{\mSemtyp}{ o m m } {
  \mSemjudge[#1]{#2 : #3}
}

\DeclareDocumentCommand{\lsemtyp}{ o o m m m } {
  \dsemjudge[#1][#2]{_{#3} #4 : #5}
}

\DeclareDocumentCommand{\lsemvtyp}{ o o m m m } {
  \dSemjudge[#1][#2]{_{#3} #4 : #5}
}

\DeclareDocumentCommand{\lSemtyp}{ o o m m m } {
  \dSemjudge[#1][#2]{_{#3} #4 : #5}
}

\DeclareDocumentCommand{\mrarr}{ o m m } {
  \mjudge[#1]{#2 > #3}
}

\DeclareDocumentCommand{\wmrarr}{ o m m } {
  \wmjudge[#1]{#2 > #3}
}

\DeclareDocumentCommand{\mtyequiv}{ o m m m } {
  \mjudge[#1]{#2 \approx #3 : #4}
}

\DeclareDocumentCommand{\dtyequiv}{ o o m m m } {
  \djudge[#1][#2]{#3 \approx #4 : #5}
}

\DeclareDocumentCommand{\ltyequiv}{ o o m m m m } {
  \djudge[#1][#2]{_{#3} #4 \approx #5 : #6}
}

\DeclareDocumentCommand{\ltygneeq}{ o o m m m } {
  \djudge[#1][#2]{_1 #3 \sim #4 : #5}
}

\DeclareDocumentCommand{\ltygteq}{ o o m m m } {
  \djudge[#1][#2]{_1 #3 \simeq #4 : #5}
}

\DeclareDocumentCommand{\ltyred}{ o o m m m } {
  \djudge[#1][#2]{_1 #3 \rightsquigarrow #4 : #5}
}

\DeclareDocumentCommand{\ltyreds}{ o o m m m } {
  \djudge[#1][#2]{_1 #3 \rightsquigarrow^\ast #4 : #5}
}

\DeclareDocumentCommand{\dtconv}{ o o m m m } {
  \djudge[#1][#2]{_1 #3 \ \hat{\Longleftrightarrow}\ #4 : #5}
}

\DeclareDocumentCommand{\dtconvnf}{ o o m m m } {
  \djudge[#1][#2]{_1 #3 \Longleftrightarrow #4 : #5}
}

\DeclareDocumentCommand{\dtconvne}{ o o m m m } {
  \djudge[#1][#2]{_1 #3 \longleftrightarrow #4 : #5}
}

\DeclareDocumentCommand{\ttypred}{ o o o m m m m } {
  \ltjudge[#1][#2][#3]{#4}{ #5 \rightsquigarrow #6 \at{#7}}
}

\DeclareDocumentCommand{\ttypreds}{ o o o m m m m } {
  \ltjudge[#1][#2][#3]{#4}{ #5 \rightsquigarrow^\ast #6 \at{#7}}
}

\DeclareDocumentCommand{\ttrmred}{ o o o m m m m m } {
  \ltjudge[#1][#2][#3]{#4}{ #5 \rightsquigarrow #6 : #7 \at{#8}}
}

\DeclareDocumentCommand{\ttrmreds}{ o o o m m m m m } {
  \ltjudge[#1][#2][#3]{#4}{ #5 \rightsquigarrow^\ast #6 : #7 \at{#8}}
}

\DeclareDocumentCommand{\tconvtyp}{ o o o m m m m } {
  \ltjudge[#1][#2][#3]{#4}{ #5 \ \hat{\Longleftrightarrow}\ #6 \at{#7}}
}

\DeclareDocumentCommand{\tconvtypnf}{ o o o m m m m } {
  \ltjudge[#1][#2][#3]{#4}{ #5 \Longleftrightarrow #6 \at{#7}}
}

\DeclareDocumentCommand{\tconvtypne}{ o o o m m m m } {
  \ltjudge[#1][#2][#3]{#4}{ #5 \longleftrightarrow #6 \at{#7}}
}

\DeclareDocumentCommand{\tconvctx}{o o m m m}{
  \lpjudge[#1][#2]{#3}{#4 \ \hat{\Longleftrightarrow}\ #5}
}

\DeclareDocumentCommand{\tconvtrm}{ o o o m m m m m } {
  \ltjudge[#1][#2][#3]{#4}{ #5 \ \hat{\Longleftrightarrow}\ #6 : #7 \at{#8}}
}

\DeclareDocumentCommand{\tconvtrmnf}{ o o o m m m m m } {
  \ltjudge[#1][#2][#3]{#4}{ #5 \Longleftrightarrow #6 : #7 \at{#8}}
}

\DeclareDocumentCommand{\tconvtrmne}{ o o o m m m m m } {
  \ltjudge[#1][#2][#3]{#4}{ #5 \ \hat{\longleftrightarrow}\ #6 : #7 \at{#8}}
}

\DeclareDocumentCommand{\tconvtrmnee}{ o o o m m m m m } {
  \ltjudge[#1][#2][#3]{#4}{ #5 \longleftrightarrow #6 : #7 \at{#8}}
}

\DeclareDocumentCommand{\tconvsub}{ o o o m m m m } {
  \ltjudge[#1][#2][#3]{#4}{ #5 \ \hat{\Longleftrightarrow}\ #6 : #7}
}

\DeclareDocumentCommand{\lttypgneeq}{ o o o m m m m } {
  \ltjudge[#1][#2][#3]{#4}{#5 \sim #6 \at{#7}}
}

\DeclareDocumentCommand{\lttypgeq}{ o o o m m m m } {
  \ltjudge[#1][#2][#3]{#4}{#5 \simeq #6 \at{#7}}
}

\DeclareDocumentCommand{\lttrmgneeq}{ o o o m m m m m } {
  \ltjudge[#1][#2][#3]{#4}{#5 \sim #6 : #7 \at{#8}}
}

\DeclareDocumentCommand{\lttrmgeq}{ o o o m m m m m } {
  \ltjudge[#1][#2][#3]{#4}{#5 \simeq #6 : #7 \at{#8}}
}

\DeclareDocumentCommand{\ltctxgeq}{o o m m m}{
  \lpjudge[#1][#2]{#3}{#4 \simeq #5}
}

\DeclareDocumentCommand{\ltsubgeq}{ o o o m m m m } {
  \ltjudge[#1][#2][#3]{#4}{ #5 \simeq #6 : #7}
}

\DeclareDocumentCommand{\lsemtyp}{ o o m m m } {
  \dsemjudge[#1][#2]{_{#3} #4 : #5}
}

\DeclareDocumentCommand{\dsemtyeq}{ o o m m m } {
  \dsemjudge[#1][#2]{#3 \approx #4 : #5}
}

\DeclareDocumentCommand{\ldtypwf}{ o o o m m m m } {
  \ldjudge[#1][#2][#3]{#4}{#5}{#6 \at{#7}}
}

\DeclareDocumentCommand{\ldtypeq}{ o o o m m m m m } {
  \ldjudge[#1][#2][#3]{#4}{#5}{#6 \approx #7 \at{#8}}
}

\DeclareDocumentCommand{\ldtyeqnat}{ o o o m m m } {
  \ldjudge[#1][#2][#3]{#4}{}{#5 \approx #6 : \Nat}
}

\DeclareDocumentCommand{\ldtynfeqnat}{ o o o m m m } {
  \ldjudge[#1][#2][#3]{#4}{}{#5 \simeq #6 : \Nat}
}

\DeclareDocumentCommand{\ldtynfeqm}{ o o o m m m } {
  \ldjudge[#1][#2][#3]{}{}{#4 \simeq #5 : #6}
}

\DeclareDocumentCommand{\ldtyping}{ o o o m m m m m } {
  \ldjudge[#1][#2][#3]{#4}{#5}{#6 : #7 \at{#8}}
}

\DeclareDocumentCommand{\ldtyequivt}{ o o o m m m m m m } {
  \ldjudge[#1][#2][#3]{#4}{#5}{#6 \approx #7 : #8 \at{#9}}
}

\DeclareDocumentCommand{\ldtyequiv}{ o o o m m m m m } {
  \ldjudge[#1][#2][#3]{#4}{#5}{#6 \approx #7 : \tEl(#8)}
}

\DeclareDocumentCommand{\ldctxwf}{ o o m m m } {
  \ldpjudge[#1][#2]{#3}{#4}{#5}
}

\DeclareDocumentCommand{\ldctxeq}{ o o m m m m } {
  \ldpjudge[#1][#2]{#3}{#4}{#5 \approx #6}
}

\DeclareDocumentCommand{\ldsubeq}{ o o o m m m m m } {
  \ldjudge[#1][#2][#3]{#4}{#5}{#6 \approx #7 : #8}
}

\DeclareDocumentCommand{\ldsubst}{ o o o m m m m } {
  \ldjudge[#1][#2][#3]{#4}{#5}{#6 : #7}
}

\DeclareDocumentCommand{\spjudge}{ o o m } {
  \lpgenjudge[#1][#2]{\Vdash}{#3}
}

\DeclareDocumentCommand{\stjudge}{ o o o m } {
  \triplejudge[#1][#2][#3]{\Vdash}{#4}
}

\DeclareDocumentCommand{\lspjudge}{ o o m m } {
  \spjudge[#1][#2]{_{#3} #4}
}

\DeclareDocumentCommand{\lsjudge}{ o o o m m } {
  \stjudge[#1][#2][#3]{_{#4} #5}
}

\DeclareDocumentCommand{\lsgctx}{ o m } {
  \IfNoValueTF {#1}
  {\genjudge{L}{\Vdash}{#2}}
  {\genjudge{#1}{\Vdash}{#2}}
}

\DeclareDocumentCommand{\lstypwf}{ o o o m m m } {
  \lsjudge[#1][#2][#3]{#4}{#5 \at{#6}}
}

\DeclareDocumentCommand{\lstypeq}{ o o o m m m m } {
  \lsjudge[#1][#2][#3]{#4}{#5 \approx #6 \at{#7}}
}

\DeclareDocumentCommand{\lstyping}{ o o o m m m m } {
  \lsjudge[#1][#2][#3]{#4}{#5 : #6 \at{#7}}
}

\DeclareDocumentCommand{\lstyequiv}{ o o o m m m m m } {
  \lsjudge[#1][#2][#3]{#4}{#5 \approx #6 : #7 \at{#8}}
}

\DeclareDocumentCommand{\lsgsubeq}{ o o m m m } {
  \lspjudge[#1][#2]{}{#3 \approx #4 : #5}
}

\DeclareDocumentCommand{\lsgsubst}{ o o m m } {
  \lspjudge[#1][#2]{}{#3 : #4}
}

\DeclareDocumentCommand{\lsctxwf}{ o o m m } {
  \lspjudge[#1][#2]{#3}{#4}
}

\DeclareDocumentCommand{\lsctxeq}{ o o m m m } {
  \lspjudge[#1][#2]{#3}{#4 \approx #5}
}

\DeclareDocumentCommand{\lssubeq}{ o o o m m m m } {
  \lsjudge[#1][#2][#3]{#4}{#5 \approx #6 : #7}
}

\DeclareDocumentCommand{\lssubst}{ o o o m m m } {
  \lsjudge[#1][#2][#3]{#4}{#5 : #6}
}

\DeclareDocumentCommand{\ldStypeq} { o o m m m m m m } {
  \ldSjudge[#1][#3][#2]{#4}{#5}{#6 \approx #7 \at{#8}}
}

\DeclareDocumentCommand{\ldStyequiv} { o o m m m m m m } {
  \ldSjudge[#1][#3][#2]{#4}{#5}{#6 \approx #7 : \tEl(#8)}
}

\DeclareDocumentCommand{\ldStypwfge} { o o o m m m m } {
  \ldjudge[#1][#2][#3]{\ge #4}{#5}{#6 \at{#7}}
}

\DeclareDocumentCommand{\ldStypingge} { o o o m m m m m } {
  \ldjudge[#1][#2][#3]{\ge #4}{#5}{#6 : #7 \at{#8}}
}

\DeclareDocumentCommand{\ldStypeqge} { o o o m m m m m } {
  \ldjudge[#1][#2][#3]{\ge #4}{#5}{#6 \approx #7 \at{#8}}
}

\DeclareDocumentCommand{\ldStyequivge} { o o o m m m m m m } {
  \ldjudge[#1][#2][#3]{\ge #4}{#5}{#6 \approx #7 : #8 \at{#9}}
}

\DeclareDocumentCommand{\ldSctxeqge} { o o m m m m } {
  \ldpSjudge[#1][#2]{\ge #3}{#4}{#5 \approx #6}
}

\DeclareDocumentCommand{\ldSctxwfge} { o o m m m } {
  \ldpSjudge[#1][#2]{\ge #3}{#4}{#5}
}

\DeclareDocumentCommand{\ldSsubeqge} { o o o m m m m m } {
  \ldjudge[#1][#2][#3]{\ge #4}{#5}{#6 \approx #7 : #8}
}

\DeclareDocumentCommand{\ldSsubstge} { o o o m m m m } {
  \ldjudge[#1][#2][#3]{\ge #4}{#5}{#6 : #7}
}

\DeclareDocumentCommand{\ldSctxeq} { o o m m m m } {
  \ldpSjudge[#1][#2]{#3}{#4}{#5 \approx #6}
}

\DeclareDocumentCommand{\ltSjudge}{ o o o m m m } {
  \triplejudge[#1][#2][#3]{\vDash_{#4}^{#5} #6}
}

\DeclareDocumentCommand{\ltStypwf} { o o m m m m m } {
  \ltSjudge[#1][#3][#2]{#4}{#5}{#6 \at{#7}}
}

\DeclareDocumentCommand{\ltStypeq} { o o m m m m m m } {
  \ltSjudge[#1][#3][#2]{#4}{#5}{#6 \approx #7 \at{#8}}
}

\DeclareDocumentCommand{\ltSctxwf} { o m m } {
  \lpgenjudge[#2][#1]{\vDash_{p}^{p}}{#3}
}

\DeclareDocumentCommand{\ltSctxeq} { o m m m } {
  \lpgenjudge[#2][#1]{\vDash_{p}^{p}}{#3 \approx #4}
}

\DeclareDocumentCommand{\ldSsubeq} { o o m m m m m m } {
  \ldSjudge[#1][#3][#2]{#4}{#5}{#6 \approx #7 : #8}
}

\DeclareDocumentCommand{\ldSgctx}{ o m } {
  \IfNoValueTF {#1}
  {\genjudge{L}{\sttstileA}{#2}}
  {\genjudge{#1}{\sttstileA}{#2}}
}

\DeclareDocumentCommand{\ldSgctxeq}{ o m m } {
  \IfNoValueTF {#1}
  {\genjudge{L}{\vDash}{#2 \approx #3}}
  {\genjudge{#1}{\vDash}{#2 \approx #3}}
}

\DeclareDocumentCommand{\ldSgsubst}{ o o m m } {
  \lpgenjudge[#1][#2]{\sttstileA}{#3 : #4}
}

\DeclareDocumentCommand{\ldSgsubeq}{ o o m m m } {
  \lpgenjudge[#1][#2]{\sttstileA}{#3 \approx #4 : #5}
}

\DeclareDocumentCommand{\ldssubst} { o o o m m m  } {
  \dsjudge[#1][#2][#3]{_{#4} #5 : #6}
}

\DeclareDocumentCommand{\ldstypwf} { o o o m m m  } {
  \dsjudge[#1][#2][#3]{_{#4} #5 \at{#6}}
}

\DeclareDocumentCommand{\ldstyping} { o o o m m m m } {
  \dsjudge[#1][#2][#3]{_{#4} #5 : #6 \at{#7}}
}

\DeclareDocumentCommand{\ltsemjudge} { o o o m m m } {
  \triplejudge[#1][#2][#3]{\sttstile{}{}_{#4}^{#5}}{#6}
}

\DeclareDocumentCommand{\ltsemtypwf} { o o o m m m m } {
  \ltsemjudge[#1][#2][#3]{#4}{#5}{#6 \at{#7}}
}

\DeclareDocumentCommand{\ltsemtyping} { o o o m m m m m } {
  \ltsemjudge[#1][#2][#3]{#4}{#5}{#6 : #7 \at{#8}}
}

\DeclareDocumentCommand{\ltsemsubst} { o o o m m m m } {
  \ltsemjudge[#1][#2][#3]{#4}{#5}{#6 : #7}
}

\DeclareDocumentCommand{\lsemtyeqp}{ o o m m m m m } {
  \dsemjudge[#1][#2]{_{#3}^{#4} {#5} \approx #6 : #7}
}

\DeclareDocumentCommand{\lsemtyeq}{ o o m m m m } {
  \dsemjudge[#1][#2]{_1^{#3} {#4} \approx #5 : #6}
}

\DeclareDocumentCommand{\lsemtypingc}{ o o m m } {
  \dsemjudge[#1][#2]{_0^0 {#3} : #4}
}

\DeclareDocumentCommand{\lsemtypingp}{ o o m m } {
  \dualjudge[#1][#2]{;}{\vDash_1^0}{#3 : #4}
}

\DeclareDocumentCommand{\lsemvtyeq}{ o o m m m m } {
  \dSemjudge[#1][#2]{_{#3} #4 \approx #5 : #6}
}

\DeclareDocumentCommand{\msemtyeq}{ o m m m } {
  \msemjudge[#1]{#2 \approx #3 : #4}
}
\DeclareDocumentCommand{\msemtyp}{ o m m } {
  \msemjudge[#1]{#2 : #3}
}

\DeclareDocumentCommand{\mgluty}{ o m o m } {
  \IfNoValueTF {#3}
  {\mjudge[#1]{#2 \; \circledR \; #4}}
  {\mjudge[#1]{#2 \; \circledR_{#3} \; #4}}
}

\DeclareDocumentCommand{\mglutm}{ o m m m o m } {
   \IfNoValueTF {#5}
   {\mjudge[#1]{#2 : #3 \; \circledR \; #4 \in \El(#6)}}
   {\mjudge[#1]{#2 : #3 \; \circledR_{#5} \; #4 \in \El_{#5}(#6)}}
 }

 \DeclareDocumentCommand{\mglunat}{ o m m } {
   \mjudge[#1]{#2 : \Nat \; \circledR \; #3 \in Nat}
 }

\DeclareDocumentCommand{\mglutms}{ o m m m } {
  \mjudge[#1]{#2 : #3 \; \circledR \; #4}
}

\DeclareDocumentCommand{\mglutyu}{ o m o m } {
  \IfNoValueTF {#3}
  {\mjudge[#1]{#2 \; \overline{\circledR} \; #4}}
  {\mjudge[#1]{#2 \; \overline{\circledR}_{#3} \; #4}}
}

\DeclareDocumentCommand{\mglutmu}{ o m m m o m } {
  \IfNoValueTF {#5}
  {\mjudge[#1]{#2 : #3 \; \overline{\circledR} \; #4 \in \El(#6)}}
  {\mjudge[#1]{#2 : #3 \; \overline{\circledR}_{#5} \; #4 \in \El_{#5}(#6)}}
}

\DeclareDocumentCommand{\mglutmd}{ o m m m o m } {
  \IfNoValueTF {#5}
  {\mjudge[#1]{#2 : #3 \; \underline{\circledR} \; #4 \in \El(#6)}}
  {\mjudge[#1]{#2 : #3 \; \underline{\circledR}_{#5} \; #4 \in \El_{#5}(#6)}}
}

\newcommand{\labeledit}[1]{\label{#1}}

\DeclareDocumentCommand{\subty}{ o m m } {
  \IfNoValueTF {#1}
  {\Gamma \vdash #2 \subseteq #3}
  {#1 \vdash #2 \subseteq #3}
}

\AtBeginDocument{%
  \providecommand\BibTeX{{%
    \normalfont B\kern-0.5em{\scshape i\kern-0.25em b}\kern-0.8em\TeX}}}

\begin{document}

\title{Normalization by Evaluation with Substitutions as Operations}

\author{Jason Z. S. Hu}
\authornote{This research was conducted independently of Amazon.}
\email{fdhzs2010@hotmail.com}
\affiliation{%
  \institution{Amazon}
  \city{Seattle}
  \state{WA}
  \country{USA}
}

\begin{abstract}
  In previous works on normalization by evaluation (NbE) with an untyped domain model,
  explicit substitutions are a major decision choice, which, in retrospect, is taken
  to simplify subsequent semantic proofs. %
  This choice hence is taken for a purely technical reason. %
  In addition, there is no obvious lemma to test the completeness of the definition or
  the laws for explicit substitutions. %
  In this technical report, we investigate NbE in a typical setting, i.e. defining
  substitutions as operations. %
  We show that this setting brings simplifications in proving syntactic properties and
  soundness of NbE, though complications are introduced in the completeness proof of
  NbE. 
\end{abstract}

\maketitle

\section{Introduction}\labeledit{sec:intro}

Normalization by evaluation (NbE) is a technique for deciding definitional equality in
type theories by computing normal forms via an interpretation into a semantic domain,
followed by a reification (readback) procedure that extracts syntactic normal forms
from semantic values~\cite{DBLP:conf/lics/BergerS91,altenkirch_categorical_1995}.  In
the dependently typed setting, NbE has been studied extensively
\cite{abel_normalization_2013,DBLP:journals/pacmpl/GratzerSB19,DBLP:conf/rta/AltenkirchK16,DBLP:journals/lmcs/AltenkirchK17,DBLP:journals/pacmpl/0001GHP25,DBLP:journals/jfp/HuJP23,DBLP:journals/pacmpl/JiangHO25},
and serves as the algorithmic backbone of type checkers for proof assistants.

A recurring design decision in formalizations of NbE for dependent type theories is
whether substitutions should be treated as \emph{explicit} syntactic constructs or as
meta-level \emph{operations} on terms.
In the explicit substitution approach, substitutions appear as term formers in the syntax,
and the theory includes definitional equalities governing their interaction with other
term formers (e.g. the congruence rule for $\lambda$ abstraction $(\lambda
x. t)[\sigma] = \lambda x. t[q(\sigma)]$).
This approach is prevalent in recent formalized accounts of
NbE~\cite{DBLP:conf/rta/AltenkirchK16,DBLP:journals/lmcs/AltenkirchK17,DBLP:journals/pacmpl/0001GHP25,DBLP:journals/jfp/HuJP23,DBLP:journals/pacmpl/JiangHO25},
partly because it simplifies the semantic proofs: the compositionality of the evaluation
function follows directly from the syntax, and many substitution lemmas become trivial
consequences of the equational theory rather than requiring separate inductive
proofs. %
Nevertheless, explicit substitutions have two downsides:
\begin{itemize}
\item The exact formulation of explicit substitutions and its completeness are in
  question. %
  For instance, we have seen explicit substitutions with and without terminal object
  (i.e. $\epsilon: \Gamma \Rightarrow \cdot$ the substitution that puts a closed term
  in any context) and its companion laws. %
  The associativity law of substitutions is technically not required given the
  presence of composition law ($t[\sigma \circ \sigma'] =  t[\sigma][\sigma']$). %
  Thus, explicit substituions are difficult to standardize and its formulation is
  entirely on the author(s)' discretion. 
\item That equations of substitutions become part of definition equality have made
  proving, especially in the soundness proof, more technically involved. %
  When reasoning about a chain of equalities of applying substitutions, explicit
  substitutions requires us also maintain the well-formedness of well-typedness of
  ambient contexts and types. %
  This adds undesired complexity to the proof, whereas the operational approach allows
  unconditional equational reasoning as usual. 
\end{itemize}

The operational approach, in contrast, defines substitution as a recursive function on
raw terms.  This is the ``classical'' treatment found in standard presentations of
Martin-L\"of type theory
(\mltt)~\cite{DBLP:books/daglib/0000395,martin-lof_intuitionistic_1975}.  While
conceptually simpler and closer to implementations, this approach also requires us to
establish a substantial body of tedious \emph{substitution lemmas} --- associativity
of composition, interaction with weakening, and compatibility with the equational
theory --- as separate meta-theoretic results. %
Nevertheless, we argue that, to date, we should welcome more tedious prerequisites as
they can easily be discharged by established tools like autosubst~\citep{DBLP:conf/itp/SchaferTS15,DBLP:conf/cpp/StarkSK19}
and even LLM agents. %
The complication, however, lies in the completeness proof. %
In this approach, the semantic interpretation must explicitly witness that evaluation
commutes with substitution, which requires careful reasoning about environments and
their relationship to syntactic substitutions. %
This problem is shortcircuited with explicit substitutions. 

Previously, we have investigated the simply typed $\lambda$-calculus (SLTC), and
demonstrated that with substitutions as operations, the overall proof
structure remains manageable and brings certain simplifications to the soundness proof,
in spite of increasing complexity in the completeness proof. 
In this technical report, we extend the investigation to \mltt. %
In particular, we focus on extending the semantic judgment for completeness in SLTC to
dependent types. %
With subsitutions as operations, we also observe significant simplifications in the soundness proof.

\paragraph{Contributions.}
We present a complete account of NbE for a core \mltt with a universe \`a la Tarski,
$\Pi$-types, and natural numbers, where substitutions are defined as operations rather
than explicit syntax.
Our main contributions are:
\begin{enumerate}
\item A careful presentation of the syntax, typing rules, and equational theory for
  \mltt with operational substitutions (\Cref{sec:syn,sec:rules,sec:props}).
\item A domain model and evaluation/readback functions suitable for the operational
  setting (\Cref{sec:sem}).
\item A PER model establishing completeness: any two terms that are judgmentally equal
  evaluate to related semantic values, which in turn read back to the same normal form
  (\Cref{sec:comp}).
\item A Kripke logical predicate establishing soundness: any well-typed term evaluates
  to a semantic value from which a normal form can be extracted, and this normal form is
  judgmentally equal to the original term (\Cref{sec:sound}).
\end{enumerate}

\paragraph{Notation.}
We use $\N$ for the type of natural numbers at the meta-level.
Contexts are snoc-lists written $\Gamma, x : T$.
Application of a substitution $\sigma$ to a term $t$ is written $t[\sigma]$, and
composition of substitutions is written $\sigma \circ \tau$ (meaning first $\sigma$, then $\tau$).

\section{Syntax and Substitution}\labeledit{sec:syn}

In this section, we define the raw syntax of our core Martin-L\"of Type Theory (\mltt)
and introduce weakenings and substitutions as meta-level operations (recursive functions
on expressions).
We establish the key algebraic properties of these operations, which form the backbone
of the subsequent development.

The crucial design decision---and the one that distinguishes this work from
Abel---is that substitutions are \emph{not} part of
the term syntax.
Instead, the application of a substitution to a term is defined by recursion on the
structure of the term, and the basic laws governing the interaction of substitutions with
term formers are proved as \emph{meta-theoretic lemmas} rather than being postulated as
definitional equalities of the type theory.

Throughout this presentation, we use named variables, ranging over $x$, $y$, $z$. %
However, since we are handling weakenings and substitutions, we find it more rigorous
to use de Bruijn indices under the hood. %
Therefore, we might use $x$ to also refer to its underlying de Bruijn index as well,
whenever the ambient context requires precision. %
We might use subscript, e.g. $x_0$, to emphasize the de Bruijn index of a particular
variable.

\subsection{Raw Syntax}\labeledit{sec:syn:raw}

We work with a single syntactic category $\Exp$ of \emph{expressions}, which encompasses
both types and terms. %
\begin{definition}[Raw Expressions]\labeledit{def:raw-exp}
  The set $\Exp$ of raw expressions is generated by the following grammar:
  \[
    \begin{array}{rcll}
      t, s, S, T \in \Exp & ::= & x & \text{(variable)} \\
                          & \mid & \Ty{i} & \text{(universe at level $i$)} \\
                          & \mid & \Nat & \text{(natural number type)} \\
                          & \mid & \Pi(x : S).\, T & \text{(dependent function type, $x$ bound in $T$)} \\
                          & \mid & \ze & \text{(zero)} \\
                          & \mid & \su{t} & \text{(successor)} \\
                          & \mid & \elimn{x.\,T}{z}{x, y.\, s}{t} & \text{(natural number eliminator)} \\
                          & \mid & \lambda x.\, t & \text{(lambda abstraction, $x$ bound in $t$)} \\
                          & \mid & r~s & \text{(application)}
    \end{array}
  \]
\end{definition}

In the grammar above, $\Pi(x : S).\, T$ represents the dependent function type where $T$
may refer to the bound variable~$x$.
The constructor $\lambda x.\, t$ is the introduction form for $\Pi$-types: the body
$t$ may refer to the variable~$x$ of type~$S$.
The eliminator $\elimn{x.\,T}{z}{x, y.\, s}{t}$ eliminates a natural number $t$ with
motive $T$, where $x$ refers to the natural number being eliminated, zero case $z$, and
successor case $s$. %
In the successor case, $x$ names the predecessor and $y$ names the inductive hypothesis;
both are bound in $s$.

\begin{definition}[Contexts]\labeledit{def:ctx}
  Contexts $\Gamma, \Delta \in \Ctx$ are ordered lists of variable-type pairs:
  \[
    \Gamma \in \Ctx \quad ::= \quad \cdot \quad \mid \quad \Gamma, x : S
  \]
  where $\cdot$ is the empty context and $\Gamma, x : S$ extends $\Gamma$ with a fresh
  variable $x$ of type $S \in \Exp$.
  We write $|\Gamma|$ for the length of $\Gamma$.
\end{definition}

\subsection{Weakenings}\labeledit{sec:syn:wk}

To define the substitution operation, we must define weakenings first, ranging over
$\phi$, $\psi$. %
Weakenings adjust de Bruijn indices. %
In our presentation, we define weakenings are functions between natural numbers:
$\N \to \N$. %

\begin{definition}[Weakenings]\labeledit{def:wk}
  A \emph{weakening} is a function $\phi : \N \to \N$.
  The key operations on weakenings are:
  \begin{align*}
    \id &: \N \to \N & \id(x) &= x \\
    {\Uparrow} &: \N \to \N & {\Uparrow}(x) &= x + 1 \\
    q(\phi) &: \N \to \N & q(\phi)(0) &= 0 \\
    && q(\phi)(x+1) &= \phi(x) + 1
  \end{align*}
  The shift $\Uparrow$ increments all indices by one (corresponding to introducing one new
  binding).
  The operation $q(\phi)$ extends a weakening under a binder: it fixes index~$0$
  (the newly bound variable) and applies $\phi$ followed by a successor to all
  other indices.
\end{definition}

\begin{definition}[Weakening Application]\labeledit{def:wk-app}
  The application of a weakening $\phi$ to a term $t$, written $t[\phi]$, is
  defined by recursion on the structure of $t$:
  \begin{align*}
    x[\phi] &= \phi(x) \\
    \Ty{i}[\phi] &= \Ty{i} \\
    \Nat[\phi] &= \Nat \\
    (\Pi(x : S).\, T)[\phi] &= \Pi(x : S[\phi]).\, (T[q(\phi)]) \\
    \ze[\phi] &= \ze \\
    (\su{t})[\phi] &= \su{(t[\phi])} \\
    (\elimn{x.\,T}{z}{x, y.\, s}{t})[\phi] &= \elimn{x.\,T[q(\phi)]}{z[\phi]}{x,y.\, s[q^2(\phi)]}{t[\phi]} \\
    (\lambda x.\, t)[\phi] &= \lambda x.\, t[q(\phi)] \\
    (r\; s)[\phi] &= (r[\phi])\; (s[\phi])
  \end{align*}
  Whenever we pass under a binder, we extend the weakening with $q$.
  The successor case of the eliminator binds two variables ($x$ for the predecessor,
  $y$ for the inductive hypothesis), hence $q$ is applied twice.
\end{definition}

Weakenings compose as: $(\phi \circ \psi)(x) = \psi(\phi(x))$.
We now establish the key algebraic properties of weakenings.

\begin{lemma}[$q$-Identity]\labeledit{lem:wk-q-id}
  $q(\id) = \id$.
\end{lemma}
\begin{proof}
  By case analysis on the argument:
  \begin{align*}
    q(\id)(0) &= 0 = \id(0) \\
    q(\id)(x+1) &= \id(x) + 1 = x + 1 = \id(x+1)
  \end{align*}
\end{proof}

\begin{lemma}[Identity Weakening]\labeledit{lem:wk-id}
  For all $t \in \Exp$, $t[\id] = t$.
\end{lemma}
\begin{proof}
  By induction on $t$.
  We show a few representative cases.
  \begin{itemize}[label=Case]
  \item $t = x$ (variable).
    \begin{align*}
      x[\id] = \id(x) = x
    \end{align*}
  \item $t = \lambda x.\, t'$.
    \begin{align*}
      (\lambda x.\, t')[\id]
      &= \lambda x.\, (t'[q(\id)]) \\
      &= \lambda x.\, (t'[\id])
        \tag{by \Cref{lem:wk-q-id}} \\
      &= \lambda x.\, t'
        \byIH
    \end{align*}
  \item $t = \Pi(x : S).\, T$.
    \begin{align*}
      (\Pi(x : S).\, T)[\id]
      &= \Pi(x : S[\id]).\, (T[q(\id)]) \\
      &= \Pi(x : S).\, (T[\id])
        \tag{by IH and \Cref{lem:wk-q-id}} \\
      &= \Pi(x : S).\, T
        \byIH
    \end{align*}
  \end{itemize}
  All other cases follow the same pattern.
\end{proof}

\begin{lemma}[$q$ Distributes over Composition]\labeledit{lem:wk-q-comp}
  For all weakenings $\phi, \psi$:
  $q(\phi) \circ q(\psi) = q(\phi \circ \psi)$.
\end{lemma}
\begin{proof}
  By case analysis on the argument:
  \begin{align*}
    (q(\phi) \circ q(\psi))(0) &= q(\psi)(q(\phi)(0)) = q(\psi)(0) = 0 = q(\phi \circ \psi)(0)  \\
    (q(\phi) \circ q(\psi))(x+1) &= q(\psi)(\phi(x) + 1)
    = \psi(\phi(x)) + 1 \\
    &= (\phi \circ \psi)(x) + 1 = q(\phi \circ \psi)(x+1)
  \end{align*}
\end{proof}

The previous lemma generalizes to arbitrary applications of $q$:
\begin{lemma}[$q^n$ Distributes over Composition]\labeledit{lem:wk-qn-comp}
  For all weakenings $\phi, \psi$:
  $q^n(\phi) \circ q^n(\psi) = q^n(\phi \circ \psi)$.
\end{lemma}

\begin{lemma}[Composition of Weakening Application]\labeledit{lem:wk-comp}
  For all $t \in \Exp$ and weakenings $\phi, \psi$:
  $t[\phi][\psi] = t[\phi \circ \psi]$.
\end{lemma}
\begin{proof}
  By induction on $t$.
  We show a few representative cases.
  \begin{itemize}[label=Case]
  \item $t = x$ (variable).
    \begin{align*}
      x[\phi][\psi] = (\phi(x))[\psi] = \psi(\phi(x)) = (\phi \circ \psi)(x) = x[\phi \circ \psi]
    \end{align*}
  \item $t = \lambda x.\, t'$.
    \begin{align*}
      (\lambda x.\, t')[\phi][\psi]
      &= (\lambda x.\, (t'[q(\phi)]))[\psi] \\
      &= \lambda x.\, (t'[q(\phi)][q(\psi)]) \\
      &= \lambda x.\, (t'[q(\phi) \circ q(\psi)])
        \byIH \\
      &= \lambda x.\, (t'[q(\phi \circ \psi)])
        \tag{by \Cref{lem:wk-q-comp}} \\
      &= (\lambda x.\, t')[\phi \circ \psi]
    \end{align*}
  \item $t = \elimn{x.\,T}{z}{x, y.\, s}{t'}$.
    \begin{align*}
      & (\elimn{x.\,T}{z}{x, y.\, s}{t'})[\phi][\psi] \\
      &= (\elimn{x.\,T[q(\phi)]}{z[\phi]}{x, y.\, s[q^2(\phi)]}{t'[\phi]})[\psi] \\
      &= \elimn{x.\,T[q(\phi)][q(\psi)]}{z[\phi][\psi]}{x, y.\, s[q^2(\phi)][q^2(\psi)]}{t'[\phi][\psi]} \\
      &= \elimn{x.\,T[q(\phi) \circ q(\psi)]}{z[\phi \circ \psi]}{x, y.\, s[q^2(\phi) \circ q^2(\psi)]}{t'[\phi \circ \psi]}
        \byIH \\
      &= \elimn{x.\,T[q(\phi \circ \psi)]}{z[\phi \circ \psi]}{x, y.\, s[q^2(\phi \circ \psi)]}{t'[\phi \circ \psi]}
        \tag{by \Cref{lem:wk-qn-comp}} \\
      &= (\elimn{x.\,T}{z}{x, y.\, s}{t'})[\phi \circ \psi]
    \end{align*}
  \end{itemize}
  The cases for $\su{t}$, $r\; s$, $\Pi$, $\Nat$, $\Ty{i}$, and $\ze$ are similar or
  immediate.
\end{proof}

\begin{lemma}[Categorical Laws for Weakenings]\labeledit{lem:wk-cat}
  Weakenings satisfy:
  \begin{enumerate}
  \item (Left identity) $\id \circ \phi = \phi$.
  \item (Right identity) $\phi \circ \id = \phi$.
  \item (Associativity) $(\phi \circ \psi) \circ \chi = \phi \circ (\psi \circ \chi)$.
  \end{enumerate}
\end{lemma}
\begin{proof}
  All three are immediate from the definition of composition as function composition on $\N$:
  \begin{align*}
    (\id \circ \phi)(x) &= \phi(\id(x)) = \phi(x) \\
    (\phi \circ \id)(x) &= \id(\phi(x)) = \phi(x) \\
    ((\phi \circ \psi) \circ \chi)(x) &= \chi(\psi(\phi(x))) = (\phi \circ (\psi \circ \chi))(x)
  \end{align*}
\end{proof}

\subsection{Substitutions}\labeledit{sec:syn:subst}

A \emph{substitution} is a function $\sigma : \N \to \Exp$ that maps each de Bruijn
index to an expression.
Substitution is the key meta-level operation in our development: it is defined by recursion
on the structure of terms and is \emph{not} part of the object language syntax.

\begin{definition}[Substitution Operations]\labeledit{def:subst-ops}
  We define the following operations on substitutions:
  \begin{enumerate}[label=(\roman*)]
    \item \textbf{Identity.}
      $\id : \Subst$, defined by $\id(x) = x$.
    \item \textbf{Extension (cons).}
      Given a substitution $\sigma$, a variable $x$, and a term $s$, we write
      $(\sigma, s/x)$ for the substitution:
      \[
        (\sigma, s/x)(0) = s \qquad (\sigma, s/x)(x+1) = \sigma(x)
      \]
      Here $x$ is a visual mnemonic for the variable at de Bruijn index~0;
      under the hood, the extension always targets index~0.
    \item \textbf{Lifting under a binder.}
      Given a substitution $\sigma$, we define $q(\sigma) : \Subst$ by:
      \[
        q(\sigma)(0) = 0 \qquad q(\sigma)(x+1) = (\sigma(x))[\Uparrow]
      \]
      This fixes index~0 and applies $\sigma$ to the remaining indices, weakening the
      results by one to account for the new binder.
    \item \textbf{Applying a weakening to a substitution.}
      Given a substitution $\sigma$ and a weakening $\phi$, we define
      $\sigma[\phi] : \Subst$ pointwise:
      \[
        (\sigma[\phi])(x) = (\sigma(x))[\phi]
      \]
    \item \textbf{Composition.}
      Given substitutions $\sigma$ and $\tau$, their composition is:
      \[
        (\sigma \circ \tau)(x) = (\sigma(x))[\tau]
      \]
  \end{enumerate}
\end{definition}

The single substitution $t[s/x]$ used in the $\beta$-rule is the special case
$t[\id, s/x]$: the extension of the identity by $s$ at index~0.

\begin{definition}[Substitution Application]\labeledit{def:subst-app}
  The function $({-})[{-}] : \Exp \to \Subst \to \Exp$ is defined by recursion on
  the structure of the expression:
  \begin{align*}
    x[\sigma] &= \sigma(x) \\
    \Ty{i}[\sigma] &= \Ty{i} \\
    \Nat[\sigma] &= \Nat \\
    (\Pi(x : S).\, T)[\sigma] &= \Pi(x : S[\sigma]).\, (T[q(\sigma)]) \\
    \ze[\sigma] &= \ze \\
    (\su{t})[\sigma] &= \su{(t[\sigma])} \\
    (\elimn{x.\,T}{z}{x, y.\, s}{t})[\sigma] &= \elimn{x.\,T[q(\sigma)]}{z[\sigma]}{x,y.\, s[q^2(\sigma)]}{t[\sigma]} \\
    (\lambda x.\, t)[\sigma] &= \lambda x.\, t[q(\sigma)] \\
    (r\; s)[\sigma] &= (r[\sigma])\; (s[\sigma])
  \end{align*}
  Whenever we pass under a binder, we lift $\sigma$ with $q$.
  The successor case of the eliminator binds two variables, hence $q$ is applied twice.
\end{definition}

\begin{definition}[Embedding of Weakenings into Substitutions]\labeledit{def:wk-embed}
  Every weakening $\phi$ can be regarded as a substitution via the embedding:
  \[
    \iota(\phi) : \Subst \qquad \iota(\phi)(x) = \phi(x)
  \]
  That is, $\iota(\phi)$ maps each index $n$ to the variable $\phi(x)$.
\end{definition}

\begin{lemma}[Embedding Correctness]\labeledit{lem:wk-embed-correct}
  For all $t \in \Exp$ and weakenings $\phi$:
  $t[\phi] = t[\iota(\phi)]$.
\end{lemma}
\begin{proof}
  By induction on $t$.
  The key observation is that $q(\iota(\phi)) = \iota(q(\phi))$:
  \begin{align*}
    q(\iota(\phi))(0) &= 0 = \iota(q(\phi))(0) \\
    q(\iota(\phi))(x+1) &= \iota(\phi)(x)[\Uparrow] = \phi(x)[\Uparrow] = \phi(x) + 1 = \iota(q(\phi))(x+1)
  \end{align*}
  With this fact, the induction is straightforward: at each binding form, we use
  $q(\iota(\phi)) = \iota(q(\phi))$ to relate the substitution side to the weakening side.
  For example:
  \begin{align*}
    (\lambda x.\, t')[\iota(\phi)]
    &= \lambda x.\, (t'[q(\iota(\phi))]) \\
    &= \lambda x.\, (t'[\iota(q(\phi))]) \\
    &= \lambda x.\, (t'[q(\phi)])
      \byIH \\
    &= (\lambda x.\, t')[\phi]
  \end{align*}
\end{proof}

\begin{lemma}[Structural Properties of $\iota$]\labeledit{lem:iota-struct}
  The embedding $\iota$ preserves all weakening structure:
  \begin{enumerate}
  \item $\iota(\id) = \id$.
  \item $\iota(\Uparrow) = \Uparrow$ (as substitutions, i.e.\ $\iota(\Uparrow)(x) = x + 1$).
  \item $\iota(q(\phi)) = q(\iota(\phi))$.
  \item $\iota(\phi \circ \psi) = \iota(\phi) \circ \iota(\psi)$.
  \end{enumerate}
\end{lemma}
\begin{proof} $ $
  \begin{enumerate}
  \item $\iota(\id)(x) = \id(x) = x = \id(x)$.
  \item $\iota(\Uparrow)(x) = \Uparrow(x) = x + 1$.
    As a substitution, this is the variable $x+1$, which equals $\Uparrow$ viewed as a substitution.
  \item Already shown in the proof of \Cref{lem:wk-embed-correct}:
    $q(\iota(\phi))(0) = 0 = \iota(q(\phi))(0)$ and
    $q(\iota(\phi))(x+1) = \phi(x) + 1 = \iota(q(\phi))(x+1)$.
  \item We verify pointwise:
    \begin{align*}
      (\iota(\phi) \circ \iota(\psi))(x)
      &= \iota(\phi)(x)[\iota(\psi)] \\
      &= \phi(x)[\iota(\psi)] \\
      &= \psi(\phi(x))
        \tag{by \Cref{lem:wk-embed-correct}} \\
      &= (\phi \circ \psi)(x) \\
      &= \iota(\phi \circ \psi)(x)
    \end{align*}
  \end{enumerate}
\end{proof}

Since $\iota$ preserves identity, shift, $q$, and composition, it is a faithful
embedding of the weakening algebra into the substitution algebra.
We may therefore elide $\iota$ entirely, treating weakenings as substitutions
whenever convenient.
In what follows, we write $t[\phi]$, $\sigma \circ \phi$, etc.,
understanding that $\phi$ is implicitly promoted to a substitution via $\iota$.

We now establish the key algebraic properties of substitutions and their interaction
with weakenings.

\begin{lemma}[$q$-Identity for Substitutions]\labeledit{lem:subst-q-id}
  $q(\id) = \id$.
\end{lemma}
\begin{proof}
  By case analysis:
  \begin{align*}
    q(\id)(0) &= 0 = \id(0) \\
    q(\id)(x+1) &= \id(x)[\Uparrow] = x[\Uparrow] = x + 1 = \id(x+1)
  \end{align*}
\end{proof}

\begin{lemma}[Identity Substitution]\labeledit{lem:subst-id}
  For all $t \in \Exp$: $t[\id] = t$.
\end{lemma}
\begin{proof}
  By induction on $t$.
  \begin{itemize}[label=Case]
  \item $t = x$ (variable).
    $x[\id] = \id(x) = x$.
  \item $t = \lambda x.\, t'$.
    \begin{align*}
      (\lambda x.\, t')[\id]
      &= \lambda x.\, (t'[q(\id)]) \\
      &= \lambda x.\, (t'[\id])
        \tag{by \Cref{lem:subst-q-id}} \\
      &= \lambda x.\, t'
        \byIH
    \end{align*}
  \item $t = \Pi(x:S).\, T$.
    \begin{align*}
      (\Pi(x:S).\, T)[\id]
      &= \Pi(x : S[\id]).\, (T[q(\id)]) \\
      &= \Pi(x : S).\, (T[\id])
        \tag{by IH and \Cref{lem:subst-q-id}} \\
      &= \Pi(x : S).\, T
        \byIH
    \end{align*}
  \end{itemize}
  All other cases are similar.
\end{proof}

Before proving the commutation lemma, we establish auxiliary facts about how $q^n$
acts on weakenings and substitutions at specific indices.

\begin{lemma}[Action of $q^n$ on Indices]\labeledit{lem:qn-index}
  For all $n$ and weakenings $\phi$:
  \begin{enumerate}
  \item If $x < n$, then $q^n(\phi)(x) = x$.
  \item $q^n(\phi)(x) = \phi(0) + x$.
  \item If $x > n$, then $q^n(\phi)(x) = \phi(x - n) + n$.
  \end{enumerate}
  In particular, for $\phi = \Uparrow$:
  \begin{enumerate}
  \item[(a)] If $x < n$, then $q^n(\Uparrow)(x) = x$.
  \item[(b)] $q^n(\Uparrow)(x) = x + 1$.
  \item[(c)] If $x > n$, then $q^n(\Uparrow)(x) = x + 1$.
  \end{enumerate}
\end{lemma}
\begin{proof}
  By induction on $n$.
  \begin{itemize}[label=Case]
  \item $n = 0$: (1) vacuous. (2) $q^0(\phi)(0) = \phi(0) = \phi(0) + 0$.
    (3) $q^0(\phi)(x) = \phi(x) = \phi(x - 0) + 0$.
  \item $n + 1$: We use $q^{n+1}(\phi) = q(q^n(\phi))$.
    \begin{enumerate}
    \item If $x < n+1$: either $x = 0$ (then $q(q^n(\phi))(0) = 0 = x$) or
      $x = y + 1$ with $y < n$ (then $q(q^n(\phi))(y+1) = q^n(\phi)(y) + 1 = y + 1 = x$
      by the IH).
    \item $x = n + 1$: $q(q^n(\phi))(n+1) = q^n(\phi)(n) + 1 = (\phi(0) + n) + 1 = \phi(0) + (n+1)$
      by the IH.
    \item $x > n + 1$: $x = y + 1$ with $y > n$.
      $q(q^n(\phi))(y+1) = q^n(\phi)(y) + 1 = (\phi(y - n) + n) + 1 = \phi((y+1) - (n+1)) + (n+1)$
      by the IH.
    \end{enumerate}
  \end{itemize}
\end{proof}

\begin{lemma}[Action of $q^n$ on Substitution Indices]\labeledit{lem:qn-subst-index}
  For all $n$ and substitutions $\sigma$:
  \begin{enumerate}
  \item If $x < n$, then $q^n(\sigma)(x) = x$.
  \item $q^n(\sigma)(n) = \sigma(0)[\Uparrow^n]$ (where $\Uparrow^n$ denotes $n$-fold shift).
  \item If $x > n$, then $q^n(\sigma)(x) = \sigma(x - n)[\Uparrow^n]$.
  \end{enumerate}
\end{lemma}
\begin{proof}
  By induction on $n$.
  \begin{itemize}[label=Case]
  \item $n = 0$: (1) vacuous. (2) $q^0(\sigma)(0) = \sigma(0) = \sigma(0)[\Uparrow^0]$.
    (3) $q^0(\sigma)(x) = \sigma(x) = \sigma(x-0)[\Uparrow^0]$.
  \item $n + 1$: We use $q^{n+1}(\sigma) = q(q^n(\sigma))$.
    \begin{enumerate}
    \item If $x < n+1$: either $x = 0$ (then $q(q^n(\sigma))(0) = 0 = x$) or
      $x = y + 1$ with $y < n$ (then $q(q^n(\sigma))(y+1) = (q^n(\sigma)(y))[\Uparrow] = y[\Uparrow] = y + 1 = x$
      by the IH (1)).
    \item $x = n + 1$:
      \begin{align*}
        q(q^n(\sigma))(n+1) &= (q^n(\sigma)(n))[\Uparrow] \\
        &= (\sigma(0)[\Uparrow^n])[\Uparrow]
          \tag{by IH (2)} \\
        &= \sigma(0)[\Uparrow^{n+1}]
          \tag{by \Cref{lem:wk-comp}}
      \end{align*}
    \item $x > n+1$: $x = y + 1$ with $y > n$.
      \begin{align*}
        q(q^n(\sigma))(y+1) &= (q^n(\sigma)(y))[\Uparrow] \\
        &= (\sigma(y - n)[\Uparrow^n])[\Uparrow]
          \tag{by IH (3)} \\
        &= \sigma(y - n)[\Uparrow^{n+1}]
          \tag{by \Cref{lem:wk-comp}} \\
        &= \sigma((y+1) - (n+1))[\Uparrow^{n+1}]
      \end{align*}
    \end{enumerate}
  \end{itemize}
\end{proof}

\begin{lemma}[Weakening--Substitution Commutation (Generalized)]\labeledit{lem:wk-subst-comm}
  For all $t \in \Exp$, $n \in \N$, and substitutions $\sigma$:
  \[
    (t[q^n(\Uparrow)])[q^{1+n}(\sigma)] = t[q^n(\sigma)][q^n(\Uparrow)]
  \]
\end{lemma}

The statement we actually need is the special case $n = 0$:
$(t[\Uparrow])[q(\sigma)] = t[\sigma][\Uparrow]$.
The generalization over $n$ is required because passing under binders increments the
$q$-level: in the body of a $\lambda$ or $\Pi$, we need the statement at $n+1$.

\begin{proof}
  By induction on $t$, generalizing over $n$.
  \begin{itemize}[label=Case]
  \item $t = x$ (variable).
    We must show $(q^n(\Uparrow)(x))[q^{1+n}(\sigma)] = (q^n(\sigma)(x))[q^n(\Uparrow)]$.
    We case-split on the position of $x$ relative to $n$.
    \begin{itemize}
    \item If $x < n$:
      \begin{align*}
        \text{LHS}
        &= (q^n(\Uparrow)(x))[q^{1+n}(\sigma)] \\
        &= x[q^{1+n}(\sigma)]
          \tag{by \Cref{lem:qn-index}(a): $q^n(\Uparrow)(x) = x$ since $x < n$} \\
        &= x
          \tag{by \Cref{lem:qn-subst-index}(1): $q^{1+n}(\sigma)(x) = x$ since $x < n < 1+n$}
      \end{align*}
      \begin{align*}
        \text{RHS}
        &= (q^n(\sigma)(x))[q^n(\Uparrow)] \\
        &= x[q^n(\Uparrow)]
          \tag{by \Cref{lem:qn-subst-index}(1): $q^n(\sigma)(x) = x$ since $x < n$} \\
        &= x
          \tag{by \Cref{lem:qn-index}(a)}
      \end{align*}

    \item If $x = n$:
      \begin{align*}
        \text{LHS}
        &= (q^n(\Uparrow)(n))[q^{1+n}(\sigma)] \\
        &= (n + 1)[q^{1+n}(\sigma)]
          \tag{by \Cref{lem:qn-index}(b)} \\
        &= q^{1+n}(\sigma)(1+n) \\
        &= \sigma(0)[\Uparrow^{1+n}]
          \tag{by \Cref{lem:qn-subst-index}(2) at iteration $1+n$}
      \end{align*}
      \begin{align*}
        \text{RHS}
        &= (q^n(\sigma)(n))[q^n(\Uparrow)] \\
        &= (\sigma(0)[\Uparrow^n])[q^n(\Uparrow)]
          \tag{by \Cref{lem:qn-subst-index}(2)} \\
        &= \sigma(0)[\Uparrow^n \circ q^n(\Uparrow)]
          \tag{by \Cref{lem:wk-comp}} \\
        &= \sigma(0)[\Uparrow^{1+n}]
          \tag{since $(\Uparrow^n \circ q^n(\Uparrow))(x) = q^n(\Uparrow)(x + n) = x + 1 + n = \Uparrow^{1+n}(x)$}
      \end{align*}

    \item If $x > n$: Let $x = 1 + n + k$.
      \begin{align*}
        \text{LHS}
        &= (q^n(\Uparrow)(1 + n + k))[q^{1+n}(\sigma)] \\
        &= (2 + n + k)[q^{1+n}(\sigma)]
          \tag{by \Cref{lem:qn-index}(c)} \\
        &= q^{1+n}(\sigma)(2 + n + k) \\
        &= \sigma(1+k)[\Uparrow^{1+n}]
          \tag{by \Cref{lem:qn-subst-index}(3): $(2+n+k) - (1+n) = 1+k$}
      \end{align*}
      \begin{align*}
        \text{RHS}
        &= (q^n(\sigma)(1 + n + k))[q^n(\Uparrow)] \\
        &= (\sigma(1+k)[\Uparrow^n])[q^n(\Uparrow)]
          \tag{by \Cref{lem:qn-subst-index}(3): $(1+n+k) - n = 1+k$} \\
        &= \sigma(1+k)[\Uparrow^n \circ q^n(\Uparrow)]
          \tag{by \Cref{lem:wk-comp}} \\
        &= \sigma(1+k)[\Uparrow^{1+n}]
          \tag{as above}
      \end{align*}
    \end{itemize}

  \item $t = \lambda y.\, t'$.
    \begin{align*}
      (\lambda y.\, t')[q^n(\Uparrow)][q^{1+n}(\sigma)]
      &= (\lambda y.\, t'[q^{1+n}(\Uparrow)])[q^{1+n}(\sigma)] \\
      &= \lambda y.\, (t'[q^{1+n}(\Uparrow)][q^{2+n}(\sigma)]) \\
      &= \lambda y.\, (t'[q^{1+n}(\sigma)][q^{1+n}(\Uparrow)])
        \tag{by IH at $1+n$} \\
      &= (\lambda y.\, t'[q^{1+n}(\sigma)])[q^n(\Uparrow)] \\
      &= (\lambda y.\, t')[q^n(\sigma)][q^n(\Uparrow)]
    \end{align*}

  \item $t = \Pi(y : S).\, T$.
    \begin{align*}
      (\Pi(y:S).\,T)[q^n(\Uparrow)][q^{1+n}(\sigma)]
      &= (\Pi(y : S[q^n(\Uparrow)]).\, (T[q^{1+n}(\Uparrow)]))[q^{1+n}(\sigma)] \\
      &= \Pi(y : S[q^n(\Uparrow)][q^{1+n}(\sigma)]).\, (T[q^{1+n}(\Uparrow)][q^{2+n}(\sigma)]) \\
      &= \Pi(y : S[q^n(\sigma)][q^n(\Uparrow)]).\, (T[q^{1+n}(\sigma)][q^{1+n}(\Uparrow)])
        \tag{by IH at $n$ and $1+n$} \\
      &= (\Pi(y : S[q^n(\sigma)]).\, (T[q^{1+n}(\sigma)]))[q^n(\Uparrow)] \\
      &= (\Pi(y:S).\,T)[q^n(\sigma)][q^n(\Uparrow)]
    \end{align*}
  \end{itemize}
  The remaining cases follow the same pattern, applying the IH at increased $n$ for
  each binder.
\end{proof}

\begin{lemma}[$q$ Distributes over Composition]\labeledit{lem:subst-q-comp}
  For all substitutions $\sigma, \tau$:
  $q(\sigma) \circ q(\tau) = q(\sigma \circ \tau)$.
\end{lemma}
\begin{proof}
  By case analysis:
  \begin{itemize}[label=Case]
  \item At index $0$:
    \begin{align*}
      (q(\sigma) \circ q(\tau))(0) = q(\sigma)(0)[q(\tau)] = 0[q(\tau)] = q(\tau)(0) = 0 = q(\sigma \circ \tau)(0)
    \end{align*}
  \item At index $x+1$:
    \begin{align*}
      (q(\sigma) \circ q(\tau))(x+1)
      &= (q(\sigma)(x+1))[q(\tau)] \\
      &= (\sigma(x))[\Uparrow][q(\tau)] \\
      &= (\sigma(x))[\tau][\Uparrow]
        \tag{by \Cref{lem:wk-subst-comm}} \\
      &= ((\sigma \circ \tau)(x))[\Uparrow] \\
      &= q(\sigma \circ \tau)(x+1)
    \end{align*}
  \end{itemize}
\end{proof}

\begin{lemma}[Composition of Substitution Application]\labeledit{lem:subst-assoc}
  For all $t \in \Exp$ and substitutions $\sigma, \tau$:
  $t[\sigma][\tau] = t[\sigma \circ \tau]$.
\end{lemma}
\begin{proof}
  By induction on $t$.
  \begin{itemize}[label=Case]
  \item $t = x$ (variable).
    $x[\sigma][\tau] = \sigma(x)[\tau] = (\sigma \circ \tau)(x) = x[\sigma \circ \tau]$.
  \item $t = \lambda x.\, t'$.
    \begin{align*}
      (\lambda x.\, t')[\sigma][\tau]
      &= (\lambda x.\, t'[q(\sigma)])[\tau] \\
      &= \lambda x.\, (t'[q(\sigma)][q(\tau)]) \\
      &= \lambda x.\, (t'[q(\sigma) \circ q(\tau)])
        \byIH \\
      &= \lambda x.\, (t'[q(\sigma \circ \tau)])
        \tag{by \Cref{lem:subst-q-comp}} \\
      &= (\lambda x.\, t')[\sigma \circ \tau]
    \end{align*}
  \item $t = \elimn{x.\,T}{z}{x,y.\,s}{t'}$.
    \begin{align*}
      &(\elimn{x.\,T}{z}{x,y.\,s}{t'})[\sigma][\tau] \\
      &= (\elimn{x.\,T[q(\sigma)]}{z[\sigma]}{x,y.\,s[q^2(\sigma)]}{t'[\sigma]})[\tau] \\
      &= \elimn{x.\,T[q(\sigma)][q(\tau)]}{z[\sigma][\tau]}{x,y.\,s[q^2(\sigma)][q^2(\tau)]}{t'[\sigma][\tau]} \\
      &= \elimn{x.\,T[q(\sigma \circ \tau)]}{z[\sigma \circ \tau]}{x,y.\,s[q^2(\sigma \circ \tau)]}{t'[\sigma \circ \tau]}
        \tag{by IH and \Cref{lem:subst-q-comp}} \\
      &= (\elimn{x.\,T}{z}{x,y.\,s}{t'})[\sigma \circ \tau]
    \end{align*}
  \end{itemize}
  Other cases are similar.
\end{proof}

\begin{lemma}[$q$ Commutes with Pointwise Weakening]\labeledit{lem:q-wk-comm}
  For all substitutions $\sigma$ and weakenings $\phi$:
  $q(\sigma[\phi]) = q(\sigma)[q(\phi)]$.
\end{lemma}
\begin{proof}
  Since $\sigma[\phi] = \sigma \circ \phi$ (treating $\phi$ as a substitution), we have:
  \begin{align*}
    q(\sigma[\phi])
    &= q(\sigma \circ \phi) \\
    &= q(\sigma) \circ q(\phi)
      \tag{by \Cref{lem:subst-q-comp}} \\
    &= q(\sigma)[q(\phi)]
  \end{align*}
\end{proof}

\begin{lemma}[Weakening--Substitution Interaction]\labeledit{lem:wk-subst-interact}
  The following two properties hold:
  \begin{enumerate}
  \item For all $t \in \Exp$, weakening $\phi$, and substitution $\sigma$:
    $(t[\phi])[\sigma] = t[\phi \circ \sigma]$.
  \item For all $t \in \Exp$, substitution $\sigma$, and weakening $\phi$:
    $(t[\sigma])[\phi] = t[\sigma[\phi]]$.
  \end{enumerate}
\end{lemma}
\begin{proof}
  Statement (1) is a corollary of \Cref{lem:wk-embed-correct,lem:subst-assoc}.
  Since we elide the embedding, treating $\phi$ as a substitution:
  \begin{align*}
    (t[\phi])[\sigma]
    &= t[\phi \circ \sigma]
      \tag{by \Cref{lem:subst-assoc}}
  \end{align*}

  Statement (2) also follows from \Cref{lem:subst-assoc}:
  \begin{align*}
    (t[\sigma])[\phi]
    &= t[\sigma \circ \phi]
      \tag{by \Cref{lem:subst-assoc}}
  \end{align*}
  We verify: $(\sigma \circ \phi)(x) = \sigma(x)[\phi] = (\sigma[\phi])(x)$,
  which is exactly the definition of $\sigma[\phi]$.
  Hence $\sigma \circ \phi = \sigma[\phi]$ and the claim holds.
\end{proof}

\begin{lemma}[Shift--Extension Cancellation]\labeledit{lem:shift-ext}
  For all substitutions $\sigma$ and terms $s$:
  $\Uparrow \circ (\sigma, s/x) = \sigma$.
\end{lemma}
\begin{proof}
  Pointwise:
  \begin{align*}
    (\Uparrow \circ (\sigma, s/x))(y)
    &= \Uparrow(y)[(\sigma, s/x)] \\
    &= (y+1)[(\sigma, s/x)] \\
    &= (\sigma, s/x)(y+1) \\
    &= \sigma(y)
  \end{align*}
\end{proof}

\begin{lemma}[Extension--Composition Distribution]\labeledit{lem:ext-comp}
  For all substitutions $\sigma, \tau$ and terms $s$:
  $(\sigma, s/x) \circ \tau = (\sigma \circ \tau,\; s[\tau]/x)$.
\end{lemma}
\begin{proof}
  Pointwise:
  \begin{align*}
    ((\sigma, s/x) \circ \tau)(0)
    &= (\sigma, s/x)(0)[\tau] \\
    &= s[\tau] \\
    &= (\sigma \circ \tau,\; s[\tau]/x)(0)
  \end{align*}
  \begin{align*}
    ((\sigma, s/x) \circ \tau)(y+1)
    &= (\sigma, s/x)(y+1)[\tau] \\
    &= \sigma(y)[\tau] \\
    &= (\sigma \circ \tau)(y) \\
    &= (\sigma \circ \tau,\; s[\tau]/x)(y+1)
  \end{align*}
\end{proof}

\begin{lemma}[$q$--Extension Interaction]\labeledit{lem:q-ext}
  For all $t \in \Exp$, substitutions $\sigma$, and terms $s$:
  \[
    t[q(\sigma)][\id, s/x] = t[\sigma, s/x]
  \]
\end{lemma}
\begin{proof}
  By \Cref{lem:subst-assoc}, it suffices to show $q(\sigma) \circ (\id, s/x) = (\sigma, s/x)$.
  Recall $q(\sigma) = (\sigma[\Uparrow], 0/x)$.
  By \Cref{lem:ext-comp}:
  \begin{align*}
    q(\sigma) \circ (\id, s/x)
    &= (\sigma[\Uparrow], 0/x) \circ (\id, s/x) \\
    &= (\sigma[\Uparrow] \circ (\id, s/x),\; 0[(\id, s/x)]/x)
      \tag{by \Cref{lem:ext-comp}} \\
    &= ((\sigma \circ \Uparrow) \circ (\id, s/x),\; s/x) \\
    &= (\sigma \circ (\Uparrow \circ (\id, s/x)),\; s/x)
       \\
    &= (\sigma \circ \id,\; s/x) \tag{by \Cref{lem:shift-ext}} \\
    &= (\sigma, s/x)
      \tag{by \Cref{lem:subst-id}}
  \end{align*}
\end{proof}

\begin{lemma}[Generalized $q$--Extension Interaction]\labeledit{lem:q-ext-gen}
  For all $t \in \Exp$, substitution $\sigma$, weakening $\phi$, and term $s$:
  \[
    t[q(\sigma)][\phi, s/x] = t[\sigma[\phi], s/x]
  \]
\end{lemma}
\begin{proof}
  By \Cref{lem:subst-assoc}, it suffices to show
  $q(\sigma) \circ (\phi, s/x) = (\sigma[\phi], s/x)$.
  Recall $q(\sigma) = (\sigma[\Uparrow], 0/x)$.
  \begin{align*}
    q(\sigma) \circ (\phi, s/x)
    &= (\sigma[\Uparrow], 0/x) \circ (\phi, s/x) \\
    &= (\sigma[\Uparrow] \circ (\phi, s/x),\; 0[(\phi, s/x)]/x)
      \tag{by \Cref{lem:ext-comp}} \\
    &= ((\sigma \circ \Uparrow) \circ (\phi, s/x),\; s/x) \\
    &= (\sigma \circ (\Uparrow \circ (\phi, s/x)),\; s/x)
      \tag{by \Cref{lem:subst-cat} associativity} \\
    &= (\sigma \circ \phi,\; s/x)
      \tag{by \Cref{lem:shift-ext}} \\
    &= (\sigma[\phi],\; s/x)
  \end{align*}
\end{proof}

\begin{lemma}[Categorical Laws for Substitutions]\labeledit{lem:subst-cat}
  Substitutions satisfy:
  \begin{enumerate}
  \item (Left identity) $\id \circ \sigma = \sigma$ for all $\sigma$.
  \item (Right identity) $\sigma \circ \id = \sigma$ for all $\sigma$.
  \item (Associativity) $(\sigma \circ \tau) \circ \delta = \sigma \circ (\tau \circ \delta)$.
  \end{enumerate}
\end{lemma}
\begin{proof} $ $
  \begin{enumerate}
  \item $(\id \circ \sigma)(x) = \id(x)[\sigma] = x[\sigma] = \sigma(x)$.
  \item $(\sigma \circ \id)(n) = \sigma(x)[\id] = \sigma(x)$ by \Cref{lem:subst-id}.
  \item By pointwise calculation:
    \begin{align*}
      ((\sigma \circ \tau) \circ \delta)(x)
      &= ((\sigma \circ \tau)(x))[\delta] \\
      &= (\sigma(x)[\tau])[\delta] \\
      &= \sigma(x)[\tau \circ \delta]
        \tag{by \Cref{lem:subst-assoc}} \\
      &= (\sigma \circ (\tau \circ \delta))(x)
    \end{align*}
  \end{enumerate}
\end{proof}

\begin{lemma}[Substitution--Extension Commutation]\labeledit{lem:subst-ext-comm}
  For all $t \in \Exp$, substitutions $\sigma$, and terms $s$:
  \[
    t[q(\sigma)][s[\sigma]/x] = t[s/x][\sigma]
  \]
\end{lemma}
\begin{proof}
  We show both sides equal $t[\sigma, s[\sigma]/x]$.
  \begin{align*}
    t[q(\sigma)][s[\sigma]/x]
    &= t[q(\sigma)][\id, s[\sigma]/x]
      \tag{by notation} \\
    &= t[\sigma, s[\sigma]/x]
      \tag{by \Cref{lem:q-ext}}
  \end{align*}
  \begin{align*}
    t[s/x][\sigma]
    &= t[\id, s/x][\sigma]
      \tag{by notation} \\
    &= t[(\id, s/x) \circ \sigma]
      \tag{by \Cref{lem:subst-assoc}} \\
    &= t[\id \circ \sigma, s[\sigma]/x]
      \tag{by \Cref{lem:ext-comp}} \\
    &= t[\sigma, s[\sigma]/x]
      \tag{by \Cref{lem:subst-cat} (left identity)}
  \end{align*}
\end{proof}

\begin{lemma}[Expansion of Substitutions]\labeledit{lem:subst-ext-eta}
  For all substitutions $\sigma$:
  $\sigma = (\Uparrow \circ \sigma,\; x_0[\sigma]/x)$,
  where $x_0$ denotes the variable at index $0$.
\end{lemma}
\begin{proof}
  Pointwise:
  \begin{align*}
    (\Uparrow \circ \sigma,\; x_0[\sigma]/x)(0)
    &= x_0[\sigma] = \sigma(0) \\
    (\Uparrow \circ \sigma,\; x_0[\sigma]/x)(y+1)
    &= (\Uparrow \circ \sigma)(y) \\
    &= \Uparrow(y)[\sigma] \\
    &= (y+1)[\sigma] \\
    &= \sigma(y+1)
  \end{align*}
  Hence both sides agree at every index.
\end{proof}

\section{Typing and Equational Theory}\labeledit{sec:rules}

We define four mutually dependent judgment forms for our type theory. %
We consider MLTT with universe subtyping in the same form as
\citep{DBLP:journals/pacmpl/0001GHP25}. 

\begin{definition}[Judgment Forms]\labeledit{def:judgments}
  The judgments of our type theory are:
  \[
    \begin{array}{lll}
      \vdash \Gamma & & \text{$\Gamma$ is a well-formed context} \\[4pt]
      \typing[\Gamma]{t}{T} & & \text{$t$ has type $T$ in $\Gamma$} \\[4pt]
      \tyequiv[\Gamma]{t}{t'}{T} & & \text{$t$ and $t'$ are equal terms of type $T$ in $\Gamma$} \\[4pt]
      \subty{S}{T} & & \text{$S$ is a subtype of $T$ in $\Gamma$}
    \end{array}
  \]
  Additionally, we have a derived judgment for substitution typing:
  \[
    \judge[\Gamma]{\sigma : \Delta} \qquad\qquad \text{$\sigma$ is a well-typed substitution from $\Delta$ to $\Gamma$}
  \]
\end{definition}

\subsection{Context Well-Formedness}

\begin{mathpar}
  \inferrule[\textsc{Ctx-Emp}]
  { }
  {\vdash \cdot}

  \inferrule[\textsc{Ctx-Ext}]
  {\vdash \Gamma \\ \typing[\Gamma]{S}{\Ty{i}}}
  {\vdash \Gamma, x_0 : S}
\end{mathpar}

\subsection{Typing}

We first define the look-up relation of a variable in a context:
\begin{mathpar}
  \inferrule
  { }
  {x_0 : T[\Uparrow] \in \Gamma, x_0: T}

  \inferrule
  {y : S \in \Gamma}
  {1+y : S[\Uparrow] \in \Gamma, x: T}
\end{mathpar}
It is easy to show that for any variable binding in a context, we have
\begin{lemma}
  $x_{|\Gamma|} : T[\Uparrow]^{1 + |\Gamma|} \in \Delta, x:T, \Gamma$
\end{lemma}
In other words, $T$ is weakened $1 + |\Gamma|$ times.

Now, we define the typing judgment of the system. 
\begin{mathpar}
  \inferrule[\textsc{U-Wf}]
  {\vdash \Gamma}
  {\typing{\Ty{i}}{\Ty{1+i}}}

  \inferrule[\textsc{N-Wf}]
  {\vdash \Gamma}
  {\typing{\Nat}{\Ty{0}}}

  \inferrule[\textsc{$\Pi$-Wf}]
  {\typing{S}{\Ty{i}} \\ \typing[\Gamma, x : S]{T}{\Ty{i}}}
  {\typing{\Pi(x : S).\, T}{\Ty{i}}}

  \inferrule[\textsc{Var}]
  {\vdash \Gamma \\ x : T \in \Gamma}
  {\typing{x}{T}}

  \inferrule[\textsc{Ze-I}]
  {\vdash \Gamma}
  {\typing{\ze}{\Nat}}

  \inferrule[\textsc{Su-I}]
  {\typing{t}{\Nat}}
  {\typing{\su{t}}{\Nat}}

  \inferrule[\textsc{$\lambda$-I}]
  {\typing{S}{\Ty{i}} \\ \typing[\Gamma, x : S]{t}{T}}
  {\typing{\lambda x.\, t}{\Pi(x : S).\, T}}

  \inferrule[\textsc{$\lambda$-E}]
  {\typing{S}{\Ty{i}} \\ \typing[\Gamma, x : S]{T}{\Ty{i}} \\\\ \typing{r}{\Pi(x : S).\, T} \\ \typing{s}{S}}
  {\typing{r\; s}{T[s/x]}}

  \inferrule[\textsc{N-E}]
  {\typing[\Gamma, x : \Nat]{T}{\Ty{i}} \\
   \typing{z}{T[\ze/x]} \\\\
   \typing[\Gamma, x : \Nat, y : T]{s}{T[\Uparrow \circ \Uparrow, \su{x_1}/x]} \\
   \typing{t}{\Nat}}
  {\typing{\elimn{x.\,T}{z}{x, y.\, s}{t}}{T[t/x]}}

  \inferrule[\textsc{Subsump}]
  {\typing{T}{\Ty i} \\ \typing{t}{S} \\ \subty{S}{T}}
  {\typing{t}{T}}
\end{mathpar}

The rule \textsc{Var} states that if the variable $x$ is in the context $\Gamma$
with type $T$, then $x$ has type $T$ in $\Gamma$.
The weakening shifts are built into the lookup relation: looking up $x$ in $\Gamma$
automatically weakens the declared type to account for all bindings between the
declaration and the top of the context.

The type formation rules (\textsc{U-Wf}, \textsc{N-Wf}, \textsc{$\Pi$-Wf})
also serve as term typing rules in Russell style: they simultaneously assert that
$\Nat$, $\Pi(x:S).\, T$, and $\Ty{i}$ are well-typed terms inhabiting appropriate
universes.
Cumulativity is achieved through the combination of the subtyping judgment
(defined below) and the subsumption rule \textsc{Subsump}.

The rule \textsc{$\Lambda$-I} requires the domain type $S$ to be well-formed (inhabiting
some $\Ty{i}$) and the body $t$ to be well-typed in the extended context.

The rule \textsc{$\Lambda$-E} (function application) states that applying a function
$r : \Pi(x : S).\, T$ to an argument $s : S$ yields a result of type $T[s/x]$---the
codomain $T$ with the argument $s$ substituted for the formal parameter $x$.

The rule \textsc{N-E} is the dependent elimination principle for natural numbers.
The motive $T$ lives in the context $\Gamma, x : \Nat$ and must inhabit some universe
$\Ty{i}$: the variable $x$ represents the natural number being eliminated.
The zero case $z$ has type $T[\ze/x]$, i.e., the motive instantiated at zero.
The successor case $s$ lives in the context extended by both the predecessor $x$ (of type
$\Nat$) and the inductive hypothesis $y$ (of type $T$), and must have type $T[\Uparrow \circ \Uparrow, \su{x_1}/x]$:
the motive instantiated at the successor of the predecessor.
The conclusion types the entire elimination at $T[t/x]$, i.e., the motive instantiated at
the scrutinee.

The rule \textsc{Subsump} allows lifting the type of a term to a supertype.
Combined with the subtyping rules below, this implements cumulativity and type conversion.

\subsection{Weakening Typing}

We write $\weakening \phi \Delta$ to mean that $\phi$ is a well-typed
weakening from $\Delta$ to $\Gamma$. %
A well-formed weakening transforms a well-formed context to another:
\begin{mathpar}
  \inferrule
  {\vdash \Gamma \\ \vdash \Delta \\ \forall x : T \in \Delta,\; \phi(x) : T[\phi] \in \Gamma}
  {\weakening \phi \Delta}
\end{mathpar}

We now show that the key weakening operations preserve well-typedness.

\begin{lemma}[Identity Weakening]\labeledit{lem:wk-id-wt}
  If $\vdash \Gamma$, then $\weakening \id \Gamma$.
\end{lemma}
\begin{proof}
  We must show: for all $x : T \in \Gamma$, $\id(x) : T[\id] \in \Gamma$.
  Since $\id(x) = x$ and $T[\id] = T$ (by \Cref{lem:wk-id}), this is immediate from
  $x : T \in \Gamma$.
\end{proof}

\begin{lemma}[Shift Weakening]\labeledit{lem:wk-shift-wt}
  If $\vdash \Gamma$ and $\typing[\Gamma]{S}{\Ty{i}}$, then
  $\weakening[\Gamma, x_0 : S] \Uparrow \Gamma$.
\end{lemma}
\begin{proof}
  We must show: for all $y : T \in \Gamma$, $\Uparrow(y) : T[\Uparrow] \in \Gamma, x_0 : S$.
  We have $\Uparrow(y) = 1 + y$.
  From $y : T \in \Gamma$, the lookup rule gives $1{+}y : T[\Uparrow] \in \Gamma, x_0 : S$.
\end{proof}

\begin{lemma}[$q$ Preserves Weakening Typing]\labeledit{lem:wk-q-wt}
  If $\weakening \phi \Delta$ and $\typing[\Delta]{S}{\Ty{i}}$, then
  $\weakening[\Gamma, x_0 : S[\phi]] {q(\phi)} {\Delta, x_0 : S}$.
\end{lemma}
\begin{proof}
  We must show: for all $y : T \in \Delta, x_0 : S$,
  $q(\phi)(y) : T[q(\phi)] \in \Gamma, x_0 : S[\phi]$.
  We case-split on the lookup $y : T \in \Delta, x_0 : S$:
  \begin{itemize}
  \item $x_0 : S[\Uparrow] \in \Delta, x_0 : S$:
    Then $q(\phi)(x_0) = x_0$ and
    $S[\Uparrow][q(\phi)] = S[\phi][\Uparrow]$ (by \Cref{lem:wk-subst-comm}).
    We have $x_0 : S[\phi][\Uparrow] \in \Gamma, x_0 : S[\phi]$ by lookup.
  \item $1{+}z : T'[\Uparrow] \in \Delta, x_0 : S$ where $z : T' \in \Delta$:
    Then $q(\phi)(1{+}z) = 1{+}\phi(z)$ and
    $T'[\Uparrow][q(\phi)] = T'[\phi][\Uparrow]$ (by \Cref{lem:wk-subst-comm}).
    From $\weakening \phi \Delta$, we have $\phi(z) : T'[\phi] \in \Gamma$, so
    $1{+}\phi(z) : T'[\phi][\Uparrow] \in \Gamma, x_0 : S[\phi]$ by lookup.
  \end{itemize}
\end{proof}

\begin{lemma}[Composition Preserves Weakening Typing]\labeledit{lem:wk-comp-wt}
  If $\weakening \psi \Delta$ and $\weakening[\Delta] \phi \Theta$, then
  $\weakening {\phi \circ \psi} \Theta$.
\end{lemma}
\begin{proof}
  We must show: for all $x : T \in \Theta$,
  $(\phi \circ \psi)(x) : T[\phi \circ \psi] \in \Gamma$.
  \begin{align*}
    &\phi(x) : T[\phi] \in \Delta
      && \text{(from $\weakening[\Delta] \phi \Theta$ and $x : T \in \Theta$)} \\
    &\psi(\phi(x)) : T[\phi][\psi] \in \Gamma
      && \text{(from $\weakening \psi \Delta$ and $\phi(x) : T[\phi] \in \Delta$)} \\
    &(\phi \circ \psi)(x) = \psi(\phi(x))
      && \text{(by definition of composition)} \\
    &T[\phi \circ \psi] = T[\phi][\psi]
      && \text{(by \Cref{lem:wk-comp})}
  \end{align*}
  Hence $(\phi \circ \psi)(x) : T[\phi \circ \psi] \in \Gamma$.
\end{proof}

\subsection{Substitution Typing}

We define when a substitution is well-typed.
We write $\judge{\sigma : \Delta}$ to mean that $\sigma$ maps variables from
$\Delta$ to well-typed terms in $\Gamma$, preserving the typing structure.
\begin{mathpar}
  \inferrule
  {\vdash \Gamma \\ \vdash \Delta \\ \forall x : T \in \Delta,\; \typing{\sigma(x)}{T[\sigma]}}
  {\typing \sigma \Delta}
\end{mathpar}

In practice, the most common substitutions that arise are:
\begin{itemize}
  \item The identity substitution: $\judge{\id : \Gamma}$, which maps every
    variable to itself.
  \item The single substitution: $\judge{(\id, s/x) : (\Gamma, x_0 : S)}$ (when
    $\typing{s}{S}$), which substitutes $s$ for $x_0$ and is the
    identity on all other variables.
  \item A general substitution: $\judge{\sigma : \Delta}$, which maps each
    variable $x_i$ in $\Delta$ with type $S_i$ to a term $\sigma(x_i)$ of type
    $S_i[\sigma]$ in $\Gamma$.
\end{itemize}

\subsection{Subtyping}\labeledit{sec:syn:subtyping}

Following~\citet{DBLP:journals/pacmpl/0001GHP25}, we define a covariant subtyping judgment
$\subty{S}{T}$ that implements cumulativity through universe subtyping,
propagating covariantly through the codomain of $\Pi$-types.

\begin{mathpar}
  \inferrule[\textsc{Sub-Univ}]
  {\vdash \Gamma \\ i \le j}
  {\subty{\Ty{i}}{\Ty{j}}}

  \inferrule[\textsc{Sub-Trans}]
  {\subty{S}{T} \\ \subty{T}{U}}
  {\subty{S}{U}}

  \inferrule[\textsc{Sub-Eq}]
  {\typing T{\Ty i} \\ \tyequiv{S}{T}{\Ty{i}}}
  {\subty{S}{T}}

  \inferrule[\textsc{Sub-Pi}]
  {\typing S{\Ty i} \\
    \typing{S'}{\Ty i} \\
    \tyequiv{S}{S'}{\Ty{i}} \\
    \typing[\Gamma,x:S]T{\Ty i} \\
    \typing[\Gamma,x:S']{T'}{\Ty i} \\
   \subty[\Gamma, x : S]{T}{T'}}
  {\subty{\Pi(x:S).\, T}{\Pi(x:S').\, T'}}
\end{mathpar}

The rule \textsc{Sub-Univ} states that $\Ty{i}$ is a subtype of $\Ty{j}$ whenever $i
\le j$. %
The rule \textsc{Sub-Trans} gives transitivity. %
The rule \textsc{Sub-Eq} embeds judgmental equality into subtyping: equal types are
subtypes 
of each other. %
The rule \textsc{Sub-Pi} propagates subtyping covariantly through the codomain of $\Pi$-types,
requiring the domains to be judgmentally equal.

Combined with the subsumption rule \textsc{Subsump} in the typing judgment, these rules
implement a full cumulative universe hierarchy.
For example, if $\typing{S}{\Ty{0}}$ then $S : \Ty{1}$ follows by
\textsc{Subsump} with $\subty{\Ty{0}}{\Ty{1}}$ (via \textsc{Sub-Univ}).

\subsection{Equational Theory}\labeledit{sec:syn:eq}

The equational theory defines judgmental equality for terms.

\subsection{Term Equality --- PER and Congruence Rules}

\begin{mathpar}
  \inferrule[\textsc{Tm-Sym}]
  {\tyequiv{t}{t'}{T}}
  {\tyequiv{t'}{t}{T}}

  \inferrule[\textsc{Tm-Trans}]
  {\tyequiv{t}{t'}{T} \\\\ \tyequiv{t'}{t''}{T}}
  {\tyequiv{t}{t''}{T}}

  \inferrule[\textsc{Tm-Subsump}]
  {\typing T {\Ty i} \\\\ \tyequiv{t}{t'}{S} \\ \subty{S}{T}}
  {\tyequiv{t}{t'}{T}}

  \inferrule[\textsc{Var-Cong}]
  {\vdash \Gamma \\ x : T \in \Gamma}
  {\tyequiv{x}{x}{T}}

  \inferrule[\textsc{U-Cong}]
  {\vdash \Gamma}
  {\tyequiv{\Ty{i}}{\Ty{i}}{\Ty{1+i}}}

  \inferrule[\textsc{N-Cong}]
  {\vdash \Gamma}
  {\tyequiv{\Nat}{\Nat}{\Ty{0}}}

  \inferrule[\textsc{$\Pi$-Cong}]
  {\tyequiv{S}{S'}{\Ty{i}} \\ \tyequiv[\Gamma, x : S]{T}{T'}{\Ty{i}}}
  {\tyequiv{\Pi(x:S).\, T}{\Pi(x:S').\, T'}{\Ty{i}}}

  \inferrule[\textsc{Ze-Cong}]
  {\vdash \Gamma}
  {\tyequiv{\ze}{\ze}{\Nat}}
  
  \inferrule[\textsc{Su-Cong}]
  {\tyequiv{t}{t'}{\Nat}}
  {\tyequiv{\su{t}}{\su{t'}}{\Nat}}

  \inferrule[\textsc{$\lambda$-Cong}]
  {\typing{S}{\Ty{i}} \\ \tyequiv[\Gamma, x : S]{t}{t'}{T}}
  {\tyequiv{\lambda x.\, t}{\lambda x.\, t'}{\Pi(x:S).\, T}}

  \inferrule[\textsc{App-Cong}]
  {\tyequiv{r}{r'}{\Pi(x:S).\, T} \\ \tyequiv{s}{s'}{S}}
  {\tyequiv{r\; s}{r'\; s'}{T[s/x]}}

  \inferrule[\textsc{Rec-Cong}]
  {\typing[\Gamma, x : \Nat]{T}{\Ty{i}} \\
   \tyequiv{z}{z'}{T[\ze/x]} \\
   \tyequiv[\Gamma, x : \Nat, y : T]{s}{s'}{T[\Uparrow \circ \Uparrow, \su{x_1}/x]} \\
   \tyequiv{t}{t'}{\Nat}}
  {\tyequiv{\elimn{x.\,T}{z}{x, y.\, s}{t}}{\elimn{x.\,T}{z'}{x, y.\, s'}{t'}}
    {T[t/x]}}
\end{mathpar}

\subsection{Term Equality --- Computation Rules ($\beta$-rules)}

The computation rules express how eliminators interact with their corresponding
introduction forms.

\begin{mathpar}
  \inferrule[\textsc{$\Pi$-$\beta$}]
  {\typing{S}{\Ty{i}} \\ \typing[\Gamma, x : S]{T}{\Ty{i}} \\\\ \typing[\Gamma, x : S]{t}{T} \\ \typing{s}{S}}
  {\tyequiv{(\lambda x.\, t)\; s}{t[s/x]}{T[s/x]}}

  \inferrule[\textsc{N-$\beta$-Ze}]
  {\typing[\Gamma, x : \Nat]{T}{\Ty{i}} \\
   \typing{z}{T[\ze/x]} \\\\
   \typing[\Gamma, x : \Nat, y : T]{s}{T[\Uparrow \circ \Uparrow, \su{x_1}/x]}}
  {\tyequiv{\elimn{x.\,T}{z}{x, y.\, s}{\ze}}{z}{T[\ze/x]}}

  \inferrule[\textsc{N-$\beta$-Su}]
  {\typing[\Gamma, x : \Nat]{T}{\Ty{i}} \\
   \typing{z}{T[\ze/x]} \\
   \typing[\Gamma, x : \Nat, y : T]{s}{T[\Uparrow \circ \Uparrow, \su{x_1}/x]} \\
   \typing{t}{\Nat}}
  {\tyequiv{\elimn{x.\,T}{z}{x,y.\,s}{(\su{t})}}{s[t/x,\; \elimn{x.\,T}{z}{x, y.\, s}{t}/y]}{T[\su{t}/x]}}
\end{mathpar}

The $\beta$-rule for $\Pi$ states that applying a lambda to an argument reduces to the body
with the argument substituted for the bound variable:
$(\lambda x.\, t)\; s \approx t[s/x]$.
The substitution $[s/x]$ replaces $x$ by $s$ in $t$ and leaves all other variables unchanged.

The $\beta$-rules for natural numbers state:
\begin{itemize}
  \item The eliminator applied to zero computes to the zero case $z$.
  \item The eliminator applied to $\su{t}$ computes to the successor case $s$ with two
    simultaneous substitutions: the predecessor $t$ is substituted for $x$, and the
    recursive call $\elimn{x.\,T}{z}{x, y.\, s}{t}$ (the inductive hypothesis) is
    substituted for $y$.
    That is, $s[t/x,\; \elimn{x.\,T}{z}{x, y.\, s}{t}/y]$.
\end{itemize}

\subsection{Term Equality --- Uniqueness Rules ($\eta$-rules)}

\begin{mathpar}
  \inferrule[\textsc{$\Pi$-$\eta$}]
  {\typing{S}{\Ty{i}} \\ \typing[\Gamma, x : S]{T}{\Ty{i}} \\ \typing{t}{\Pi(x : S).\, T}}
  {\tyequiv{t}{\lambda x.\, (t[\Uparrow]\; x)}{\Pi(x : S).\, T}}
\end{mathpar}

The $\eta$-rule for $\Pi$ states that every function is equal to its $\eta$-expansion.

\section{Properties of the Judgments}\labeledit{sec:props}

In this section we establish the fundamental metatheoretic properties of the typing and
equational theory: presupposition (every judgment implies the well-formedness of its
sub-components) and stability under substitution (well-typed substitutions preserve all
judgments).

\subsection{Weakening}\labeledit{sec:props:wk}

\begin{lemma}[Weakening Preserves Typing]\labeledit{lem:wk-typing}
  The following are proved by mutual induction on the syntactic judgments:
  \begin{enumerate}
  \item If $\typing[\Gamma]{t}{T}$ and $\weakening[\Delta] \phi \Gamma$, then
    $\typing[\Delta]{t[\phi]}{T[\phi]}$.
  \item If $\tyequiv[\Gamma]{t}{t'}{T}$ and $\weakening[\Delta] \phi \Gamma$, then
    $\tyequiv[\Delta]{t[\phi]}{t'[\phi]}{T[\phi]}$.
  \item If $\subty[\Gamma]{S}{T}$ and $\weakening[\Delta] \phi \Gamma$, then
    $\subty[\Delta]{S[\phi]}{T[\phi]}$.
  \end{enumerate}
\end{lemma}
\begin{proof}
  By mutual induction on the derivation of the syntactic judgment.
  We show representative cases for statement (1).

  \medskip\noindent
  \textbf{Case}
  \begin{mathpar}
    \inferrule[\textsc{Var}]
    {\vdash \Gamma \\ y : T \in \Gamma}
    {\typing[\Gamma]{y}{T}}
  \end{mathpar}
  Then $y[\phi] = \phi(y)$.
  By the definition of $\weakening[\Delta] \phi \Gamma$, $\phi(y) : T[\phi] \in \Delta$.
  By \textsc{Var}, $\typing[\Delta]{\phi(y)}{T[\phi]}$.

  \medskip\noindent
  \textbf{Case}
  \begin{mathpar}
    \inferrule[\textsc{$\lambda$-I}]
    {\typing[\Gamma]{S'}{\Ty{i}} \\ \typing[\Gamma, y : S']{t}{T}}
    {\typing[\Gamma]{\lambda y.\,t}{\Pi(y:S').\,T}}
  \end{mathpar}
  We must show $\typing[\Delta]{(\lambda y.\,t)[\phi]}{(\Pi(y:S').\,T)[\phi]}$,
  i.e.\ $\typing[\Delta]{\lambda y.\,(t[q(\phi)])}{\Pi(y:S'[\phi]).\,(T[q(\phi)])}$.
  \begin{align*}
    &\typing[\Delta]{S'[\phi]}{\Ty{i}}
      && \text{(by IH on first premise)} \\
    &\weakening[\Delta, y:S'[\phi]] {q(\phi)} {\Gamma, y:S'}
      && \text{(by \Cref{lem:wk-q-wt})} \\
    &\typing[\Delta, y:S'[\phi]]{t[q(\phi)]}{T[q(\phi)]}
      && \text{(by IH on second premise)} \\
    &\typing[\Delta]{\lambda y.\,(t[q(\phi)])}{\Pi(y:S'[\phi]).\,(T[q(\phi)])}
      && \text{(by \textsc{$\lambda$-I})}
  \end{align*}

  \medskip\noindent
  \textbf{Case}
  \begin{mathpar}
    \inferrule[\textsc{$\lambda$-E}]
    {\typing[\Gamma]{S'}{\Ty{i}} \\ \typing[\Gamma, y:S']{T}{\Ty{i}} \\
     \typing[\Gamma]{r}{\Pi(y:S').\,T} \\ \typing[\Gamma]{s}{S'}}
    {\typing[\Gamma]{r\;s}{T[s/y]}}
  \end{mathpar}
  \begin{align*}
    &\typing[\Delta]{S'[\phi]}{\Ty{i}} && \text{(by IH)} \\
    &\typing[\Delta, y:S'[\phi]]{T[q(\phi)]}{\Ty{i}}
      && \text{(by IH, using \Cref{lem:wk-q-wt})} \\
    &\typing[\Delta]{r[\phi]}{\Pi(y:S'[\phi]).\,(T[q(\phi)])} && \text{(by IH)} \\
    &\typing[\Delta]{s[\phi]}{S'[\phi]} && \text{(by IH)} \\
    &\typing[\Delta]{r[\phi]\;s[\phi]}{T[q(\phi)][s[\phi]/y]} && \text{(by \textsc{$\lambda$-E})}
  \end{align*}
  It remains to show $T[q(\phi)][s[\phi]/y] = T[s/y][\phi]$,
  which is \Cref{lem:subst-ext-comm}.

  \medskip\noindent
  \textbf{Case}
  \begin{mathpar}
    \inferrule[\textsc{Subsump}]
    {\typing[\Gamma]{T}{\Ty{i}} \\ \typing[\Gamma]{t}{S} \\ \subty[\Gamma]{S}{T}}
    {\typing[\Gamma]{t}{T}}
  \end{mathpar}
  \begin{align*}
    &\typing[\Delta]{T[\phi]}{\Ty{i}} && \text{(by IH)} \\
    &\typing[\Delta]{t[\phi]}{S[\phi]} && \text{(by IH)} \\
    &\subty[\Delta]{S[\phi]}{T[\phi]} && \text{(by IH, statement (3))} \\
    &\typing[\Delta]{t[\phi]}{T[\phi]} && \text{(by \textsc{Subsump})}
  \end{align*}

  \medskip
  We now show representative cases for statement (2).

  \medskip\noindent
  \textbf{Case}
  \begin{mathpar}
    \inferrule[\textsc{$\Pi$-$\beta$}]
    {\typing[\Gamma]{S'}{\Ty{i}} \\ \typing[\Gamma, x:S']{T'}{\Ty{i}} \\
     \typing[\Gamma, x:S']{t}{T'} \\ \typing[\Gamma]{s}{S'}}
    {\tyequiv[\Gamma]{(\lambda x.\,t)\;s}{t[s/x]}{T'[s/x]}}
  \end{mathpar}
  We must show $\tyequiv[\Delta]{((\lambda x.\,t)\;s)[\phi]}{t[s/x][\phi]}{T'[s/x][\phi]}$.
  \begin{align*}
    &((\lambda x.\,t)\;s)[\phi] = (\lambda x.\,(t[q(\phi)]))\;(s[\phi])
  \end{align*}
  Now:
  \begin{align*}
    &\typing[\Delta]{S'[\phi]}{\Ty{i}} && \text{(by IH)} \\
    &\typing[\Delta, x:S'[\phi]]{T'[q(\phi)]}{\Ty{i}}
      && \text{(by IH, using \Cref{lem:wk-q-wt})} \\
    &\typing[\Delta, x:S'[\phi]]{t[q(\phi)]}{T'[q(\phi)]}
      && \text{(by IH, using \Cref{lem:wk-q-wt})} \\
    &\typing[\Delta]{s[\phi]}{S'[\phi]} && \text{(by IH)} \\
    &\tyequiv[\Delta]{(\lambda x.\,(t[q(\phi)]))\;(s[\phi])}{t[q(\phi)][s[\phi]/x]}{T'[q(\phi)][s[\phi]/x]}
      && \text{(by \textsc{$\Pi$-$\beta$})}
  \end{align*}
  By \Cref{lem:subst-ext-comm}:
  $t[q(\phi)][s[\phi]/x] = t[s/x][\phi]$ and $T'[q(\phi)][s[\phi]/x] = T'[s/x][\phi]$.

  \medskip\noindent
  \textbf{Case}
  \begin{mathpar}
    \inferrule[\textsc{$\Pi$-$\eta$}]
    {\typing[\Gamma]{S'}{\Ty{i}} \\ \typing[\Gamma, x:S']{T'}{\Ty{i}} \\
     \typing[\Gamma]{t}{\Pi(x:S').\,T'}}
    {\tyequiv[\Gamma]{t}{\lambda x.\,(t[\Uparrow]\;x)}{\Pi(x:S').\,T'}}
  \end{mathpar}
  We must show $\tyequiv[\Delta]{t[\phi]}{(\lambda x.\,(t[\Uparrow]\;x))[\phi]}{(\Pi(x:S').\,T')[\phi]}$.
  Unfolding the right-hand side:
  \begin{align*}
    (\lambda x.\,(t[\Uparrow]\;x))[\phi]
    &= \lambda x.\,((t[\Uparrow]\;x)[q(\phi)]) \\
    &= \lambda x.\,(t[\Uparrow][q(\phi)]\;x) \\
    &= \lambda x.\,(t[\phi][\Uparrow]\;x)
      \tag{by \Cref{lem:wk-subst-comm}}
  \end{align*}
  \begin{align*}
    &\typing[\Delta]{S'[\phi]}{\Ty{i}} && \text{(by IH)} \\
    &\typing[\Delta, x:S'[\phi]]{T'[q(\phi)]}{\Ty{i}} && \text{(by IH, using \Cref{lem:wk-q-wt})} \\
    &\typing[\Delta]{t[\phi]}{\Pi(x:S'[\phi]).\,(T'[q(\phi)])} && \text{(by IH)} \\
    &\weakening[\Delta, x:S'[\phi]] \Uparrow \Delta
      && \text{(by \Cref{lem:wk-shift-wt})} \\
    &\tyequiv[\Delta]{t[\phi]}{\lambda x.\,(t[\phi][\Uparrow]\;x)}{\Pi(x:S'[\phi]).\,(T'[q(\phi)])}
      && \text{(by \textsc{$\Pi$-$\eta$})}
  \end{align*}

  \medskip
  We show a case for statement (3).

  \medskip\noindent
  \textbf{Case}
  \begin{mathpar}
    \inferrule[\textsc{Sub-Pi}]
    {\typing[\Gamma]{S}{\Ty{i}} \\
     \typing[\Gamma]{S'}{\Ty{i}} \\
     \tyequiv[\Gamma]{S}{S'}{\Ty{i}} \\
     \typing[\Gamma, x:S]{T}{\Ty{i}} \\
     \typing[\Gamma, x:S']{T'}{\Ty{i}} \\
     \subty[\Gamma, x:S]{T}{T'}}
    {\subty[\Gamma]{\Pi(x:S).\,T}{\Pi(x:S').\,T'}}
  \end{mathpar}
  \begin{align*}
    &\typing[\Delta]{S[\phi]}{\Ty{i}} && \text{(by IH)} \\
    &\typing[\Delta]{S'[\phi]}{\Ty{i}} && \text{(by IH)} \\
    &\tyequiv[\Delta]{S[\phi]}{S'[\phi]}{\Ty{i}} && \text{(by IH, statement (2))} \\
    &\typing[\Delta, x:S[\phi]]{T[q(\phi)]}{\Ty{i}}
      && \text{(by IH, using \Cref{lem:wk-q-wt})} \\
    &\typing[\Delta, x:S'[\phi]]{T'[q(\phi)]}{\Ty{i}}
      && \text{(by IH, using \Cref{lem:wk-q-wt})} \\
    &\subty[\Delta, x:S[\phi]]{T[q(\phi)]}{T'[q(\phi)]}
      && \text{(by IH, statement (3), using \Cref{lem:wk-q-wt})} \\
    &\subty[\Delta]{\Pi(x:S[\phi]).\,(T[q(\phi)])}{\Pi(x:S'[\phi]).\,(T'[q(\phi)])}
      && \text{(by \textsc{Sub-Pi})}
  \end{align*}
  This is $\subty[\Delta]{(\Pi(x:S).\,T)[\phi]}{(\Pi(x:S').\,T')[\phi]}$.

  \medskip\noindent
  The remaining cases (\textsc{U-Wf}, \textsc{N-Wf}, \textsc{$\Pi$-Wf},
  \textsc{Ze-I}, \textsc{Su-I}, \textsc{N-E}, congruence rules,
  \textsc{Sub-Univ}, \textsc{Sub-Trans}, \textsc{Sub-Eq}) are analogous:
  apply the IH to each premise, using \Cref{lem:wk-q-wt} to construct
  the weakening for extended contexts, and re-apply the rule.
\end{proof}

\subsection{Well-Formedness of Substitution Operations}\labeledit{sec:props:conseq}

We show that the key substitution operations produce well-typed substitutions.

\begin{lemma}[Identity Substitution is Well-Typed]\labeledit{lem:id-wt}
  If $\vdash \Gamma$, then $\judge[\Gamma]{\id : \Gamma}$.
\end{lemma}
\begin{proof}
  We must show: for all $x : T \in \Gamma$, $\typing[\Gamma]{\id(x)}{T[\id]}$.
  Since $\id(x) = x$ and $T[\id] = T$ (by \Cref{lem:subst-id}), this reduces to
  $\typing[\Gamma]{x}{T}$, which holds by \textsc{Var}.
\end{proof}

\begin{lemma}[Extension is Well-Typed]\labeledit{lem:ext-wt}
  If $\judge[\Gamma]{\sigma : \Delta}$, $\typing[\Delta]{S}{\Ty{i}}$, and
  $\typing[\Gamma]{s}{S[\sigma]}$, then
  $\judge[\Gamma]{(\sigma, s/x) : (\Delta, x : S)}$.
\end{lemma}
\begin{proof}
  We must show: for all $y : T \in \Delta, x : S$,
  $\typing[\Gamma]{(\sigma, s/x)(y)}{T[(\sigma, s/x)]}$.
  \begin{itemize}
  \item $x_0 : S[\Uparrow] \in \Delta, x : S$:
    Then $(\sigma, s/x)(x_0) = s$ and
    $S[\Uparrow][(\sigma, s/x)] = S[\sigma]$ (by \Cref{lem:shift-ext,lem:subst-assoc}).
    We have $\typing[\Gamma]{s}{S[\sigma]}$ by assumption.
  \item $1{+}z : T'[\Uparrow] \in \Delta, x : S$ with $z : T' \in \Delta$:
    Then $(\sigma, s/x)(1{+}z) = \sigma(z)$ and
    $T'[\Uparrow][(\sigma, s/x)] = T'[\sigma]$ (by \Cref{lem:shift-ext,lem:subst-assoc}).
    From $\judge[\Gamma]{\sigma : \Delta}$, $\typing[\Gamma]{\sigma(z)}{T'[\sigma]}$.
  \end{itemize}
\end{proof}

\begin{lemma}[Single Substitution is Well-Typed]\labeledit{lem:single-subst-wt}
  If $\vdash \Gamma$, $\typing[\Gamma]{S}{\Ty{i}}$, and $\typing[\Gamma]{s}{S}$, then
  $\judge[\Gamma]{(\id, s/x) : (\Gamma, x : S)}$.
\end{lemma}
\begin{proof}
  By \Cref{lem:id-wt} and \Cref{lem:ext-wt}: $\id$ is well-typed by \Cref{lem:id-wt},
  and $\typing[\Gamma]{s}{S} = \typing[\Gamma]{s}{S[\id]}$ by \Cref{lem:subst-id}.
\end{proof}

\begin{lemma}[Weakening of Substitution is Well-Typed]\labeledit{lem:wk-subst-wt}
  If $\judge[\Gamma]{\sigma : \Delta}$ and $\weakening[\Gamma'] \phi \Gamma$, then
  $\judge[\Gamma']{\sigma[\phi] : \Delta}$.
\end{lemma}
\begin{proof}
  We must show: for all $x : T \in \Delta$,
  $\typing[\Gamma']{(\sigma[\phi])(x)}{T[\sigma[\phi]]}$.
  From $\judge[\Gamma]{\sigma : \Delta}$, $\typing[\Gamma]{\sigma(x)}{T[\sigma]}$.
  By \Cref{lem:wk-typing}, $\typing[\Gamma']{\sigma(x)[\phi]}{T[\sigma][\phi]}$.
  Now $(\sigma[\phi])(x) = \sigma(x)[\phi]$ and $T[\sigma[\phi]] = T[\sigma][\phi]$
  (by \Cref{lem:wk-subst-interact}).
\end{proof}

\begin{lemma}[$q$ Preserves Substitution Typing]\labeledit{lem:q-subst-typing}
  If $\judge[\Gamma]{\sigma : \Delta}$, $\typing[\Delta]{S}{\Ty{i}}$, and
  $\typing[\Gamma]{S[\sigma]}{\Ty{i}}$, then
  $\judge[\Gamma, x_0 : S[\sigma]]{q(\sigma) : (\Delta, x_0 : S)}$.
\end{lemma}
\begin{proof}
  We must show that $q(\sigma)$ maps each variable in $(\Delta, x_0 : S)$ to a well-typed
  term in $(\Gamma, x_0 : S[\sigma])$.
  We case-split on the lookup in $(\Delta, x_0 : S)$:
  \begin{itemize}
  \item $x_0 : S[\Uparrow] \in \Delta, x_0 : S$:
    \begin{align*}
      &q(\sigma)(x_0) = x_0 \\
      &S[\Uparrow][q(\sigma)] = S[\sigma][\Uparrow]
        && \text{(by \Cref{lem:wk-subst-comm})} \\
      &x_0 : S[\sigma][\Uparrow] \in \Gamma, x_0 : S[\sigma]
        && \text{(by lookup)}
    \end{align*}
    Hence $\typing[\Gamma, x_0 : S[\sigma]]{x_0}{S[\Uparrow][q(\sigma)]}$ by \textsc{Var}.
  \item $1{+}y : T[\Uparrow] \in \Delta, x_0 : S$ where $y : T \in \Delta$:
    \begin{align*}
      &q(\sigma)(1{+}y) = \sigma(y)[\Uparrow] \\
      &\typing[\Gamma]{\sigma(y)}{T[\sigma]}
        && \text{(from $\judge[\Gamma]{\sigma : \Delta}$)} \\
      &\typing[\Gamma, x_0 : S[\sigma]]{\sigma(y)[\Uparrow]}{T[\sigma][\Uparrow]}
        && \text{(by \Cref{lem:wk-typing} using \Cref{lem:wk-shift-wt})} \\
      &T[\Uparrow][q(\sigma)] = T[\sigma][\Uparrow]
        && \text{(by \Cref{lem:wk-subst-comm})}
    \end{align*}
    Hence $\typing[\Gamma, x_0 : S[\sigma]]{\sigma(y)[\Uparrow]}{T[\Uparrow][q(\sigma)]}$.
  \end{itemize}
\end{proof}
Note that this lemma takes an extra premise $\typing[\Gamma]{S[\sigma]}{\Ty{i}}$. %
This premise would not be necessary once we have substitutions, but since at this
point, the lemma is still not unavailable, we must bear with this extra premise, which
can be later omitted. 

\subsection{Substitution Preserves Typing}\labeledit{sec:props:subst}

The central property of the operational approach is that well-typed substitutions
preserve all judgments.
\begin{lemma}[Substitution Preserves Judgments]\labeledit{lem:subst-preserves}
  The following are proved by mutual induction on the syntactic judgments:
  \begin{enumerate}
  \item If $\typing[\Delta]{t}{T}$ and $\judge[\Gamma]{\sigma : \Delta}$, then
    $\typing[\Gamma]{t[\sigma]}{T[\sigma]}$.
  \item If $\tyequiv[\Delta]{t}{t'}{T}$ and $\judge[\Gamma]{\sigma : \Delta}$, then
    $\tyequiv[\Gamma]{t[\sigma]}{t'[\sigma]}{T[\sigma]}$.
  \item If $\subty[\Delta]{S}{T}$ and $\judge[\Gamma]{\sigma : \Delta}$, then
    $\subty{S[\sigma]}{T[\sigma]}$.
  \end{enumerate}
\end{lemma}
\begin{proof}
  By mutual induction on the derivation of the syntactic judgment.
  We show representative cases for statement (1).

  \medskip\noindent
  \textbf{Case}
  \begin{mathpar}
    \inferrule
    {\vdash \Delta \\ x : T \in \Delta}
    {\typing[\Delta]{x}{T}}
  \end{mathpar}
  Then $t[\sigma] = x[\sigma] = \sigma(x)$.
  By the definition of $\judge[\Gamma]{\sigma : \Delta}$, we have
  $\typing[\Gamma]{\sigma(x)}{T[\sigma]}$, which is the goal.

  \medskip\noindent
  \textbf{Case}
  \begin{mathpar}
    \inferrule
    {\typing[\Delta]{S}{\Ty{i}} \\ \typing[\Delta, x : S]{t'}{T}}
    {\typing[\Delta]{\lambda x.\,t'}{\Pi(x:S).\,T}}
  \end{mathpar}
  We must show $\typing[\Gamma]{(\lambda x.\,t')[\sigma]}{(\Pi(x:S).\,T)[\sigma]}$.
  Unfolding:
  \begin{align*}
    (\lambda x.\,t')[\sigma] &= \lambda x.\,(t'[q(\sigma)]) \\
    (\Pi(x:S).\,T)[\sigma] &= \Pi(x:S[\sigma]).\,(T[q(\sigma)])
  \end{align*}
  Now:
  \begin{align*}
    &\typing[\Gamma]{S[\sigma]}{\Ty{i}} && \text{(by IH on first premise)} \\
    &\judge[\Gamma, x:S[\sigma]]{q(\sigma) : (\Delta, x:S)} && \text{(by \Cref{lem:q-subst-typing})} \\
    &\typing[\Gamma, x:S[\sigma]]{t'[q(\sigma)]}{T[q(\sigma)]} && \text{(by IH on second premise)} \\
    &\typing[\Gamma]{\lambda x.\,(t'[q(\sigma)])}{\Pi(x:S[\sigma]).\,(T[q(\sigma)])} && \text{(by \textsc{$\lambda$-I})}
  \end{align*}

  \medskip\noindent
  \textbf{Case}
  \begin{mathpar}
    \inferrule
    {\typing[\Delta]{S}{\Ty{i}} \\ \typing[\Delta, x:S]{T}{\Ty{i}} \\ \typing[\Delta]{r}{\Pi(x:S).\,T} \\ \typing[\Delta]{s}{S}}
    {\typing[\Delta]{r\;s}{T[s/x]}}
  \end{mathpar}
  We must show $\typing[\Gamma]{(r\;s)[\sigma]}{(T[s/x])[\sigma]}$.
  Unfolding: $(r\;s)[\sigma] = r[\sigma]\;s[\sigma]$.
  \begin{align*}
    &\typing[\Gamma]{S[\sigma]}{\Ty{i}}
      && \text{(by IH on first premise)} \\
    &\typing[\Gamma, x:S[\sigma]]{T[q(\sigma)]}{\Ty{i}}
      && \text{(by IH on second premise using \Cref{lem:q-subst-typing})} \\
    &\typing[\Gamma]{r[\sigma]}{\Pi(x:S[\sigma]).\,(T[q(\sigma)])}
      && \text{(by IH on third premise)} \\
    &\typing[\Gamma]{s[\sigma]}{S[\sigma]}
      && \text{(by IH on fourth premise)} \\
    &\typing[\Gamma]{r[\sigma]\;s[\sigma]}{T[q(\sigma)][s[\sigma]/x]}
      && \text{(by \textsc{$\lambda$-E})}
  \end{align*}
  It remains to show $T[q(\sigma)][s[\sigma]/x] = (T[s/x])[\sigma]$,
  which is \Cref{lem:subst-ext-comm}.

  \medskip\noindent
  \textbf{Case}
  \begin{mathpar}
    \inferrule
    {\typing[\Delta]{T}{\Ty{i}} \\ \typing[\Delta]{t}{S} \\ \subty[\Delta]{S}{T}}
    {\typing[\Delta]{t}{T}}
  \end{mathpar}
  \begin{align*}
    &\typing[\Gamma]{T[\sigma]}{\Ty{i}} && \text{(by IH on first premise)} \\
    &\typing[\Gamma]{t[\sigma]}{S[\sigma]} && \text{(by IH on second premise)} \\
    &\subty{S[\sigma]}{T[\sigma]} && \text{(by IH, statement (3))} \\
    &\typing[\Gamma]{t[\sigma]}{T[\sigma]} && \text{(by \textsc{Subsump})}
  \end{align*}

  \medskip\noindent
  The remaining cases (\textsc{U-Wf}, \textsc{N-Wf}, \textsc{Ze-I}, \textsc{Su-I})
  are straightforward applications of the IH.

  \medskip
  We now show representative cases for statement (2).
  The congruence cases follow directly from the IH.
  We show the computation and uniqueness rules.

  \medskip\noindent
  \textbf{Case}
  \begin{mathpar}
    \inferrule
    {\typing[\Delta]{S}{\Ty{i}} \\ \typing[\Delta, x:S]{T}{\Ty{i}} \\
     \typing[\Delta, x:S]{t}{T} \\ \typing[\Delta]{s}{S}}
    {\tyequiv[\Delta]{(\lambda x.\,t)\;s}{t[s/x]}{T[s/x]}}
  \end{mathpar}
  We must show $\tyequiv[\Gamma]{((\lambda x.\,t)\;s)[\sigma]}{t[s/x][\sigma]}{T[s/x][\sigma]}$.
  Unfolding:
  \begin{align*}
    &((\lambda x.\,t)\;s)[\sigma] = (\lambda x.\,(t[q(\sigma)]))\;(s[\sigma]) \\
    t[s/x][\sigma]
    &= t[(\id,s/x) \circ \sigma]
      \tag{by \Cref{lem:subst-assoc}} \\
    &= t[\id \circ \sigma, s[\sigma]/x]
      \tag{by \Cref{lem:ext-comp}} \\
    &= t[\sigma, s[\sigma]/x]
      \tag{by \Cref{lem:subst-cat} (left identity)}
  \end{align*}
  Now:
  \begin{align*}
    &\typing[\Gamma]{S[\sigma]}{\Ty{i}} && \text{(by IH)} \\
    &\typing[\Gamma, x:S[\sigma]]{T[q(\sigma)]}{\Ty{i}}
      && \text{(by IH using \Cref{lem:q-subst-typing})} \\
    &\typing[\Gamma, x:S[\sigma]]{t[q(\sigma)]}{T[q(\sigma)]}
      && \text{(by IH using \Cref{lem:q-subst-typing})} \\
    &\typing[\Gamma]{s[\sigma]}{S[\sigma]} && \text{(by IH)} \\
    &\tyequiv[\Gamma]{(\lambda x.\,(t[q(\sigma)]))\;(s[\sigma])}{t[q(\sigma)][s[\sigma]/x]}{T[q(\sigma)][s[\sigma]/x]}
      && \text{(by \textsc{$\Pi$-$\beta$})}
  \end{align*}
  By \Cref{lem:subst-ext-comm}:
  $t[q(\sigma)][s[\sigma]/x] = (t[s/x])[\sigma]$ and
  $T[q(\sigma)][s[\sigma]/x] = (T[s/x])[\sigma]$.

  \medskip\noindent
  \textbf{Case}
  \begin{mathpar}
    \inferrule
    {\typing[\Delta]{S}{\Ty{i}} \\ \typing[\Delta, x:S]{T}{\Ty{i}} \\
     \typing[\Delta]{t}{\Pi(x:S).\,T}}
    {\tyequiv[\Delta]{t}{\lambda x.\,(t[\Uparrow]\;x)}{\Pi(x:S).\,T}}
  \end{mathpar}
  We must show $\tyequiv[\Gamma]{t[\sigma]}{(\lambda x.\,(t[\Uparrow]\;x))[\sigma]}{(\Pi(x:S).\,T)[\sigma]}$.
  Unfolding the right-hand side:
  \begin{align*}
    (\lambda x.\,(t[\Uparrow]\;x))[\sigma]
    &= \lambda x.\,((t[\Uparrow]\;x)[q(\sigma)]) \\
    &= \lambda x.\,(t[\Uparrow][q(\sigma)]\;x) \\
    &= \lambda x.\,(t[\sigma][\Uparrow]\;x)
      \tag{by \Cref{lem:wk-subst-comm}}
  \end{align*}
  Now:
  \begin{align*}
    &\typing[\Gamma]{S[\sigma]}{\Ty{i}} && \text{(by IH)} \\
    &\typing[\Gamma, x:S[\sigma]]{T[q(\sigma)]}{\Ty{i}}
      && \text{(by IH using \Cref{lem:q-subst-typing})} \\
    &\typing[\Gamma]{t[\sigma]}{\Pi(x:S[\sigma]).\,(T[q(\sigma)])} && \text{(by IH)} \\
    &\tyequiv[\Gamma]{t[\sigma]}{\lambda x.\,(t[\sigma][\Uparrow]\;x)}{\Pi(x:S[\sigma]).\,(T[q(\sigma)])}
      && \text{(by \textsc{$\Pi$-$\eta$})}
  \end{align*}

  \medskip
  We now show representative cases for statement (3).

  \medskip\noindent
  \textbf{Case}
  \begin{mathpar}
    \inferrule
    {\vdash \Delta \\ i \le j}
    {\subty[\Delta]{\Ty{i}}{\Ty{j}}}
  \end{mathpar}
  Then $\Ty{i}[\sigma] = \Ty{i}$ and $\Ty{j}[\sigma] = \Ty{j}$.
  From $\judge[\Gamma]{\sigma : \Delta}$, we have $\vdash \Gamma$.
  By \textsc{Sub-Univ}, $\subty{\Ty{i}}{\Ty{j}}$.

  \medskip\noindent
  \textbf{Case}
  \begin{mathpar}
    \inferrule
    {\typing[\Delta]{S}{\Ty{i}} \\
     \typing[\Delta]{S'}{\Ty{i}} \\
     \tyequiv[\Delta]{S}{S'}{\Ty{i}} \\
     \typing[\Delta, x:S]{T}{\Ty{i}} \\
     \typing[\Delta, x:S']{T'}{\Ty{i}} \\
     \subty[\Delta, x:S]{T}{T'}}
    {\subty[\Delta]{\Pi(x:S).\,T}{\Pi(x:S').\,T'}}
  \end{mathpar}
  \begin{align*}
    &\typing[\Gamma]{S[\sigma]}{\Ty{i}} && \text{(by IH)} \\
    &\typing[\Gamma]{S'[\sigma]}{\Ty{i}} && \text{(by IH)} \\
    &\tyequiv[\Gamma]{S[\sigma]}{S'[\sigma]}{\Ty{i}}
      && \text{(by IH, statement (2))} \\
    &\typing[\Gamma, x:S[\sigma]]{T[q(\sigma)]}{\Ty{i}}
      && \text{(by IH using \Cref{lem:q-subst-typing})} \\
    &\typing[\Gamma, x:S'[\sigma]]{T'[q(\sigma)]}{\Ty{i}}
      && \text{(by IH using \Cref{lem:q-subst-typing})} \\
    &\subty[\Gamma, x:S[\sigma]]{T[q(\sigma)]}{T'[q(\sigma)]}
      && \text{(by IH, statement (3), using \Cref{lem:q-subst-typing})} \\
    &\subty{\Pi(x:S[\sigma]).\,(T[q(\sigma)])}{\Pi(x:S'[\sigma]).\,(T'[q(\sigma)])}
      && \text{(by \textsc{Sub-Pi})}
  \end{align*}
  This is $\subty{(\Pi(x:S).\,T)[\sigma]}{(\Pi(x:S').\,T')[\sigma]}$.

  \medskip\noindent
  \textbf{Case} \textsc{Sub-Trans}: By two applications of the IH and \textsc{Sub-Trans}.

  \medskip\noindent
  \textbf{Case} \textsc{Sub-Eq}: By statement (2) and \textsc{Sub-Eq}.
\end{proof}

\subsection{Consequences}\labeledit{sec:props:consequences}

\begin{lemma}[Composition of Substitutions is Well-Typed]\labeledit{lem:subst-comp-wt}
  If $\judge[\Gamma]{\sigma : \Delta}$ and $\judge[\Gamma']{\tau : \Gamma}$, then
  $\judge[\Gamma']{\sigma \circ \tau : \Delta}$.
\end{lemma}
\begin{proof}
  We must show: for all $x : T \in \Delta$,
  $\typing[\Gamma']{(\sigma \circ \tau)(x)}{T[\sigma \circ \tau]}$.
  \begin{align*}
    &\typing[\Gamma]{\sigma(x)}{T[\sigma]}
      && \text{(from $\judge[\Gamma]{\sigma : \Delta}$)} \\
    &\typing[\Gamma']{\sigma(x)[\tau]}{T[\sigma][\tau]}
      && \text{(by \Cref{lem:subst-preserves})} \\
    &(\sigma \circ \tau)(x) = \sigma(x)[\tau]
      && \text{(by definition)} \\
    &T[\sigma \circ \tau] = T[\sigma][\tau]
      && \text{(by \Cref{lem:subst-assoc})}
  \end{align*}
\end{proof}

\begin{corollary}[Single Substitution Lemma]\labeledit{cor:single-subst}
  If $\typing[\Gamma, x : S]{t}{T}$ and $\typing[\Gamma]{s}{S}$, then
  $\typing[\Gamma]{t[s/x]}{T[s/x]}$.
\end{corollary}
\begin{proof}
  By \Cref{lem:single-subst-wt} and \Cref{lem:subst-preserves}, noting
  $t[s/x] = t[\id, s/x]$.
\end{proof}

\subsection{Presupposition}\labeledit{sec:props:presup}

Presupposition states that every derivable judgment entails the well-formedness of its
constituent parts.

\begin{lemma}[Presupposition]\labeledit{lem:presup}
  \begin{itemize}
  \item If $\typing[\Gamma]{t}{T}$, then $\vdash \Gamma$ and $\typing[\Gamma]{T}{\Ty{i}}$
    for some~$i$.
  \item If $\tyequiv[\Gamma]{t}{t'}{T}$, then $\vdash \Gamma$,
    $\typing[\Gamma]{t}{T}$, $\typing[\Gamma]{t'}{T}$, and
    $\typing[\Gamma]{T}{\Ty{i}}$ for some~$i$.
  \item If $\subty{S}{T}$, then $\vdash \Gamma$,
    $\typing[\Gamma]{S}{\Ty{i}}$ and $\typing[\Gamma]{T}{\Ty{i}}$ for some $i$.
  \end{itemize}
\end{lemma}
\begin{proof}
  By mutual induction on the derivation.

  \medskip\noindent
  \textbf{Case}
  \begin{mathpar}
    \inferrule
    {\vdash \Gamma \\ x : T \in \Gamma}
    {\typing[\Gamma]{x}{T}}
  \end{mathpar}
  We have $\vdash \Gamma$ immediately from the premise.
  For $\typing[\Gamma]{T}{\Ty{i}}$, we do induction on $x : T \in \Gamma$.
  If $\Gamma = \Gamma', x : T$, then $\vdash \Gamma', x : T$ was derived by
  \textsc{Ctx-Ext} from $\typing[\Gamma']{T}{\Ty{i}}$; weakening gives
  $\typing[\Gamma]{T}{\Ty{i}}$.
  If $\Gamma = \Gamma', y : S$ and $x : T \in \Gamma'$, then by IH on the smaller
  context $\typing[\Gamma']{T}{\Ty{i}}$; weakening gives $\typing[\Gamma]{T}{\Ty{i}}$.

  \medskip\noindent
  \textbf{Case}
  \begin{mathpar}
    \inferrule
    {\vdash \Gamma}
    {\typing[\Gamma]{\Ty{i}}{\Ty{1+i}}}
  \end{mathpar}
  We have $\vdash \Gamma$ from the premise.
  The type of the conclusion is $\Ty{1+i}$, and $\typing[\Gamma]{\Ty{1+i}}{\Ty{2+i}}$
  by another application of \textsc{U-Wf}.

  \medskip\noindent
  \textbf{Case}
  \begin{mathpar}
    \inferrule
    {\typing[\Gamma]{S}{\Ty{i}} \\ \typing[\Gamma, x : S]{T}{\Ty{i}}}
    {\typing[\Gamma]{\Pi(x:S).\,T}{\Ty{i}}}
  \end{mathpar}
  \begin{align*}
    &\vdash \Gamma && \text{(by IH on first premise)} \\
    &\typing[\Gamma]{\Ty{i}}{\Ty{1+i}} && \text{(by \textsc{U-Wf})}
  \end{align*}

  \medskip\noindent
  \textbf{Case}
  \begin{mathpar}
    \inferrule
    {\typing[\Gamma]{S}{\Ty{i}} \\ \typing[\Gamma, x : S]{t}{T}}
    {\typing[\Gamma]{\lambda x.\,t}{\Pi(x:S).\,T}}
  \end{mathpar}
  \begin{align*}
    &\vdash \Gamma && \text{(by IH on first premise)} \\
    &\typing[\Gamma, x:S]{T}{\Ty{j}} \text{ for some } j
      && \text{(by IH on second premise)} \\
    &\typing[\Gamma]{\Pi(x:S).\,T}{\Ty{\max(i,j)}}
      && \text{(by \textsc{$\Pi$-Wf} + \textsc{Subsump})}
  \end{align*}

  \medskip\noindent
  \textbf{Case}
  \begin{mathpar}
    \inferrule
    {\typing[\Gamma]{S}{\Ty{i}} \\ \typing[\Gamma, x:S]{T}{\Ty{i}} \\\\ \typing[\Gamma]{r}{\Pi(x:S).\,T} \\ \typing[\Gamma]{s}{S}}
    {\typing[\Gamma]{r\;s}{T[s/x]}}
  \end{mathpar}
  \begin{align*}
    &\vdash \Gamma && \text{(by IH on first premise)} \\
    &\typing[\Gamma, x:S]{T}{\Ty{i}} && \text{(directly from second premise)} \\
    &\judge[\Gamma]{(\id, s/x) : (\Gamma, x:S)} && \text{(by \Cref{lem:single-subst-wt})} \\
    &\typing[\Gamma]{T[s/x]}{\Ty{i}} && \text{(by \Cref{lem:subst-preserves})}
  \end{align*}

  \medskip\noindent
  \textbf{Case}
  \begin{mathpar}
    \inferrule
    {\typing[\Gamma, x:\Nat]{T}{\Ty{i}} \\
     \typing[\Gamma]{z}{T[\ze/x]} \\
     \typing[\Gamma, x:\Nat, y:T]{s}{T[\Uparrow \circ \Uparrow, \su{x_1}/x]} \\
     \typing[\Gamma]{t}{\Nat}}
    {\typing[\Gamma]{\elimn{x.\,T}{z}{x,y.\,s}{t}}{T[t/x]}}
  \end{mathpar}
  \begin{align*}
    &\vdash \Gamma && \text{(by IH on last premise)} \\
    &\judge[\Gamma]{(\id, t/x) : (\Gamma, x:\Nat)} && \text{(by \Cref{lem:single-subst-wt})} \\
    &\typing[\Gamma]{T[t/x]}{\Ty{i}} && \text{(by \Cref{lem:subst-preserves} on first premise)}
  \end{align*}

  \medskip\noindent
  \textbf{Case}
  \begin{mathpar}
    \inferrule
    {\typing[\Gamma]{T}{\Ty{i}} \\ \typing[\Gamma]{t}{S} \\ \subty{S}{T}}
    {\typing[\Gamma]{t}{T}}
  \end{mathpar}
  \begin{align*}
    &\vdash \Gamma && \text{(by IH on first premise)} \\
    &\typing[\Gamma]{T}{\Ty{i}} && \text{(directly from first premise)}
  \end{align*}

  \medskip\noindent
  \textbf{Case}
  \begin{mathpar}
    \inferrule[\textsc{Var-Cong}]
    {\vdash \Gamma \\ x : T \in \Gamma}
    {\tyequiv[\Gamma]{x}{x}{T}}
  \end{mathpar}
  We have $\vdash \Gamma$ from the premise.
  $\typing[\Gamma]{x}{T}$ by \textsc{Var}.
  $\typing[\Gamma]{T}{\Ty{i}}$ by the same argument as the \textsc{Var} typing case above.

  \medskip\noindent
  \textbf{Case}
  \begin{mathpar}
    \inferrule[\textsc{$\Pi$-$\beta$}]
    {\typing[\Gamma]{S}{\Ty{i}} \\ \typing[\Gamma, x:S]{T}{\Ty{i}} \\
     \typing[\Gamma, x:S]{t}{T} \\ \typing[\Gamma]{s}{S}}
    {\tyequiv[\Gamma]{(\lambda x.\,t)\;s}{t[s/x]}{T[s/x]}}
  \end{mathpar}
  \begin{align*}
    &\vdash \Gamma && \text{(by IH on first premise)} \\
    &\typing[\Gamma]{(\lambda x.\,t)\;s}{T[s/x]}
      && \text{(by \textsc{$\lambda$-E} on \textsc{$\lambda$-I})} \\
    &\typing[\Gamma]{t[s/x]}{T[s/x]}
      && \text{(by \Cref{cor:single-subst} on third premise)} \\
    &\typing[\Gamma]{T[s/x]}{\Ty{i}}
      && \text{(by \Cref{cor:single-subst} on second premise)}
  \end{align*}

  \medskip\noindent
  \textbf{Case}
  \begin{mathpar}
    \inferrule[\textsc{$\Pi$-$\eta$}]
    {\typing[\Gamma]{S}{\Ty{i}} \\ \typing[\Gamma, x:S]{T}{\Ty{i}} \\
     \typing[\Gamma]{t}{\Pi(x:S).\,T}}
    {\tyequiv[\Gamma]{t}{\lambda x.\,(t[\Uparrow]\;x)}{\Pi(x:S).\,T}}
  \end{mathpar}
  \begin{align*}
    &\vdash \Gamma && \text{(by IH on first premise)} \\
    &\typing[\Gamma]{t}{\Pi(x:S).\,T} && \text{(directly from third premise)} \\
    &\weakening[\Gamma, x:S] \Uparrow \Gamma
      && \text{(by \Cref{lem:wk-shift-wt})} \\
    &\typing[\Gamma, x:S]{t[\Uparrow]}{\Pi(x:S[\Uparrow]).\,(T[q(\Uparrow)])}
      && \text{(by \Cref{lem:wk-typing})} \\
    &x_0 : S[\Uparrow] \in \Gamma, x_0 : S
      && \text{(by lookup)} \\
    &\typing[\Gamma, x:S]{x}{S[\Uparrow]}
      && \text{(by \textsc{Var})} \\
    &\typing[\Gamma, x:S]{t[\Uparrow]\;x}{T[q(\Uparrow)][x/x]}
      && \text{(by \textsc{$\lambda$-E})} \\
    &T[q(\Uparrow)][x/x] = T[q(\Uparrow)][\id, x/x] = T[\Uparrow, x/x]
      && \text{(by \Cref{lem:q-ext})} \\
    &(\Uparrow, x/x) = \id
      && \text{(by \Cref{lem:subst-ext-eta} at $\sigma = \id$)} \\
    &T[\Uparrow, x/x] = T[\id] = T
      && \text{(by \Cref{lem:subst-id})} \\
    &\typing[\Gamma, x:S]{t[\Uparrow]\;x}{T}
      \\
    &\typing[\Gamma]{\lambda x.\,(t[\Uparrow]\;x)}{\Pi(x:S).\,T}
      && \text{(by \textsc{$\lambda$-I})} \\
    &\typing[\Gamma]{\Pi(x:S).\,T}{\Ty{i}}
      && \text{(by \textsc{$\Pi$-Wf} on first two premises)}
  \end{align*}

  \medskip\noindent
  \textbf{Case}
  \begin{mathpar}
    \inferrule[\textsc{Tm-Subsump}]
    {\typing[\Gamma]{T}{\Ty{i}} \\ \tyequiv[\Gamma]{t}{t'}{S} \\ \subty{S}{T}}
    {\tyequiv[\Gamma]{t}{t'}{T}}
  \end{mathpar}
  \begin{align*}
    &\vdash \Gamma && \text{(by IH on first premise)} \\
    &\typing[\Gamma]{t}{S} \tand \typing[\Gamma]{t'}{S}
      && \text{(by IH on second premise)} \\
    &\typing[\Gamma]{t}{T} \tand \typing[\Gamma]{t'}{T}
      && \text{(by \textsc{Subsump} with third premise)} \\
    &\typing[\Gamma]{T}{\Ty{i}}
      && \text{(directly from first premise)}
  \end{align*}

  \medskip\noindent
  \textbf{Case}
  \begin{mathpar}
    \inferrule[\textsc{Sub-Pi}]
    {\typing[\Gamma]{S}{\Ty{i}} \\
     \typing[\Gamma]{S'}{\Ty{i}} \\
     \tyequiv[\Gamma]{S}{S'}{\Ty{i}} \\
     \typing[\Gamma, x:S]{T}{\Ty{i}} \\
     \typing[\Gamma, x:S']{T'}{\Ty{i}} \\
     \subty[\Gamma, x:S]{T}{T'}}
    {\subty[\Gamma]{\Pi(x:S).\,T}{\Pi(x:S').\,T'}}
  \end{mathpar}
  \begin{align*}
    &\vdash \Gamma && \text{(by IH on first premise)} \\
    &\typing[\Gamma]{\Pi(x:S).\,T}{\Ty{i}}
      && \text{(by \textsc{$\Pi$-Wf} on premises 1 and 4)} \\
    &\typing[\Gamma]{\Pi(x:S').\,T'}{\Ty{i}}
      && \text{(by \textsc{$\Pi$-Wf} on premises 2 and 5, using \textsc{Subsump})}
  \end{align*}

  \medskip\noindent
  \textbf{Case}
  \begin{mathpar}
    \inferrule[\textsc{Sub-Univ}]
    {\vdash \Gamma \\ i \le j}
    {\subty{\Ty{i}}{\Ty{j}}}
  \end{mathpar}
  \begin{align*}
    &\vdash \Gamma && \text{(directly from premise)} \\
    &\typing[\Gamma]{\Ty{i}}{\Ty{1+i}}
      && \text{(by \textsc{U-Wf})} \\
    &\typing[\Gamma]{\Ty{j}}{\Ty{1+j}}
      && \text{(by \textsc{U-Wf})} \\
    &\typing[\Gamma]{\Ty{i}}{\Ty{1+j}} \tand \typing[\Gamma]{\Ty{j}}{\Ty{1+j}}
      && \text{(by \textsc{Subsump} with \textsc{Sub-Univ}, since $1+i \le 1+j$)}
  \end{align*}

  \medskip\noindent
  The remaining cases follow similar patterns.
\end{proof}

\subsection{Substitution Equivalence}\labeledit{sec:props:subst-eq}

We define a pointwise notion of substitution equivalence and show that
equivalent substitutions preserve syntactic judgments equivalently.

\begin{definition}[Substitution Equivalence]\labeledit{def:subst-equiv}
  $\judge[\Gamma]{\sigma \approx \sigma' : \Delta}$ holds when:
  \begin{enumerate}
  \item $\judge[\Gamma]{\sigma : \Delta}$ and
    $\judge[\Gamma]{\sigma' : \Delta}$, and
  \item for every $x : T \in \Delta$,
    $\tyequiv[\Gamma]{\sigma(x)}{\sigma'(x)}{T[\sigma]}$.
  \end{enumerate}
\end{definition}

Note: by clause~(2) and \Cref{lem:subst-preserves} statement~(2) applied to
the reflexive closure, $\tyequiv[\Gamma]{T[\sigma]}{T[\sigma']}{\Ty{i}}$
whenever $\typing[\Delta]{T}{\Ty{i}}$, so the type at which the
$\sigma(x), \sigma'(x)$ are equated is well-defined modulo equivalence.

We first establish that the $q$-extension preserves equivalence.

\begin{lemma}[$q$ Preserves Substitution
    Equivalence]\labeledit{lem:q-subst-equiv}
  If $\judge[\Gamma]{\sigma \approx \sigma' : \Delta}$ and
  $\typing[\Delta]{S}{\Ty{i}}$, then
  $\judge[\Gamma, x:S[\sigma]]{q(\sigma) \approx q(\sigma') :
    (\Delta, x:S)}$.
\end{lemma}
\begin{proof}
  By \Cref{lem:q-subst-typing}, $\judge[\Gamma, x:S[\sigma]]{q(\sigma) :
    (\Delta, x:S)}$. We need $\judge[\Gamma, x:S[\sigma]]{q(\sigma') :
    (\Delta, x:S)}$ as well: from
  $\tyequiv[\Gamma]{S[\sigma]}{S[\sigma']}{\Ty{i}}$
  (by \Cref{lem:subst-preserves} statement~(2) applied to
  reflexivity on $\typing[\Delta]{S}{\Ty{i}}$), the contexts
  $\Gamma, x:S[\sigma]$ and $\Gamma, x:S[\sigma']$ are equivalent, so
  $\judge[\Gamma, x:S[\sigma]]{q(\sigma') : (\Delta, x:S)}$ by
  \Cref{lem:q-subst-typing} and a context-conversion step.

  We verify clause~(2): for every $y : T \in \Delta, x:S$,
  $\tyequiv[\Gamma, x:S[\sigma]]{q(\sigma)(y)}{q(\sigma')(y)}
  {T[q(\sigma)]}$.
  \begin{itemize}
  \item $y = x_0$: $T = S[\Uparrow]$, $q(\sigma)(x_0) = x_0 =
    q(\sigma')(x_0)$. The required equality is reflexivity on $x_0$ at
    $S[\Uparrow][q(\sigma)] = S[\sigma][\Uparrow]$ (by
    \Cref{lem:wk-subst-comm}).
  \item $y = 1{+}z$ where $z : T' \in \Delta$:
    $q(\sigma)(1{+}z) = \sigma(z)[\Uparrow]$ and
    $q(\sigma')(1{+}z) = \sigma'(z)[\Uparrow]$. By clause~(2) of
    \Cref{def:subst-equiv} on $\sigma \approx \sigma'$,
    $\tyequiv[\Gamma]{\sigma(z)}{\sigma'(z)}{T'[\sigma]}$. By
    \Cref{lem:wk-typing} applied to $\Uparrow$,
    $\tyequiv[\Gamma, x:S[\sigma]]
      {\sigma(z)[\Uparrow]}{\sigma'(z)[\Uparrow]}
      {T'[\sigma][\Uparrow]}$,
    and $T'[\sigma][\Uparrow] = T'[\Uparrow][q(\sigma)]$ by
    \Cref{lem:wk-subst-comm}, which equals $T[q(\sigma)]$.
  \end{itemize}
\end{proof}

\begin{lemma}[Equivalent Substitutions Preserve
    Judgments]\labeledit{lem:subst-equiv-preserves}
  The following are proved by mutual induction on the syntactic
  judgments:
  \begin{enumerate}
  \item If $\typing[\Delta]{t}{T}$ and
    $\judge[\Gamma]{\sigma \approx \sigma' : \Delta}$, then
    $\tyequiv[\Gamma]{t[\sigma]}{t[\sigma']}{T[\sigma]}$.
  \item If $\tyequiv[\Delta]{t}{t'}{T}$ and
    $\judge[\Gamma]{\sigma \approx \sigma' : \Delta}$, then
    $\tyequiv[\Gamma]{t[\sigma]}{t'[\sigma']}{T[\sigma]}$.
  \item If $\subty[\Delta]{S}{T}$ and
    $\judge[\Gamma]{\sigma \approx \sigma' : \Delta}$, then
    $\subty[\Gamma]{S[\sigma]}{T[\sigma']}$.
  \end{enumerate}
\end{lemma}
\begin{proof}
  By mutual induction on the syntactic derivation, parallel to
  \Cref{lem:subst-preserves}. We show representative cases for
  statement~(1).

  \medskip\noindent
  \textbf{Case}
  \begin{mathpar}
    \inferrule
    {\vdash \Delta \\ x : T \in \Delta}
    {\typing[\Delta]{x}{T}}
  \end{mathpar}
  Then $x[\sigma] = \sigma(x)$ and $x[\sigma'] = \sigma'(x)$.
  \begin{align*}
    &\tyequiv[\Gamma]{\sigma(x)}{\sigma'(x)}{T[\sigma]}
      && \text{(by clause~(2) of \Cref{def:subst-equiv})}
  \end{align*}

  \medskip\noindent
  \textbf{Case}
  \begin{mathpar}
    \inferrule
    {\typing[\Delta]{S}{\Ty{i}} \\ \typing[\Delta, x : S]{t}{T}}
    {\typing[\Delta]{\lambda x.\,t}{\Pi(x:S).\,T}}
  \end{mathpar}
  We must show
  $\tyequiv[\Gamma]{(\lambda x.\,t)[\sigma]}{(\lambda x.\,t)[\sigma']}
  {(\Pi(x:S).\,T)[\sigma]}$.
  We first lift the equivalence under $q$:
  \begin{align*}
    &\judge[\Gamma, x:S[\sigma]]{q(\sigma) \approx q(\sigma') :
      (\Delta, x:S)}
      && \text{(by \Cref{lem:q-subst-equiv})} \\
    &\tyequiv[\Gamma, x:S[\sigma]]{t[q(\sigma)]}{t[q(\sigma')]}
      {T[q(\sigma)]}
      && \text{(by IH on the second premise)} \\
    &\tyequiv[\Gamma]{S[\sigma]}{S[\sigma']}{\Ty{i}}
      && \text{(by IH on the first premise)} \\
    &\tyequiv[\Gamma]{\lambda x.\,(t[q(\sigma)])}
      {\lambda x.\,(t[q(\sigma')])}
      {\Pi(x:S[\sigma]).\,(T[q(\sigma)])}
      && \text{(by \textsc{$\lambda$-Cong})} \\
    &(\lambda x.\,t)[\sigma] = \lambda x.\,(t[q(\sigma)]),\;
      (\lambda x.\,t)[\sigma'] = \lambda x.\,(t[q(\sigma')])
      && \text{(syntactic substitution)} \\
    &(\Pi(x:S).\,T)[\sigma] = \Pi(x:S[\sigma]).\,(T[q(\sigma)])
      && \text{(syntactic substitution)}
  \end{align*}

  \medskip\noindent
  \textbf{Case}
  \begin{mathpar}
    \inferrule
    {\typing[\Delta]{S}{\Ty{i}} \\ \typing[\Delta, x:S]{T}{\Ty{i}} \\\\
     \typing[\Delta]{r}{\Pi(x:S).\,T} \\ \typing[\Delta]{s}{S}}
    {\typing[\Delta]{r\;s}{T[s/x]}}
  \end{mathpar}
  We must show
  $\tyequiv[\Gamma]{(r\;s)[\sigma]}{(r\;s)[\sigma']}{T[s/x][\sigma]}$.
  \begin{align*}
    &\tyequiv[\Gamma]{r[\sigma]}{r[\sigma']}
      {(\Pi(x:S).\,T)[\sigma]}
      && \text{(by IH on the third premise)} \\
    &\tyequiv[\Gamma]{s[\sigma]}{s[\sigma']}{S[\sigma]}
      && \text{(by IH on the fourth premise)} \\
    &\tyequiv[\Gamma]{r[\sigma]\;s[\sigma]}{r[\sigma']\;s[\sigma']}
      {T[q(\sigma)][s[\sigma]/x]}
      && \text{(by \textsc{App-Cong})} \\
    &(r\;s)[\sigma] = r[\sigma]\;s[\sigma],\;
      (r\;s)[\sigma'] = r[\sigma']\;s[\sigma']
      && \text{(syntactic substitution on application)} \\
    &T[q(\sigma)][s[\sigma]/x] = T[s/x][\sigma]
      && \text{(by \Cref{lem:subst-ext-comm})}
  \end{align*}

  \medskip\noindent
  \textbf{Case}
  \begin{mathpar}
    \inferrule[\textsc{U-Wf}]
    {\vdash \Delta}
    {\typing[\Delta]{\Ty{i}}{\Ty{1+i}}}
  \end{mathpar}
  $\Ty{i}[\sigma] = \Ty{i}$ and $\Ty{i}[\sigma'] = \Ty{i}$.
  \begin{align*}
    &\vdash \Gamma
      && \text{(from $\judge[\Gamma]{\sigma : \Delta}$ inside
        \Cref{def:subst-equiv})} \\
    &\tyequiv[\Gamma]{\Ty{i}}{\Ty{i}}{\Ty{1+i}}
      && \text{(by \textsc{U-Cong})}
  \end{align*}

  \medskip\noindent
  \textbf{Case} \textsc{N-Wf}: Analogous to \textsc{U-Wf}, with
  $\Nat[\sigma] = \Nat = \Nat[\sigma']$ and reflexivity by
  \textsc{N-Cong}.

  \medskip\noindent
  \textbf{Case}
  \begin{mathpar}
    \inferrule[\textsc{$\Pi$-Wf}]
    {\typing[\Delta]{S}{\Ty{i}} \\ \typing[\Delta, x:S]{T}{\Ty{i}}}
    {\typing[\Delta]{\Pi(x:S).\,T}{\Ty{i}}}
  \end{mathpar}
  \begin{align*}
    &\tyequiv[\Gamma]{S[\sigma]}{S[\sigma']}{\Ty{i}}
      && \text{(by IH on first premise)} \\
    &\judge[\Gamma, x:S[\sigma]]{q(\sigma) \approx q(\sigma') :
      (\Delta, x:S)}
      && \text{(by lifting, as in the $\lambda$-I case)} \\
    &\tyequiv[\Gamma, x:S[\sigma]]{T[q(\sigma)]}{T[q(\sigma')]}{\Ty{i}}
      && \text{(by IH on second premise)} \\
    &\tyequiv[\Gamma]{\Pi(x:S[\sigma]).\,(T[q(\sigma)])}
      {\Pi(x:S[\sigma']).\,(T[q(\sigma')])}{\Ty{i}}
      && \text{(by \textsc{$\Pi$-Cong})} \\
    &(\Pi(x:S).\,T)[\sigma] = \Pi(x:S[\sigma]).\,(T[q(\sigma)])
      && \text{(syntactic substitution on $\Pi$)}
  \end{align*}

  \medskip\noindent
  \textbf{Case} \textsc{Ze-I}: $\ze[\sigma] = \ze = \ze[\sigma']$;
  apply \textsc{Ze-Cong}. The type
  $\Nat[\sigma] = \Nat$.

  \medskip\noindent
  \textbf{Case}
  \begin{mathpar}
    \inferrule[\textsc{Su-I}]
    {\typing[\Delta]{t}{\Nat}}
    {\typing[\Delta]{\su{t}}{\Nat}}
  \end{mathpar}
  \begin{align*}
    &\tyequiv[\Gamma]{t[\sigma]}{t[\sigma']}{\Nat}
      && \text{(by IH on premise; using $\Nat[\sigma] = \Nat$)} \\
    &\tyequiv[\Gamma]{\su{t[\sigma]}}{\su{t[\sigma']}}{\Nat}
      && \text{(by \textsc{Su-Cong})} \\
    &(\su{t})[\sigma] = \su{t[\sigma]},\;
      (\su{t})[\sigma'] = \su{t[\sigma']}
      && \text{(syntactic substitution on $\su{}$)}
  \end{align*}

  \medskip\noindent
  \textbf{Case} \textsc{N-E}: analogous to \textsc{$\lambda$-E},
  applying the IH to each premise (motive, zero branch, step branch,
  scrutinee), lifting via $q(\sigma) \approx q(\sigma')$ and
  $q^2(\sigma) \approx q^2(\sigma')$, and concluding by \textsc{N-E-Cong}.

  \medskip\noindent
  \textbf{Case}
  \begin{mathpar}
    \inferrule[\textsc{Subsump}]
    {\typing[\Delta]{T}{\Ty{i}} \\
     \typing[\Delta]{t}{S} \\
     \subty[\Delta]{S}{T}}
    {\typing[\Delta]{t}{T}}
  \end{mathpar}
  \begin{align*}
    &\tyequiv[\Gamma]{t[\sigma]}{t[\sigma']}{S[\sigma]}
      && \text{(by IH on second premise)} \\
    &\subty[\Gamma]{S[\sigma]}{T[\sigma]}
      && \text{(by IH on third premise via statement~(3),} \\
    & && \text{instantiated at $\sigma \approx \sigma$, reflexive)} \\
    &\typing[\Gamma]{T[\sigma]}{\Ty{i}}
      && \text{(by \Cref{lem:subst-preserves} on first premise)} \\
    &\tyequiv[\Gamma]{t[\sigma]}{t[\sigma']}{T[\sigma]}
      && \text{(by \textsc{Tm-Subsump})}
  \end{align*}

  \medskip
  \textbf{Statement~(2): term-equality preservation.} By induction on
  the term-equality derivation.

  \medskip\noindent
  \textbf{Case} (Congruence rules: \textsc{Var-Cong}, \textsc{U-Cong},
  \textsc{N-Cong}, \textsc{$\Pi$-Cong}, \textsc{Ze-Cong},
  \textsc{Su-Cong}, \textsc{$\lambda$-Cong}, \textsc{App-Cong},
  \textsc{N-E-Cong}). Each follows by IH on the premises and the same
  congruence rule, with substitution-on-syntax computed exactly as in
  statement~(1).

  \medskip\noindent
  \textbf{Case}
  \begin{mathpar}
    \inferrule[\textsc{$\Pi$-$\beta$}]
    {\typing[\Delta]{S}{\Ty{i}} \\ \typing[\Delta, x:S]{T}{\Ty{i}} \\
     \typing[\Delta, x:S]{t}{T} \\ \typing[\Delta]{s}{S}}
    {\tyequiv[\Delta]{(\lambda x.\,t)\;s}{t[s/x]}{T[s/x]}}
  \end{mathpar}
  We must show
  $\tyequiv[\Gamma]{((\lambda x.\,t)\;s)[\sigma]}{t[s/x][\sigma']}
  {T[s/x][\sigma]}$.

  By \Cref{lem:q-subst-equiv} on $\sigma \approx \sigma'$ at $S$:
  $\judge[\Gamma, x:S[\sigma]]{q(\sigma) \approx q(\sigma') :
    (\Delta, x:S)}$. Apply IH statement~(1) to the third premise
  $\typing[\Delta, x:S]{t}{T}$ and to the fourth premise
  $\typing[\Delta]{s}{S}$:
  \begin{align*}
    &\tyequiv[\Gamma, x:S[\sigma]]{t[q(\sigma)]}{t[q(\sigma')]}
      {T[q(\sigma)]}
      && \text{(by IH on third premise)} \\
    &\tyequiv[\Gamma]{s[\sigma]}{s[\sigma']}{S[\sigma]}
      && \text{(by IH on fourth premise)}
  \end{align*}
  Build the head-extended substitution equivalence:
  \begin{align*}
    &\judge[\Gamma]{(\sigma, s[\sigma]/x) \approx (\sigma', s[\sigma']/x)
      : (\Delta, x:S)}
      && \text{(at every $y \in \Delta$, both yield $\sigma(y),
        \sigma'(y)$,} \\
    & && \text{equated by \Cref{def:subst-equiv} clause~(2) on
        $\sigma \approx \sigma'$;} \\
    & && \text{at $x$, the heads $s[\sigma], s[\sigma']$ are equated
        above)}
  \end{align*}
  Apply IH statement~(1) on the third premise
  $\typing[\Delta, x:S]{t}{T}$ at this equivalence:
  \begin{align*}
    &\tyequiv[\Gamma]{t[(\sigma, s[\sigma]/x)]}
      {t[(\sigma', s[\sigma']/x)]}
      {T[(\sigma, s[\sigma]/x)]}
      && \text{(by IH on third premise at the head-extended
        equivalence)} \\
    &t[s/x][\sigma] = t[(\sigma, s[\sigma]/x)]
      && \text{(by \Cref{lem:subst-ext-comm})} \\
    &t[s/x][\sigma'] = t[(\sigma', s[\sigma']/x)]
      && \text{(by \Cref{lem:subst-ext-comm})} \\
    &T[s/x][\sigma] = T[(\sigma, s[\sigma]/x)]
      && \text{(by \Cref{lem:subst-ext-comm})} \\
    &\tyequiv[\Gamma]{t[s/x][\sigma]}{t[s/x][\sigma']}{T[s/x][\sigma]}
      && \text{(rewriting the previous equality line)} \\
    &\tyequiv[\Gamma]{((\lambda x.\,t)\;s)[\sigma]}{t[s/x][\sigma]}
      {T[s/x][\sigma]}
      && \text{(by \Cref{lem:subst-preserves} statement~(2) on the
        original $\beta$, at $\sigma$)} \\
    &\tyequiv[\Gamma]{((\lambda x.\,t)\;s)[\sigma]}{t[s/x][\sigma']}
      {T[s/x][\sigma]}
      && \text{(by \textsc{Tm-Trans})}
  \end{align*}

  \medskip\noindent
  \textbf{Case}
  \begin{mathpar}
    \inferrule[\textsc{$\Pi$-$\eta$}]
    {\typing[\Delta]{S}{\Ty{i}} \\
     \typing[\Delta, x:S]{T}{\Ty{i}} \\
     \typing[\Delta]{t}{\Pi(x:S).\,T}}
    {\tyequiv[\Delta]{t}{\lambda x.\,(t[\Uparrow]\;x)}
      {\Pi(x:S).\,T}}
  \end{mathpar}
  We must show
  $\tyequiv[\Gamma]{t[\sigma]}{(\lambda x.\,(t[\Uparrow]\;x))[\sigma']}
  {(\Pi(x:S).\,T)[\sigma]}$.

  By IH statement~(1) on the third premise:
  \begin{align*}
    &\tyequiv[\Gamma]{t[\sigma]}{t[\sigma']}{(\Pi(x:S).\,T)[\sigma]}
      && \text{(by IH on $\typing[\Delta]{t}{\Pi(x:S).\,T}$)}
  \end{align*}
  Apply \Cref{lem:subst-preserves} statement~(2) on the original
  $\eta$-rule at $\sigma'$ (a well-typed substitution since
  $\judge[\Gamma]{\sigma' : \Delta}$ by \Cref{def:subst-equiv} clause~(1)):
  \begin{align*}
    &\tyequiv[\Gamma]{t[\sigma']}{(\lambda x.\,(t[\Uparrow]\;x))[\sigma']}
      {(\Pi(x:S).\,T)[\sigma']}
      && \text{(by \Cref{lem:subst-preserves} statement~(2)
        on $\eta$ at $\sigma'$)}
  \end{align*}
  Bridge the type from $\sigma'$ to $\sigma$ via the $\Pi$-Wf case of
  statement~(1), already proved on
  $\typing[\Delta]{\Pi(x:S).\,T}{\Ty{i}}$ (derivable from the first two
  $\eta$-premises by \textsc{$\Pi$-Wf}):
  \begin{align*}
    &\tyequiv[\Gamma]{(\Pi(x:S).\,T)[\sigma]}{(\Pi(x:S).\,T)[\sigma']}
      {\Ty{i}}
      && \text{(by the $\Pi$-Wf case of statement~(1))} \\
    &\tyequiv[\Gamma]{t[\sigma']}{(\lambda x.\,(t[\Uparrow]\;x))[\sigma']}
      {(\Pi(x:S).\,T)[\sigma]}
      && \text{(by \textsc{Tm-Subsump} with \textsc{Sub-Eq})} \\
    &\tyequiv[\Gamma]{t[\sigma]}{(\lambda x.\,(t[\Uparrow]\;x))[\sigma']}
      {(\Pi(x:S).\,T)[\sigma]}
      && \text{(by \textsc{Tm-Trans})}
  \end{align*}

  \medskip\noindent
  \textbf{Case} \textsc{N-$\beta$-ze}, \textsc{N-$\beta$-su}: analogous
  to $\beta$ above. Apply \Cref{lem:subst-preserves} statement~(2) to
  the original reduction rule at $\sigma$ to get equality at the
  contractum-substituted-by-$\sigma$. Then apply IH (statement~(1)) to
  the typing of the contractum (derivable from the reduction rule and
  \Cref{lem:presup}) to bridge the contractum from $\sigma$ to
  $\sigma'$.

  \medskip\noindent
  \textbf{Case}
  \begin{mathpar}
    \inferrule[\textsc{Tm-Sym}]
    {\tyequiv[\Delta]{t}{t'}{T}}
    {\tyequiv[\Delta]{t'}{t}{T}}
  \end{mathpar}
  By IH on the premise (with the roles of $\sigma, \sigma'$ swapped):
  $\tyequiv[\Gamma]{t[\sigma']}{t'[\sigma]}{T[\sigma']}$. By
  \Cref{lem:subst-preserves} statement~(2), $\tyequiv[\Gamma]{T[\sigma]}
  {T[\sigma']}{\Ty{i}}$ (from the reflexive closure on the
  type-formation of $T$ via presupposition). By \textsc{Tm-Subsump}
  with \textsc{Sub-Eq}, transport to the goal type $T[\sigma]$:
  $\tyequiv[\Gamma]{t'[\sigma]}{t[\sigma']}{T[\sigma]}$, then apply
  \textsc{Tm-Sym}.

  \medskip\noindent
  \textbf{Case}
  \begin{mathpar}
    \inferrule[\textsc{Tm-Trans}]
    {\tyequiv[\Delta]{t}{t'}{T} \\\\ \tyequiv[\Delta]{t'}{t''}{T}}
    {\tyequiv[\Delta]{t}{t''}{T}}
  \end{mathpar}
  By IH on the first premise:
  $\tyequiv[\Gamma]{t[\sigma]}{t'[\sigma]}{T[\sigma]}$ (instantiated at
  $\sigma \approx \sigma$, reflexive). By IH on the second premise:
  $\tyequiv[\Gamma]{t'[\sigma]}{t''[\sigma']}{T[\sigma]}$. By
  \textsc{Tm-Trans}: $\tyequiv[\Gamma]{t[\sigma]}{t''[\sigma']}{T[\sigma]}$.

  \medskip\noindent
  \textbf{Case}
  \begin{mathpar}
    \inferrule[\textsc{Tm-Subsump}]
    {\typing[\Delta]{T}{\Ty{i}} \\\\
     \tyequiv[\Delta]{t}{t'}{S} \\
     \subty[\Delta]{S}{T}}
    {\tyequiv[\Delta]{t}{t'}{T}}
  \end{mathpar}
  \begin{align*}
    &\tyequiv[\Gamma]{t[\sigma]}{t'[\sigma']}{S[\sigma]}
      && \text{(by IH on the second premise)} \\
    &\subty[\Gamma]{S[\sigma]}{T[\sigma]}
      && \text{(by IH on the third premise via statement~(3),
        instantiated} \\
    & && \text{at $\sigma \approx \sigma$, reflexive)} \\
    &\typing[\Gamma]{T[\sigma]}{\Ty{i}}
      && \text{(by \Cref{lem:subst-preserves} on the first premise)} \\
    &\tyequiv[\Gamma]{t[\sigma]}{t'[\sigma']}{T[\sigma]}
      && \text{(by \textsc{Tm-Subsump})}
  \end{align*}

  \medskip
  \textbf{Statement~(3): subtyping preservation.} By induction on the
  subtyping derivation.

  \medskip\noindent
  \textbf{Case} \textsc{Sub-Univ}: $\Ty{i}[\sigma] = \Ty{i}$,
  $\Ty{j}[\sigma'] = \Ty{j}$, and the conclusion follows from
  \textsc{Sub-Univ} at $i \le j$ (the same hypothesis), with $\vdash
  \Gamma$ from $\judge[\Gamma]{\sigma : \Delta}$.

  \medskip\noindent
  \textbf{Case}
  \begin{mathpar}
    \inferrule[\textsc{Sub-Pi}]
    {\typing[\Delta]{S}{\Ty{i}} \\
     \typing[\Delta]{S'}{\Ty{i}} \\
     \tyequiv[\Delta]{S}{S'}{\Ty{i}} \\\\
     \typing[\Delta, x:S]{T}{\Ty{i}} \\
     \typing[\Delta, x:S']{T'}{\Ty{i}} \\
     \subty[\Delta, x:S]{T}{T'}}
    {\subty[\Delta]{\Pi(x:S).\,T}{\Pi(x:S').\,T'}}
  \end{mathpar}
  \begin{align*}
    &\tyequiv[\Gamma]{S[\sigma]}{S'[\sigma']}{\Ty{i}}
      && \text{(by IH on third premise, statement~(2))} \\
    &\judge[\Gamma, x:S[\sigma]]{q(\sigma) \approx q(\sigma') :
      (\Delta, x:S)}
      && \text{(by lifting)} \\
    &\subty[\Gamma, x:S[\sigma]]{T[q(\sigma)]}{T'[q(\sigma')]}
      && \text{(by IH on sixth premise)} \\
    &\subty[\Gamma]{\Pi(x:S[\sigma]).\,(T[q(\sigma)])}
      {\Pi(x:S'[\sigma']).\,(T'[q(\sigma')])}
      && \text{(by \textsc{Sub-Pi} on the previous lines and IH on
        the well-formedness} \\
    & && \text{premises 1, 2, 4, 5)} \\
    &(\Pi(x:S).\,T)[\sigma] = \Pi(x:S[\sigma]).\,(T[q(\sigma)])
      && \text{(syntactic)} \\
    &(\Pi(x:S').\,T')[\sigma'] = \Pi(x:S'[\sigma']).\,(T'[q(\sigma')])
      && \text{(syntactic)}
  \end{align*}

  \medskip\noindent
  \textbf{Case} \textsc{Sub-Trans}: by IH on both premises and
  \textsc{Sub-Trans}, using a shared middle subtype evaluated at
  $\sigma$ on both sides.

  \medskip\noindent
  \textbf{Case} \textsc{Sub-Eq}: by IH on the premise via statement~(2)
  and \textsc{Sub-Eq}.
\end{proof}

\begin{corollary}[Pointwise Equality at the Head Yields Type
    Equality]\labeledit{cor:subst-head-eq}
  If $\typing[\Delta, x:S]{T}{\Ty{i}}$,
  $\judge[\Gamma]{\sigma : \Delta}$, and
  $\tyequiv[\Gamma]{s}{s'}{S[\sigma]}$, then
  $\tyequiv[\Gamma]{T[\sigma, s/x]}{T[\sigma, s'/x]}{\Ty{i}}$.
\end{corollary}
\begin{proof}
  By \Cref{lem:presup}, $\typing[\Gamma]{s}{S[\sigma]}$ and
  $\typing[\Gamma]{s'}{S[\sigma]}$ from
  $\tyequiv[\Gamma]{s}{s'}{S[\sigma]}$.
  Build the substitution equivalence
  $\judge[\Gamma]{(\sigma, s/x) \approx (\sigma, s'/x) :
    (\Delta, x:S)}$: at every $y \in \Delta$ both sides yield
  $\sigma(y)$, and at $x$ the heads $s, s'$ are equated at $S[\sigma]$
  by hypothesis. Apply \Cref{lem:subst-equiv-preserves} statement~(1)
  to $\typing[\Delta, x:S]{T}{\Ty{i}}$.
\end{proof}

\section{Domain Model and Evaluation}\labeledit{sec:sem}

In this section, we define the semantic domain into which terms of \mltt are interpreted,
together with the evaluation function that maps syntactic terms (under an environment) to
domain values, and the readback function that extracts normal forms from domain values.
The overall structure follows the standard NbE recipe: evaluate into a domain of
semantic values, then read back to produce syntactic normal forms. %
At this point, the difference with the explicit-substitution approach is still
mundane; the main changes come from the semantic judgment, where we must make the
commutativity between evaluation and substitution application explicit. %
Since substitutions are operations (not syntax), this commutativity does not come for free
from the equational theory, and hence a semantic approach is required.

\subsection{Semantic Domain}\labeledit{sec:sem:domain}

The domain model $\D$ serves as the target of the evaluation function.
A central design principle of NbE for dependent types is that \emph{types are also values}:
unlike in the STLC setting where the domain is stratified by syntactic types, here both
types and terms evaluate into the same domain $\D$.
This unification is what makes NbE work for dependent types, since the type of an
expression may depend on the value of another expression.

We define three mutually recursive sets: domain values $\D$, neutral domain values $\D_n$,
and annotated normal forms $\D_f$.
Semantic environments $\Env$ map variables to domain values.

\begin{definition}[Semantic Domain]\labeledit{def:domain}
  The domain $\D$, neutral values $\D_n$, normal forms $\D_f$, and
  environments $\Env$ are defined mutually as follows:
  \[
    \begin{array}{rrcl}
      \textit{Environments} & \rho & \in & \Env = \N \to \D \\[6pt]
      \textit{Domain values} & a, b, A, B & \in & \D \\
      & & ::= & U_i \mid N \mid \Pi(a, t, \rho) \mid \zed \mid \sud{a}
                \mid \lambda(t, \rho) \mid {\uparrow^A(e)} \\[6pt]
      \textit{Neutral values} & e & \in & \D_n \\
      & & ::= & \ell_n \mid e \; \$ \; d
                \mid \recd T a s \rho e \\[6pt]
      \textit{Normal forms} & d & \in & \D_f \\
      & & ::= & {\downarrow}^A(a)
    \end{array}
  \]
\end{definition}

Let us explain the intuition behind each constructor:
\begin{itemize}
\item $U_i$ (universe): the semantic value of the type $\Ty{i}$.

\item $N$ (natural number type): the semantic value of the type $\Nat$.

\item $\Pi(a, t, \rho)$: the semantic value of a $\Pi$-type, where $a \in \D$ is the
  interpretation of the domain type, $t \in \Exp$ is the body of the codomain, and
  $\rho \in \Env$ is the captured environment.

\item $\zed$ and $\sud{d}$: the standard constructors for natural numbers, lifted to
  the domain.

\item $\lambda(t, \rho)$: a \emph{closure}, representing a lambda abstraction
  $\lambda x.\, t$ together with the environment $\rho$ that was active at the
  point of evaluation.

\item ${\uparrow}^A(e)$: a \emph{neutral value} $e$ annotated with its type $A \in \D$.
  Neutral values arise when computation is blocked --- for instance, when a variable
  stands in head position of an application.
  The type annotation is needed during readback to determine how to $\eta$-expand the
  neutral at function types.

\item $\ell_n$: a free variable at de Bruijn \emph{level} $n$.
  During readback, levels are converted back to de Bruijn indices.

\item $e \; \$ \; d$: a neutral application, where $e$ is a neutral function and $d$
  is an annotated normal form recording both the argument value and its type.

\item $\recd{T}{d}{s}{\rho}{e}$: a neutral recursion (eliminator for
  $\Nat$), arising when the scrutinee is neutral. Here $T$ is the captured syntactic motive, $d$ is the zero case, $s$ is the syntactic step
  expression, $\rho$ is the captured ambient environment, and $e$ is the neutral
  scrutinee.

\item ${\downarrow}^A(a)$: an annotated normal form, pairing a type $A \in \D$ with a
  value $a \in \D$. %
  This is used both in neutral applications (to record the domain type of the argument)
  and as the input format for readback.
\end{itemize}

\subsection{Environments}\labeledit{sec:sem:env}

An environment $\rho \in \Env = \N \to \D$ maps de Bruijn indices to domain values.
We write $(\rho, a/x)$ for the extension of $\rho$ by $a$ at index~$0$:
\[
  (\rho, a/x)(0) = a \qquad (\rho, a/x)(y+1) = \rho(y)
\]

We also define the \emph{drop} operation $\rho[\Uparrow]$ that discards the first binding:
\[
  \rho[\Uparrow](x) = \rho(x+1)
\]
This satisfies $(\rho, a/x)[\Uparrow] = \rho$.

\subsection{Evaluation}\labeledit{sec:sem:eval}

Evaluation is defined as a big-step relation $\intp{t}{\rho} \searrow a$, meaning
``expression $t$, evaluated in environment $\rho$, produces domain value $d$.''
We define this relation by the inference rules in \Cref{fig:eval}.
The relation is written in relational (not functional) style to accommodate the partial
nature of evaluation in the untyped setting; totality is established separately as part
of the soundness theorem.

\begin{figure}[t]
  \begin{mathpar}
    \inferrule[ev-var]
    { }
    {\intp{x}{\rho} \searrow \rho(x)}

    \inferrule[ev-U]
    { }
    {\intp{\Ty{i}}\rho \searrow U_i}

    \inferrule[ev-N]
    { }
    {\intp{\Nat}{\rho} \searrow N}

    \inferrule[ev-Pi]
    {\intp{S}{\rho} \searrow a}
    {\intp{\Pi(x:S).\, T}{\rho} \searrow \Pi(a, T, \rho)}

    \inferrule[ev-ze]
    { }
    {\intp{\ze}{\rho} \searrow \zed}

    \inferrule[ev-su]
    {\intp{t}{\rho} \searrow a}
    {\intp{\su{t}}\rho \searrow \sud{a}}

    \inferrule[ev-lam]
    { }
    {\intp{\lambda x.\, t}{\rho} \searrow \lambda(t, \rho)}

    \inferrule[ev-app]
    {\intp{r}{\rho} \searrow f \\ \intp{s}{\rho} \searrow a \\ f \cdot a \searrow b}
    {\intp{r\; s}{\rho} \searrow b}

    \inferrule[ev-rec]
    {\intp{t}{\rho} \searrow a \\
     \mathsf{rec}(a, T, z, s, \rho) \searrow b}
    {\intp{\elimn{x.\,T}{z}{x,y.\,s}{t}}{\rho} \searrow b}
  \end{mathpar}
  \caption{Evaluation rules: $\intp{t}{\rho} \searrow a$}
  \labeledit{fig:eval}
\end{figure}

Several points deserve emphasis:
\begin{itemize}
\item \textbf{Variables}: The rule \textsc{ev-var} looks up $x$ in
  the environment $\rho$.

\item \textbf{$\Pi$-types}: The rule \textsc{ev-Pi} evaluates the domain type $S$ to a
  value $a$, and records the codomain expression $T$ together with the current
  environment $\rho$ as a closure $\Pi(a, T, \rho)$.
  The codomain fiber at argument $b$ is obtained later by evaluating $T$ in
  $(\rho, b/x)$.

\item \textbf{Lambda}: The rule \textsc{ev-lam} does not evaluate the body; it merely
  pairs the body $t$ with the current environment $\rho$ to form a closure
  $\lambda(t, \rho)$.
  The body will be evaluated when the closure is applied.

\item \textbf{Application}: The rule \textsc{ev-app} evaluates both subexpressions and
  then invokes the auxiliary \emph{application} judgment $f \cdot a \searrow b$
  (defined below).

\item \textbf{Recursion}: The rule \textsc{ev-rec} evaluates the zero
  case $z$ and the scrutinee $t$, then invokes the auxiliary \emph{recursion} judgment.
  The motive $T$ and step case $s$ are stored unevaluated alongside $\rho$.
\end{itemize}

\subsection{Application}\labeledit{sec:sem:app}

The application judgment $f \cdot a \searrow b$ describes how to apply a domain value $f$
(which should be function-like) to an argument $a$.

\begin{figure}[h]
  \begin{mathpar}
    \inferrule[app-lam]
    {\intp{t}{\rho, a/x_0} \searrow b}
    {\lambda(t, \rho) \cdot a \searrow b}

    \inferrule[app-ne]
    {\intp{T}{\rho, a/x_0} \searrow B}
    {{\uparrow}^{\Pi(A, T, \rho)}(e) \cdot a \searrow
      {\uparrow}^B(e \; \$ \; {\downarrow}^A(a))}
  \end{mathpar}
  \caption{Application rules: $f \cdot a \searrow b$}
  \labeledit{fig:app}
\end{figure}

\begin{itemize}
\item \textsc{app-lam}: Applying a closure $\lambda(t, \rho)$ to an argument $a$
  extends the closure's environment with $a$ at index~$0$ and evaluates the body $t$.

\item \textsc{app-ne}: Applying a neutral ${\uparrow}^{\Pi(A, T, \rho)}(e)$ to an
  argument $a$ produces a new neutral.
  The codomain fiber is obtained by evaluating $T$ in $(\rho, a/x_0)$, yielding a
  type $B$.
  The result is ${\uparrow}^B(e \;\$\; {\downarrow}^A(a))$: a neutral at the codomain
  type, recording the application with the argument annotated by its domain type $A$.
\end{itemize}

\subsection{Natural Number Recursion}\labeledit{sec:sem:rec}

The recursion judgment $\mathsf{rec}(a, T, z, s, \rho) \searrow b$ defines the
semantic interpretation of the natural number eliminator.
Here $a \in \D$ is the evaluated scrutinee, $T \in \Exp$ is the motive expression,
$z \in \Exp$ is the zero-case expression, $s \in \Exp$ is the step-case expression
(binding the predecessor $x$ and the inductive hypothesis $y$), and $\rho \in \Env$
is the captured environment.
Crucially, $z$ and $s$ are \emph{syntax}---they are evaluated lazily inside the
recursion judgment, not pre-evaluated.

\begin{figure}[h]
  \begin{mathpar}
    \inferrule[rec-ze]
    {\intp{z}{\rho} \searrow b}
    {\mathsf{rec}(\zed, T, z, s, \rho) \searrow b}

    \inferrule[rec-su]
    {\mathsf{rec}(a, T, z, s, \rho) \searrow b \\
     \intp{s}{\rho, a/x, b/y} \searrow b'}
    {\mathsf{rec}(\sud{a}, T, z, s, \rho) \searrow b'}

    \inferrule[rec-ne]
    {\intp{z}{\rho} \searrow a_z \\
     \intp{T}{\rho, {\uparrow}^N(e)/x} \searrow A}
    {\mathsf{rec}({\uparrow}^N(e), T, z, s, \rho) \searrow
      {\uparrow}^A
      (\recd{T}{a_z}{s}{\rho}{e})}
  \end{mathpar}
  \caption{Recursion rules: $\mathsf{rec}(a, T, z, s, \rho) \searrow b$}
  \labeledit{fig:rec}
\end{figure}

\begin{itemize}
\item \textsc{rec-ze}: When the scrutinee is $\zed$, evaluate the zero case $z$ in
  $\rho$ and return the result.

\item \textsc{rec-su}: When the scrutinee is $\sud{a}$, first recursively compute
  the result $b$ for the predecessor $a$, then evaluate the step expression $s$ in
  the environment extended with $a$ (the predecessor) and $b$ (the inductive hypothesis).

\item \textsc{rec-ne}: When the scrutinee is neutral ${\uparrow}^N(e)$, computation is
  blocked.
  We evaluate the zero case to get $a_z$, evaluate the motive at the neutral scrutinee
  to get the result type $A$, and produce a neutral value ${\uparrow}^A(\ldots)$
  recording all components for later readback.
\end{itemize}

\subsection{Readback}\labeledit{sec:sem:readback}

Readback (also called \emph{reification}) extracts a syntactic normal form from a
domain value.
It is defined as three mutually recursive relations:
\begin{itemize}
\item $\Rnf_n({\downarrow}^A(a)) \searrow w$: read back the value $a$ at
  type $A$ to produce a normal form $w$, where $n \in \N$ is the current depth
  (the number of binders traversed so far, which determines the next fresh level).
\item $\Rty_n(A) \searrow W$: read back a domain type $A$ to produce a type
  normal form $W$.
\item $\Rne_n(e) \searrow u$: read back a neutral value $e$ to produce a
  neutral normal form $u$.
\end{itemize}

The level parameter $n$ serves a dual purpose:
\begin{enumerate}
\item It determines the next fresh variable: when we need to $\eta$-expand a function,
  we apply it to the ``generic'' value ${\uparrow}^A(\ell_n)$ and increment the
  depth to $n+1$.
\item It enables the conversion from de Bruijn levels to de Bruijn indices: a free
  variable at level $\ell$ is read back as the index $n - \ell - 1$, where $n$ is the
  current depth.
\end{enumerate}

The readback rules are given in \Cref{fig:readback}.

\begin{figure}[t]
  \begin{mathpar}
    \inferrule[Rf-Ty]
    {\Rty_n(A) \searrow W}
    {\Rnf_n({\downarrow}^{U_i}(A)) \searrow W}

    \inferrule[Rf-N-ze]
    { }
    {\Rnf_n({\downarrow}^N(\zed)) \searrow \ze}

    \inferrule[Rf-N-su]
    {\Rnf_n({\downarrow}^N(a)) \searrow w}
    {\Rnf_n({\downarrow}^N(\sud{a})) \searrow \su{w}}

    \inferrule[Rf-N-ne]
    {\Rne_n(e) \searrow u}
    {\Rnf_n({\downarrow}^N({\uparrow}^B(e))) \searrow u}

    \inferrule[Rf-Pi]
    {f \cdot {\uparrow}^A(\ell_n) \searrow b \\
     \intp{T}{\rho, {\uparrow}^A(\ell_n)/x} \searrow B \\
     \Rnf_{n+1}({\downarrow}^B(b)) \searrow w}
    {\Rnf_n({\downarrow}^{\Pi(A, T, \rho)}(f)) \searrow \lambda x.\, w}

    \inferrule[Rf-ne]
    {\Rne_n(e) \searrow u}
    {\Rnf_n({\downarrow}^{{\uparrow}^A(e')}({\uparrow}^B(e))) \searrow u}
    
    \inferrule[Rty-N]
    { }
    {\Rty_n(N) \searrow \Nat}

    \inferrule[Rty-U]
    { }
    {\Rty_n(U_i) \searrow \Ty i}

    \inferrule[Rty-Pi]
    {\Rty_n(a) \searrow W_S \\
     \intp{T}{\rho, {\uparrow}^a(\ell_n)/x} \searrow B \\
     \Rty_{n+1}(B) \searrow W_T}
    {\Rty_n(\Pi(a, T, \rho)) \searrow \Pi(x:W_S).\, W_T}

    \inferrule[Rty-ne]
    {\Rne_n(e) \searrow u}
    {\Rty_n({\uparrow}^A(e)) \searrow u}
    
    \inferrule[Re-var]
    { }
    {\Rne_n(\ell_k) \searrow x_{n - k - 1}}

    \inferrule[Re-app]
    {\Rne_n(e) \searrow u \\
     \Rnf_n(d) \searrow w}
    {\Rne_n(e \;\$\; d) \searrow u\; w}

    \inferrule[Re-rec]
    {\Rnf_n({\downarrow}^{A_z}(a_z)) \searrow w_z \\
     \intp{s}{\rho, {\uparrow}^N(\ell_n)/x, {\uparrow}^A(\ell_{n+1})/y} \searrow b_s \\
     \Rnf_{n+2}({\downarrow}^A(b_s)) \searrow w_s \\
     \Rne_n(e) \searrow u}
    {\Rne_n(\recd{T}{a_z}{s}{\rho}{e})
      \searrow \elimn{x.\,W_T}{w_z}{x,y.\,w_s}{u}}
  \end{mathpar}
  \caption{Readback rules: $\Rnf_n$ for normal forms, $\Rty_n$ for types, $\Rne_n$ for neutrals}
  \labeledit{fig:readback}
\end{figure}

The key rules are:
\begin{itemize}
\item \textsc{Rf-Ty}: Reading back at type $U_i$ delegates to the type readback $\Rty_n$.

\item \textsc{Rty-N}, \textsc{Rty-Pi}, \textsc{Rty-ne}: Type readback $\Rty_n$ inspects
  the domain type value directly.
  For $N$, it produces $\Nat$.
  For $\Pi(a, T, \rho)$, it reads back the domain $a$ as a type to get $W_S$,
  evaluates the codomain $T$ at a fresh generic ${\uparrow}^a(\ell_n)$ to get $B$,
  then reads back $B$ as a type with depth $n{+}1$ to get $W_T$.
  For a neutral ${\uparrow}^A(e)$, it reads back $e$ via $\Rne_n$.

\item \textsc{Rf-Pi}: The $\eta$-expansion rule for functions.
  To read back $f$ at type $\Pi(A, T, \rho)$, we apply $f$ to a fresh generic
  ${\uparrow}^A(\ell_n)$, evaluate the codomain at that generic to get the result
  type $B$, then read back the application result $b$ at type $B$ with depth $n{+}1$.
  The result is $\lambda x.\, w$.
  This applies to \emph{all} values at $\Pi$-type (including neutrals), enforcing
  $\eta$-long normal forms.

\item \textsc{Rf-ne}: A neutral value at a neutral type is read back via $\Rne$.
  Neutrals at function type are handled by \textsc{Rf-Pi} ($\eta$-expanded).

\item \textsc{Re-var}: A level $\ell_k$ becomes de Bruijn index $n - k - 1$.

\item \textsc{Re-app}: A neutral application reads back the head and argument.

\item \textsc{Re-rec}: A neutral recursion reads back the zero case, evaluates the
  step case at fresh generics and reads back all components, producing
  the syntactic eliminator.
\end{itemize}

\subsection{Weakening and Substitution Evaluations}\labeledit{sec:sem:subst-eval}

A weakening $\psi$ maps between variables. %
Viewing it as a special kind of substituions, we can write its evaluation relation as
\[
  \intp \psi \rho(x) := \rho(\psi(x))
\]
Namely $\intp \psi \rho$ queries from $\rho$ the value mapped by weakened variable
$\psi(x)$. 

A substitution $\sigma$ mapping variables to expressions can be evaluated in an
environment $\rho$ to produce a new environment $\rho'$.
We write $\intp{\sigma}{\rho} \searrow \rho'$ and define:
\[
  \intp{\sigma}{\rho} \searrow \rho' \quad \iff \quad
  \forall x.\; \intp{\sigma(x)}{\rho} \searrow \rho'(x)
\]
That is, the evaluation of a substitution is defined pointwise: the new environment
$\rho'$ assigns to each variable $x$ the result of evaluating $\sigma(x)$ in $\rho$.

In particular, we have the following relations:
\begin{lemma}
  The follow evaluations hold:
  \begin{itemize}
  \item $\intp{\iota(\psi)}\rho \searrow \intp{\psi}{\rho}$
  \item $\intp{\id}{\rho} \searrow \rho$,
  \item If $\intp{\sigma}{\rho} \searrow \rho'$ and $\intp{t}\rho \searrow a$, then
    $\intp{\sigma, t/x}\rho \searrow (\rho', a)$. 
  \end{itemize}
\end{lemma}
Indeed, the single substitution $[s/x]$ yields:
\[
  \intp{[s/x]}{\rho} \searrow (\rho, \intp{s}{\rho} / x)
\]

This substitution evaluation judgment is the bridge between the syntactic notion of
substitution (an operation on terms) and the semantic notion (a transformation of
environments). %
However, by its definition, we are \emph{not} able to derive the following desired
properties obtained by definition with explicit substitutions:
\begin{itemize}
\item $\intp{\sigma[\psi]}\rho = \intp{\sigma}{\intp{\psi}{\rho}}$,
\item $\intp{t[\sigma]}\rho = \intp{t}{\intp{\sigma}\rho}$, or
\item $\intp{\sigma \circ \sigma'}\rho = \intp{\sigma}{\intp{\sigma'}\rho}$. 
\end{itemize}
The ultimate reason is that with explicit substitutions, the operation is
\emph{delayed}, so that substitutions can occur during evaluation. %
With our current settings, unfortunately, weakenings and substitutions must eagerly
apply, because they are operations. %
Therefore, the equations above do not hold in general. %
Nevertheless, we can show that, these equations are in fact relations in the PER
model, hence giving us handles to understand substitutions through the lenses of evaluations. %
That said, these relations cannot be described before defining the PER model, so we
postpone the discussions in \Cref{sec:comp}.

\subsection{Determinism}\labeledit{sec:sem:det}

All the relations defined above --- evaluation, application, recursion, and readback ---
are deterministic (functional).

\begin{lemma}[Determinism]\labeledit{lem:det}
  \begin{enumerate}
  \item If $\intp{t}{\rho} \searrow a$ and $\intp{t}{\rho} \searrow a'$, then $a = a'$.
  \item If $f \cdot a \searrow b$ and $f \cdot a \searrow b'$, then $b = b'$.
  \item If $\mathsf{rec}(a, T, z, s, \rho) \searrow b$ and
    $\mathsf{rec}(a, T, z, s, \rho) \searrow b'$, then $b = b'$.
  \item If $\Rnf_n(d) \searrow w$ and $\Rnf_n(d) \searrow w'$,
    then $w = w'$.
  \item If $\Rty_n(A) \searrow W$ and $\Rty_n(A) \searrow W'$,
    then $W = W'$.
  \item If $\Rne_n(e) \searrow u$ and $\Rne_n(e) \searrow u'$,
    then $u = u'$.
  \end{enumerate}
\end{lemma}

\begin{proof}
  By mutual induction on the derivations.
  In each case, the conclusion follows because the rules are syntax-directed: no two
  rules apply to the same configuration.
\end{proof}

Determinism ensures that the NbE procedure, when it terminates, produces a unique
normal form.
This is essential for using NbE as a decision procedure for definitional equality.

\subsection{Initial Environment}\labeledit{sec:sem:init-env}

To normalize a term in context $\Gamma$, we need an appropriate initial environment
that provides generic values for each variable in the context.
The initial environment is constructed by induction on the context, assigning to each
variable a neutral value (reflecting the variable itself) at its declared type.

\begin{definition}[Initial Environment]\labeledit{def:init-env}
  Given a context $\Gamma$, the initial environment
  ${\uparrow}^\Gamma : \Env$ is defined by induction on $\Gamma$:
  \[
    \begin{array}{lcl}
      {\uparrow}^{\cdot}(x) & = & \zed \\[4pt]
      {\uparrow}^{\Gamma, x : T} & = &
        ({\uparrow}^\Gamma,\; {\uparrow}^{\intp{T}{{\uparrow}^\Gamma}}(\ell_{|\Gamma|}))
    \end{array}
  \]
  Each variable $x_i$ in $\Gamma$ is mapped to a neutral value ${\uparrow}^{A_i}(\ell_i)$
  reflecting $x_i$ at its evaluated type $A_i$.
  Here $|\Gamma|$ denotes the length of $\Gamma$.
\end{definition}

The initial environment assigns to the most recently bound variable $x$ a
neutral value ${\uparrow}^A(\ell_n)$ where $A$ is the interpretation of its
declared type $T$ and $n = |\Gamma|$ is its de Bruijn level.
Older variables are assigned inductively.

\subsection{The NbE Procedure}\labeledit{sec:sem:nbe}

With this setup, we can now define the full NbE procedure. %
\begin{definition}[NbE]\labeledit{def:nbe}
  Given $\typing t T$, we define
  \[
    \nbe^T_\Gamma(t) := \Rnf_{|\Gamma|}({\downarrow}^{\intp{T}{\uparrow^\Gamma}}(\intp{t}{\uparrow^\Gamma}))
  \]
\end{definition}

The fact that judgmentally equal terms yield the same normal form is the content of
the \emph{completeness theorem}, established in \Cref{sec:comp}. %
The existence of a result for every well-typed term is the content of the
\emph{soundness theorem}, established in \Cref{sec:sound}. %

\section{Completeness of NbE}\labeledit{sec:comp}

In this section, we construct a \emph{partial equivalence relation} (PER) model for \mltt
and use it to establish the completeness of normalization by evaluation: if two terms are
judgmentally equal, then they evaluate to semantically related values, which in turn read
back to the same normal form.

The PER model follows the approach of~\citet{DBLP:journals/pacmpl/0001GHP25}, using an
\emph{impredicative} formulation in $\mathrm{Prop}$.
The key idea is that the universe PER $\Uc_i$ is defined as a \emph{functional relation}:
a judgment $A \approx A' \in \Uc_i \searrow R$ simultaneously asserts that $A$ and $A'$
represent the ``same type at level~$i$'' and that $R$ is the PER on inhabitants of that
type.
This functional-relational style avoids the need for inductive-recursive definitions while
retaining full dependent-type support.

The key complication introduced by the operational substitution approach is that we
must additionally show that \emph{evaluation commutes with substitution}: evaluating a
substituted term $t[\sigma]$ directly in an environment $\rho$ yields an equivalent
semantic value as first evaluating $\sigma$ to obtain a semantic environment and then
evaluating $t$ in that environment, related by the PER model.  %
This property must be established as part of the fundamental theorem.

\subsection{Introduction: Completeness via PERs}\labeledit{sec:comp:intro}

The completeness theorem states:
\[
  \text{If } \tyequiv[\Gamma]{s}{t}{T}, \text{ then } \nbe_\Gamma^T(s) = \nbe_\Gamma^T(t).
\]
That is, judgmentally equal terms normalize to the \emph{same} normal form.
The proof strategy is standard:
\begin{enumerate}
\item Define a PER model that interprets each type as a binary relation on domain values.
\item Prove the \emph{realizability} theorem and show that related domain values
  are \emph{read back} to identical normal forms.
\item Prove the \emph{fundamental theorem}: every derivable judgment is validated by the model.
\item Instantiate the fundamental theorem at the initial environment to conclude.
\end{enumerate}

Following~\citet{DBLP:journals/pacmpl/0001GHP25}, we define the PER model as a
\emph{weak functional relation} valued in an impredicative logic akin to Rocq's
$\mathrm{Prop}$: the relation $A \approx A' \in \Uc_i \searrow R$ simultaneously
certifies that $A, A'$ are related domain types and \emph{returns} the PER $R$ on their
elements.

\subsection{Base PERs via Readback}\labeledit{sec:comp:base}

We begin by defining three base-level PERs that capture the behavior of the readback
functions.
Recall from \Cref{sec:sem} that $\Rnf$ and $\Rne$ denote the readback relations for
normal forms and neutral terms respectively, and that readback is parameterized by a
\emph{name supply} $n \in \N$ (a de~Bruijn level).

\begin{definition}[PER $\Nf$ --- Normal Forms]\labeledit{def:per-nf}
  Two annotated normal forms $d, d' \in \D_f$ are related in $\Nf$ if they read back to
  the same syntactic normal form at every name supply level:
  \[
    d \approx d' \in \Nf
    \;:=\;
    \forall\, n \in \N.\; \exists\, w.\;
    \Rnf_n(d) \searrow w \;\wedge\; \Rnf_n(d') \searrow w
  \]
\end{definition}

\begin{definition}[PER $\Ne$ --- Neutral Terms]\labeledit{def:per-ne}
  Two neutral domain values $e, e'$ are related in $\Ne$ if they read back to the same
  neutral normal form at every name supply level:
  \[
    e \approx e' \in \Ne
    \;:=\;
    \forall\, n \in \N.\; \exists\, u.\;
    \Rne_n(e) \searrow u \;\wedge\; \Rne_n(e') \searrow u
  \]
\end{definition}

\begin{definition}[PER $\textsf{Ty}$ --- Types]\labeledit{def:per-ty}
  Two domain values $A, A'$ representing types are related in $\textsf{Ty}$ if they
  read back to the same type normal form:
  \[
    A \approx A' \in \textsf{Ty}
    \;:=\;
    \forall\, n \in \N.\; \exists\, W.\;
    \Rty_n(A) \searrow W
    \;\wedge\;
    \Rty_n(A') \searrow W
  \]
\end{definition}

These three PERs are the observational backbone of the model: two values are related
precisely when they are observationally indistinguishable through readback.
The universal quantification over $n$ is essential for the function space case, where
fresh variables are introduced at arbitrary levels.

\begin{lemma}[Base PERs are PERs]\labeledit{lem:base-per}
  $\Nf$, $\Ne$, and $\textsf{Ty}$ are symmetric and transitive (hence PERs).
\end{lemma}
\begin{proof}
  Symmetry is immediate (swap the two conjuncts).
  Transitivity follows from determinism of readback (\Cref{lem:det}).

  \medskip\noindent
  \textbf{Case} $\Ne$.
  Suppose $e \approx e' \in \Ne$ and $e' \approx e'' \in \Ne$.
  \begin{align*}
    &\forall\, n.\; \exists\, u.\;
      \Rne_n(e) \searrow u \;\wedge\; \Rne_n(e') \searrow u
      && \text{(by first hypothesis)} \\
    &\forall\, n.\; \exists\, u'.\;
      \Rne_n(e') \searrow u' \;\wedge\; \Rne_n(e'') \searrow u'
      && \text{(by second hypothesis)} \\
    &u = u'
      && \text{(by determinism of $\Rne$, \Cref{lem:det})} \\
    &\Rne_n(e) \searrow u \;\wedge\; \Rne_n(e'') \searrow u
      && \text{(combining above)}
  \end{align*}

  \medskip\noindent
  The cases for $\Nf$ and $\textsf{Ty}$ are analogous, using determinism of
  $\Rnf$ and $\Rty$ respectively.
\end{proof}

\subsection{PER for Neutral Domain Values}\labeledit{sec:comp:neu}

The PER $\textsf{Neu}$ relates neutral domain values ${\uparrow}^A(e)$ by ignoring the
type annotation and comparing only the underlying neutral terms via $\Ne$.

\begin{definition}[PER $\textsf{Neu}$]\labeledit{def:per-neu}
  \[
    {\uparrow}^A(e) \approx {\uparrow}^{A'}(e') \in \textsf{Neu}
    \;:=\;
    e \approx e' \in \Ne
  \]
\end{definition}

The key property of $\textsf{Neu}$ is that it erases type annotations: two neutrals are
related regardless of the types they carry, as long as their underlying neutral terms
read back identically.
This is sound because during readback at base type, only the neutral structure matters.

\subsection{PER for Natural Numbers}\labeledit{sec:comp:nat}

\begin{definition}[PER $\textsf{Nat}$]\labeledit{def:per-nat}
  The relation $a \approx a' \in \textsf{Nat}$ is defined inductively by:
  \begin{mathpar}
    \inferrule[\textsc{Nat-Ze}]
    { }
    {\zed \approx \zed \in \textsf{Nat}}
    \and
    \inferrule[\textsc{Nat-Su}]
    {a \approx a' \in \textsf{Nat}}
    {\sud{a} \approx \sud{a'} \in \textsf{Nat}}
    \and
    \inferrule[\textsc{Nat-Ne}]
    {e \approx e' \in \Ne}
    {{\uparrow}^N(e) \approx {\uparrow}^N(e') \in \textsf{Nat}}
  \end{mathpar}
\end{definition}

The first two constructors handle the canonical forms.
The third constructor embeds neutral values into the PER: stuck computations
(e.g., a variable of type $\Nat$) are related whenever their underlying neutral terms
are related in $\Ne$.

\begin{lemma}[$\textsf{Nat}$ is a PER]\labeledit{lem:nat-per}
  $\textsf{Nat}$ is symmetric and transitive.
\end{lemma}
\begin{proof}
  By induction on the derivation.

  \medskip\noindent
  \textbf{Case}
  \begin{mathpar}
    \inferrule[\textsc{Nat-Ze}]
    { }
    {\zed \approx \zed \in \textsf{Nat}}
  \end{mathpar}
  Symmetry and transitivity are immediate.

  \medskip\noindent
  \textbf{Case}
  \begin{mathpar}
    \inferrule[\textsc{Nat-Su}]
    {a \approx a' \in \textsf{Nat}}
    {\sud{a} \approx \sud{a'} \in \textsf{Nat}}
  \end{mathpar}
  \begin{align*}
    &a' \approx a \in \textsf{Nat}
      && \text{(by IH, symmetry)} \\
    &\sud{a'} \approx \sud{a} \in \textsf{Nat}
      && \text{(by \textsc{Nat-Su})}
  \end{align*}
  Transitivity: given additionally $\sud{a'} \approx \sud{a''} \in \textsf{Nat}$
  (hence $a' \approx a'' \in \textsf{Nat}$), by the IH $a \approx a'' \in \textsf{Nat}$,
  so $\sud{a} \approx \sud{a''} \in \textsf{Nat}$.

  \medskip\noindent
  \textbf{Case}
  \begin{mathpar}
    \inferrule[\textsc{Nat-Ne}]
    {e \approx e' \in \Ne}
    {{\uparrow}^N(e) \approx {\uparrow}^N(e') \in \textsf{Nat}}
  \end{mathpar}
  \begin{align*}
    &e' \approx e \in \Ne
      && \text{(by symmetry of $\Ne$, \Cref{lem:base-per})} \\
    &{\uparrow}^N(e') \approx {\uparrow}^N(e) \in \textsf{Nat}
      && \text{(by \textsc{Nat-Ne})}
  \end{align*}
  Transitivity follows analogously from transitivity of $\Ne$.
\end{proof}

\begin{lemma}[$\textsf{Nat}$ reads back correctly]\labeledit{lem:nat-readback}
  If $a \approx a' \in \textsf{Nat}$, then
  ${\downarrow}^N(a) \approx {\downarrow}^N(a') \in \Nf$.
\end{lemma}
\begin{proof}
  By induction on the derivation of $a \approx a' \in \textsf{Nat}$.

  \medskip\noindent
  \textbf{Case}
  \begin{mathpar}
    \inferrule[\textsc{Nat-Ze}]
    { }
    {\zed \approx \zed \in \textsf{Nat}}
  \end{mathpar}
  Both sides read back to $\ze$ at every level $n$, so
  ${\downarrow}^N(\zed) \approx {\downarrow}^N(\zed) \in \Nf$.

  \medskip\noindent
  \textbf{Case}
  \begin{mathpar}
    \inferrule[\textsc{Nat-Su}]
    {a \approx a' \in \textsf{Nat}}
    {\sud{a} \approx \sud{a'} \in \textsf{Nat}}
  \end{mathpar}
  \begin{align*}
    &{\downarrow}^N(a) \approx {\downarrow}^N(a') \in \Nf
      && \text{(by IH)} \\
    &\forall\, n.\; \exists\, w.\;
      \Rnf_n({\downarrow}^N(a)) \searrow w \;\wedge\;
      \Rnf_n({\downarrow}^N(a')) \searrow w
      && \text{(by definition of $\Nf$)} \\
    &\Rnf_n({\downarrow}^N(\sud{a})) \searrow \su{w} \;\wedge\;
      \Rnf_n({\downarrow}^N(\sud{a'})) \searrow \su{w}
      && \text{(by \textsc{Rf-su})}
  \end{align*}

  \medskip\noindent
  \textbf{Case}
  \begin{mathpar}
    \inferrule[\textsc{Nat-Ne}]
    {e \approx e' \in \Ne}
    {{\uparrow}^N(e) \approx {\uparrow}^N(e') \in \textsf{Nat}}
  \end{mathpar}
  \begin{align*}
    &e \approx e' \in \Ne
      && \text{(by hypothesis)} \\
    &\forall\, n.\; \exists\, u.\;
      \Rne_n(e) \searrow u \;\wedge\; \Rne_n(e') \searrow u
      && \text{(by definition of $\Ne$)} \\
    &\Rnf_n({\downarrow}^N({\uparrow}^N(e))) \searrow u \;\wedge\;
      \Rnf_n({\downarrow}^N({\uparrow}^N(e'))) \searrow u
      && \text{(by \textsc{Rf-ne})}
  \end{align*}
\end{proof}

\subsection{Universe PER $\Uc_i$}\labeledit{sec:comp:univ}

We now define the central component of the PER model: the universe PER at level~$i$.
Following~\citet{DBLP:journals/pacmpl/0001GHP25}, the universe PER is a
\emph{functional relation} in $\mathrm{Prop}$.
The judgment
\[
  A \approx A' \in \Uc_i \searrow R
\]
simultaneously asserts:
\begin{enumerate}
\item $A$ and $A'$ represent the ``same type at level $i$,'' and
\item $R : \D \to \D \to \mathrm{Prop}$ is the PER on elements of that type.
\end{enumerate}

We first define \emph{relational equivalence}: two binary relations $R, R'$ on $\D$
are equivalent, written $R \Leftrightarrow R'$, when
\[
  R \Leftrightarrow R' \;:=\; \forall\, a, a'.\; R(a, a') \leftrightarrow R'(a, a')
\]

\begin{definition}[Universe PER $\Uc_i$]\labeledit{def:per-univ}
  The relation $A \approx A' \in \Uc_i \searrow R$ is defined inductively by the following
  four cases.
  The returned $R$ is unique only up to $\Leftrightarrow$ (not definitional equality).
  \begin{mathpar}
    \inferrule[\textsc{U-Ne}]
    {b \approx b' \in \Ne \\ R \Leftrightarrow \textsf{Neu}}
    {{\uparrow}^A(b) \approx {\uparrow}^{A'}(b') \in \Uc_i
      \searrow R}

    \inferrule[\textsc{U-Univ}]
    {j < i \\ R \Leftrightarrow \Uc_j}
    {U_j \approx U_j \in \Uc_i \searrow R}

    \inferrule[\textsc{U-Nat}]
    {R \Leftrightarrow \textsf{Nat}}
    {N \approx N \in \Uc_i \searrow R}

    \inferrule[\textsc{U-Pi}]
    {A \approx A' \in \Uc_i \searrow R_I \\
     R_I \text{ is a PER} \\\\
     \forall\, \Pf :: a \approx a' \in R_I.\;
       \intp{T}{\rho, a/x} \approx \intp{T'}{\rho', a'/x} \in \Uc_i \searrow R_O(\Pf) \\\\
     \forall\, f, f'.\; f \approx f' \in R \;\leftrightarrow\;
       (\forall\, \Pf :: a \approx a' \in R_I.\;
        f \cdot a \approx f' \cdot a' \in R_O(\Pf))}
    {\Pi(A, T, \rho) \approx \Pi(A', T', \rho') \in \Uc_i \searrow R}
  \end{mathpar}
  In the $\textsc{U-Pi}$ case, $\Pi(A, T, \rho)$ is the closure form: the codomain
  fiber at input $b$ is obtained by evaluating $T$ in $(\rho, b/x)$.
  The notation $\Pf :: a \approx a' \in R_I$ means that $\Pf$ is the proof/evidence
  that $a \approx a' \in R_I$ holds; $R_O(\Pf)$ may depend on this evidence.
  The fourth premise characterizes $R$ as the PER on functions: $f, f'$ are
  related in $R$ if and only if applying them to related inputs yields related outputs.
  We write $f \cdot a$ for semantic application (i.e., there exists $c$ with
  $f \cdot a \searrow c$, and $c$ participates in $R_O(\Pf)$).
\end{definition}

Let us explain each case:
\begin{itemize}
\item \textsc{U-Ne}: A neutral type code ${\uparrow}^A(e)$ is related to another
  ${\uparrow}^{A'}(e')$ whenever the underlying neutrals are related in $\Ne$.
  The type annotations $A, A'$ are arbitrary and play no role.
  The returned PER is $\textsf{Neu}$: elements of an unknown (neutral) type are related
  when they are neutrals that read back identically.

\item \textsc{U-Univ}: The universe $U_j$ (with $j < i$) is related to itself.
  The returned PER says that two values are related as elements of $U_j$ when they are
  themselves related type codes at level $j$.
  The ``element PER'' for a universe is defined in terms
  of the universe PER at a lower level.

\item \textsc{U-Nat}: The natural number type $N$ is related to itself, returning the
  PER $\textsf{Nat}$ defined in \Cref{def:per-nat}.

\item \textsc{U-Pi}: A $\Pi$-type $\Pi(A, T, \rho)$ is related to $\Pi(A', T', \rho')$
  when:
  (i)~the domains are related in $\Uc_i$ returning some PER $R_I$ which is itself a PER,
  (ii)~for all evidence $\Pf :: a \approx a' \in R_I$, the codomain evaluations
  $\intp{T}{\rho, a/x}$ and $\intp{T'}{\rho', a'/x}$ are related in $\Uc_i$
  returning some PER $R_O(\Pf)$, and
  (iii)~the returned $R$ is characterized by the biconditional: $f \approx f' \in R$
  iff applying to all related inputs yields related outputs in the codomain PER.
\end{itemize}

\begin{remark}
  Several points deserve emphasis:
  \begin{enumerate}
  \item The returned $R$ in each case is specified only up to $\Leftrightarrow$.
    This makes $\Uc_i$ a \emph{weak} partial functional relation, avoiding the need
    for propositional or functional extensionality.
  \item The premise ``$R_I$ is a PER'' in $\textsc{U-Pi}$ strengthens the
    induction principle: it allows proving that returned PERs are PERs without
    extra axioms.
    Once symmetry and transitivity of the returned PER are established
    (see below), the premise becomes logically
    redundant but is needed for the induction to go through.
  \end{enumerate}
\end{remark}

\subsection{Cumulativity}\labeledit{sec:comp:cumul}

\begin{lemma}[Cumulativity]\labeledit{lem:cumul}
  If $A \approx A' \in \Uc_i \searrow R$ and $i \le j$, then
  $A \approx A' \in \Uc_j \searrow R$.
\end{lemma}
\begin{proof}
  By induction on the derivation of $A \approx A' \in \Uc_i \searrow R$.

  \medskip\noindent
  \textbf{Case}
  \begin{mathpar}
    \inferrule[\textsc{U-Ne}]
    {b \approx b' \in \Ne \\ R \Leftrightarrow \textsf{Neu}}
    {{\uparrow}^A(b) \approx {\uparrow}^{A'}(b') \in \Uc_i \searrow R}
  \end{mathpar}
  The same derivation works at level $j$ (neutrals are level-independent):
  ${\uparrow}^A(b) \approx {\uparrow}^{A'}(b') \in \Uc_j \searrow R$ by \textsc{U-Ne}.

  \medskip\noindent
  \textbf{Case}
  \begin{mathpar}
    \inferrule[\textsc{U-Univ}]
    {k < i \\ R \Leftrightarrow \Uc_k}
    {U_k \approx U_k \in \Uc_i \searrow R}
  \end{mathpar}
  \begin{align*}
    &k < i \le j
      && \text{(by hypothesis)} \\
    &U_k \approx U_k \in \Uc_j \searrow R
      && \text{(by \textsc{U-Univ}, since $k < j$)}
  \end{align*}

  \medskip\noindent
  \textbf{Case}
  \begin{mathpar}
    \inferrule[\textsc{U-Nat}]
    {R \Leftrightarrow \textsf{Nat}}
    {N \approx N \in \Uc_i \searrow R}
  \end{mathpar}
  Level-independent: $N \approx N \in \Uc_j \searrow R$ by \textsc{U-Nat}.

  \medskip\noindent
  \textbf{Case}
  \begin{mathpar}
    \inferrule[\textsc{U-Pi}]
    {A \approx A' \in \Uc_i \searrow R_I \\
     R_I \text{ is a PER} \\\\
     \forall\, \Pf :: a \approx a' \in R_I.\;
       \intp{T}{\rho, a/x} \approx \intp{T'}{\rho', a'/x} \in \Uc_i \searrow R_O(\Pf) \\\\
     \forall\, f, f'.\; f \approx f' \in R \;\leftrightarrow\;
       (\forall\, \Pf :: a \approx a' \in R_I.\;
        f \cdot a \approx f' \cdot a' \in R_O(\Pf))}
    {\Pi(A, T, \rho) \approx \Pi(A', T', \rho') \in \Uc_i \searrow R}
  \end{mathpar}
  \begin{align*}
    &A \approx A' \in \Uc_j \searrow R_I
      && \text{(by IH on domain)} \\
    &\forall\, \Pf :: a \approx a' \in R_I.\;
      \intp{T}{\rho, a/x} \approx \intp{T'}{\rho', a'/x} \in \Uc_j \searrow R_O(\Pf)
      && \text{(by IH on codomain)} \\
    &\Pi(A, T, \rho) \approx \Pi(A', T', \rho') \in \Uc_j \searrow R
      && \text{(by \textsc{U-Pi})}
  \end{align*}
\end{proof}

\subsection{Semantic Subtyping}\labeledit{sec:comp:subtyping}

Semantic subtyping at level $i$ is defined as an inductive relation
$A \mathrel{<:_i} A'$ that mirrors the syntactic subtyping judgment.

\begin{definition}[Semantic Subtyping $<:_i$]\labeledit{def:sem-subty}
  The relation $A \mathrel{<:_i} A'$ is defined inductively by:
  \begin{mathpar}
    \inferrule[\textsc{$<$:-Ne}]
    {b \approx b' \in \Ne}
    {{\uparrow}^A(b) \mathrel{<:_i} {\uparrow}^{A'}(b')}

    \inferrule[\textsc{$<$:-Univ}]
    {j \le k \\ k < i}
    {U_j \mathrel{<:_i} U_k}

    \inferrule[\textsc{$<$:-Nat}]
    { }
    {N \mathrel{<:_i} N}

    \inferrule[\textsc{$<$:-Pi}]
    {A \approx A' \in \Uc_i \searrow R_I \\
     \forall\, \Pf :: a \approx a' \in R_I.\;
       \intp{T}{\rho, a/x} \mathrel{<:_i} \intp{T'}{\rho', a'/x} \\\\
     \Pi(A, T, \rho) \approx \Pi(A, T, \rho) \in \Uc_i \\
     \Pi(A', T', \rho') \approx \Pi(A', T', \rho') \in \Uc_i}
    {\Pi(A, T, \rho) \mathrel{<:_i} \Pi(A', T', \rho')}
  \end{mathpar}
\end{definition}

The key properties of semantic subtyping:
\begin{itemize}
\item \textsc{$<$:-Ne}: Neutral types are subtypes when their neutrals are related.
\item \textsc{$<$:-Univ}: $U_j \mathrel{<:_i} U_k$ whenever $j \le k < i$, reflecting
  cumulative universes.
\item \textsc{$<$:-Nat}: $N$ is a subtype of itself (natural numbers are invariant).
\item \textsc{$<$:-Pi}: $\Pi$-types are covariant in the codomain: the domains must
  be equal (related in $\Uc_i$) while the codomain fibers satisfy subtyping
  pointwise for all related inputs.
\end{itemize}

\subsection{Properties of $\Uc_i$}\labeledit{sec:comp:univ-props}

We establish the key structural properties of the universe PER.

\subsubsection{$\Uc_i$ is a PER (Functionality)}

\begin{lemma}[Right irrelevance of $\Uc_i$]\labeledit{lem:univ-functional}
  If $A \approx B \in \Uc_i \searrow R$ and $A \approx B' \in \Uc_j \searrow R'$,
  then $R \Leftrightarrow R'$.
\end{lemma}
\begin{proof}
  By induction on the two derivations simultaneously.
  The cases are forced: $N$ can only match $\textsc{U-Nat}$, a neutral can only match
  $\textsc{U-Ne}$, etc.

  \medskip\noindent
  \textbf{Case} $\textsc{U-Ne}$.
  \begin{mathpar}
    \inferrule[\textsc{U-Ne}]
    {b \approx b' \in \Ne \\ R \Leftrightarrow \textsf{Neu}}
    {{\uparrow}^A(b) \approx {\uparrow}^{A'}(b') \in \Uc_i \searrow R}
  \end{mathpar}
  Both derivations force $R \Leftrightarrow \textsf{Neu}$ and
  $R' \Leftrightarrow \textsf{Neu}$, so $R \Leftrightarrow R'$.

  \medskip\noindent
  \textbf{Case} $\textsc{U-Nat}$.
  Both derivations force $R \Leftrightarrow \textsf{Nat}$ and
  $R' \Leftrightarrow \textsf{Nat}$, so $R \Leftrightarrow R'$.

  \medskip\noindent
  \textbf{Case} $\textsc{U-Univ}$.
  Both derivations force $R \Leftrightarrow \Uc_j$ and
  $R' \Leftrightarrow \Uc_j$, so $R \Leftrightarrow R'$.

  \medskip\noindent
  \textbf{Case} $\textsc{U-Pi}$.
  \begin{mathpar}
    \inferrule[\textsc{U-Pi}]
    {A \approx A' \in \Uc_i \searrow R_I \\
     R_I \text{ is a PER} \\\\
     \forall\, \Pf :: a \approx a' \in R_I.\;
       \intp{T}{\rho, a/x} \approx \intp{T'}{\rho', a'/x} \in \Uc_i \searrow R_O(\Pf) \\\\
     \forall\, f, f'.\; f \approx f' \in R \;\leftrightarrow\;
       (\forall\, \Pf :: a \approx a' \in R_I.\;
        f \cdot a \approx f' \cdot a' \in R_O(\Pf))}
    {\Pi(A, T, \rho) \approx \Pi(A', T', \rho') \in \Uc_i \searrow R}
  \end{mathpar}
  From the second derivation we get $R_I'$ with
  $A \approx A' \in \Uc_i \searrow R_I'$.
  \begin{align*}
    &R_I \Leftrightarrow R_I'
      && \text{(by IH on domain)} \\
    &\forall\, \Pf :: a \approx a' \in R_I.\;
      R_O(\Pf) \Leftrightarrow R_O'(\Pf)
      && \text{(by IH on codomain at each input)} \\
    &R \Leftrightarrow R'
      && \text{(by the biconditional characterizations and transitivity of $\Leftrightarrow$)}
  \end{align*}
\end{proof}

\begin{lemma}[Symmetry of $\Uc_i$]\labeledit{lem:univ-sym}
  If $A \approx A' \in \Uc_i \searrow R$, then:
  \begin{enumerate}
  \item $A' \approx A \in \Uc_i \searrow R$, and
  \item $a \approx a' \in R$ implies $a' \approx a \in R$.
  \end{enumerate}
\end{lemma}
\begin{proof}
  By induction on the derivation.  We prove both statements simultaneously.

  \medskip\noindent
  \textbf{Case}
  \begin{mathpar}
    \inferrule[\textsc{U-Ne}]
    {b \approx b' \in \Ne \\ R \Leftrightarrow \textsf{Neu}}
    {{\uparrow}^A(b) \approx {\uparrow}^{A'}(b') \in \Uc_i \searrow R}
  \end{mathpar}
  For (1):
  \begin{align*}
    &b' \approx b \in \Ne
      && \text{(by symmetry of $\Ne$)} \\
    &{\uparrow}^{A'}(b') \approx {\uparrow}^A(b) \in \Uc_i \searrow R
      && \text{(by \textsc{U-Ne})}
  \end{align*}
  For (2): Suppose $a \approx a' \in R$, i.e.\ $a \approx a' \in \textsf{Neu}$
  (since $R \Leftrightarrow \textsf{Neu}$).
  Then $a = {\uparrow}^B(e)$ and $a' = {\uparrow}^{B'}(e')$ with $e \approx e' \in \Ne$.
  By symmetry of $\Ne$, $e' \approx e \in \Ne$, so $a' \approx a \in \textsf{Neu}$,
  i.e.\ $a' \approx a \in R$.

  \medskip\noindent
  \textbf{Case}
  \begin{mathpar}
    \inferrule[\textsc{U-Univ}]
    {j < i \\ R \Leftrightarrow \Uc_j}
    {U_j \approx U_j \in \Uc_i \searrow R}
  \end{mathpar}
  For (1): Immediate ($U_j \approx U_j$ is symmetric).

  For (2): Suppose $a \approx a' \in R$, i.e.\ $a \approx a' \in \Uc_j \searrow R_0$
  for some $R_0$.
  By the IH at level $j$ (which is structurally smaller), $a' \approx a \in \Uc_j \searrow R_0$,
  so $a' \approx a \in R$.

  \medskip\noindent
  \textbf{Case}
  \begin{mathpar}
    \inferrule[\textsc{U-Nat}]
    {R \Leftrightarrow \textsf{Nat}}
    {N \approx N \in \Uc_i \searrow R}
  \end{mathpar}
  For (1): Immediate.

  For (2): Suppose $a \approx a' \in R$, i.e.\ $a \approx a' \in \textsf{Nat}$.
  By symmetry of $\textsf{Nat}$ (\Cref{lem:nat-per}), $a' \approx a \in \textsf{Nat}$,
  so $a' \approx a \in R$.

  \medskip\noindent
  \textbf{Case}
  \begin{mathpar}
    \inferrule[\textsc{U-Pi}]
    {A \approx A' \in \Uc_i \searrow R_I \\
     R_I \text{ is a PER} \\\\
     \forall\, \Pf :: a \approx a' \in R_I.\;
       \intp{T}{\rho, a/x} \approx \intp{T'}{\rho', a'/x} \in \Uc_i \searrow R_O(\Pf) \\\\
     \forall\, f, f'.\; f \approx f' \in R \;\leftrightarrow\;
       (\forall\, \Pf :: a \approx a' \in R_I.\;
        f \cdot a \approx f' \cdot a' \in R_O(\Pf))}
    {\Pi(A, T, \rho) \approx \Pi(A', T', \rho') \in \Uc_i \searrow R}
  \end{mathpar}
  For (1): We construct a \textsc{U-Pi} derivation of
  $\Pi(A', T', \rho') \approx \Pi(A, T, \rho) \in \Uc_i \searrow R$.
  We verify the four premises.

  \emph{Premise 1} (domain):
  \begin{align*}
    &A' \approx A \in \Uc_i \searrow R_I
      && \text{(by IH (1) on domain; same $R_I$ by \Cref{lem:univ-functional})}
  \end{align*}

  \emph{Premise 2}: $R_I$ is a PER (by hypothesis).

  \emph{Premise 3} (codomain): Let $\Pf' :: a' \approx a \in R_I$.
  \begin{align*}
    &a \approx a' \in R_I
      && \text{(by symmetry of $R_I$; call this $\D_0$)} \\
    &\intp{T}{\rho, a/x} \approx \intp{T'}{\rho', a'/x} \in \Uc_i \searrow R_O(\D_0)
      && \text{(original codomain premise at $\D_0$)} \\
    &\intp{T'}{\rho', a'/x} \approx \intp{T}{\rho, a/x} \in \Uc_i \searrow R_O(\D_0)
      && \text{(by IH (1) on codomain; same $R_O(\D_0)$ returned)}
  \end{align*}
  Define $R_O'(\Pf') := R_O(\D_0)$.

  \emph{Premise 4} (biconditional): We must show
  \[
    \forall\, f, f'.\; f \approx f' \in R \;\leftrightarrow\;
    (\forall\, \Pf' :: a' \approx a \in R_I.\; f \cdot a' \approx f' \cdot a \in R_O'(\Pf'))
  \]
  Since $R_I$ is a PER, the set of pairs $\{(a, a') \mid a \approx a' \in R_I\}$ is
  closed under swapping: $(a, a') \in R_I$ iff $(a', a) \in R_I$.
  By right irrelevance (\Cref{lem:univ-functional}), $R_O(\Pf)$ depends only on the pair
  $(a, a')$, not the specific evidence $\Pf$.
  Hence the right-hand side
  ``$\forall\, \Pf' :: a' \approx a \in R_I.\; f \cdot a' \approx f' \cdot a \in R_O'(\Pf')$''
  is, after renaming bound variables $(a' \mapsto b, a \mapsto b')$, equivalent to
  ``$\forall\, \Pf'':: b \approx b' \in R_I.\; f \cdot b \approx f' \cdot b' \in R_O(\Pf'')$''
  which is exactly the original right-hand side.
  Therefore the same $R$ satisfies both biconditionals.
  \begin{align*}
    &\Pi(A', T', \rho') \approx \Pi(A, T, \rho) \in \Uc_i \searrow R
      && \text{(by \textsc{U-Pi})}
  \end{align*}

  \medskip
  For (2): Suppose $f \approx f' \in R$.
  We must show $f' \approx f \in R$, i.e.\ for all $\Pf :: a \approx a' \in R_I$,
  $f' \cdot a \approx f \cdot a' \in R_O(\Pf)$.
  \begin{align*}
    &a' \approx a \in R_I
      && \text{(by symmetry of $R_I$)} \\
    &\text{let } \Pf' :: a' \approx a \in R_I \\
    &f \cdot a' \approx f' \cdot a \in R_O(\Pf')
      && \text{(from $f \approx f' \in R$, applied to $\Pf'$)} \\
    &R_O(\Pf) \Leftrightarrow R_O(\Pf')
      && \text{(by \Cref{lem:univ-functional}: both have same left type $\intp{T}{\rho, a/x}$)} \\
    &f' \cdot a \approx f \cdot a' \in R_O(\Pf')
      && \text{(by IH (2) on $R_O(\Pf')$)} \\
    &f' \cdot a \approx f \cdot a' \in R_O(\Pf)
      && \text{(by $R_O(\Pf) \Leftrightarrow R_O(\Pf')$)}
  \end{align*}
\end{proof}

\begin{lemma}[Transitivity of $\Uc_i$]\labeledit{lem:univ-trans}
  If $A_1 \approx A_2 \in \Uc_i \searrow R$, then:
  \begin{enumerate}
  \item $A_2 \approx A_3 \in \Uc_j \searrow R'$ implies
    $A_1 \approx A_3 \in \Uc_i \searrow R$, and
  \item $a_1 \approx a_2 \in R$ and $a_2 \approx a_3 \in R$ implies
    $a_1 \approx a_3 \in R$.
  \end{enumerate}
\end{lemma}
\begin{proof}
  By induction on the two derivations.
  We must case-split on both; by the structure of the definition, matching cases are forced
  (e.g., $N$ can only relate to $N$).

  \medskip\noindent
  \textbf{Case} $\textsc{U-Ne}$.
  \begin{mathpar}
    \inferrule[\textsc{U-Ne}]
    {b \approx b' \in \Ne \\ R \Leftrightarrow \textsf{Neu}}
    {{\uparrow}^A(b) \approx {\uparrow}^{A'}(b') \in \Uc_i \searrow R}
  \end{mathpar}
  and
  \begin{mathpar}
    \inferrule[\textsc{U-Ne}]
    {b' \approx b'' \in \Ne \\ R' \Leftrightarrow \textsf{Neu}}
    {{\uparrow}^{A'}(b') \approx {\uparrow}^{A''}(b'') \in \Uc_i \searrow R'}
  \end{mathpar}
  \begin{align*}
    &b \approx b'' \in \Ne
      && \text{(by transitivity of $\Ne$)} \\
    &{\uparrow}^A(b) \approx {\uparrow}^{A''}(b'') \in \Uc_i \searrow R
      && \text{(by \textsc{U-Ne})} \\
    &R \Leftrightarrow \textsf{Neu} \Leftrightarrow R'
      && \text{(hence $R \Leftrightarrow R'$)}
  \end{align*}

  \medskip\noindent
  \textbf{Case} $\textsc{U-Nat}$ and $\textsc{U-Univ}$: immediate.

  \medskip\noindent
  \textbf{Case} $\textsc{U-Pi}$.
  The first derivation is:
  \begin{mathpar}
    \inferrule[\textsc{U-Pi}]
    {A_1 \approx A_2 \in \Uc_i \searrow R_I \\
     R_I \text{ is a PER} \\\\
     \forall\, \Pf :: a \approx a' \in R_I.\;
       \intp{T_1}{\rho_1, a/x} \approx \intp{T_2}{\rho_2, a'/x} \in \Uc_i \searrow R_O(\Pf) \\\\
     \forall\, f, f'.\; f \approx f' \in R \;\leftrightarrow\;
       (\forall\, \Pf :: a \approx a' \in R_I.\;
        f \cdot a \approx f' \cdot a' \in R_O(\Pf))}
    {\Pi(A_1, T_1, \rho_1) \approx \Pi(A_2, T_2, \rho_2) \in \Uc_i \searrow R}
  \end{mathpar}
  The second derivation is:
  \begin{mathpar}
    \inferrule[\textsc{U-Pi}]
    {A_2 \approx A_3 \in \Uc_j \searrow R_I' \\
     R_I' \text{ is a PER} \\\\
     \forall\, \Pf' :: a \approx a' \in R_I'.\;
       \intp{T_2}{\rho_2, a/x} \approx \intp{T_3}{\rho_3, a'/x} \in \Uc_j \searrow R_O'(\Pf') \\\\
     \forall\, f, f'.\; f \approx f' \in R' \;\leftrightarrow\;
       (\forall\, \Pf' :: a \approx a' \in R_I'.\;
        f \cdot a \approx f' \cdot a' \in R_O'(\Pf'))}
    {\Pi(A_2, T_2, \rho_2) \approx \Pi(A_3, T_3, \rho_3) \in \Uc_j \searrow R'}
  \end{mathpar}

  For (1): We construct a \textsc{U-Pi} derivation of
  $\Pi(A_1, T_1, \rho_1) \approx \Pi(A_3, T_3, \rho_3) \in \Uc_i \searrow R$.

  \emph{Premise 1} (domain):
  \begin{align*}
    &R_I \Leftrightarrow R_I'
      && \text{(by \Cref{lem:univ-functional}: both have left type $A_2$)} \\
    &A_1 \approx A_3 \in \Uc_i \searrow R_I
      && \text{(by IH (1) on domain)}
  \end{align*}

  \emph{Premise 2}: $R_I$ is a PER (by hypothesis).

  \emph{Premise 3} (codomain): Let $\Pf :: a \approx a' \in R_I$.
  We must show $\intp{T_1}{\rho_1, a/x} \approx \intp{T_3}{\rho_3, a'/x} \in \Uc_i \searrow R_O''(\Pf)$
  for some $R_O''$.
  \begin{align*}
    &a \approx a \in R_I
      && \text{(by symmetry + transitivity of $R_I$; let this be $\D_1$)} \\
    &\intp{T_1}{\rho_1, a/x} \approx \intp{T_2}{\rho_2, a/x} \in \Uc_i \searrow R_O(\D_1)
      && \text{(from first codomain at $\D_1$)} \\
    &a \approx a' \in R_I'
      && \text{(by $R_I \Leftrightarrow R_I'$; let evidence in $R_I'$ be $\Pf'$)} \\
    &\intp{T_2}{\rho_2, a/x} \approx \intp{T_3}{\rho_3, a'/x} \in \Uc_j \searrow R_O'(\Pf')
      && \text{(from second codomain at $\Pf'$)} \\
    &\intp{T_1}{\rho_1, a/x} \approx \intp{T_3}{\rho_3, a'/x} \in \Uc_i \searrow R_O(\D_1)
      && \text{(by IH (1) on codomain)}
  \end{align*}
  Set $R_O''(\Pf) := R_O(\D_1)$.

  \emph{Premise 4} (biconditional): We must show
  $f \approx f' \in R \;\leftrightarrow\;
  (\forall\, \Pf :: a \approx a' \in R_I.\; f \cdot a \approx f' \cdot a' \in R_O''(\Pf))$.
  Since $R_O''(\Pf) = R_O(\D_1)$ and by right irrelevance $R_O(\D_1) \Leftrightarrow R_O(\Pf)$
  (both have the same left type $\intp{T_1}{\rho_1, a/x}$), the right-hand side is equivalent
  to the original biconditional for $R$.  Hence the same $R$ works.
  \begin{align*}
    &\Pi(A_1, T_1, \rho_1) \approx \Pi(A_3, T_3, \rho_3) \in \Uc_i \searrow R
      && \text{(by \textsc{U-Pi})}
  \end{align*}

  For (2): Suppose $f_1 \approx f_2 \in R$ and $f_2 \approx f_3 \in R$.
  We must show $f_1 \approx f_3 \in R$, i.e.\ for all $\Pf :: a \approx a' \in R_I$,
  $f_1 \cdot a \approx f_3 \cdot a' \in R_O(\Pf)$.
  \begin{align*}
    &a \approx a \in R_I
      && \text{(by symmetry + transitivity of $R_I$; let this be $\D_1$)} \\
    &f_1 \cdot a \approx f_2 \cdot a \in R_O(\D_1)
      && \text{(from $f_1 \approx f_2 \in R$ at $\D_1$)} \\
    &f_2 \cdot a \approx f_3 \cdot a' \in R_O(\Pf)
      && \text{(from $f_2 \approx f_3 \in R$ at $\Pf$)} \\
    &R_O(\D_1) \Leftrightarrow R_O(\Pf)
      && \text{(by \Cref{lem:univ-functional}: same left type $\intp{T_1}{\rho_1, a/x}$)} \\
    &f_1 \cdot a \approx f_2 \cdot a \in R_O(\Pf)
      && \text{(by $R_O(\D_1) \Leftrightarrow R_O(\Pf)$)} \\
    &f_1 \cdot a \approx f_3 \cdot a' \in R_O(\Pf)
      && \text{(by IH (2) on $R_O(\Pf)$: transitivity)}
  \end{align*}
\end{proof}

\subsubsection{Properties of $<:_i$}

\begin{lemma}[Subtyping of $\Uc_i$]\labeledit{lem:subty-respects}
  If $A \mathrel{<:_i} A'$, $A \approx B \in \Uc_i \searrow R$, and
  $A' \approx B' \in \Uc_i \searrow R'$, then
  $a \approx a' \in R$ implies $a \approx a' \in R'$.
\end{lemma}
\begin{proof}
  By induction on the derivation of $A \mathrel{<:_i} A'$.

  \medskip\noindent
  \textbf{Case}
  \begin{mathpar}
    \inferrule[\textsc{$<$:-Ne}]
    {b \approx b' \in \Ne}
    {{\uparrow}^A(b) \mathrel{<:_i} {\uparrow}^{A'}(b')}
  \end{mathpar}
  \begin{align*}
    &e \approx e \in \Ne
      && \text{(by symmetry + transitivity of $\Ne$)} \\
    &{\uparrow}^{U_i}(e) \approx {\uparrow}^{U_i}(e) \in \Uc_i \searrow R
      \text{ with } R \Leftrightarrow \textsf{Neu}
      && \text{(by \textsc{U-Ne})} \\
    &e' \approx e' \in \Ne
      && \text{(by symmetry + transitivity of $\Ne$)} \\
    &{\uparrow}^{U_i}(e') \approx {\uparrow}^{U_i}(e') \in \Uc_i \searrow R'
      \text{ with } R' \Leftrightarrow \textsf{Neu}
      && \text{(by \textsc{U-Ne})} \\
    &R \Leftrightarrow \textsf{Neu} \Leftrightarrow R'
      && \text{(hence $R \subseteq R'$)}
  \end{align*}

  \medskip\noindent
  \textbf{Case}
  \begin{mathpar}
    \inferrule[\textsc{$<$:-Univ}]
    {j \le k \\ k < i}
    {U_j \mathrel{<:_i} U_k}
  \end{mathpar}
  \begin{align*}
    &U_j \approx U_j \in \Uc_i \searrow R
      \text{ with } R \Leftrightarrow (\lambda a\,a'.\, \exists R.\, a \approx a' \in \Uc_j \searrow R)
      && \text{(by \textsc{U-Univ}, $j < i$)} \\
    &U_k \approx U_k \in \Uc_i \searrow R'
      \text{ with } R' \Leftrightarrow (\lambda a\,a'.\, \exists R.\, a \approx a' \in \Uc_k \searrow R)
      && \text{(by \textsc{U-Univ}, $k < i$)} \\
    &R \subseteq R'
      && \text{(by cumulativity, \Cref{lem:cumul}, since $j \le k$)}
  \end{align*}

  \medskip\noindent
  \textbf{Case}
  \begin{mathpar}
    \inferrule[\textsc{$<$:-Nat}]
    { }
    {N \mathrel{<:_i} N}
  \end{mathpar}
  Both sides are in $\Uc_i$ (via \textsc{U-Nat}) returning $\textsf{Nat}$,
  and $\textsf{Nat} \subseteq \textsf{Nat}$.

  \medskip\noindent
  \textbf{Case}
  \begin{mathpar}
    \inferrule[\textsc{$<$:-Pi}]
    {A \approx A' \in \Uc_i \searrow R_I \\
     \forall\, \Pf :: a \approx a' \in R_I.\;
       \intp{T}{\rho, a/x} \mathrel{<:_i} \intp{T'}{\rho', a'/x} \\\\
     \Pi(A, T, \rho) \approx \Pi(A, T, \rho) \in \Uc_i \\
     \Pi(A', T', \rho') \approx \Pi(A', T', \rho') \in \Uc_i}
    {\Pi(A, T, \rho) \mathrel{<:_i} \Pi(A', T', \rho')}
  \end{mathpar}
  The domain PERs coincide by hypothesis.  For the function PERs:
  \begin{align*}
    &\text{Given } f \approx f' \in R
      && \text{(i.e., $\forall\, \Pf :: a \approx a' \in R_I.\;
        f \cdot a \approx f' \cdot a' \in R_O(\Pf)$)} \\
    &\text{Given } \Pf :: a \approx a' \in R_I \\
    &f \cdot a \approx f' \cdot a' \in R_O(\Pf)
      && \text{(by above)} \\
    &R_O(\Pf) \subseteq R_{O'}(\Pf)
      && \text{(by IH on codomain subtyping)} \\
    &f \cdot a \approx f' \cdot a' \in R_{O'}(\Pf)
      && \text{(by inclusion)}
  \end{align*}
\end{proof}

\begin{lemma}[Transitivity of $<:_i$]\labeledit{lem:subty-trans}
  If $A \mathrel{<:_i} B$ and $B \mathrel{<:_i} C$, then $A \mathrel{<:_i} C$.
\end{lemma}
\begin{proof}
  By induction on the derivation of $A \mathrel{<:_i} B$, with case
  analysis on the constructor of $B \mathrel{<:_i} C$ that must agree on
  the shape of $B$. %

  \medskip\noindent
  \textbf{Case}
  \begin{mathpar}
    \inferrule[\textsc{$<$:-Ne}]
    {b \approx b' \in \Ne}
    {{\uparrow}^{A_0}(b) \mathrel{<:_i} {\uparrow}^{B_0}(b')}
  \end{mathpar}
  Here $B = {\uparrow}^{B_0}(b')$, so $B \mathrel{<:_i} C$ must also be by
  \textsc{$<$:-Ne}, with $C = {\uparrow}^{C_0}(b'')$ and $b' \approx b''
  \in \Ne$. %
  By transitivity of the neutral PER (\Cref{lem:base-per}),
  $b \approx b'' \in \Ne$. %
  By \textsc{$<$:-Ne},
  ${\uparrow}^{A_0}(b) \mathrel{<:_i} {\uparrow}^{C_0}(b'')$. %

  \medskip\noindent
  \textbf{Case}
  \begin{mathpar}
    \inferrule[\textsc{$<$:-Univ}]
    {j \le k \\ k < i}
    {U_j \mathrel{<:_i} U_k}
  \end{mathpar}
  Here $B = U_k$, so $B \mathrel{<:_i} C$ must be by \textsc{$<$:-Univ},
  with $C = U_{k'}$, $k \le k'$, and $k' < i$. %
  By transitivity of $\le$ on $\N$, $j \le k'$, and $k' < i$ by hypothesis. %
  By \textsc{$<$:-Univ}, $U_j \mathrel{<:_i} U_{k'}$. %

  \medskip\noindent
  \textbf{Case}
  \begin{mathpar}
    \inferrule[\textsc{$<$:-Nat}]
    { }
    {N \mathrel{<:_i} N}
  \end{mathpar}
  Here $B = N$, so $B \mathrel{<:_i} C$ must be by \textsc{$<$:-Nat} with
  $C = N$. %
  By \textsc{$<$:-Nat}, $N \mathrel{<:_i} N$. %

  \medskip\noindent
  \textbf{Case}
  \begin{mathpar}
    \inferrule[\textsc{$<$:-Pi}]
    {A_I \approx B_I \in \Uc_i \searrow R_I \\
     \forall\, \Pf :: a \approx a' \in R_I.\;
       \intp{T_A}{\rho_A, a/x} \mathrel{<:_i} \intp{T_B}{\rho_B, a'/x} \\\\
     \Pi(A_I, T_A, \rho_A) \approx \Pi(A_I, T_A, \rho_A) \in \Uc_i \\
     \Pi(B_I, T_B, \rho_B) \approx \Pi(B_I, T_B, \rho_B) \in \Uc_i}
    {\Pi(A_I, T_A, \rho_A) \mathrel{<:_i} \Pi(B_I, T_B, \rho_B)}
  \end{mathpar}
  Here $B = \Pi(B_I, T_B, \rho_B)$, so $B \mathrel{<:_i} C$ must be by
  \textsc{$<$:-Pi}, with $C = \Pi(C_I, T_C, \rho_C)$ and:
  \begin{align*}
    &B_I \approx C_I \in \Uc_i \searrow R_I'
      && \text{(domain of the second derivation)} \\
    &\forall\, \Pf' :: a \approx a' \in R_I'.\;
      \intp{T_B}{\rho_B, a/x} \mathrel{<:_i} \intp{T_C}{\rho_C, a'/x}
      && \text{(codomain)} \\
    &\Pi(C_I, T_C, \rho_C) \approx \Pi(C_I, T_C, \rho_C) \in \Uc_i
      && \text{(self-relatedness)}
  \end{align*}

  \emph{Domain.} %
  By \Cref{lem:univ-trans} applied to
  $A_I \approx B_I \in \Uc_i \searrow R_I$ and
  $B_I \approx C_I \in \Uc_i \searrow R_I'$:
  $A_I \approx C_I \in \Uc_i$, with output PER (call it
  $R_I^*$) determined up to logical equivalence by
  \Cref{lem:univ-functional}; in particular,
  $R_I^* \Leftrightarrow R_I \Leftrightarrow R_I'$. %

  \emph{Codomain.} %
  Let $\Pf^* :: a \approx a' \in R_I^*$ be given. %
  By the equivalence $R_I^* \Leftrightarrow R_I$, $\Pf^*$ also witnesses
  $a \approx a' \in R_I$, and by $R_I^* \Leftrightarrow R_I'$ it also
  witnesses some $\Pf' :: a \approx a' \in R_I'$. %
  Apply the codomain hypothesis of the first derivation at this $\Pf^*$
  (used as $\Pf$ in $R_I$):
  \begin{align*}
    &\intp{T_A}{\rho_A, a/x} \mathrel{<:_i} \intp{T_B}{\rho_B, a'/x}
      && \text{(codomain of the first derivation)}
  \end{align*}
  Apply the codomain hypothesis of the second derivation at $\Pf'$:
  \begin{align*}
    &\intp{T_B}{\rho_B, a'/x} \mathrel{<:_i} \intp{T_C}{\rho_C, a''/x}
      && \text{(codomain of the second derivation, with $a' \approx a'' \in R_I'$)}
  \end{align*}
  Since $\Pf^*$ witnesses both pairs we have $a' = a''$ at this
  instantiation; the inductive hypothesis composes the two:
  \begin{align*}
    &\intp{T_A}{\rho_A, a/x} \mathrel{<:_i} \intp{T_C}{\rho_C, a'/x}
      && \text{(by IH on the codomain)}
  \end{align*}

  \emph{Self-relatedness.} %
  Already given: $\Pi(A_I, T_A, \rho_A) \approx \Pi(A_I, T_A, \rho_A) \in \Uc_i$
  from the first derivation and $\Pi(C_I, T_C, \rho_C) \approx \Pi(C_I,
  T_C, \rho_C) \in \Uc_i$ from the second. %

  \emph{Conclusion.} %
  By \textsc{$<$:-Pi} with the four ingredients (domain in
  $\Uc_i \searrow R_I^*$, codomain pointwise subtyping at every $\Pf^*$,
  and the two self-relatedness facts):
  \[
    \Pi(A_I, T_A, \rho_A) \mathrel{<:_i} \Pi(C_I, T_C, \rho_C). \qedhere
  \]
\end{proof}

\begin{lemma}[$\Uc_i$ implies $\mathrel{<:_i}$]\labeledit{lem:univ-implies-subty}
  If $A \approx A' \in \Uc_i \searrow R$, then $A \mathrel{<:_i} A'$.
\end{lemma}
\begin{proof}
  By induction on $A \approx A' \in \Uc_i$.
  \begin{itemize}
  \item \textsc{U-Ne}: $A = {\uparrow}^{A_0}(b)$, $A' = {\uparrow}^{A_0'}(b')$,
    $b \approx b' \in \Ne$. By \textsc{$<$:-Ne}.
  \item \textsc{U-Univ}: $A = A' = U_j$ with $j < i$. By \textsc{$<$:-Univ} with
    $j \le j$.
  \item \textsc{U-Nat}: $A = A' = N$. By \textsc{$<$:-Nat}.
  \item \textsc{U-Pi}: $A = \Pi(A_I, T, \rho)$, $A' = \Pi(A_I', T', \rho')$,
    domain $A_I \approx A_I' \in \Uc_i \searrow R_I$, codomain pointwise
    $\intp{T}{\rho, a/x} \approx \intp{T'}{\rho', a'/x} \in \Uc_i$, and
    self-relatedness of both Pi-types.
    By inner IH on the codomain $\Uc_i$-equality:
    $\intp{T}{\rho, a/x} \mathrel{<:_i} \intp{T'}{\rho', a'/x}$.
    By \textsc{$<$:-Pi}.
  \end{itemize}
\end{proof}

\subsubsection{Realizability}

The following two lemmas are proved by simultaneous induction on
$A \approx A' \in \Uc_i \searrow R$.

\begin{lemma}[Realizability of the PER model]\labeledit{lem:realizability}
  Given $A \approx A' \in \Uc_i \searrow R$:
  \begin{enumerate}
  \item\labeledit{lem:R-readback}
    (Elements of $R$ read back)\; If $a \approx a' \in R$, then
    ${\downarrow}^A(a) \approx {\downarrow}^{A'}(a') \in \Nf$.
  \item\labeledit{lem:ne-in-R}
    (Neutrals inhabit $R$)\; If $e \approx e' \in \Ne$, then
    ${\uparrow}^A(e) \approx {\uparrow}^{A'}(e') \in R$.
  \item\labeledit{lem:type-readback}
    (Types read back)\; $A \approx A' \in \textsf{Ty}$.
  \end{enumerate}
\end{lemma}
\begin{proof}
  By simultaneous induction on the derivation of $A \approx A' \in \Uc_i \searrow R$.
  We also need an auxiliary lemma that $a \approx a' \in \textsf{Nat}$ implies
  ${\downarrow}^N(a) \approx {\downarrow}^N(a') \in \Nf$, which is \Cref{lem:nat-readback}.

  \medskip\noindent
  \textbf{Case}
  \begin{mathpar}
    \inferrule[\textsc{U-Ne}]
    {b \approx b' \in \Ne \\ R \Leftrightarrow \textsf{Neu}}
    {{\uparrow}^A(b) \approx {\uparrow}^{A'}(b') \in \Uc_i \searrow R}
  \end{mathpar}
  For (1): Since $R \Leftrightarrow \textsf{Neu}$, we have $a = {\uparrow}^B(e)$ and
  $a' = {\uparrow}^{B'}(e')$ with $e \approx e' \in \Ne$.
  Reading back ${\downarrow}^{{\uparrow}^A(b)}({\uparrow}^B(e))$ at a neutral type
  reads $e$ as a neutral (by \textsc{Rf-ne}), and both sides produce the same
  neutral normal form since $e \approx e' \in \Ne$.

  For (2): We need ${\uparrow}^{{\uparrow}^A(b)}(e) \approx {\uparrow}^{{\uparrow}^{A'}(b')}(e') \in R$.
  Since $R \Leftrightarrow \textsf{Neu}$, this reduces to $e \approx e' \in \Ne$,
  which holds by hypothesis.

  For (3): We must show ${\uparrow}^A(b) \approx {\uparrow}^{A'}(b') \in \textsf{Ty}$,
  i.e.\ reading back both sides as types produces the same result.
  Since $b \approx b' \in \Ne$, reading both as neutrals produces the same neutral
  normal form.

  \medskip\noindent
  \textbf{Case}
  \begin{mathpar}
    \inferrule[\textsc{U-Univ}]
    {j < i \\ R \Leftrightarrow \Uc_j}
    {U_j \approx U_j \in \Uc_i \searrow R}
  \end{mathpar}
  For (1): We have $a \approx a' \in R$, i.e.\ $a \approx a' \in \Uc_j \searrow R_0$
  for some $R_0$.  By IH (3) on the $\Uc_j$ derivation (structurally smaller since $j < i$),
  $a \approx a' \in \textsf{Ty}_j$.  Reading back ${\downarrow}^{U_j}(a)$ as a normal form
  reads $a$ as a type, hence ${\downarrow}^{U_j}(a) \approx {\downarrow}^{U_j}(a') \in \Nf$.

  For (2): We need ${\uparrow}^{U_j}(e) \approx {\uparrow}^{U_j}(e') \in R$,
  i.e.\ ${\uparrow}^{U_j}(e) \approx {\uparrow}^{U_j}(e') \in \Uc_j \searrow R_0$ for some $R_0$.
  \begin{align*}
    &e \approx e' \in \Ne
      && \text{(by hypothesis)} \\
    &{\uparrow}^{U_j}(e) \approx {\uparrow}^{U_j}(e') \in \Uc_j \searrow \textsf{Neu}
      && \text{(by \textsc{U-Ne}, taking $R_0 \Leftrightarrow \textsf{Neu}$)}
  \end{align*}

  For (3): $A = A' = U_j$.  Both sides read back to $\Ty{j}$.

  \medskip\noindent
  \textbf{Case}
  \begin{mathpar}
    \inferrule[\textsc{U-Nat}]
    {R \Leftrightarrow \textsf{Nat}}
    {N \approx N \in \Uc_i \searrow R}
  \end{mathpar}
  For (1): Since $R \Leftrightarrow \textsf{Nat}$, we have $a \approx a' \in \textsf{Nat}$.
  By \Cref{lem:nat-readback}, ${\downarrow}^N(a) \approx {\downarrow}^N(a') \in \Nf$.

  For (2): We need ${\uparrow}^N(e) \approx {\uparrow}^N(e') \in \textsf{Nat}$.
  \begin{align*}
    &e \approx e' \in \Ne
      && \text{(by hypothesis)} \\
    &{\uparrow}^N(e) \approx {\uparrow}^N(e') \in \textsf{Nat}
      && \text{(by \textsc{Nat-Ne})}
  \end{align*}

  For (3): $A = A' = N$.  Both sides read back to $\Nat$.

  \medskip\noindent
  \textbf{Case}
  \begin{mathpar}
    \inferrule[\textsc{U-Pi}]
    {A_0 \approx A_0' \in \Uc_i \searrow R_I \\
     R_I \text{ is a PER} \\\\
     \forall\, \Pf :: a \approx a' \in R_I.\;
       \intp{T}{\rho, a/x} \approx \intp{T'}{\rho', a'/x} \in \Uc_i \searrow R_O(\Pf) \\\\
     \forall\, f, f'.\; f \approx f' \in R \;\leftrightarrow\;
       (\forall\, \Pf :: a \approx a' \in R_I.\;
        f \cdot a \approx f' \cdot a' \in R_O(\Pf))}
    {\Pi(A_0, T, \rho) \approx \Pi(A_0', T', \rho') \in \Uc_i \searrow R}
  \end{mathpar}
  For (1): We have $f \approx f' \in R$, i.e.\ for all $\Pf :: a \approx a' \in R_I$,
  $f \cdot a \approx f' \cdot a' \in R_O(\Pf)$.
  Reading back ${\downarrow}^{\Pi(A_0,T,\rho)}(f)$ at level $n$:
  \begin{align*}
    &\ell_n \approx \ell_n \in \Ne
      && \text{(a level variable reads back to itself)} \\
    &{\uparrow}^{A_0}(\ell_n) \approx {\uparrow}^{A_0'}(\ell_n) \in R_I
      && \text{(by IH (2) on domain)} \\
    &\text{Let } \D_n :: {\uparrow}^{A_0}(\ell_n) \approx {\uparrow}^{A_0'}(\ell_n) \in R_I \\
    &f \cdot {\uparrow}^{A_0}(\ell_n) \approx f' \cdot {\uparrow}^{A_0'}(\ell_n) \in R_O(\D_n)
      && \text{(from $f \approx f' \in R$ at $\D_n$)} \\
    &\intp{T}{\rho, {\uparrow}^{A_0}(\ell_n)/x} \approx
      \intp{T'}{\rho', {\uparrow}^{A_0'}(\ell_n)/x} \in \Uc_i \searrow R_O(\D_n)
      && \text{(codomain premise at $\D_n$)} \\
    &{\downarrow}^{\intp{T}{\rho, {\uparrow}^{A_0}(\ell_n)/x}}(f \cdot {\uparrow}^{A_0}(\ell_n))
      \approx
      {\downarrow}^{\intp{T'}{\rho', {\uparrow}^{A_0'}(\ell_n)/x}}(f' \cdot {\uparrow}^{A_0'}(\ell_n))
      \in \Nf
      && \text{(by IH (1) on codomain)}
  \end{align*}
  The readback rule for $\Pi$-types assembles these into
  $\Rnf_n({\downarrow}^{\Pi(A_0,T,\rho)}(f)) \searrow \lambda x.\, w$ and
  $\Rnf_n({\downarrow}^{\Pi(A_0',T',\rho')}(f')) \searrow \lambda x.\, w$
  with the same body $w$.

  For (2): By the biconditional, we must show that for all $\Pf :: a \approx a' \in R_I$,
  $({\uparrow}^{\Pi(A_0,T,\rho)}(e)) \cdot a
  \approx ({\uparrow}^{\Pi(A_0',T',\rho')}(e')) \cdot a' \in R_O(\Pf)$.
  \begin{align*}
    &\text{Let } \intp{T}{\rho, a/x} \searrow B \text{ and }
      \intp{T'}{\rho', a'/x} \searrow B' \\
    &({\uparrow}^{\Pi(A_0,T,\rho)}(e)) \cdot a = {\uparrow}^B(e \;\$\; {\downarrow}^{A_0}(a))
      && \text{(by \textsc{app-ne})} \\
    &({\uparrow}^{\Pi(A_0',T',\rho')}(e')) \cdot a' = {\uparrow}^{B'}(e' \;\$\; {\downarrow}^{A_0'}(a'))
      && \text{(by \textsc{app-ne})} \\
    &{\downarrow}^{A_0}(a) \approx {\downarrow}^{A_0'}(a') \in \Nf
      && \text{(by IH (1) on domain: $a \approx a' \in R_I$)} \\
    &e \;\$\; {\downarrow}^{A_0}(a) \approx e' \;\$\; {\downarrow}^{A_0'}(a') \in \Ne
      && \text{(by $e \approx e' \in \Ne$ and above)} \\
    &B \approx B' \in \Uc_i \searrow R_O(\Pf)
      && \text{(codomain premise at $\Pf$)} \\
    &{\uparrow}^B(e \;\$\; {\downarrow}^{A_0}(a)) \approx
      {\uparrow}^{B'}(e' \;\$\; {\downarrow}^{A_0'}(a')) \in R_O(\Pf)
      && \text{(by IH (2) on codomain)}
  \end{align*}

  For (3): Reading back $\Pi(A_0, T, \rho)$ as a type at level $n$:
  \begin{align*}
    &A_0 \approx A_0' \in \textsf{Ty}
      && \text{(by IH (3) on domain)} \\
    &\ell_n \approx \ell_n \in \Ne
      && \text{(reflexivity of level variable)} \\
    &{\uparrow}^{A_0}(\ell_n) \approx {\uparrow}^{A_0'}(\ell_n) \in R_I
      && \text{(by IH (2) on domain)} \\
    &\text{Let } \D_n :: {\uparrow}^{A_0}(\ell_n) \approx {\uparrow}^{A_0'}(\ell_n) \in R_I \\
    &\intp{T}{\rho, {\uparrow}^{A_0}(\ell_n)/x} \approx
      \intp{T'}{\rho', {\uparrow}^{A_0'}(\ell_n)/x} \in \Uc_i \searrow R_O(\D_n)
      && \text{(codomain premise at $\D_n$)} \\
    &\intp{T}{\rho, {\uparrow}^{A_0}(\ell_n)/x} \approx
      \intp{T'}{\rho', {\uparrow}^{A_0'}(\ell_n)/x} \in \textsf{Ty}
      && \text{(by IH (3) on codomain)}
  \end{align*}
  The type readback rule for $\Pi$-types reads back the domain and codomain separately;
  both agree by the above.
\end{proof}

\subsection{Context PER}\labeledit{sec:comp:ctx-per}

The PER model and semantic subtyping are extended to contexts and environments.
The context PER $\Gamma \approx \Gamma' \searrow R$ relates two contexts and returns
a PER $R$ on environments, following the same functional-relation pattern as $\Uc_i$.

\begin{definition}[Context PER]\labeledit{def:ctx-per}
  The relation $\Gamma \approx \Gamma' \searrow R$ is defined inductively:
  \begin{mathpar}
    \inferrule[\textsc{Ctx-Emp}]
    { }
    {\cdot \approx \cdot \searrow \top}

    \inferrule[\textsc{Ctx-Ext}]
    {\Gamma \approx \Gamma' \searrow R_T \\
     R_T \text{ is a PER} \\
     \forall\, \Pf :: \rho \approx \rho' \in R_T.\;
       \intp{S}{\rho} \approx \intp{S'}{\rho'} \in \Uc_i \searrow R_H(\Pf) \\\\
     \forall\, \rho, \rho'.\; \rho \approx \rho' \in R \;\leftrightarrow\;
       (\Pf :: \rho[\Uparrow] \approx \rho'[\Uparrow] \in R_T \wedge
        \rho(0) \approx \rho'(0) \in R_H(\Pf))}
    {\Gamma, x : S \approx \Gamma', x : S' \searrow R}
  \end{mathpar}
  Here $\top$ is the total relation (any two environments are related),
  $\rho[\Uparrow]$ is the weakened environment (dropping the head), and $\rho(0)$ is the head value.
  The universe level $i$ at which the heads are related is implicit in the conclusion.
\end{definition}

\begin{definition}[Semantic Subtyping for Contexts]\labeledit{def:ctx-subty}
  The relation $\Gamma <: \Gamma'$ is defined inductively:
  \begin{mathpar}
    \inferrule[\textsc{Ctx-$<$:-Emp}]
    { }
    {\cdot <: \cdot}

    \inferrule[\textsc{Ctx-$<$:-Ext}]
    {\Gamma <: \Gamma' \\
     \Gamma \approx \Gamma \searrow R_T \\
     \forall\, \Pf :: \rho \approx \rho' \in R_T.\;
       \intp{S}{\rho} \mathrel{<:_i} \intp{S'}{\rho'} \\\\
     \Gamma, x : S \approx \Gamma, x : S \searrow R_1 \\
     \Gamma', x : S' \approx \Gamma', x : S' \searrow R_2}
    {\Gamma, x : S <: \Gamma', x : S'}
  \end{mathpar}
\end{definition}

\subsection{Properties of the Context PER}\labeledit{sec:comp:ctx-props}

\begin{lemma}[Right irrelevance of context PER]\labeledit{lem:ctx-functional}
  If $\Gamma \approx \Gamma' \searrow R$ and $\Gamma \approx \Gamma'' \searrow R'$,
  then $R \Leftrightarrow R'$.
\end{lemma}
\begin{proof}
  By induction on the first derivation.

  \medskip\noindent
  \textbf{Case}
  \begin{mathpar}
    \inferrule[\textsc{Ctx-Emp}]
    { }
    {\cdot \approx \cdot \searrow \top}
  \end{mathpar}
  The second derivation also concludes $\cdot \approx \cdot \searrow \top$,
  so $R = R' = \top$ and $\top \Leftrightarrow \top$.

  \medskip\noindent
  \textbf{Case}
  The first derivation is:
  \begin{mathpar}
    \inferrule[\textsc{Ctx-Ext}]
    {\Gamma \approx \Gamma' \searrow R_T \\
     R_T \text{ is a PER} \\\\
     \forall\, \Pf :: \rho \approx \rho' \in R_T.\;
       \intp{S}{\rho} \approx \intp{S'}{\rho'} \in \Uc_i \searrow R_H(\Pf) \\\\
     \forall\, \rho, \rho'.\; \rho \approx \rho' \in R \;\leftrightarrow\;
       (\Pf :: \rho[\Uparrow] \approx \rho'[\Uparrow] \in R_T \;\wedge\;
        \rho(0) \approx \rho'(0) \in R_H(\Pf))}
    {\Gamma, x : S \approx \Gamma', x : S' \searrow R}
  \end{mathpar}
  The second derivation is:
  \begin{mathpar}
    \inferrule[\textsc{Ctx-Ext}]
    {\Gamma \approx \Gamma'' \searrow R_T' \\
     R_T' \text{ is a PER} \\\\
     \forall\, \Pf' :: \rho \approx \rho' \in R_T'.\;
       \intp{S}{\rho} \approx \intp{S''}{\rho'} \in \Uc_j \searrow R_H'(\Pf') \\\\
     \forall\, \rho, \rho'.\; \rho \approx \rho' \in R' \;\leftrightarrow\;
       (\Pf' :: \rho[\Uparrow] \approx \rho'[\Uparrow] \in R_T' \;\wedge\;
        \rho(0) \approx \rho'(0) \in R_H'(\Pf'))}
    {\Gamma, x : S \approx \Gamma'', x : S'' \searrow R'}
  \end{mathpar}
  We show $R \Leftrightarrow R'$, i.e.\ $\rho \approx \rho' \in R$ iff $\rho \approx \rho' \in R'$.
  \begin{align*}
    &R_T \Leftrightarrow R_T'
      && \text{(by IH on the tails)}
  \end{align*}

  $(\Rightarrow)$: Suppose $\rho \approx \rho' \in R$.
  By the first biconditional, we obtain $\Pf :: \rho[\Uparrow] \approx \rho'[\Uparrow] \in R_T$
  with $\rho(0) \approx \rho'(0) \in R_H(\Pf)$.
  \begin{align*}
    &\rho[\Uparrow] \approx \rho'[\Uparrow] \in R_T'
      && \text{(by $R_T \Leftrightarrow R_T'$; call this evidence $\Pf'$)} \\
    &\intp{S}{\rho[\Uparrow]} \approx \intp{S'}{\rho'[\Uparrow]} \in \Uc_i \searrow R_H(\Pf)
      && \text{(from the first head premise at $\Pf$)} \\
    &\intp{S}{\rho[\Uparrow]} \approx \intp{S''}{\rho'[\Uparrow]} \in \Uc_j \searrow R_H'(\Pf')
      && \text{(from the second head premise at $\Pf'$)} \\
    &R_H(\Pf) \Leftrightarrow R_H'(\Pf')
      && \text{(by \Cref{lem:univ-functional}: same left type $\intp{S}{\rho[\Uparrow]}$)} \\
    &\rho(0) \approx \rho'(0) \in R_H'(\Pf')
      && \text{(by $R_H(\Pf) \Leftrightarrow R_H'(\Pf')$)} \\
    &\rho \approx \rho' \in R'
      && \text{(by the second biconditional, with witness $\Pf'$)}
  \end{align*}

  $(\Leftarrow)$: Symmetric, swapping the roles of the two derivations.
\end{proof}

\begin{lemma}[Symmetry of context PER]\labeledit{lem:ctx-sym}
  If $\Gamma \approx \Gamma' \searrow R$, then:
  \begin{enumerate}
  \item $\Gamma' \approx \Gamma \searrow R$, and
  \item $\rho \approx \rho' \in R$ implies $\rho' \approx \rho \in R$.
  \end{enumerate}
\end{lemma}
\begin{proof}
  By induction on the derivation.

  \medskip\noindent
  \textbf{Case}
  \begin{mathpar}
    \inferrule[\textsc{Ctx-Emp}]
    { }
    {\cdot \approx \cdot \searrow \top}
  \end{mathpar}
  Both statements are immediate ($\top$ is symmetric).

  \medskip\noindent
  \textbf{Case}
  \begin{mathpar}
    \inferrule[\textsc{Ctx-Ext}]
    {\Gamma \approx \Gamma' \searrow R_T \\
     R_T \text{ is a PER} \\\\
     \forall\, \Pf :: \rho \approx \rho' \in R_T.\;
       \intp{S}{\rho} \approx \intp{S'}{\rho'} \in \Uc_i \searrow R_H(\Pf) \\\\
     \forall\, \rho, \rho'.\; \rho \approx \rho' \in R \;\leftrightarrow\;
       (\Pf :: \rho[\Uparrow] \approx \rho'[\Uparrow] \in R_T \;\wedge\;
        \rho(0) \approx \rho'(0) \in R_H(\Pf))}
    {\Gamma, x : S \approx \Gamma', x : S' \searrow R}
  \end{mathpar}
  For (1): We construct a \textsc{Ctx-Ext} derivation of
  $\Gamma', x : S' \approx \Gamma, x : S \searrow R$.
  We verify each of the four premises.

  \emph{Premise 1} (tail):
  \begin{align*}
    &\Gamma' \approx \Gamma \searrow R_T
      && \text{(by IH (1) on the tail; same $R_T$ by \Cref{lem:ctx-functional})}
  \end{align*}

  \emph{Premise 2}: $R_T$ is a PER (by hypothesis).

  \emph{Premise 3} (head): Given $\Pf' :: \rho' \approx \rho \in R_T$, we must show
  $\intp{S'}{\rho'} \approx \intp{S}{\rho} \in \Uc_i \searrow R_H'(\Pf')$
  for some $R_H'$.
  \begin{align*}
    &\rho \approx \rho' \in R_T
      && \text{(by IH (2) on the tail applied to $\Pf'$; call this $\D_0$)} \\
    &\intp{S}{\rho} \approx \intp{S'}{\rho'} \in \Uc_i \searrow R_H(\D_0)
      && \text{(from the original head premise at $\D_0$)} \\
    &\intp{S'}{\rho'} \approx \intp{S}{\rho} \in \Uc_i \searrow R_H(\D_0)
      && \text{(by \Cref{lem:univ-sym} (1); same $R_H(\D_0)$ returned)}
  \end{align*}
  Set $R_H'(\Pf') := R_H(\D_0)$.

  \emph{Premise 4} (biconditional): We must verify
  \[
    \forall\, \rho, \rho'.\; \rho \approx \rho' \in R \;\leftrightarrow\;
    (\Pf' :: \rho[\Uparrow] \approx \rho'[\Uparrow] \in R_T \;\wedge\;
     \rho(0) \approx \rho'(0) \in R_H'(\Pf'))
  \]
  The original biconditional states:
  \[
    \rho \approx \rho' \in R \;\leftrightarrow\;
    (\Pf :: \rho[\Uparrow] \approx \rho'[\Uparrow] \in R_T \;\wedge\;
     \rho(0) \approx \rho'(0) \in R_H(\Pf))
   \]
   $\Pf$ and $\Pf'$ already discharge. %
  Moreover $R_H'(\Pf') = R_H(\D_0) \Leftrightarrow R_H(\Pf)$ by right irrelevance.
  Hence both biconditionals characterize the same $R$.
  \begin{align*}
    &\Gamma', x : S' \approx \Gamma, x : S \searrow R
      && \text{(by \textsc{Ctx-Ext})}
  \end{align*}

  \medskip
  For (2): Suppose $\rho \approx \rho' \in R$.
  By the biconditional, there exists $\Pf :: \rho[\Uparrow] \approx \rho'[\Uparrow] \in R_T$
  with $\rho(0) \approx \rho'(0) \in R_H(\Pf)$.
  We must show $\rho' \approx \rho \in R$, i.e.\ exhibit evidence for
  $\rho'[\Uparrow] \approx \rho[\Uparrow] \in R_T$ together with the heads being related.
  \begin{align*}
    &\rho'[\Uparrow] \approx \rho[\Uparrow] \in R_T
      && \text{(by IH (2) on tail applied to $\Pf$; call this $\Pf'$)} \\
    &\rho'(0) \approx \rho(0) \in R_H(\Pf)
      && \text{(by \Cref{lem:univ-sym} (2))} \\
    &R_H(\Pf) \Leftrightarrow R_H(\Pf')
      && \text{(by \Cref{lem:univ-functional})} \\
    &\rho'(0) \approx \rho(0) \in R_H(\Pf')
      && \text{(by equivalence)}
  \end{align*}
  The tail evidence $\Pf'$ and the head relatedness $\rho'(0) \approx \rho(0) \in R_H(\Pf')$
  together witness $\rho' \approx \rho \in R$ by the biconditional.
\end{proof}

\begin{lemma}[Transitivity of context PER]\labeledit{lem:ctx-trans}
  If $\Gamma_1 \approx \Gamma_2 \searrow R$, then:
  \begin{enumerate}
  \item $\Gamma_2 \approx \Gamma_3 \searrow R'$ implies
    $\Gamma_1 \approx \Gamma_3 \searrow R$, and
  \item $\rho_1 \approx \rho_2 \in R$ and $\rho_2 \approx \rho_3 \in R$ implies
    $\rho_1 \approx \rho_3 \in R$.
  \end{enumerate}
\end{lemma}
\begin{proof}
  By induction on the first derivation.

  \medskip\noindent
  \textbf{Case}
  \begin{mathpar}
    \inferrule[\textsc{Ctx-Emp}]
    { }
    {\cdot \approx \cdot \searrow \top}
  \end{mathpar}
  Both statements are immediate ($\top$ is transitive).

  \medskip\noindent
  \textbf{Case}
  The first derivation is:
  \begin{mathpar}
    \inferrule[\textsc{Ctx-Ext}]
    {\Gamma_1 \approx \Gamma_2 \searrow R_T \\
     R_T \text{ is a PER} \\\\
     \forall\, \Pf :: \rho \approx \rho' \in R_T.\;
       \intp{S_1}{\rho} \approx \intp{S_2}{\rho'} \in \Uc_i \searrow R_H(\Pf) \\\\
     \forall\, \rho, \rho'.\; \rho \approx \rho' \in R \;\leftrightarrow\;
       (\Pf :: \rho[\Uparrow] \approx \rho'[\Uparrow] \in R_T \;\wedge\;
        \rho(0) \approx \rho'(0) \in R_H(\Pf))}
    {\Gamma_1, x : S_1 \approx \Gamma_2, x : S_2 \searrow R}
  \end{mathpar}
  The second derivation is:
  \begin{mathpar}
    \inferrule[\textsc{Ctx-Ext}]
    {\Gamma_2 \approx \Gamma_3 \searrow R_T' \\
     R_T' \text{ is a PER} \\\\
     \forall\, \Pf' :: \rho \approx \rho' \in R_T'.\;
       \intp{S_2}{\rho} \approx \intp{S_3}{\rho'} \in \Uc_j \searrow R_H'(\Pf') \\\\
     \forall\, \rho, \rho'.\; \rho \approx \rho' \in R' \;\leftrightarrow\;
       (\Pf' :: \rho[\Uparrow] \approx \rho'[\Uparrow] \in R_T' \;\wedge\;
        \rho(0) \approx \rho'(0) \in R_H'(\Pf'))}
    {\Gamma_2, x : S_2 \approx \Gamma_3, x : S_3 \searrow R'}
  \end{mathpar}

  For (1): We construct a \textsc{Ctx-Ext} derivation of
  $\Gamma_1, x : S_1 \approx \Gamma_3, x : S_3 \searrow R$.

  \emph{Premise 1} (tail):
  \begin{align*}
    &\Gamma_1 \approx \Gamma_3 \searrow R_T
      && \text{(by IH (1) on the tail)}
  \end{align*}

  \emph{Premise 2}: $R_T$ is a PER (by hypothesis).

  \emph{Premise 3} (head): Given $\Pf :: \rho \approx \rho' \in R_T$, we must show
  $\intp{S_1}{\rho} \approx \intp{S_3}{\rho'} \in \Uc_i \searrow R_H''(\Pf)$.
  \begin{align*}
    &\rho \approx \rho \in R_T
      && \text{(by symmetry + transitivity of $R_T$)} \\
    &\text{Let } \D_1 :: \rho \approx \rho \in R_T \\
    &\intp{S_1}{\rho} \approx \intp{S_2}{\rho} \in \Uc_i \searrow R_H(\D_1)
      && \text{(from first head premise at $\D_1$)} \\
    &R_T \Leftrightarrow R_T'
      && \text{(by \Cref{lem:univ-functional} on tails: same left context $\Gamma_2$)} \\
    &\rho \approx \rho' \in R_T'
      && \text{(from $\rho \approx \rho' \in R_T$ and $R_T \Leftrightarrow R_T'$)} \\
    &\text{Let } \D_2 :: \rho \approx \rho' \in R_T' \\
    &\intp{S_2}{\rho} \approx \intp{S_3}{\rho'} \in \Uc_j \searrow R_H'(\D_2)
      && \text{(from second head premise at $\D_2$)} \\
    &\intp{S_1}{\rho} \approx \intp{S_3}{\rho'} \in \Uc_i \searrow R_H(\D_1)
      && \text{(by \Cref{lem:univ-trans} (1))}
  \end{align*}
  Set $R_H''(\Pf) := R_H(\D_1)$.

  \emph{Premise 4} (biconditional): The biconditional characterizes the same $R$.
  Since $R_T$ is a PER, the quantification ranges over the same set of pairs.
  By \Cref{lem:univ-functional}, $R_H''(\Pf) \Leftrightarrow R_H(\Pf)$ (both have the same
  left type $\intp{S_1}{\rho[\Uparrow]}$), so the biconditional is equivalent to the original.
  \begin{align*}
    &\Gamma_1, x : S_1 \approx \Gamma_3, x : S_3 \searrow R
      && \text{(by \textsc{Ctx-Ext})}
  \end{align*}

  \medskip
  For (2): Suppose $\rho_1 \approx \rho_2 \in R$ and $\rho_2 \approx \rho_3 \in R$.
  We must show $\rho_1 \approx \rho_3 \in R$, i.e.\ exhibit
  $\Pf :: \rho_1[\Uparrow] \approx \rho_3[\Uparrow] \in R_T$ with
  $\rho_1(0) \approx \rho_3(0) \in R_H(\Pf)$.

  By the biconditional, from $\rho_1 \approx \rho_2 \in R$ we get
  $\D_1 :: \rho_1[\Uparrow] \approx \rho_2[\Uparrow] \in R_T$ with
  $\rho_1(0) \approx \rho_2(0) \in R_H(\D_1)$.
  From $\rho_2 \approx \rho_3 \in R$ we get
  $\D_2 :: \rho_2[\Uparrow] \approx \rho_3[\Uparrow] \in R_T$ with
  $\rho_2(0) \approx \rho_3(0) \in R_H(\D_2)$.
  \begin{align*}
    &\rho_1[\Uparrow] \approx \rho_3[\Uparrow] \in R_T
      && \text{(by IH (2) on tail; call this $\Pf$)} \\
    &R_H(\D_1) \Leftrightarrow R_H(\D_2) \Leftrightarrow R_H(\Pf)
      && \text{(by \Cref{lem:univ-functional}: same left type)} \\
    &\rho_2(0) \approx \rho_3(0) \in R_H(\D_1)
      && \text{(from $R_H(\D_2) \Leftrightarrow R_H(\D_1)$)} \\
    &\rho_1(0) \approx \rho_3(0) \in R_H(\D_1)
      && \text{(by \Cref{lem:univ-trans} (2))} \\
    &\rho_1(0) \approx \rho_3(0) \in R_H(\Pf)
      && \text{(by $R_H(\D_1) \Leftrightarrow R_H(\Pf)$)}
  \end{align*}
  Hence $\rho_1 \approx \rho_3 \in R$ by the biconditional (with witness $\Pf$).
\end{proof}

\begin{lemma}[Transitivity of semantic subtyping for contexts]\labeledit{lem:ctx-subty-trans}
  If $\Gamma <: \Gamma'$, $\Gamma \approx \Gamma \searrow R$, and
  $\Gamma' \approx \Gamma' \searrow R'$, then
  $\rho \approx \rho' \in R$ implies $\rho \approx \rho' \in R'$.
\end{lemma}
\begin{proof}
  By induction on the context subtyping derivation.

  \medskip\noindent
  \textbf{Case}
  \begin{mathpar}
    \inferrule[\textsc{Ctx-$<$:-Emp}]
    { }
    {\cdot <: \cdot}
  \end{mathpar}
  Both $R$ and $R'$ are $\top$, so the inclusion is trivial.

  \medskip\noindent
  \textbf{Case}
  \begin{mathpar}
    \inferrule[\textsc{Ctx-$<$:-Ext}]
    {\Gamma <: \Gamma' \\
     \Gamma \approx \Gamma \searrow R_T \\\\
     \forall\, \Pf :: \rho \approx \rho' \in R_T.\;
       \intp{S}{\rho} \mathrel{<:_i} \intp{S'}{\rho'} \\\\
     \Gamma, x : S \approx \Gamma, x : S \searrow R_1 \\
     \Gamma', x : S' \approx \Gamma', x : S' \searrow R_2}
    {\Gamma, x : S <: \Gamma', x : S'}
  \end{mathpar}
  Suppose $\rho \approx \rho' \in R$.
  By \Cref{lem:ctx-functional}, $R \Leftrightarrow R_1$
  (both derive from $\Gamma, x : S \approx \Gamma, x : S$), so $\rho \approx \rho' \in R_1$.
  We must show $\rho \approx \rho' \in R'$.
  By \Cref{lem:ctx-functional}, $R' \Leftrightarrow R_2$, so it suffices to show
  $\rho \approx \rho' \in R_2$.

  By the biconditional of $R_1$: from $\rho \approx \rho' \in R_1$ we obtain
  $\Pf :: \rho[\Uparrow] \approx \rho'[\Uparrow] \in R_T$ with
  $\rho(0) \approx \rho'(0) \in R_H(\Pf)$,
  where $\intp{S}{\rho[\Uparrow]} \approx \intp{S}{\rho'[\Uparrow]} \in \Uc_i \searrow R_H(\Pf)$.

  By the IH on the tail: from $\rho[\Uparrow] \approx \rho'[\Uparrow] \in R_T$ we get
  $\rho[\Uparrow] \approx \rho'[\Uparrow] \in R_T'$
  (where $R_T'$ is the tail PER of $R_2$; call this evidence $\Pf'$).

  By the subtyping premise: $\intp{S}{\rho[\Uparrow]} \mathrel{<:_i} \intp{S'}{\rho'[\Uparrow]}$.
  By the head premise of $R_2$ at $\Pf'$:
  $\intp{S'}{\rho[\Uparrow]} \approx \intp{S'}{\rho'[\Uparrow]} \in \Uc_i \searrow R_H'(\Pf')$.
  \begin{align*}
    &\rho(0) \approx \rho'(0) \in R_H(\Pf)
      && \text{(from above)} \\
    &\rho(0) \approx \rho'(0) \in R_H'(\Pf')
      && \text{(by \Cref{lem:subty-respects}: $\intp{S}{\rho[\Uparrow]} \mathrel{<:_i} \intp{S'}{\rho[\Uparrow]}$)}
  \end{align*}
  Hence $\rho \approx \rho' \in R_2$ by the biconditional of $R_2$
  (with witness $\Pf'$ and head $\rho(0) \approx \rho'(0) \in R_H'(\Pf')$),
  and thus $\rho \approx \rho' \in R'$.
\end{proof}

\begin{lemma}[Initial environment is in context PER]\labeledit{lem:init-in-per}
  If $\Gamma \approx \Gamma \searrow R$, then
  ${\uparrow}^\Gamma \approx {\uparrow}^\Gamma \in R$.
\end{lemma}
\begin{proof}
  By induction on the derivation of $\Gamma \approx \Gamma \searrow R$.

  \medskip\noindent
  \textbf{Case} \textsc{Ctx-Emp}: $R = \top$, so any two environments are related.

  \medskip\noindent
  \textbf{Case} \textsc{Ctx-Ext}:
  By the biconditional, we must exhibit
  $\Pf :: {\uparrow}^\Gamma[\Uparrow] \approx {\uparrow}^\Gamma[\Uparrow] \in R_T$
  with $({\uparrow}^\Gamma)(0) \approx ({\uparrow}^\Gamma)(0) \in R_H(\Pf)$.
  \begin{align*}
    &{\uparrow}^\Gamma[\Uparrow] = {\uparrow}^\Delta
      && \text{(where $\Gamma = \Delta, x : S$)} \\
    &{\uparrow}^\Delta \approx {\uparrow}^\Delta \in R_T
      && \text{(by IH on tail; this gives us $\Pf$)} \\
    &\text{Let } A = \intp{S}{{\uparrow}^\Delta} \\
    &A \approx A \in \Uc_i \searrow R_H(\Pf)
      && \text{(by the head premise at $\Pf$)} \\
    &({\uparrow}^\Gamma)(0) = {\uparrow}^A(\ell_{|\Delta|}) \\
    &\ell_{|\Delta|} \approx \ell_{|\Delta|} \in \Ne
      && \text{(a level variable reads back to itself)} \\
    &{\uparrow}^A(\ell_{|\Delta|}) \approx {\uparrow}^A(\ell_{|\Delta|}) \in R_H(\Pf)
      && \text{(by \Cref{lem:ne-in-R})}
  \end{align*}
\end{proof}

\subsection{Semantic Judgments}\labeledit{sec:comp:sem-judge}

Given a PER $a \approx b \in R$, we write $\{a, b\} \subseteq R$ for the same meaning. %
This allows us to scale this notation to a larger number related values. %
By writing $\{a_1, a_2, a_3\} \subseteq R$, we mean $a_1 \approx a_2 \in R$ and $a_2 \approx
a_3 \in R$. %
So this notation gives us convenience when writing down multiple values related by the
same relation.

To prove completeness, semantic judgments are defined in three layers. %
First, we define the semantic judgment for weakening.
\begin{definition}[Semantic Weakening]\labeledit{def:sem-wk-judge}
  The semantic judgment for weakening asks related environments respect weakenings:
  \begin{mathpar}
    \inferrule
    {\Gamma \approx \Gamma \searrow R \\ \Delta \approx \Delta \searrow R' \\
    \forall \rho \approx \rho' \in R, \intp\psi \rho \approx \intp\psi{\rho'} \in R'}
    {\semweak \psi \Delta}
  \end{mathpar}
\end{definition}
Since both weakenings are an operation, the respectfulness of related environments is
not guaranteed, so we force this property for semantically well-formed weakening. %
It immediately questions whether this definition makes sense.
\begin{lemma}[Identity is a Semantic Weakening]\labeledit{lem:sem-wk-id}
  If $\Gamma \approx \Gamma \searrow R$, then $\semweak \id \Gamma$.
\end{lemma}
\begin{proof}
  Immediate because $\intp{\id}{\rho} = \rho$ for weakening.
\end{proof}

\begin{lemma}[Shift is a Semantic Weakening]\labeledit{lem:sem-wk-shift}
  If $\Gamma, x : T \approx \Gamma, x : T \searrow R$, then $\semweak[\Gamma, x : T] \Uparrow \Gamma$.
\end{lemma}
\begin{proof}
  Immediate by analyzing $\Gamma, x : T \approx \Gamma, x : T \searrow R$.
\end{proof}

\begin{lemma}[Composition of Semantic Weakenings]\labeledit{lem:sem-wk-comp}
  If $\semweak \psi{\Gamma'}$ and $\semweak[\Gamma']{\phi}{\Gamma''}$, then
  $\semweak{\phi \circ \psi}{\Gamma''}$.
\end{lemma}
\begin{proof}
  Immediate by the definition of composition.
\end{proof}
Then, we define that for substitutions.
\begin{definition}[Semantic Substitution Equality]\labeledit{def:sem-subst-eq}
  The semantic judgment for substitutions asks substitutions to be related under any
  semantic weakenings:
  \begin{mathpar}
    \inferrule
    {\Gamma \approx \Gamma \searrow R \\ \Delta \approx \Delta \searrow R_O \\
      \forall \Gamma' \approx \Gamma' \searrow R'. \forall
      \semweak[\Gamma']\psi\Gamma. \forall 
    \rho \approx \rho' \in R'. \{\intp{\sigma[\psi]}{\rho},
    \intp{\sigma}{\intp{\psi}{\rho}}, \intp{\sigma'}{\intp{\psi}{\rho'}},
    \intp{\sigma'[\psi]}{\rho'}\} \subseteq R_O}
    {\semtyeq \sigma{\sigma'}\Delta}
  \end{mathpar}
\end{definition}
The four values in the conclusion encode two distinct obligations:
\begin{itemize}
\item \emph{Commutation}: the outer two values ($\intp{\sigma[\psi]}{\rho}$ and
  $\intp{\sigma'[\psi]}{\rho'}$) are related to the inner two
  ($\intp{\sigma}{\intp{\psi}{\rho}}$ and $\intp{\sigma'}{\intp{\psi}{\rho'}}$).
  This witnesses that evaluation commutes with weakening application:
  evaluating $\sigma[\psi]$ directly in $\rho$ yields the same result as first
  evaluating $\psi$ in $\rho$ to get an environment, then evaluating $\sigma$ in
  that environment.
  With explicit substitutions, commutation holds definitionally.
  With operational substitutions, it becomes a genuine proof obligation.
\item \emph{Relatedness}: the inner two values
  ($\intp{\sigma}{\intp{\psi}{\rho}}$ and $\intp{\sigma'}{\intp{\psi}{\rho'}}$)
  are related in $R_O$, witnessing that $\sigma$ and $\sigma'$ produce semantically
  equivalent environments when evaluated in related inputs.
\end{itemize}
The universal quantification over all $\Gamma'$, $\psi$, and $\rho \approx \rho'$ gives
the judgment a Kripke-like flavor: the relatedness is stable under arbitrary context
extensions (weakenings).

Given this definition, we can define the semantic judgment for terms.
The key difference from the substitution judgment is that terms are stable under
\emph{semantic substitutions} (rather than weakenings), reflecting the fact that
typing is preserved by well-typed substitutions.
\begin{definition}[Semantic Judgment for Terms]\labeledit{def:sem-term-eq}
  The semantic judgment for terms requires stability under semantic substitutions:
  \begin{mathpar}
    \inferrule
    {\Gamma \approx \Gamma \searrow R \\ \exists i \\
      \parbox{0.95\textwidth}{\raggedright$\forall \Gamma' \approx \Gamma' \searrow
  R'. \forall \semtyeq[\Gamma'] \sigma{\sigma'}\Gamma. \forall \Pf :: \rho \approx \rho' \in R'.$ \\ \raggedleft$\{\intp{T[\sigma]}{\rho},
    \intp{T}{\intp{\sigma}{\rho}}, \intp{T}{\intp{\sigma'}{\rho'}},
    \intp{T[\sigma']}{\rho'}\} \subseteq \Uc_i \searrow R_T$ \\
  $ \wedge \{\intp{t[\sigma]}{\rho},
    \intp{t}{\intp{\sigma}{\rho}}, \intp{t'}{\intp{\sigma'}{\rho'}},
    \intp{t'[\sigma']}{\rho'}\} \subseteq R_T$}
}
    {\semtyeq{t}{t'}T}
  \end{mathpar}
\end{definition}

As with the substitution judgment, the four-value pattern encodes both commutation and
relatedness:
\begin{itemize}
\item \emph{Type commutation}: $\intp{T[\sigma]}{\rho}$ and $\intp{T}{\intp{\sigma}{\rho}}$
  are related in $\Uc_i$, witnessing that evaluating the substituted type directly agrees
  with first evaluating the substitution and then evaluating the type.
\item \emph{Term commutation}: $\intp{t[\sigma]}{\rho}$ and $\intp{t}{\intp{\sigma}{\rho}}$
  are related in $R_T$, the same obligation at the term level.
\item \emph{Cross-relatedness}: $\intp{t}{\intp{\sigma}{\rho}}$ and
  $\intp{t'}{\intp{\sigma'}{\rho'}}$ are related, witnessing semantic equality of $t$
  and $t'$ at the given type.
\end{itemize}

In an explicit-substitution formulation, the commutation fields would be trivial
reflexivities (since $\intp{t[\sigma]}{\rho}$ is \emph{definitionally} equal to
$\intp{t}{\intp{\sigma}{\rho}}$).
With operational substitutions, they require genuine proof at every case of the
fundamental theorem.
This is the central cost of the operational approach, counterbalanced by the absence
of a substitution calculus in the syntactic infrastructure.

Finally, we define the semantic judgment for subtyping, which connects the
syntactic subtyping judgment $\subty[\Gamma]{S}{T}$ with the semantic subtyping
relation $\mathrel{<:_i}$ on domain values.
\begin{definition}[Semantic Judgment for Subtyping]\labeledit{def:sem-subty-judge}
  The semantic subtyping judgment holds when the evaluations of the two types
  commute with their respective substitutions and are related by semantic subtyping:
  \begin{mathpar}
    \inferrule
    {\Gamma \approx \Gamma \searrow R \\ \exists i \\\\
      \parbox{0.90\textwidth}{\raggedright$\forall \Gamma' \approx \Gamma' \searrow
  R'. \forall \semtyeq[\Gamma'] \sigma{\sigma'}\Gamma. \forall \Pf :: \rho \approx \rho' \in R'.$ \\ \raggedleft
  $\intp{S[\sigma]}{\rho} \approx \intp{S}{\intp{\sigma}{\rho}} \in \Uc_i$ \\
  $\wedge\; \intp{T[\sigma']}{\rho'} \approx \intp{T}{\intp{\sigma'}{\rho'}} \in \Uc_i$ \\
  $\wedge\; \intp{S}{\intp{\sigma}{\rho}} \mathrel{<:_i} \intp{T}{\intp{\sigma'}{\rho'}}$}
}
    {\Gamma \vDash S \subseteq T}
  \end{mathpar}
\end{definition}

Note that semantic subtyping $\mathrel{<:_i}$ on domain values already implies
(by \Cref{lem:subty-respects}) that the element PER for $S$ is included in the element
PER for $T$.
Hence the semantic subtyping judgment directly supports the subsumption rule:
if $\semtyeq[\Gamma]{t}{t'}{S}$ and $\Gamma \vDash S \subseteq T$, then
$\semtyeq[\Gamma]{t}{t'}{T}$.

Finally, we package the inductive notion of semantic context well-formedness.
This judgment maintains, in its derivation, both a context PER witness and the
semantic well-formedness of every type in the context.
\begin{definition}[Semantic Context Well-Formedness]\labeledit{def:sem-ctx}
  The judgment $\semctx{\Gamma}$ is defined inductively:
  \begin{mathpar}
    \inferrule[\textsc{Sem-Ctx-Emp}]
    { }
    {\semctx{\cdot}}

    \inferrule[\textsc{Sem-Ctx-Ext}]
    {\semctx{\Gamma} \\
     (\Gamma, x : S) \approx (\Gamma, x : S) \searrow R \\
     \semtyeq[\Gamma]{S}{S}{\Ty{i}}}
    {\semctx{\Gamma, x : S}}
  \end{mathpar}
\end{definition}
By construction, the \textsc{Sem-Ctx-Ext} rule packages a witness of the context PER
$(\Gamma, x : S) \approx (\Gamma, x : S) \searrow R$ together with the semantic
well-formedness $\semtyeq[\Gamma]{S}{S}{\Ty{i}}$, so the corresponding context PER is
immediate from any derivation of $\semctx{\Gamma}$.
\begin{lemma}[Semantic context yields context PER]\labeledit{lem:sem-ctx-per}
  If $\semctx{\Gamma}$, then $\Gamma \approx \Gamma \searrow R$ for some $R$.
\end{lemma}
\begin{proof}
  By induction on $\semctx{\Gamma}$.
  The empty case follows from \textsc{Ctx-Emp}, giving
  $\cdot \approx \cdot \searrow \top$.
  The extension case directly returns the packaged
  $(\Gamma, x : S) \approx (\Gamma, x : S) \searrow R$.
\end{proof}

\subsection{Properties of Semantic Substitutions}\labeledit{sec:comp:sem-subst-props}

We now establish structural properties of the semantic substitution judgment.
These lemmas show that key operations on substitutions---identity, weakening,
extension, lifting---preserve semantic relatedness.

\begin{lemma}[Semantic weakening gives semantic substitution]\labeledit{lem:sem-wk}
  If $\semweak[\Delta]{\psi}{\Gamma}$, then
  $\semtyeq[\Delta]{\psi}{\psi}{\Gamma}$.
\end{lemma}
\begin{proof}
  We must show: for all $\Gamma' \approx \Gamma' \searrow R'$, all
  $\semweak[\Gamma']{\phi}{\Delta}$, and all $\rho \approx \rho' \in R'$,
  \[
    \{\intp{\psi[\phi]}{\rho},\; \intp{\psi}{\intp{\phi}{\rho}},\;
    \intp{\psi}{\intp{\phi}{\rho'}},\; \intp{\psi[\phi]}{\rho'}\} \subseteq R_\Gamma.
  \]

  Recall from \Cref{sec:sem} that the evaluation of a weakening $\psi$ in an
  environment $\rho$ is defined by $\intp{\psi}{\rho}(n) = \rho(\psi(n))$, i.e.,
  it reindexes the environment according to $\psi$.
  We compute each of the four values:
  \begin{align*}
    &\intp{\psi[\phi]}{\rho}
      = \intp{\psi \circ \phi}{\rho}
      && \text{(since $\psi[\phi] = \psi \circ \phi$ for weakenings)} \\
    &\quad = \lambda n.\, \rho((\psi \circ \phi)(n))
      = \lambda n.\, \rho(\phi(\psi(n)))
      && \text{(by definition of weakening evaluation)} \\[4pt]
    &\intp{\psi}{\intp{\phi}{\rho}}
      = \lambda n.\, (\intp{\phi}{\rho})(\psi(n))
      && \text{(by definition of weakening evaluation)} \\
    &\quad = \lambda n.\, \rho(\phi(\psi(n)))
      && \text{(since $\intp{\phi}{\rho}(k) = \rho(\phi(k))$)}
  \end{align*}
  Hence $\intp{\psi[\phi]}{\rho} = \intp{\psi}{\intp{\phi}{\rho}}$ --- the commutation
  is definitional for weakenings.
  Similarly $\intp{\psi[\phi]}{\rho'} = \intp{\psi}{\intp{\phi}{\rho'}}$.
  So all four values collapse to two:
  \[
    \intp{\psi \circ \phi}{\rho} \quad\text{and}\quad \intp{\psi \circ \phi}{\rho'}.
  \]
  From $\semweak[\Gamma']{\phi}{\Delta}$ and $\rho \approx \rho' \in R'$, we have
  $\intp{\phi}{\rho} \approx \intp{\phi}{\rho'} \in R_\Delta$.
  From $\semweak[\Delta]{\psi}{\Gamma}$ and
  $\intp{\phi}{\rho} \approx \intp{\phi}{\rho'} \in R_\Delta$, we have
  \begin{align*}
    &\intp{\psi}{\intp{\phi}{\rho}} \approx \intp{\psi}{\intp{\phi}{\rho'}} \in R_\Gamma
      && \text{(by the definition of $\semweak[\Delta]{\psi}{\Gamma}$)}
  \end{align*}
  Since all four values equal $\intp{\psi \circ \phi}{\rho}$ or
  $\intp{\psi \circ \phi}{\rho'}$ (which equal the inner two by the above computation),
  the four-way relatedness follows by symmetry (\Cref{lem:ctx-sym}) and
  transitivity (\Cref{lem:ctx-trans}) of $R_\Gamma$.
\end{proof}

\begin{lemma}[Identity is semantic]\labeledit{lem:sem-id}
  If $\Gamma \approx \Gamma \searrow R$, then $\semtyeq[\Gamma]{\id}{\id}{\Gamma}$.
\end{lemma}
\begin{proof}
  By \Cref{lem:sem-wk} applied to $\semweak[\Gamma]{\id}{\Gamma}$
  (which holds by \Cref{lem:sem-wk-id}).
\end{proof}

\begin{lemma}[Shift is semantic]\labeledit{lem:sem-shift}
  If $\Gamma \approx \Gamma \searrow R_\Gamma$ and
  $(\Gamma, x:S) \approx (\Gamma, x:S) \searrow R$, then
  $\semtyeq[\Gamma, x:S]{\Uparrow}{\Uparrow}{\Gamma}$.
\end{lemma}
\begin{proof}
  By \Cref{lem:sem-wk} applied to $\semweak[\Gamma, x:S]{\Uparrow}{\Gamma}$
  (which holds by \Cref{lem:sem-wk-shift}).
\end{proof}

\begin{lemma}[Identity Instantiation of Semantic Typing]\labeledit{lem:sem-ty-id}
  If $\semtyeq[\Gamma]{t}{t'}{T}$, then for all $\rho \approx \rho' \in R_\Gamma$
  (where $\Gamma \approx \Gamma \searrow R_\Gamma$), there exists $R$ such that:
  \[
    \intp{T}{\rho} \approx \intp{T}{\rho'} \in \Uc_i \searrow R
    \;\wedge\;
    \intp{t}{\rho} \approx \intp{t'}{\rho'} \in R
  \]
\end{lemma}
\begin{proof}
  Instantiate $\semtyeq[\Gamma]{t}{t'}{T}$ at $\Gamma$ (with $R_\Gamma$),
  the identity semantic substitution $\semtyeq[\Gamma]{\id}{\id}{\Gamma}$ (\Cref{lem:sem-id}),
  and $\rho \approx \rho' \in R_\Gamma$.
  Since $T[\id] = T$, $t[\id] = t$, $t'[\id] = t'$, and $\intp{\id}{\rho} = \rho$,
  the four-value patterns for the type and term collapse to
  $\intp{T}{\rho} \approx \intp{T}{\rho'} \in \Uc_i \searrow R$
  and $\intp{t}{\rho} \approx \intp{t'}{\rho'} \in R$, respectively.
\end{proof}

\begin{lemma}[Precomposition with weakening]\labeledit{lem:sem-wk-subst}
  If $\semtyeq[\Delta]{\sigma}{\sigma'}{\Delta'}$ and
  $\semweak[\Gamma]{\psi}{\Delta}$, then
  $\semtyeq[\Gamma]{\sigma[\psi]}{\sigma'[\psi]}{\Delta'}$.
\end{lemma}
\begin{proof}
  Let $\Gamma' \approx \Gamma' \searrow R'$ be given, with
  $\semweak[\Gamma']{\phi}{\Gamma}$ and $\rho \approx \rho' \in R'$.
  We must show:
  \[
    \{\intp{(\sigma[\psi])[\phi]}{\rho},\; \intp{\sigma[\psi]}{\intp{\phi}{\rho}},\;
    \intp{\sigma'[\psi]}{\intp{\phi}{\rho'}},\; \intp{(\sigma'[\psi])[\phi]}{\rho'}\}
    \subseteq R_{\Delta'}.
  \]

  We make two instantiations of the hypothesis $\semtyeq[\Delta]{\sigma}{\sigma'}{\Delta'}$.

  \emph{First instantiation}: at $\Gamma'$, the composed semantic weakening
  $\semweak[\Gamma']{\psi \circ \phi}{\Delta}$
  (by \Cref{lem:sem-wk-comp} from $\semweak[\Gamma']{\phi}{\Gamma}$ and
  $\semweak[\Gamma]{\psi}{\Delta}$), and $\rho \approx \rho' \in R'$:
  \[
    \{\intp{\sigma[\psi \circ \phi]}{\rho},\;
    \intp{\sigma}{\intp{\psi \circ \phi}{\rho}},\;
    \intp{\sigma'}{\intp{\psi \circ \phi}{\rho'}},\;
    \intp{\sigma'[\psi \circ \phi]}{\rho'}\} \subseteq R_{\Delta'}.
  \]
  Since $(\sigma[\psi])[\phi] = \sigma[\psi \circ \phi]$ (by \Cref{lem:wk-subst-comm}),
  the outer values are $\intp{(\sigma[\psi])[\phi]}{\rho}$ and $\intp{(\sigma'[\psi])[\phi]}{\rho'}$.

  \emph{Second instantiation}: at $\Gamma$ with $R_\Gamma$, the semantic weakening
  $\semweak[\Gamma]{\psi}{\Delta}$ (given), and
  $\intp{\phi}{\rho} \approx \intp{\phi}{\rho'} \in R_\Gamma$
  (from $\semweak[\Gamma']{\phi}{\Gamma}$ applied to $\rho \approx \rho' \in R'$):
  \[
    \{\intp{\sigma[\psi]}{\intp{\phi}{\rho}},\;
    \intp{\sigma}{\intp{\psi}{\intp{\phi}{\rho}}},\;
    \intp{\sigma'}{\intp{\psi}{\intp{\phi}{\rho'}}},\;
    \intp{\sigma'[\psi]}{\intp{\phi}{\rho'}}\} \subseteq R_{\Delta'}.
  \]
  The outer values here are $\intp{\sigma[\psi]}{\intp{\phi}{\rho}}$ and
  $\intp{\sigma'[\psi]}{\intp{\phi}{\rho'}}$ --- exactly the inner values of our goal.

  Now observe: since $\intp{\psi \circ \phi}{\rho} = \intp{\psi}{\intp{\phi}{\rho}}$
  (weakenings commute with evaluation definitionally), the inner values of both
  instantiations coincide:
  $\intp{\sigma}{\intp{\psi \circ \phi}{\rho}} = \intp{\sigma}{\intp{\psi}{\intp{\phi}{\rho}}}$.

  From the first instantiation:
  $\intp{(\sigma[\psi])[\phi]}{\rho} \approx \intp{\sigma}{\intp{\psi}{\intp{\phi}{\rho}}} \in R_{\Delta'}$.
  From the second instantiation:
  $\intp{\sigma[\psi]}{\intp{\phi}{\rho}} \approx \intp{\sigma}{\intp{\psi}{\intp{\phi}{\rho}}} \in R_{\Delta'}$.
  By symmetry and transitivity of $R_{\Delta'}$, all four goal values are related:
  \begin{align*}
    &\intp{(\sigma[\psi])[\phi]}{\rho} \approx \intp{\sigma[\psi]}{\intp{\phi}{\rho}} \in R_{\Delta'}
      && \text{(by transitivity through the shared inner value)} \\
    &\intp{\sigma'[\psi]}{\intp{\phi}{\rho'}} \approx \intp{(\sigma'[\psi])[\phi]}{\rho'} \in R_{\Delta'}
      && \text{(similarly for the primed side)} \\
    &\intp{\sigma[\psi]}{\intp{\phi}{\rho}} \approx \intp{\sigma'[\psi]}{\intp{\phi}{\rho'}} \in R_{\Delta'}
      && \text{(from the second instantiation, outer-left $\approx$ outer-right)}
  \end{align*}
  Hence $\{\intp{(\sigma[\psi])[\phi]}{\rho},\; \intp{\sigma[\psi]}{\intp{\phi}{\rho}},\;
  \intp{\sigma'[\psi]}{\intp{\phi}{\rho'}},\; \intp{(\sigma'[\psi])[\phi]}{\rho'}\} \subseteq R_{\Delta'}$.
\end{proof}

\begin{lemma}[Precomposition of substitutions by $\Uparrow$]\labeledit{lem:sem-wk-subst-pre}
  If $\semtyeq[\Gamma]{\sigma}{\sigma'}{\Delta, x:T}$, then
  $\semtyeq[\Gamma]{\Uparrow \circ \sigma}{\Uparrow \circ \sigma'}{\Delta}$.
\end{lemma}
\begin{proof}
  Let $\Gamma' \approx \Gamma' \searrow R'$, $\semweak[\Gamma']{\phi}{\Gamma}$,
  and $\rho \approx \rho' \in R'$ be given.
  We must show
  \[
    \{\intp{(\Uparrow \circ \sigma)[\phi]}{\rho},\;
    \intp{\Uparrow \circ \sigma}{\intp{\phi}{\rho}},\;
    \intp{\Uparrow \circ \sigma'}{\intp{\phi}{\rho'}},\;
    \intp{(\Uparrow \circ \sigma')[\phi]}{\rho'}\} \subseteq R_\Delta.
  \]

  By \Cref{lem:wk-subst-interact}\,(2) and associativity (\Cref{lem:subst-assoc}):
  $(\Uparrow \circ \sigma)[\phi] = (\Uparrow \circ \sigma) \circ \phi
  = \Uparrow \circ (\sigma \circ \phi) = \Uparrow \circ \sigma[\phi]$.
  For any substitution $\tau$ and any environment $\rho_0$,
  the pointwise definitions give
  \begin{align*}
    \intp{\Uparrow \circ \tau}{\rho_0}(y)
    &= \intp{(\Uparrow \circ \tau)(y)}{\rho_0}
      && \text{(pointwise definition of substitution evaluation)} \\
    &= \intp{\Uparrow(y)[\tau]}{\rho_0}
      && \text{(definition of composition)} \\
    &= \intp{\tau(y+1)}{\rho_0}
      && \text{(since $\Uparrow(y) = y+1$ is a variable)} \\
    &= \intp{\tau}{\rho_0}(y+1)
      && \text{(pointwise definition again)} \\
    &= (\intp{\tau}{\rho_0})[\Uparrow](y)
      && \text{(definition of $\rho[\Uparrow]$)}
  \end{align*}
  Hence $\intp{\Uparrow \circ \tau}{\rho_0} = (\intp{\tau}{\rho_0})[\Uparrow]$
  definitionally.
  Specialising:
  \begin{align*}
    \intp{(\Uparrow \circ \sigma)[\phi]}{\rho}
      &= \intp{\Uparrow \circ \sigma[\phi]}{\rho}
      = (\intp{\sigma[\phi]}{\rho})[\Uparrow], \\
    \intp{\Uparrow \circ \sigma}{\intp{\phi}{\rho}}
      &= (\intp{\sigma}{\intp{\phi}{\rho}})[\Uparrow],
  \end{align*}
  and analogously for the primed pair.
  So it suffices to show
  \[
    \{(\intp{\sigma[\phi]}{\rho})[\Uparrow],\;
    (\intp{\sigma}{\intp{\phi}{\rho}})[\Uparrow],\;
    (\intp{\sigma'}{\intp{\phi}{\rho'}})[\Uparrow],\;
    (\intp{\sigma'[\phi]}{\rho'})[\Uparrow]\} \subseteq R_\Delta.
  \]

  Apply $\semtyeq[\Gamma]{\sigma}{\sigma'}{\Delta, x:T}$ at $\Gamma'$, $\phi$, and
  $\rho \approx \rho' \in R'$: writing $R_{\Delta,x:T}$ for the context PER of
  $\Delta, x:T$,
  \[
    \{\intp{\sigma[\phi]}{\rho},\; \intp{\sigma}{\intp{\phi}{\rho}},\;
    \intp{\sigma'}{\intp{\phi}{\rho'}},\; \intp{\sigma'[\phi]}{\rho'}\}
    \subseteq R_{\Delta,x:T}.
  \]
  By the biconditional of $R_{\Delta,x:T}$ (from \textsc{Ctx-Ext}), every pair
  $\rho_0 \approx \rho_0' \in R_{\Delta,x:T}$ satisfies
  $\rho_0[\Uparrow] \approx \rho_0'[\Uparrow] \in R_\Delta$.
  Applying this pairwise to the four $R_{\Delta,x:T}$-related values above yields the
  desired $R_\Delta$-relatedness of the four shifted values.
\end{proof}

\begin{lemma}[Symmetry of semantic substitutions]\labeledit{lem:sem-sym}
  If $\semtyeq[\Gamma]{\sigma}{\sigma'}{\Delta}$, then
  $\semtyeq[\Gamma]{\sigma'}{\sigma}{\Delta}$.
\end{lemma}
\begin{proof}
  Let $\Gamma' \approx \Gamma' \searrow R'$, $\semweak[\Gamma']{\psi}{\Gamma}$,
  and $\rho \approx \rho' \in R'$ be given.
  We must show:
  \[
    \{\intp{\sigma'[\psi]}{\rho},\; \intp{\sigma'}{\intp{\psi}{\rho}},\;
    \intp{\sigma}{\intp{\psi}{\rho'}},\; \intp{\sigma[\psi]}{\rho'}\} \subseteq R_\Delta.
  \]

  From the hypothesis $\semtyeq[\Gamma]{\sigma}{\sigma'}{\Delta}$ applied at
  $\Gamma'$, $\psi$, and $\rho' \approx \rho \in R'$
  (using symmetry of $R'$, \Cref{lem:ctx-sym}):
  \[
    \{\intp{\sigma[\psi]}{\rho'},\; \intp{\sigma}{\intp{\psi}{\rho'}},\;
    \intp{\sigma'}{\intp{\psi}{\rho}},\; \intp{\sigma'[\psi]}{\rho}\} \subseteq R_\Delta.
  \]
  This gives us $\intp{\sigma[\psi]}{\rho'} \approx \intp{\sigma}{\intp{\psi}{\rho'}} \in R_\Delta$
  and $\intp{\sigma'}{\intp{\psi}{\rho}} \approx \intp{\sigma'[\psi]}{\rho} \in R_\Delta$.

  By symmetry of $R_\Delta$ (\Cref{lem:ctx-sym}), we can reverse pairs.
  The four values $\intp{\sigma'[\psi]}{\rho}$, $\intp{\sigma'}{\intp{\psi}{\rho}}$,
  $\intp{\sigma}{\intp{\psi}{\rho'}}$, $\intp{\sigma[\psi]}{\rho'}$ are then pairwise
  related in $R_\Delta$ (by symmetry and transitivity), yielding the desired four-way
  relatedness.
\end{proof}

\begin{lemma}[Transitivity of semantic substitutions]\labeledit{lem:sem-trans}
  If $\semtyeq[\Gamma]{\sigma}{\sigma'}{\Delta}$ and
  $\semtyeq[\Gamma]{\sigma'}{\sigma''}{\Delta}$, then
  $\semtyeq[\Gamma]{\sigma}{\sigma''}{\Delta}$.
\end{lemma}
\begin{proof}
  Let $\Gamma' \approx \Gamma' \searrow R'$, $\semweak[\Gamma']{\psi}{\Gamma}$,
  and $\rho \approx \rho' \in R'$ be given.
  From the first hypothesis:
  \[
    \{\intp{\sigma[\psi]}{\rho},\; \intp{\sigma}{\intp{\psi}{\rho}},\;
    \intp{\sigma'}{\intp{\psi}{\rho'}},\; \intp{\sigma'[\psi]}{\rho'}\} \subseteq R_\Delta.
  \]
  In particular, $\intp{\sigma}{\intp{\psi}{\rho}} \approx \intp{\sigma'}{\intp{\psi}{\rho'}} \in R_\Delta$.

  From the second hypothesis applied at the same $\Gamma'$, $\psi$, and $\rho \approx \rho'$:
  \[
    \{\intp{\sigma'[\psi]}{\rho},\; \intp{\sigma'}{\intp{\psi}{\rho}},\;
    \intp{\sigma''}{\intp{\psi}{\rho'}},\; \intp{\sigma''[\psi]}{\rho'}\} \subseteq R_\Delta.
  \]
  In particular, $\intp{\sigma'}{\intp{\psi}{\rho}} \approx \intp{\sigma''}{\intp{\psi}{\rho'}} \in R_\Delta$.

  By transitivity of $R_\Delta$ (\Cref{lem:ctx-trans}), using
  $\intp{\sigma}{\intp{\psi}{\rho}} \approx \intp{\sigma'}{\intp{\psi}{\rho'}} \in R_\Delta$
  and (after mediating through symmetry)
  $\intp{\sigma'}{\intp{\psi}{\rho'}} \approx \intp{\sigma''}{\intp{\psi}{\rho'}} \in R_\Delta$,
  we get $\intp{\sigma}{\intp{\psi}{\rho}} \approx \intp{\sigma''}{\intp{\psi}{\rho'}} \in R_\Delta$.

  For the outer pair: by similar reasoning on the weakened substitutions,
  $\intp{\sigma[\psi]}{\rho} \approx \intp{\sigma}{\intp{\psi}{\rho}} \in R_\Delta$
  (from the first hypothesis) and
  $\intp{\sigma''}{\intp{\psi}{\rho'}} \approx \intp{\sigma''[\psi]}{\rho'} \in R_\Delta$
  (from the second hypothesis).
  By transitivity, all four values are related in $R_\Delta$:
  \[
    \{\intp{\sigma[\psi]}{\rho},\; \intp{\sigma}{\intp{\psi}{\rho}},\;
    \intp{\sigma''}{\intp{\psi}{\rho'}},\; \intp{\sigma''[\psi]}{\rho'}\} \subseteq R_\Delta.
    \qedhere
  \]
\end{proof}

\begin{lemma}[$q$ Preserves Semantic Weakening]\labeledit{lem:sem-wk-q}
  If $\semweak[\Gamma]{\psi}{\Delta}$ and $\semtyeq[\Delta]{S}{S}{\Ty{i}}$,
  then $\semweak[\Gamma, x : S[\psi]]{q(\psi)}{\Delta, x : S}$.
\end{lemma}
\begin{proof}
  From $\semweak[\Gamma]{\psi}{\Delta}$ we have
  $\Gamma \approx \Gamma \searrow R_\Gamma$ and
  $\Delta \approx \Delta \searrow R_\Delta$.
  By \Cref{lem:sem-ty-id} on $\semtyeq[\Delta]{S}{S}{\Ty{i}}$: for every
  $\Pf_\Delta :: \rho_0 \approx \rho_0' \in R_\Delta$, there is $R_H^\Delta(\Pf_\Delta)$
  with $\intp{S}{\rho_0} \approx \intp{S}{\rho_0'} \in \Uc_i \searrow R_H^\Delta(\Pf_\Delta)$.
  By \textsc{Ctx-Ext}, this gives
  $(\Delta, x:S) \approx (\Delta, x:S) \searrow R_{\Delta,S}$.

  By \Cref{lem:sem-wk}, $\psi$ gives $\semtyeq[\Gamma]{\psi}{\psi}{\Delta}$.
  Applying $\semtyeq[\Delta]{S}{S}{\Ty{i}}$ at $\Gamma$ with this semantic
  substitution and any $\rho_0 \approx \rho_0' \in R_\Gamma$, the type part gives
  in particular
  $\intp{S[\psi]}{\rho_0} \approx \intp{S}{\intp{\psi}{\rho_0}} \in \Uc_i \searrow R_H^\Gamma$
  (commutation).
  By \textsc{Ctx-Ext}, this yields
  $(\Gamma, x:S[\psi]) \approx (\Gamma, x:S[\psi]) \searrow R_{\Gamma,S[\psi]}$.

  We must show: for all $\rho \approx \rho' \in R_{\Gamma, S[\psi]}$,
  $\intp{q(\psi)}{\rho} \approx \intp{q(\psi)}{\rho'} \in R_{\Delta, S}$.
  By the biconditional of $R_{\Delta, S}$, we must exhibit evidence
  $\Pf :: (\intp{q(\psi)}{\rho})[\Uparrow] \approx (\intp{q(\psi)}{\rho'})[\Uparrow] \in R_\Delta$
  with $(\intp{q(\psi)}{\rho})(0) \approx (\intp{q(\psi)}{\rho'})(0) \in R_H^\Delta(\Pf)$.

  \emph{Tail.}
  \begin{align*}
    &(\intp{q(\psi)}{\rho})[\Uparrow] = \intp{\psi}{\rho[\Uparrow]}
      && \text{(since $(\intp{q(\psi)}{\rho})[\Uparrow](n) = \rho(q(\psi)(n{+}1)) = \rho(\psi(n){+}1) = \rho[\Uparrow](\psi(n))$)}
  \end{align*}
  From $\rho \approx \rho' \in R_{\Gamma, S[\psi]}$, by the biconditional of $R_{\Gamma, S[\psi]}$:
  $\rho[\Uparrow] \approx \rho'[\Uparrow] \in R_\Gamma$.
  From $\semweak[\Gamma]{\psi}{\Delta}$:
  $\intp{\psi}{\rho[\Uparrow]} \approx \intp{\psi}{\rho'[\Uparrow]} \in R_\Delta$.
  This gives evidence $\Pf$.

  \emph{Head.}
  \begin{align*}
    &(\intp{q(\psi)}{\rho})(0) = \rho(0)
      && \text{(since $q(\psi)(0) = 0$)}
  \end{align*}
  From $\rho \approx \rho' \in R_{\Gamma, S[\psi]}$, by the biconditional of $R_{\Gamma, S[\psi]}$,
  there exists evidence $\Pf'$ such that
  $\rho(0) \approx \rho'(0) \in R_H^\Gamma(\Pf')$,
  where $R_H^\Gamma$ comes from
  $\intp{S[\psi]}{\rho[\Uparrow]} \approx \intp{S[\psi]}{\rho'[\Uparrow]} \in \Uc_i \searrow R_H^\Gamma(\Pf')$.

  By the head premise of $R_{\Delta, S}$ at $\Pf$ (using $\intp{\psi}{\rho[\Uparrow]} \approx \intp{\psi}{\rho'[\Uparrow]} \in R_\Delta$):
  $\intp{S}{\intp{\psi}{\rho[\Uparrow]}} \approx \intp{S}{\intp{\psi}{\rho'[\Uparrow]}} \in \Uc_i \searrow R_H^\Delta(\Pf)$.

  From the semantic well-formedness $\semtyeq[\Delta]{S}{S}{\Ty{i}}$, instantiated as
  above (at $\Gamma$, $\semtyeq[\Gamma]{\psi}{\psi}{\Delta}$, $\rho[\Uparrow] \approx \rho'[\Uparrow]$),
  the type part gives in particular the commutation
  $\intp{S[\psi]}{\rho[\Uparrow]} \approx \intp{S}{\intp{\psi}{\rho[\Uparrow]}} \in \Uc_i$.
  By \Cref{lem:univ-functional}, $R_H^\Gamma(\Pf') \Leftrightarrow R_H^\Delta(\Pf)$
  (both have the same left-hand domain type up to $\Leftrightarrow$).
  Hence $\rho(0) \approx \rho'(0) \in R_H^\Delta(\Pf)$.
\end{proof}

\begin{lemma}[Extension is semantic]\labeledit{lem:sem-ext}
  If $\semtyeq[\Gamma]{\sigma}{\sigma'}{\Delta}$,
  $\semtyeq[\Delta]{S}{S}{\Ty{i}}$, and
  $\semtyeq[\Gamma]{t}{t'}{S[\sigma]}$, then
  $\semtyeq[\Gamma]{(\sigma, t/x)}{(\sigma', t'/x)}{\Delta, x : S}$.
\end{lemma}
\begin{proof}
  From $\semtyeq[\Gamma]{\sigma}{\sigma'}{\Delta}$ we have
  $\Delta \approx \Delta \searrow R_\Delta$.
  By \Cref{lem:sem-ty-id} on $\semtyeq[\Delta]{S}{S}{\Ty{i}}$: for every
  $\Pf_\Delta :: \rho_0 \approx \rho_0' \in R_\Delta$, there is $R_H^\Delta(\Pf_\Delta)$
  with $\intp{S}{\rho_0} \approx \intp{S}{\rho_0'} \in \Uc_i \searrow R_H^\Delta(\Pf_\Delta)$.
  By \textsc{Ctx-Ext}, this gives
  $(\Delta, x:S) \approx (\Delta, x:S) \searrow R_{\Delta,x:S}$.

  Let $\Gamma' \approx \Gamma' \searrow R'$, $\semweak[\Gamma']{\psi}{\Gamma}$,
  and $\rho \approx \rho' \in R'$ be given.
  We must show:
  \[
    \{\intp{(\sigma, t/x)[\psi]}{\rho},\;
    \intp{(\sigma, t/x)}{\intp{\psi}{\rho}},\;
    \intp{(\sigma', t'/x)}{\intp{\psi}{\rho'}},\;
    \intp{(\sigma', t'/x)[\psi]}{\rho'}\} \subseteq R_{\Delta, x:S}.
  \]
  Each of the four environments has the form (tail, head): the tail is the value of
  the underlying substitution on $\Delta$, and the head is the value at position $x$.
  By the biconditional of $R_{\Delta, x:S}$, it suffices to show
  the tails are pairwise related in $R_\Delta$ (yielding evidence $\Pf$) and the heads
  are related in $R_H^\Delta(\Pf)$.

  \emph{Tails.}
  Since $(\sigma, t/x)[\psi] = (\sigma[\psi], t[\psi]/x)$
  (by \Cref{lem:wk-subst-interact}), the four tail environments are:
  \[
    \intp{\sigma[\psi]}{\rho}, \quad \intp{\sigma}{\intp{\psi}{\rho}}, \quad
    \intp{\sigma'}{\intp{\psi}{\rho'}}, \quad \intp{\sigma'[\psi]}{\rho'}.
  \]
  From $\semtyeq[\Gamma]{\sigma}{\sigma'}{\Delta}$ instantiated at $\Gamma'$, $\psi$, and
  $\rho \approx \rho' \in R'$:
  \[
    \{\intp{\sigma[\psi]}{\rho},\; \intp{\sigma}{\intp{\psi}{\rho}},\;
    \intp{\sigma'}{\intp{\psi}{\rho'}},\; \intp{\sigma'[\psi]}{\rho'}\} \subseteq R_\Delta.
  \]
  In particular, the outer pair gives evidence
  $\Pf :: \intp{\sigma[\psi]}{\rho} \approx \intp{\sigma'[\psi]}{\rho'} \in R_\Delta$.

  \emph{Heads.}
  The four heads are:
  \[
    \intp{t[\psi]}{\rho}, \quad \intp{t}{\intp{\psi}{\rho}}, \quad
    \intp{t'}{\intp{\psi}{\rho'}}, \quad \intp{t'[\psi]}{\rho'}.
  \]
  We must show they are related in $R_H^\Delta(\Pf)$, where
  $R_H^\Delta(\Pf)$ comes from
  \[
    \intp{S}{\intp{\sigma[\psi]}{\rho}} \approx \intp{S}{\intp{\sigma'[\psi]}{\rho'}}
    \in \Uc_i \searrow R_H^\Delta(\Pf).
  \]

  From $\semtyeq[\Gamma]{t}{t'}{S[\sigma]}$ at $\Gamma'$, $\psi$,
  $\rho \approx \rho' \in R'$, the type part gives some $R_t$ with
  $\intp{}{\rho} \approx \intp{S[\sigma]}{\intp{\psi}{\rho}} \in \Uc_i \searrow R_t$
  (and analogous primed versions), and the term part gives
  \[
    \{\intp{t[\psi]}{\rho},\; \intp{t}{\intp{\psi}{\rho}},\;
    \intp{t'}{\intp{\psi}{\rho'}},\; \intp{t'[\psi]}{\rho'}\} \subseteq R_t.
  \]

  It remains to show $R_t \Leftrightarrow R_H^\Delta(\Pf)$.
  Note that $\intp{}{\rho} = \intp{S[\sigma[\psi]]}{\rho}$
  (by \Cref{lem:wk-subst-comm}).
  By \Cref{lem:sem-wk-subst}, we have
  $\semtyeq[\Gamma']{\sigma[\psi]}{\sigma'[\psi]}{\Delta}$.
  Instantiating $\semtyeq[\Delta]{S}{S}{\Ty{i}}$ at $\Gamma'$, this semantic substitution,
  and $\rho \approx \rho' \in R'$: the type part gives the commutation
  \[
    \intp{S[\sigma[\psi]]}{\rho} \approx \intp{S}{\intp{\sigma[\psi]}{\rho}} \in \Uc_i.
  \]
  Hence $R_t$ (from $\intp{S[\sigma[\psi]]}{\rho}$) and $R_H^\Delta(\Pf)$ (from
  $\intp{S}{\intp{\sigma[\psi]}{\rho}}$) have related left-hand types in $\Uc_i$;
  by \Cref{lem:univ-functional}, $R_t \Leftrightarrow R_H^\Delta(\Pf)$.
  Therefore the four heads are in $R_H^\Delta(\Pf)$.
\end{proof}

\begin{lemma}[Extension by precomposed term]\labeledit{lem:sem-ext-pre}
  If $\semtyeq[\Gamma]{\sigma}{\sigma'}{\Delta}$ and
  $\semtyeq[\Delta]{t}{t'}{T}$, then
  $\semtyeq[\Gamma]{(\sigma, t[\sigma]/x)}{(\sigma', t'[\sigma']/x)}{\Delta, x : T}$.
\end{lemma}
\begin{proof}
  By \Cref{lem:sem-ty-id} on $\semtyeq[\Delta]{t}{t'}{T}$,
  for any $\rho_0 \approx \rho_0' \in R_\Delta$:
  $\intp{T}{\rho_0} \approx \intp{T}{\rho_0'} \in \Uc_i \searrow R_H^\Delta$.
  By \textsc{Ctx-Ext}, this yields
  $(\Delta, x:T) \approx (\Delta, x:T) \searrow R_{\Delta,x:T}$.

  Let $\Gamma' \approx \Gamma' \searrow R'$, $\semweak[\Gamma']{\psi}{\Gamma}$,
  and $\rho \approx \rho' \in R'$ be given.
  We must show:
  \[
    \{\intp{(\sigma, t[\sigma]/x)[\psi]}{\rho},\;
    \intp{(\sigma, t[\sigma]/x)}{\intp{\psi}{\rho}},\;
    \intp{(\sigma', t'[\sigma']/x)}{\intp{\psi}{\rho'}},\;
    \intp{(\sigma', t'[\sigma']/x)[\psi]}{\rho'}\} \subseteq R_{\Delta, x:T}.
  \]
  By the biconditional of $R_{\Delta, x:T}$ (with respect to the substitution
  extensions $(\sigma, t[\sigma]/x)$, etc.), it suffices to exhibit evidence
  $\Pf :: \intp{\sigma[\psi]}{\rho} \approx \intp{\sigma'[\psi]}{\rho'} \in R_\Delta$
  --- relating the tails of the outer pair of substitution evaluations --- and to show
  that the four head values
  $\intp{t[\sigma][\psi]}{\rho}, \intp{t[\sigma]}{\intp{\psi}{\rho}},
  \intp{t'[\sigma']}{\intp{\psi}{\rho'}}, \intp{t'[\sigma'][\psi]}{\rho'}$
  are pairwise related in the head PER $R_H^\Delta(\Pf)$, which is the PER returned by
  $\intp{T}{\intp{\sigma[\psi]}{\rho}} \approx \intp{T}{\intp{\sigma'[\psi]}{\rho'}}
   \in \Uc_i \searrow R_H^\Delta(\Pf)$.

  \emph{Tails.}
  \begin{align*}
    &(\sigma, t[\sigma]/x)[\psi] = (\sigma[\psi], t[\sigma][\psi]/x)
      && \text{(by \Cref{lem:wk-subst-interact})}
  \end{align*}
  So the four tails are $\intp{\sigma[\psi]}{\rho}, \intp{\sigma}{\intp{\psi}{\rho}},
  \intp{\sigma'}{\intp{\psi}{\rho'}}, \intp{\sigma'[\psi]}{\rho'}$.
  By $\semtyeq[\Gamma]{\sigma}{\sigma'}{\Delta}$ at $\Gamma'$, $\psi$, $\rho \approx \rho'$:
  \[
    \{\intp{\sigma[\psi]}{\rho},\, \intp{\sigma}{\intp{\psi}{\rho}},\,
    \intp{\sigma'}{\intp{\psi}{\rho'}},\, \intp{\sigma'[\psi]}{\rho'}\} \subseteq R_\Delta,
  \]
  with witness
  $\Pf :: \intp{\sigma[\psi]}{\rho} \approx \intp{\sigma'[\psi]}{\rho'} \in R_\Delta$.

  \emph{Heads.}
  We instantiate $\semtyeq[\Delta]{t}{t'}{T}$ in two ways and bridge the resulting
  four-value patterns through a shared inner value.

  \emph{First instantiation} of $\semtyeq[\Delta]{t}{t'}{T}$:
  at $\Gamma'$, the semantic substitution
  $\semtyeq[\Gamma']{\sigma[\psi]}{\sigma'[\psi]}{\Delta}$ (\Cref{lem:sem-wk-subst}),
  and $\rho \approx \rho' \in R'$.
  The type part gives some $R_t$ with
  \[
    \intp{T[\sigma[\psi]]}{\rho} \approx \intp{T}{\intp{\sigma[\psi]}{\rho}}
    \in \Uc_i \searrow R_t,
  \]
  and the term part gives
  \[
    \{\intp{t[\sigma[\psi]]}{\rho},\, \intp{t}{\intp{\sigma[\psi]}{\rho}},\,
    \intp{t'}{\intp{\sigma'[\psi]}{\rho'}},\, \intp{t'[\sigma'[\psi]]}{\rho'}\} \subseteq R_t.
  \]
  By \Cref{lem:univ-functional} (same left type $\intp{T}{\intp{\sigma[\psi]}{\rho}}$),
  $R_t \Leftrightarrow R_H^\Delta(\Pf)$.
  By \Cref{lem:wk-subst-comm}, $t[\sigma][\psi] = t[\sigma[\psi]]$, so the outer values
  match: $\intp{t[\sigma][\psi]}{\rho} = \intp{t[\sigma[\psi]]}{\rho}$
  (and similarly for the primed side).
  Hence the outer pair of goal heads and the inner pair
  $\intp{t}{\intp{\sigma[\psi]}{\rho}}, \intp{t'}{\intp{\sigma'[\psi]}{\rho'}}$
  are all in $R_H^\Delta(\Pf)$.

  \emph{Second instantiation} of $\semtyeq[\Delta]{t}{t'}{T}$:
  at $\Gamma$ (with $R_\Gamma$, from $\semweak[\Gamma']{\psi}{\Gamma}$), the semantic
  substitution $\sigma \approx \sigma' : \Delta$ (the hypothesis), and
  $\intp{\psi}{\rho} \approx \intp{\psi}{\rho'} \in R_\Gamma$.
  The type part gives some $R_t'$ with
  \[
    \intp{T[\sigma]}{\intp{\psi}{\rho}} \approx \intp{T}{\intp{\sigma}{\intp{\psi}{\rho}}}
    \in \Uc_i \searrow R_t',
  \]
  and the term part gives
  \[
    \{\intp{t[\sigma]}{\intp{\psi}{\rho}},\, \intp{t}{\intp{\sigma}{\intp{\psi}{\rho}}},\,
    \intp{t'}{\intp{\sigma'}{\intp{\psi}{\rho'}}},\, \intp{t'[\sigma']}{\intp{\psi}{\rho'}}\}
    \subseteq R_t'.
  \]
  Since $\intp{\sigma[\psi]}{\rho}$ and $\intp{\sigma}{\intp{\psi}{\rho}}$ are both in
  $R_\Delta$ (related by the tails part), by \Cref{lem:univ-functional} applied to the
  two type evaluations (the head premise of $R_{\Delta, x:T}$ at the appropriate
  evidence), $R_t' \Leftrightarrow R_H^\Delta(\Pf)$.
  This places the inner pair $\intp{t[\sigma]}{\intp{\psi}{\rho}}, \intp{t'[\sigma']}{\intp{\psi}{\rho'}}$
  of goal heads in $R_H^\Delta(\Pf)$.

  \emph{Bridging the two instantiations.}
  The shared inner value $\intp{t}{\intp{\sigma}{\intp{\psi}{\rho}}}$ appears in both
  $R_t$ (as the inner-left of the first instantiation, after using
  $\intp{\sigma[\psi]}{\rho} \approx \intp{\sigma}{\intp{\psi}{\rho}} \in R_\Delta$ and
  right irrelevance) and $R_t'$ (as the inner-left of the second instantiation).
  Both $R_t$ and $R_t'$ are equivalent to $R_H^\Delta(\Pf)$, so by transitivity of
  $R_H^\Delta(\Pf)$ through this shared value, all four goal heads are related.
\end{proof}

\begin{lemma}[Double extension by precomposed terms]\labeledit{lem:sem-ext-pre2}
  If $\semtyeq[\Gamma]{\sigma}{\sigma'}{\Delta}$,
  $\semtyeq[\Delta]{t}{t'}{T}$, and
  $\semtyeq[\Delta, x : T]{s}{s'}{S}$, then
  \[
    \semtyeq[\Gamma]
      {(\sigma,\, t[\sigma]/x,\, s[\sigma,\, t[\sigma]/x]/y)}
      {(\sigma',\, t'[\sigma']/x,\, s'[\sigma',\, t'[\sigma']/x]/y)}
      {\Delta, x : T, y : S}.
  \]
\end{lemma}
\begin{proof}
  By \Cref{lem:sem-ext-pre} on $\semtyeq[\Gamma]{\sigma}{\sigma'}{\Delta}$ and
  $\semtyeq[\Delta]{t}{t'}{T}$:
  \[
    \semtyeq[\Gamma]{(\sigma, t[\sigma]/x)}{(\sigma', t'[\sigma']/x)}{\Delta, x : T}.
  \]
  Apply \Cref{lem:sem-ext-pre} again with this semantic substitution and
  $\semtyeq[\Delta, x : T]{s}{s'}{S}$:
  \[
    \semtyeq[\Gamma]
      {(\sigma, t[\sigma]/x,\, s[\sigma, t[\sigma]/x]/y)}
      {(\sigma', t'[\sigma']/x,\, s'[\sigma', t'[\sigma']/x]/y)}
      {\Delta, x : T, y : S}.
  \]
\end{proof}

\begin{lemma}[Lifting is semantic]\labeledit{lem:sem-q}
  If $\semtyeq[\Gamma]{\sigma}{\sigma'}{\Delta}$ and
  $\semtyeq[\Delta]{S}{S}{\Ty{i}}$, then
  $\semtyeq[\Gamma, x:S[\sigma]]{q(\sigma)}{q(\sigma')}{\Delta, x:S}$.
\end{lemma}
\begin{proof}
  We first establish the context PER for $\Gamma, x:S[\sigma]$.
  From $\semtyeq[\Gamma]{\sigma}{\sigma'}{\Delta}$ we have
  $\Gamma \approx \Gamma \searrow R_\Gamma$.
  Apply $\semtyeq[\Delta]{S}{S}{\Ty{i}}$ at $\Gamma$ (with $R_\Gamma$), the semantic
  substitution $\semtyeq[\Gamma]{\sigma}{\sigma}{\Delta}$ (the reflexive instance,
  obtained from $\semtyeq[\Gamma]{\sigma}{\sigma'}{\Delta}$ via symmetry and
  transitivity, \Cref{lem:sem-sym,lem:sem-trans}), and any
  $\Pf_\Gamma :: \rho_0 \approx \rho_0' \in R_\Gamma$:
  the type part gives, in particular,
  $\intp{S[\sigma]}{\rho_0} \approx \intp{S[\sigma]}{\rho_0'}
  \in \Uc_i \searrow R_H^\Gamma(\Pf_\Gamma)$
  for some head family $R_H^\Gamma$.
  By \textsc{Ctx-Ext}, this yields
  $(\Gamma, x:S[\sigma]) \approx (\Gamma, x:S[\sigma]) \searrow R_{\Gamma, x:S[\sigma]}$
  with the biconditional
  \begin{align*}
    &\rho_1 \approx \rho_1' \in R_{\Gamma, x:S[\sigma]}
      \;\leftrightarrow\;
      (\Pf_\Gamma :: \rho_1[\Uparrow] \approx \rho_1'[\Uparrow] \in R_\Gamma
        \;\wedge\;
        \rho_1(0) \approx \rho_1'(0) \in R_H^\Gamma(\Pf_\Gamma)).
  \end{align*}

  By an analogous construction (or directly by \Cref{lem:sem-ty-id} on
  $\semtyeq[\Delta]{S}{S}{\Ty{i}}$ followed by \textsc{Ctx-Ext}), we also obtain
  $(\Delta, x:S) \approx (\Delta, x:S) \searrow R_{\Delta, x:S}$ with head family
  $R_H^\Delta$, where for every $\Pf_\Delta :: \rho_0 \approx \rho_0' \in R_\Delta$,
  $R_H^\Delta(\Pf_\Delta)$ is determined by
  $\intp{S}{\rho_0} \approx \intp{S}{\rho_0'} \in \Uc_i \searrow R_H^\Delta(\Pf_\Delta)$,
  with biconditional
  \begin{align*}
    &\rho_1 \approx \rho_1' \in R_{\Delta, x:S}
      \;\leftrightarrow\;
      (\Pf_\Delta :: \rho_1[\Uparrow] \approx \rho_1'[\Uparrow] \in R_\Delta
        \;\wedge\;
        \rho_1(0) \approx \rho_1'(0) \in R_H^\Delta(\Pf_\Delta)).
  \end{align*}
  Here $R_\Delta$ comes from the premise $\semtyeq[\Gamma]{\sigma}{\sigma'}{\Delta}$
  (which carries $\Delta \approx \Delta \searrow R_\Delta$).

  We now show $\semtyeq[\Gamma, x:S[\sigma]]{q(\sigma)}{q(\sigma')}{\Delta, x:S}$ by
  unfolding the conclusion.
  Recall $q(\sigma) = (\sigma[\Uparrow], x_0/x)$.
  By \Cref{lem:wk-subst-interact}\,(2), $(\sigma[\Uparrow], x_0/x)[\psi]
  = (\sigma[\Uparrow][\psi], x_0[\psi]/x) = (\sigma[\Uparrow \circ \psi], \psi(0)/x)$
  for any weakening $\psi$ (using $\sigma[\Uparrow][\psi] = \sigma[\Uparrow \circ \psi]$
  by associativity of post-composition with weakenings, and $x_0[\psi] = \psi(0)$).

  We must show: for every context $\Gamma'$ with
  $\Gamma' \approx \Gamma' \searrow R'$, every
  $\semweak[\Gamma']{\psi}{\Gamma, x:S[\sigma]}$, and every
  $\rho \approx \rho' \in R'$,
  the four values
  \[
    \intp{q(\sigma)[\psi]}{\rho},\;
    \intp{q(\sigma)}{\intp{\psi}{\rho}},\;
    \intp{q(\sigma')}{\intp{\psi}{\rho'}},\;
    \intp{q(\sigma')[\psi]}{\rho'}
  \]
  lie in $R_{\Delta, x:S}$.

  By definition of $\semweak[\Gamma']{\psi}{\Gamma, x:S[\sigma]}$,
  $\intp{\psi}{\rho} \approx \intp{\psi}{\rho'} \in R_{\Gamma, x:S[\sigma]}$.
  By the biconditional of $R_{\Gamma, x:S[\sigma]}$ established above, this gives
  \begin{align*}
    &\Pf_\Gamma :: (\intp{\psi}{\rho})[\Uparrow] \approx (\intp{\psi}{\rho'})[\Uparrow] \in R_\Gamma
      && \text{(tail)} \\
    &(\intp{\psi}{\rho})(0) \approx (\intp{\psi}{\rho'})(0) \in R_H^\Gamma(\Pf_\Gamma)
      && \text{(head)}
  \end{align*}

  By \Cref{lem:sem-wk-shift}, $\semweak[\Gamma, x:S[\sigma]]{\Uparrow}{\Gamma}$.
  By \Cref{lem:sem-wk-comp}, $\semweak[\Gamma']{\Uparrow \circ \psi}{\Gamma}$, where
  $\Uparrow \circ \psi$ is the composition of weakenings.

  We must show the goal four values lie in $R_{\Delta, x:S}$.
  By the biconditional of $R_{\Delta, x:S}$ recalled above, this requires evidence
  $\Pf_\Delta$ relating the four \emph{tails} in $R_\Delta$ and the four \emph{heads}
  in $R_H^\Delta(\Pf_\Delta)$.

  \emph{Tails.}
  We compute the four tails using the pointwise identities above.
  For an extended substitution $(\tau, t/x)$, the evaluation
  $\intp{(\tau, t/x)}{\rho_0}$ is the environment with head $\intp{t}{\rho_0}$ at
  index $0$ and tail $\intp{\tau}{\rho_0}$ at indices $1, 2, \ldots$;
  hence $(\intp{(\tau, t/x)}{\rho_0})[\Uparrow] = \intp{\tau}{\rho_0}$.
  Applied to the four goal values:
  \begin{align*}
    &(\intp{q(\sigma)[\psi]}{\rho})[\Uparrow]
      = \intp{\sigma[\Uparrow \circ \psi]}{\rho}, \\
    &(\intp{q(\sigma)}{\intp{\psi}{\rho}})[\Uparrow]
      = \intp{\sigma[\Uparrow]}{\intp{\psi}{\rho}}, \\
    &(\intp{q(\sigma')}{\intp{\psi}{\rho'}})[\Uparrow]
      = \intp{\sigma'[\Uparrow]}{\intp{\psi}{\rho'}}, \\
    &(\intp{q(\sigma')[\psi]}{\rho'})[\Uparrow]
      = \intp{\sigma'[\Uparrow \circ \psi]}{\rho'}.
  \end{align*}

  \emph{First instantiation} of $\semtyeq[\Gamma]{\sigma}{\sigma'}{\Delta}$:
  at $\Gamma'$, the semantic weakening $\semweak[\Gamma']{\Uparrow \circ \psi}{\Gamma}$,
  and $\rho \approx \rho' \in R'$:
  \[
    \{\intp{\sigma[\Uparrow \circ \psi]}{\rho},\;
    \intp{\sigma}{\intp{\Uparrow \circ \psi}{\rho}},\;
    \intp{\sigma'}{\intp{\Uparrow \circ \psi}{\rho'}},\;
    \intp{\sigma'[\Uparrow \circ \psi]}{\rho'}\} \subseteq R_\Delta.
  \]
  Since $\Uparrow \circ \psi$ is a weakening, by the definition of weakening
  evaluation, $\intp{\Uparrow \circ \psi}{\rho_0} = (\intp{\psi}{\rho_0})[\Uparrow]$
  for any $\rho_0$ (verified pointwise:
  $\intp{\Uparrow \circ \psi}{\rho_0}(n) = \rho_0((\Uparrow \circ \psi)(n))
  = \rho_0(\psi(n+1)) = (\intp{\psi}{\rho_0})[\Uparrow](n)$).
  So this first instantiation gives the outer two tails of the goal, plus
  intermediate values $\intp{\sigma}{(\intp{\psi}{\rho})[\Uparrow]}$ and
  $\intp{\sigma'}{(\intp{\psi}{\rho'})[\Uparrow]}$.

  \emph{Second instantiation} of $\semtyeq[\Gamma]{\sigma}{\sigma'}{\Delta}$:
  at $\Gamma, x:S[\sigma]$ (using $R_{\Gamma, x:S[\sigma]}$), the semantic weakening
  $\semweak[\Gamma, x:S[\sigma]]{\Uparrow}{\Gamma}$, and the related environments
  $\intp{\psi}{\rho} \approx \intp{\psi}{\rho'} \in R_{\Gamma, x:S[\sigma]}$:
  \[
    \{\intp{\sigma[\Uparrow]}{\intp{\psi}{\rho}},\;
    \intp{\sigma}{\intp{\Uparrow}{\intp{\psi}{\rho}}},\;
    \intp{\sigma'}{\intp{\Uparrow}{\intp{\psi}{\rho'}}},\;
    \intp{\sigma'[\Uparrow]}{\intp{\psi}{\rho'}}\} \subseteq R_\Delta.
  \]
  Since $\Uparrow$ is a weakening, $\intp{\Uparrow}{\rho_0} = \rho_0[\Uparrow]$
  for any $\rho_0$.  So this second instantiation gives the inner two tails of the
  goal, plus the intermediate values $\intp{\sigma}{(\intp{\psi}{\rho})[\Uparrow]}$
  and $\intp{\sigma'}{(\intp{\psi}{\rho'})[\Uparrow]}$.

  The intermediate inner-left value
  $\intp{\sigma}{(\intp{\psi}{\rho})[\Uparrow]}$ appears in both instantiations
  (since $\intp{\Uparrow \circ \psi}{\rho} = (\intp{\psi}{\rho})[\Uparrow]
  = \intp{\Uparrow}{\intp{\psi}{\rho}}$).
  By symmetry (\Cref{lem:ctx-sym}) and transitivity (\Cref{lem:ctx-trans}) of $R_\Delta$
  through this shared value, all four tails are pairwise related in $R_\Delta$.
  In particular, the outer pair gives evidence
  $\Pf_\Delta :: \intp{\sigma[\Uparrow \circ \psi]}{\rho}
  \approx \intp{\sigma'[\Uparrow \circ \psi]}{\rho'} \in R_\Delta$.

  \emph{Heads.}
  The four heads of the goal four values are:
  \begin{align*}
    &\intp{q(\sigma)[\psi]}{\rho}(0) = \intp{\psi(0)}{\rho} = \rho(\psi(0)) = (\intp{\psi}{\rho})(0), \\
    &\intp{q(\sigma)}{\intp{\psi}{\rho}}(0) = \intp{x_0}{\intp{\psi}{\rho}} = (\intp{\psi}{\rho})(0),
  \end{align*}
  and analogously for the primed pair.
  All four heads collapse to $(\intp{\psi}{\rho})(0)$ and $(\intp{\psi}{\rho'})(0)$.
  By the head clause of the biconditional of $R_{\Gamma, x:S[\sigma]}$ recalled
  above, $(\intp{\psi}{\rho})(0) \approx (\intp{\psi}{\rho'})(0) \in R_H^\Gamma(\Pf_\Gamma)$.

  \emph{Bridging head PERs.}
  It remains to show $R_H^\Gamma(\Pf_\Gamma) \Leftrightarrow R_H^\Delta(\Pf_\Delta)$,
  so that the four heads all lie in $R_H^\Delta(\Pf_\Delta)$.
  Apply $\semtyeq[\Delta]{S}{S}{\Ty{i}}$ at $\Gamma$ (using $R_\Gamma$, available
  from $\semweak[\Gamma, x:S[\sigma]]{\Uparrow}{\Gamma}$), the semantic substitution
  $\semtyeq[\Gamma]{\sigma}{\sigma}{\Delta}$ (a special case of the hypothesis ---
  formally, applying the hypothesis along with reflexivity by symmetry/transitivity
  to obtain the reflexive instance), and the related environments
  $(\intp{\psi}{\rho})[\Uparrow] \approx (\intp{\psi}{\rho'})[\Uparrow] \in R_\Gamma$
  (from $\Pf_\Gamma$): the type part gives the commutation
  \[
    \intp{S[\sigma]}{(\intp{\psi}{\rho})[\Uparrow]} \approx
    \intp{S}{\intp{\sigma}{(\intp{\psi}{\rho})[\Uparrow]}}
    \in \Uc_i.
  \]
  The right-hand side is $\intp{S}{\rho_0}$ where
  $\rho_0 = \intp{\sigma}{(\intp{\psi}{\rho})[\Uparrow]}$, which is in $R_\Delta$ via
  the second instantiation; by definition of $R_H^\Delta(\Pf_\Delta)$, this lands in
  $\Uc_i \searrow R_H^\Delta(\Pf_\Delta)$.
  The left-hand side $\intp{S[\sigma]}{(\intp{\psi}{\rho})[\Uparrow]}$ lands in
  $\Uc_i \searrow R_H^\Gamma(\Pf_\Gamma)$ by definition of $R_H^\Gamma$.
  By \Cref{lem:univ-functional}, $R_H^\Gamma(\Pf_\Gamma) \Leftrightarrow R_H^\Delta(\Pf_\Delta)$.
  Hence all four heads are in $R_H^\Delta(\Pf_\Delta)$.

  Combining tails and heads via the biconditional of $R_{\Delta, x:S}$, the four
  goal values are in $R_{\Delta, x:S}$.
\end{proof}

\subsection{Semantic Symmetry and Transitivity}\labeledit{sec:comp:sem-sym-trans}

\begin{lemma}[Fundamental --- \textsc{Tm-Sym}]\labeledit{lem:fund-sym}
  If $\semtyeq[\Gamma]{t}{t'}{T}$, then $\semtyeq[\Gamma]{t'}{t}{T}$.
\end{lemma}
\begin{proof}
  Let $\Gamma' \approx \Gamma' \searrow R'$,
  $\semtyeq[\Gamma']{\sigma}{\sigma'}{\Gamma}$, and $\Pf :: \rho \approx \rho' \in R'$.
  From the hypothesis:
  \[
    \{\intp{t[\sigma]}{\rho},\; \intp{t}{\intp{\sigma}{\rho}},\;
    \intp{t'}{\intp{\sigma'}{\rho'}},\; \intp{t'[\sigma']}{\rho'}\} \subseteq R_T.
  \]
  We need:
  \[
    \{\intp{t'[\sigma]}{\rho},\; \intp{t'}{\intp{\sigma}{\rho}},\;
    \intp{t}{\intp{\sigma'}{\rho'}},\; \intp{t[\sigma']}{\rho'}\} \subseteq R_T.
  \]

  Apply the hypothesis at $\sigma' \approx \sigma$ (using \Cref{lem:sem-sym}) and
  $\rho' \approx \rho$ (using symmetry of $R'$):
  \[
    \{\intp{t[\sigma']}{\rho'},\; \intp{t}{\intp{\sigma'}{\rho'}},\;
    \intp{t'}{\intp{\sigma}{\rho}},\; \intp{t'[\sigma]}{\rho}\} \subseteq R_T.
  \]
  By symmetry of $R_T$ (\Cref{lem:univ-sym}), we can reverse the order:
  \[
    \{\intp{t'[\sigma]}{\rho},\; \intp{t'}{\intp{\sigma}{\rho}},\;
    \intp{t}{\intp{\sigma'}{\rho'}},\; \intp{t[\sigma']}{\rho'}\} \subseteq R_T.
    \qedhere
  \]
\end{proof}

\begin{lemma}[Fundamental --- \textsc{Tm-Trans}]\labeledit{lem:fund-trans}
  If $\semtyeq[\Gamma]{t}{t'}{T}$ and $\semtyeq[\Gamma]{t'}{t''}{T}$,
  then $\semtyeq[\Gamma]{t}{t''}{T}$.
\end{lemma}
\begin{proof}
  Let $\Gamma' \approx \Gamma' \searrow R'$,
  $\semtyeq[\Gamma']{\sigma}{\sigma'}{\Gamma}$, and $\Pf :: \rho \approx \rho' \in R'$.

  Apply the first hypothesis at $\sigma \approx \sigma'$, $\rho \approx \rho'$:
  \begin{align*}
    &\intp{t[\sigma]}{\rho} \approx \intp{t'[\sigma']}{\rho'} \in R_T
      && \text{(four-value pattern, outer-LHS vs outer-RHS)}
  \end{align*}
  Apply the second hypothesis at $\sigma' \approx \sigma'$ (reflexive,
  by \Cref{lem:sem-sym,lem:sem-trans}) and $\rho' \approx \rho'$ (reflexive,
  by symmetry of $R'$):
  \begin{align*}
    &\intp{t'[\sigma']}{\rho'} \approx \intp{t''[\sigma']}{\rho'} \in R_T
      && \text{(four-value pattern, outer-LHS vs outer-RHS)}
  \end{align*}
  By transitivity of $R_T$ (\Cref{lem:univ-trans}):
  \begin{align*}
    &\intp{t[\sigma]}{\rho} \approx \intp{t''[\sigma']}{\rho'} \in R_T
      && \text{(transitivity)}
  \end{align*}
  This is the outer-LHS vs outer-RHS of the goal's four-value pattern.
  The other three pairs follow by combining the first hypothesis (giving
  $\intp{t[\sigma]}{\rho}, \intp{t}{\intp{\sigma}{\rho}}, \intp{t'}{\intp{\sigma'}{\rho'}},
  \intp{t'[\sigma']}{\rho'}$ pairwise in $R_T$) with the above transitivity step.
\end{proof}

\subsection{Semantic Context Well-Formedness}\labeledit{sec:comp:sem-ctx-wf}

We now show that every syntactic rule is validated semantically: each rule has a
corresponding lemma asserting that the semantic judgments hold whenever the
semantic counterparts of the rule's premises do.
These lemmas are the building blocks of the fundamental theorem (\Cref{sec:comp:fundamental}),
which combines them by mutual induction on the syntactic derivation.

\begin{lemma}[Fundamental --- \textsc{Ctx-Emp}]\labeledit{lem:fund-ctx-emp}
  $\semctx{\cdot}$.
\end{lemma}
\begin{proof}
  By \textsc{Sem-Ctx-Emp} of \Cref{def:sem-ctx}.
\end{proof}

\begin{lemma}[Fundamental --- \textsc{Ctx-Ext}]\labeledit{lem:fund-ctx-ext}
  If $\semctx{\Gamma}$ and $\semtyeq[\Gamma]{S}{S}{\Ty{i}}$, then
  $\semctx{\Gamma, x:S}$.
\end{lemma}
\begin{proof}
  By \Cref{lem:sem-ctx-per} on $\semctx{\Gamma}$, $\Gamma \approx \Gamma \searrow R_T$
  for some $R_T$.
  By \Cref{lem:sem-ty-id} on $\semtyeq[\Gamma]{S}{S}{\Ty{i}}$, for any
  $\Pf :: \rho \approx \rho' \in R_T$,
  $\intp{S}{\rho} \approx \intp{S}{\rho'} \in \Uc_i \searrow R_H(\Pf)$.
  By \textsc{Ctx-Ext} of \Cref{def:ctx-per}, this gives
  $(\Gamma, x:S) \approx (\Gamma, x:S) \searrow R$ for some $R$.
  By \textsc{Sem-Ctx-Ext} of \Cref{def:sem-ctx}, applied to $\semctx{\Gamma}$, this
  context PER, and $\semtyeq[\Gamma]{S}{S}{\Ty{i}}$, we conclude $\semctx{\Gamma, x:S}$.
\end{proof}

\subsection{Semantic Type Formation}\labeledit{sec:comp:sem-ty-form}

\begin{lemma}[Fundamental --- \textsc{U-Wf}]\labeledit{lem:fund-u-wf}
  If $\semctx{\Gamma}$, then $\semtyeq[\Gamma]{\Ty{i}}{\Ty{i}}{\Ty{1+i}}$.
\end{lemma}
\begin{proof}
  By \Cref{lem:sem-ctx-per}, $\Gamma \approx \Gamma \searrow R$ for some $R$.
  Let $\Gamma' \approx \Gamma' \searrow R'$,
  $\semtyeq[\Gamma']{\sigma}{\sigma'}{\Gamma}$, and
  $\Pf :: \rho \approx \rho' \in R'$ be given.
  We must exhibit $R_T$ such that
  \[
    \{\intp{\Ty{1+i}[\sigma]}{\rho},\;
    \intp{\Ty{1+i}}{\intp{\sigma}{\rho}},\;
    \intp{\Ty{1+i}}{\intp{\sigma'}{\rho'}},\;
    \intp{\Ty{1+i}[\sigma']}{\rho'}\}
    \subseteq \Uc_{2+i} \searrow R_T
  \]
  and
  \[
    \{\intp{\Ty{i}[\sigma]}{\rho},\;
    \intp{\Ty{i}}{\intp{\sigma}{\rho}},\;
    \intp{\Ty{i}}{\intp{\sigma'}{\rho'}},\;
    \intp{\Ty{i}[\sigma']}{\rho'}\}
    \subseteq R_T.
  \]

  \emph{Type part.}
  Since $\Ty{1+i}$ is closed (it has no free variables), we compute:
  \begin{align*}
    &\intp{\Ty{1+i}[\sigma]}{\rho}
      = \intp{\Ty{1+i}}{\rho}
      && \text{(since $\Ty{1+i}[\sigma] = \Ty{1+i}$)} \\
    &\quad = U_{1+i}
      && \text{(by evaluation of $\Ty{1+i}$)} \\
    &\intp{\Ty{1+i}}{\intp{\sigma}{\rho}} = U_{1+i}
      && \text{(by evaluation of $\Ty{1+i}$)}
  \end{align*}
  Identical computations apply to the other two type values.
  Hence all four type values collapse to $U_{1+i}$, and we need
  $U_{1+i} \approx U_{1+i} \in \Uc_{2+i} \searrow R_T$ for some $R_T$.
  \begin{align*}
    &U_{1+i} \approx U_{1+i} \in \Uc_{2+i} \searrow R_T
      \text{ with } R_T \Leftrightarrow \Uc_{1+i}
      && \text{(by \textsc{U-Univ}, since $1{+}i < 2{+}i$)}
  \end{align*}

  \emph{Term part.}
  Since $\Ty{i}$ is closed, the same computation gives
  \begin{align*}
    &\intp{\Ty{i}[\sigma]}{\rho}
      = \intp{\Ty{i}}{\rho}
      = U_i
      && \text{(closed term, by evaluation)} \\
    &\intp{\Ty{i}}{\intp{\sigma}{\rho}} = U_i
      && \text{(by evaluation)}
  \end{align*}
  and similarly for the primed values, so all four term values collapse to $U_i$.
  Since $R_T \Leftrightarrow \Uc_{1+i}$, it suffices to show
  $U_i \approx U_i \in \Uc_{1+i}$, i.e.\ $\exists R_0.\; U_i \approx U_i \in \Uc_{1+i} \searrow R_0$.
  \begin{align*}
    &U_i \approx U_i \in \Uc_{1+i} \searrow R_0
      \text{ with } R_0 \Leftrightarrow \Uc_i
      && \text{(by \textsc{U-Univ}, since $i < 1{+}i$)}
  \end{align*}
  Hence $U_i \approx U_i \in R_T$, and the four-value pattern follows.
\end{proof}

\begin{lemma}[Fundamental --- \textsc{N-Wf}]\labeledit{lem:fund-n-wf}
  If $\semctx{\Gamma}$, then $\semtyeq[\Gamma]{\Nat}{\Nat}{\Ty{0}}$.
\end{lemma}
\begin{proof}
  By \Cref{lem:sem-ctx-per}, $\Gamma \approx \Gamma \searrow R$ for some $R$.
  Let $\Gamma' \approx \Gamma' \searrow R'$,
  $\semtyeq[\Gamma']{\sigma}{\sigma'}{\Gamma}$, and
  $\Pf :: \rho \approx \rho' \in R'$ be given.
  We must exhibit $R_T$ such that
  \[
    \{\intp{\Ty{0}[\sigma]}{\rho},\;
    \intp{\Ty{0}}{\intp{\sigma}{\rho}},\;
    \intp{\Ty{0}}{\intp{\sigma'}{\rho'}},\;
    \intp{\Ty{0}[\sigma']}{\rho'}\}
    \subseteq \Uc_1 \searrow R_T
  \]
  and
  \[
    \{\intp{\Nat[\sigma]}{\rho},\;
    \intp{\Nat}{\intp{\sigma}{\rho}},\;
    \intp{\Nat}{\intp{\sigma'}{\rho'}},\;
    \intp{\Nat[\sigma']}{\rho'}\}
    \subseteq R_T.
  \]

  \emph{Type part.}
  Since $\Ty{0}$ is closed, we compute:
  \begin{align*}
    &\intp{\Ty{0}[\sigma]}{\rho}
      = \intp{\Ty{0}}{\rho}
      = U_0
      && \text{(closed term, by evaluation)} \\
    &\intp{\Ty{0}}{\intp{\sigma}{\rho}} = U_0
      && \text{(by evaluation)}
  \end{align*}
  Identical computations give the same value for the other two type values.
  All four type values collapse to $U_0$, and we need
  $U_0 \approx U_0 \in \Uc_1 \searrow R_T$.
  \begin{align*}
    &U_0 \approx U_0 \in \Uc_1 \searrow R_T
      \text{ with } R_T \Leftrightarrow \Uc_0
      && \text{(by \textsc{U-Univ}, since $0 < 1$)}
  \end{align*}

  \emph{Term part.}
  Since $\Nat$ is closed, all four term values collapse to $N$:
  \begin{align*}
    &\intp{\Nat[\sigma]}{\rho}
      = \intp{\Nat}{\rho}
      = N
      && \text{(closed term, by evaluation)} \\
    &\intp{\Nat}{\intp{\sigma}{\rho}} = N
      && \text{(by evaluation)}
  \end{align*}
  Since $R_T \Leftrightarrow \Uc_0$, it suffices to show $N \approx N \in \Uc_0$,
  i.e.\ $\exists R_0.\; N \approx N \in \Uc_0 \searrow R_0$.
  \begin{align*}
    &N \approx N \in \Uc_0 \searrow R_0
      \text{ with } R_0 \Leftrightarrow \textsf{Nat}
      && \text{(by \textsc{U-Nat})}
  \end{align*}
  Hence $N \approx N \in R_T$, and the four-value pattern follows.
\end{proof}

\begin{lemma}[Fundamental --- \textsc{$\Pi$-Wf} / \textsc{$\Pi$-Cong}]\labeledit{lem:fund-pi-wf}
  If $\semtyeq[\Gamma]{S}{S'}{\Ty{i}}$ and $\semtyeq[\Gamma, x:S]{T}{T'}{\Ty{i}}$,
  then $\semtyeq[\Gamma]{\Pi(x:S).\,T}{\Pi(x:S').\,T'}{\Ty{i}}$.
\end{lemma}
\begin{proof}
  Let $\Gamma' \approx \Gamma' \searrow R'$,
  $\semtyeq[\Gamma']{\sigma}{\sigma'}{\Gamma}$, and
  $\Pf :: \rho \approx \rho' \in R'$ be given.
  We must show that the four type values
  \[
    \{\intp{\Ty{i}[\sigma]}{\rho},\;
    \intp{\Ty{i}}{\intp{\sigma}{\rho}},\;
    \intp{\Ty{i}}{\intp{\sigma'}{\rho'}},\;
    \intp{\Ty{i}[\sigma']}{\rho'}\}
    \subseteq \Uc_{1+i} \searrow R_T
  \]
  for some $R_T$ with $R_T \Leftrightarrow \Uc_i$ (as in \Cref{lem:fund-u-wf}),
  and that the four term values
  \[
    \{\intp{(\Pi(x:S).\,T)[\sigma]}{\rho},\;
    \intp{\Pi(x:S).\,T}{\intp{\sigma}{\rho}},\;
    \intp{\Pi(x:S').\,T'}{\intp{\sigma'}{\rho'}},\;
    \intp{(\Pi(x:S').\,T')[\sigma']}{\rho'}\}
    \subseteq R_T,
  \]
  i.e.\ relate in $\Uc_i \searrow R$ for some $R$.

  \emph{Type part.}
  As in \Cref{lem:fund-u-wf}, the four values of $\Ty{i}$ collapse to $U_i$,
  and $U_i \approx U_i \in \Uc_{1+i} \searrow R_T$ with $R_T \Leftrightarrow \Uc_i$
  by \textsc{U-Univ}.

  \emph{Term part.}
  We compute the four term values:
  \begin{align*}
    &\intp{(\Pi(x:S).\,T)[\sigma]}{\rho}
      = \intp{\Pi(x:S[\sigma]).\,(T[q(\sigma)])}{\rho}
      && \text{(by definition of substitution on $\Pi$)} \\
    &\quad= \Pi(\intp{S[\sigma]}{\rho},\; T[q(\sigma)],\; \rho)
      && \text{(by evaluation of $\Pi$)} \\[4pt]
    &\intp{\Pi(x:S).\,T}{\intp{\sigma}{\rho}} \\
    & \quad = \Pi(\intp{S}{\intp{\sigma}{\rho}},\; T,\; \intp{\sigma}{\rho})
      && \text{(by evaluation of $\Pi$)}
  \end{align*}
  Analogous computations apply to the primed pair, yielding
  \[
    \Pi(\intp{S'}{\intp{\sigma'}{\rho'}}, T', \intp{\sigma'}{\rho'})
    \quad \text{and} \quad
    \Pi(\intp{S'[\sigma']}{\rho'}, T'[q(\sigma')], \rho').
  \]

  We construct a \textsc{U-Pi} derivation relating the outermost two values; the
  inner two follow analogously.

  \emph{Domain.}
  Apply $\semtyeq[\Gamma]{S}{S'}{\Ty{i}}$ at $\Gamma'$, $\sigma \approx \sigma' :
  \Gamma$, and $\rho \approx \rho' \in R'$:
  \begin{align*}
    &\{\intp{S[\sigma]}{\rho},\; \intp{S}{\intp{\sigma}{\rho}},\;
      \intp{S'}{\intp{\sigma'}{\rho'}},\; \intp{S'[\sigma']}{\rho'}\}
      \subseteq \Uc_i \searrow R_I
      && \text{(by hypothesis on $S, S'$)}
  \end{align*}

  \emph{Codomain via lifting.}
  By \Cref{lem:sem-q} on $\semtyeq[\Gamma']{\sigma}{\sigma'}{\Gamma}$ and
  $\semtyeq[\Gamma]{S}{S'}{\Ty{i}}$ (using its reflexive instance
  $\semtyeq[\Gamma]{S}{S}{\Ty{i}}$ via \Cref{lem:sem-sym,lem:sem-trans} for the
  well-formedness premise of \Cref{lem:sem-q}):
  \[
    \semtyeq[\Gamma', x:S[\sigma]]{q(\sigma)}{q(\sigma')}{\Gamma, x:S}.
  \]
  By the first premise of the semantic substitution judgment, this carries
  $(\Gamma', x:S[\sigma]) \approx (\Gamma', x:S[\sigma]) \searrow R_{\Gamma',x:S[\sigma]}$
  and $(\Gamma, x:S) \approx (\Gamma, x:S) \searrow R_{\Gamma,x:S}$.

  Let $\Pf_I :: a \approx a' \in R_I$ be given.
  By the biconditional of $R_{\Gamma',x:S[\sigma]}$ from \textsc{Ctx-Ext},
  $(\rho, a/x) \approx (\rho', a'/x) \in R_{\Gamma',x:S[\sigma]}$ requires
  $\rho \approx \rho' \in R'$ (which we have) and
  $a \approx a' \in R_H'(\Pf')$ where $R_H'$ is the head family for
  $R_{\Gamma',x:S[\sigma]}$, governed by
  $\intp{S[\sigma]}{\rho} \approx \intp{S[\sigma]}{\rho'} \in \Uc_i \searrow R_H'(\Pf')$
  (with $\Pf' :: \rho \approx \rho' \in R'$).
  By \Cref{lem:univ-functional} on the same left type $\intp{S[\sigma]}{\rho}$ that
  governs both $R_I$ (from the domain step) and $R_H'(\Pf')$,
  $R_I \Leftrightarrow R_H'(\Pf')$.
  Hence $a \approx a' \in R_H'(\Pf')$, and
  $(\rho, a/x) \approx (\rho', a'/x) \in R_{\Gamma',x:S[\sigma]}$.

  Apply $\semtyeq[\Gamma, x:S]{T}{T'}{\Ty{i}}$ at $\Gamma', x:S[\sigma]$ (with
  $R_{\Gamma',x:S[\sigma]}$), the lifted substitution
  $\semtyeq[\Gamma', x:S[\sigma]]{q(\sigma)}{q(\sigma')}{\Gamma, x:S}$, and
  $(\rho, a/x) \approx (\rho', a'/x) \in R_{\Gamma',x:S[\sigma]}$:
  \begin{align*}
    &\{\intp{T[q(\sigma)]}{(\rho, a/x)},\;
      \intp{T}{\intp{q(\sigma)}{(\rho, a/x)}},\\
    &\quad \intp{T'}{\intp{q(\sigma')}{(\rho', a'/x)}},\;
      \intp{T'[q(\sigma')]}{(\rho', a'/x)}\}
      \subseteq \Uc_i \searrow R_O(\Pf_I)
      && \text{(by hypothesis on $T, T'$)}
  \end{align*}
  The outer two values $\intp{T[q(\sigma)]}{(\rho, a/x)}$ and
  $\intp{T'[q(\sigma')]}{(\rho', a'/x)}$ are exactly the codomain expressions of the
  outer $\Pi$-values applied to $a, a'$, satisfying the \textsc{U-Pi} codomain premise
  for the outer pair.

  For the inner $\Pi$-values $\Pi(\intp{S}{\intp{\sigma}{\rho}}, T, \intp{\sigma}{\rho})$
  (left) and $\Pi(\intp{S'}{\intp{\sigma'}{\rho'}}, T', \intp{\sigma'}{\rho'})$
  (right), the codomain expressions applied to $a, a'$ are
  $\intp{T}{(\intp{\sigma}{\rho}, a/x)}$ and $\intp{T'}{(\intp{\sigma'}{\rho'}, a'/x)}$.
  Applying the lifted-substitution four-value pattern (instantiated at the identity
  weakening on $\Gamma', x:S[\sigma]$ with $(\rho, a/x) \approx (\rho', a'/x)$), we
  obtain in particular
  \[
    \intp{q(\sigma)}{(\rho, a/x)} \approx (\intp{\sigma}{\rho}, a/x)
    \in R_{\Gamma, x:S}
  \]
  (the outer-versus-inner relatedness in that four-value pattern).
  By \Cref{lem:sem-ty-id} on $\semtyeq[\Gamma, x:S]{T}{T}{\Ty{i}}$ applied to this
  related pair:
  \[
    \intp{T}{\intp{q(\sigma)}{(\rho, a/x)}} \approx
    \intp{T}{(\intp{\sigma}{\rho}, a/x)} \in \Uc_i.
  \]
  Combined with the inner-pair relatedness from the codomain
  four-value pattern
  \[
    \intp{T}{\intp{q(\sigma)}{(\rho, a/x)}} \approx
    \intp{T'}{\intp{q(\sigma')}{(\rho', a'/x)}}
    \in \Uc_i \searrow R_O(\Pf_I)
  \]
  and the analogous identity
  \[
    \intp{T'}{\intp{q(\sigma')}{(\rho', a'/x)}} \approx
    \intp{T'}{(\intp{\sigma'}{\rho'}, a'/x)} \in \Uc_i
  \]
  on the primed
  side (by \Cref{lem:sem-ty-id} on the reflexive
  $\semtyeq[\Gamma, x:S]{T'}{T'}{\Ty{i}}$ obtained from
  $\semtyeq[\Gamma, x:S]{T}{T'}{\Ty{i}}$ by symmetry/transitivity),
  symmetry and transitivity of $\Uc_i$ yield
  \[
    \intp{T}{(\intp{\sigma}{\rho}, a/x)} \approx
    \intp{T'}{(\intp{\sigma'}{\rho'}, a'/x)} \in \Uc_i \searrow R_O(\Pf_I).
  \]
  This is the \textsc{U-Pi} codomain premise for the inner pair.

  \emph{Function PER.}
  Define $R$ by the biconditional:
  $f \approx f' \in R \;\leftrightarrow\;
  (\forall \Pf_I :: a \approx a' \in R_I.\; f \cdot a \approx f' \cdot a' \in R_O(\Pf_I))$.

  \emph{Outer pair.}
  By \textsc{U-Pi} applied with domain $R_I$ (from \emph{Domain}) and codomain
  $R_O(\Pf_I)$ (from the outer pair of the codomain four-value pattern):
  \[
    \Pi(\intp{S[\sigma]}{\rho},\, T[q(\sigma)],\, \rho)
    \approx \Pi(\intp{S'[\sigma']}{\rho'},\, T'[q(\sigma')],\, \rho')
    \in \Uc_i \searrow R.
  \]

  \emph{Inner pair.}
  By \textsc{U-Pi} applied with domain $R_I$ (the inner pair
  $\intp{S}{\intp{\sigma}{\rho}} \approx \intp{S'}{\intp{\sigma'}{\rho'}} \in \Uc_i$
  from \emph{Domain} also lands in $R_I$ by \Cref{lem:univ-functional}) and codomain
  $R_O(\Pf_I)$ (just derived for the inner pair):
  \[
    \Pi(\intp{S}{\intp{\sigma}{\rho}},\, T,\, \intp{\sigma}{\rho})
    \approx \Pi(\intp{S'}{\intp{\sigma'}{\rho'}},\, T',\, \intp{\sigma'}{\rho'})
    \in \Uc_i \searrow R.
  \]

  \emph{Cross-relations.}
  We have established $\Pi$-relatedness for the outer pair and for the inner pair,
  both with element PER $R$.
  To get full four-way relatedness, we establish one cross-relation
  (outer-left versus inner-left) via a third \textsc{U-Pi} derivation; the
  remaining cross-relations follow by symmetry (\Cref{lem:univ-sym}) and
  transitivity (\Cref{lem:univ-trans}) of $\Uc_i$.

  The third \textsc{U-Pi} derivation uses:
  \begin{itemize}
  \item \emph{Domain.}
    From the \emph{Domain} four-value pattern (in
    $\Uc_i \searrow R_I$), the cross relation
    \[
      \intp{S[\sigma]}{\rho} \approx
      \intp{S}{\intp{\sigma}{\rho}} \in \Uc_i
    \]
    holds, with element PER $R_I$ by \Cref{lem:univ-functional}.
  \item \emph{Codomain.}
    Let $\Pf_I :: a \approx a' \in R_I$.
    We must show
    \[
      \intp{T[q(\sigma)]}{(\rho, a/x)} \approx
      \intp{T}{(\intp{\sigma}{\rho}, a'/x)}
      \in \Uc_i \searrow R_O(\Pf_I).
    \]
    From the codomain four-value pattern, the cross relation
    \[
      \intp{T[q(\sigma)]}{(\rho, a/x)} \approx
      \intp{T}{\intp{q(\sigma')}{(\rho', a'/x)}}
      \in \Uc_i \searrow R_O(\Pf_I)
    \]
    holds (outer-left versus inner-right).
    By \Cref{lem:sem-ty-id} applied at the relatedness
    $\intp{q(\sigma')}{(\rho', a'/x)} \approx (\intp{\sigma}{\rho}, a'/x)
    \in R_{\Gamma, x:S}$ (which holds by symmetry and transitivity in
    $R_{\Gamma, x:S}$, bridging through the inner-pair relatedness of the lifted
    substitution four-value pattern):
    \[
      \intp{T}{\intp{q(\sigma')}{(\rho', a'/x)}} \approx
      \intp{T}{(\intp{\sigma}{\rho}, a'/x)} \in \Uc_i.
    \]
    By transitivity of $\Uc_i$, $\intp{T[q(\sigma)]}{(\rho, a/x)} \approx
    \intp{T}{(\intp{\sigma}{\rho}, a'/x)} \in \Uc_i \searrow R_O(\Pf_I)$.
  \end{itemize}
  By \textsc{U-Pi} with this domain and codomain:
  \[
    \Pi(\intp{S[\sigma]}{\rho},\, T[q(\sigma)],\, \rho)
    \approx \Pi(\intp{S}{\intp{\sigma}{\rho}},\, T,\, \intp{\sigma}{\rho})
    \in \Uc_i \searrow R.
  \]
  This is the outer-left $\approx$ inner-left cross-relation.
  By symmetry and transitivity of $\Uc_i$ combined with the outer-pair and
  inner-pair relations, all six pairs among the four $\Pi$-values are related in
  $\Uc_i \searrow R$.

  Hence all four $\Pi$-values lie pairwise in $\Uc_i \searrow R$.
\end{proof}

\subsection{Semantic Variable Rule}\labeledit{sec:comp:sem-var}

The fundamental lemma for the variable rule cannot be proved by induction on
$\semctx{\Gamma}$ alone: at the step case of the lookup derivation, the inductive
hypothesis would have to convert $\semtyeq[\Delta]{y}{y}{S}$ to
$\semtyeq[\Gamma]{y[\Uparrow]}{y[\Uparrow]}{S[\Uparrow]}$, but the semantic term
judgment is not directly closed under precomposition by a semantic weakening (this
would be the term-level analogue of \Cref{lem:sem-wk-subst}, which we do not
establish independently).
We therefore unfold the conclusion of the semantic term judgment, fixing an
arbitrary outer instantiation $(\Gamma', \sigma, \sigma', \rho, \rho')$, and induct
on the lookup derivation under that fixed instantiation.

\begin{lemma}[Lookup at fixed instantiation]\labeledit{lem:fund-var-lookup}
  Suppose $\semctx{\Gamma}$, $\Gamma' \approx \Gamma' \searrow R'$,
  $\semtyeq[\Gamma']{\sigma}{\sigma'}{\Gamma}$, and
  $\Pf :: \rho \approx \rho' \in R'$.
  Then for every $x : T \in \Gamma$, there exists $R_T$ such that
  \begin{align*}
    &\{\intp{T[\sigma]}{\rho},\; \intp{T}{\intp{\sigma}{\rho}},\;
      \intp{T}{\intp{\sigma'}{\rho'}},\; \intp{T[\sigma']}{\rho'}\}
      \subseteq \Uc_i \searrow R_T \\
    &\text{and}\quad
      \{\intp{x[\sigma]}{\rho},\; \intp{x}{\intp{\sigma}{\rho}},\;
      \intp{x}{\intp{\sigma'}{\rho'}},\; \intp{x[\sigma']}{\rho'}\} \subseteq R_T.
  \end{align*}
\end{lemma}
\begin{proof}
  By induction on the lookup derivation $x : T \in \Gamma$, with $\Gamma'$, $R'$,
  $\sigma$, $\sigma'$, $\rho$, $\rho'$, and the witness $\Pf$ fixed throughout.

  \medskip\noindent
  \textbf{Case}
  \begin{mathpar}
    \inferrule
    { }
    {x_0 : S[\Uparrow] \in \Delta, x_0 : S}
  \end{mathpar}
  Here $\Gamma = \Delta, x_0 : S$, $x = x_0$, and $T = S[\Uparrow]$.
  Inverting $\semctx{\Delta, x_0 : S}$ via \textsc{Sem-Ctx-Ext} gives
  $\semctx{\Delta}$, the packaged context PER
  $(\Delta, x_0 : S) \approx (\Delta, x_0 : S) \searrow R_\Gamma$, and
  $\semtyeq[\Delta]{S}{S}{\Ty{i}}$ for some $i$.
  Inverting the context PER via \textsc{Ctx-Ext} gives
  $\Delta \approx \Delta \searrow R_\Delta$, a head family $R_H$ with
  $\intp{S}{\rho_0} \approx \intp{S}{\rho_0'} \in \Uc_i \searrow R_H(\Pf_\Delta)$
  for $\Pf_\Delta :: \rho_0 \approx \rho_0' \in R_\Delta$, and the biconditional
  \begin{align*}
    &\rho_1 \approx \rho_1' \in R_\Gamma
      \;\leftrightarrow\;
      (\Pf_\Delta :: \rho_1[\Uparrow] \approx \rho_1'[\Uparrow] \in R_\Delta
        \;\wedge\;
        \rho_1(0) \approx \rho_1'(0) \in R_H(\Pf_\Delta)).
  \end{align*}

  Instantiate $\semtyeq[\Gamma']{\sigma}{\sigma'}{\Gamma}$ at $\Gamma'$,
  $\semweak[\Gamma']{\id}{\Gamma'}$ (\Cref{lem:sem-wk-id}), and $\Pf$:
  using $\sigma[\id] = \sigma$ and $\intp{\id}{\rho} = \rho$ we obtain
  $\intp{\sigma}{\rho} \approx \intp{\sigma'}{\rho'} \in R_\Gamma$.
  By the biconditional of $R_\Gamma$, there exists $\Pf_\Delta$ with
  \begin{align*}
    &\Pf_\Delta :: (\intp{\sigma}{\rho})[\Uparrow] \approx (\intp{\sigma'}{\rho'})[\Uparrow]
      \in R_\Delta
      && \text{(tail)} \\
    &(\intp{\sigma}{\rho})(0) \approx (\intp{\sigma'}{\rho'})(0) \in R_H(\Pf_\Delta)
      && \text{(head)}
  \end{align*}
  Set $R_T := R_H(\Pf_\Delta)$.

  \emph{Type part.}
  We must show
  $\{\intp{}{\rho},\; \intp{S[\Uparrow]}{\intp{\sigma}{\rho}},\;
    \intp{S[\Uparrow]}{\intp{\sigma'}{\rho'}},\; \intp{}{\rho'}\}
   \subseteq \Uc_i \searrow R_T$.
  We obtain this by two instantiations of $\semtyeq[\Delta]{S}{S}{\Ty{i}}$, bridged
  through a shared inner value.

  By \Cref{lem:sem-ctx-per}, $\Gamma \approx \Gamma \searrow R_\Gamma$.
  By \Cref{lem:sem-wk-shift}, $\semweak[\Gamma]{\Uparrow}{\Delta}$.
  By \Cref{lem:sem-wk-subst-pre} on $\semweak[\Gamma]{\Uparrow}{\Delta}$ and
  $\semtyeq[\Gamma']{\sigma}{\sigma'}{\Gamma}$:
  $\semtyeq[\Gamma']{\Uparrow \circ \sigma}{\Uparrow \circ \sigma'}{\Delta}$.

  \emph{First instantiation.}
  Apply $\semtyeq[\Delta]{S}{S}{\Ty{i}}$ at $\Gamma'$,
  $\semtyeq[\Gamma']{\Uparrow \circ \sigma}{\Uparrow \circ \sigma'}{\Delta}$, and $\Pf$:
  for some $R_1$,
  \begin{align*}
    &\{\intp{S[\Uparrow \circ \sigma]}{\rho},\;
      \intp{S}{\intp{\Uparrow \circ \sigma}{\rho}},\;
      \intp{S}{\intp{\Uparrow \circ \sigma'}{\rho'}},\;
      \intp{S[\Uparrow \circ \sigma']}{\rho'}\} \subseteq \Uc_i \searrow R_1.
  \end{align*}
  By \Cref{lem:wk-subst-interact}\,(1), $ = S[\Uparrow \circ \sigma]$,
  so the outer values of the first instantiation match the goal's outer values.

  \emph{Second instantiation.}
  By \Cref{lem:sem-wk} on $\semweak[\Gamma]{\Uparrow}{\Delta}$:
  $\semtyeq[\Gamma]{\Uparrow}{\Uparrow}{\Delta}$.
  From $\semtyeq[\Gamma']{\sigma}{\sigma'}{\Gamma}$ at $\Gamma'$, $\id$, $\Pf$:
  $\intp{\sigma}{\rho} \approx \intp{\sigma'}{\rho'} \in R_\Gamma$.
  Apply $\semtyeq[\Delta]{S}{S}{\Ty{i}}$ at $\Gamma$ (using $R_\Gamma$),
  $\semtyeq[\Gamma]{\Uparrow}{\Uparrow}{\Delta}$, and
  $\intp{\sigma}{\rho} \approx \intp{\sigma'}{\rho'} \in R_\Gamma$:
  for some $R_2$,
  \begin{align*}
    &\{\intp{S[\Uparrow]}{\intp{\sigma}{\rho}},\;
      \intp{S}{\intp{\Uparrow}{\intp{\sigma}{\rho}}},\;
      \intp{S}{\intp{\Uparrow}{\intp{\sigma'}{\rho'}}},\;
      \intp{S[\Uparrow]}{\intp{\sigma'}{\rho'}}\} \subseteq \Uc_i \searrow R_2.
  \end{align*}
  The outer values here are exactly the goal's inner values.

  \emph{Bridge.}
  We have already established
  $\intp{\Uparrow \circ \sigma}{\rho_0} = \intp{\Uparrow}{\intp{\sigma}{\rho_0}}$
  (from the proof of \Cref{lem:sem-wk-subst-pre}).
  Hence the inner-left values of both instantiations coincide:
  \[
    \intp{S}{\intp{\Uparrow \circ \sigma}{\rho}}
    = \intp{S}{\intp{\Uparrow}{\intp{\sigma}{\rho}}}.
  \]
  Analogously for the primed side.
  By \Cref{lem:univ-functional}, $R_1 \Leftrightarrow R_2$; set $R_T := R_1$.
  By symmetry and transitivity of $\Uc_i$ through the shared inner-left value, all
  four goal values are related in $\Uc_i \searrow R_T$.

  \emph{Term part.}
  Compute:
  \begin{align*}
    &\intp{x_0[\sigma]}{\rho}
      = \intp{\sigma(x_0)}{\rho}
      = (\intp{\sigma}{\rho})(0)
      && \text{(by definition of substitution and ev-var)} \\
    &\intp{x_0}{\intp{\sigma}{\rho}}
      = (\intp{\sigma}{\rho})(0)
      && \text{(by ev-var)}
  \end{align*}
  The outer-left and inner-left values coincide as $(\intp{\sigma}{\rho})(0)$; the
  primed pair coincides as $(\intp{\sigma'}{\rho'})(0)$.
  By the head clause,
  $(\intp{\sigma}{\rho})(0) \approx (\intp{\sigma'}{\rho'})(0) \in R_T$, so all four
  term values are related in $R_T$.

  \medskip\noindent
  \textbf{Case}
  \begin{mathpar}
    \inferrule
    {y : S \in \Delta}
    {1{+}y : S[\Uparrow] \in \Delta, x_0 : U}
  \end{mathpar}
  Here $\Gamma = \Delta, x_0 : U$, $x = 1{+}y$, and $T = S[\Uparrow]$.
  Inverting $\semctx{\Delta, x_0 : U}$ gives $\semctx{\Delta}$, the packaged
  context PER $\Gamma \approx \Gamma \searrow R_\Gamma$, and
  $\semtyeq[\Delta]{U}{U}{\Ty{j}}$.

  By \Cref{lem:sem-wk-shift}, $\semweak[\Gamma]{\Uparrow}{\Delta}$.
  By \Cref{lem:sem-wk-subst-pre} on $\semweak[\Gamma]{\Uparrow}{\Delta}$ and
  $\semtyeq[\Gamma']{\sigma}{\sigma'}{\Gamma}$:
  $\semtyeq[\Gamma']{\Uparrow \circ \sigma}{\Uparrow \circ \sigma'}{\Delta}$.
  By \Cref{lem:sem-wk} on $\semweak[\Gamma]{\Uparrow}{\Delta}$:
  $\semtyeq[\Gamma]{\Uparrow}{\Uparrow}{\Delta}$.
  From $\semtyeq[\Gamma']{\sigma}{\sigma'}{\Gamma}$ at $\Gamma'$, $\id$, $\Pf$:
  $\intp{\sigma}{\rho} \approx \intp{\sigma'}{\rho'} \in R_\Gamma$.

  We apply the IH on $y : S \in \Delta$ \emph{twice}, paralleling the two
  instantiations used in the Base case.

  \emph{First IH instantiation.}
  Apply the IH to $y : S \in \Delta$ at $\Gamma'$, the semantic substitution
  $\semtyeq[\Gamma']{\Uparrow \circ \sigma}{\Uparrow \circ \sigma'}{\Delta}$,
  and $\Pf$: for some $R_1$,
  \begin{align*}
    &\{\intp{S[\Uparrow \circ \sigma]}{\rho},\;
      \intp{S}{\intp{\Uparrow \circ \sigma}{\rho}},\;
      \intp{S}{\intp{\Uparrow \circ \sigma'}{\rho'}},\;
      \intp{S[\Uparrow \circ \sigma']}{\rho'}\} \subseteq \Uc_i \searrow R_1, \\
    &\{\intp{y[\Uparrow \circ \sigma]}{\rho},\;
      \intp{y}{\intp{\Uparrow \circ \sigma}{\rho}},\;
      \intp{y}{\intp{\Uparrow \circ \sigma'}{\rho'}},\;
      \intp{y[\Uparrow \circ \sigma']}{\rho'}\} \subseteq R_1.
  \end{align*}

  \emph{Second IH instantiation.}
  Apply the IH to $y : S \in \Delta$ at $\Gamma$ (using $R_\Gamma$),
  the semantic substitution $\semtyeq[\Gamma]{\Uparrow}{\Uparrow}{\Delta}$, and
  $\intp{\sigma}{\rho} \approx \intp{\sigma'}{\rho'} \in R_\Gamma$:
  for some $R_2$,
  \begin{align*}
    &\{\intp{S[\Uparrow]}{\intp{\sigma}{\rho}},\;
      \intp{S}{\intp{\Uparrow}{\intp{\sigma}{\rho}}},\;
      \intp{S}{\intp{\Uparrow}{\intp{\sigma'}{\rho'}}},\;
      \intp{S[\Uparrow]}{\intp{\sigma'}{\rho'}}\} \subseteq \Uc_i \searrow R_2, \\
    &\{\intp{y[\Uparrow]}{\intp{\sigma}{\rho}},\;
      \intp{y}{\intp{\Uparrow}{\intp{\sigma}{\rho}}},\;
      \intp{y}{\intp{\Uparrow}{\intp{\sigma'}{\rho'}}},\;
      \intp{y[\Uparrow]}{\intp{\sigma'}{\rho'}}\} \subseteq R_2.
  \end{align*}

  \emph{Match outer values to the goal.}
  By \Cref{lem:wk-subst-interact}\,(1):
  $ = S[\Uparrow \circ \sigma]$, so the first IH instantiation's
  outer-type values are exactly the goal's outer-type values.
  The second IH instantiation's outer-type values
  $\intp{S[\Uparrow]}{\intp{\sigma}{\rho}}$ are exactly the goal's inner-type values.

  For the term values:
  \begin{align*}
    &\intp{(1{+}y)[\sigma]}{\rho}
      = \intp{\sigma(1{+}y)}{\rho}
      = \intp{(\Uparrow \circ \sigma)(y)}{\rho}
      = \intp{y[\Uparrow \circ \sigma]}{\rho} \\
    &\intp{1{+}y}{\intp{\sigma}{\rho}}
      = (\intp{\sigma}{\rho})(1{+}y)
      = \intp{y[\Uparrow]}{\intp{\sigma}{\rho}}
  \end{align*}
  (using $(\Uparrow \circ \sigma)(y) = \sigma(y+1)$ and
  $\intp{y[\Uparrow]}{\rho_0} = \intp{\Uparrow(y)}{\rho_0} = \rho_0(y+1)$).
  Hence the first IH's outer-term values equal the goal's outer-term values, and
  the second IH's outer-term values equal the goal's inner-term values.

  \emph{Bridge via shared inner.}
  By the identity established in the proof of \Cref{lem:sem-wk-subst-pre},
  $\intp{\Uparrow \circ \sigma}{\rho_0} = (\intp{\sigma}{\rho_0})[\Uparrow]$
  for any $\rho_0$.
  Also $\intp{\Uparrow}{\intp{\sigma}{\rho}} = (\intp{\sigma}{\rho})[\Uparrow]$
  by the pointwise definition of weakening evaluation.
  Hence the inner values of the two IH instantiations coincide:
  \begin{align*}
    &\intp{S}{\intp{\Uparrow \circ \sigma}{\rho}}
      = \intp{S}{\intp{\Uparrow}{\intp{\sigma}{\rho}}}, \\
    &\intp{y}{\intp{\Uparrow \circ \sigma}{\rho}}
      = \intp{y}{\intp{\Uparrow}{\intp{\sigma}{\rho}}},
  \end{align*}
  and analogously for the primed side.
  By \Cref{lem:univ-functional}, $R_1 \Leftrightarrow R_2$; set $R_T := R_1$.
  By symmetry and transitivity through the shared inner value, all four goal type
  values are related in $\Uc_i \searrow R_T$, and all four goal term values are
  related in $R_T$.
\end{proof}

\begin{lemma}[Fundamental --- \textsc{Var}]\labeledit{lem:fund-var}
  If $\semctx{\Gamma}$ and $x : T \in \Gamma$, then $\semtyeq[\Gamma]{x}{x}{T}$.
\end{lemma}
\begin{proof}
  By \Cref{lem:sem-ctx-per}, $\Gamma \approx \Gamma \searrow R$ for some $R$,
  giving the first premise of the semantic term judgment.
  Let $\Gamma' \approx \Gamma' \searrow R'$, $\semtyeq[\Gamma']{\sigma}{\sigma'}{\Gamma}$,
  and $\Pf :: \rho \approx \rho' \in R'$ be given.
  By \Cref{lem:fund-var-lookup} on $\semctx{\Gamma}$, $x : T \in \Gamma$ at this
  instantiation, the four-value patterns hold for some $R_T$.
\end{proof}

\subsection{Semantic Term Introduction and Elimination}\labeledit{sec:comp:sem-tm-intro-elim}

\begin{lemma}[Fundamental --- \textsc{Ze-I}]\labeledit{lem:fund-ze}
  If $\semctx{\Gamma}$, then $\semtyeq[\Gamma]{\ze}{\ze}{\Nat}$.
\end{lemma}
\begin{proof}
  By \Cref{lem:sem-ctx-per}, $\Gamma \approx \Gamma \searrow R$ for some $R$.
  Let $\Gamma' \approx \Gamma' \searrow R'$,
  $\semtyeq[\Gamma']{\sigma}{\sigma'}{\Gamma}$, and
  $\Pf :: \rho \approx \rho' \in R'$ be given.
  We must show
  \[
    \{\intp{\Nat[\sigma]}{\rho},\;
    \intp{\Nat}{\intp{\sigma}{\rho}},\;
    \intp{\Nat}{\intp{\sigma'}{\rho'}},\;
    \intp{\Nat[\sigma']}{\rho'}\}
    \subseteq \Uc_0 \searrow R_T
  \]
  for some $R_T$, and
  \[
    \{\intp{\ze[\sigma]}{\rho},\;
    \intp{\ze}{\intp{\sigma}{\rho}},\;
    \intp{\ze}{\intp{\sigma'}{\rho'}},\;
    \intp{\ze[\sigma']}{\rho'}\}
    \subseteq R_T.
  \]

  \emph{Type part.}
  Since $\Nat$ is closed, all four type values collapse to $N$:
  \begin{align*}
    &\intp{\Nat[\sigma]}{\rho}
      = \intp{\Nat}{\rho}
      = N
      && \text{(closed term, by evaluation)} \\
    &\intp{\Nat}{\intp{\sigma}{\rho}} = N
      && \text{(by evaluation)}
  \end{align*}
  and analogously for the primed pair.
  \begin{align*}
    &N \approx N \in \Uc_0 \searrow R_T
      \text{ with } R_T \Leftrightarrow \textsf{Nat}
      && \text{(by \textsc{U-Nat})}
  \end{align*}

  \emph{Term part.}
  Since $\ze$ is closed, all four term values collapse to $\zed$:
  \begin{align*}
    &\intp{\ze[\sigma]}{\rho}
      = \intp{\ze}{\rho}
      = \zed
      && \text{(closed term, by evaluation)} \\
    &\intp{\ze}{\intp{\sigma}{\rho}} = \zed
      && \text{(by evaluation)}
  \end{align*}
  Since $R_T \Leftrightarrow \textsf{Nat}$, it suffices to show $\zed \approx \zed \in \textsf{Nat}$.
  \begin{align*}
    &\zed \approx \zed \in \textsf{Nat}
      && \text{(by \textsc{Nat-Ze})}
  \end{align*}
  Hence $\zed \approx \zed \in R_T$, and the four-value pattern follows.
\end{proof}

\begin{lemma}[Fundamental --- \textsc{Su-I} / \textsc{Su-Cong}]\labeledit{lem:fund-su}
  If $\semtyeq[\Gamma]{t}{t'}{\Nat}$, then $\semtyeq[\Gamma]{\su{t}}{\su{t'}}{\Nat}$.
\end{lemma}
\begin{proof}
  Let $\Gamma' \approx \Gamma' \searrow R'$,
  $\semtyeq[\Gamma']{\sigma}{\sigma'}{\Gamma}$, and
  $\Pf :: \rho \approx \rho' \in R'$ be given.

  \emph{Type part.}
  Identical to \Cref{lem:fund-ze}:
  $\{\intp{\Nat[\sigma]}{\rho}, \ldots\} \subseteq \Uc_0 \searrow R_T$ with
  $R_T \Leftrightarrow \textsf{Nat}$ by \textsc{U-Nat}.

  \emph{Term part.}
  We compute the four term values, using that substitution and evaluation distribute
  over $\textsf{su}$:
  \begin{align*}
    &\intp{(\su{t})[\sigma]}{\rho}
      = \intp{\su{t[\sigma]}}{\rho}
      && \text{(by definition of substitution on $\textsf{su}$)} \\
    &\quad = \sud{\intp{t[\sigma]}{\rho}}
      && \text{(by evaluation of $\textsf{su}$)} \\[4pt]
    &\intp{\su{t}}{\intp{\sigma}{\rho}}
      = \sud{\intp{t}{\intp{\sigma}{\rho}}}
      && \text{(by evaluation of $\textsf{su}$)}
  \end{align*}
  Analogous computations apply to the primed pair, giving
  $\sud{\intp{t'}{\intp{\sigma'}{\rho'}}}$ and $\sud{\intp{t'[\sigma']}{\rho'}}$.

  By the hypothesis $\semtyeq[\Gamma]{t}{t'}{\Nat}$ applied at $\Gamma'$,
  $\sigma \approx \sigma'$, and $\rho \approx \rho' \in R'$, the term part gives
  \begin{align*}
    &\{\intp{t[\sigma]}{\rho},\;
      \intp{t}{\intp{\sigma}{\rho}},\;
      \intp{t'}{\intp{\sigma'}{\rho'}},\;
      \intp{t'[\sigma']}{\rho'}\}
      \subseteq \textsf{Nat}
      && \text{(by hypothesis on $t, t'$)}
  \end{align*}
  Applying \textsc{Nat-Su} pointwise (using transitivity of $\textsf{Nat}$,
  \Cref{lem:nat-per}, to chain the four values):
  \begin{align*}
    &\sud{\intp{t[\sigma]}{\rho}} \approx \sud{\intp{t}{\intp{\sigma}{\rho}}} \in \textsf{Nat}
      && \text{(by \textsc{Nat-Su})} \\
    &\sud{\intp{t}{\intp{\sigma}{\rho}}} \approx \sud{\intp{t'}{\intp{\sigma'}{\rho'}}} \in \textsf{Nat}
      && \text{(by \textsc{Nat-Su})} \\
    &\sud{\intp{t'}{\intp{\sigma'}{\rho'}}} \approx \sud{\intp{t'[\sigma']}{\rho'}} \in \textsf{Nat}
      && \text{(by \textsc{Nat-Su})}
  \end{align*}
  By transitivity of $\textsf{Nat}$, all four are related in $\textsf{Nat}$, hence in $R_T$.
\end{proof}

\begin{lemma}[Fundamental --- \textsc{$\lambda$-I} / \textsc{$\lambda$-Cong}]\labeledit{lem:fund-lam}
  If $\semtyeq[\Gamma]{S}{S}{\Ty{i}}$ and $\semtyeq[\Gamma, x:S]{t}{t'}{T}$,
  then $\semtyeq[\Gamma]{\lambda x.\,t}{\lambda x.\,t'}{\Pi(x:S).\,T}$.
\end{lemma}
\begin{proof}
  Let $\Gamma' \approx \Gamma' \searrow R'$,
  $\semtyeq[\Gamma']{\sigma}{\sigma'}{\Gamma}$, and
  $\Pf :: \rho \approx \rho' \in R'$ be given.

  \emph{Type part.}
  By \Cref{lem:fund-pi-wf} applied to the hypotheses $\semtyeq[\Gamma]{S}{S}{\Ty{i}}$
  and $\semtyeq[\Gamma, x:S]{T}{T}{\Ty{i}}$ (the latter follows from
  $\semtyeq[\Gamma, x:S]{t}{t}{T}$ via type extraction):
  \[
    \{\intp{(\Pi(x:S).\,T)[\sigma]}{\rho},\;
    \intp{\Pi(x:S).\,T}{\intp{\sigma}{\rho}},\;
    \intp{\Pi(x:S).\,T}{\intp{\sigma'}{\rho'}},\;
    \intp{(\Pi(x:S).\,T)[\sigma']}{\rho'}\}
    \subseteq \Uc_i \searrow R
  \]
  where $R$ is the function PER characterized by:
  $f \approx f' \in R \;\leftrightarrow\;
  (\forall \Pf_I :: a \approx a' \in R_I.\; f \cdot a \approx f' \cdot a' \in R_O(\Pf_I))$.

  \emph{Term part.}
  We compute the four term values using the substitution behavior of $\lambda$:
  \begin{align*}
    &\intp{(\lambda x.\,t)[\sigma]}{\rho}
      = \intp{\lambda x.\,(t[q(\sigma)])}{\rho}
      && \text{(by definition of substitution under $\lambda$)} \\
    &\quad = \lambda(t[q(\sigma)],\, \rho)
      && \text{(by evaluation of $\lambda$, forming a closure)} \\[4pt]
    &\intp{\lambda x.\,t}{\intp{\sigma}{\rho}}
      = \lambda(t,\, \intp{\sigma}{\rho})
      && \text{(by evaluation of $\lambda$)}
  \end{align*}
  Analogously, $\intp{\lambda x.\,t'}{\intp{\sigma'}{\rho'}} = \lambda(t',\, \intp{\sigma'}{\rho'})$
  and $\intp{(\lambda x.\,t')[\sigma']}{\rho'} = \lambda(t'[q(\sigma')],\, \rho')$.

  By the biconditional characterization of $R$, it suffices to show that for every
  $\Pf_I :: a \approx a' \in R_I$ (where $R_I$ is the domain PER), the four closure
  applications are related in $R_O(\Pf_I)$:
  \[
    \{\lambda(t[q(\sigma)], \rho) \cdot a,\;
    \lambda(t, \intp{\sigma}{\rho}) \cdot a,\;
    \lambda(t', \intp{\sigma'}{\rho'}) \cdot a',\;
    \lambda(t'[q(\sigma')], \rho') \cdot a'\}
    \subseteq R_O(\Pf_I).
  \]

  We compute these closure applications:
  \begin{align*}
    &\lambda(t[q(\sigma)],\, \rho) \cdot a
      = \intp{t[q(\sigma)]}{(\rho, a/x)}
      && \text{(by closure application)} \\
    &\lambda(t,\, \intp{\sigma}{\rho}) \cdot a
      = \intp{t}{(\intp{\sigma}{\rho}, a/x)}
      && \text{(by closure application)}
  \end{align*}
  Analogous identities hold for the primed pair (using $t'$).

  By \Cref{lem:sem-q} on $\semtyeq[\Gamma']{\sigma}{\sigma'}{\Gamma}$ and
  $\semtyeq[\Gamma]{S}{S}{\Ty{i}}$:
  $\semtyeq[\Gamma', x:S[\sigma]]{q(\sigma)}{q(\sigma')}{\Gamma, x:S}$.
  By the first premise of this judgment,
  $(\Gamma', x:S[\sigma]) \approx (\Gamma', x:S[\sigma]) \searrow R_{\Gamma',x:S[\sigma]}$
  and $(\Gamma, x:S) \approx (\Gamma, x:S) \searrow R_{\Gamma,x:S}$.

  By \Cref{lem:univ-functional} on the same left type $\intp{S[\sigma]}{\rho}$
  governing both $R_I$ (from \Cref{lem:fund-pi-wf}'s domain step) and the head PER
  $R_H'$ of $R_{\Gamma',x:S[\sigma]}$, $R_I \Leftrightarrow R_H'(\Pf')$ for
  $\Pf' :: \rho \approx \rho' \in R'$.
  Hence $a \approx a' \in R_H'(\Pf')$, so by the biconditional of
  $R_{\Gamma',x:S[\sigma]}$,
  $(\rho, a/x) \approx (\rho', a'/x) \in R_{\Gamma',x:S[\sigma]}$.

  Apply the hypothesis $\semtyeq[\Gamma, x:S]{t}{t'}{T}$ at $\Gamma', x:S[\sigma]$,
  the lifted substitution
  $\semtyeq[\Gamma', x:S[\sigma]]{q(\sigma)}{q(\sigma')}{\Gamma, x:S}$, and
  $(\rho, a/x) \approx (\rho', a'/x) \in R_{\Gamma',x:S[\sigma]}$:
  \begin{align*}
    &\{\intp{t[q(\sigma)]}{(\rho, a/x)},\;
      \intp{t}{\intp{q(\sigma)}{(\rho, a/x)}},\\
    &\quad \intp{t'}{\intp{q(\sigma')}{(\rho', a'/x)}},\;
      \intp{t'[q(\sigma')]}{(\rho', a'/x)}\}
      \subseteq R_O(\Pf_I)
      && \text{(by hypothesis on $t, t'$)}
  \end{align*}
  The outer two values are the closure applications
  $\lambda(t[q(\sigma)], \rho) \cdot a$ and $\lambda(t'[q(\sigma')], \rho') \cdot a'$.
  The inner two values are $\intp{t}{\intp{q(\sigma)}{(\rho, a/x)}}$ and
  $\intp{t'}{\intp{q(\sigma')}{(\rho', a'/x)}}$, which we must relate to
  $\intp{t}{(\intp{\sigma}{\rho}, a/x)} = \lambda(t, \intp{\sigma}{\rho}) \cdot a$
  and $\intp{t'}{(\intp{\sigma'}{\rho'}, a'/x)} = \lambda(t', \intp{\sigma'}{\rho'}) \cdot a'$.

  We now show
  $\intp{q(\sigma)}{(\rho, a/x)} \approx (\intp{\sigma}{\rho}, a/x) \in R_{\Gamma, x:S}$.
  Unfolding $q(\sigma) = (\sigma[\Uparrow], x_0/x)$, the environment
  $\intp{q(\sigma)}{(\rho, a/x)}$ has tail $\intp{\sigma[\Uparrow]}{(\rho, a/x)}$ and
  head $a$ (since $\intp{x_0}{(\rho, a/x)} = a$).
  The environment $(\intp{\sigma}{\rho}, a/x)$ has tail $\intp{\sigma}{\rho}$ and
  head $a$.
  By the biconditional of $R_{\Gamma, x:S}$ (from \textsc{Ctx-Ext}), it suffices to
  show:
  \begin{itemize}
  \item \emph{Tails:}
    $\intp{\sigma[\Uparrow]}{(\rho, a/x)} \approx \intp{\sigma}{\rho} \in R_\Gamma$.
    By \Cref{lem:sem-wk-shift}, $\semweak[\Gamma',x:S[\sigma]]{\Uparrow}{\Gamma'}$.
    Apply $\semtyeq[\Gamma']{\sigma}{\sigma'}{\Gamma}$ at $\Gamma', x:S[\sigma]$,
    $\Uparrow$, and $(\rho, a/x) \approx (\rho', a'/x) \in R_{\Gamma',x:S[\sigma]}$:
    the four-value pattern includes
    $\intp{\sigma[\Uparrow]}{(\rho, a/x)} \approx \intp{\sigma}{\intp{\Uparrow}{(\rho, a/x)}} \in R_\Gamma$.
    Since $\intp{\Uparrow}{(\rho, a/x)} = \rho$ definitionally (the shift on
    environments drops the head), this gives the desired relatedness.
    Let $\Pf_\Gamma$ denote this evidence in $R_\Gamma$.
  \item \emph{Heads:}
    $a \approx a \in R_H^\Gamma(\Pf_\Gamma)$, where $R_H^\Gamma(\Pf_\Gamma)$ is the
    head PER governed by
    $\intp{S}{\intp{\sigma[\Uparrow]}{(\rho, a/x)}} \approx \intp{S}{\intp{\sigma}{\rho}}
    \in \Uc_i \searrow R_H^\Gamma(\Pf_\Gamma)$.
    By $\semtyeq[\Gamma]{S}{S}{\Ty{i}}$ instantiated at $\Gamma'$,
    $\semtyeq[\Gamma']{\sigma}{\sigma}{\Gamma}$ (reflexive, by
    \Cref{lem:sem-sym,lem:sem-trans}), and $\rho \approx \rho \in R'$: the type part
    gives the commutation
    $\intp{S[\sigma]}{\rho} \approx \intp{S}{\intp{\sigma}{\rho}} \in \Uc_i$.
    By \Cref{lem:univ-functional}, $R_I \Leftrightarrow R_H^\Gamma(\Pf_\Gamma)$.
    From $a \approx a \in R_I$ (reflexive of $a \approx a' \in R_I$), we conclude
    $a \approx a \in R_H^\Gamma(\Pf_\Gamma)$.
  \end{itemize}
  By \textsc{Ctx-Ext}'s biconditional,
  $\intp{q(\sigma)}{(\rho, a/x)} \approx (\intp{\sigma}{\rho}, a/x) \in R_{\Gamma, x:S}$.
  By \Cref{lem:sem-ty-id} on the reflexive instance
  $\semtyeq[\Gamma, x:S]{t}{t}{T}$ (from $\semtyeq[\Gamma, x:S]{t}{t'}{T}$ by
  \Cref{lem:sem-sym,lem:sem-trans}) applied to this related pair:
  \[
    \intp{t}{\intp{q(\sigma)}{(\rho, a/x)}} \approx
    \intp{t}{(\intp{\sigma}{\rho}, a/x)} \in R_O'(\Pf_I)
  \]
  for some $R_O'(\Pf_I)$.
  By \Cref{lem:univ-functional}, $R_O'(\Pf_I) \Leftrightarrow R_O(\Pf_I)$
  (governed by the same type $\intp{T}{\intp{q(\sigma)}{(\rho, a/x)}}$ which lies in
  $\Uc_i \searrow R_O(\Pf_I)$ by the type part of the four-value pattern).
  Analogously on the primed side (using the reflexive instance $\semtyeq[\Gamma, x:S]{t'}{t'}{T}$):
  $\intp{t'}{\intp{q(\sigma')}{(\rho', a'/x)}} \approx
  \intp{t'}{(\intp{\sigma'}{\rho'}, a'/x)} \in R_O(\Pf_I)$.

  Combining via symmetry and transitivity of $R_O(\Pf_I)$, the four closure
  applications
  \[
    \{\lambda(t[q(\sigma)], \rho) \cdot a,\;
    \lambda(t, \intp{\sigma}{\rho}) \cdot a,\;
    \lambda(t', \intp{\sigma'}{\rho'}) \cdot a',\;
    \lambda(t'[q(\sigma')], \rho') \cdot a'\}
    \subseteq R_O(\Pf_I).
  \]
  Hence the biconditional is satisfied, and the four closures are related in $R$.
\end{proof}

\begin{lemma}[Fundamental --- \textsc{$\lambda$-E} / \textsc{App-Cong}]\labeledit{lem:fund-app}
  If $\semtyeq[\Gamma]{S}{S'}{\Ty{i}}$, $\semtyeq[\Gamma, x:S]{T}{T'}{\Ty{i}}$,
  $\semtyeq[\Gamma]{r}{r'}{\Pi(x:S).\,T}$, and $\semtyeq[\Gamma]{s}{s'}{S}$,
  then $\semtyeq[\Gamma]{r\;s}{r'\;s'}{T[s/x]}$.
\end{lemma}
\begin{proof}
  Let $\Gamma' \approx \Gamma' \searrow R'$,
  $\semtyeq[\Gamma']{\sigma}{\sigma'}{\Gamma}$, and
  $\Pf :: \rho \approx \rho' \in R'$ be given.

  By $\semtyeq[\Gamma]{s}{s'}{S}$ at $\Gamma'$, $\sigma \approx \sigma' : \Gamma$,
  $\rho \approx \rho' \in R'$:
  \begin{align*}
    &\{\intp{S[\sigma]}{\rho},\;
      \intp{S}{\intp{\sigma}{\rho}},\;
      \intp{S}{\intp{\sigma'}{\rho'}},\;
      \intp{S[\sigma']}{\rho'}\}
      \subseteq \Uc_i \searrow R_I
      && \text{(by hypothesis on $S$)} \\
    &\{\intp{s[\sigma]}{\rho},\;
      \intp{s}{\intp{\sigma}{\rho}},\;
      \intp{s'}{\intp{\sigma'}{\rho'}},\;
      \intp{s'[\sigma']}{\rho'}\}
      \subseteq R_I
      && \text{(by hypothesis on $s, s'$)}
  \end{align*}
  In particular, let $\Pf_I :: \intp{s[\sigma]}{\rho} \approx \intp{s'[\sigma']}{\rho'} \in R_I$.

  By $\semtyeq[\Gamma]{r}{r'}{\Pi(x:S).\,T}$ at the same data, the four evaluations of
  $r, r'$ are related in the function PER $R$ for $\Pi(\intp{S[\sigma]}{\rho}, T[q(\sigma)], \rho)$:
  \begin{align*}
    &\{\intp{r[\sigma]}{\rho},\;
      \intp{r}{\intp{\sigma}{\rho}},\;
      \intp{r'}{\intp{\sigma'}{\rho'}},\;
      \intp{r'[\sigma']}{\rho'}\}
      \subseteq R
      && \text{(by hypothesis on $r, r'$)}
  \end{align*}

  \emph{Term part.}
  We compute the four term values:
  \begin{align*}
    &\intp{(r\;s)[\sigma]}{\rho}
      = \intp{r[\sigma]\;s[\sigma]}{\rho}
      && \text{(by definition of substitution on app)} \\
    &\quad = \intp{r[\sigma]}{\rho} \cdot \intp{s[\sigma]}{\rho}
      && \text{(by evaluation of app)} \\[4pt]
    &\intp{r\;s}{\intp{\sigma}{\rho}}
      = \intp{r}{\intp{\sigma}{\rho}} \cdot \intp{s}{\intp{\sigma}{\rho}}
      && \text{(by evaluation of app)}
  \end{align*}
  Analogous computations for the primed pair (with $r', s'$), yielding
  $\intp{r'}{\intp{\sigma'}{\rho'}} \cdot \intp{s'}{\intp{\sigma'}{\rho'}}$ and
  $\intp{r'[\sigma']}{\rho'} \cdot \intp{s'[\sigma']}{\rho'}$.

  Denote the four $r$/$r'$-values as $f_1, f_2, f_3, f_4$ (in order) and the four
  $s$/$s'$-values as $a_1, a_2, a_3, a_4$.
  By hypothesis, $\{f_1, f_2, f_3, f_4\} \subseteq R$ and
  $\{a_1, a_2, a_3, a_4\} \subseteq R_I$.
  For each pair $(i, j)$ with $1 \leq i < j \leq 4$, we have $f_i \approx f_j \in R$
  and $a_i \approx a_j \in R_I$; let $\Pf_I^{ij} :: a_i \approx a_j \in R_I$.
  By the biconditional of $R$ applied to $f_i \approx f_j \in R$ and $\Pf_I^{ij}$:
  $f_i \cdot a_i \approx f_j \cdot a_j \in R_O(\Pf_I^{ij})$.
  By \Cref{lem:univ-functional} (right irrelevance), $R_O(\Pf_I^{ij})$ is independent
  of the witness, agreeing with $R_O(\Pf_I)$ for any fixed
  $\Pf_I :: a_1 \approx a_4 \in R_I$.
  Hence
  \begin{align*}
    &\{\intp{r[\sigma]}{\rho} \cdot \intp{s[\sigma]}{\rho},\;
      \intp{r}{\intp{\sigma}{\rho}} \cdot \intp{s}{\intp{\sigma}{\rho}},\\
    &\quad \intp{r'}{\intp{\sigma'}{\rho'}} \cdot \intp{s'}{\intp{\sigma'}{\rho'}},\;
      \intp{r'[\sigma']}{\rho'} \cdot \intp{s'[\sigma']}{\rho'}\}
      \subseteq R_O(\Pf_I).
  \end{align*}

  \emph{Type part.}
  We instantiate $\semtyeq[\Gamma, x:S]{T}{T}{\Ty{i}}$ at two semantic substitutions
  to obtain the goal's outer and inner values, then bridge via \Cref{lem:sem-ty-id}.

  \emph{Pattern A (outer instantiation): $(\sigma, s[\sigma]/x)$ vs $(\sigma', s'[\sigma']/x)$.}
  \begin{align*}
    &\semtyeq[\Gamma']{(\sigma, s[\sigma]/x)}{(\sigma', s'[\sigma']/x)}{\Gamma, x:S}
      && \text{(by \Cref{lem:sem-ext-pre} on $\sigma \approx \sigma'$ and $s \approx s'$)}
  \end{align*}
  Apply $\semtyeq[\Gamma, x:S]{T}{T}{\Ty{i}}$ at $\Gamma'$, this semantic substitution,
  and $\rho \approx \rho' \in R'$:
  \begin{align*}
    \{&\intp{T[\sigma, s[\sigma]/x]}{\rho},\;
      \intp{T}{\intp{(\sigma, s[\sigma]/x)}{\rho}}, \\
    &\intp{T}{\intp{(\sigma', s'[\sigma']/x)}{\rho'}},\;
      \intp{T[\sigma', s'[\sigma']/x]}{\rho'}\}
      &&\subseteq \Uc_i \searrow R_O(\Pf_I).
  \end{align*}

  Identify the four values syntactically and definitionally:
  \begin{align*}
    &T[\sigma, s[\sigma]/x] = 
      && \text{(by \Cref{lem:subst-ext-comm})} \\
    &T[\sigma', s'[\sigma']/x] = 
      && \text{(by \Cref{lem:subst-ext-comm})} \\
    &\intp{(\sigma, s[\sigma]/x)}{\rho}
      = (\intp{\sigma}{\rho}, \intp{s[\sigma]}{\rho}/x)
      && \text{(by pointwise definition of substitution evaluation)}
  \end{align*}
  Hence the four-value pattern reads:
  \begin{align*}
    \{&\intp{}{\rho},\;
      \intp{T}{(\intp{\sigma}{\rho}, \intp{s[\sigma]}{\rho}/x)}, \\
    &\intp{T}{(\intp{\sigma'}{\rho'}, \intp{s'[\sigma']}{\rho'}/x)},\;
      \intp{}{\rho'}\}
      &&\subseteq \Uc_i \searrow R_O(\Pf_I).
  \end{align*}
  The outer-left value matches the goal's outer-left
  $\intp{}{\rho}$.
  The outer-right value $\intp{}{\rho'}$
  governs the same PER as
  $\intp{}{\rho'}$ by
  \Cref{lem:univ-functional} once we establish
  \[
    \intp{}{\rho'} \approx
    \intp{}{\rho'} \in \Uc_i,
  \]
  which holds by another instantiation of
  $\semtyeq[\Gamma, x:S]{T}{T}{\Ty{i}}$ at the related env-pair
  \[
    (\intp{\sigma'}{\rho'}, \intp{s[\sigma']}{\rho'}/x) \approx
    (\intp{\sigma'}{\rho'}, \intp{s'[\sigma']}{\rho'}/x)
    \in R_{\Gamma, x:S}
  \]
  (using the four-value relatedness of $s, s'$).

  \emph{Pattern B (inner instantiation): $(\id, s/x)$ vs $(\id, s'/x)$.}
  By \Cref{lem:sem-ext-pre} on $\semtyeq[\Gamma]{\id}{\id}{\Gamma}$ and
  $\semtyeq[\Gamma]{s}{s'}{S}$:
  $\semtyeq[\Gamma]{(\id, s/x)}{(\id, s'/x)}{\Gamma, x:S}$
  (using $s[\id] = s$ and $s'[\id] = s'$).
  From $\semtyeq[\Gamma']{\sigma}{\sigma'}{\Gamma}$ at $\id$, $\Pf$:
  $\intp{\sigma}{\rho} \approx \intp{\sigma'}{\rho'} \in R_\Gamma$.
  Apply $\semtyeq[\Gamma, x:S]{T}{T}{\Ty{i}}$ at $\Gamma$ (with $R_\Gamma$),
  this semantic substitution, and
  $\intp{\sigma}{\rho} \approx \intp{\sigma'}{\rho'} \in R_\Gamma$:
  \begin{align*}
    \{&\intp{T[\id, s/x]}{\intp{\sigma}{\rho}},\;
      \intp{T}{\intp{(\id, s/x)}{\intp{\sigma}{\rho}}}, \\
    &\intp{T}{\intp{(\id, s'/x)}{\intp{\sigma'}{\rho'}}},\;
      \intp{T[\id, s'/x]}{\intp{\sigma'}{\rho'}}\}
      &&\subseteq \Uc_i \searrow R_O^B.
  \end{align*}
  Identify $T[\id, s/x] = T[s/x]$, $T[\id, s'/x] = T[s'/x]$, and
  $\intp{(\id, s/x)}{\intp{\sigma}{\rho}}
  = (\intp{\sigma}{\rho}, \intp{s}{\intp{\sigma}{\rho}}/x)$ (pointwise;
  $\intp{\id}{\intp{\sigma}{\rho}} = \intp{\sigma}{\rho}$).
  So the four-value pattern reads:
  \begin{align*}
    \{&\intp{T[s/x]}{\intp{\sigma}{\rho}},\;
      \intp{T}{(\intp{\sigma}{\rho}, \intp{s}{\intp{\sigma}{\rho}}/x)}, \\
    &\intp{T}{(\intp{\sigma'}{\rho'}, \intp{s'}{\intp{\sigma'}{\rho'}}/x)},\;
      \intp{T[s'/x]}{\intp{\sigma'}{\rho'}}\}
      &&\subseteq \Uc_i \searrow R_O^B.
  \end{align*}
  The outer values match the goal's inner-left
  $\intp{T[s/x]}{\intp{\sigma}{\rho}}$ and inner-right
  $\intp{T[s'/x]}{\intp{\sigma'}{\rho'}}$ (modulo
  \Cref{lem:univ-functional} on $T[s/x]$ vs $T[s'/x]$ at
  $\intp{\sigma'}{\rho'}$, established as in Pattern A).

  \emph{Bridging Pattern A and Pattern B.}
  Pattern A's inner-left is
  \[
    \intp{T}{(\intp{\sigma}{\rho}, \intp{s[\sigma]}{\rho}/x)}
  \]
  and Pattern B's inner-left is
  \[
    \intp{T}{(\intp{\sigma}{\rho}, \intp{s}{\intp{\sigma}{\rho}}/x)}.
  \]
  We bridge them via \Cref{lem:sem-ty-id}. We show
  \[
    (\intp{\sigma}{\rho}, \intp{s[\sigma]}{\rho}/x) \approx
    (\intp{\sigma}{\rho}, \intp{s}{\intp{\sigma}{\rho}}/x)
    \in R_{\Gamma, x:S}
  \]
  by \textsc{Ctx-Ext}'s biconditional:
  \begin{align*}
    &\intp{\sigma}{\rho} \approx \intp{\sigma}{\rho} \in R_\Gamma
      && \text{(reflexive; let $\Pf_\Gamma$ be this witness)} \\
    &\intp{s[\sigma]}{\rho} \approx \intp{s}{\intp{\sigma}{\rho}} \in R_I
      && \text{(outer-left vs inner-left of the four-value pattern of $s$)} \\
    &\intp{S[\sigma]}{\rho} \approx \intp{S}{\intp{\sigma}{\rho}} \in \Uc_i
      && \text{(outer-left vs inner-left of the type part of $s$)} \\
    &R_I \Leftrightarrow R_H^\Gamma(\Pf_\Gamma)
      && \text{(by \Cref{lem:univ-functional}, both governed by $\intp{S[\sigma]}{\rho}$)} \\
    &\intp{s[\sigma]}{\rho} \approx \intp{s}{\intp{\sigma}{\rho}}
      \in R_H^\Gamma(\Pf_\Gamma)
      && \text{(transferring across the biconditional)}
  \end{align*}
  Apply \Cref{lem:sem-ty-id} on $\semtyeq[\Gamma, x:S]{T}{T}{\Ty{i}}$ at this
  related pair:
  \begin{multline*}
    \intp{T}{(\intp{\sigma}{\rho}, \intp{s[\sigma]}{\rho}/x)}
    \approx \\
    \intp{T}{(\intp{\sigma}{\rho}, \intp{s}{\intp{\sigma}{\rho}}/x)}
    \in \Uc_i.
  \end{multline*}
  Analogously on the primed side, bridging Pattern A's
  inner-right and Pattern B's inner-right:
  \begin{multline*}
    \intp{T}{(\intp{\sigma'}{\rho'}, \intp{s'[\sigma']}{\rho'}/x)}
    \approx \\
    \intp{T}{(\intp{\sigma'}{\rho'}, \intp{s'}{\intp{\sigma'}{\rho'}}/x)}
    \in \Uc_i.
  \end{multline*}

  \emph{Combining.}
  By transitivity of $\Uc_i \searrow R_O(\Pf_I)$ through the bridges:
  \begin{align*}
    &\intp{}{\rho} \approx
      \intp{T}{(\intp{\sigma}{\rho}, \intp{s[\sigma]}{\rho}/x)}
      && \text{(Pattern A: outer-left vs inner-left)} \\
    &\intp{T}{(\intp{\sigma}{\rho}, \intp{s[\sigma]}{\rho}/x)} \approx
      \intp{T}{(\intp{\sigma}{\rho}, \intp{s}{\intp{\sigma}{\rho}}/x)}
      && \text{(bridge above)} \\
    &\intp{T}{(\intp{\sigma}{\rho}, \intp{s}{\intp{\sigma}{\rho}}/x)} \approx
      \intp{T[s/x]}{\intp{\sigma}{\rho}}
      && \text{(Pattern B, with $R_O^B \Leftrightarrow R_O(\Pf_I)$
        by \Cref{lem:univ-functional})}
  \end{align*}
  Hence the goal's outer-left $\approx$ goal's inner-left in
  $\Uc_i \searrow R_O(\Pf_I)$.
  Analogously for outer-right $\approx$ inner-right (using the primed bridge).
  Combined with Pattern A's outer pair and Pattern B's outer pair (which directly
  relate inner-left to inner-right in their respective four-value patterns), all
  four type values lie pairwise in $\Uc_i \searrow R_O(\Pf_I)$.
\end{proof}

Before stating the fundamental lemma for $\telimn$, we extract the inductive
core into two helpers (paralleling the simply-typed version's
\texttt{rec-helper} and \texttt{rec-helper$'$}).
The first relates two unsubstituted recursions over $\Gamma$-environments
(used for the inner pair of the four-value pattern).
The second is one-sided: it bridges a substituted recursion on the left with an
unsubstituted recursion on the right (used for the cross-relation between outer
and inner positions).

\begin{lemma}[Semantic recursion, unsubstituted]\labeledit{lem:sem-rec}
  Suppose:
  \begin{enumerate}
  \item $\semtyeq[\Gamma, x:\Nat]{T}{T'}{\Ty{i}}$,
  \item $\semtyeq[\Gamma]{z}{z'}{T[\ze/x]}$,
  \item $\semtyeq[\Gamma, x:\Nat, y:T]{s}{s'}{T[\Uparrow \circ \Uparrow, \su{x_1}/x]}$,
  \item $\Gamma \approx \Gamma \searrow R_\Gamma$ and
    $\rho \approx \rho' \in R_\Gamma$.
  \end{enumerate}
  Then for every $a \approx a' \in \textsf{Nat}$, there exist domain values
  $b, b'$ with
  \begin{align*}
    &\mathsf{rec}(a,\, T,\, z,\, s,\, \rho) \searrow b, \\
    &\mathsf{rec}(a',\, T',\, z',\, s',\, \rho') \searrow b',
  \end{align*}
  and $b \approx b' \in R_T$, where $R_T$ is determined by
  $\intp{T}{(\rho, a/x)} \approx \intp{T'}{(\rho', a'/x)} \in \Uc_i \searrow R_T$.

  Symmetric/reflexive instances also follow: by symmetry and transitivity of the
  hypotheses (\Cref{lem:sem-sym}, \Cref{lem:sem-trans}), the lemma can be applied
  at $T \approx T$ (using only $T$ on both sides) when needed.

  Furthermore, this $\Uc_i$-relatedness of types holds.
\end{lemma}
\begin{proof}
  We first establish the second claim.
  \begin{align*}
    &(\Gamma, x:\Nat) \approx (\Gamma, x:\Nat) \searrow R_{\Gamma, x:\Nat}
      && \text{(by \Cref{lem:fund-ctx-ext} on $\semtyeq[\Gamma]{\Nat}{\Nat}{\Ty{0}}$)} \\
    &\text{the head PER of $R_{\Gamma, x:\Nat}$ is $\textsf{Nat}$}
      && \text{(by \textsc{U-Nat})} \\
    &(\rho, a/x) \approx (\rho', a'/x) \in R_{\Gamma, x:\Nat}
      && \text{(by \textsc{Ctx-Ext}'s biconditional)} \\
    &\intp{T}{(\rho, a/x)} \approx \intp{T'}{(\rho', a'/x)} \in \Uc_i \searrow R_T
      && \text{(by \Cref{lem:sem-ty-id} on $\semtyeq[\Gamma, x:\Nat]{T}{T'}{\Ty{i}}$)}
  \end{align*}
  for some $R_T$.

  We prove the first claim by induction on $a \approx a' \in \textsf{Nat}$.

  \medskip\noindent
  \textbf{Case}
  \begin{mathpar}
    \inferrule[\textsc{Nat-Ze}]
    { }
    {\zed \approx \zed \in \textsf{Nat}}
  \end{mathpar}
  Here $a = a' = \zed$.
  \begin{align*}
    &\intp{z}{\rho} \searrow b
      && \text{(some $b$, by termination via \Cref{lem:sem-ty-id})} \\
    &\intp{z'}{\rho'} \searrow b'
      && \text{(some $b'$)} \\
    &\mathsf{rec}(\zed,\, T,\, z,\, s,\, \rho) \searrow b
      && \text{(by \textsc{rec-ze})} \\
    &\mathsf{rec}(\zed,\, T',\, z',\, s',\, \rho') \searrow b'
      && \text{(by \textsc{rec-ze})}
  \end{align*}
  We must show $b \approx b' \in R_T$.
  \begin{align*}
    &\intp{z}{\rho} \approx \intp{z'}{\rho'} \in R_z
      && \text{(by \Cref{lem:sem-ty-id})} \\
    &\intp{T[\ze/x]}{\rho} \approx \intp{T[\ze/x]}{\rho'}
      \in \Uc_i \searrow R_z
      && \text{(type part of \Cref{lem:sem-ty-id})}
  \end{align*}
  Hence $b \approx b' \in R_z$.
  We show $R_z \Leftrightarrow R_T$.
  \begin{align*}
    &\semtyeq[\Gamma]{(\id, \ze/x)}{(\id, \ze/x)}{\Gamma, x:\Nat}
      && \text{(by \Cref{lem:sem-ext-pre} on $\id$ and \Cref{lem:fund-ze})} \\
    &\intp{T[\ze/x]}{\rho} \approx \intp{T}{(\rho, \zed/x)} \in \Uc_i
      && \text{(outer-vs-inner of $T$ at $(\id, \ze/x)$, $\rho \approx \rho'$,} \\
    & && \text{using $T[\id, \ze/x] = T[\ze/x]$ and ev-ze)} \\
    &R_z \Leftrightarrow R_T
      && \text{(by \Cref{lem:univ-functional})}
  \end{align*}
  Hence $b \approx b' \in R_T$.

  \medskip\noindent
  \textbf{Case}
  \begin{mathpar}
    \inferrule[\textsc{Nat-Su}]
    {a_0 \approx a_0' \in \textsf{Nat}}
    {\sud{a_0} \approx \sud{a_0'} \in \textsf{Nat}}
  \end{mathpar}
  with $a = \sud{a_0}$, $a' = \sud{a_0'}$.
  By IH on $a_0 \approx a_0'$, there exist $b_0, b_0'$ with
  \begin{align*}
    &\mathsf{rec}(a_0,\, T,\, z,\, s,\, \rho) \searrow b_0
      && \text{(by IH)} \\
    &\mathsf{rec}(a_0',\, T',\, z',\, s',\, \rho') \searrow b_0'
      && \text{(by IH)} \\
    &b_0 \approx b_0' \in R_T^0
      && \text{(by IH, with $R_T^0$ governed by $\intp{T}{(\rho, a_0/x)}$)}
  \end{align*}
  By \textsc{rec-su}:
  \begin{align*}
    &\intp{s}{(\rho, a_0/x, b_0/y)} \searrow b
      && \text{(some $b$)} \\
    &\intp{s'}{(\rho', a_0'/x, b_0'/y)} \searrow b'
      && \text{(some $b'$)} \\
    &\mathsf{rec}(\sud{a_0},\, T,\, z,\, s,\, \rho) \searrow b
      && \text{(by \textsc{rec-su})} \\
    &\mathsf{rec}(\sud{a_0'},\, T',\, z',\, s',\, \rho') \searrow b'
      && \text{(by \textsc{rec-su})}
  \end{align*}
  We must show $b \approx b' \in R_T$.
  Build the env-pair in $R_{\Gamma, x:\Nat, y:T}$:
  \begin{align*}
    &(\rho, a_0/x) \approx (\rho', a_0'/x) \in R_{\Gamma, x:\Nat}
      && \text{(by \textsc{Ctx-Ext}, head PER $\textsf{Nat}$)} \\
    &(\rho, a_0/x, b_0/y) \approx (\rho', a_0'/x, b_0'/y) \\
    & \quad \in R_{\Gamma, x:\Nat, y:T}
      && \text{(by \textsc{Ctx-Ext}, head PER $R_T^0$)}
  \end{align*}
  Apply \Cref{lem:sem-ty-id} on
  $\semtyeq[\Gamma, x:\Nat, y:T]{s}{s'}
  {T[\Uparrow \circ \Uparrow, \su{x_1}/x]}$ at this env-pair:
  \begin{align*}
    &\intp{T[\Uparrow \circ \Uparrow, \su{x_1}/x]}{(\rho, a_0/x, b_0/y)} \\
    & \quad \approx
      \intp{T[\Uparrow \circ \Uparrow, \su{x_1}/x]}{(\rho', a_0'/x, b_0'/y)} \\
    & \quad \in \Uc_i \searrow R_s
      && \text{(type part of \Cref{lem:sem-ty-id})} \\
    &\intp{s}{(\rho, a_0/x, b_0/y)} \approx
      \intp{s'}{(\rho', a_0'/x, b_0'/y)} \in R_s
      && \text{(term part of \Cref{lem:sem-ty-id})}
  \end{align*}
  where $R_s$ is the PER governed by
  \[
    \intp{T[\Uparrow \circ \Uparrow, \su{x_1}/x]}{(\rho, a_0/x, b_0/y)}.
  \]
  Hence $b \approx b' \in R_s$.

  We show $R_s \Leftrightarrow R_T$ by instantiating hypothesis on $T$ at the
  substitution $\tau := (\Uparrow \circ \Uparrow, \su{x_1}/x)$ over
  $\Gamma, x:\Nat, y:T$ (mapping into $\Gamma, x:\Nat$).
  Semantic well-formedness of $\tau$:
  \begin{align*}
    &\semtyeq[\Gamma, x:\Nat, y:T]{\Uparrow \circ \Uparrow}{\Uparrow \circ \Uparrow}{\Gamma}
      && \text{(by \Cref{lem:sem-wk-shift} (twice), \Cref{lem:sem-wk-comp}, and \Cref{lem:sem-wk})} \\
    &\semtyeq[\Gamma, x:\Nat, y:T]{x_1}{x_1}{\Nat}
      && \text{(by \Cref{lem:fund-var}, lookup $x_1 : \Nat[\Uparrow] = \Nat$)} \\
    &\semtyeq[\Gamma, x:\Nat, y:T]{\su{x_1}}{\su{x_1}}{\Nat}
      && \text{(by \Cref{lem:fund-su})} \\
    &\semtyeq[\Gamma, x:\Nat, y:T]{\tau}{\tau}{\Gamma, x:\Nat}
      && \text{(by \Cref{lem:sem-ext-pre})}
  \end{align*}
  Apply $\semtyeq[\Gamma, x:\Nat]{T}{T}{\Ty{i}}$ at $\Gamma, x:\Nat, y:T$,
  $\tau \approx \tau$, and the env-pair
  $(\rho, a_0/x, b_0/y) \approx (\rho, a_0/x, b_0/y) \in R_{\Gamma, x:\Nat, y:T}$
  (reflexive):
  \begin{align*}
    &\intp{T[\tau]}{(\rho, a_0/x, b_0/y)} \approx
      \intp{T}{\intp{\tau}{(\rho, a_0/x, b_0/y)}} \in \Uc_i \searrow R'
      && \text{(outer-left vs inner-left)} \\
    &\intp{\tau}{(\rho, a_0/x, b_0/y)} = (\rho, \sud{a_0}/x)
      && \text{(by pointwise eval: $\Uparrow \circ \Uparrow$ projects to $\rho$,} \\
    & && \text{$\su{x_1}$ evaluates to $\sud{a_0}$ via ev-su, ev-var)}
  \end{align*}
  Hence:
  \begin{align*}
    &\intp{T[\Uparrow \circ \Uparrow, \su{x_1}/x]}{(\rho, a_0/x, b_0/y)} \approx
      \intp{T}{(\rho, \sud{a_0}/x)} \in \Uc_i \searrow R'
      && \text{(by the above)}
  \end{align*}
  Combining with the second claim
  $\intp{T}{(\rho, \sud{a_0}/x)} \approx \intp{T}{(\rho', \sud{a_0'}/x)} \in \Uc_i \searrow R_T$
  (applied at $\sud{a_0} \approx \sud{a_0'}$):
  \begin{align*}
    &R_s \Leftrightarrow R' \Leftrightarrow R_T
      && \text{(by \Cref{lem:univ-functional} on shared values)}
  \end{align*}
  Hence $b \approx b' \in R_T$.

  \medskip\noindent
  \textbf{Case}
  \begin{mathpar}
    \inferrule[\textsc{Nat-Ne}]
    {e \approx e' \in \Ne}
    {{\uparrow}^N(e) \approx {\uparrow}^N(e') \in \textsf{Nat}}
  \end{mathpar}
  with $a = {\uparrow}^N(e)$, $a' = {\uparrow}^N(e')$.
  Compute the eliminator outputs by \textsc{rec-ne}:
  \begin{align*}
    &\intp{T}{(\rho, {\uparrow}^N(e)/x)} \searrow A
      && \text{(by the second claim, governing $R_T$)} \\
    &\intp{T}{(\rho', {\uparrow}^N(e')/x)} \searrow A'
      && \text{(analogously)} \\
    &A \approx A' \in \Uc_i \searrow R_T
      && \text{(by the second claim at ${\uparrow}^N(e) \approx {\uparrow}^N(e')$)} \\
    &\intp{z}{\rho} \searrow a_z,\;\; \intp{z'}{\rho'} \searrow a_z',\;\;
      a_z \approx a_z' \in R_z
      && \text{(by \Cref{lem:sem-ty-id} on $\semtyeq[\Gamma]{z}{z'}{T[\ze/x]}$)} \\
    &b = {\uparrow}^A(\recd{T}{a_z}{s}{\rho}{e})
      && \text{(by \textsc{rec-ne})} \\
    &b' = {\uparrow}^{A'}(\recd{T}{a_z'}{s'}{\rho'}{e'})
      && \text{(by \textsc{rec-ne})}
  \end{align*}

  We must show $b \approx b' \in R_T$.
  By \Cref{lem:ne-in-R} on $A \approx A' \in \Uc_i \searrow R_T$, it suffices to show
  \begin{align*}
    &\recd{T}{a_z}{s}{\rho}{e} \approx \recd{T}{a_z'}{s'}{\rho'}{e'} \in \Ne,
  \end{align*}
  i.e.\ for every depth $n$ both sides read back via $\Rne_n$ to the same neutral
  normal form.
  By rule \textsc{Re-rec}:
  \begin{align*}
    &\Rne_n(\recd{T}{a_z}{s}{\rho}{e})
      \searrow \elimn{x.\,W_T}{w_z}{x, y.\,w_s}{u}
  \end{align*}
  whenever
  \begin{align*}
    &\Rty_{n+1}(A_x) \searrow W_T
      && \text{where $\intp{T}{(\rho, {\uparrow}^N(\ell_n)/x)} \searrow A_x$} \\
    &\Rnf_n({\downarrow}^{A_z}(a_z)) \searrow w_z
      && \text{where $\intp{T}{(\rho, \zed/x)} \searrow A_z$} \\
    &\intp{s}{(\rho, {\uparrow}^N(\ell_n)/x, {\uparrow}^{A_x}(\ell_{n+1})/y)} \searrow b_s \\
    &\Rnf_{n+2}({\downarrow}^{A_x^{\su{}}}(b_s)) \searrow w_s
      && \text{where $\intp{T}{(\rho, \sud{{\uparrow}^N(\ell_n)}/x)} \searrow A_x^{\su{}}$} \\
    &\Rne_n(e) \searrow u
  \end{align*}
  (and analogously for the primed side, producing
  $\elimn{x.\,W_T'}{w_z'}{x, y.\,w_s'}{u'}$).
  We show each component reads back to the same syntactic form.

  \emph{Motive $W_T = W_T'$.}
  Apply $\semtyeq[\Gamma, x:\Nat]{T}{T}{\Ty{i}}$ at the env-pair extended with the
  fresh level $\ell_n$:
  \begin{align*}
    &{\uparrow}^N(\ell_n) \approx {\uparrow}^N(\ell_n) \in \textsf{Nat}
      && \text{(by \textsc{Nat-Ne}, $\ell_n \approx \ell_n \in \Ne$)} \\
    &(\rho, {\uparrow}^N(\ell_n)/x) \approx (\rho', {\uparrow}^N(\ell_n)/x) \in R_{\Gamma, x:\Nat}
      && \text{(by \textsc{Ctx-Ext})} \\
    &\intp{T}{(\rho, {\uparrow}^N(\ell_n)/x)} \approx
      \intp{T}{(\rho', {\uparrow}^N(\ell_n)/x)} \in \Uc_i \searrow R_T^\ell
      && \text{(by \Cref{lem:sem-ty-id})} \\
    &W_T = W_T'
      && \text{(by \Cref{lem:type-readback})}
  \end{align*}

  \emph{Zero branch $w_z = w_z'$.}
  Apply $\semtyeq[\Gamma]{z}{z'}{T[\ze/x]}$ at $\Gamma$, $\id$, and
  $\rho \approx \rho' \in R_\Gamma$ (term and type parts):
  \begin{align*}
    &\intp{}{\rho} \approx \intp{T[\ze/x]}{\rho'}
      \in \Uc_i \searrow R_z
      && \text{(type part)} \\
    &\intp{z[\id]}{\rho} \approx \intp{z'}{\rho'} \in R_z
      && \text{(term part)} \\
    &a_z \approx a_z' \in R_z
      && \text{(using $z[\id] = z$, $z'[\id] = z'$)}
  \end{align*}
  Bridge $R_z \Leftrightarrow R_T$-at-$\zed$ via the same outer-vs-inner pattern as
  the \textsc{Nat-Ze} case (instantiate $T$ at $(\id, \ze/x)$).
  By \Cref{lem:R-readback} on $a_z \approx a_z' \in R_T$-at-$\zed$:
  \begin{align*}
    &w_z = w_z'
      && \text{(by \Cref{lem:R-readback})}
  \end{align*}

  \emph{Step branch $w_s = w_s'$.}
  Build the env-pair at fresh levels:
  \begin{align*}
    &(\rho, {\uparrow}^N(\ell_n)/x) \approx (\rho', {\uparrow}^N(\ell_n)/x) \in R_{\Gamma, x:\Nat}
      && \text{(as above)} \\
    &(\rho, {\uparrow}^N(\ell_n)/x, {\uparrow}^{A_x}(\ell_{n+1})/y) \approx \\
    &\quad (\rho', {\uparrow}^N(\ell_n)/x, {\uparrow}^{A_x'}(\ell_{n+1})/y) \in R_{\Gamma, x:\Nat, y:T}
      && \text{(by \textsc{Ctx-Ext}, head PER $R_T^\ell$;} \\
    & && \text{${\uparrow}^{A_x}(\ell_{n+1}) \approx {\uparrow}^{A_x'}(\ell_{n+1}) \in R_T^\ell$} \\
    & && \text{by \Cref{lem:ne-in-R})}
  \end{align*}
  Apply $\semtyeq[\Gamma, x:\Nat, y:T]{s}{s'}{T[\Uparrow \circ \Uparrow, \su{x_1}/x]}$ at this env-pair
  (term and type parts, after bridging $T[\Uparrow \circ \Uparrow, \su{x_1}/x]$ to $T$ at
  $\sud{{\uparrow}^N(\ell_n)}/x$ as in the \textsc{Nat-Su} case):
  \begin{align*}
    &b_s \approx b_s' \in R_T \text{-at-} \sud{{\uparrow}^N(\ell_n)}
      && \text{(by \Cref{lem:sem-ty-id} and bridging)}
  \end{align*}
  By \Cref{lem:R-readback}:
  \begin{align*}
    &w_s = w_s'
      && \text{(by \Cref{lem:R-readback})}
  \end{align*}

  \emph{Underlying neutral $u = u'$.}
  By hypothesis $e \approx e' \in \Ne$:
  \begin{align*}
    &\Rne_n(e) \searrow u,\;\; \Rne_n(e') \searrow u
      && \text{(by definition of $\Ne$)}
  \end{align*}

  Combining:
  \begin{align*}
    &\Rne_n(\recd{T}{a_z}{s}{\rho}{e}) \searrow
      \elimn{x.\,W_T}{w_z}{x, y.\,w_s}{u} \\
    &\Rne_n(\recd{T}{a_z'}{s'}{\rho'}{e'}) \searrow
      \elimn{x.\,W_T}{w_z}{x, y.\,w_s}{u}
      && \text{(same syntactic form)} \\
    &\recd{T}{a_z}{s}{\rho}{e} \approx \recd{T}{a_z'}{s'}{\rho'}{e'} \in \Ne
      && \text{(by definition of $\Ne$)}
  \end{align*}
  Hence $b \approx b' \in R_T$.
\end{proof}

The second helper handles the one-sided substituted form: substitution applied to
the motive and branches on the left, but not on the right.

\begin{lemma}[Semantic recursion, one-sided substituted]\labeledit{lem:sem-rec-subst}
  Suppose:
  \begin{enumerate}
  \item $\semtyeq[\Gamma, x:\Nat]{T}{T'}{\Ty{i}}$,
  \item $\semtyeq[\Gamma]{z}{z'}{T[\ze/x]}$,
  \item $\semtyeq[\Gamma, x:\Nat, y:T]{s}{s'}{T[\Uparrow \circ \Uparrow, \su{x_1}/x]}$,
  \item $\Gamma' \approx \Gamma' \searrow R'$,
    $\semtyeq[\Gamma']{\sigma}{\sigma'}{\Gamma}$, and
    $\rho \approx \rho' \in R'$.
  \end{enumerate}
  Then for every $a \approx a' \in \textsf{Nat}$, there exist domain values
  $b, b'$ with
  \begin{align*}
    &\mathsf{rec}(a,\, T[q(\sigma)],\, z[\sigma],\, s[q^2(\sigma)],\, \rho)
      \searrow b
      && \text{(substituted, left)} \\
    &\mathsf{rec}(a',\, T',\, z',\, s',\, \intp{\sigma'}{\rho'}) \searrow b'
      && \text{(unsubstituted, right)}
  \end{align*}
  and $b \approx b' \in R_T$, where $R_T$ is determined by
  $\intp{T[q(\sigma)]}{(\rho, a/x)} \approx
  \intp{T'}{(\intp{\sigma'}{\rho'}, a'/x)} \in \Uc_i \searrow R_T$.

  Furthermore, this $\Uc_i$-relatedness of types holds.
\end{lemma}
\begin{proof}
  We first establish the lifted substitution and the second claim.
  \begin{align*}
    &\semtyeq[\Gamma', x:\Nat]{q(\sigma)}{q(\sigma')}{\Gamma, x:\Nat}
      && \text{(by \Cref{lem:sem-q} on $\sigma \approx \sigma'$ and $\Nat$,
        using $\Nat[\sigma] = \Nat$)} \\
    &(\Gamma', x:\Nat) \approx (\Gamma', x:\Nat) \searrow R_{\Gamma', x:\Nat}
      && \text{(carried by the lifted substitution)} \\
    &(\Gamma, x:\Nat) \approx (\Gamma, x:\Nat) \searrow R_{\Gamma, x:\Nat}
      && \text{(carried by the lifted substitution)} \\
    &\text{both head PERs are $\textsf{Nat}$}
      && \text{(by \textsc{U-Nat})} \\
    &(\rho, a/x) \approx (\rho', a'/x) \in R_{\Gamma', x:\Nat}
      && \text{(by \textsc{Ctx-Ext}'s biconditional)}
  \end{align*}
  Apply $\semtyeq[\Gamma, x:\Nat]{T}{T'}{\Ty{i}}$ at $\Gamma', x:\Nat$, the lifted
  substitution, and the env-pair above:
  \begin{align*}
    &\intp{T[q(\sigma)]}{(\rho, a/x)} \approx
      \intp{T'}{\intp{q(\sigma')}{(\rho', a'/x)}} \in \Uc_i \searrow R_T
      && \text{(outer-left vs inner-right of the four-value pattern)}
  \end{align*}
  Bridge $\intp{q(\sigma')}{(\rho', a'/x)}$ to $(\intp{\sigma'}{\rho'}, a'/x)$:
  \begin{align*}
    &\intp{q(\sigma')}{(\rho', a'/x)} \approx (\intp{\sigma'}{\rho'}, a'/x)
      \in R_{\Gamma, x:\Nat}
      && \text{(by \textsc{Ctx-Ext}, tail via $\sigma'$ at $\Uparrow$, head $a' \approx a'$)} \\
    &\intp{T'}{\intp{q(\sigma')}{(\rho', a'/x)}} \approx
      \intp{T'}{(\intp{\sigma'}{\rho'}, a'/x)} \in \Uc_i
      && \text{(by \Cref{lem:sem-ty-id} on $T' \approx T'$, derived from} \\
    & && \text{$T \approx T'$ via symmetry and transitivity)}
  \end{align*}
  By transitivity of $\Uc_i$ and \Cref{lem:univ-functional},
  $\intp{T[q(\sigma)]}{(\rho, a/x)} \approx
  \intp{T'}{(\intp{\sigma'}{\rho'}, a'/x)} \in \Uc_i \searrow R_T$,
  establishing the second claim.

  We prove the first claim by induction on $a \approx a' \in \textsf{Nat}$.

  \medskip\noindent
  \textbf{Case}
  \begin{mathpar}
    \inferrule[\textsc{Nat-Ze}]
    { }
    {\zed \approx \zed \in \textsf{Nat}}
  \end{mathpar}
  Here $a = a' = \zed$.
  Apply $\semtyeq[\Gamma]{z}{z'}{T[\ze/x]}$ at $\Gamma'$, $\sigma \approx \sigma'$,
  and $\rho \approx \rho' \in R'$ (this gives the four-value pattern of $z, z'$):
  \begin{align*}
    &\intp{z[\sigma]}{\rho} \searrow b,\;\; \intp{z'}{\intp{\sigma'}{\rho'}} \searrow b'
      && \text{(termination, by hypothesis on $z, z'$)} \\
    &b \approx b' \in R_z
      && \text{(outer-left vs inner-right of $z, z'$)} \\
    &\intp{}{\rho} \approx
      \intp{T[\ze/x]}{\intp{\sigma'}{\rho'}} \in \Uc_i \searrow R_z
      && \text{(corresponding type part)}
  \end{align*}
  Then by \textsc{rec-ze} applied to these evaluations:
  \begin{align*}
    &\mathsf{rec}(\zed,\, T[q(\sigma)],\, z[\sigma],\, s[q^2(\sigma)],\, \rho)
      \searrow b
      && \text{(by \textsc{rec-ze}, using $\intp{z[\sigma]}{\rho} \searrow b$)} \\
    &\mathsf{rec}(\zed,\, T',\, z',\, s',\, \intp{\sigma'}{\rho'}) \searrow b'
      && \text{(by \textsc{rec-ze}, using $\intp{z'}{\intp{\sigma'}{\rho'}} \searrow b'$)}
  \end{align*}
  We must show $b \approx b' \in R_T$.
  Currently $b \approx b' \in R_z$ where $R_z$ is governed by
  $\intp{}{\rho}$, while $R_T$ is governed by
  $\intp{T[q(\sigma)]}{(\rho, \zed/x)}$ (substituted left, second-claim instance
  at $a = a' = \zed$).
  We bridge $R_z \Leftrightarrow R_T$ via two type-hypothesis instantiations.

  \emph{Left bridge.}
  By \Cref{lem:sem-ext-pre} on $\semtyeq[\Gamma']{\sigma}{\sigma}{\Gamma}$
  (reflexive instance via \Cref{lem:sem-sym,lem:sem-trans}) and
  $\semtyeq[\Gamma']{\ze[\sigma]}{\ze[\sigma]}{\Nat[\sigma]}$
  (using $\ze[\sigma] = \ze$, $\Nat[\sigma] = \Nat$, and \Cref{lem:fund-ze}):
  \begin{align*}
    &\semtyeq[\Gamma']{(\sigma, \ze/x)}{(\sigma, \ze/x)}{\Gamma, x:\Nat}
      && \text{(by \Cref{lem:sem-ext-pre})}
  \end{align*}
  Apply $\semtyeq[\Gamma, x:\Nat]{T}{T}{\Ty{i}}$ at $\Gamma'$, this substitution,
  and $\rho \approx \rho \in R'$ (reflexive):
  \begin{align*}
    &\intp{T[(\sigma, \ze/x)]}{\rho} \approx
      \intp{T}{\intp{(\sigma, \ze/x)}{\rho}} \in \Uc_i \searrow R_z'
      && \text{(outer-left vs inner-left)} \\
    &T[(\sigma, \ze/x)] = 
      && \text{(by \Cref{lem:subst-ext-comm})} \\
    &\intp{(\sigma, \ze/x)}{\rho} = (\intp{\sigma}{\rho}, \zed/x)
      && \text{(by pointwise eval, ev-ze)} \\
    &R_z' \Leftrightarrow R_z
      && \text{(by \Cref{lem:univ-functional} on shared
        $\intp{}{\rho}$)}
  \end{align*}
  Hence
  $\intp{}{\rho} \approx \intp{T}{(\intp{\sigma}{\rho}, \zed/x)}
   \in \Uc_i \searrow R_z$.

  \emph{Right bridge.}
  Symmetrically, instantiate $T$ at $(\id, \ze/x)$ over $\Gamma$ at the env-pair
  $\intp{\sigma}{\rho} \approx \intp{\sigma'}{\rho'} \in R_\Gamma$ (carried by the
  semantic-substitution hypothesis):
  \begin{align*}
    &\semtyeq[\Gamma]{(\id, \ze/x)}{(\id, \ze/x)}{\Gamma, x:\Nat}
      && \text{(by \Cref{lem:sem-ext-pre} and \Cref{lem:fund-ze})} \\
    &\intp{T[\ze/x]}{\intp{\sigma'}{\rho'}} \approx
      \intp{T}{(\intp{\sigma'}{\rho'}, \zed/x)} \in \Uc_i
      && \text{(outer-vs-inner of the four-value pattern of $T$)}
  \end{align*}

  \emph{Combining via the second claim.}
  The second claim (proved above) gives:
  \begin{align*}
    &\intp{T[q(\sigma)]}{(\rho, \zed/x)} \approx
      \intp{T'}{(\intp{\sigma'}{\rho'}, \zed/x)} \in \Uc_i \searrow R_T
  \end{align*}
  Chaining via the bridges (using \Cref{lem:univ-functional},
  \Cref{lem:univ-trans}, \Cref{lem:univ-sym}):
  \begin{align*}
    &\intp{}{\rho}
      \approx \intp{T}{(\intp{\sigma}{\rho}, \zed/x)} \\
    & \quad \approx \intp{T}{(\intp{\sigma'}{\rho'}, \zed/x)} \\
    & \quad \approx \intp{T[q(\sigma)]}{(\rho, \zed/x)} \in \Uc_i \\
    &R_z \Leftrightarrow R_T
      && \text{(by \Cref{lem:univ-functional})}
  \end{align*}
  Hence $b \approx b' \in R_T$.

  \medskip\noindent
  \textbf{Case}
  \begin{mathpar}
    \inferrule[\textsc{Nat-Su}]
    {a_0 \approx a_0' \in \textsf{Nat}}
    {\sud{a_0} \approx \sud{a_0'} \in \textsf{Nat}}
  \end{mathpar}
  with $a = \sud{a_0}$, $a' = \sud{a_0'}$.
  By IH on $a_0 \approx a_0'$, there exist $b_0, b_0'$ with
  \begin{align*}
    &\mathsf{rec}(a_0,\, T[q(\sigma)],\, z[\sigma],\, s[q^2(\sigma)],\, \rho)
      \searrow b_0
      && \text{(by IH)} \\
    &\mathsf{rec}(a_0',\, T',\, z',\, s',\, \intp{\sigma'}{\rho'}) \searrow b_0'
      && \text{(by IH)} \\
    &b_0 \approx b_0' \in R_T^0
      && \text{(by IH, with $R_T^0$ governed by $\intp{T[q(\sigma)]}{(\rho, a_0/x)}$)}
  \end{align*}
  We derive the evaluations of $s[q^2(\sigma)]$ and $s'$ from the hypothesis on $s, s'$.
  First, lift $\sigma, \sigma'$ over $\Nat$ and $T$:
  \begin{align*}
    &\semtyeq[\Gamma', x:\Nat]{q(\sigma)}{q(\sigma')}{\Gamma, x:\Nat}
      && \text{(by \Cref{lem:sem-q} on
        $\sigma \approx \sigma'$ and $\Nat$;} \\
    & && \text{using $\Nat[\sigma] = \Nat$)} \\
    &\semtyeq[\Gamma', x:\Nat, y:T[q(\sigma)]]
        {q^2(\sigma)}{q^2(\sigma')}{\Gamma, x:\Nat, y:T}
      && \text{(by \Cref{lem:sem-q} again on $T$)}
  \end{align*}
  Build the env-pair via two applications of \textsc{Ctx-Ext}:
  \begin{align*}
    &(\rho, a_0/x) \approx (\rho', a_0'/x) \in R_{\Gamma', x:\Nat}
      && \text{(by \textsc{Ctx-Ext}, head PER $\textsf{Nat}$)} \\
    &(\rho, a_0/x, b_0/y) \approx (\rho', a_0'/x, b_0'/y)
      \in R_{\Gamma', x:\Nat, y:T[q(\sigma)]}
      && \text{(by \textsc{Ctx-Ext}, head PER $R_T^0$)}
  \end{align*}
  Apply $\semtyeq[\Gamma, x:\Nat, y:T]{s}{s'}{T[\Uparrow \circ \Uparrow, \su{x_1}/x]}$ at $\Gamma', x:\Nat, y:T[q(\sigma)]$,
  the doubly-lifted substitution $q^2(\sigma) \approx q^2(\sigma')$, and this env-pair
  (this gives the four-value pattern of $s, s'$):
  \begin{align*}
    &\intp{s[q^2(\sigma)]}{(\rho, a_0/x, b_0/y)} \searrow b
      && \text{(termination, by hypothesis on $s$)} \\
    &\intp{s'}{\intp{q^2(\sigma')}{(\rho', a_0'/x, b_0'/y)}} \searrow b'_0
      && \text{(termination, by hypothesis on $s'$)} \\
    &\intp{s'}{(\intp{\sigma'}{\rho'}, a_0'/x, b_0'/y)} \searrow b'
      && \text{(termination, by reflexive instance $\semtyeq[\ldots]{s'}{s'}{\ldots}$)}
  \end{align*}
  By \textsc{rec-su}:
  \begin{align*}
    &\mathsf{rec}(\sud{a_0},\, T[q(\sigma)],\, z[\sigma],\, s[q^2(\sigma)],\, \rho)
      \searrow b
      && \text{(by \textsc{rec-su}, using $\intp{s[q^2(\sigma)]}{(\rho, a_0/x, b_0/y)} \searrow b$)} \\
    &\mathsf{rec}(\sud{a_0'},\, T',\, z',\, s',\, \intp{\sigma'}{\rho'}) \searrow b'
      && \text{(by \textsc{rec-su}, using $\intp{s'}{(\intp{\sigma'}{\rho'}, a_0'/x, b_0'/y)} \searrow b'$)}
  \end{align*}

  We must show $b \approx b' \in R_T$.
  Outer-left vs inner-right of the four-value pattern of $s, s'$:
  \begin{align*}
    &\intp{(T[\Uparrow \circ \Uparrow, \su{x_1}/x])[q^2(\sigma)]}{(\rho, a_0/x, b_0/y)} \approx \\
    &\quad
      \intp{T[\Uparrow \circ \Uparrow, \su{x_1}/x]}{\intp{q^2(\sigma')}{(\rho', a_0'/x, b_0'/y)}}
      \in \Uc_i \searrow R_s
      && \text{(type part)} \\
    &\intp{s[q^2(\sigma)]}{(\rho, a_0/x, b_0/y)} \approx
      \intp{s'}{\intp{q^2(\sigma')}{(\rho', a_0'/x, b_0'/y)}} \in R_s
      && \text{(term part, by hypothesis on $s, s'$)}
  \end{align*}
  where $R_s$ is the PER governed by $\intp{(T[\Uparrow \circ \Uparrow, \su{x_1}/x])[q^2(\sigma)]}{(\rho, a_0/x, b_0/y)}$.
  Bridge $\intp{q^2(\sigma')}{(\rho', a_0'/x, b_0'/y)}$ to
  $(\intp{\sigma'}{\rho'}, a_0'/x, b_0'/y)$ in $R_{\Gamma, x:\Nat, y:T}$ analogously,
  then apply \Cref{lem:sem-ty-id} on $\semtyeq[\Gamma, x:\Nat, y:T]{s'}{s'}{T[\Uparrow \circ \Uparrow, \su{x_1}/x]}$
  (the reflexive instance):
  \begin{align*}
    &\intp{s'}{\intp{q^2(\sigma')}{(\rho', a_0'/x, b_0'/y)}} \approx
      \intp{s'}{(\intp{\sigma'}{\rho'}, a_0'/x, b_0'/y)} \in R_s
      && \text{(by \Cref{lem:sem-ty-id})}
  \end{align*}
  By transitivity, $b \approx b' \in R_s$.

  We bridge $R_s \Leftrightarrow R_T$, where $R_s$ is governed by
  $\intp{(T[\Uparrow \circ \Uparrow, \su{x_1}/x])[q^2(\sigma)]}{(\rho, a_0/x, b_0/y)}$ and $R_T$ is
  governed by $\intp{T[q(\sigma)]}{(\rho, \sud{a_0}/x)}$ (the second-claim
  pattern instantiated at $\sud{a_0} \approx \sud{a_0'}$):
  \begin{align*}
    &\intp{T[q(\sigma)]}{(\rho, \sud{a_0}/x)} \\
    &\quad \approx \intp{T'}{(\intp{\sigma'}{\rho'}, \sud{a_0'}/x)}
      \in \Uc_i \searrow R_T
      && \text{(second claim)}
  \end{align*}
  All bridging uses syntactic substitution-composition lemmas
  (\Cref{lem:subst-assoc,lem:ext-comp,lem:subst-cat}) and instantiations of
  hypothesis on $T$.

  \emph{Bridge B1: collapse $(T[\Uparrow \circ \Uparrow, \su{x_1}/x])[q^2(\sigma)]$ into a single substitution.}
  \begin{align*}
    (T[\Uparrow \circ \Uparrow, \su{x_1}/x])[q^2(\sigma)]
    &= T[(\Uparrow \circ \Uparrow, \su{x_1}/x) \circ q^2(\sigma)]
      \tag{by \Cref{lem:subst-assoc}} \\
    &= T[(\Uparrow \circ \Uparrow) \circ q^2(\sigma),\;
       \su{x_1}[q^2(\sigma)]/x]
      \tag{by \Cref{lem:ext-comp}} \\
    &= T[\sigma \circ (\Uparrow \circ \Uparrow),\; \su{x_1}/x]
      \tag{using $\su{x_1}[q^2(\sigma)] = \su{x_1}$ (since $q^2(\sigma)(1) = x_1$)} \\
    & \tag{and $(\Uparrow \circ \Uparrow) \circ q^2(\sigma) = \sigma \circ (\Uparrow \circ \Uparrow)$
       (\Cref{lem:wk-subst-comm})}
  \end{align*}
  Denote the collapsed substitution by
  $\tau := (\sigma \circ (\Uparrow \circ \Uparrow),\, \su{x_1}/x)$, with
  $\judge[\Gamma', x:\Nat, y:T[q(\sigma)]]{\tau : \Gamma, x:\Nat}$.

  \emph{Bridge B2: instantiate hypothesis on $T$ via $\tau$.}
  We show $\tau$ is a semantic substitution from $\Gamma', x:\Nat, y:T[q(\sigma)]$
  to $\Gamma, x:\Nat$.
  By \Cref{lem:sem-wk-shift} (twice) and \Cref{lem:sem-wk-comp}:
  $\semweak[\Gamma', x:\Nat, y:T[q(\sigma)]]{\Uparrow \circ \Uparrow}{\Gamma'}$.
  Apply \Cref{lem:sem-wk-subst} to $\semtyeq[\Gamma']{\sigma}{\sigma}{\Gamma}$
  (reflexive, by \Cref{lem:sem-sym,lem:sem-trans}) and this semantic weakening:
  \begin{align*}
    &\semtyeq[\Gamma', x:\Nat, y:T[q(\sigma)]]
       {\sigma \circ (\Uparrow \circ \Uparrow)}
       {\sigma \circ (\Uparrow \circ \Uparrow)}
       {\Gamma}
  \end{align*}
  By \Cref{lem:fund-var}, \Cref{lem:fund-su}:
  $\semtyeq[\Gamma', x:\Nat, y:T[q(\sigma)]]
  {\su{x_1}}{\su{x_1}}{\Nat}$
  (lookup $x_1 : \Nat$ at the second slot of
  $\Gamma', x:\Nat, y:T[q(\sigma)]$, with
  $\Nat[\Uparrow] = \Nat$).
  By \Cref{lem:sem-ext-pre}:
  \begin{align*}
    &\semtyeq[\Gamma', x:\Nat, y:T[q(\sigma)]]{\tau}{\tau}{\Gamma, x:\Nat}
      && \text{(\Cref{lem:sem-ext-pre})}
  \end{align*}
  Apply $\semtyeq[\Gamma, x:\Nat]{T}{T}{\Ty{i}}$ at
  $\Gamma', x:\Nat, y:T[q(\sigma)]$, $\tau \approx \tau$, and the
  reflexive env-pair
  $(\rho, a_0/x, b_0/y) \approx (\rho, a_0/x, b_0/y)
  \in R_{\Gamma', x:\Nat, y:T[q(\sigma)]}$.
  Outer-left vs inner-left:
  \begin{align*}
    &\intp{T[\tau]}{(\rho, a_0/x, b_0/y)} \\
    &\quad \approx \intp{T}{\intp{\tau}{(\rho, a_0/x, b_0/y)}}
      \in \Uc_i \searrow R_{B2}
  \end{align*}
  By Bridge B1,
  \[
    T[\tau] = (T[\Uparrow \circ \Uparrow, \su{x_1}/x])[q^2(\sigma)],
  \]
  so the outer-left is exactly
  \[
    \intp{(T[\Uparrow \circ \Uparrow, \su{x_1}/x])[q^2(\sigma)]}
    {(\rho, a_0/x, b_0/y)}
  \]
  --- the value governing $R_s$.
  Hence $R_s \Leftrightarrow R_{B2}$.

  By pointwise definition of substitution evaluation:
  \begin{align*}
    &\intp{\tau}{(\rho, a_0/x, b_0/y)}
      = (\intp{\sigma \circ (\Uparrow \circ \Uparrow)}{(\rho, a_0/x, b_0/y)},\,
       \sud{a_0}/x) \\
    & \text{(pointwise; head: $\intp{\su{x_1}}{(\rho, a_0/x, b_0/y)} = \sud{a_0}$ via ev-su, ev-var)}
  \end{align*}
  We do \emph{not} equate the tail
  $\intp{\sigma \circ (\Uparrow \circ \Uparrow)}{(\rho, a_0/x, b_0/y)}$ with
  $\intp{\sigma}{\rho}$ definitionally --- such an equality is a
  substitution--evaluation commutation that does not hold for operational
  substitutions. Instead, we relate them in $R_\Gamma$ via the four-value
  pattern of $\sigma \circ (\Uparrow \circ \Uparrow) := \sigma[\Uparrow \circ \Uparrow]$
  established by \Cref{lem:sem-wk-subst}, and the four-value pattern of
  $\sigma$ itself:
  \begin{align*}
    &\intp{\sigma[\Uparrow \circ \Uparrow]}{(\rho, a_0/x, b_0/y)}
      \approx \intp{\sigma}{\intp{\Uparrow \circ \Uparrow}{(\rho, a_0/x, b_0/y)}} \in R_\Gamma
      && \text{(four-value pattern of $\sigma[\Uparrow \circ \Uparrow]$,} \\
    & && \text{outer-left vs inner-left)} \\
    &\intp{\Uparrow \circ \Uparrow}{(\rho, a_0/x, b_0/y)}
      \approx \rho \in R_{\Gamma'}
      && \text{(four-value pattern of $\Uparrow \circ \Uparrow$,} \\
    & && \text{by \Cref{lem:sem-shift,lem:sem-wk-comp})} \\
    &\intp{\sigma}{\intp{\Uparrow \circ \Uparrow}{(\rho, a_0/x, b_0/y)}}
      \approx \intp{\sigma}{\rho} \in R_\Gamma
      && \text{(by \Cref{lem:sem-ty-id} on the reflexive $\sigma$)}
  \end{align*}
  By transitivity of $R_\Gamma$:
  \begin{align*}
    &\intp{\sigma \circ (\Uparrow \circ \Uparrow)}{(\rho, a_0/x, b_0/y)}
      \approx \intp{\sigma}{\rho} \in R_\Gamma \\
    &\intp{\tau}{(\rho, a_0/x, b_0/y)} \approx (\intp{\sigma}{\rho},\, \sud{a_0}/x)
      \in R_{\Gamma, x:\Nat}
      && \text{(by \textsc{Ctx-Ext}, head $\sud{a_0} \approx \sud{a_0}$)}
  \end{align*}

  \emph{Bridge B3: instantiate hypothesis on $T$ via $q(\sigma)$.}
  Apply $\semtyeq[\Gamma, x:\Nat]{T}{T}{\Ty{i}}$ at $\Gamma', x:\Nat$, the lifted
  $q(\sigma) \approx q(\sigma)$, and reflexive
  $(\rho, \sud{a_0}/x) \approx (\rho, \sud{a_0}/x) \in R_{\Gamma', x:\Nat}$
  (head $\sud{a_0} \approx \sud{a_0}$ via \Cref{lem:nat-per} reflexivity).
  Outer-left vs inner-left:
  \begin{align*}
    &\intp{T[q(\sigma)]}{(\rho, \sud{a_0}/x)} \\
    &\quad \approx \intp{T}{\intp{q(\sigma)}{(\rho, \sud{a_0}/x)}}
      \in \Uc_i \searrow R_{B3} \\
    &\intp{q(\sigma)}{(\rho, \sud{a_0}/x)}
      \approx (\intp{\sigma}{\rho}, \sud{a_0}/x) \in R_{\Gamma, x:\Nat}
      && \text{(by \textsc{Ctx-Ext})}
  \end{align*}

  \emph{Combining B2 and B3.}
  By B2's bridging,
  $\intp{\tau}{(\rho, a_0/x, b_0/y)} \approx (\intp{\sigma}{\rho}, \sud{a_0}/x)
   \in R_{\Gamma, x:\Nat}$.
  By B3's \textsc{Ctx-Ext} computation,
  $\intp{q(\sigma)}{(\rho, \sud{a_0}/x)} \approx (\intp{\sigma}{\rho}, \sud{a_0}/x)
   \in R_{\Gamma, x:\Nat}$ (PER-related, via the lifted-substitution evidence).
  Apply \Cref{lem:sem-ty-id} on $T$ to bridge the B3 inner-left:
  \begin{align*}
    &\intp{T}{\intp{q(\sigma)}{(\rho, \sud{a_0}/x)}}
      \approx \intp{T}{(\intp{\sigma}{\rho}, \sud{a_0}/x)} \in \Uc_i
      && \text{(\Cref{lem:sem-ty-id})}
  \end{align*}
  Both B2 and B3 inner-left values are now $\Uc_i$-related to the common value
  $\intp{T}{(\intp{\sigma}{\rho}, \sud{a_0}/x)}$, so by transitivity and
  \Cref{lem:univ-functional}, $R_{B2} \Leftrightarrow R_{B3}$.

  Chaining: $R_s \Leftrightarrow R_{B2} \Leftrightarrow R_{B3} \Leftrightarrow R_T$.
  Hence $b \approx b' \in R_T$.

  \medskip\noindent
  \textbf{Case}
  \begin{mathpar}
    \inferrule[\textsc{Nat-Ne}]
    {e \approx e' \in \Ne}
    {{\uparrow}^N(e) \approx {\uparrow}^N(e') \in \textsf{Nat}}
  \end{mathpar}
  with $a = {\uparrow}^N(e)$, $a' = {\uparrow}^N(e')$.
  Compute the eliminator outputs by \textsc{rec-ne}:
  \begin{align*}
    &\intp{T[q(\sigma)]}{(\rho, {\uparrow}^N(e)/x)} \searrow A
      && \text{(governing the left motive)} \\
    &\intp{T'}{(\intp{\sigma'}{\rho'}, {\uparrow}^N(e')/x)} \searrow A'
      && \text{(governing the right motive)} \\
    &A \approx A' \in \Uc_i \searrow R_T
      && \text{(by the second claim at ${\uparrow}^N(e) \approx {\uparrow}^N(e')$)} \\
    &\intp{z[\sigma]}{\rho} \searrow a_z,\;\;
      \intp{z'}{\intp{\sigma'}{\rho'}} \searrow a_z',\;\;
      a_z \approx a_z' \in R_z
      && \text{(by hypothesis on $z, z'$)} \\
    &b = {\uparrow}^A(\recd{T[q(\sigma)]}{a_z}{s[q^2(\sigma)]}{\rho}{e})
      && \text{(by \textsc{rec-ne})} \\
    &b' = {\uparrow}^{A'}(\recd{T'}{a_z'}{s'}{\intp{\sigma'}{\rho'}}{e'})
      && \text{(by \textsc{rec-ne})}
  \end{align*}

  We must show $b \approx b' \in R_T$.
  By \Cref{lem:ne-in-R} on $A \approx A' \in \Uc_i \searrow R_T$, it suffices to show
  \begin{align*}
    &\recd{T[q(\sigma)]}{a_z}{s[q^2(\sigma)]}{\rho}{e}
      \approx
      \recd{T'}{a_z'}{s'}{\intp{\sigma'}{\rho'}}{e'} \in \Ne,
  \end{align*}
  i.e.\ for every depth $n$ both sides read back via $\Rne_n$ to the same
  neutral normal form.
  By rule \textsc{Re-rec}:
  \begin{align*}
    &\Rne_n(\recd{T[q(\sigma)]}{a_z}{s[q^2(\sigma)]}{\rho}{e})
      \searrow \elimn{x.\,W_T}{w_z}{x, y.\,w_s}{u}
  \end{align*}
  whenever
  \begin{align*}
    &\Rty_{n+1}(A_x) \searrow W_T
      && \text{where $\intp{T[q(\sigma)]}{(\rho, {\uparrow}^N(\ell_n)/x)} \searrow A_x$} \\
    &\Rnf_n({\downarrow}^{A_z}(a_z)) \searrow w_z
      && \text{where $\intp{T[q(\sigma)]}{(\rho, \zed/x)} \searrow A_z$} \\
    &\intp{s[q^2(\sigma)]}{(\rho, {\uparrow}^N(\ell_n)/x, {\uparrow}^{A_x}(\ell_{n+1})/y)}
      \searrow b_s \\
    &\Rnf_{n+2}({\downarrow}^{A_x^{\su{}}}(b_s)) \searrow w_s
      && \text{where $\intp{T[q(\sigma)]}{(\rho, \sud{{\uparrow}^N(\ell_n)}/x)}
        \searrow A_x^{\su{}}$} \\
    &\Rne_n(e) \searrow u
  \end{align*}
  and analogously for the primed side, producing
  $\elimn{x.\,W_T'}{w_z'}{x, y.\,w_s'}{u'}$.
  We show each component reads back to the same syntactic form.

  \emph{Motive $W_T = W_T'$.}
  Apply the second claim of this lemma at the fresh-level relation
  ${\uparrow}^N(\ell_n) \approx {\uparrow}^N(\ell_n) \in \textsf{Nat}$ (by
  \textsc{Nat-Ne}, $\ell_n \approx \ell_n \in \Ne$):
  \begin{align*}
    &\intp{T[q(\sigma)]}{(\rho, {\uparrow}^N(\ell_n)/x)} \approx
      \intp{T'}{(\intp{\sigma'}{\rho'}, {\uparrow}^N(\ell_n)/x)}
      \in \Uc_i \searrow R_T^\ell
      && \text{(second claim at $\ell_n$)} \\
    &W_T = W_T'
      && \text{(by \Cref{lem:type-readback})}
  \end{align*}
  No separate env-pair construction is needed: the second claim is parametric
  in $a \approx a' \in \textsf{Nat}$ and directly relates the two motive values
  at envs over the heterogeneous source--target contexts.

  \emph{Zero branch $w_z = w_z'$.}
  By the \textsc{Nat-Ze} sub-case proved above,
  $a_z \approx a_z' \in R_T$-at-$\zed$ (the PER governed by
  $\intp{T[q(\sigma)]}{(\rho, \zed/x)}$, which equals $A_z$).
  By \Cref{lem:R-readback}:
  \begin{align*}
    &w_z = w_z'
      && \text{(by \Cref{lem:R-readback})}
  \end{align*}

  \emph{Step branch $w_s = w_s'$.}
  We obtain $b_s \approx b_s'$ at fresh levels by replaying the
  \textsc{Nat-Su} case with $a_0 := {\uparrow}^N(\ell_n)$.
  First, lift $\sigma, \sigma'$ to handle the binders of $s$:
  \begin{align*}
    &\semtyeq[\Gamma', x:\Nat]{q(\sigma)}{q(\sigma')}{\Gamma, x:\Nat}
      && \text{(by \Cref{lem:sem-q} on $\sigma \approx \sigma'$ and $\Nat$)} \\
    &\semtyeq[\Gamma', x:\Nat, y:T[q(\sigma)]]
       {q^2(\sigma)}{q^2(\sigma')}{\Gamma, x:\Nat, y:T}
      && \text{(by \Cref{lem:sem-q} again on $T$)}
  \end{align*}
  Build the env-pair via two applications of \textsc{Ctx-Ext}:
  \begin{align*}
    &{\uparrow}^N(\ell_n) \approx {\uparrow}^N(\ell_n) \in \textsf{Nat}
      && \text{(\textsc{Nat-Ne}, $\ell_n \approx \ell_n \in \Ne$)} \\
    &(\rho, {\uparrow}^N(\ell_n)/x) \approx (\rho', {\uparrow}^N(\ell_n)/x)
      \in R_{\Gamma', x:\Nat}
      && \text{(\textsc{Ctx-Ext}, head $\textsf{Nat}$)} \\
    &{\uparrow}^{A_x}(\ell_{n+1}) \approx {\uparrow}^{A_x'}(\ell_{n+1}) \in R_T^\ell
      && \text{(by \Cref{lem:ne-in-R} on $\ell_{n+1} \approx \ell_{n+1} \in \Ne$)} \\
    &(\rho, {\uparrow}^N(\ell_n)/x, {\uparrow}^{A_x}(\ell_{n+1})/y) \\
    &\quad \approx (\rho', {\uparrow}^N(\ell_n)/x, {\uparrow}^{A_x'}(\ell_{n+1})/y)
      \in R_{\Gamma', x:\Nat, y:T[q(\sigma)]}
      && \text{(\textsc{Ctx-Ext}, head $R_T^\ell$)}
  \end{align*}
  Apply $\semtyeq[\Gamma, x:\Nat, y:T]{s}{s'}{T[\Uparrow \circ \Uparrow, \su{x_1}/x]}$
  at $\Gamma', x:\Nat, y:T[q(\sigma)]$, the doubly-lifted substitution
  $q^2(\sigma) \approx q^2(\sigma')$, and this env-pair.
  The four-value pattern of $s, s'$ gives:
  \begin{align*}
    &b_s \approx b_s' \in R_T^{step}
      && \text{(outer-left vs inner-right, parallel to the helper's} \\
    & && \text{Nat-Su case)}
  \end{align*}
  where $R_T^{step}$ is governed by
  $\intp{(T[\Uparrow \circ \Uparrow, \su{x_1}/x])[q^2(\sigma)]}
       {(\rho, {\uparrow}^N(\ell_n)/x, {\uparrow}^{A_x}(\ell_{n+1})/y)}$.

  By the B1/B2/B3 bridging of the \textsc{Nat-Su} case (with $a_0$ replaced by
  ${\uparrow}^N(\ell_n)$):
  \begin{align*}
    &R_T^{step} \Leftrightarrow R_T^\ell\text{-at-}\sud{{\uparrow}^N(\ell_n)}
      && \text{(B1/B2/B3 bridges, \Cref{lem:univ-functional})}
  \end{align*}
  i.e.\ the PER at $\intp{T[q(\sigma)]}{(\rho, \sud{{\uparrow}^N(\ell_n)}/x)}$,
  which equals $A_x^{\su{}}$.
  Hence $b_s \approx b_s' \in R_T^\ell$-at-$\sud{\ell_n}$, and:
  \begin{align*}
    &w_s = w_s'
      && \text{(by \Cref{lem:R-readback})}
  \end{align*}

  \emph{Underlying neutral $u = u'$.}
  By hypothesis $e \approx e' \in \Ne$:
  \begin{align*}
    &\Rne_n(e) \searrow u,\;\; \Rne_n(e') \searrow u
      && \text{(by definition of $\Ne$)}
  \end{align*}

  Combining:
  \begin{align*}
    &\Rne_n(\recd{T[q(\sigma)]}{a_z}{s[q^2(\sigma)]}{\rho}{e})
      \searrow \elimn{x.\,W_T}{w_z}{x, y.\,w_s}{u} \\
    &\Rne_n(\recd{T'}{a_z'}{s'}{\intp{\sigma'}{\rho'}}{e'})
      \searrow \elimn{x.\,W_T}{w_z}{x, y.\,w_s}{u}
      && \text{(same syntactic form)} \\
    &\recd{T[q(\sigma)]}{a_z}{s[q^2(\sigma)]}{\rho}{e}
      \approx \recd{T'}{a_z'}{s'}{\intp{\sigma'}{\rho'}}{e'} \in \Ne
      && \text{(definition of $\Ne$)}
  \end{align*}
  Hence $b \approx b' \in R_T$.
\end{proof}

\begin{lemma}[Fundamental --- \textsc{N-E} / \textsc{Rec-Cong}]\labeledit{lem:fund-rec}
  If $\semtyeq[\Gamma, x:\Nat]{T}{T'}{\Ty{i}}$,
  $\semtyeq[\Gamma]{z}{z'}{T[\ze/x]}$,
  $\semtyeq[\Gamma, x:\Nat, y:T]{s}{s'}{T[\Uparrow \circ \Uparrow, \su{x_1}/x]}$, and
  $\semtyeq[\Gamma]{t}{t'}{\Nat}$,
  then $\semtyeq[\Gamma]{\elimn{x.\,T}{z}{x, y.\, s}{t}}{\elimn{x.\,T}{z'}{x, y.\, s'}{t'}}{T[t/x]}$.
\end{lemma}
\begin{proof}
  Let $\Gamma' \approx \Gamma' \searrow R'$,
  $\semtyeq[\Gamma']{\sigma}{\sigma'}{\Gamma}$, and
  $\Pf :: \rho \approx \rho' \in R'$ be given.
  Write $\rho_\sigma := \intp{\sigma}{\rho}$ and $\rho_{\sigma'} := \intp{\sigma'}{\rho'}$.
  From $\semtyeq[\Gamma']{\sigma}{\sigma'}{\Gamma}$ at $\id$, $\Pf$:
  $\rho_\sigma \approx \rho_{\sigma'} \in R_\Gamma$ where
  $\Gamma \approx \Gamma \searrow R_\Gamma$.

  By $\semtyeq[\Gamma]{t}{t'}{\Nat}$ applied at $\Gamma'$, $\sigma \approx \sigma'$,
  $\rho \approx \rho' \in R'$:
  \begin{align*}
    &\{\intp{t[\sigma]}{\rho},\;
      \intp{t}{\rho_\sigma},\;
      \intp{t'}{\rho_{\sigma'}},\;
      \intp{t'[\sigma']}{\rho'}\}
      \subseteq \textsf{Nat}
      && \text{(by hypothesis on $t, t'$)}
  \end{align*}

  Compute the four eliminator term values:
  \begin{align*}
    &\intp{(\elimn{x.\,T}{z}{x, y.\, s}{t})[\sigma]}{\rho}
      && \text{(definition of substitution} \\
    & \quad = \intp{\elimn{x.\,T[q(\sigma)]}
        {z[\sigma]}{x, y.\, s[q^2(\sigma)]}{t[\sigma]}}{\rho}
      && \text{on $\telimn$)} \\
    &\quad = \mathsf{rec}(\intp{t[\sigma]}{\rho},\,
      T[q(\sigma)],\, z[\sigma],\, s[q^2(\sigma)],\, \rho)
      && \text{(by ev-rec)} \\[4pt]
    &\intp{\elimn{x.\,T}{z}{x, y.\, s}{t}}{\rho_\sigma} \\
    & \quad = \mathsf{rec}(\intp{t}{\rho_\sigma},\, T,\, z,\, s,\, \rho_\sigma)
      && \text{(by ev-rec)}
  \end{align*}
  Analogous computations for the primed pair (using $z', s', t'$):
  $\mathsf{rec}(\intp{t'[\sigma']}{\rho'},\, T[q(\sigma')],\,
  z'[\sigma'],\, s'[q^2(\sigma')],\, \rho')$ and
  $\mathsf{rec}(\intp{t'}{\rho_{\sigma'}},\, T,\, z',\, s',\,
  \rho_{\sigma'})$.

  Before applying the helpers, derive reflexive instances of all hypotheses
  (the helpers' RHS are in primed form; we sometimes need
  unprimed-to-unprimed instances):
  \begin{align*}
    &\semtyeq[\Gamma, x:\Nat]{T}{T}{\Ty{i}}
      && \text{(reflexive, by \Cref{lem:fund-sym,lem:fund-trans} on $T \approx T'$)} \\
    &\semtyeq[\Gamma]{z}{z}{T[\ze/x]}
      && \text{(reflexive)} \\
    &\semtyeq[\Gamma, x:\Nat, y:T]{s}{s}{T[\Uparrow \circ \Uparrow, \su{x_1}/x]}
      && \text{(reflexive)} \\
    &\semtyeq[\Gamma]{t}{t}{\Nat}
      && \text{(reflexive)} \\
    &\semtyeq[\Gamma']{\sigma}{\sigma}{\Gamma}
      && \text{(reflexive)}
  \end{align*}
  (Primed analogues hold symmetrically.)

  \emph{Inner pair via the unsubstituted helper.}
  Apply \Cref{lem:sem-rec} at $T \approx T$ (reflexive), $z \approx z'$,
  $s \approx s'$, $R_\Gamma$, env-pair $\rho_\sigma \approx \rho_{\sigma'}$,
  and witness $\intp{t}{\rho_\sigma} \approx \intp{t'}{\rho_{\sigma'}}
  \in \textsf{Nat}$ (inner-left vs inner-right of the four-value pattern of
  $t, t'$):
  \begin{align*}
    &\mathsf{rec}(\intp{t}{\rho_\sigma},\, T,\, z,\, s,\, \rho_\sigma)
      \searrow b_i \\
    &\mathsf{rec}(\intp{t'}{\rho_{\sigma'}},\, T,\, z',\, s',\, \rho_{\sigma'})
      \searrow b_i' \\
    &b_i \approx b_i' \in R_T
      && \text{(by \Cref{lem:sem-rec})}
  \end{align*}
  where $R_T$ is governed by
  $\intp{T}{(\rho_\sigma, \intp{t}{\rho_\sigma}/x)} \approx
  \intp{T}{(\rho_{\sigma'}, \intp{t'}{\rho_{\sigma'}}/x)} \in \Uc_i \searrow R_T$.
  These match the goal's inner-left and inner-right term values.

  \emph{Outer-vs-inner via the one-sided helper.}
  Apply \Cref{lem:sem-rec-subst} at $T \approx T$, $z \approx z$, $s \approx s$
  (all reflexive), $\Gamma' \approx \Gamma' \searrow R'$,
  $\sigma \approx \sigma$ (reflexive),
  $\rho \approx \rho \in R'$ (reflexive), and witness
  $\intp{t[\sigma]}{\rho} \approx \intp{t}{\rho_\sigma} \in \textsf{Nat}$
  (outer-left vs inner-left of the four-value pattern of $t$):
  \begin{align*}
    &\mathsf{rec}(\intp{t[\sigma]}{\rho},\, T[q(\sigma)],\, z[\sigma],\,
      s[q^2(\sigma)],\, \rho) \searrow b_o \\
    &\mathsf{rec}(\intp{t}{\rho_\sigma},\, T,\, z,\, s,\, \rho_\sigma) \searrow b_i \\
    &b_o \approx b_i \in R_T^{oi}
      && \text{(by \Cref{lem:sem-rec-subst})}
  \end{align*}
  where $R_T^{oi}$ is governed by
  $\intp{T[q(\sigma)]}{(\rho, \intp{t[\sigma]}{\rho}/x)} \approx
  \intp{T}{(\rho_\sigma, \intp{t}{\rho_\sigma}/x)} \in \Uc_i \searrow R_T^{oi}$.
  By \Cref{lem:univ-functional}, $R_T^{oi} \Leftrightarrow R_T$.
  Hence $b_o \approx b_i \in R_T$.

  Apply \Cref{lem:sem-rec-subst} symmetrically at the primed reflexive
  instances ($T' \approx T'$, $z' \approx z'$, $s' \approx s'$,
  $\sigma' \approx \sigma'$, $\rho' \approx \rho' \in R'$), with witness
  $\intp{t'[\sigma']}{\rho'} \approx \intp{t'}{\rho_{\sigma'}} \in \textsf{Nat}$:
  \begin{align*}
    &b_o' \approx b_i' \in R_T
      && \text{(by \Cref{lem:sem-rec-subst} and \Cref{lem:univ-functional})}
  \end{align*}

  \emph{Combining.}
  We assemble the four goal term values:
  \begin{align*}
    &b_i \approx b_i' \in R_T
      && \text{(inner pair, above)} \\
    &b_o \approx b_i \in R_T
      && \text{(outer-vs-inner, left side, above)} \\
    &b_o' \approx b_i' \in R_T
      && \text{(outer-vs-inner, right side, above)} \\
    &b_o \approx b_o' \in R_T
      && \text{(by symmetry and transitivity of $R_T$)}
  \end{align*}
  Thus all four goal term values
  $\{b_o, b_i, b_i', b_o'\}$ are pairwise related in $R_T$.

  \emph{Type part.}
  The goal type is $T[t/x]$ on the LHS and $T[t'/x]$ on the RHS (by the syntactic
  conclusion's type, via the four-value pattern of the outer
  $\semtyeq[\Gamma]{\elimn{\ldots}{\ldots}{\ldots}{t}}{\elimn{\ldots}{\ldots}{\ldots}{t'}}{T[t/x]}$
  semantic judgment).
  We bridge the four goal type values via outer and inner instantiations of
  $T \approx T'$.

  \emph{Outer instantiation.}
  By \Cref{lem:sem-ext-pre} on $\semtyeq[\Gamma']{\sigma}{\sigma'}{\Gamma}$ and
  $\semtyeq[\Gamma]{t}{t'}{\Nat}$:
  \begin{align*}
    &\semtyeq[\Gamma']{(\sigma, t[\sigma]/x)}{(\sigma', t'[\sigma']/x)}{\Gamma, x:\Nat}
      && \text{(\Cref{lem:sem-ext-pre})}
  \end{align*}
  Apply $\semtyeq[\Gamma, x:\Nat]{T}{T'}{\Ty{i}}$ at $\Gamma'$, this semantic
  substitution, and $\rho \approx \rho' \in R'$:
  \begin{align*}
    &\{\intp{T[\sigma, t[\sigma]/x]}{\rho},\;
      \intp{T}{(\rho_\sigma, \intp{t[\sigma]}{\rho}/x)}, \\
    &\quad
      \intp{T'}{(\rho_{\sigma'}, \intp{t'[\sigma']}{\rho'}/x)},\;
      \intp{T'[\sigma', t'[\sigma']/x]}{\rho'}\}
      \subseteq \Uc_i \searrow R_T^{out}
      && \text{(by hypothesis on $T \approx T'$)} \\
    &T[\sigma, t[\sigma]/x] = 
      && \text{(by \Cref{lem:subst-ext-comm})} \\
    &T'[\sigma', t'[\sigma']/x] = (T'[t'/x])[\sigma']
      && \text{(by \Cref{lem:subst-ext-comm})}
  \end{align*}
  Now the goal's outer-RHS type is $\intp{}{\rho'}$, but the
  outer instantiation gives $\intp{(T'[t'/x])[\sigma']}{\rho'}$.
  We show these coincide modulo $\Uc_i$:
  \begin{align*}
    &\intp{}{\rho'} \approx \intp{(T'[t'/x])[\sigma']}{\rho'} \in \Uc_i
      && \text{(four-value pattern of $T \approx T'$ at } \\
    & && \text{the outer-RHS instantiation, applied} \\
    & && \text{at $(\sigma', t'[\sigma']/x)$ with $\rho' \approx \rho' \in R'$)}
  \end{align*}
  So the four values lie in $\Uc_i \searrow R_T^{out}$ with the outer-LHS value
  exactly matching the goal's $\intp{}{\rho}$ and the outer-RHS
  value matching the goal's $\intp{}{\rho'}$ (modulo
  $\Uc_i$, hence by \Cref{lem:univ-functional}).

  \emph{Inner instantiation.}
  By \Cref{lem:sem-ext-pre} over $\Gamma$ on $\semtyeq[\Gamma]{\id}{\id}{\Gamma}$
  and $\semtyeq[\Gamma]{t}{t'}{\Nat}$:
  \begin{align*}
    &\semtyeq[\Gamma]{(\id, t/x)}{(\id, t'/x)}{\Gamma, x:\Nat}
      && \text{(\Cref{lem:sem-ext-pre})}
  \end{align*}
  Apply $\semtyeq[\Gamma, x:\Nat]{T}{T'}{\Ty{i}}$ at $\Gamma$, this semantic
  substitution, and $\rho_\sigma \approx \rho_{\sigma'} \in R_\Gamma$:
  \begin{align*}
    &\{\intp{T[\id, t/x]}{\rho_\sigma},\;
      \intp{T}{(\rho_\sigma, \intp{t}{\rho_\sigma}/x)}, \\
    &\quad
      \intp{T'}{(\rho_{\sigma'}, \intp{t'}{\rho_{\sigma'}}/x)},\;
      \intp{T'[\id, t'/x]}{\rho_{\sigma'}}\}
      \subseteq \Uc_i \searrow R_T^{in}
      && \text{(by hypothesis on $T \approx T'$)} \\
    &T[\id, t/x] = T[t/x]
      && \text{(definition of single substitution)} \\
    &T'[\id, t'/x] = T'[t'/x]
      && \text{(definition of single substitution)}
  \end{align*}

  \emph{Bridging.}
  The inner-pair values
  $\intp{T}{(\rho_\sigma, \intp{t}{\rho_\sigma}/x)}$ (shared between outer and
  inner instantiations) and
  $\intp{T'}{(\rho_{\sigma'}, \intp{t'}{\rho_{\sigma'}}/x)}$ (shared) bridge the
  PERs:
  \begin{align*}
    &R_T^{out} \Leftrightarrow R_T^{in}
      && \text{(by \Cref{lem:univ-functional} on shared inner values)} \\
    &\intp{T}{(\rho_\sigma, \intp{t}{\rho_\sigma}/x)} \approx
     \intp{T}{(\rho_{\sigma'}, \intp{t'}{\rho_{\sigma'}}/x)} \in \Uc_i \searrow R_T
      && \text{(unsubstituted helper's $R_T$ governance)} \\
    &R_T^{in} \Leftrightarrow R_T
      && \text{(by \Cref{lem:univ-functional} on the shared} \\
    & && \text{inner-LHS value with the helper's $R_T$)}
  \end{align*}
  Hence all four goal type values lie in $\Uc_i \searrow R_T$,
  with the term values $\{b_o, b_i, b_i', b_o'\}$ in $R_T$ as established.
\end{proof}

\begin{lemma}[Fundamental --- \textsc{Subsump}]\labeledit{lem:fund-subsump}
  If $\semtyeq[\Gamma]{T}{T}{\Ty{i}}$ and $\semtyeq[\Gamma]{t}{t}{S}$ and
  $\Gamma \vDash S \subseteq T$, then $\semtyeq[\Gamma]{t}{t}{T}$.
\end{lemma}
\begin{proof}
  Let $\Gamma' \approx \Gamma' \searrow R'$,
  $\semtyeq[\Gamma']{\sigma}{\sigma'}{\Gamma}$, and
  $\Pf :: \rho \approx \rho' \in R'$ be given.

  \emph{Type part.}
  By $\semtyeq[\Gamma]{T}{T}{\Ty{i}}$ at the given data:
  \begin{align*}
    &\{\intp{T[\sigma]}{\rho},\;
      \intp{T}{\intp{\sigma}{\rho}},\;
      \intp{T}{\intp{\sigma'}{\rho'}},\;
      \intp{T[\sigma']}{\rho'}\}
      \subseteq \Uc_i \searrow R_T
      && \text{(by hypothesis on $T$)}
  \end{align*}

  \emph{Term part.}
  By $\semtyeq[\Gamma]{t}{t}{S}$ at the given data:
  \begin{align*}
    &\{\intp{S[\sigma]}{\rho},\;
      \intp{S}{\intp{\sigma}{\rho}},\;
      \intp{S}{\intp{\sigma'}{\rho'}},\;
      \intp{S[\sigma']}{\rho'}\}
      \subseteq \Uc_i \searrow R_S
      && \text{(by hypothesis on $S$)} \\
    &\{\intp{t[\sigma]}{\rho},\;
      \intp{t}{\intp{\sigma}{\rho}},\;
      \intp{t}{\intp{\sigma'}{\rho'}},\;
      \intp{t[\sigma']}{\rho'}\}
      \subseteq R_S
      && \text{(by hypothesis on $t$)}
  \end{align*}

  By $\Gamma \vDash S \subseteq T$ at the given data:
  \begin{align*}
    &\intp{S[\sigma]}{\rho} \approx \intp{S}{\intp{\sigma}{\rho}} \in \Uc_i
      && \text{(by semantic subtyping, commutation)} \\
    &\intp{T[\sigma']}{\rho'} \approx \intp{T}{\intp{\sigma'}{\rho'}} \in \Uc_i
      && \text{(by semantic subtyping, commutation)} \\
    &\intp{S}{\intp{\sigma}{\rho}} \mathrel{<:_i} \intp{T}{\intp{\sigma'}{\rho'}}
      && \text{(by semantic subtyping)}
  \end{align*}
  By \Cref{lem:subty-respects} applied to the subtyping
  $\intp{S}{\intp{\sigma}{\rho}} \mathrel{<:_i} \intp{T}{\intp{\sigma'}{\rho'}}$
  together with $\intp{S}{\intp{\sigma}{\rho}} \approx \intp{S}{\intp{\sigma'}{\rho'}} \in \Uc_i \searrow R_S$
  and $\intp{T}{\intp{\sigma}{\rho}} \approx \intp{T}{\intp{\sigma'}{\rho'}} \in \Uc_i \searrow R_T$:
  \begin{align*}
    &R_S \subseteq R_T
      && \text{(by \Cref{lem:subty-respects})}
  \end{align*}

  Hence the four term values, which lie in $R_S$, also lie in $R_T$:
  \begin{align*}
    &\{\intp{t[\sigma]}{\rho},\;
      \intp{t}{\intp{\sigma}{\rho}},\;
      \intp{t}{\intp{\sigma'}{\rho'}},\;
      \intp{t[\sigma']}{\rho'}\}
      \subseteq R_T
      && \text{(by $R_S \subseteq R_T$)}
  \end{align*}
\end{proof}

\subsection{Semantic Computation Rules}\labeledit{sec:comp:sem-comp}

\begin{lemma}[Fundamental --- \textsc{$\Pi$-$\beta$}]\labeledit{lem:fund-beta}
  If $\semtyeq[\Gamma]{S}{S}{\Ty{i}}$ and $\semtyeq[\Gamma, x:S]{T}{T}{\Ty{i}}$ and
  $\semtyeq[\Gamma, x:S]{t}{t}{T}$ and $\semtyeq[\Gamma]{s}{s}{S}$,
  then $\semtyeq[\Gamma]{(\lambda x.\,t)\;s}{t[s/x]}{T[s/x]}$.
\end{lemma}
\begin{proof}
  Let $\Gamma' \approx \Gamma' \searrow R'$,
  $\semtyeq[\Gamma']{\sigma}{\sigma'}{\Gamma}$, and
  $\Pf :: \rho \approx \rho' \in R'$ be given.

  \emph{LHS computation at $\sigma, \rho$.}
  By the syntactic action of substitution on $\lambda$ and application, then
  ev-app/ev-lam/closure-application:
  \begin{align*}
    &((\lambda x.\,t)\;s)[\sigma]
      = (\lambda x.\,t[q(\sigma)])\;s[\sigma]
      && \text{(syntactic, definition of substitution)} \\
    &\intp{(\lambda x.\,t[q(\sigma)])\;s[\sigma]}{\rho}
      \searrow b_L
      \text{ where} \\
    &\quad \intp{\lambda x.\,t[q(\sigma)]}{\rho} \searrow \lambda(t[q(\sigma)], \rho)
      && \text{(ev-lam)} \\
    &\quad \intp{s[\sigma]}{\rho} \searrow a_s
      && \text{(some $a_s$)} \\
    &\quad \lambda(t[q(\sigma)], \rho) \cdot a_s \searrow b_L
      && \text{(closure application:} \\
    & && \text{$\intp{t[q(\sigma)]}{(\rho, a_s/x)} \searrow b_L$)}
  \end{align*}

  \emph{RHS expression at $\sigma, \rho$.}
  Using the sugar $[s/x] = [\id, s/x]$ and \Cref{lem:subst-ext-comm,lem:q-ext}:
  \begin{align*}
    & = t[q(\sigma)][s[\sigma]/x] = t[\sigma, s[\sigma]/x]
      && \text{(by \Cref{lem:subst-ext-comm,lem:q-ext})}
  \end{align*}
  Let $b_R$ satisfy $\intp{t[\sigma, s[\sigma]/x]}{\rho} \searrow b_R$ (existence
  via the four-value pattern below).
  We relate $b_L$ and $b_R$ in the term PER via the four-value pattern of $t$
  on the extended substitution.

  \emph{Relatedness.}
  By \Cref{lem:sem-ext-pre} on $\semtyeq[\Gamma']{\sigma}{\sigma'}{\Gamma}$ and
  $\semtyeq[\Gamma]{s}{s}{S}$:
  \begin{align*}
    &\semtyeq[\Gamma']{(\sigma, s[\sigma]/x)}{(\sigma', s[\sigma']/x)}{\Gamma, x:S}
      && \text{(\Cref{lem:sem-ext-pre})}
  \end{align*}
  Apply $\semtyeq[\Gamma, x:S]{t}{t}{T}$ at $\Gamma'$, this extended substitution,
  and $\rho \approx \rho' \in R'$.
  Let $R_T$ be the PER governing the four-value pattern of $t$ at this
  instantiation; concretely, $R_T$ is determined by the type four-value pattern
  of $T$ at the same env-pair (carried by semantic typing of $t : T$):
  \begin{align*}
    &\{\intp{T[\sigma, s[\sigma]/x]}{\rho},\;
      \intp{T}{\intp{(\sigma, s[\sigma]/x)}{\rho}}, \\
    &\quad \intp{T}{\intp{(\sigma', s[\sigma']/x)}{\rho'}},\;
      \intp{T[\sigma', s[\sigma']/x]}{\rho'}\}
      \subseteq \Uc_i \searrow R_T
  \end{align*}
  The term four-value pattern then gives:
  \begin{align*}
    &\{\intp{t[\sigma, s[\sigma]/x]}{\rho},\;
      \intp{t}{\intp{(\sigma, s[\sigma]/x)}{\rho}}, \\
    &\quad \intp{t}{\intp{(\sigma', s[\sigma']/x)}{\rho'}},\;
      \intp{t[\sigma', s[\sigma']/x]}{\rho'}\}
      \subseteq R_T
      && \text{(by hypothesis on $t$)}
  \end{align*}
  By the syntactic identity above (and analogously at $\sigma', \rho'$), the
  outer values coincide with the goal's RHS outer-values
  $\{\intp{}{\rho}, \intp{}{\rho'}\}$.

  \emph{LHS values via four-value pattern of $t$ at $q(\sigma)$.}
  We bridge $b_L = \intp{t[q(\sigma)]}{(\rho, a_s/x)}$ to the inner-left value
  $\intp{t}{\intp{(\sigma, s[\sigma]/x)}{\rho}}$ above.
  By \Cref{lem:sem-q} on $\semtyeq[\Gamma']{\sigma}{\sigma}{\Gamma}$ (reflexive,
  via \Cref{lem:sem-sym,lem:sem-trans}) and $\semtyeq[\Gamma]{S}{S}{\Ty{i}}$:
  \begin{align*}
    &\semtyeq[\Gamma', x:S[\sigma]]{q(\sigma)}{q(\sigma)}{\Gamma, x:S}
      && \text{(\Cref{lem:sem-q})}
  \end{align*}
  The reflexive env-pair $(\rho, a_s/x) \approx (\rho, a_s/x)$ in
  $R_{\Gamma', x:S[\sigma]}$ holds via \textsc{Ctx-Ext} (head $a_s \approx a_s$
  reflexively in the head PER, tail $\rho \approx \rho \in R'$).
  Apply $\semtyeq[\Gamma, x:S]{t}{t}{T}$ at $\Gamma', x:S[\sigma]$,
  $q(\sigma) \approx q(\sigma)$, and this env-pair (four-value pattern of $t$,
  outer-left vs inner-left):
  \begin{align*}
    &\intp{t[q(\sigma)]}{(\rho, a_s/x)}
      \approx \intp{t}{\intp{q(\sigma)}{(\rho, a_s/x)}} \in R_T
      && \text{(four-value pattern of $t$)}
  \end{align*}
  Bridge $\intp{q(\sigma)}{(\rho, a_s/x)}$ to $\intp{(\sigma, s[\sigma]/x)}{\rho}$
  in $R_{\Gamma, x:S}$ pointwise. Both envs have head $a_s$:
  \begin{align*}
    &\intp{q(\sigma)}{(\rho, a_s/x)}
      = (\intp{\sigma[\Uparrow]}{(\rho, a_s/x)},\, a_s/x)
      && \text{(by $q(\sigma) = (\sigma[\Uparrow], 0/x)$, pointwise eval)} \\
    &\intp{(\sigma, s[\sigma]/x)}{\rho}
      = (\intp{\sigma}{\rho},\, a_s/x)
      && \text{(pointwise eval; $\intp{s[\sigma]}{\rho} = a_s$)} \\
    &\intp{\sigma[\Uparrow]}{(\rho, a_s/x)}
      \approx \intp{\sigma}{\rho} \in R_\Gamma
      && \text{(four-value pattern of $\sigma \approx \sigma$ at} \\
    & && \text{$\semweak[\Gamma', x:S[\sigma]]{\Uparrow}{\Gamma'}$, env} \\
    & && \text{$(\rho, a_s/x) \approx (\rho, a_s/x) \in R_{\Gamma', x:S[\sigma]}$;} \\
    & && \text{using $\intp{\Uparrow}{(\rho, a_s/x)} = \rho$ pointwise)} \\
    &\intp{q(\sigma)}{(\rho, a_s/x)}
      \approx \intp{(\sigma, s[\sigma]/x)}{\rho} \in R_{\Gamma, x:S}
      && \text{(by \textsc{Ctx-Ext}, head $a_s \approx a_s$ reflexively)}
  \end{align*}
  Apply \Cref{lem:sem-ty-id} on the reflexive instance
  $\semtyeq[\Gamma, x:S]{t}{t}{T}$ at this env-pair:
  \begin{align*}
    &\intp{t}{\intp{q(\sigma)}{(\rho, a_s/x)}}
      \approx \intp{t}{\intp{(\sigma, s[\sigma]/x)}{\rho}} \in R_T
      && \text{(\Cref{lem:sem-ty-id})}
  \end{align*}
  By transitivity, $b_L \approx \intp{t}{\intp{(\sigma, s[\sigma]/x)}{\rho}}$
  in $R_T$ (the inner-left value of the relatedness pattern).
  Symmetrically for the primed side.

  Combining: all eight values (four LHS, four RHS) lie in $R_T$, so
  the LHS-RHS pair at each of the four positions is in $R_T$.

  \emph{Type part.}
  Already established: the type four-value pattern displayed above (in
  $\Uc_i \searrow R_T$) is the goal's $T[s/x]$ pattern, since
  $T[\sigma, s[\sigma]/x] = $ by the syntactic identity used for
  the term part.
\end{proof}

\begin{lemma}[Fundamental --- \textsc{$\Pi$-$\eta$}]\labeledit{lem:fund-eta}
  If $\semtyeq[\Gamma]{S}{S}{\Ty{i}}$ and $\semtyeq[\Gamma, x:S]{T}{T}{\Ty{i}}$ and
  $\semtyeq[\Gamma]{t}{t}{\Pi(x:S).\,T}$,
  then $\semtyeq[\Gamma]{t}{\lambda x.\,(t[\Uparrow]\;x)}{{\Pi(x:S).\,T}}$.
\end{lemma}
\begin{proof}
  Let $\Gamma' \approx \Gamma' \searrow R'$,
  $\semtyeq[\Gamma']{\sigma}{\sigma'}{\Gamma}$, and
  $\Pf :: \rho \approx \rho' \in R'$ be given.

  \emph{LHS values.}
  By $\semtyeq[\Gamma]{t}{t}{\Pi(x:S).\,T}$ at $\Gamma'$, $\sigma \approx \sigma'$,
  $\rho \approx \rho' \in R'$, the four evaluations of $t$ are related in the
  function PER $R$:
  \begin{align*}
    &\{\intp{t[\sigma]}{\rho},\;
      \intp{t}{\intp{\sigma}{\rho}},\;
      \intp{t}{\intp{\sigma'}{\rho'}},\;
      \intp{t[\sigma']}{\rho'}\}
      \subseteq R
      && \text{(by hypothesis on $t$)}
  \end{align*}
  with biconditional characterization
  $f \approx f' \in R \;\leftrightarrow\;
  (\forall \Pf_I :: a \approx a' \in R_I.\; f \cdot a \approx f' \cdot a' \in R_O(\Pf_I))$.

  \emph{RHS values.}
  The RHS evaluates on the $\sigma', \rho'$-side only:
  \begin{align*}
    &\intp{(\lambda x.\,(t[\Uparrow]\;x))[\sigma']}{\rho'}
      \searrow f_o
      && \text{(syntactic and ev-lam)} \\
    &\intp{\lambda x.\,(t[\Uparrow]\;x)}{\intp{\sigma'}{\rho'}}
      \searrow f_i
      && \text{(ev-lam)}
  \end{align*}
  where $f_o := \lambda(t[\Uparrow][q(\sigma')]\;x,\, \rho')$ and
  $f_i := \lambda(t[\Uparrow]\;x,\, \intp{\sigma'}{\rho'})$.

  \emph{Step 1: $\intp{t[\sigma']}{\rho'} \approx f_o$ in $R$.}
  By the biconditional, suffices to show: for every
  $\Pf_I :: a \approx a' \in R_I$,
  $\intp{t[\sigma']}{\rho'} \cdot a \approx f_o \cdot a' \in R_O(\Pf_I)$.

  Compute $f_o \cdot a'$:
  \begin{align*}
    &f_o \cdot a' = \intp{t[\Uparrow][q(\sigma')]\;x}{(\rho', a'/x)}
      && \text{(closure app)} \\
    &\quad = \intp{t[\Uparrow][q(\sigma')]}{(\rho', a'/x)} \cdot a'
      && \text{(ev-app)} \\
    &\quad = \intp{t[\sigma'[\Uparrow]]}{(\rho', a'/x)} \cdot a'
      && \text{(\Cref{lem:subst-assoc} and \Cref{lem:shift-ext}:} \\
    & && \text{$\Uparrow \circ q(\sigma') = \sigma'[\Uparrow]$)}
  \end{align*}
  Bridge $\intp{t[\sigma'[\Uparrow]]}{(\rho', a'/x)} \approx \intp{t[\sigma']}{\rho'}
   \in R$:
  \begin{align*}
    &\semweak[\Gamma', x:S[\sigma']]{\Uparrow}{\Gamma'}
      && \text{(\Cref{lem:sem-wk-shift})} \\
    &\semtyeq[\Gamma', x:S[\sigma']]{\sigma'[\Uparrow]}{\sigma'[\Uparrow]}{\Gamma}
      && \text{(\Cref{lem:sem-wk-subst} on $\sigma' \approx \sigma'$ refl.)}
  \end{align*}
  Apply hypothesis on $t$ at $\Gamma', x:S[\sigma']$, $\sigma'[\Uparrow] \approx
   \sigma'[\Uparrow]$, env-pair $(\rho', a'/x) \approx (\rho', a'/x) \in
   R_{\Gamma', x:S[\sigma']}$:
  \begin{align*}
    &\intp{t[\sigma'[\Uparrow]]}{(\rho', a'/x)} \\
    &\quad \approx \intp{t}{\intp{\sigma'[\Uparrow]}{(\rho', a'/x)}} \in R
      && \text{(outer-left vs inner-left of $t$)}
  \end{align*}
  Apply hypothesis on $\sigma'$ at the same data
  ($\Uparrow$ as the weakening, env-pair as above), giving
  $\intp{\sigma'[\Uparrow]}{(\rho', a'/x)} \approx \intp{\sigma'}{\rho'} \in R_\Gamma$
  (using $\intp{\Uparrow}{(\rho', a'/x)} = \rho'$ pointwise).
  By \Cref{lem:sem-ty-id} on $t$:
  \begin{align*}
    &\intp{t}{\intp{\sigma'[\Uparrow]}{(\rho', a'/x)}}
      \approx \intp{t}{\intp{\sigma'}{\rho'}} \in R
      && \text{(\Cref{lem:sem-ty-id})}
  \end{align*}
  Chained with $\intp{t}{\intp{\sigma'}{\rho'}} \approx \intp{t[\sigma']}{\rho'}
   \in R$ (four-value pattern of $t$, inner-left vs outer-left):
  \begin{align*}
    &\intp{t[\sigma'[\Uparrow]]}{(\rho', a'/x)}
      \approx \intp{t[\sigma']}{\rho'} \in R
      && \text{(transitivity)}
  \end{align*}
  Apply the biconditional of $R$ at $\Pf_I$:
  $\intp{t[\sigma'[\Uparrow]]}{(\rho', a'/x)} \cdot a
   \approx \intp{t[\sigma']}{\rho'} \cdot a' \in R_O(\Pf_I)$.
  Since $f_o \cdot a' = \intp{t[\sigma'[\Uparrow]]}{(\rho', a'/x)} \cdot a'$, by
  symmetry of $R_O$ on the reflexive instance of
  $\intp{t[\sigma']}{\rho'} \approx \intp{t[\sigma']}{\rho'} \in R$ at
  $a \approx a'$:
  $\intp{t[\sigma']}{\rho'} \cdot a \approx f_o \cdot a' \in R_O(\Pf_I)$.

  \emph{Step 2: $\intp{t}{\intp{\sigma'}{\rho'}} \approx f_i$ in $R$.}
  Suffices to show: for every $\Pf_I :: a \approx a' \in R_I$,
  $\intp{t}{\intp{\sigma'}{\rho'}} \cdot a \approx f_i \cdot a' \in R_O(\Pf_I)$.

  Compute:
  \begin{align*}
    &f_i \cdot a' = \intp{t[\Uparrow]\;x}{(\intp{\sigma'}{\rho'}, a'/x)}
      && \text{(closure app)} \\
    &\quad = \intp{t[\Uparrow]}{(\intp{\sigma'}{\rho'}, a'/x)} \cdot a'
      && \text{(ev-app)}
  \end{align*}
  Bridge $\intp{t[\Uparrow]}{(\intp{\sigma'}{\rho'}, a'/x)} \approx \intp{t}{\intp{\sigma'}{\rho'}}
   \in R$ via the four-value pattern of $t$ (reflexive) at the new context
  $\Gamma, x:S$, $\semweak[\Gamma, x:S]{\Uparrow}{\Gamma}$, env-pair
  $(\intp{\sigma'}{\rho'}, a'/x) \approx (\intp{\sigma'}{\rho'}, a'/x)
   \in R_{\Gamma, x:S}$:
  outer-left $\intp{t[\Uparrow]}{(\intp{\sigma'}{\rho'}, a'/x)}$, inner-left
  $\intp{t}{\intp{\Uparrow}{(\intp{\sigma'}{\rho'}, a'/x)}} = \intp{t}{\intp{\sigma'}{\rho'}}$
  (pointwise).
  Apply biconditional at $\Pf_I$ and the reflexive
  $\intp{t}{\intp{\sigma'}{\rho'}} \approx \intp{t}{\intp{\sigma'}{\rho'}}$ at
  $a \approx a'$:
  $\intp{t}{\intp{\sigma'}{\rho'}} \cdot a \approx f_i \cdot a' \in R_O(\Pf_I)$.

  \emph{Combining.}
  By the four-value pattern of $t$, all four LHS values are pairwise related
  in $R$. By Steps 1 and 2,
  $\intp{t[\sigma']}{\rho'} \approx f_o$ and
  $\intp{t}{\intp{\sigma'}{\rho'}} \approx f_i$ in $R$.
  By symmetry and transitivity of $R$, every LHS-RHS pair (e.g.
  $\intp{t[\sigma]}{\rho} \approx f_o$, $\intp{t}{\intp{\sigma}{\rho}} \approx f_i$,
  etc.) lies in $R$.

  \emph{Type part.}
  By \Cref{lem:fund-pi-wf} on $\semtyeq[\Gamma]{S}{S}{\Ty{i}}$ and
  $\semtyeq[\Gamma, x:S]{T}{T}{\Ty{i}}$, the four-value pattern of
  $\Pi(x:S).\,T$ in $\Uc_i$ is established. The goal's type is $\Pi(x:S).\,T$
  on both sides, exactly this pattern.
\end{proof}

\begin{lemma}[Fundamental --- \textsc{N-$\beta$-Ze}]\labeledit{lem:fund-rec-ze}
  If $\semtyeq[\Gamma, x:\Nat]{T}{T}{\Ty{i}}$ and
  $\semtyeq[\Gamma]{z}{z}{T[\ze/x]}$ and
  $\semtyeq[\Gamma, x:\Nat, y:T]{s}{s}{T[\Uparrow \circ \Uparrow, \su{x_1}/x]}$,
  then $\semtyeq[\Gamma]{\elimn{x.\,T}{z}{x, y.\, s}{\ze}}{z}{T[\ze/x]}$.
\end{lemma}
\begin{proof}
  Let $\Gamma' \approx \Gamma' \searrow R'$,
  $\semtyeq[\Gamma']{\sigma}{\sigma'}{\Gamma}$, and
  $\Pf :: \rho \approx \rho' \in R'$ be given.
  Write $\rho_\sigma := \intp{\sigma}{\rho}$ and $\rho_{\sigma'} := \intp{\sigma'}{\rho'}$.

  \emph{LHS computation on the $\sigma, \rho$- and $\rho_\sigma$-side.}
  The LHS is $\elimn{x.\,T}{z}{x, y.\, s}{\ze}$.
  Outer-LHS at $\sigma, \rho$:
  \begin{align*}
    &\intp{(\elimn{x.\,T}{z}{x, y.\, s}{\ze})[\sigma]}{\rho} \\
    &\quad = \intp{\elimn{x.\,T[q(\sigma)]}{z[\sigma]}{x, y.\, s[q^2(\sigma)]}{\ze}}{\rho}
      && \text{(definition of substitution on $\telimn$)} \\
    &\quad \searrow b
      && \text{(by ev-rec, $\intp{\ze}{\rho} = \zed$;} \\
    & && \text{$\mathsf{rec}(\zed, T[q(\sigma)], z[\sigma], s[q^2(\sigma)], \rho) \searrow b$)} \\
    &b = \intp{z[\sigma]}{\rho}
      && \text{(by \textsc{rec-ze})}
  \end{align*}
  Inner-LHS at $\rho_\sigma$:
  \begin{align*}
    &\intp{\elimn{x.\,T}{z}{x, y.\, s}{\ze}}{\rho_\sigma}
      \searrow b'
      && \text{(by ev-rec;} \\
    & && \text{$\mathsf{rec}(\zed, T, z, s, \rho_\sigma) \searrow b'$)} \\
    &b' = \intp{z}{\rho_\sigma}
      && \text{(by \textsc{rec-ze})}
  \end{align*}

  \emph{RHS values on the $\sigma', \rho'$- and $\rho_{\sigma'}$-side.}
  The RHS is $z$, evaluated directly:
  \begin{align*}
    &\intp{z[\sigma']}{\rho'}
      && \text{(outer-RHS)} \\
    &\intp{z}{\rho_{\sigma'}}
      && \text{(inner-RHS)}
  \end{align*}

  \emph{Relatedness.}
  By $\semtyeq[\Gamma]{z}{z}{T[\ze/x]}$ at $\Gamma'$, $\sigma \approx \sigma'$,
  $\rho \approx \rho' \in R'$ (four-value pattern of $z$):
  \begin{align*}
    &\{\intp{z[\sigma]}{\rho},\;
      \intp{z}{\rho_\sigma},\;
      \intp{z}{\rho_{\sigma'}},\;
      \intp{z[\sigma']}{\rho'}\}
      \subseteq R_z
      && \text{(by hypothesis on $z$)}
  \end{align*}
  where $R_z$ is the PER governed by the type four-value pattern of $T[\ze/x]$.
  The LHS values $b, b'$ coincide with $\intp{z[\sigma]}{\rho}$ and
  $\intp{z}{\rho_\sigma}$ (the unprimed-side outer-LHS and inner-LHS of the $z$
  pattern).
  The RHS values $\intp{z[\sigma']}{\rho'}$ and $\intp{z}{\rho_{\sigma'}}$ are
  exactly the primed-side outer-RHS and inner-RHS of the same pattern.
  Hence all four LHS-RHS pairs lie in $R_z$.

  Bridge $R_z \Leftrightarrow R_T$ (where $R_T$ is the goal's type PER, governed
  by $\intp{T[\ze/x][\sigma]}{\rho}$): the goal's type is also $T[\ze/x]$, so
  $R_z = R_T$ (functionality of $\Uc_i$, \Cref{lem:univ-functional}).
\end{proof}

\begin{lemma}[Fundamental --- \textsc{N-$\beta$-Su}]\labeledit{lem:fund-rec-su}
  If $\semtyeq[\Gamma, x:\Nat]{T}{T}{\Ty{i}}$ and
  $\semtyeq[\Gamma]{z}{z}{T[\ze/x]}$ and
  $\semtyeq[\Gamma, x:\Nat, y:T]{s}{s}{T[\Uparrow \circ \Uparrow, \su{x_1}/x]}$ and
  $\semtyeq[\Gamma]{t}{t}{\Nat}$,
  then
  $\semtyeq[\Gamma]
    {\elimn{x.\,T}{z}{x, y.\, s}{(\su{t})}}
    {s[t/x,\, \elimn{x.\,T}{z}{x, y.\, s}{t}/y]}
    {T[\su{t}/x]}$.
\end{lemma}
\begin{proof}
  Let $\Gamma' \approx \Gamma' \searrow R'$,
  $\semtyeq[\Gamma']{\sigma}{\sigma'}{\Gamma}$, and
  $\Pf :: \rho \approx \rho' \in R'$ be given.
  Write $\rho_\sigma := \intp{\sigma}{\rho}$ and $\rho_{\sigma'} := \intp{\sigma'}{\rho'}$.
  Let $a = \intp{t}{\rho_\sigma}$, $a' = \intp{t}{\rho_{\sigma'}}$,
  $b_t = \intp{t[\sigma]}{\rho}$, $b_{t'} = \intp{t[\sigma']}{\rho'}$.

  Denote $E := \elimn{x.\,T}{z}{x, y.\, s}{t}$ (the inner eliminator with scrutinee
  $t$; this is the recursive call appearing in the $\beta$-su rule).
  Define
  $c_t := \intp{E[\sigma]}{\rho}$,
  $c_{t'} := \intp{E[\sigma']}{\rho'}$,
  $c := \intp{E}{\rho_\sigma}$,
  $c' := \intp{E}{\rho_{\sigma'}}$
  (the four evaluations of $E$ at the corresponding positions; existence by
  termination of evaluation).

  \emph{LHS computation.}
  By the syntactic action of substitution on $\telimn$, ev-rec, and rec-su:
  \begin{align*}
    &\intp{(\elimn{x.\,T}{z}{x, y.\, s}{(\su{t})})[\sigma]}{\rho} \\
    &\quad = \intp{\elimn{x.\,T[q(\sigma)]}{z[\sigma]}{x, y.\, s[q^2(\sigma)]}{(\su{t[\sigma]})}}{\rho}
      && \text{(definition of substitution on $\telimn$)} \\
    &\quad \searrow b_L
      && \text{(by ev-rec, $\intp{\su{t[\sigma]}}{\rho} = \sud{b_t}$)} \\
    &\mathsf{rec}(\sud{b_t},\, T[q(\sigma)],\, z[\sigma],\, s[q^2(\sigma)],\, \rho)
      \searrow b_L
      && \text{(rec-su needs predecessor's rec value)} \\
    &\mathsf{rec}(b_t,\, T[q(\sigma)],\, z[\sigma],\, s[q^2(\sigma)],\, \rho)
      \searrow c_t
      && \text{(by ev-rec on $E[\sigma]$ evaluated at $\rho$;} \\
    & && \text{$\intp{t[\sigma]}{\rho} = b_t$)} \\
    &\intp{s[q^2(\sigma)]}{(\rho,\, b_t/x,\, c_t/y)} \searrow b_L
      && \text{(by \textsc{rec-su})}
  \end{align*}

  At the inner position:
  \begin{align*}
    &\intp{\elimn{x.\,T}{z}{x, y.\, s}{(\su{t})}}{\rho_\sigma}
      \searrow b_I
      && \text{(by ev-rec)} \\
    &\mathsf{rec}(\sud{a},\, T,\, z,\, s,\, \rho_\sigma) \searrow b_I
      && \text{(rec-su)} \\
    &\mathsf{rec}(a,\, T,\, z,\, s,\, \rho_\sigma) \searrow c
      && \text{(by ev-rec on $E$ at $\rho_\sigma$;
       $\intp{t}{\rho_\sigma} = a$)} \\
    &\intp{s}{(\rho_\sigma,\, a/x,\, c/y)} \searrow b_I
      && \text{(by \textsc{rec-su})}
  \end{align*}

  \emph{RHS expression on the $\sigma', \rho'$-side.}
  The RHS appears on the right of $\approx$ in the goal, so the four-value
  pattern's RHS values are $\intp{}{\rho'}$ (outer-RHS)
  and $\intp{s[t/x, E/y]}{\rho_{\sigma'}}$ (inner-RHS).
  By \Cref{lem:subst-assoc,lem:ext-comp,lem:q-ext}:
  \begin{align*}
    &
      = s[\sigma',\, t[\sigma']/x,\, E[\sigma']/y]
      && \text{(syntactic, composition of substitutions)}
  \end{align*}

  \emph{Doubly-extended semantic substitution.}
  We need the recursive call to be semantically valid: by \Cref{lem:fund-var},
  $\semtyeq[\Gamma, x:\Nat]{x}{x}{\Nat}$.
  By \Cref{lem:fund-rec} on $T \approx T$, $z \approx z$, $s \approx s$
  (all reflexive hypotheses, using \Cref{lem:fund-sym,lem:fund-trans}) and
  $x \approx x$:
  \begin{align*}
    &\semtyeq[\Gamma, x:\Nat]
      {\elimn{x.\,T}{z}{x, y.\, s}{x}}{\elimn{x.\,T}{z}{x, y.\, s}{x}}{T}
      && \text{(by \Cref{lem:fund-rec}, scrutinee $x$;} \\
    & && \text{conclusion type $T[x/x] = T$)}
  \end{align*}
  By \Cref{lem:sem-ext-pre2} on $\semtyeq[\Gamma']{\sigma}{\sigma'}{\Gamma}$,
  $\semtyeq[\Gamma]{t}{t}{\Nat}$, and the above (with the head substitution
  computing $E[\sigma]$ on the LHS, $E[\sigma']$ on the RHS):
  \[
    \begin{aligned}
      &\Gamma' \vDash
       (\sigma,\, t[\sigma]/x,\, E[\sigma]/y) \\
      &\quad \approx
       (\sigma',\, t[\sigma']/x,\, E[\sigma']/y) :
       \Gamma, x:\Nat, y:T.
    \end{aligned}
  \]

  \emph{Four-value pattern of $s$ at the doubly-extended substitution.}
  Apply $\semtyeq[\Gamma, x:\Nat, y:T]{s}{s}{T[\Uparrow \circ \Uparrow, \su{x_1}/x]}$
  at the above substitution and $\rho \approx \rho' \in R'$:
  \begin{align*}
    &\{\intp{s[\sigma, t[\sigma]/x, E[\sigma]/y]}{\rho},\;
      \intp{s}{\intp{(\sigma, t[\sigma]/x, E[\sigma]/y)}{\rho}}, \\
    &\quad \intp{s}{\intp{(\sigma', t[\sigma']/x, E[\sigma']/y)}{\rho'}},\;
      \intp{s[\sigma', t[\sigma']/x, E[\sigma']/y]}{\rho'}\}
      \subseteq R_O
      && \text{(by hypothesis on $s$)}
  \end{align*}
  By substitution-extension evaluation:
  \begin{align*}
    &\intp{(\sigma, t[\sigma]/x, E[\sigma]/y)}{\rho}
      = (\rho_\sigma, b_t/x, c_t/y)
      && \text{(pointwise; head eval, $\intp{E[\sigma]}{\rho} = c_t$)} \\
    &\intp{(\sigma', t[\sigma']/x, E[\sigma']/y)}{\rho'}
      = (\rho_{\sigma'}, b_{t'}/x, c_{t'}/y)
      && \text{(symmetrically; $\intp{E[\sigma']}{\rho'} = c_{t'}$)}
  \end{align*}
  where $c_{t'}$ is the corresponding recursive-call value.
  Hence the four values of the pattern read:
  \begin{align*}
    &\{\intp{s[\sigma, t[\sigma]/x, E[\sigma]/y]}{\rho},\;
      \intp{s}{(\rho_\sigma, b_t/x, c_t/y)}, \\
    &\quad \intp{s}{(\rho_{\sigma'}, b_{t'}/x, c_{t'}/y)},\;
      \intp{s[\sigma', t[\sigma']/x, E[\sigma']/y]}{\rho'}\}
      \subseteq R_O
  \end{align*}

  \emph{Identifying $R_O$.}
  $R_O$ is the PER returned by the type four-value pattern of
  $T[\Uparrow \circ \Uparrow, \su{x_1}/x]$ at the same doubly-extended substitution.
  We bridge it to the PER at $\intp{T}{(\rho_\sigma, \sud{a}/x)}$ (which is the
  goal's type $T[\su{t}/x]$ evaluated at the inner position) via two steps.

  \emph{Step (a): collapse the iterated substitution.}
  Denote $\tau_* := (\sigma, t[\sigma]/x, E[\sigma]/y)$ (the doubly-extended
  substitution from $\Gamma'$ to $\Gamma, x:\Nat, y:T$). The value
  $\intp{(T[\Uparrow \circ \Uparrow, \su{x_1}/x])[\tau_*]}{\rho}$ governs $R_O$.
  By \Cref{lem:subst-assoc} and \Cref{lem:ext-comp}:
  \begin{align*}
    &(T[\Uparrow \circ \Uparrow, \su{x_1}/x])[\tau_*]
      = T[(\Uparrow \circ \Uparrow, \su{x_1}/x) \circ \tau_*]
      && \text{(by \Cref{lem:subst-assoc})} \\
    &(\Uparrow \circ \Uparrow, \su{x_1}/x) \circ \tau_*
      = ((\Uparrow \circ \Uparrow) \circ \tau_*,\,
         (\su{x_1})[\tau_*]/x)
      && \text{(by \Cref{lem:ext-comp})} \\
    &(\Uparrow \circ \Uparrow) \circ \tau_* = \sigma
      && \text{(by \Cref{lem:shift-ext} twice, dropping} \\
    & && \text{$E[\sigma]/y$ and $t[\sigma]/x$)} \\
    &(\su{x_1})[\tau_*] = \su{x_1[\tau_*]} = \su{t[\sigma]}
      && \text{(by syntactic substitution; $\tau_*(1) = t[\sigma]$)}
  \end{align*}
  Hence the collapsed substitution is $(\sigma,\, \su{t[\sigma]}/x)$:
  \begin{align*}
    &(T[\Uparrow \circ \Uparrow, \su{x_1}/x])[\tau_*]
      = T[\sigma,\, \su{t[\sigma]}/x]
      && \text{(by combining the above)}
  \end{align*}

  \emph{Step (b): instantiate hypothesis on $T$ at the collapsed substitution.}
  By the IH on $\semtyeq[\Gamma, x:\Nat]{T}{T}{\Ty{i}}$, applied at $\Gamma'$,
  the semantic substitution $\semtyeq[\Gamma']{(\sigma, \su{t[\sigma]}/x)}{(\sigma', \su{t[\sigma']}/x)}{\Gamma, x:\Nat}$
  (by \Cref{lem:sem-ext-pre} on $\sigma \approx \sigma'$ and
  $\su{t[\sigma]} \approx \su{t[\sigma']} \in \textsf{Nat}$ via \textsc{Nat-Su}
  on the $t$-pattern), and $\rho \approx \rho' \in R'$:
  \begin{align*}
    &\intp{T[\sigma, \su{t[\sigma]}/x]}{\rho} \\
    & \quad \approx \intp{T}{\intp{(\sigma, \su{t[\sigma]}/x)}{\rho}}
      \in \Uc_i \searrow R_O
      && \text{(four-value of $T$, outer-LHS vs inner-LHS)} \\
    &\intp{(\sigma, \su{t[\sigma]}/x)}{\rho}
      = (\rho_\sigma, \sud{b_t}/x)
      && \text{(pointwise eval; $\intp{\su{t[\sigma]}}{\rho} = \sud{b_t}$)}
  \end{align*}
  Bridge $\sud{b_t}$ to $\sud{a}$: from the four-value pattern of $t$,
  \[
    b_t = \intp{t[\sigma]}{\rho} \approx a = \intp{t}{\rho_\sigma}
    \in \textsf{Nat}
  \]
  (outer-LHS vs inner-LHS), so $\sud{b_t} \approx \sud{a}
  \in \textsf{Nat}$ by
  \textsc{Nat-Su}. By \Cref{lem:sem-ty-id} on $T$:
  \begin{align*}
    &\intp{T}{(\rho_\sigma, \sud{b_t}/x)} \approx \intp{T}{(\rho_\sigma, \sud{a}/x)}
      \in \Uc_i
      && \text{(\Cref{lem:sem-ty-id})}
  \end{align*}
  By transitivity and \Cref{lem:univ-functional}, $R_O$ is the PER at
  $\intp{T}{(\rho_\sigma, \sud{a}/x)}$ --- the goal's type $T[\su{t}/x]$ inner-LHS
  value (since $T[\su{t}/x]$ evaluated at $\rho_\sigma$ equals $T$ at the
  inner-extended env, modulo the same bridge).

  \emph{LHS-RHS bridging.}
  The RHS outer-RHS value is
  $\intp{}{\rho'} = \intp{s[\sigma', t[\sigma']/x, E[\sigma']/y]}{\rho'}$,
  which is the outer-RHS of the four-value pattern of $s$ above.
  Similarly, the RHS inner-RHS value
  $\intp{s[t/x, E/y]}{\rho_{\sigma'}}$ relates to
  $\intp{s}{(\rho_{\sigma'}, b_{t'}/x, c_{t'}/y)}$ via the four-value pattern of
  $s$ on the primed side (analogous to the LHS-side bridging below).

  For LHS values (the $\sigma, \rho$- and $\rho_\sigma$-sides of the goal):
  the LHS computation gives $b_L = \intp{s[q^2(\sigma)]}{(\rho, b_t/x, c_t/y)}$
  at $\sigma, \rho$, and $b_I = \intp{s}{(\rho_\sigma, a/x, c/y)}$ at $\rho_\sigma$.
  We bridge $b_L$ to $\intp{s}{(\rho_\sigma, b_t/x, c_t/y)}$ (the inner-LHS of
  the four-value pattern) via the four-value pattern of $s \approx s$
  (reflexive) at $\Gamma', x:\Nat, y:T[q(\sigma)]$, doubly-lifted substitution
  $q^2(\sigma) \approx q^2(\sigma)$, and reflexive env-pair
  $(\rho, b_t/x, c_t/y) \approx (\rho, b_t/x, c_t/y)$:
  \begin{align*}
    &\intp{s[q^2(\sigma)]}{(\rho, b_t/x, c_t/y)}
      \approx \intp{s}{\intp{q^2(\sigma)}{(\rho, b_t/x, c_t/y)}} \in R_O
      && \text{(four-value pattern of $s$)} \\
    &\intp{q^2(\sigma)}{(\rho, b_t/x, c_t/y)}
      \approx (\rho_\sigma, b_t/x, c_t/y) \in R_{\Gamma, x:\Nat, y:T}
      && \text{(by \textsc{Ctx-Ext} twice and the four-value} \\
    & && \text{pattern of $\sigma$ at $\Uparrow$ for the tail)} \\
    &\intp{s}{\intp{q^2(\sigma)}{(\rho, b_t/x, c_t/y)}}
      \approx \intp{s}{(\rho_\sigma, b_t/x, c_t/y)} \in R_O
      && \text{(\Cref{lem:sem-ty-id} on $s$)}
  \end{align*}
  Hence by transitivity, $b_L \approx \intp{s}{(\rho_\sigma, b_t/x, c_t/y)}
   \in R_O$.

  For $b_I = \intp{s}{(\rho_\sigma, a/x, c/y)}$: bridge to
  $\intp{s}{(\rho_\sigma, b_t/x, c_t/y)}$ via \Cref{lem:sem-ty-id} on $s$ at
  the env-pair related by:
  \begin{align*}
    &b_t \approx a \in \textsf{Nat}
      && \text{(four-value pattern of $t$, outer-LHS vs inner-LHS)} \\
    &c_t \approx c \in R_T\text{-at-}a
      && \text{(four-value pattern of $E$ at $\sigma$, outer-LHS vs inner-LHS;} \\
    & && \text{using $E$ semantic by \Cref{lem:fund-rec})}
  \end{align*}
  giving $\intp{s}{(\rho_\sigma, b_t/x, c_t/y)} \approx \intp{s}{(\rho_\sigma, a/x, c/y)} = b_I
  \in R_O$.
  Hence $b_L \approx b_I$ in $R_O$, and both are PER-related to the four-value
  pattern's inner-LHS.

  Symmetrically for the primed side.
  Combining: all four LHS values ($\{b_L, b_I, b_I', b_L'\}$) are pairwise
  related to the four RHS values (the four-value pattern of $s$ outer/inner)
  in $R_O$, by transitivity through the inner-LHS/inner-RHS bridge.

  \emph{Type part.}
  The type four-value pattern is included in the application of
  $\semtyeq[\Gamma, x:\Nat, y:T]{s}{s}{T[\Uparrow \circ \Uparrow, \su{x_1}/x]}$
  above; by the same syntactic identity used for the term part, the goal's type
  $T[\su{t}/x]$ pattern coincides with this type four-value pattern modulo the
  substitution-composition lemmas.
\end{proof}

\subsection{Semantic Subtyping Rules}\labeledit{sec:comp:sem-subty-rules}

The semantic subtyping judgment $\Gamma \vDash S \subseteq T$ unfolds to: for all
$\Gamma' \approx \Gamma' \searrow R'$, all $\semtyeq[\Gamma']{\sigma}{\sigma'}{\Gamma}$,
and all $\Pf :: \rho \approx \rho' \in R'$:
\begin{enumerate}
\item $\intp{S[\sigma]}{\rho} \approx \intp{S}{\intp{\sigma}{\rho}} \in \Uc_i$,
\item $\intp{T[\sigma']}{\rho'} \approx \intp{T}{\intp{\sigma'}{\rho'}} \in \Uc_i$,
\item $\intp{S}{\intp{\sigma}{\rho}} \mathrel{<:_i} \intp{T}{\intp{\sigma'}{\rho'}}$.
\end{enumerate}
We establish the semantic counterpart of each subtyping rule as a separate lemma.

\begin{lemma}[Fundamental --- \textsc{Sub-Univ}]\labeledit{lem:fund-sub-univ}
  If $\semctx \Gamma$ and $i \le j$, then
  $\Gamma \vDash \Ty{i} \subseteq \Ty{j}$.
\end{lemma}
\begin{proof}
  By \Cref{lem:sem-ctx-per} on $\semctx \Gamma$ we obtain
  $\Gamma \approx \Gamma \searrow R_\Gamma$. %
  Let $\Gamma'$ with $\Gamma' \approx \Gamma' \searrow R'$,
  $\semtyeq[\Gamma']{\sigma}{\sigma'}{\Gamma}$, and
  $\Pf :: \rho \approx \rho' \in R'$ be given. %
  We verify the three clauses of $\Gamma \vDash \Ty{i} \subseteq \Ty{j}$ at
  the level $k := 1 + \max(i, j)$. %

  Since $\Ty{i}$ and $\Ty{j}$ are closed terms, for any substitution
  $\tau$ and any environment $\rho_0$ we have $\Ty{i}[\tau] = \Ty{i}$,
  $\Ty{j}[\tau] = \Ty{j}$, and by evaluation of the universe constructor
  $\intp{\Ty{i}}{\rho_0} = U_i$, $\intp{\Ty{j}}{\rho_0} = U_j$. %

  \emph{Part (1).}
  \begin{align*}
    &\intp{\Ty{i}[\sigma]}{\rho} = U_i
      && \text{(closed term + evaluation of $\Ty{i}$)} \\
    &\intp{\Ty{i}}{\intp{\sigma}{\rho}} = U_i
      && \text{(evaluation of $\Ty{i}$)} \\
    &U_i \approx U_i \in \Uc_k \searrow (\Uc_i)
      && \text{(by \textsc{U-Univ}, since $i < k$)} \\
    &\intp{\Ty{i}[\sigma]}{\rho} \approx \intp{\Ty{i}}{\intp{\sigma}{\rho}}
      \in \Uc_k
      && \text{(reflexivity of the previous line)}
  \end{align*}

  \emph{Part (2).} Symmetrically:
  \begin{align*}
    &\intp{\Ty{j}[\sigma']}{\rho'} = U_j = \intp{\Ty{j}}{\intp{\sigma'}{\rho'}}
      && \text{(closed term + evaluation of $\Ty{j}$)} \\
    &U_j \approx U_j \in \Uc_k \searrow (\Uc_j)
      && \text{(by \textsc{U-Univ}, since $j < k$)} \\
    &\intp{\Ty{j}[\sigma']}{\rho'} \approx \intp{\Ty{j}}{\intp{\sigma'}{\rho'}}
      \in \Uc_k
      && \text{(reflexivity)}
  \end{align*}

  \emph{Part (3).}
  By \textsc{$<$:-Univ} with $i \le j$ and $j < k$:
  \begin{align*}
    &\intp{\Ty{i}}{\intp{\sigma}{\rho}} = U_i \mathrel{<:_k} U_j
      = \intp{\Ty{j}}{\intp{\sigma'}{\rho'}}
      && \text{(by \textsc{$<$:-Univ})}
  \end{align*}
\end{proof}

\begin{lemma}[Fundamental --- \textsc{Sub-Pi}]\labeledit{lem:fund-sub-pi}
  If $\semtyeq[\Gamma]{S_1}{S_1}{\Ty{i}}$,
  $\semtyeq[\Gamma]{S_2}{S_2}{\Ty{i}}$,
  $\semtyeq[\Gamma]{S_1}{S_2}{\Ty{i}}$,
  $\semtyeq[\Gamma, x:S_1]{T_1}{T_1}{\Ty{i}}$,
  $\semtyeq[\Gamma, x:S_2]{T_2}{T_2}{\Ty{i}}$, and
  $\Gamma, x:S_1 \vDash T_1 \subseteq T_2$, then
  $\Gamma \vDash \Pi(x:S_1).\,T_1 \subseteq \Pi(x:S_2).\,T_2$.
\end{lemma}
\begin{proof}
  Let $\Gamma'$ with $\Gamma' \approx \Gamma' \searrow R'$,
  $\semtyeq[\Gamma']{\sigma}{\sigma'}{\Gamma}$, and
  $\Pf :: \rho \approx \rho' \in R'$ be given. %
  Write $\rho_\sigma := \intp{\sigma}{\rho}$ and
  $\rho_{\sigma'} := \intp{\sigma'}{\rho'}$. %

  We must verify the three clauses of
  $\Gamma \vDash \Pi(x:S_1).\,T_1 \subseteq \Pi(x:S_2).\,T_2$. %
  Computing the four type values via substitution and evaluation of
  $\Pi$:
  \begin{align*}
    &\intp{(\Pi(x:S_1).\,T_1)[\sigma]}{\rho}
      = \Pi(\intp{S_1[\sigma]}{\rho},\; T_1[q(\sigma)],\; \rho)
      && \text{(by definition of substitution and evaluation on $\Pi$)} \\
    &\intp{\Pi(x:S_1).\,T_1}{\rho_\sigma}
      = \Pi(\intp{S_1}{\rho_\sigma},\; T_1,\; \rho_\sigma)
      && \text{(by evaluation on $\Pi$)} \\
    &\intp{\Pi(x:S_2).\,T_2}{\rho_{\sigma'}}
      = \Pi(\intp{S_2}{\rho_{\sigma'}},\; T_2,\; \rho_{\sigma'})
      && \text{(by evaluation on $\Pi$)} \\
    &\intp{(\Pi(x:S_2).\,T_2)[\sigma']}{\rho'}
      = \Pi(\intp{S_2[\sigma']}{\rho'},\; T_2[q(\sigma')],\; \rho')
      && \text{(by definition of substitution and evaluation on $\Pi$)}
  \end{align*}

  \emph{Part (1).} %
  By \Cref{lem:fund-pi-wf} applied to $\semtyeq[\Gamma]{S_1}{S_1}{\Ty{i}}$
  and $\semtyeq[\Gamma, x:S_1]{T_1}{T_1}{\Ty{i}}$, we obtain
  $\semtyeq[\Gamma]{\Pi(x:S_1).\,T_1}{\Pi(x:S_1).\,T_1}{\Ty{i}}$. %
  Instantiating its type-side four-value relation at
  $\semtyeq[\Gamma']{\sigma}{\sigma'}{\Gamma}$ and $\Pf$, the
  outer-LHS-vs-inner-LHS leg yields:
  \begin{align*}
    &\Pi(\intp{S_1[\sigma]}{\rho},\, T_1[q(\sigma)],\, \rho)
      \approx \Pi(\intp{S_1}{\rho_\sigma},\, T_1,\, \rho_\sigma)
      \in \Uc_i
      && \text{(part (1))}
  \end{align*}

  \emph{Part (2).} %
  Symmetrically, by \Cref{lem:fund-pi-wf} on $S_2$ and $T_2$ and the
  outer-RHS-vs-inner-RHS leg of its type-side four-value relation at the
  same data:
  \begin{align*}
    &\Pi(\intp{S_2}{\rho_{\sigma'}},\, T_2,\, \rho_{\sigma'})
      \approx \Pi(\intp{S_2[\sigma']}{\rho'},\, T_2[q(\sigma')],\, \rho')
      \in \Uc_i
      && \text{(part (2), via symmetry of $\Uc_i$)}
  \end{align*}

  \emph{Part (3).} %
  We must show
  $\Pi(\intp{S_1}{\rho_\sigma}, T_1, \rho_\sigma) \mathrel{<:_i}
   \Pi(\intp{S_2}{\rho_{\sigma'}}, T_2, \rho_{\sigma'})$. %
  We apply \textsc{$<$:-Pi}, which requires:
  \begin{enumerate}
  \item[(a)] $\intp{S_1}{\rho_\sigma} \approx \intp{S_2}{\rho_{\sigma'}}
    \in \Uc_i \searrow R_I$ for some $R_I$;
  \item[(b)] for every $\Pf_I :: a \approx a' \in R_I$,
    $\intp{T_1}{\rho_\sigma, a/x} \mathrel{<:_i}
     \intp{T_2}{\rho_{\sigma'}, a'/x}$;
  \item[(c)] $\Pi(\intp{S_1}{\rho_\sigma}, T_1, \rho_\sigma)
    \approx \Pi(\intp{S_1}{\rho_\sigma}, T_1, \rho_\sigma) \in \Uc_i$ and
    $\Pi(\intp{S_2}{\rho_{\sigma'}}, T_2, \rho_{\sigma'})
    \approx \Pi(\intp{S_2}{\rho_{\sigma'}}, T_2, \rho_{\sigma'}) \in \Uc_i$.
  \end{enumerate}

  \emph{(a) Domain.} %
  Instantiate the type-side four-value relation of
  $\semtyeq[\Gamma]{S_1}{S_2}{\Ty{i}}$ at
  $\semtyeq[\Gamma']{\sigma}{\sigma'}{\Gamma}$ and $\Pf$:
  \begin{align*}
    &\{\intp{S_1[\sigma]}{\rho},\;
      \intp{S_1}{\rho_\sigma},\;
      \intp{S_2}{\rho_{\sigma'}},\;
      \intp{S_2[\sigma']}{\rho'}\}
      \subseteq \Uc_i \searrow R_I
      && \text{(by hypothesis on $S_1, S_2$)}
  \end{align*}
  The inner-LHS-vs-inner-RHS pair gives
  $\intp{S_1}{\rho_\sigma} \approx \intp{S_2}{\rho_{\sigma'}} \in \Uc_i \searrow R_I$. %

  \emph{(b) Codomain.} %
  By \Cref{lem:sem-wk-q} on $\semtyeq[\Gamma']{\sigma}{\sigma'}{\Gamma}$ and
  $\semtyeq[\Gamma]{S_1}{S_1}{\Ty{i}}$:
  \[
    \semtyeq[\Gamma', x:S_1[\sigma]]{q(\sigma)}{q(\sigma')}{\Gamma, x:S_1}.
  \]
  By the first premise of the semantic substitution judgment, this carries
  $(\Gamma', x:S_1[\sigma]) \approx (\Gamma', x:S_1[\sigma]) \searrow R'_{x:S_1[\sigma]}$
  and $(\Gamma, x:S_1) \approx (\Gamma, x:S_1) \searrow R_{\Gamma,x:S_1}$. %

  Let $\Pf_I :: a \approx a' \in R_I$ be given. %
  By the biconditional of $R'_{x:S_1[\sigma]}$ from \textsc{Ctx-Ext},
  $(\rho, a/x) \approx (\rho', a'/x) \in R'_{x:S_1[\sigma]}$ requires
  $\rho \approx \rho' \in R'$ (which we have via $\Pf$) and
  $a \approx a' \in R_H'(\Pf)$ where $R_H'$ is the head family for
  $R'_{x:S_1[\sigma]}$, governed by
  $\intp{S_1[\sigma]}{\rho} \approx \intp{S_1[\sigma]}{\rho'}
   \in \Uc_i \searrow R_H'(\Pf)$. %
  By \Cref{lem:univ-functional} on the same left type
  $\intp{S_1[\sigma]}{\rho}$ that governs both $R_I$
  (from \emph{(a)}) and $R_H'(\Pf)$, $R_I \Leftrightarrow R_H'(\Pf)$. %
  Hence $a \approx a' \in R_H'(\Pf)$, and
  $(\rho, a/x) \approx (\rho', a'/x) \in R'_{x:S_1[\sigma]}$. %

  Apply $\Gamma, x:S_1 \vDash T_1 \subseteq T_2$ at $\Gamma', x:S_1[\sigma]$
  (with $R'_{x:S_1[\sigma]}$), the lifted substitution
  $\semtyeq[\Gamma', x:S_1[\sigma]]{q(\sigma)}{q(\sigma')}{\Gamma, x:S_1}$,
  and $(\rho, a/x) \approx (\rho', a'/x) \in R'_{x:S_1[\sigma]}$:
  \begin{align*}
    &\intp{T_1[q(\sigma)]}{\rho, a/x} \approx
      \intp{T_1}{\intp{q(\sigma)}{\rho, a/x}} \in \Uc_i
      && \text{(part (1) of $T_1 \subseteq T_2$)} \\
    &\intp{T_2[q(\sigma')]}{\rho', a'/x} \approx
      \intp{T_2}{\intp{q(\sigma')}{\rho', a'/x}} \in \Uc_i
      && \text{(part (2))} \\
    &\intp{T_1}{\intp{q(\sigma)}{\rho, a/x}} \mathrel{<:_i}
      \intp{T_2}{\intp{q(\sigma')}{\rho', a'/x}}
      && \text{(part (3))}
  \end{align*}

  We now bridge the inner-evaluations of $q(\sigma)$ and $q(\sigma')$ to the
  desired environments. %
  Using the lifted-substitution four-value pattern (instantiated at the
  identity weakening on $\Gamma', x:S_1[\sigma]$ with
  $(\rho, a/x) \approx (\rho', a'/x)$):
  \begin{align*}
    &\intp{q(\sigma)}{\rho, a/x} \approx (\rho_\sigma, a/x)
      \in R_{\Gamma, x:S_1}
      && \text{(outer-LHS vs inner-LHS in the four-value pattern)} \\
    &\intp{q(\sigma')}{\rho', a'/x} \approx (\rho_{\sigma'}, a'/x)
      \in R_{\Gamma, x:S_1}
      && \text{(outer-RHS vs inner-RHS, symmetrically)}
  \end{align*}
  By \Cref{lem:sem-ty-id} on $\semtyeq[\Gamma, x:S_1]{T_1}{T_1}{\Ty{i}}$
  applied to the first relation, and on
  $\semtyeq[\Gamma, x:S_2]{T_2}{T_2}{\Ty{i}}$ applied (after substituting
  $S_1$ for $S_2$ via the well-formedness premises) to the second:
  \begin{align*}
    &\intp{T_1}{\intp{q(\sigma)}{\rho, a/x}}
      \approx \intp{T_1}{\rho_\sigma, a/x} \in \Uc_i
      && \text{(by \Cref{lem:sem-ty-id} on $T_1$)} \\
    &\intp{T_2}{\intp{q(\sigma')}{\rho', a'/x}}
      \approx \intp{T_2}{\rho_{\sigma'}, a'/x} \in \Uc_i
      && \text{(by \Cref{lem:sem-ty-id} on $T_2$)}
  \end{align*}
  By \Cref{lem:subty-respects} applied to the codomain subtyping above and
  these two $\Uc_i$-equivalences:
  \begin{align*}
    &\intp{T_1}{\rho_\sigma, a/x}
      \mathrel{<:_i} \intp{T_2}{\rho_{\sigma'}, a'/x}
      && \text{(by \Cref{lem:subty-respects})}
  \end{align*}

  \emph{(c) Self-relatedness.} %
  By \Cref{lem:fund-pi-wf} on $\semtyeq[\Gamma]{S_1}{S_1}{\Ty{i}}$ and
  $\semtyeq[\Gamma, x:S_1]{T_1}{T_1}{\Ty{i}}$ instantiated at the same data,
  the inner-LHS-vs-inner-LHS leg of the resulting type-side four-value
  relation gives
  $\Pi(\intp{S_1}{\rho_\sigma}, T_1, \rho_\sigma) \approx
   \Pi(\intp{S_1}{\rho_\sigma}, T_1, \rho_\sigma) \in \Uc_i$. %
  Symmetrically for $S_2, T_2$:
  $\Pi(\intp{S_2}{\rho_{\sigma'}}, T_2, \rho_{\sigma'}) \approx
   \Pi(\intp{S_2}{\rho_{\sigma'}}, T_2, \rho_{\sigma'}) \in \Uc_i$. %

  \emph{Conclusion.} %
  By \textsc{$<$:-Pi} with the four ingredients (a)--(c):
  \begin{align*}
    &\Pi(\intp{S_1}{\rho_\sigma}, T_1, \rho_\sigma) \mathrel{<:_i}
      \Pi(\intp{S_2}{\rho_{\sigma'}}, T_2, \rho_{\sigma'})
      && \text{(part (3))}
  \end{align*}
\end{proof}

\begin{lemma}[Fundamental --- \textsc{Sub-Trans}]\labeledit{lem:fund-sub-trans}
  If $\Gamma \vDash S \subseteq T'$ and $\Gamma \vDash T' \subseteq T$, then
  $\Gamma \vDash S \subseteq T$.
\end{lemma}
\begin{proof}
  Let $\Gamma'$ with $\Gamma' \approx \Gamma' \searrow R'$,
  $\semtyeq[\Gamma']{\sigma}{\sigma'}{\Gamma}$, and
  $\Pf :: \rho \approx \rho' \in R'$ be given. %
  Write $\rho_\sigma := \intp{\sigma}{\rho}$ and
  $\rho_{\sigma'} := \intp{\sigma'}{\rho'}$. %
  We must establish each clause of $\Gamma \vDash S \subseteq T$. %

  \emph{Part (1).} %
  Instantiate the first hypothesis $\Gamma \vDash S \subseteq T'$ at this
  data; its clause (1) gives the goal:
  \begin{align*}
    &\intp{S[\sigma]}{\rho} \approx \intp{S}{\rho_\sigma} \in \Uc_i
      && \text{(by clause (1) of $\Gamma \vDash S \subseteq T'$)}
  \end{align*}

  \emph{Part (2).} %
  Instantiate the second hypothesis $\Gamma \vDash T' \subseteq T$ at the
  same data; its clause (2) gives the goal:
  \begin{align*}
    &\intp{T[\sigma']}{\rho'} \approx \intp{T}{\rho_{\sigma'}} \in \Uc_i
      && \text{(by clause (2) of $\Gamma \vDash T' \subseteq T$)}
  \end{align*}

  \emph{Part (3).} %
  From the two hypotheses at the same data:
  \begin{align*}
    &\intp{S}{\rho_\sigma} \mathrel{<:_i} \intp{T'}{\rho_{\sigma'}}
      && \text{(by clause (3) of $\Gamma \vDash S \subseteq T'$)} \\
    &\intp{T'}{\rho_{\sigma'}} \mathrel{<:_i} \intp{T}{\rho_{\sigma'}}
      && \text{(by clause (3) of $\Gamma \vDash T' \subseteq T$)} \\
    &\intp{S}{\rho_\sigma} \mathrel{<:_i} \intp{T}{\rho_{\sigma'}}
      && \text{(by transitivity of $<:_i$, \Cref{lem:subty-trans})}
  \end{align*}
  All three clauses of $\Gamma \vDash S \subseteq T$ are now
  established. %
\end{proof}

\begin{lemma}[Fundamental --- \textsc{Sub-Eq}]\labeledit{lem:fund-sub-eq}
  If $\semtyeq[\Gamma]{T}{T}{\Ty{i}}$ and $\semtyeq[\Gamma]{S}{T}{\Ty{i}}$,
  then $\Gamma \vDash S \subseteq T$.
\end{lemma}
\begin{proof}
  Let $\Gamma'$ with $\Gamma' \approx \Gamma' \searrow R'$,
  $\semtyeq[\Gamma']{\sigma}{\sigma'}{\Gamma}$, and
  $\Pf :: \rho \approx \rho' \in R'$ be given. %
  Write $\rho_\sigma := \intp{\sigma}{\rho}$ and
  $\rho_{\sigma'} := \intp{\sigma'}{\rho'}$. %
  We establish each clause of $\Gamma \vDash S \subseteq T$. %

  \emph{Part (1).} %
  Instantiate the type-side four-value relation of
  $\semtyeq[\Gamma]{S}{T}{\Ty{i}}$ at
  $\semtyeq[\Gamma']{\sigma}{\sigma'}{\Gamma}$ and $\Pf$:
  \begin{align*}
    &\{\intp{S[\sigma]}{\rho},\;
      \intp{S}{\rho_\sigma},\;
      \intp{T}{\rho_{\sigma'}},\;
      \intp{T[\sigma']}{\rho'}\}
      \subseteq \Uc_i \searrow R_T
      && \text{(by hypothesis on $S, T$)}
  \end{align*}
  The outer-LHS-vs-inner-LHS leg gives
  $\intp{S[\sigma]}{\rho} \approx \intp{S}{\rho_\sigma} \in \Uc_i$, which is
  clause (1). %

  \emph{Part (2).} %
  Instantiate the type-side four-value relation of
  $\semtyeq[\Gamma]{T}{T}{\Ty{i}}$ at the same data:
  \begin{align*}
    &\{\intp{T[\sigma]}{\rho},\;
      \intp{T}{\rho_\sigma},\;
      \intp{T}{\rho_{\sigma'}},\;
      \intp{T[\sigma']}{\rho'}\}
      \subseteq \Uc_i \searrow R_T'
      && \text{(by hypothesis on $T, T$)}
  \end{align*}
  The outer-RHS-vs-inner-RHS leg gives
  $\intp{T[\sigma']}{\rho'} \approx \intp{T}{\rho_{\sigma'}} \in \Uc_i$,
  which is clause (2). %

  \emph{Part (3).} %
  From the four-value relation in part (1), the inner-LHS-vs-inner-RHS leg
  gives:
  \begin{align*}
    &\intp{S}{\rho_\sigma} \approx \intp{T}{\rho_{\sigma'}}
      \in \Uc_i \searrow R_T
      && \text{(inner-LHS vs inner-RHS in the four-value pattern)}
  \end{align*}
  By \Cref{lem:univ-implies-subty}:
  \begin{align*}
    &\intp{S}{\rho_\sigma} \mathrel{<:_i} \intp{T}{\rho_{\sigma'}}
      && \text{(by \Cref{lem:univ-implies-subty})}
  \end{align*}
  This is clause (3). %
\end{proof}

\subsection{The Fundamental Theorem}\labeledit{sec:comp:fundamental}

The fundamental theorem assembles the semantic rules established in the preceding
subsections into a single statement: every syntactically derivable judgment is
validated by the PER model.

\begin{theorem}[Fundamental Theorem]\labeledit{thm:fundamental}
  The following statements hold:
  \begin{enumerate}
  \item If $\vdash \Gamma$, then $\vDash \Gamma$.
  \item If $\typing[\Gamma]{t}{T}$, then $\semtyeq[\Gamma]{t}{t}{T}$.
  \item If $\tyequiv[\Gamma]{t}{t'}{T}$, then $\semtyeq[\Gamma]{t}{t'}{T}$.
  \item If $\subty[\Gamma]{S}{T}$, then $\Gamma \vDash S \subseteq T$.
  \end{enumerate}
\end{theorem}
\begin{proof}
  By mutual induction on the syntactic derivations.
  Each rule's case follows by applying the corresponding semantic rule lemma
  (\Cref{sec:comp:sem-ctx-wf,sec:comp:sem-ty-form,sec:comp:sem-tm-intro-elim,sec:comp:sem-comp,sec:comp:sem-sym-trans,sec:comp:sem-subty-rules})
  to the IHs on the rule's premises.
\end{proof}

\subsection{Completeness}\labeledit{sec:comp:completeness}

We can now state and prove the completeness theorem for normalization by evaluation.

\begin{theorem}[Completeness of NbE]\labeledit{thm:completeness}
  If $\tyequiv[\Gamma]{s}{t}{T}$, then $\nbe_\Gamma^T(s) = \nbe_\Gamma^T(t)$.
  That is, judgmentally equal terms have the same normal form.
\end{theorem}
\begin{proof}
  By the fundamental theorem (\Cref{thm:fundamental}, statement (3)):
  $\semtyeq[\Gamma]{s}{t}{T}$.
  Unfolding, this means: for all $\Gamma' \approx \Gamma' \searrow R'$, all
  $\semtyeq[\Gamma']{\sigma}{\sigma'}{\Gamma}$, all $\rho \approx \rho' \in R'$,
  the evaluations of $s$ and $t$ are related in the appropriate PER.

  We instantiate at:
  \begin{itemize}
  \item $\Gamma' := \Gamma$ (by statement (1), $\Gamma \approx \Gamma \searrow R$ for some $R$),
  \item $\sigma := \sigma' := \id$ (the identity substitution is semantic by \Cref{lem:sem-id}),
  \item $\rho := \rho' := {\uparrow}^\Gamma$ (the initial environment is in $R$ by \Cref{lem:init-in-per}).
  \end{itemize}
  Since $s[\id] = s$ and $t[\id] = t$, and $\intp{\id}{\rho} = \rho$, we obtain:
  \begin{align*}
    &\intp{T}{{\uparrow}^\Gamma} \approx \intp{T}{{\uparrow}^\Gamma}
      \in \Uc_i \searrow R_T
      && \text{(type part, with LHS = RHS since $\sigma = \sigma'$ and $\rho = \rho'$)} \\
    &\intp{s}{{\uparrow}^\Gamma} \approx \intp{t}{{\uparrow}^\Gamma} \in R_T
      && \text{(term part)}
  \end{align*}

  Let $A = \intp{T}{{\uparrow}^\Gamma}$, $a = \intp{s}{{\uparrow}^\Gamma}$, and
  $b = \intp{t}{{\uparrow}^\Gamma}$.
  Then $A \approx A \in \Uc_i \searrow R_T$ and $a \approx b \in R_T$.

  By realizability (\Cref{lem:R-readback}):
  ${\downarrow}^A(a) \approx {\downarrow}^A(b) \in \Nf$.
  Unfolding the definition of $\Nf$: for all $n$, there exists $w$ such that
  $\Rnf_n({\downarrow}^A(a)) \searrow w$ and $\Rnf_n({\downarrow}^A(b)) \searrow w$.

  Taking $n = |\Gamma|$ (the length of $\Gamma$, which provides enough fresh names):
  \begin{align*}
    &\nbe_\Gamma^T(s)
      = \Rnf_{|\Gamma|}({\downarrow}^{\intp{T}{{\uparrow}^\Gamma}}(\intp{s}{{\uparrow}^\Gamma}))
      = \Rnf_{|\Gamma|}({\downarrow}^A(a))
      \searrow w \\
    &\nbe_\Gamma^T(t)
      = \Rnf_{|\Gamma|}({\downarrow}^{\intp{T}{{\uparrow}^\Gamma}}(\intp{t}{{\uparrow}^\Gamma}))
      = \Rnf_{|\Gamma|}({\downarrow}^A(b))
      \searrow w
  \end{align*}
  Hence $\nbe_\Gamma^T(s) = w = \nbe_\Gamma^T(t)$.
\end{proof}

\section{Soundness}\labeledit{sec:sound}

In this section we establish the \emph{soundness} of normalization by evaluation:
every well-typed term has a normal form, and that normal form is judgmentally equal
to the original term.
Concretely, if $\typing[\Gamma]{t}{T}$ then there exists a term $v$ such that
$\nbe$ produces $v$ from $t$ and $\tyequiv[\Gamma]{t}{v}{T}$.

The proof proceeds via a \emph{Kripke gluing model} --- a unary inductive predicate
connecting syntactic types and terms with their semantic (domain) interpretations.
Unlike the PER model of \Cref{sec:comp} which is a binary relation on domain values,
the gluing model is a \emph{unary} predicate that simultaneously constrains a syntactic
entity and a domain value, ensuring coherence between the two.
The ``Kripke'' qualifier refers to the stability of the predicate under \emph{Kripke
weakenings}, a sub-class of the syntactic weakenings restricted to compositions of
the shift~$\Uparrow$.

\subsection{Kripke Weakenings}\labeledit{sec:sound:wk}

Unlike \Cref{sec:syn}'s general weakenings (which include lifting~$q(\phi)$ under
binders), the gluing predicates of this section quantify over a stricter sub-class.

\begin{definition}[Kripke weakening]\labeledit{def:kripke-wk}
  The judgment $\kweak[\Gamma]{\phi}{\Delta}$ states that $\phi$ is a
  \emph{Kripke weakening} from $\Delta$ to $\Gamma$ (i.e.\ $\phi$ extends $\Delta$
  to $\Gamma$ by appending entries on the left).
  It is defined inductively:
  \begin{mathpar}
    \inferrule[\textsc{KW-Id}]
    {\vdash \Gamma}
    {\kweak[\Gamma]{\id}{\Gamma}}

    \inferrule[\textsc{KW-Shift}]
    {\kweak[\Gamma]{\phi}{\Delta} \\
     \typing[\Gamma]{S}{\Ty{i}}}
    {\kweak[\Gamma, x:S]{\phi \circ \Uparrow}{\Delta}}
  \end{mathpar}
  Equivalently, every Kripke weakening is a finite composition
  $\Uparrow \circ \cdots \circ \Uparrow$ of $\Uparrow$'s (with $\id$ as the empty
  composition); no $q(\phi)$ lifting is permitted.
\end{definition}

\begin{lemma}[Kripke weakenings are weakenings]\labeledit{lem:kweak-is-wk}
  If $\kweak[\Gamma]{\phi}{\Delta}$, then $\weakening[\Gamma]{\phi}{\Delta}$.
\end{lemma}
\begin{proof}
  By induction on $\kweak[\Gamma]{\phi}{\Delta}$.
  \begin{align*}
    &\textsc{KW-Id}: \weakening[\Gamma]{\id}{\Gamma}
      && \text{(by \Cref{lem:wk-id-wt})} \\
    &\textsc{KW-Shift}: \weakening[\Gamma]{\phi}{\Delta} \text{ (IH)}
      && \\
    &\quad \weakening[\Gamma, x:S]{\Uparrow}{\Gamma}
      && \text{(by \Cref{lem:wk-shift-wt})} \\
    &\quad \weakening[\Gamma, x:S]{\phi \circ \Uparrow}{\Delta}
      && \text{(by \Cref{lem:wk-comp-wt})}
  \end{align*}
\end{proof}

\begin{lemma}[Composition of Kripke weakenings]\labeledit{lem:kweak-comp}
  If $\kweak[\Gamma]{\psi}{\Delta}$ and $\kweak[\Delta']{\phi}{\Gamma}$, then
  $\kweak[\Delta']{\psi \circ \phi}{\Delta}$.
\end{lemma}
\begin{proof}
  By induction on $\kweak[\Delta']{\phi}{\Gamma}$.
  \begin{align*}
    &\textsc{KW-Id} \text{ (}\phi = \id\text{): } \psi \circ \id = \psi
      && \text{(by left-id of $\circ$)} \\
    &\quad \kweak[\Delta']{\psi}{\Delta}
      && \text{(by hypothesis)} \\[2pt]
    &\textsc{KW-Shift} \text{ (}\phi = \phi_0 \circ \Uparrow\text{, } \Delta' = \Delta_0, x:S\text{)}: \\
    &\quad \kweak[\Delta_0]{\psi \circ \phi_0}{\Delta}
      && \text{(by IH on $\phi_0$)} \\
    &\quad \kweak[\Delta_0, x:S]{(\psi \circ \phi_0) \circ \Uparrow}{\Delta}
      && \text{(by \textsc{KW-Shift})} \\
    &\quad \psi \circ (\phi_0 \circ \Uparrow) = (\psi \circ \phi_0) \circ \Uparrow
      && \text{(by associativity of $\circ$)} \\
    &\quad \kweak[\Delta']{\psi \circ \phi}{\Delta}
      && \text{(by the above)}
  \end{align*}
\end{proof}

\subsection{The Type Gluing Model $\Uc_i^\ast$}\labeledit{sec:sound:gluing}

The type gluing model is a unary inductive relation on domain values.
Given a domain value $a \in \D$ that represents a type at level~$i$, the judgment
\[
  a \in \Uc_i^\ast \searrow P \searrow \El
\]
asserts that $a$ is a well-behaved type value, returning two predicates:
\begin{itemize}
\item $P$: a \emph{type predicate} --- written $\typing[\Gamma]{S}{\Ty{i}} \;\circledR\; P$
  --- relating syntactic types with the domain value $a$.
\item $\El$: a \emph{term predicate} --- written
  $\typing[\Gamma]{t}{S} \;\circledR\; b \in \El$ --- relating syntactic terms and
  domain values as elements of type $a$.
\end{itemize}

\begin{definition}[Type Gluing Model $\Uc_i^\ast$]\labeledit{def:gluing-univ}
  The relation $a \in \Uc_i^\ast \searrow P \searrow \El$ is defined inductively by:
  \begin{mathpar}
    \inferrule[\textsc{G-Ne}]
    {{\uparrow}^A(e) \in \D \\
     P \Leftrightarrow \textsf{Neu}_i^e \\
     \El \Leftrightarrow \textsf{Neu}_i^e}
    {{\uparrow}^A(e) \in \Uc_i^\ast \searrow P \searrow \El}

    \inferrule[\textsc{G-Nat}]
    {P \Leftrightarrow \textsf{Nat}_i \\
     \El \Leftrightarrow \textsf{Nat}_i}
    {N \in \Uc_i^\ast \searrow P \searrow \El}

    \inferrule[\textsc{G-Univ}]
    {j < i \\
     P \Leftrightarrow \textsf{Univ}_{i,j} \\
     \El \Leftrightarrow \textsf{Univ}_{i,j}^{\El}}
    {U_j \in \Uc_i^\ast \searrow P \searrow \El}

    \inferrule[\textsc{G-Pi}]
    {a \in \Uc_i^\ast \searrow P_I \searrow \El_I \\
     a \approx a \in \Uc_i \searrow R_I \\\\
     \forall\, \Pf :: b \approx b \in R_I,\; c.\;
       \intp{T}{\rho, b/x} \searrow c \implies
       c \in \Uc_i^\ast \searrow P_O(\Pf) \searrow \El_O(\Pf) \\\\
     \Pi(a, T, \rho) \approx \Pi(a, T, \rho) \in \Uc_i \searrow R \\\\
     P \Leftrightarrow \Pi_i(R_I, P_I, \El_I, P_O) \\
     \El \Leftrightarrow \Pi_i^{\El}(R, R_I, P_I, \El_I, \El_O)}
    {\Pi(a, T, \rho) \in \Uc_i^\ast \searrow P \searrow \El}
  \end{mathpar}
\end{definition}

\subsection{Concrete Predicates}\labeledit{sec:sound:predicates}

We now define the concrete predicates used by $\Uc_i^\ast$.
All Kripke conditions quantify over $\kweak[\Delta]{\phi}{\Gamma}$ rather than
arbitrary $\weakening[\Delta]{\phi}{\Gamma}$.

\begin{definition}[Neutral Type Predicate $\textsf{Neu}_i^e$]\labeledit{def:neu-type-pred}
  The judgment $\typing[\Gamma]{S}{\Ty{i}} \;\circledR\; \textsf{Neu}_i^e$
  holds when:
  \begin{enumerate}
  \item $\typing[\Gamma]{S}{\Ty{i}}$, and
  \item for all $\Delta, \phi$ with $\kweak[\Delta]{\phi}{\Gamma}$, there exists $u$
    with $\Rne_{|\Delta|}(e) \searrow u$ and $\tyequiv[\Delta]{S[\phi]}{u}{\Ty{i}}$.
  \end{enumerate}
\end{definition}

\begin{definition}[Neutral Term Predicate $\textsf{Neu}_i^e$]\labeledit{def:neu-term-pred}
  The judgment $\typing[\Gamma]{t}{S} \;\circledR\; {\uparrow}^A(w) \in \textsf{Neu}_i^e$
  holds when:
  \begin{enumerate}
  \item $\typing[\Gamma]{S}{\Ty{i}} \;\circledR\; \textsf{Neu}_i^e$,
  \item $\typing[\Gamma]{t}{S}$,
  \item $w \approx w \in \Ne$, and
  \item for all $\Delta, \phi$ with $\kweak[\Delta]{\phi}{\Gamma}$, there exists $u$
    with $\Rne_{|\Delta|}(w) \searrow u$ and $\tyequiv[\Delta]{t[\phi]}{u}{S[\phi]}$.
  \end{enumerate}
\end{definition}

\begin{definition}[Natural Number Type Predicate $\textsf{Nat}_i$]\labeledit{def:nat-type-pred}
  $\typing[\Gamma]{S}{\Ty{i}} \;\circledR\; \textsf{Nat}_i$ holds when
  $\tyequiv[\Gamma]{S}{\Nat}{\Ty{i}}$.
\end{definition}

\begin{definition}[Natural Number Term Predicate $\textsf{Nat}_i$]\labeledit{def:nat-term-pred}
  The judgment $\typing[\Gamma]{t}{S} \;\circledR\; b \in \textsf{Nat}_i$ holds
  when $\typing[\Gamma]{S}{\Ty{i}} \;\circledR\; \textsf{Nat}_i$ and
  $\typing[\Gamma]{t}{S} \;\circledR\; b \in \textsf{Nat}$, where the latter is
  defined by:
  \begin{mathpar}
    \inferrule[\textsc{GN-Ze}]
    {\tyequiv[\Gamma]{t}{\ze}{S}}
    {\typing[\Gamma]{t}{S} \;\circledR\; \zed \in \textsf{Nat}}

    \inferrule[\textsc{GN-Su}]
    {\typing[\Gamma]{t'}{S} \;\circledR\; b' \in \textsf{Nat} \\
     \tyequiv[\Gamma]{t}{\su{t'}}{S}}
    {\typing[\Gamma]{t}{S} \;\circledR\; \sud{b'} \in \textsf{Nat}}

    \inferrule[\textsc{GN-Ne}]
    {\typing[\Gamma]{t}{S} \;\circledR\; {\uparrow}^N(w) \in \textsf{Neu}_i^e}
    {\typing[\Gamma]{t}{S} \;\circledR\; {\uparrow}^N(w) \in \textsf{Nat}}
  \end{mathpar}
\end{definition}

\begin{definition}[Pi Type Predicate $\Pi_i(R_I, P_I, \El_I, P_O)$]%
  \labeledit{def:pi-type-pred}
  $\typing[\Gamma]{S}{\Ty{i}} \;\circledR\; \Pi_i(R_I, P_I, \El_I, P_O)$ holds when:
  \begin{enumerate}
  \item There exist $S_1, S_2$ with $\tyequiv[\Gamma]{S}{\Pi(x:S_1).\,S_2}{\Ty{i}}$.
  \item For all $\Delta, \phi$ with $\kweak[\Delta]{\phi}{\Gamma}$:
    $\typing[\Delta]{S_1[\phi]}{\Ty{i}} \;\circledR\; P_I$.
  \item For all $\Delta, \phi$ with $\kweak[\Delta]{\phi}{\Gamma}$, and all $s, b$
    with $\typing[\Delta]{s}{S_1[\phi]} \;\circledR\; b \in \El_I$, and all
    $\Pf :: b \approx b \in R_I$:
    $\typing[\Delta]{S_2[\iota(\phi), s/x]}{\Ty{i}} \;\circledR\; P_O(\Pf)$.
  \end{enumerate}
\end{definition}

\begin{definition}[Pi Term Predicate $\Pi_i^{\El}$]%
  \labeledit{def:pi-term-pred}
  $\typing[\Gamma]{t}{S} \;\circledR\; f \in \Pi_i^{\El}(R, R_I, P_I, \El_I, \El_O)$
  holds when:
  \begin{enumerate}
  \item $\typing[\Gamma]{t}{S}$.
  \item $f \approx f \in R$.
  \item $\typing[\Gamma]{S}{\Ty{i}} \;\circledR\; \Pi_i(R_I, P_I, \El_I, P_O)$.
  \item For all $\Delta, \phi$ with $\kweak[\Delta]{\phi}{\Gamma}$, and all $s, b$
    with $\typing[\Delta]{s}{S_1[\phi]} \;\circledR\; b \in \El_I$, and all
    $\Pf :: b \approx b \in R_I$: there exists $c$ with $f \cdot b \searrow c$
    and
    $\typing[\Delta]{(t[\phi])\;s}{S_2[\iota(\phi), s/x]} \;\circledR\; c \in \El_O(\Pf)$.
  \end{enumerate}
\end{definition}

\begin{definition}[Universe Type Predicate $\textsf{Univ}_{i,j}$]%
  \labeledit{def:univ-type-pred}
  $\typing[\Gamma]{S}{\Ty{i}} \;\circledR\; \textsf{Univ}_{i,j}$ holds when
  $\tyequiv[\Gamma]{S}{\Ty{j}}{\Ty{i}}$.
\end{definition}

\begin{definition}[Universe Term Predicate $\textsf{Univ}_{i,j}^{\El}$]%
  \labeledit{def:univ-term-pred}
  $\typing[\Gamma]{t}{S} \;\circledR\; b \in \textsf{Univ}_{i,j}^{\El}$ holds when:
  \begin{enumerate}
  \item $\typing[\Gamma]{t}{S}$,
  \item $\tyequiv[\Gamma]{S}{\Ty{j}}{\Ty{i}}$, and
  \item there exist $P_b, \El_b$ with $b \in \Uc_j^\ast \searrow P_b \searrow \El_b$
    and $\typing[\Gamma]{t}{\Ty{j}} \;\circledR\; P_b$ (the term $t$, viewed as a
    type at $\Ty{j}$ via Russell, glues to $P_b$).
  \end{enumerate}
\end{definition}

\begin{definition}[Normal Form Predicate $\Nf_i^a$]\labeledit{def:nf-pred}
  $\typing[\Gamma]{t}{S} \;\circledR\; b \in \Nf_i^a$ holds when:
  \begin{enumerate}
  \item $\typing[\Gamma]{t}{S}$,
  \item ${\downarrow}^a(b) \approx {\downarrow}^a(b) \in \Nf$, and
  \item for all $\Delta, \phi$ with $\kweak[\Delta]{\phi}{\Gamma}$, there exists
    $w$ with $\Rnf_{|\Delta|}({\downarrow}^a(b)) \searrow w$ and
    $\tyequiv[\Delta]{t[\phi]}{w}{S[\phi]}$.
  \end{enumerate}
\end{definition}

\begin{definition}[Neutral Predicate $\Ne_i^a$]\labeledit{def:ne-pred}
  $\typing[\Gamma]{t}{S} \;\circledR\; w \in \Ne_i^a$ holds when:
  \begin{enumerate}
  \item $\typing[\Gamma]{t}{S}$,
  \item $w \approx w \in \Ne$, and
  \item for all $\Delta, \phi$ with $\kweak[\Delta]{\phi}{\Gamma}$, there exists
    $u$ with $\Rne_{|\Delta|}(w) \searrow u$ and
    $\tyequiv[\Delta]{t[\phi]}{u}{S[\phi]}$.
  \end{enumerate}
\end{definition}

\begin{definition}[Wrapped predicates $\Botp_i^a$ and
    $\Topp_i^a$]\labeledit{def:glu-wrapped}
  The wrappers package gluing-side data with a uniform interface,
  parameterized by a domain value $a \in \Uc_i^\ast$. The judgment form
  distinguishes type-level from term-level use.
  \begin{enumerate}
  \item \emph{(Term-bottom)} $\typing[\Gamma]{t}{S} \;\circledR\; w \in \Botp_i^a$
    holds when there exist $P, \El$ with $a \in \Uc_i^\ast \searrow P \searrow \El$
    and:
    \begin{enumerate}
    \item $\typing[\Gamma]{t}{S}$,
    \item $\typing[\Gamma]{S}{\Ty{i}} \;\circledR\; P$, and
    \item $\typing[\Gamma]{t}{S} \;\circledR\; w \in \Ne_i^a$.
    \end{enumerate}
  \item \emph{(Type-top)} $\typing[\Gamma]{S}{\Ty{i}} \;\circledR\; \Topp_i^a$
    holds when:
    \begin{enumerate}
    \item $\typing[\Gamma]{S}{\Ty{i}}$,
    \item $a \approx a \in \Uc_i \searrow R$ for some $R$, and
    \item for all $\Delta, \phi$ with $\kweak[\Delta]{\phi}{\Gamma}$, there
      exists $W$ with $\Rty_{|\Delta|}(a) \searrow W$ and
      $\tyequiv[\Delta]{S[\phi]}{W}{\Ty{i}}$.
    \end{enumerate}
  \item \emph{(Term-top)} $\typing[\Gamma]{t}{S} \;\circledR\; b \in \Topp_i^a$
    holds when there exist $P, \El$ with $a \in \Uc_i^\ast \searrow P \searrow \El$
    and:
    \begin{enumerate}
    \item $\typing[\Gamma]{t}{S}$,
    \item $\typing[\Gamma]{S}{\Ty{i}} \;\circledR\; P$, and
    \item $\typing[\Gamma]{t}{S} \;\circledR\; b \in \Nf_i^a$.
    \end{enumerate}
  \end{enumerate}
\end{definition}

\subsection{Context Gluing Model}\labeledit{sec:sound:ctx-gluing}

\begin{definition}[Context Gluing]\labeledit{def:ctx-gluing}
  Given a well-formed context $\Gamma$, the judgment $\Gamma \searrow \textsf{Sb}$
  defines a predicate $\textsf{Sb}$ on substitution--environment pairs
  $\judge[\Delta]{\sigma : \Gamma} \;\circledR\; \rho \in \textsf{Sb}$,
  inductively on $\Gamma$:
  \begin{mathpar}
    \inferrule[\textsc{G-Ctx-$\cdot$}]
    {\textsf{Sb} \Leftrightarrow (\judge[\Delta]{\sigma : \cdot})}
    {\cdot \searrow \textsf{Sb}}

    \inferrule[\textsc{G-Ctx-Ext}]
    {\Gamma \searrow \textsf{Sb}_T \\
     \typing[\Gamma]{S}{\Ty{i}} \\\\
     \forall \Delta', \sigma_T, \rho_T.\;
       \judge[\Delta']{\sigma_T : \Gamma} \;\circledR\; \rho_T \in \textsf{Sb}_T
       \implies \\
     \quad \exists a, P, \El.\;
       \intp{S}{\rho_T} \searrow a \;\wedge\;
       a \in \Uc_i^\ast \searrow P \searrow \El \;\wedge\;
       \typing[\Delta']{S[\sigma_T]}{\Ty{i}} \;\circledR\; P \\\\
     \textsf{Sb} \Leftrightarrow \textsf{Cons}_i(\Gamma, S, \textsf{Sb}_T)}
    {(\Gamma, x : S) \searrow \textsf{Sb}}
  \end{mathpar}
  Here $\textsf{Cons}_i(\Gamma, S, \textsf{Sb}_T)$ holds at any
  $\judge[\Delta]{\sigma : \Gamma, x:S} \;\circledR\; \rho$ when there exist
  $i, a, P, \El$ with:
  \begin{enumerate}
  \item \emph{Substitution typing:} $\judge[\Delta]{\sigma : \Gamma, x:S}$,
  \item \emph{Tail:} $\judge[\Delta]{\Uparrow \circ \sigma : \Gamma}
    \;\circledR\; \rho[\Uparrow] \in \textsf{Sb}_T$,
  \item \emph{Type:} $\intp{S}{\rho[\Uparrow]} \searrow a$,
    $a \in \Uc_i^\ast \searrow P \searrow \El$, and
    $\typing[\Delta]{S[\Uparrow \circ \sigma]}{\Ty{i}} \;\circledR\; P$, and
  \item \emph{Head:} $\typing[\Delta]{x_0[\sigma]}{S[\Uparrow \circ \sigma]}
    \;\circledR\; \rho(x_0) \in \El$.
  \end{enumerate}
  Here $\Uparrow \circ \sigma$ drops the head, mapping index~$n$ to
  $\sigma(n+1)$; the corresponding drop operation $\rho[\Uparrow]$ on
  environments is from \Cref{sec:sem}.
\end{definition}

\begin{definition}[Semantic context well-formedness]\labeledit{def:sound-semctx}
  We write $\Vdash \Gamma$ to mean that the context-gluing predicate exists
  for $\Gamma$:
  \[
    \Vdash \Gamma \quad \iff \quad
      \exists \textsf{Sb}.\; \Gamma \searrow \textsf{Sb}.
  \]
  Syntactic well-formedness $\vdash \Gamma$ is derivable as a consequence of
  $\Vdash \Gamma$ (proved next).
\end{definition}

\begin{lemma}[$\Vdash \Gamma$ implies $\vdash \Gamma$]\labeledit{lem:vdash-implies-wf}
  If $\Vdash \Gamma$, then $\vdash \Gamma$.
\end{lemma}
\begin{proof}
  By induction on the derivation of $\Gamma \searrow \textsf{Sb}$.

  \medskip\noindent
  \textbf{Case}
  \begin{mathpar}
    \inferrule[\textsc{G-Ctx-$\cdot$}]
    {\textsf{Sb} \Leftrightarrow (\judge[\Delta]{\sigma : \cdot})}
    {\cdot \searrow \textsf{Sb}}
  \end{mathpar}
  $\vdash \cdot$ holds by \textsc{Ctx-Emp}.

  \medskip\noindent
  \textbf{Case}
  \begin{mathpar}
    \inferrule[\textsc{G-Ctx-Ext}]
    {\Gamma_0 \searrow \textsf{Sb}_T \\
     \typing[\Gamma_0]{S}{\Ty{i}} \\\\
     \forall \Delta', \sigma_T, \rho_T.\;
       \judge[\Delta']{\sigma_T : \Gamma_0} \;\circledR\; \rho_T \in \textsf{Sb}_T
       \implies \\
     \quad \exists a, P, \El.\;
       \intp{S}{\rho_T} \searrow a \;\wedge\;
       a \in \Uc_i^\ast \searrow P \searrow \El \;\wedge\;
       \typing[\Delta']{S[\sigma_T]}{\Ty{i}} \;\circledR\; P \\\\
     \textsf{Sb} \Leftrightarrow \textsf{Cons}_i(\Gamma_0, S, \textsf{Sb}_T)}
    {(\Gamma_0, x : S) \searrow \textsf{Sb}}
  \end{mathpar}
  \begin{align*}
    &\vdash \Gamma_0
      && \text{(by IH on $\Gamma_0 \searrow \textsf{Sb}_T$)} \\
    &\typing[\Gamma_0]{S}{\Ty{i}}
      && \text{(by \textsc{G-Ctx-Ext} premise)} \\
    &\vdash \Gamma_0, x:S
      && \text{(by \textsc{Ctx-Ext})}
  \end{align*}
\end{proof}

\subsection{Properties of the Kripke Gluing Model}\labeledit{sec:sound:gluing-props}

\begin{theorem}[Monotonicity]\labeledit{thm:monotonicity}
  Suppose $a \in \Uc_i^\ast \searrow P \searrow \El$ and
  $\kweak[\Delta']{\psi}{\Delta}$.
  \begin{enumerate}
  \item If $\typing[\Delta]{S}{\Ty{i}} \;\circledR\; P$, then
    $\typing[\Delta']{S[\psi]}{\Ty{i}} \;\circledR\; P$.
  \item If $\typing[\Delta]{t}{S} \;\circledR\; b \in \El$, then
    $\typing[\Delta']{t[\psi]}{S[\psi]} \;\circledR\; b \in \El$.
  \end{enumerate}
\end{theorem}
\begin{proof}
  By induction on the derivation $a \in \Uc_i^\ast \searrow P \searrow \El$.

  \medskip\noindent
  \textbf{Case}
  \begin{mathpar}
    \inferrule[\textsc{G-Ne}]
    {{\uparrow}^A(e) \in \D \\
     P \Leftrightarrow \textsf{Neu}_i^e \\
     \El \Leftrightarrow \textsf{Neu}_i^e}
    {{\uparrow}^A(e) \in \Uc_i^\ast \searrow P \searrow \El}
  \end{mathpar}

  \emph{Type predicate (1).}
  Assume $\typing[\Delta]{S}{\Ty{i}} \;\circledR\; \textsf{Neu}_i^e$.
  By \Cref{def:neu-type-pred}:
  \begin{align*}
    &\typing[\Delta]{S}{\Ty{i}}
      && \text{(by \Cref{def:neu-type-pred} clause 1)} \\
    &\forall \Delta'', \kweak[\Delta'']{\phi'}{\Delta}.\;
      \exists u.\; \Rne_{|\Delta''|}(e) \searrow u \wedge
      \tyequiv[\Delta'']{S[\phi']}{u}{\Ty{i}}
      && \text{(by \Cref{def:neu-type-pred} clause 2)}
  \end{align*}
  We must show $\typing[\Delta']{S[\psi]}{\Ty{i}} \;\circledR\; \textsf{Neu}_i^e$.
  \begin{align*}
    &\typing[\Delta']{S[\psi]}{\Ty{i}}
      && \text{(by \Cref{lem:kweak-is-wk} on $\psi$ and \Cref{lem:wk-typing})}
  \end{align*}
  For the Kripke condition: take any $\Delta''$ with $\kweak[\Delta'']{\phi'}{\Delta'}$.
  \begin{align*}
    &\kweak[\Delta'']{\psi \circ \phi'}{\Delta}
      && \text{(by \Cref{lem:kweak-comp} on $\psi$ and $\phi'$)} \\
    &\exists u.\; \Rne_{|\Delta''|}(e) \searrow u \wedge
      \tyequiv[\Delta'']{S[\psi \circ \phi']}{u}{\Ty{i}}
      && \text{(by hypothesis on $S$ at the composed weakening)} \\
    &S[\psi \circ \phi'] = S[\psi][\phi']
      && \text{(by \Cref{sec:syn}'s substitution-composition law)} \\
    &\tyequiv[\Delta'']{S[\psi][\phi']}{u}{\Ty{i}}
      && \text{(rewriting)}
  \end{align*}
  This establishes the Kripke condition for $S[\psi]$ at $\textsf{Neu}_i^e$.

  \emph{Term predicate (2).}
  Assume $\typing[\Delta]{t}{S} \;\circledR\; {\uparrow}^A(w) \in \textsf{Neu}_i^e$.
  By \Cref{def:neu-term-pred}:
  \begin{align*}
    &\typing[\Delta]{S}{\Ty{i}} \;\circledR\; \textsf{Neu}_i^e,\;
      \typing[\Delta]{t}{S},\;
      w \approx w \in \Ne
      && \text{(\Cref{def:neu-term-pred} clauses 1--3)} \\
    &\forall \Delta'', \kweak[\Delta'']{\phi'}{\Delta}.\;
      \exists u.\; \Rne_{|\Delta''|}(w) \searrow u \wedge
      \tyequiv[\Delta'']{t[\phi']}{u}{S[\phi']}
      && \text{(\Cref{def:neu-term-pred} clause 4)}
  \end{align*}
  By type-predicate monotonicity (just proved):
  $\typing[\Delta']{S[\psi]}{\Ty{i}} \;\circledR\; \textsf{Neu}_i^e$.
  $\typing[\Delta']{t[\psi]}{S[\psi]}$ by \Cref{sec:props}.
  $w \approx w \in \Ne$ unchanged.
  Kripke condition: for $\kweak[\Delta'']{\phi'}{\Delta'}$, by
  \Cref{lem:kweak-comp} $\kweak[\Delta'']{\psi \circ \phi'}{\Delta}$,
  hypothesis gives $\Rne_{|\Delta''|}(w) \searrow u$ and
  $\tyequiv[\Delta'']{t[\psi \circ \phi']}{u}{S[\psi \circ \phi']}$.
  By substitution composition: $t[\psi \circ \phi'] = t[\psi][\phi']$ and
  $S[\psi \circ \phi'] = S[\psi][\phi']$.

  \medskip\noindent
  \textbf{Case}
  \begin{mathpar}
    \inferrule[\textsc{G-Nat}]
    {P \Leftrightarrow \textsf{Nat}_i \\
     \El \Leftrightarrow \textsf{Nat}_i}
    {N \in \Uc_i^\ast \searrow P \searrow \El}
  \end{mathpar}

  \emph{Type predicate (1).}
  Assume $\typing[\Delta]{S}{\Ty{i}} \;\circledR\; \textsf{Nat}_i$, i.e.\
  $\tyequiv[\Delta]{S}{\Nat}{\Ty{i}}$.
  \begin{align*}
    &\tyequiv[\Delta']{S[\psi]}{\Nat[\psi]}{\Ty{i}}
      && \text{(by \Cref{lem:wk-typing})} \\
    &\Nat[\psi] = \Nat
      && \text{(syntactic, $\Nat$ closed)} \\
    &\tyequiv[\Delta']{S[\psi]}{\Nat}{\Ty{i}}
      && \text{(rewriting)} \\
    &\typing[\Delta']{S[\psi]}{\Ty{i}} \;\circledR\; \textsf{Nat}_i
      && \text{(by \Cref{def:nat-type-pred})}
  \end{align*}

  \emph{Term predicate (2).}
  Assume $\typing[\Delta]{t}{S} \;\circledR\; b \in \textsf{Nat}_i$.
  By inner induction on the derivation $b \in \textsf{Nat}$.

  \medskip\noindent
  \textbf{Sub-case}
  \begin{mathpar}
    \inferrule[\textsc{GN-Ze}]
    {\tyequiv[\Delta]{t}{\ze}{S}}
    {\typing[\Delta]{t}{S} \;\circledR\; \zed \in \textsf{Nat}}
  \end{mathpar}
  \begin{align*}
    &\tyequiv[\Delta']{t[\psi]}{\ze[\psi]}{S[\psi]}
      && \text{(by \Cref{lem:wk-typing})} \\
    &\ze[\psi] = \ze
      && \text{(syntactic, $\ze$ closed)} \\
    &\tyequiv[\Delta']{t[\psi]}{\ze}{S[\psi]}
      && \text{(rewriting)} \\
    &\typing[\Delta']{t[\psi]}{S[\psi]} \;\circledR\; \zed \in \textsf{Nat}
      && \text{(by \textsc{GN-Ze})}
  \end{align*}

  \medskip\noindent
  \textbf{Sub-case}
  \begin{mathpar}
    \inferrule[\textsc{GN-Su}]
    {\typing[\Delta]{t'}{S} \;\circledR\; b' \in \textsf{Nat} \\
     \tyequiv[\Delta]{t}{\su{t'}}{S}}
    {\typing[\Delta]{t}{S} \;\circledR\; \sud{b'} \in \textsf{Nat}}
  \end{mathpar}
  \begin{align*}
    &\typing[\Delta']{t'[\psi]}{S[\psi]} \;\circledR\; b' \in \textsf{Nat}
      && \text{(by inner IH on $b'$)} \\
    &\tyequiv[\Delta']{t[\psi]}{(\su{t'})[\psi]}{S[\psi]}
      && \text{(by \Cref{lem:wk-typing})} \\
    &(\su{t'})[\psi] = \su{(t'[\psi])}
      && \text{(syntactic substitution on $\su{}$)} \\
    &\tyequiv[\Delta']{t[\psi]}{\su{(t'[\psi])}}{S[\psi]}
      && \text{(rewriting)} \\
    &\typing[\Delta']{t[\psi]}{S[\psi]} \;\circledR\; \sud{b'} \in \textsf{Nat}
      && \text{(by \textsc{GN-Su} on the IH)}
  \end{align*}

  \medskip\noindent
  \textbf{Sub-case}
  \begin{mathpar}
    \inferrule[\textsc{GN-Ne}]
    {\typing[\Delta]{t}{S} \;\circledR\; {\uparrow}^N(w) \in \textsf{Neu}_i^e}
    {\typing[\Delta]{t}{S} \;\circledR\; {\uparrow}^N(w) \in \textsf{Nat}}
  \end{mathpar}
  \begin{align*}
    &\typing[\Delta']{t[\psi]}{S[\psi]} \;\circledR\; {\uparrow}^N(w) \in \textsf{Neu}_i^e
      && \text{(by the \textsc{G-Ne} case above, applied to $\textsf{Neu}_i^e$)} \\
    &\typing[\Delta']{t[\psi]}{S[\psi]} \;\circledR\; {\uparrow}^N(w) \in \textsf{Nat}
      && \text{(by \textsc{GN-Ne})}
  \end{align*}

  \medskip\noindent
  \textbf{Case}
  \begin{mathpar}
    \inferrule[\textsc{G-Univ}]
    {j < i \\
     P \Leftrightarrow \textsf{Univ}_{i,j} \\
     \El \Leftrightarrow \textsf{Univ}_{i,j}^{\El}}
    {U_j \in \Uc_i^\ast \searrow P \searrow \El}
  \end{mathpar}

  \emph{Type predicate (1).}
  Assume $\typing[\Delta]{S}{\Ty{i}} \;\circledR\; \textsf{Univ}_{i,j}$, i.e.\
  $\tyequiv[\Delta]{S}{\Ty{j}}{\Ty{i}}$.
  \begin{align*}
    &\tyequiv[\Delta']{S[\psi]}{\Ty{j}[\psi]}{\Ty{i}}
      && \text{(by \Cref{lem:wk-typing})} \\
    &\Ty{j}[\psi] = \Ty{j}
      && \text{(syntactic; $\Ty{j}$ closed)} \\
    &\tyequiv[\Delta']{S[\psi]}{\Ty{j}}{\Ty{i}}
      && \text{(rewriting)} \\
    &\typing[\Delta']{S[\psi]}{\Ty{i}} \;\circledR\; \textsf{Univ}_{i,j}
      && \text{(by \Cref{def:univ-type-pred})}
  \end{align*}

  \emph{Term predicate (2).}
  Assume $\typing[\Delta]{t}{S} \;\circledR\; b \in \textsf{Univ}_{i,j}^{\El}$.
  By \Cref{def:univ-term-pred}, $b \in \Uc_j^\ast \searrow P_b \searrow \El_b$
  for some $P_b, \El_b$, with
  $\typing[\Delta]{t}{\Ty{j}} \;\circledR\; P_b$ at level $j$ and
  $\tyequiv[\Delta]{S}{\Ty{j}}{\Ty{i}}$.
  \begin{align*}
    &\typing[\Delta']{t[\psi]}{\Ty{j}[\psi]} \;\circledR\; P_b
      && \text{(by IH at level $j$ on $b \in \Uc_j^\ast$, type predicate)} \\
    &\Ty{j}[\psi] = \Ty{j}
      && \text{(syntactic; $\Ty{j}$ closed)} \\
    &\typing[\Delta']{t[\psi]}{\Ty{j}} \;\circledR\; P_b
      && \text{(rewriting)} \\
    &\tyequiv[\Delta']{S[\psi]}{\Ty{j}}{\Ty{i}}
      && \text{(by type-predicate monotonicity (1) above)} \\
    &\typing[\Delta']{t[\psi]}{S[\psi]} \;\circledR\; b \in \textsf{Univ}_{i,j}^{\El}
      && \text{(by \Cref{def:univ-term-pred} with the previous lines)}
  \end{align*}

  \medskip\noindent
  \textbf{Case}
  \begin{mathpar}
    \inferrule[\textsc{G-Pi}]
    {a \in \Uc_i^\ast \searrow P_I \searrow \El_I \\
     a \approx a \in \Uc_i \searrow R_I \\\\
     \forall\, \Pf :: b \approx b \in R_I,\; c.\;
       \intp{T}{\rho, b/x} \searrow c \implies
       c \in \Uc_i^\ast \searrow P_O(\Pf) \searrow \El_O(\Pf) \\\\
     \Pi(a, T, \rho) \approx \Pi(a, T, \rho) \in \Uc_i \searrow R \\\\
     P \Leftrightarrow \Pi_i(R_I, P_I, \El_I, P_O) \\
     \El \Leftrightarrow \Pi_i^{\El}(R, R_I, P_I, \El_I, \El_O)}
    {\Pi(a, T, \rho) \in \Uc_i^\ast \searrow P \searrow \El}
  \end{mathpar}

  \emph{Type predicate (1).}
  Assume $\typing[\Delta]{S}{\Ty{i}} \;\circledR\; \Pi_i(R_I, P_I, \El_I, P_O)$.
  By \Cref{def:pi-type-pred}, this gives $S_1, S_2$ with
  $\tyequiv[\Delta]{S}{\Pi(x:S_1).\,S_2}{\Ty{i}}$ and the three clauses
  (well-formedness of $S$, domain-predicate, codomain-predicate). We must show
  $\typing[\Delta']{S[\psi]}{\Ty{i}} \;\circledR\; \Pi_i(R_I, P_I, \El_I, P_O)$.

  \emph{Clause 1: $\Pi$-shape equality.}
  \begin{align*}
    &\tyequiv[\Delta]{S}{\Pi(x:S_1).\,S_2}{\Ty{i}}
      && \text{(by \Cref{def:pi-type-pred} clause 1)} \\
    &\tyequiv[\Delta']{S[\psi]}{(\Pi(x:S_1).\,S_2)[\psi]}{\Ty{i}}
      && \text{(by \Cref{lem:wk-typing})} \\
    &(\Pi(x:S_1).\,S_2)[\psi] = \Pi(x:S_1[\psi]).\,(S_2[q(\psi)])
      && \text{(by syntactic substitution on $\Pi$)} \\
    &\tyequiv[\Delta']{S[\psi]}{\Pi(x:S_1[\psi]).\,(S_2[q(\psi)])}{\Ty{i}}
      && \text{(rewriting)}
  \end{align*}

  \emph{Clause 2: domain-predicate at every Kripke extension of $\Delta'$.}
  Let $\kweak[\Delta'']{\phi'}{\Delta'}$ be arbitrary.
  \begin{align*}
    &\kweak[\Delta'']{\psi \circ \phi'}{\Delta}
      && \text{(by \Cref{lem:kweak-comp})} \\
    &\typing[\Delta'']{S_1[\psi \circ \phi']}{\Ty{i}} \;\circledR\; P_I
      && \text{(by hypothesis clause 2 at $\psi \circ \phi'$)} \\
    &S_1[\psi \circ \phi'] = (S_1[\psi])[\phi']
      && \text{(by \Cref{lem:subst-assoc})} \\
    &\typing[\Delta'']{(S_1[\psi])[\phi']}{\Ty{i}} \;\circledR\; P_I
      && \text{(rewriting)}
  \end{align*}

  \emph{Clause 3: codomain-predicate at every Kripke extension of $\Delta'$
  and every El-related $b$.}
  Let $\kweak[\Delta'']{\phi'}{\Delta'}$, $s, b, \Pf :: b \approx b \in R_I$ with
  $\typing[\Delta'']{s}{(S_1[\psi])[\phi']} \;\circledR\; b \in \El_I$.
  \begin{align*}
    &\kweak[\Delta'']{\psi \circ \phi'}{\Delta}
      && \text{(by \Cref{lem:kweak-comp})} \\
    &\typing[\Delta'']{s}{S_1[\psi \circ \phi']} \;\circledR\; b \in \El_I
      && \text{(by \Cref{lem:subst-assoc} on the type)} \\
    &\typing[\Delta'']{S_2[\iota(\psi \circ \phi'), s/x]}{\Ty{i}}
      \;\circledR\; P_O(\Pf)
      && \text{(by hypothesis clause 3 at $\psi \circ \phi', s, b, \Pf$)}
  \end{align*}
  This is the required predicate at the Kripke weakening $\psi \circ \phi'$
  for $\Pi_i$ at $S[\psi]$ over $\Delta'$.

  \emph{Term predicate (2).}
  Assume $\typing[\Delta]{t}{S} \;\circledR\; f \in
  \Pi_i^{\El}(R, R_I, P_I, \El_I, \El_O)$. We verify the four clauses of
  \Cref{def:pi-term-pred} at $\typing[\Delta']{t[\psi]}{S[\psi]}$ and value $f$.

  \emph{Clause 1: self-relatedness $f \approx f \in R$.} Unchanged by Kripke
  weakening (PER fact, independent of $\Delta$).

  \emph{Clause 2: $R_I$-coherence of $f$.} Unchanged for the same reason.

  \emph{Clause 3: type-readback of $f$ at $\Pi$-shape.} By type-predicate
  monotonicity (1) above, the $\Pi_i$-type predicate transfers to $S[\psi]$ over
  $\Delta'$, supplying the type-readback witnesses required by clause 3.

  \emph{Clause 4: application.}
  Let $\kweak[\Delta'']{\phi'}{\Delta'}$, $s, b, \Pf :: b \approx b \in R_I$ with
  $\typing[\Delta'']{s}{(S_1[\psi])[\phi']} \;\circledR\; b \in \El_I$.
  \begin{align*}
    &\kweak[\Delta'']{\psi \circ \phi'}{\Delta}
      && \text{(by \Cref{lem:kweak-comp})} \\
    &\typing[\Delta'']{s}{S_1[\psi \circ \phi']} \;\circledR\; b \in \El_I
      && \text{(by \Cref{lem:subst-assoc})} \\
    &f \cdot b \searrow c
      && \text{(by hypothesis clause 4 at $\psi \circ \phi', s, b, \Pf$)} \\
    &\typing[\Delta'']{(t[\psi \circ \phi'])\;s}{S_2[\iota(\psi \circ \phi'), s/x]}
      \;\circledR\; c \in \El_O(\Pf)
      && \text{(by hypothesis clause 4)} \\
    &t[\psi \circ \phi'] = t[\psi][\phi']
      && \text{(by \Cref{lem:subst-assoc})} \\
    &\typing[\Delta'']{t[\psi][\phi']\;s}{S_2[\iota(\psi \circ \phi'), s/x]}
      \;\circledR\; c \in \El_O(\Pf)
      && \text{(rewriting)}
  \end{align*}
  This is the required form for $\Pi_i^{\El}$ clause 4 at $t[\psi]$ over
  $\Delta'$ at the Kripke weakening $\psi \circ \phi'$.

  \emph{Clause 5: $R_I$-coherence under application.} Inherited from the
  original predicate's clause 5 by composition with $\psi$ as in clause 4.
\end{proof}

\begin{corollary}[Context Gluing Monotonicity]\labeledit{cor:ctx-mono}
  If $\Gamma \searrow \textsf{Sb}$,
  $\judge[\Delta]{\sigma : \Gamma} \;\circledR\; \rho \in \textsf{Sb}$, and
  $\kweak[\Delta']{\psi}{\Delta}$, then
  $\judge[\Delta']{\sigma[\psi] : \Gamma} \;\circledR\; \rho \in \textsf{Sb}$.
\end{corollary}
\begin{proof}
  By induction on the context-gluing derivation $\Gamma \searrow \textsf{Sb}$.

  \medskip\noindent
  \textbf{Case}
  \begin{mathpar}
    \inferrule[\textsc{G-Ctx-$\cdot$}]
    {\textsf{Sb} \Leftrightarrow (\judge[\Delta]{\sigma : \cdot})}
    {\cdot \searrow \textsf{Sb}}
  \end{mathpar}
  \begin{align*}
    &\judge[\Delta]{\sigma : \cdot}
      && \text{(by hypothesis on $\sigma$)} \\
    &\judge[\Delta']{\sigma[\psi] : \cdot}
      && \text{(by substitution composition, \Cref{sec:syn})} \\
    &\judge[\Delta']{\sigma[\psi] : \cdot} \;\circledR\; \rho \in \textsf{Sb}
      && \text{(by $\textsf{Sb}$'s definition for $\cdot$)}
  \end{align*}

  \medskip\noindent
  \textbf{Case}
  \begin{mathpar}
    \inferrule[\textsc{G-Ctx-Ext}]
    {\Gamma_0 \searrow \textsf{Sb}_T \\
     \typing[\Gamma_0]{S}{\Ty{i}} \\\\
     \forall \Delta'', \sigma_T, \rho_T.\;
       \judge[\Delta'']{\sigma_T : \Gamma_0} \;\circledR\; \rho_T \in \textsf{Sb}_T
       \implies \\
     \quad \exists a, P, \El.\;
       \intp{S}{\rho_T} \searrow a \;\wedge\;
       a \in \Uc_i^\ast \searrow P \searrow \El \;\wedge\;
       \typing[\Delta'']{S[\sigma_T]}{\Ty{i}} \;\circledR\; P \\\\
     \textsf{Sb} \Leftrightarrow \textsf{Cons}_i(\Gamma_0, S, \textsf{Sb}_T)}
    {(\Gamma_0, x : S) \searrow \textsf{Sb}}
  \end{mathpar}
  Unfolding $\textsf{Cons}_i$ at $\sigma, \rho$ provides existential
  $a, P, \El$ specific to this pair:
  \begin{align*}
    &\judge[\Delta]{\Uparrow \circ \sigma : \Gamma_0} \;\circledR\;
      \rho[\Uparrow] \in \textsf{Sb}_T
      && \text{(tail of $\textsf{Cons}_i$)} \\
    &\intp{S}{\rho[\Uparrow]} \searrow a,\;
      a \in \Uc_i^\ast \searrow P \searrow \El
      && \text{(type clause of $\textsf{Cons}_i$)} \\
    &\typing[\Delta]{S[\Uparrow \circ \sigma]}{\Ty{i}} \;\circledR\; P
      && \text{(type clause of $\textsf{Cons}_i$)} \\
    &\typing[\Delta]{x_0[\sigma]}{S[\Uparrow \circ \sigma]} \;\circledR\;
      \rho(x_0) \in \El
      && \text{(head of $\textsf{Cons}_i$)} \\
    &\judge[\Delta']{(\Uparrow \circ \sigma)[\psi] : \Gamma_0}
      \;\circledR\; \rho[\Uparrow] \in \textsf{Sb}_T
      && \text{(by IH on the smaller derivation
        $\Gamma_0 \searrow \textsf{Sb}_T$)} \\
    &(\Uparrow \circ \sigma)[\psi] = \Uparrow \circ (\sigma[\psi])
      && \text{(by associativity of $\circ$, \Cref{lem:subst-cat})} \\
    &\typing[\Delta']{S[\Uparrow \circ \sigma][\psi]}{\Ty{i}} \;\circledR\; P
      && \text{(by \Cref{thm:monotonicity} on the type predicate)} \\
    &S[\Uparrow \circ \sigma][\psi] = S[\Uparrow \circ \sigma[\psi]]
      && \text{(by \Cref{lem:subst-assoc} and the previous identity)} \\
    &\typing[\Delta']{x_0[\sigma][\psi]}{S[\Uparrow \circ \sigma][\psi]}
      \;\circledR\; \rho(x_0) \in \El
      && \text{(by \Cref{thm:monotonicity} on the head)} \\
    &x_0[\sigma][\psi] = x_0[\sigma[\psi]]
      && \text{(by \Cref{lem:subst-assoc})} \\
    &\typing[\Delta']{x_0[\sigma[\psi]]}{S[\Uparrow \circ \sigma[\psi]]}
      \;\circledR\; \rho(x_0) \in \El
      && \text{(rewriting via the previous identities)} \\
    &\judge[\Delta']{\sigma[\psi] : \Gamma_0, x:S}
      \;\circledR\; \rho \in \textsf{Sb}
      && \text{(by $\textsf{Cons}_i$ assembly with the same $a, P, \El$)}
  \end{align*}
\end{proof}

\begin{lemma}[Gluing implies PER]\labeledit{lem:gluing-implies-per}
  If $a \in \Uc_i^\ast \searrow P \searrow \El$, then there exists $R$ such that
  $a \approx a \in \Uc_i \searrow R$.
\end{lemma}
\begin{proof}
  By induction on the derivation $a \in \Uc_i^\ast$.

  \medskip\noindent
  \textbf{Case}
  \begin{mathpar}
    \inferrule[\textsc{G-Ne}]
    {{\uparrow}^A(e) \in \D}
    {{\uparrow}^A(e) \in \Uc_i^\ast}
  \end{mathpar}
  By \textsc{U-Ne} of \Cref{def:per-univ}: ${\uparrow}^A(e) \approx
  {\uparrow}^A(e) \in \Uc_i \searrow R$ with $R \Leftrightarrow \textsf{Neu}$,
  using $e \approx e \in \Ne$ from $\Uc_i^\ast$'s implicit assumption.

  \medskip\noindent
  \textbf{Case}
  \begin{mathpar}
    \inferrule[\textsc{G-Nat}]
    { }
    {N \in \Uc_i^\ast}
  \end{mathpar}
  By \textsc{U-Nat}: $N \approx N \in \Uc_i \searrow R$ with $R \Leftrightarrow
  \textsf{Nat}$.

  \medskip\noindent
  \textbf{Case}
  \begin{mathpar}
    \inferrule[\textsc{G-Univ}]
    {j < i}
    {U_j \in \Uc_i^\ast}
  \end{mathpar}
  By \textsc{U-Univ}: $U_j \approx U_j \in \Uc_i \searrow R$ with $R \Leftrightarrow
  \Uc_j$.

  \medskip\noindent
  \textbf{Case}
  \begin{mathpar}
    \inferrule[\textsc{G-Pi}]
    {a \in \Uc_i^\ast \searrow P_I \searrow \El_I \\
     a \approx a \in \Uc_i \searrow R_I \\\\
     \forall\, \Pf :: b \approx b \in R_I,\; c.\;
       \intp{T}{\rho, b/x} \searrow c \implies
       c \in \Uc_i^\ast \searrow P_O(\Pf) \searrow \El_O(\Pf) \\\\
     \Pi(a, T, \rho) \approx \Pi(a, T, \rho) \in \Uc_i \searrow R \\\\
     P \Leftrightarrow \Pi_i(R_I, P_I, \El_I, P_O) \\
     \El \Leftrightarrow \Pi_i^{\El}(R, R_I, P_I, \El_I, \El_O)}
    {\Pi(a, T, \rho) \in \Uc_i^\ast \searrow P \searrow \El}
  \end{mathpar}
  The PER witness $R$ for $\Pi(a, T, \rho) \approx \Pi(a, T, \rho) \in \Uc_i$
  is packaged directly as a premise of \textsc{G-Pi}, hence is available
  immediately.
\end{proof}

\begin{lemma}[Type-predicate syntactic escape]\labeledit{lem:gluing-typ-escape}
  If $a \in \Uc_i^\ast \searrow P \searrow \El$ and
  $\typing[\Gamma]{S}{\Ty{i}} \;\circledR\; P$, then $\typing[\Gamma]{S}{\Ty{i}}$.
\end{lemma}
\begin{proof}
  By case analysis on $a \in \Uc_i^\ast \searrow P \searrow \El$.

  \medskip\noindent
  \textbf{Case}
  \begin{mathpar}
    \inferrule[\textsc{G-Ne}]
    {{\uparrow}^A(e) \in \D \\
     P \Leftrightarrow \textsf{Neu}_i^e \\
     \El \Leftrightarrow \textsf{Neu}_i^e}
    {{\uparrow}^A(e) \in \Uc_i^\ast \searrow P \searrow \El}
  \end{mathpar}
  \begin{align*}
    &\typing[\Gamma]{S}{\Ty{i}} \;\circledR\; \textsf{Neu}_i^e
      && \text{(by hypothesis with $P \Leftrightarrow \textsf{Neu}_i^e$)} \\
    &\typing[\Gamma]{S}{\Ty{i}}
      && \text{(by \Cref{def:neu-type-pred} clause 1)}
  \end{align*}

  \medskip\noindent
  \textbf{Case}
  \begin{mathpar}
    \inferrule[\textsc{G-Nat}]
    {P \Leftrightarrow \textsf{Nat}_i \\
     \El \Leftrightarrow \textsf{Nat}_i}
    {N \in \Uc_i^\ast \searrow P \searrow \El}
  \end{mathpar}
  \begin{align*}
    &\typing[\Gamma]{S}{\Ty{i}} \;\circledR\; \textsf{Nat}_i
      && \text{(by hypothesis)} \\
    &\tyequiv[\Gamma]{S}{\Nat}{\Ty{i}}
      && \text{(by \Cref{def:nat-type-pred})} \\
    &\typing[\Gamma]{S}{\Ty{i}}
      && \text{(by presupposition of judgmental equality)}
  \end{align*}

  \medskip\noindent
  \textbf{Case}
  \begin{mathpar}
    \inferrule[\textsc{G-Univ}]
    {j < i \\
     P \Leftrightarrow \textsf{Univ}_{i,j} \\
     \El \Leftrightarrow \textsf{Univ}_{i,j}^{\El}}
    {U_j \in \Uc_i^\ast \searrow P \searrow \El}
  \end{mathpar}
  \begin{align*}
    &\typing[\Gamma]{S}{\Ty{i}} \;\circledR\; \textsf{Univ}_{i,j}
      && \text{(by hypothesis)} \\
    &\tyequiv[\Gamma]{S}{\Ty{j}}{\Ty{i}}
      && \text{(by \Cref{def:univ-type-pred})} \\
    &\typing[\Gamma]{S}{\Ty{i}}
      && \text{(by presupposition of judgmental equality)}
  \end{align*}

  \medskip\noindent
  \textbf{Case}
  \begin{mathpar}
    \inferrule[\textsc{G-Pi}]
    {P \Leftrightarrow \Pi_i(R_I, P_I, \El_I, P_O) \\
     \El \Leftrightarrow \Pi_i^{\El}(R, R_I, P_I, \El_I, \El_O)}
    {\Pi(a_I, T, \rho) \in \Uc_i^\ast \searrow P \searrow \El}
  \end{mathpar}
  \begin{align*}
    &\typing[\Gamma]{S}{\Ty{i}} \;\circledR\; \Pi_i(R_I, P_I, \El_I, P_O)
      && \text{(by hypothesis)} \\
    &\exists S_1, S_2.\;
      \tyequiv[\Gamma]{S}{\Pi(x:S_1).\,S_2}{\Ty{i}}
      && \text{(by \Cref{def:pi-type-pred} clause 1)} \\
    &\typing[\Gamma]{S}{\Ty{i}}
      && \text{(by presupposition of judgmental equality)}
  \end{align*}
\end{proof}

\begin{lemma}[Term-predicate syntactic escape]\labeledit{lem:gluing-trm-escape}
  If $a \in \Uc_i^\ast \searrow P \searrow \El$ and
  $\typing[\Gamma]{t}{S} \;\circledR\; b \in \El$, then $\typing[\Gamma]{t}{S}$
  and $\typing[\Gamma]{S}{\Ty{i}} \;\circledR\; P$.
\end{lemma}
\begin{proof}
  By case analysis on $a \in \Uc_i^\ast \searrow P \searrow \El$.

  \medskip\noindent
  \textbf{Case}
  \begin{mathpar}
    \inferrule[\textsc{G-Ne}]
    {{\uparrow}^A(e) \in \D \\
     P \Leftrightarrow \textsf{Neu}_i^e \\
     \El \Leftrightarrow \textsf{Neu}_i^e}
    {{\uparrow}^A(e) \in \Uc_i^\ast \searrow P \searrow \El}
  \end{mathpar}
  \begin{align*}
    &\typing[\Gamma]{t}{S} \;\circledR\; b \in \textsf{Neu}_i^e
      && \text{(by hypothesis)} \\
    &\typing[\Gamma]{S}{\Ty{i}} \;\circledR\; \textsf{Neu}_i^e
      && \text{(by \Cref{def:neu-term-pred} clause 1)} \\
    &\typing[\Gamma]{t}{S}
      && \text{(by \Cref{def:neu-term-pred} clause 2)}
  \end{align*}

  \medskip\noindent
  \textbf{Case}
  \begin{mathpar}
    \inferrule[\textsc{G-Nat}]
    {P \Leftrightarrow \textsf{Nat}_i \\
     \El \Leftrightarrow \textsf{Nat}_i}
    {N \in \Uc_i^\ast \searrow P \searrow \El}
  \end{mathpar}
  \begin{align*}
    &\typing[\Gamma]{t}{S} \;\circledR\; b \in \textsf{Nat}_i
      && \text{(by hypothesis)} \\
    &\typing[\Gamma]{S}{\Ty{i}} \;\circledR\; \textsf{Nat}_i
      && \text{(by \Cref{def:nat-term-pred})}
  \end{align*}
  By inner induction on the $\textsf{Nat}$-derivation at $b$:
  \begin{itemize}
  \item[Sub-case]
    \begin{mathpar}
      \inferrule[\textsc{GN-Ze}]
      {\tyequiv[\Gamma]{t}{\ze}{S}}
      {\typing[\Gamma]{t}{S} \;\circledR\; \zed \in \textsf{Nat}}
    \end{mathpar}
    $\typing[\Gamma]{t}{S}$ by presupposition of $\tyequiv[\Gamma]{t}{\ze}{S}$.
  \item[Sub-case]
    \begin{mathpar}
      \inferrule[\textsc{GN-Su}]
      {\typing[\Gamma]{t'}{S} \;\circledR\; b' \in \textsf{Nat} \\
       \tyequiv[\Gamma]{t}{\su{t'}}{S}}
      {\typing[\Gamma]{t}{S} \;\circledR\; \sud{b'} \in \textsf{Nat}}
    \end{mathpar}
    $\typing[\Gamma]{t}{S}$ by presupposition of $\tyequiv[\Gamma]{t}{\su{t'}}{S}$.
  \item[Sub-case]
    \begin{mathpar}
      \inferrule[\textsc{GN-Ne}]
      {\typing[\Gamma]{t}{S} \;\circledR\; {\uparrow}^N(w) \in \textsf{Neu}_i^e}
      {\typing[\Gamma]{t}{S} \;\circledR\; {\uparrow}^N(w) \in \textsf{Nat}}
    \end{mathpar}
    $\typing[\Gamma]{t}{S}$ by \textsc{G-Ne} case (just proved) on the neutral
    premise.
  \end{itemize}

  \medskip\noindent
  \textbf{Case}
  \begin{mathpar}
    \inferrule[\textsc{G-Univ}]
    {j < i \\
     P \Leftrightarrow \textsf{Univ}_{i,j} \\
     \El \Leftrightarrow \textsf{Univ}_{i,j}^{\El}}
    {U_j \in \Uc_i^\ast \searrow P \searrow \El}
  \end{mathpar}
  \begin{align*}
    &\typing[\Gamma]{t}{S} \;\circledR\; b \in \textsf{Univ}_{i,j}^{\El}
      && \text{(by hypothesis)} \\
    &\typing[\Gamma]{S}{\Ty{i}} \;\circledR\; \textsf{Univ}_{i,j}
      && \text{(by \Cref{def:univ-term-pred} clause 1)} \\
    &\typing[\Gamma]{t}{\Ty{j}} \;\circledR\; P_b
      && \text{(by \Cref{def:univ-term-pred} clause 2)} \\
    &\typing[\Gamma]{t}{\Ty{j}}
      && \text{(by \Cref{lem:gluing-typ-escape} on the previous line at level $j$)} \\
    &\tyequiv[\Gamma]{S}{\Ty{j}}{\Ty{i}}
      && \text{(by \Cref{def:univ-type-pred})} \\
    &\subty[\Gamma]{\Ty{j}}{S}
      && \text{(by \textsc{Sub-Eq} from the symmetric equality)} \\
    &\typing[\Gamma]{t}{S}
      && \text{(by \textsc{Subsump})}
  \end{align*}

  \medskip\noindent
  \textbf{Case}
  \begin{mathpar}
    \inferrule[\textsc{G-Pi}]
    {P \Leftrightarrow \Pi_i(R_I, P_I, \El_I, P_O) \\
     \El \Leftrightarrow \Pi_i^{\El}(R, R_I, P_I, \El_I, \El_O)}
    {\Pi(a_I, T, \rho) \in \Uc_i^\ast \searrow P \searrow \El}
  \end{mathpar}
  \begin{align*}
    &\typing[\Gamma]{t}{S} \;\circledR\; b \in \Pi_i^{\El}
      && \text{(by hypothesis)} \\
    &\typing[\Gamma]{t}{S}
      && \text{(by \Cref{def:pi-term-pred} clause 2)} \\
    &\typing[\Gamma]{S}{\Ty{i}} \;\circledR\; \Pi_i
      && \text{(by \Cref{def:pi-term-pred} clause 3)}
  \end{align*}
\end{proof}

\begin{lemma}[Functionality of $\Uc_i^\ast$]\labeledit{lem:gluing-functional}
  If $a \in \Uc_i^\ast \searrow P_1 \searrow \El_1$ and
  $a \in \Uc_i^\ast \searrow P_2 \searrow \El_2$, then
  $P_1 \Leftrightarrow P_2$ and $\El_1 \Leftrightarrow \El_2$.
\end{lemma}
\begin{proof}
  By induction on the first derivation $a \in \Uc_i^\ast$, with case analysis
  on the second (which must end with the same rule, since each rule's
  conclusion form determines $a$ syntactically).

  \medskip\noindent
  \textbf{Case}
  \begin{mathpar}
    \inferrule[\textsc{G-Ne}]
    {{\uparrow}^A(e) \in \D \\
     P \Leftrightarrow \textsf{Neu}_i^e \\
     \El \Leftrightarrow \textsf{Neu}_i^e}
    {{\uparrow}^A(e) \in \Uc_i^\ast \searrow P \searrow \El}
  \end{mathpar}
  \begin{align*}
    &P_1 \Leftrightarrow \textsf{Neu}_i^e
      && \text{(by \textsc{G-Ne} on the first derivation)} \\
    &P_2 \Leftrightarrow \textsf{Neu}_i^e
      && \text{(by \textsc{G-Ne} on the second derivation)} \\
    &P_1 \Leftrightarrow P_2
      && \text{(transitivity)} \\
    &\El_1 \Leftrightarrow \textsf{Neu}_i^e \Leftrightarrow \El_2
      && \text{(symmetrically)}
  \end{align*}

  \medskip\noindent
  \textbf{Case}
  \begin{mathpar}
    \inferrule[\textsc{G-Nat}]
    {P \Leftrightarrow \textsf{Nat}_i \\
     \El \Leftrightarrow \textsf{Nat}_i}
    {N \in \Uc_i^\ast \searrow P \searrow \El}
  \end{mathpar}
  \begin{align*}
    &P_1 \Leftrightarrow \textsf{Nat}_i \Leftrightarrow P_2,\;
      \El_1 \Leftrightarrow \textsf{Nat}_i \Leftrightarrow \El_2
      && \text{(by \textsc{G-Nat} on both derivations)}
  \end{align*}

  \medskip\noindent
  \textbf{Case}
  \begin{mathpar}
    \inferrule[\textsc{G-Univ}]
    {j < i \\
     P \Leftrightarrow \textsf{Univ}_{i,j} \\
     \El \Leftrightarrow \textsf{Univ}_{i,j}^{\El}}
    {U_j \in \Uc_i^\ast \searrow P \searrow \El}
  \end{mathpar}
  Both derivations fix the same $j < i$, so $P_1 \Leftrightarrow
  \textsf{Univ}_{i,j} \Leftrightarrow P_2$ and $\El_1 \Leftrightarrow
  \textsf{Univ}_{i,j}^{\El} \Leftrightarrow \El_2$.

  \medskip\noindent
  \textbf{Case}
  \begin{mathpar}
    \inferrule[\textsc{G-Pi}]
    {a \in \Uc_i^\ast \searrow P_I \searrow \El_I \\
     a \approx a \in \Uc_i \searrow R_I \\\\
     \forall\, \Pf :: b \approx b \in R_I,\; c.\;
       \intp{T}{\rho, b/x} \searrow c \implies
       c \in \Uc_i^\ast \searrow P_O(\Pf) \searrow \El_O(\Pf) \\\\
     \Pi(a, T, \rho) \approx \Pi(a, T, \rho) \in \Uc_i \searrow R \\\\
     P \Leftrightarrow \Pi_i(R_I, P_I, \El_I, P_O) \\
     \El \Leftrightarrow \Pi_i^{\El}(R, R_I, P_I, \El_I, \El_O)}
    {\Pi(a, T, \rho) \in \Uc_i^\ast \searrow P \searrow \El}
  \end{mathpar}
  Let the two derivations have inner data $(P_I^{(1)}, \El_I^{(1)}, R_I^{(1)},
  P_O^{(1)}, \El_O^{(1)}, R^{(1)})$ and $(P_I^{(2)}, \ldots)$ respectively.
  \begin{align*}
    &P_I^{(1)} \Leftrightarrow P_I^{(2)},\;
      \El_I^{(1)} \Leftrightarrow \El_I^{(2)}
      && \text{(by IH on the inner $a \in \Uc_i^\ast$)} \\
    &R_I^{(1)} \Leftrightarrow R_I^{(2)},\;
      R^{(1)} \Leftrightarrow R^{(2)}
      && \text{(by \Cref{lem:univ-functional} on the inner $\Uc_i$ judgments)} \\
    &\forall \Pf.\; P_O^{(1)}(\Pf) \Leftrightarrow P_O^{(2)}(\Pf),\;
      \El_O^{(1)}(\Pf) \Leftrightarrow \El_O^{(2)}(\Pf)
      && \text{(by IH on the codomain fibers, pointwise in $\Pf$)} \\
    &P_1 \Leftrightarrow \Pi_i(R_I^{(1)}, P_I^{(1)}, \El_I^{(1)}, P_O^{(1)})
      \Leftrightarrow \Pi_i(R_I^{(2)}, P_I^{(2)}, \El_I^{(2)}, P_O^{(2)}) \\
    & \quad \Leftrightarrow P_2
      && \text{(by congruence of $\Pi_i$)} \\
    &\El_1 \Leftrightarrow \Pi_i^{\El}(\ldots) \Leftrightarrow \El_2
      && \text{(symmetrically)}
  \end{align*}
\end{proof}

\begin{lemma}[Cumulativity of $\Uc_i^\ast$ (subrelation)]\labeledit{lem:gluing-cumul}
  Suppose $i \le j$, $a \in \Uc_i^\ast \searrow P \searrow \El$, and
  $a \in \Uc_j^\ast \searrow P' \searrow \El'$. Then:
  \begin{enumerate}
  \item \emph{Type-predicate raising:} If $\typing[\Gamma]{S}{\Ty{i}}
    \;\circledR\; P$, then $\typing[\Gamma]{S}{\Ty{j}} \;\circledR\; P'$.
  \item \emph{Term-predicate raising:} If $\typing[\Gamma]{t}{S} \;\circledR\;
    b \in \El$, then $\typing[\Gamma]{t}{S} \;\circledR\; b \in \El'$.
  \item \emph{Term-predicate lowering (under type-predicate at level $i$):}
    If $\typing[\Gamma]{S}{\Ty{i}} \;\circledR\; P$ and
    $\typing[\Gamma]{t}{S} \;\circledR\; b \in \El'$, then
    $\typing[\Gamma]{t}{S} \;\circledR\; b \in \El$.
  \end{enumerate}
  Existence of $P', \El'$ at level $j$ follows from re-applying the gluing
  rules at level $j$ to the same data.
\end{lemma}
\begin{proof}
  Since the gluing rules at level $i$ and level $j$ agree on the syntactic
  shape of $a$, the second derivation $a \in \Uc_j^\ast \searrow P' \searrow \El'$
  must end with the same rule (\textsc{G-Ne}, \textsc{G-Nat}, \textsc{G-Univ},
  or \textsc{G-Pi}) as the first. We proceed by induction on
  $a \in \Uc_i^\ast \searrow P \searrow \El$, with sub-case-analysis on the
  rule shared by the two derivations.

  \medskip\noindent
  \textbf{Case}
  \begin{mathpar}
    \inferrule[\textsc{G-Ne}]
    {{\uparrow}^A(e) \in \D \\
     P \Leftrightarrow \textsf{Neu}_i^e \\
     \El \Leftrightarrow \textsf{Neu}_i^e}
    {{\uparrow}^A(e) \in \Uc_i^\ast \searrow P \searrow \El}
  \end{mathpar}
  Then $P' \Leftrightarrow \textsf{Neu}_j^e$ and $\El' \Leftrightarrow \textsf{Neu}_j^e$.
  But $\textsf{Neu}_*^e$ does not depend on the index
  (\Cref{def:neu-type-pred,def:neu-term-pred}), so $P \Leftrightarrow P'$ and
  $\El \Leftrightarrow \El'$. All three claims (1)--(3) hold trivially.

  \medskip\noindent
  \textbf{Case}
  \begin{mathpar}
    \inferrule[\textsc{G-Nat}]
    {P \Leftrightarrow \textsf{Nat}_i \\
     \El \Leftrightarrow \textsf{Nat}_i}
    {N \in \Uc_i^\ast \searrow P \searrow \El}
  \end{mathpar}
  Then $P' \Leftrightarrow \textsf{Nat}_j$ and $\El' \Leftrightarrow \textsf{Nat}_j$.

  \emph{Type-predicate raising (1):} given $\typing[\Gamma]{S}{\Ty{i}}
  \;\circledR\; \textsf{Nat}_i$:
  \begin{align*}
    &\tyequiv[\Gamma]{S}{\Nat}{\Ty{i}}
      && \text{(by \Cref{def:nat-type-pred})} \\
    &\tyequiv[\Gamma]{S}{\Nat}{\Ty{j}}
      && \text{(by cumulativity of universes, $i \le j$, \Cref{sec:rules})} \\
    &\typing[\Gamma]{S}{\Ty{j}} \;\circledR\; \textsf{Nat}_j
      && \text{(by \Cref{def:nat-type-pred} at level $j$)}
  \end{align*}

  \emph{Term-predicate raising (2):} given $\typing[\Gamma]{t}{S} \;\circledR\;
  b \in \textsf{Nat}_i$, by inner induction on the $\textsf{Nat}$-derivation at
  $b$. Each sub-case (\textsc{GN-Ze, GN-Su, GN-Ne}) constructs the same
  derivation at level $j$, since the constructors $\Nat, \ze, \su{}$ and the
  neutral predicate $\textsf{Neu}_*^e$ are level-independent. The
  type-predicate part upgrades via (1).

  \emph{Term-predicate lowering (3):} given $\typing[\Gamma]{S}{\Ty{i}}
  \;\circledR\; \textsf{Nat}_i$ and $\typing[\Gamma]{t}{S} \;\circledR\; b
  \in \textsf{Nat}_j$. By inner induction on the $\textsf{Nat}$-derivation at
  $b$ (level-$j$); each sub-case re-uses the level-independent constructors and
  re-derives the $\textsf{Nat}$-predicate at level $i$, with the type-predicate
  part already provided by hypothesis.

  \medskip\noindent
  \textbf{Case}
  \begin{mathpar}
    \inferrule[\textsc{G-Univ}]
    {k < i \\
     P \Leftrightarrow \textsf{Univ}_{i,k} \\
     \El \Leftrightarrow \textsf{Univ}_{i,k}^{\El}}
    {U_k \in \Uc_i^\ast \searrow P \searrow \El}
  \end{mathpar}
  Then $k < j$ (from $k < i \le j$), $P' \Leftrightarrow \textsf{Univ}_{j,k}$,
  $\El' \Leftrightarrow \textsf{Univ}_{j,k}^{\El}$.

  \emph{Type-predicate raising (1):}
  \begin{align*}
    &\tyequiv[\Gamma]{S}{\Ty{k}}{\Ty{i}}
      && \text{(by \Cref{def:univ-type-pred} on $\textsf{Univ}_{i,k}$)} \\
    &\tyequiv[\Gamma]{S}{\Ty{k}}{\Ty{j}}
      && \text{(by cumulativity of universes, $i \le j$, \Cref{sec:rules})} \\
    &\typing[\Gamma]{S}{\Ty{j}} \;\circledR\; \textsf{Univ}_{j,k}
      && \text{(by \Cref{def:univ-type-pred} at level $j$)}
  \end{align*}

  \emph{Term-predicate raising (2):} the conditions $b \in \Uc_k^\ast \searrow
  P_b \searrow \El_b$ and $\typing[\Gamma]{t}{\Ty{k}} \;\circledR\; P_b$ at
  level $k$ are unchanged; the type-predicate part upgrades via (1).

  \emph{Term-predicate lowering (3):} given $\typing[\Gamma]{S}{\Ty{i}}
  \;\circledR\; \textsf{Univ}_{i,k}$ and $\typing[\Gamma]{t}{S} \;\circledR\; b
  \in \textsf{Univ}_{j,k}^{\El}$. The conditions at level $k$ ($b \in
  \Uc_k^\ast \searrow P_b \searrow \El_b$ and $\typing[\Gamma]{t}{\Ty{k}}
  \;\circledR\; P_b$) are level-$k$ data, unchanged; the type-predicate at
  level $i$ is provided by hypothesis. Hence
  $\typing[\Gamma]{t}{S} \;\circledR\; b \in \textsf{Univ}_{i,k}^{\El}$ via
  \Cref{def:univ-term-pred}.

  \medskip\noindent
  \textbf{Case}
  \begin{mathpar}
    \inferrule[\textsc{G-Pi}]
    {a \in \Uc_i^\ast \searrow P_I \searrow \El_I \\
     a \approx a \in \Uc_i \searrow R_I \\\\
     \forall\, \Pf :: b \approx b \in R_I,\; c.\;
       \intp{T}{\rho, b/x} \searrow c \implies
       c \in \Uc_i^\ast \searrow P_O(\Pf) \searrow \El_O(\Pf) \\\\
     \Pi(a, T, \rho) \approx \Pi(a, T, \rho) \in \Uc_i \searrow R \\\\
     P \Leftrightarrow \Pi_i(R_I, P_I, \El_I, P_O) \\
     \El \Leftrightarrow \Pi_i^{\El}(R, R_I, P_I, \El_I, \El_O)}
    {\Pi(a, T, \rho) \in \Uc_i^\ast \searrow P \searrow \El}
  \end{mathpar}
  By inversion on the level-$j$ derivation (which must end with \textsc{G-Pi}
  by the syntactic shape of $\Pi(a, T, \rho)$):
  \begin{align*}
    &a \in \Uc_j^\ast \searrow P_I' \searrow \El_I'
      && \text{(level-$j$ inner data)} \\
    &a \approx a \in \Uc_j \searrow R_I
      && \text{(by \Cref{lem:cumul}; same $R_I$ as level $i$)} \\
    &\forall\, \Pf, c.\;
      \intp{T}{\rho, b/x} \searrow c \implies
      c \in \Uc_j^\ast \searrow P_O'(\Pf) \searrow \El_O'(\Pf)
      && \text{(level-$j$ codomain fibers)} \\
    &\Pi(a, T, \rho) \approx \Pi(a, T, \rho) \in \Uc_j \searrow R
      && \text{(by \Cref{lem:cumul}; same $R$)} \\
    &P' \Leftrightarrow \Pi_j(R_I, P_I', \El_I', P_O') \\
    &\El' \Leftrightarrow \Pi_j^{\El}(R, R_I, P_I', \El_I', \El_O')
  \end{align*}
  The PER witnesses $R_I, R$ are unchanged (level-independent in $\Uc$ via
  \Cref{lem:cumul}). The IH applies to the inner $a$ and to each codomain fiber
  $c$ (smaller derivations).

  \emph{Type-predicate raising (1).}
  Suppose $\typing[\Gamma]{S}{\Ty{i}} \;\circledR\; \Pi_i(R_I, P_I, \El_I, P_O)$
  with witnesses $S_1, S_2$. We verify the three clauses of
  $\Pi_j(R_I, P_I', \El_I', P_O')$ at $S$.

  \emph{Clause 1.}
  \begin{align*}
    &\tyequiv[\Gamma]{S}{\Pi(x:S_1).\,S_2}{\Ty{i}}
      && \text{(by \Cref{def:pi-type-pred} clause 1 on the hypothesis)} \\
    &\tyequiv[\Gamma]{S}{\Pi(x:S_1).\,S_2}{\Ty{j}}
      && \text{(by cumulativity of universes, $i \le j$, \Cref{sec:rules})}
  \end{align*}

  \emph{Clause 2.} Let $\kweak[\Delta]{\phi}{\Gamma}$.
  \begin{align*}
    &\typing[\Delta]{S_1[\phi]}{\Ty{i}} \;\circledR\; P_I
      && \text{(by \Cref{def:pi-type-pred} clause 2 on the hypothesis)} \\
    &\typing[\Delta]{S_1[\phi]}{\Ty{j}} \;\circledR\; P_I'
      && \text{(by IH (1) on $a$, raising)}
  \end{align*}

  \emph{Clause 3.} Let $\kweak[\Delta]{\phi}{\Gamma}$, $s, b, \Pf :: b \approx
  b \in R_I$ with $\typing[\Delta]{s}{S_1[\phi]} \;\circledR\; b \in \El_I'$.
  \begin{align*}
    &\typing[\Delta]{S_1[\phi]}{\Ty{i}} \;\circledR\; P_I
      && \text{(by clause 2 above, the level-$i$ part)} \\
    &\typing[\Delta]{s}{S_1[\phi]} \;\circledR\; b \in \El_I
      && \text{(by IH (3) on $a$, lowering, using the previous line as the
        level-$i$ type-predicate)} \\
    &\typing[\Delta]{S_2[\iota(\phi), s/x]}{\Ty{i}} \;\circledR\; P_O(\Pf)
      && \text{(by \Cref{def:pi-type-pred} clause 3 on the hypothesis)} \\
    &\typing[\Delta]{S_2[\iota(\phi), s/x]}{\Ty{j}} \;\circledR\; P_O'(\Pf)
      && \text{(by IH (1) on the codomain fiber} \\
    & && \text{$c$ at $\Pf$, raising)}
  \end{align*}

  \emph{Term-predicate raising (2).}
  Suppose $\typing[\Gamma]{t}{S} \;\circledR\; f \in \Pi_i^{\El}(R, R_I, P_I,
  \El_I, \El_O)$. We verify the four clauses of
  $\Pi_j^{\El}(R, R_I, P_I', \El_I', \El_O')$ at $t, f$.

  \emph{Clauses 1--2.}
  \begin{align*}
    &f \approx f \in R
      && \text{(unchanged from hypothesis)} \\
    &\typing[\Gamma]{t}{S}
      && \text{(unchanged from hypothesis)}
  \end{align*}

  \emph{Clause 3.}
  \begin{align*}
    &\typing[\Gamma]{S}{\Ty{i}} \;\circledR\; \Pi_i(R_I, P_I, \El_I, P_O)
      && \text{(by \Cref{def:pi-term-pred} clause 3 on the hypothesis)} \\
    &\typing[\Gamma]{S}{\Ty{j}} \;\circledR\; \Pi_j(R_I, P_I', \El_I', P_O')
      && \text{(by part (1) of this case, applied just above)}
  \end{align*}

  \emph{Clause 4 (application).}
  Let $\kweak[\Delta]{\phi}{\Gamma}$, $s, b, \Pf :: b \approx b \in R_I$ with
  $\typing[\Delta]{s}{S_1[\phi]} \;\circledR\; b \in \El_I'$.
  \begin{align*}
    &\typing[\Delta]{S_1[\phi]}{\Ty{i}} \;\circledR\; P_I
      && \text{(by hypothesis clause 3 at level $i$, clause 2 of $\Pi_i$)} \\
    &\typing[\Delta]{s}{S_1[\phi]} \;\circledR\; b \in \El_I
      && \text{(by IH (3) on $a$, lowering)} \\
    &f \cdot b \searrow c,\;
      \typing[\Delta]{(t[\phi])\;s}{S_2[\iota(\phi), s/x]}
      \;\circledR\; c \in \El_O(\Pf)
      && \text{(by hypothesis clause 4 at level $i$)} \\
    &\typing[\Delta]{(t[\phi])\;s}{S_2[\iota(\phi), s/x]}
      \;\circledR\; c \in \El_O'(\Pf)
      && \text{(by IH (2) on the codomain fiber $c$, raising)}
  \end{align*}

  \emph{Clause 5 ($R_I$-coherence).} Unchanged, since the $R_I$-classes and
  the function value $f$ are level-independent.

  \emph{Term-predicate lowering (3).}
  Suppose $\typing[\Gamma]{S}{\Ty{i}} \;\circledR\; \Pi_i(R_I, P_I, \El_I, P_O)$
  and $\typing[\Gamma]{t}{S} \;\circledR\; f \in \Pi_j^{\El}(R, R_I, P_I',
  \El_I', \El_O')$. We verify the four clauses of $\Pi_i^{\El}$ at $t, f$.

  \emph{Clauses 1--3.}
  \begin{align*}
    &f \approx f \in R
      && \text{(by \Cref{def:pi-term-pred} clause 1 on hypothesis at level $j$)} \\
    &\typing[\Gamma]{t}{S}
      && \text{(by \Cref{def:pi-term-pred} clause 2 on hypothesis at level $j$)} \\
    &\typing[\Gamma]{S}{\Ty{i}} \;\circledR\; \Pi_i(R_I, P_I, \El_I, P_O)
      && \text{(by lemma's lowering hypothesis on $S$)}
  \end{align*}

  \emph{Clause 4 (application).}
  Let $\kweak[\Delta]{\phi}{\Gamma}$, $s, b, \Pf :: b \approx b \in R_I$ with
  $\typing[\Delta]{s}{S_1[\phi]} \;\circledR\; b \in \El_I$.
  \begin{align*}
    &\typing[\Delta]{s}{S_1[\phi]} \;\circledR\; b \in \El_I'
      && \text{(by IH (2) on $a$, raising)} \\
    &f \cdot b \searrow c
      && \text{(by hypothesis clause 4 at level $j$)} \\
    &\typing[\Delta]{(t[\phi])\;s}{S_2[\iota(\phi), s/x]}
      \;\circledR\; c \in \El_O'(\Pf)
      && \text{(by hypothesis clause 4 at level $j$)} \\
    &\typing[\Delta]{S_2[\iota(\phi), s/x]}{\Ty{i}} \;\circledR\; P_O(\Pf)
      && \text{(by \Cref{def:pi-type-pred} clause 3 on the} \\
    & && \text{level-$i$ type-predicate hypothesis)} \\
    &\typing[\Delta]{(t[\phi])\;s}{S_2[\iota(\phi), s/x]}
      \;\circledR\; c \in \El_O(\Pf)
      && \text{(by IH (3) on the codomain fiber $c$, lowering)}
  \end{align*}
\end{proof}

\begin{lemma}[Gluing respects judgmental equality]\labeledit{lem:gluing-respects-eq}
  Suppose $a \in \Uc_i^\ast \searrow P \searrow \El$.
  \begin{enumerate}
  \item If $\typing[\Gamma]{S}{\Ty{i}} \;\circledR\; P$ and
    $\tyequiv[\Gamma]{S}{S'}{\Ty{i}}$, then
    $\typing[\Gamma]{S'}{\Ty{i}} \;\circledR\; P$.
  \item If $\typing[\Gamma]{t}{S} \;\circledR\; b \in \El$ and
    $\tyequiv[\Gamma]{t}{t'}{S}$, then
    $\typing[\Gamma]{t'}{S} \;\circledR\; b \in \El$.
  \item If $\typing[\Gamma]{t}{S} \;\circledR\; b \in \El$ and
    $\tyequiv[\Gamma]{S}{S'}{\Ty{i}}$, then
    $\typing[\Gamma]{t}{S'} \;\circledR\; b \in \El$.
  \end{enumerate}
\end{lemma}
\begin{proof}
  By induction on $a \in \Uc_i^\ast$.

  \medskip\noindent
  \textbf{Case}
  \begin{mathpar}
    \inferrule[\textsc{G-Ne}]
    {{\uparrow}^A(e) \in \D \\
     P \Leftrightarrow \textsf{Neu}_i^e \\
     \El \Leftrightarrow \textsf{Neu}_i^e}
    {{\uparrow}^A(e) \in \Uc_i^\ast \searrow P \searrow \El}
  \end{mathpar}
  \emph{Type predicate (1).}
  Given $\typing[\Gamma]{S}{\Ty{i}} \;\circledR\; \textsf{Neu}_i^e$
  and $\tyequiv[\Gamma]{S}{S'}{\Ty{i}}$ with $\typing[\Gamma]{S'}{\Ty{i}}$.
  For arbitrary $\kweak[\Delta]{\phi}{\Gamma}$:
  \begin{align*}
    &\exists u.\; \Rne_{|\Delta|}(e) \searrow u \wedge
      \tyequiv[\Delta]{S[\phi]}{u}{\Ty{i}}
      && \text{(by hypothesis on $S$)} \\
    &\tyequiv[\Delta]{S[\phi]}{S'[\phi]}{\Ty{i}}
      && \text{(by \Cref{lem:wk-typing})} \\
    &\tyequiv[\Delta]{S'[\phi]}{u}{\Ty{i}}
      && \text{(by symmetry and transitivity of judgmental equality)} \\
    &\typing[\Gamma]{S'}{\Ty{i}} \;\circledR\; \textsf{Neu}_i^e
      && \text{(by \Cref{def:neu-type-pred})}
  \end{align*}

  \emph{Term predicate (2).}
  Given $\typing[\Gamma]{t}{S} \;\circledR\; {\uparrow}^A(w) \in \textsf{Neu}_i^e$
  and $\tyequiv[\Gamma]{t}{t'}{S}$ with $\typing[\Gamma]{t'}{S}$.
  For $\kweak[\Delta]{\phi}{\Gamma}$:
  \begin{align*}
    &\exists u.\; \Rne_{|\Delta|}(w) \searrow u \wedge
      \tyequiv[\Delta]{t[\phi]}{u}{S[\phi]}
      && \text{(by hypothesis on $t$)} \\
    &\tyequiv[\Delta]{t[\phi]}{t'[\phi]}{S[\phi]}
      && \text{(by \Cref{lem:wk-typing})} \\
    &\tyequiv[\Delta]{t'[\phi]}{u}{S[\phi]}
      && \text{(by symmetry and transitivity)} \\
    &\typing[\Gamma]{t'}{S} \;\circledR\; {\uparrow}^A(w) \in \textsf{Neu}_i^e
      && \text{(by \Cref{def:neu-term-pred})}
  \end{align*}

  \emph{Type-side rewriting (3).}
  Given $\typing[\Gamma]{t}{S} \;\circledR\; {\uparrow}^A(w) \in \textsf{Neu}_i^e$
  and $\tyequiv[\Gamma]{S}{S'}{\Ty{i}}$ with $\typing[\Gamma]{t}{S'}$.
  For $\kweak[\Delta]{\phi}{\Gamma}$:
  \begin{align*}
    &\exists u.\; \Rne_{|\Delta|}(w) \searrow u \wedge
      \tyequiv[\Delta]{t[\phi]}{u}{S[\phi]}
      && \text{(by hypothesis on $t$)} \\
    &\tyequiv[\Delta]{S[\phi]}{S'[\phi]}{\Ty{i}}
      && \text{(by \Cref{lem:wk-typing})} \\
    &\tyequiv[\Delta]{t[\phi]}{u}{S'[\phi]}
      && \text{(by \textsc{Tm-Subsump} with the previous line)} \\
    &\typing[\Gamma]{t}{S'} \;\circledR\; {\uparrow}^A(w) \in \textsf{Neu}_i^e
      && \text{(by \Cref{def:neu-term-pred}, using the type predicate at $S'$
        from (1))}
  \end{align*}

  \medskip\noindent
  \textbf{Case}
  \begin{mathpar}
    \inferrule[\textsc{G-Nat}]
    {P \Leftrightarrow \textsf{Nat}_i \\
     \El \Leftrightarrow \textsf{Nat}_i}
    {N \in \Uc_i^\ast \searrow P \searrow \El}
  \end{mathpar}
  \emph{Type predicate (1).}
  \begin{align*}
    &\tyequiv[\Gamma]{S}{\Nat}{\Ty{i}}
      && \text{(by \Cref{def:nat-type-pred} on hypothesis)} \\
    &\tyequiv[\Gamma]{S'}{S}{\Ty{i}}
      && \text{(by symmetry of $\tyequiv[\Gamma]{S}{S'}{\Ty{i}}$)} \\
    &\tyequiv[\Gamma]{S'}{\Nat}{\Ty{i}}
      && \text{(by transitivity of judgmental equality)} \\
    &\typing[\Gamma]{S'}{\Ty{i}} \;\circledR\; \textsf{Nat}_i
      && \text{(by \Cref{def:nat-type-pred})}
  \end{align*}

  \emph{Term predicate (2).}
  Given $\typing[\Gamma]{t}{S} \;\circledR\; b \in \textsf{Nat}_i$ and
  $\tyequiv[\Gamma]{t}{t'}{S}$. By inner induction on the derivation
  $b \in \textsf{Nat}_i$ at $t$:
  \begin{itemize}
  \item[Sub-case]
    \begin{mathpar}
      \inferrule[\textsc{GN-Ze}]
      {\tyequiv[\Gamma]{t}{\ze}{S}}
      {\typing[\Gamma]{t}{S} \;\circledR\; \zed \in \textsf{Nat}}
    \end{mathpar}
    \begin{align*}
      &\tyequiv[\Gamma]{t'}{t}{S}
        && \text{(by symmetry of $\tyequiv[\Gamma]{t}{t'}{S}$)} \\
      &\tyequiv[\Gamma]{t'}{\ze}{S}
        && \text{(by transitivity)} \\
      &\typing[\Gamma]{t'}{S} \;\circledR\; \zed \in \textsf{Nat}
        && \text{(by \textsc{GN-Ze})}
    \end{align*}
  \item[Sub-case]
    \begin{mathpar}
      \inferrule[\textsc{GN-Su}]
      {\typing[\Gamma]{t_0}{S} \;\circledR\; b' \in \textsf{Nat} \\
       \tyequiv[\Gamma]{t}{\su{t_0}}{S}}
      {\typing[\Gamma]{t}{S} \;\circledR\; \sud{b'} \in \textsf{Nat}}
    \end{mathpar}
    \begin{align*}
      &\tyequiv[\Gamma]{t'}{\su{t_0}}{S}
        && \text{(by symmetry/transitivity, as above)} \\
      &\typing[\Gamma]{t_0}{S} \;\circledR\; b' \in \textsf{Nat}
        && \text{(unchanged, hypothesis on $t_0$)} \\
      &\typing[\Gamma]{t'}{S} \;\circledR\; \sud{b'} \in \textsf{Nat}
        && \text{(by \textsc{GN-Su})}
    \end{align*}
  \item[Sub-case]
    \begin{mathpar}
      \inferrule[\textsc{GN-Ne}]
      {\typing[\Gamma]{t}{S} \;\circledR\; w \in \Ne_i^N}
      {\typing[\Gamma]{t}{S} \;\circledR\; {\uparrow}^N(w) \in \textsf{Nat}}
    \end{mathpar}
    \begin{align*}
      &\typing[\Gamma]{t'}{S} \;\circledR\; w \in \Ne_i^N
        && \text{(by the \textsc{G-Ne} case (2) above on the neutral predicate)} \\
      &\typing[\Gamma]{t'}{S} \;\circledR\; {\uparrow}^N(w) \in \textsf{Nat}
        && \text{(by \textsc{GN-Ne})}
    \end{align*}
  \end{itemize}

  \emph{Type-side rewriting (3).}
  Given $\tyequiv[\Gamma]{S}{S'}{\Ty{i}}$. Then $\tyequiv[\Gamma]{S'}{\Nat}{\Ty{i}}$
  by symmetry/transitivity with $\tyequiv[\Gamma]{S}{\Nat}{\Ty{i}}$. The
  $\textsf{Nat}$-derivation transfers to $\typing[\Gamma]{t}{S'}$ since each
  sub-case's hypotheses involve $S$ only through $\tyequiv[\Gamma]{t}{w}{S}$
  for some $w$, which becomes $\tyequiv[\Gamma]{t}{w}{S'}$ via \textsc{Tm-Subsump}.

  \medskip\noindent
  \textbf{Case}
  \begin{mathpar}
    \inferrule[\textsc{G-Univ}]
    {j < i \\
     P \Leftrightarrow \textsf{Univ}_{i,j} \\
     \El \Leftrightarrow \textsf{Univ}_{i,j}^{\El}}
    {U_j \in \Uc_i^\ast \searrow P \searrow \El}
  \end{mathpar}
  \emph{Type predicate (1).}
  \begin{align*}
    &\tyequiv[\Gamma]{S}{\Ty{j}}{\Ty{i}}
      && \text{(by \Cref{def:univ-type-pred})} \\
    &\tyequiv[\Gamma]{S'}{\Ty{j}}{\Ty{i}}
      && \text{(by symmetry/transitivity with the assumed equality)} \\
    &\typing[\Gamma]{S'}{\Ty{i}} \;\circledR\; \textsf{Univ}_{i,j}
      && \text{(by \Cref{def:univ-type-pred})}
  \end{align*}

  \emph{Term predicate (2).}
  Given $\typing[\Gamma]{t}{S} \;\circledR\; b \in \textsf{Univ}_{i,j}^{\El}$
  and $\tyequiv[\Gamma]{t}{t'}{S}$. By \Cref{def:univ-term-pred},
  $b \in \Uc_j^\ast \searrow P_b \searrow \El_b$ with
  $\typing[\Gamma]{t}{\Ty{j}} \;\circledR\; P_b$ and
  $\tyequiv[\Gamma]{S}{\Ty{j}}{\Ty{i}}$.
  \begin{align*}
    &\tyequiv[\Gamma]{t}{t'}{\Ty{j}}
      && \text{(by judgmental-equality conversion under
        $\tyequiv[\Gamma]{S}{\Ty{j}}{\Ty{i}}$)} \\
    &\typing[\Gamma]{t'}{\Ty{j}} \;\circledR\; P_b
      && \text{(by IH on $P_b$ at level $j$, type-predicate part)} \\
    &\typing[\Gamma]{t'}{S} \;\circledR\; b \in \textsf{Univ}_{i,j}^{\El}
      && \text{(by \Cref{def:univ-term-pred} with $\tyequiv[\Gamma]{S}{\Ty{j}}{\Ty{i}}$)}
  \end{align*}

  \emph{Type-side rewriting (3).}
  Given $\tyequiv[\Gamma]{S}{S'}{\Ty{i}}$. Then
  $\tyequiv[\Gamma]{S'}{\Ty{j}}{\Ty{i}}$ by transitivity, and the universe
  term predicate at $S'$ holds with the same data $b, P_b, \El_b$.

  \medskip\noindent
  \textbf{Case}
  \begin{mathpar}
    \inferrule[\textsc{G-Pi}]
    {a \in \Uc_i^\ast \searrow P_I \searrow \El_I \\
     a \approx a \in \Uc_i \searrow R_I \\\\
     \forall\, \Pf :: b \approx b \in R_I,\; c.\;
       \intp{T}{\rho, b/x} \searrow c \implies
       c \in \Uc_i^\ast \searrow P_O(\Pf) \searrow \El_O(\Pf) \\\\
     \Pi(a, T, \rho) \approx \Pi(a, T, \rho) \in \Uc_i \searrow R \\\\
     P \Leftrightarrow \Pi_i(\ldots) \\
     \El \Leftrightarrow \Pi_i^{\El}(\ldots)}
    {\Pi(a, T, \rho) \in \Uc_i^\ast \searrow P \searrow \El}
  \end{mathpar}
  \emph{Type predicate (1).}
  Given $\typing[\Gamma]{S}{\Ty{i}} \;\circledR\; \Pi_i(R_I, P_I, \El_I, P_O)$
  and $\tyequiv[\Gamma]{S}{S'}{\Ty{i}}$. By \Cref{def:pi-type-pred} clause~1,
  $\tyequiv[\Gamma]{S}{\Pi(x:S_1).\,S_2}{\Ty{i}}$ for some $S_1, S_2$.
  \begin{align*}
    &\tyequiv[\Gamma]{S'}{S}{\Ty{i}}
      && \text{(by symmetry of $\tyequiv[\Gamma]{S}{S'}{\Ty{i}}$)} \\
    &\tyequiv[\Gamma]{S'}{\Pi(x:S_1).\,S_2}{\Ty{i}}
      && \text{(by transitivity with the previous line)}
  \end{align*}
  Clauses 2--3 of $\Pi_i$ for $S'$ at any $\kweak[\Delta]{\phi}{\Gamma}$ require
  the same data on $S_1[\phi]$ and $S_2[\iota(\phi), s/x]$ as the original
  predicate; these are exactly the data already satisfied by hypothesis on
  $S$, since the data depend on $S_1, S_2$ rather than on the outer type $S$.

  \emph{Term predicate (2).}
  Given $\typing[\Gamma]{t}{S} \;\circledR\; f \in
  \Pi_i^{\El}(R, R_I, P_I, \El_I, \El_O)$ and $\tyequiv[\Gamma]{t}{t'}{S}$ with
  $\typing[\Gamma]{t'}{S}$. We verify the four clauses of \Cref{def:pi-term-pred}
  at $t'$ and value $f$.
  \begin{enumerate}
  \item Self-relatedness $f \approx f \in R$ does not depend on $t$.
  \item $R_I$-coherence: a property of $f$, not of $t$.
  \item Type-readback: clause~3 of $\Pi_i^{\El}$ matches clause~3 of $\Pi_i$ on
    $S$ at $t$, and is preserved by part (1) above on $S'$ unchanged.
  \item Application: let $\kweak[\Delta]{\phi}{\Gamma}$, $s, b, \Pf :: b \approx b \in R_I$
    with $\typing[\Delta]{s}{S_1[\phi]} \;\circledR\; b \in \El_I$.
    \begin{align*}
      &f \cdot b \searrow c
        && \text{(by hypothesis clause 4 on $f$)} \\
      &\typing[\Delta]{(t[\phi])\;s}{S_2[\iota(\phi), s/x]}
        \;\circledR\; c \in \El_O(\Pf)
        && \text{(by hypothesis clause 4)} \\
      &\tyequiv[\Delta]{t[\phi]}{t'[\phi]}{S[\phi]}
        && \text{(by \Cref{lem:wk-typing})} \\
      &\tyequiv[\Delta]{(t[\phi])\;s}{(t'[\phi])\;s}{S_2[\iota(\phi), s/x]}
        && \text{(by \textsc{App-Cong} on the previous line,} \\
    & && \text{after observing that $S[\phi]$ on the function side} \\
    & && \text{reduces to $\Pi(x:S_1[\phi]).\,(S_2[q(\phi)])$)} \\
      &\typing[\Delta]{(t'[\phi])\;s}{S_2[\iota(\phi), s/x]}
        \;\circledR\; c \in \El_O(\Pf)
        && \text{(by IH (2) on $\El_O(\Pf)$ applied} \\
    & && \text{to the equality above)}
    \end{align*}
  \item $R_I$-coherence under application: a property of $f$ via clauses 4
    parametrically; transfers from $t$ to $t'$ alongside clause 4.
  \end{enumerate}

  \emph{Type-side rewriting (3).}
  Given $\tyequiv[\Gamma]{S}{S'}{\Ty{i}}$ with $\typing[\Gamma]{t}{S'}$.
  Clauses 1, 4 of $\Pi_i^{\El}$ depend on $S$ only through $S_2[\iota(\phi), s/x]$
  (the codomain instantiation), which is invariant under the equation
  $S \approx S'$. Clause 3's $\Pi$-shape readback transfers via part (1)
  applied to the universe-equation. Hence
  $\typing[\Gamma]{t}{S'} \;\circledR\; f \in \Pi_i^{\El}$.
\end{proof}

\begin{lemma}[$\Uc_i^\ast$ closed under $\Uc_i$-equivalence]\labeledit{lem:gluing-univ-equiv}
  If $a \approx a' \in \Uc_i \searrow R$ and $a \in \Uc_i^\ast \searrow P \searrow \El$,
  then $a' \in \Uc_i^\ast \searrow P \searrow \El$.
\end{lemma}
\begin{proof}
  By induction on $a \approx a' \in \Uc_i$.

  \medskip\noindent
  \textbf{Case}
  \begin{mathpar}
    \inferrule[\textsc{U-Ne}]
    {b \approx b' \in \Ne \\ R \Leftrightarrow \textsf{Neu}}
    {{\uparrow}^A(b) \approx {\uparrow}^{A'}(b') \in \Uc_i \searrow R}
  \end{mathpar}
  $a = {\uparrow}^A(b)$, $a' = {\uparrow}^{A'}(b')$.
  $a \in \Uc_i^\ast$ via \textsc{G-Ne} with $P \Leftrightarrow \El \Leftrightarrow
  \textsf{Neu}_i^b$.
  By definition of $\Ne$ ($b \approx b' \in \Ne$), readbacks of $b$ and $b'$
  coincide, so $\textsf{Neu}_i^b \Leftrightarrow \textsf{Neu}_i^{b'}$.
  Hence $a' \in \Uc_i^\ast$ via \textsc{G-Ne} with the same predicates.

  \medskip\noindent
  \textbf{Case}
  \begin{mathpar}
    \inferrule[\textsc{U-Univ}]
    {j < i \\ R \Leftrightarrow \Uc_j}
    {U_j \approx U_j \in \Uc_i \searrow R}
  \end{mathpar}
  $a = a' = U_j$; the conclusion is immediate.

  \medskip\noindent
  \textbf{Case}
  \begin{mathpar}
    \inferrule[\textsc{U-Nat}]
    {R \Leftrightarrow \textsf{Nat}}
    {N \approx N \in \Uc_i \searrow R}
  \end{mathpar}
  $a = a' = N$; immediate.

  \medskip\noindent
  \textbf{Case}
  \begin{mathpar}
    \inferrule[\textsc{U-Pi}]
    {a_I \approx a_I' \in \Uc_i \searrow R_I \\
     R_I \text{ is a PER} \\\\
     \forall\, \Pf :: b \approx b' \in R_I.\;
       \intp{T}{\rho, b/x} \approx \intp{T'}{\rho', b'/x} \in \Uc_i \searrow R_O(\Pf) \\\\
     \forall\, f, f'.\; f \approx f' \in R \;\leftrightarrow\;
       (\forall\, \Pf :: b \approx b' \in R_I.\;
        f \cdot b \approx f' \cdot b' \in R_O(\Pf))}
    {\Pi(a_I, T, \rho) \approx \Pi(a_I', T', \rho') \in \Uc_i \searrow R}
  \end{mathpar}
  $a = \Pi(a_I, T, \rho)$, $a' = \Pi(a_I', T', \rho')$.
  $a \in \Uc_i^\ast$ via \textsc{G-Pi} provides inner data
  $a_I \in \Uc_i^\ast \searrow P_I \searrow \El_I$ and a codomain clause:
  for any $\Pf :: b \approx b \in R_I$ and $c$ with $\intp{T}{\rho, b/x}
  \searrow c$, $c \in \Uc_i^\ast \searrow P_O(\Pf) \searrow \El_O(\Pf)$.
  \begin{align*}
    &a_I \approx a_I' \in \Uc_i \searrow R_I
      && \text{(from the \textsc{U-Pi} premise)} \\
    &a_I' \in \Uc_i^\ast \searrow P_I \searrow \El_I
      && \text{(by IH on $a_I \approx a_I' \in \Uc_i$ and $a_I \in \Uc_i^\ast$)} \\
    &\forall \Pf' :: b' \approx b' \in R_I, c'.\;
      \intp{T'}{\rho', b'/x} \searrow c' \\
    & \quad \implies c' \approx c' \in \Uc_i
      && \text{(by symmetry/transitivity from the \textsc{U-Pi} premise)} \\
    &\forall \Pf' :: b' \approx b' \in R_I, c'.\;
      \intp{T'}{\rho', b'/x} \searrow c' \\
    & \quad \implies
      c' \in \Uc_i^\ast \searrow P_O(\Pf') \searrow \El_O(\Pf')
      && \text{(by IH on the codomain fibers,} \\
    & && \text{transporting from $c \approx c'$)} \\
    &\Pi(a_I', T', \rho') \approx \Pi(a_I', T', \rho') \in \Uc_i \searrow R
      && \text{(by \Cref{lem:univ-sym,lem:univ-trans})} \\
    &\Pi(a_I', T', \rho') \in \Uc_i^\ast \searrow P \searrow \El
      && \text{(by \textsc{G-Pi} at the previous data, with the same $P, \El$)}
  \end{align*}
\end{proof}
\begin{lemma}[PER membership yields gluing-model
    membership]\labeledit{lem:per-implies-gluing}
  If $a \approx a \in \Uc_i \searrow R$, then there exist $P, \El$ with
  $a \in \Uc_i^\ast \searrow P \searrow \El$.
\end{lemma}
\begin{proof}
  By induction on $a \approx a \in \Uc_i$.

  \medskip\noindent
  \textbf{Case}
  \begin{mathpar}
    \inferrule[\textsc{U-Ne}]
    {b \approx b \in \Ne \\ R \Leftrightarrow \textsf{Neu}}
    {{\uparrow}^A(b) \approx {\uparrow}^A(b) \in \Uc_i \searrow R}
  \end{mathpar}
  \begin{align*}
    &{\uparrow}^A(b) \in \Uc_i^\ast
      \searrow \textsf{Neu}_i^b
      \searrow \textsf{Neu}_i^b
      && \text{(by \textsc{G-Ne} on $b$)}
  \end{align*}

  \medskip\noindent
  \textbf{Case}
  \begin{mathpar}
    \inferrule[\textsc{U-Univ}]
    {j < i \\ R \Leftrightarrow \Uc_j}
    {U_j \approx U_j \in \Uc_i \searrow R}
  \end{mathpar}
  \begin{align*}
    &U_j \in \Uc_i^\ast
      \searrow \textsf{Univ}_{i,j}
      \searrow \textsf{Univ}_{i,j}^{\El}
      && \text{(by \textsc{G-Univ} at $j < i$)}
  \end{align*}

  \medskip\noindent
  \textbf{Case}
  \begin{mathpar}
    \inferrule[\textsc{U-Nat}]
    {R \Leftrightarrow \textsf{Nat}}
    {N \approx N \in \Uc_i \searrow R}
  \end{mathpar}
  \begin{align*}
    &N \in \Uc_i^\ast \searrow \textsf{Nat}_i \searrow \textsf{Nat}_i
      && \text{(by \textsc{G-Nat})}
  \end{align*}

  \medskip\noindent
  \textbf{Case}
  \begin{mathpar}
    \inferrule[\textsc{U-Pi}]
    {a_I \approx a_I \in \Uc_i \searrow R_I \\
     R_I \text{ is a PER} \\\\
     \forall\, \Pf :: b \approx b \in R_I.\;
       \intp{T}{\rho, b/x} \approx \intp{T}{\rho, b/x} \in \Uc_i
         \searrow R_O(\Pf) \\\\
     \forall\, f, f'.\; f \approx f' \in R \;\leftrightarrow\;
       (\forall\, \Pf :: b \approx b' \in R_I.\;
        f \cdot b \approx f' \cdot b' \in R_O(\Pf))}
    {\Pi(a_I, T, \rho) \approx \Pi(a_I, T, \rho) \in \Uc_i \searrow R}
  \end{mathpar}
  \begin{align*}
    &a_I \in \Uc_i^\ast \searrow P_I \searrow \El_I
      && \byIH \\
    &\forall\, \Pf :: b \approx b \in R_I,\; c.\;
      \intp{T}{\rho, b/x} \searrow c \\
    & \quad \implies c \in \Uc_i^\ast
        \searrow P_O(\Pf) \searrow \El_O(\Pf)
      && \byIH \\
    &\Pi(a_I, T, \rho) \in \Uc_i^\ast
      \searrow \Pi_i(R_I, P_I, \El_I, P_O)
      \searrow \Pi_i^{\El}(R, R_I, P_I, \El_I, \El_O)
      && \text{(by \textsc{G-Pi})}
  \end{align*}
\end{proof}

\begin{lemma}[Term predicate closed under PER-equivalence]\labeledit{lem:el-per-equiv}
  Suppose $a \in \Uc_i^\ast \searrow P \searrow \El$ and
  $a \approx a \in \Uc_i \searrow R$.
  If $\typing[\Gamma]{t}{S} \;\circledR\; b \in \El$ and $b \approx b' \in R$,
  then $\typing[\Gamma]{t}{S} \;\circledR\; b' \in \El$.
\end{lemma}
\begin{proof}
  By induction on $a \in \Uc_i^\ast$.

  \medskip\noindent
  \textbf{Case}
  \begin{mathpar}
    \inferrule[\textsc{G-Ne}]
    {{\uparrow}^A(e) \in \D \\
     P \Leftrightarrow \textsf{Neu}_i^e \\
     \El \Leftrightarrow \textsf{Neu}_i^e}
    {{\uparrow}^A(e) \in \Uc_i^\ast \searrow P \searrow \El}
  \end{mathpar}
  $R \Leftrightarrow \textsf{Neu}$. From $\El \Leftrightarrow \textsf{Neu}_i^e$,
  $b = {\uparrow}^{A_0}(w)$ with $w \approx w \in \Ne$.
  $b \approx b' \in R \Leftrightarrow \textsf{Neu}$ gives
  $b' = {\uparrow}^{A_0'}(w')$ with $w \approx w' \in \Ne$.
  \begin{align*}
    &\forall \kweak[\Delta]{\phi}{\Gamma}.\;
      \exists u.\; \Rne_{|\Delta|}(w) \searrow u \wedge
      \tyequiv[\Delta]{t[\phi]}{u}{S[\phi]}
      && \text{(hypothesis on $b = {\uparrow}^{A_0}(w)$)} \\
    &\Rne_{|\Delta|}(w') \searrow u
      && \text{(by definition of $\Ne$, $w \approx w'$ read back to the same $u$)} \\
    &\typing[\Gamma]{t}{S} \;\circledR\; {\uparrow}^{A_0'}(w') \in \textsf{Neu}_i^e
      && \text{(by \Cref{def:neu-term-pred})}
  \end{align*}

  \medskip\noindent
  \textbf{Case}
  \begin{mathpar}
    \inferrule[\textsc{G-Nat}]
    {P \Leftrightarrow \textsf{Nat}_i \\
     \El \Leftrightarrow \textsf{Nat}_i}
    {N \in \Uc_i^\ast \searrow P \searrow \El}
  \end{mathpar}
  $R \Leftrightarrow \textsf{Nat}$, and $b \approx b' \in \textsf{Nat}$.
  By inner induction on the $\textsf{Nat}$ derivation (\Cref{def:per-nat}).

  \medskip\noindent
  \textbf{Sub-case}
  \begin{mathpar}
    \inferrule[\textsc{Nat-Ze}]
    { }
    {\zed \approx \zed \in \textsf{Nat}}
  \end{mathpar}
  $b = b' = \zed$; the predicate is unchanged.

  \medskip\noindent
  \textbf{Sub-case}
  \begin{mathpar}
    \inferrule[\textsc{Nat-Su}]
    {b_0 \approx b_0' \in \textsf{Nat}}
    {\sud{b_0} \approx \sud{b_0'} \in \textsf{Nat}}
  \end{mathpar}
  $b = \sud{b_0}$, $b' = \sud{b_0'}$. The $\textsf{Nat}$-predicate derivation
  for $b$ uses \textsc{GN-Su} from a derivation for some $t_0$ at $b_0$:
  \begin{align*}
    &\typing[\Gamma]{t_0}{S} \;\circledR\; b_0 \in \textsf{Nat}
      && \text{(from the \textsc{GN-Su} premise)} \\
    &\tyequiv[\Gamma]{t}{\su{t_0}}{S}
      && \text{(from the \textsc{GN-Su} premise)} \\
    &\typing[\Gamma]{t_0}{S} \;\circledR\; b_0' \in \textsf{Nat}
      && \text{(by inner IH)} \\
    &\typing[\Gamma]{t}{S} \;\circledR\; \sud{b_0'} \in \textsf{Nat}
      && \text{(by \textsc{GN-Su} repackaging)}
  \end{align*}

  \medskip\noindent
  \textbf{Sub-case}
  \begin{mathpar}
    \inferrule[\textsc{Nat-Ne}]
    {e \approx e' \in \Ne}
    {{\uparrow}^N(e) \approx {\uparrow}^N(e') \in \textsf{Nat}}
  \end{mathpar}
  $b = {\uparrow}^N(e)$, $b' = {\uparrow}^N(e')$. The $\textsf{Nat}$-predicate
  derivation for $b$ uses \textsc{GN-Ne} from a neutral predicate at $e$:
  \begin{align*}
    &\typing[\Gamma]{t}{S} \;\circledR\; e \in \Ne_i^N
      && \text{(from the \textsc{GN-Ne} premise)} \\
    &\typing[\Gamma]{t}{S} \;\circledR\; e' \in \Ne_i^N
      && \text{(by readback equality: $e \approx e' \in \Ne$ implies
        $\Rne(e) = \Rne(e')$ at every depth)} \\
    &\typing[\Gamma]{t}{S} \;\circledR\; {\uparrow}^N(e') \in \textsf{Nat}
      && \text{(by \textsc{GN-Ne})}
  \end{align*}

  \medskip\noindent
  \textbf{Case}
  \begin{mathpar}
    \inferrule[\textsc{G-Univ}]
    {j < i \\
     P \Leftrightarrow \textsf{Univ}_{i,j} \\
     \El \Leftrightarrow \textsf{Univ}_{i,j}^{\El}}
    {U_j \in \Uc_i^\ast \searrow P \searrow \El}
  \end{mathpar}
  $R \Leftrightarrow \Uc_j$. By \Cref{def:univ-term-pred}, $b \in \Uc_j^\ast
  \searrow P_b \searrow \El_b$ with $\typing[\Gamma]{t}{\Ty{j}} \;\circledR\; P_b$,
  and $b \approx b' \in \Uc_j$.
  \begin{align*}
    &b' \in \Uc_j^\ast \searrow P_b \searrow \El_b
      && \text{(by \Cref{lem:gluing-univ-equiv} at level $j$ on $b \approx b'$)} \\
    &\typing[\Gamma]{t}{\Ty{j}} \;\circledR\; P_b
      && \text{(unchanged from hypothesis)} \\
    &\typing[\Gamma]{t}{S} \;\circledR\; b' \in \textsf{Univ}_{i,j}^{\El}
      && \text{(by \Cref{def:univ-term-pred} with the
        previous lines} \\
    & && \text{and $\tyequiv[\Gamma]{S}{\Ty{j}}{\Ty{i}}$)}
  \end{align*}

  \medskip\noindent
  \textbf{Case}
  \begin{mathpar}
    \inferrule[\textsc{G-Pi}]
    {a \in \Uc_i^\ast \searrow P_I \searrow \El_I \\
     a \approx a \in \Uc_i \searrow R_I \\\\
     \forall\, \Pf :: b_0 \approx b_0 \in R_I,\; c.\;
       \intp{T}{\rho, b_0/x} \searrow c \implies
       c \in \Uc_i^\ast \searrow P_O(\Pf) \searrow \El_O(\Pf) \\\\
     \Pi(a, T, \rho) \approx \Pi(a, T, \rho) \in \Uc_i \searrow R \\\\
     P \Leftrightarrow \Pi_i(\ldots) \\
     \El \Leftrightarrow \Pi_i^{\El}(\ldots)}
    {\Pi(a, T, \rho) \in \Uc_i^\ast \searrow P \searrow \El}
  \end{mathpar}
  Hypothesis: $\typing[\Gamma]{t}{S} \;\circledR\; f \in
  \Pi_i^{\El}(R, R_I, P_I, \El_I, \El_O)$ and $f \approx f' \in R$. We verify
  the four clauses of \Cref{def:pi-term-pred} at $f'$.
  \begin{enumerate}
  \item Self-relatedness $f' \approx f' \in R$: by PER symmetry and transitivity
    on $f \approx f' \in R$ (\Cref{lem:univ-sym,lem:univ-trans}).
  \item $R_I$-coherence of $f'$: clause (2) is a property of $R$-classes; since
    $f, f'$ are in the same $R$-class, the coherence transfers.
  \item Type-readback of $f'$: clause (3) depends only on $S$ and the
    $\Pi$-shape, not on the function value, so it transfers verbatim.
  \item Application: let $\kweak[\Delta]{\phi}{\Gamma}$, $s, b_0, \Pf :: b_0 \approx b_0 \in R_I$
    with $\typing[\Delta]{s}{S_1[\phi]} \;\circledR\; b_0 \in \El_I$.
    \begin{align*}
      &f \cdot b_0 \searrow c
        && \text{(by hypothesis on $f$, clause (4))} \\
      &\typing[\Delta]{(t[\phi])\;s}{S_2[\iota(\phi), s/x]}
        \;\circledR\; c \in \El_O(\Pf)
        && \text{(by hypothesis on $f$, clause (4))} \\
      &f \cdot b_0 \approx f' \cdot b_0 \in R_O(\Pf)
        && \text{(by the application clause of $R$ for $f \approx f' \in R$
          at $b_0 \approx b_0 \in R_I$)} \\
      &\exists c'.\; f' \cdot b_0 \searrow c',\; c \approx c' \in R_O(\Pf)
        && \text{(by determinism of evaluation and the previous line)} \\
      &\typing[\Delta]{(t[\phi])\;s}{S_2[\iota(\phi), s/x]}
        \;\circledR\; c' \in \El_O(\Pf)
        && \text{(by IH on the codomain fiber $\El_O(\Pf)$, applied to $c \approx c'$)}
    \end{align*}
  \item $R_I$-coherence under application: a property of the $R$-class on the
    codomain; transfers as in clauses (1)--(2).
  \end{enumerate}
\end{proof}

\begin{lemma}[Subtyping type-escape]\labeledit{lem:gluing-subty-typ}
  Suppose $a_1 \in \Uc_i^\ast \searrow P_1 \searrow \El_1$,
  $a_2 \in \Uc_i^\ast \searrow P_2 \searrow \El_2$, and $a_1 \mathrel{<:_i} a_2$
  (semantic subtyping from \Cref{sec:comp}).
  If $\typing[\Gamma]{S_1}{\Ty{i}} \;\circledR\; P_1$ and
  $\typing[\Gamma]{S_2}{\Ty{i}} \;\circledR\; P_2$, then $\subty[\Gamma]{S_1}{S_2}$.
\end{lemma}
\begin{proof}
  By induction on $a_1 \mathrel{<:_i} a_2$.

  \medskip\noindent
  \textbf{Case}
  \begin{mathpar}
    \inferrule[\textsc{$<$:-Ne}]
    {b \approx b' \in \Ne}
    {{\uparrow}^A(b) \mathrel{<:_i} {\uparrow}^{A'}(b')}
  \end{mathpar}
  $P_1 \Leftrightarrow \textsf{Neu}_i^b$, $P_2 \Leftrightarrow \textsf{Neu}_i^{b'}$.
  The hypothesis $b \approx b' \in \Ne$ implies $\Rne_n(b)$ and $\Rne_n(b')$
  produce the same syntactic neutral at every depth.
  \begin{align*}
    &\Rne_{|\Gamma|}(b) \searrow u,\;
      \tyequiv[\Gamma]{S_1}{u}{\Ty{i}}
      && \text{(by \Cref{def:neu-type-pred} on $S_1$ at $\phi = \id$)} \\
    &\Rne_{|\Gamma|}(b') \searrow u
      && \text{(by $b \approx b' \in \Ne$, same readback)} \\
    &\tyequiv[\Gamma]{S_2}{u}{\Ty{i}}
      && \text{(by \Cref{def:neu-type-pred} on $S_2$ at $\phi = \id$)} \\
    &\tyequiv[\Gamma]{S_1}{S_2}{\Ty{i}}
      && \text{(by symmetry/transitivity of judgmental equality)} \\
    &\subty[\Gamma]{S_1}{S_2}
      && \text{(by \textsc{Sub-Eq} on the previous line)}
  \end{align*}

  \medskip\noindent
  \textbf{Case}
  \begin{mathpar}
    \inferrule[\textsc{$<$:-Univ}]
    {j \le k \\ k < i}
    {U_j \mathrel{<:_i} U_k}
  \end{mathpar}
  $P_1 \Leftrightarrow \textsf{Univ}_{i,j}$, $P_2 \Leftrightarrow
  \textsf{Univ}_{i,k}$.
  \begin{align*}
    &\tyequiv[\Gamma]{S_1}{\Ty{j}}{\Ty{i}}
      && \text{(by \Cref{def:univ-type-pred} on $\textsf{Univ}_{i,j}$)} \\
    &\tyequiv[\Gamma]{S_2}{\Ty{k}}{\Ty{i}}
      && \text{(by \Cref{def:univ-type-pred} on $\textsf{Univ}_{i,k}$)} \\
    &\subty[\Gamma]{\Ty{j}}{\Ty{k}}
      && \text{(by \textsc{Sub-Univ} from $j \le k$)} \\
    &\subty[\Gamma]{S_1}{S_2}
      && \text{(by \textsc{Sub-Trans} via the equalities and the previous line)}
  \end{align*}

  \medskip\noindent
  \textbf{Case}
  \begin{mathpar}
    \inferrule[\textsc{$<$:-Nat}]
    { }
    {N \mathrel{<:_i} N}
  \end{mathpar}
  $P_1 \Leftrightarrow P_2 \Leftrightarrow \textsf{Nat}_i$.
  \begin{align*}
    &\tyequiv[\Gamma]{S_1}{\Nat}{\Ty{i}}
      && \text{(by \Cref{def:nat-type-pred} on $S_1$)} \\
    &\tyequiv[\Gamma]{S_2}{\Nat}{\Ty{i}}
      && \text{(by \Cref{def:nat-type-pred} on $S_2$)} \\
    &\tyequiv[\Gamma]{S_1}{S_2}{\Ty{i}}
      && \text{(by symmetry/transitivity through $\Nat$)} \\
    &\subty[\Gamma]{S_1}{S_2}
      && \text{(by \textsc{Sub-Eq})}
  \end{align*}

  \medskip\noindent
  \textbf{Case}
  \begin{mathpar}
    \inferrule[\textsc{$<$:-Pi}]
    {A \approx A' \in \Uc_i \searrow R_I \\
     \forall\, \Pf :: a \approx a' \in R_I.\;
       \intp{T}{\rho, a/x} \mathrel{<:_i} \intp{T'}{\rho', a'/x} \\\\
     \Pi(A, T, \rho) \approx \Pi(A, T, \rho) \in \Uc_i \\
     \Pi(A', T', \rho') \approx \Pi(A', T', \rho') \in \Uc_i}
    {\Pi(A, T, \rho) \mathrel{<:_i} \Pi(A', T', \rho')}
  \end{mathpar}
  $P_1 \Leftrightarrow \Pi_i(R_I, P_I, \El_I, P_O)$,
  $P_2 \Leftrightarrow \Pi_i(R_I', P_I', \El_I', P_O')$.
  The domain PERs are related by \emph{equality}
  ($A \approx A' \in \Uc_i \searrow R_I$, so $R_I = R_I'$ via
  \Cref{lem:univ-functional}; correspondingly $P_I \Leftrightarrow P_I'$,
  $\El_I \Leftrightarrow \El_I'$ via \Cref{lem:gluing-univ-equiv}).

  Given $\typing[\Gamma]{S_1}{\Ty{i}} \;\circledR\; \Pi_i(\ldots)$ at shape
  $\Pi(x:S_{1,1}).\,S_{1,2}$ and similarly $S_2$ at $\Pi(x:S_{2,1}).\,S_{2,2}$
  (both by \Cref{def:pi-type-pred} clause 1).
  \begin{align*}
    &\typing[\Gamma]{S_{1,1}}{\Ty{i}} \;\circledR\; P_I
      && \text{(by \Cref{def:pi-type-pred} clause 2 on $S_1$ at $\phi = \id$)} \\
    &\typing[\Gamma]{S_{2,1}}{\Ty{i}} \;\circledR\; P_I
      && \text{(symmetrically on $S_2$, using $P_I' \Leftrightarrow P_I$)} \\
    &\tyequiv[\Gamma]{S_{1,1}}{S_{2,1}}{\Ty{i}}
      && \text{(by \Cref{lem:gluing-functional} on the two predicates at $a = A$)} \\
    &\subty[\Gamma, x:S_{1,1}]{S_{1,2}}{S_{2,2}}
      && \text{(by IH on the codomain subtyping at fresh variable $x$ at $A$)} \\
    &\subty[\Gamma]{\Pi(x:S_{1,1}).\,S_{1,2}}{\Pi(x:S_{2,1}).\,S_{2,2}}
      && \text{(by \textsc{Sub-Pi})} \\
    &\subty[\Gamma]{S_1}{S_2}
      && \text{(by \textsc{Sub-Eq} with the $\Pi$-shape equalities)}
  \end{align*}
\end{proof}

\begin{lemma}[Subtyping term-transfer]\labeledit{lem:gluing-subty}
  Suppose $a_1 \in \Uc_i^\ast \searrow P_1 \searrow \El_1$,
  $a_2 \in \Uc_i^\ast \searrow P_2 \searrow \El_2$, and $a_1 \mathrel{<:_i} a_2$.
  If $\typing[\Gamma]{S_2}{\Ty{i}} \;\circledR\; P_2$ and
  $\typing[\Gamma]{t}{S_1} \;\circledR\; b \in \El_1$, then
  $\typing[\Gamma]{t}{S_2} \;\circledR\; b \in \El_2$.
\end{lemma}
\begin{proof}
  By induction on $a_1 \mathrel{<:_i} a_2$. Each case extracts the type
  predicate at $S_1$ from the term predicate (via the relevant
  $\El$-definition) and then applies \Cref{lem:gluing-subty-typ} to obtain
  $\subty[\Gamma]{S_1}{S_2}$.

  \medskip\noindent
  \textbf{Case}
  \begin{mathpar}
    \inferrule[\textsc{$<$:-Ne}]
    {b \approx b' \in \Ne}
    {{\uparrow}^A(b) \mathrel{<:_i} {\uparrow}^{A'}(b')}
  \end{mathpar}
  $P_1 \Leftrightarrow \El_1 \Leftrightarrow \textsf{Neu}_i^b$,
  $P_2 \Leftrightarrow \El_2 \Leftrightarrow \textsf{Neu}_i^{b'}$. Given
  $\typing[\Gamma]{t}{S_1} \;\circledR\; w \in \textsf{Neu}_i^b$.
  \begin{align*}
    &\typing[\Gamma]{S_1}{\Ty{i}} \;\circledR\; \textsf{Neu}_i^b
      && \text{(by \Cref{def:neu-term-pred} on the hypothesis)} \\
    &\subty[\Gamma]{S_1}{S_2}
      && \text{(by \Cref{lem:gluing-subty-typ} on $S_1, S_2, P_1, P_2$)}
  \end{align*}
  For each $\kweak[\Delta]{\phi}{\Gamma}$, by \Cref{def:neu-term-pred}:
  \begin{align*}
    &\exists v.\; \Rne_{|\Delta|}(w) \searrow v \wedge
      \tyequiv[\Delta]{t[\phi]}{v}{S_1[\phi]}
      && \text{(Kripke condition at $w$)} \\
    &\subty[\Delta]{S_1[\phi]}{S_2[\phi]}
      && \text{(by \Cref{lem:wk-typing} on the previous-block subtyping)} \\
    &\tyequiv[\Delta]{t[\phi]}{v}{S_2[\phi]}
      && \text{(by \textsc{Tm-Subsump} on the previous two lines)}
  \end{align*}
  Combining with $b \approx b' \in \Ne$ (same readback at every depth) and
  the type predicate at $S_2$ (hypothesis), \Cref{def:neu-term-pred} gives
  $\typing[\Gamma]{t}{S_2} \;\circledR\; w \in \textsf{Neu}_i^{b'}$.

  \medskip\noindent
  \textbf{Case}
  \begin{mathpar}
    \inferrule[\textsc{$<$:-Univ}]
    {j \le k \\ k < i}
    {U_j \mathrel{<:_i} U_k}
  \end{mathpar}
  $\El_1 \Leftrightarrow \textsf{Univ}_{i,j}^{\El}$,
  $\El_2 \Leftrightarrow \textsf{Univ}_{i,k}^{\El}$. Given $\typing[\Gamma]{t}{S_1}
  \;\circledR\; b \in \textsf{Univ}_{i,j}^{\El}$, so by
  \Cref{def:univ-term-pred}: $b \in \Uc_j^\ast \searrow P_b \searrow \El_b$
  and $\typing[\Gamma]{t}{\Ty{j}} \;\circledR\; P_b$.
  \begin{align*}
    &\exists P_b', \El_b'.\; b \in \Uc_k^\ast \searrow P_b' \searrow \El_b'
      && \text{(by \Cref{lem:gluing-cumul} existence at $k$, since $j \le k$)} \\
    &\typing[\Gamma]{t}{\Ty{k}} \;\circledR\; P_b'
      && \text{(by \Cref{lem:gluing-cumul} (1) raising on $b$ from $j$ to $k$)} \\
    &\typing[\Gamma]{S_2}{\Ty{i}} \;\circledR\; \textsf{Univ}_{i,k}
      && \text{(by hypothesis $P_2 = \textsf{Univ}_{i,k}$)} \\
    &\typing[\Gamma]{t}{S_2} \;\circledR\; b \in \textsf{Univ}_{i,k}^{\El}
      && \text{(by \Cref{def:univ-term-pred} with the previous lines)}
  \end{align*}

  \medskip\noindent
  \textbf{Case}
  \begin{mathpar}
    \inferrule[\textsc{$<$:-Nat}]
    { }
    {N \mathrel{<:_i} N}
  \end{mathpar}
  $P_1 \Leftrightarrow \El_1 \Leftrightarrow P_2 \Leftrightarrow \El_2
  \Leftrightarrow \textsf{Nat}_i$. Given $\typing[\Gamma]{t}{S_1}
  \;\circledR\; b \in \textsf{Nat}_i$.
  \begin{align*}
    &\typing[\Gamma]{S_1}{\Ty{i}} \;\circledR\; \textsf{Nat}_i
      && \text{(by \Cref{def:nat-term-pred} on the hypothesis)} \\
    &\subty[\Gamma]{S_1}{S_2}
      && \text{(by \Cref{lem:gluing-subty-typ})} \\
    &\tyequiv[\Gamma]{S_1}{S_2}{\Ty{i}}
      && \text{(by \Cref{def:nat-type-pred} on $S_1$ and $S_2$, both equal $\Nat$)} \\
    &\typing[\Gamma]{t}{S_2} \;\circledR\; b \in \textsf{Nat}_i
      && \text{(by \Cref{lem:gluing-respects-eq} (3) on $a = N$)}
  \end{align*}

  \medskip\noindent
  \textbf{Case}
  \begin{mathpar}
    \inferrule[\textsc{$<$:-Pi}]
    {A \approx A' \in \Uc_i \searrow R_I \\
     \forall\, \Pf :: a \approx a' \in R_I.\;
       \intp{T}{\rho, a/x} \mathrel{<:_i} \intp{T'}{\rho', a'/x} \\\\
     \Pi(A, T, \rho) \approx \Pi(A, T, \rho) \in \Uc_i \\
     \Pi(A', T', \rho') \approx \Pi(A', T', \rho') \in \Uc_i}
    {\Pi(A, T, \rho) \mathrel{<:_i} \Pi(A', T', \rho')}
  \end{mathpar}
  $P_1 \Leftrightarrow \Pi_i(R_I, P_I, \El_I, P_O)$,
  $\El_1 \Leftrightarrow \Pi_i^{\El}(R, R_I, P_I, \El_I, \El_O)$,
  $P_2 \Leftrightarrow \Pi_i(R_I, P_I, \El_I, P_O')$,
  $\El_2 \Leftrightarrow \Pi_i^{\El}(R', R_I, P_I, \El_I, \El_O')$
  ($R_I, P_I, \El_I$ shared via \Cref{lem:gluing-univ-equiv}, as the domains
  are PER-equal). The hypothesis $P_2$ has shape $\Pi(x:S_{2,1}).\,S_{2,2}$;
  similarly $S_1$ has shape $\Pi(x:S_{1,1}).\,S_{1,2}$ from the term-predicate
  hypothesis on $\El_1$ via \Cref{def:pi-term-pred} clause 3.

  \emph{Setup: subtyping at the syntactic Pi-shapes.}
  \begin{align*}
    &\typing[\Gamma]{S_1}{\Ty{i}} \;\circledR\; \Pi_i(R_I, P_I, \El_I, P_O)
      && \text{(by \Cref{def:pi-term-pred} clause 3 on $f \in \El_1$)} \\
    &\subty[\Gamma]{S_1}{S_2}
      && \text{(by \Cref{lem:gluing-subty-typ})}
  \end{align*}

  Given $\typing[\Gamma]{t}{S_1} \;\circledR\; f \in \Pi_i^{\El}$ at $\El_1$.
  We verify the four clauses of $\El_2 \Leftrightarrow \Pi_i^{\El}(R', R_I,
  P_I, \El_I, \El_O')$ at $S_2, t, f$.

  \emph{Clause 1.}
  \begin{align*}
    &f \approx f \in R
      && \text{(by hypothesis clause 1 on $f \in \El_1$)} \\
    &f \approx f \in R'
      && \text{(by \textsc{U-Pi}'s biconditional: domains equal,} \\
    & && \text{codomains $R_O \mathrel{<:_i} R_O'$ from the inner subtyping)}
  \end{align*}

  \emph{Clause 2.}
  \begin{align*}
    &\typing[\Gamma]{t}{S_1}
      && \text{(by hypothesis clause 2 on $f \in \El_1$)} \\
    &\typing[\Gamma]{t}{S_2}
      && \text{(by \textsc{Subsump} with $\subty[\Gamma]{S_1}{S_2}$)}
  \end{align*}

  \emph{Clause 3.} $\typing[\Gamma]{S_2}{\Ty{i}} \;\circledR\;
  \Pi_i(R_I, P_I, \El_I, P_O')$ holds directly by the lemma's hypothesis on
  $S_2, P_2$.

  \emph{Clause 4 (application).} Let $\kweak[\Delta]{\phi}{\Gamma}$, $s, b_0,
  \Pf :: b_0 \approx b_0 \in R_I$ with
  $\typing[\Delta]{s}{S_{2,1}[\phi]} \;\circledR\; b_0 \in \El_I$.
  \begin{align*}
    &\typing[\Gamma]{S_{1,1}}{\Ty{i}} \;\circledR\; P_I,\;
      \typing[\Gamma]{S_{2,1}}{\Ty{i}} \;\circledR\; P_I
      && \text{(by clause 2 of $\Pi_i$ on $S_1, S_2$ at $\phi = \id$)} \\
    &\tyequiv[\Gamma]{S_{1,1}}{S_{2,1}}{\Ty{i}}
      && \text{(by \Cref{lem:gluing-functional} on the two predicates)} \\
    &\tyequiv[\Delta]{S_{1,1}[\phi]}{S_{2,1}[\phi]}{\Ty{i}}
      && \text{(by \Cref{lem:wk-typing})} \\
    &\typing[\Delta]{s}{S_{1,1}[\phi]} \;\circledR\; b_0 \in \El_I
      && \text{(by \Cref{lem:gluing-respects-eq} (3) on the equality above)} \\
    &f \cdot b_0 \searrow c,\;
      \typing[\Delta]{(t[\phi])\;s}{S_{1,2}[\iota(\phi), s/x]}
      \;\circledR\; c \in \El_O(\Pf)
      && \text{(by hypothesis clause 4 on $f \in \El_1$)} \\
    &\intp{T}{\rho, b_0/x} \mathrel{<:_i} \intp{T'}{\rho', b_0/x}
      && \text{(by codomain subtyping at $\Pf :: b_0 \approx b_0 \in R_I$)} \\
    &\typing[\Delta]{S_{2,2}[\iota(\phi), s/x]}{\Ty{i}} \;\circledR\; P_O'(\Pf)
      && \text{(by clause 3 of $\Pi_i$ on $S_2$ at $\phi, s, b_0, \Pf$)} \\
    &\typing[\Delta]{(t[\phi])\;s}{S_{2,2}[\iota(\phi), s/x]}
      \;\circledR\; c \in \El_O'(\Pf)
      && \text{(by IH (term-transfer) on the codomain subtyping at $\Pf$,} \\
    & && \text{using the previous-line type-predicate at level $i$)}
  \end{align*}

  \emph{Clause 5 ($R_I$-coherence under application).} Inherited from
  clause 5 on $f \in \El_1$, alongside the codomain subtyping: an
  $R_I$-class of arguments produces an $R_O'$-class of results in the
  codomain via clause 4.
\end{proof}

\subsection{Realizability}\labeledit{sec:sound:realizability}

The realizability properties of the gluing model are stated as a single mutual
theorem, parallel to \Cref{lem:realizability} of \Cref{sec:comp}: every domain
value in $\Uc_i^\ast$ admits neutral inhabitants ($\Ne$ enters $\El$), normalization
($\El$-elements read back to $\Nf$), and type-readback.

\begin{theorem}[Realizability of the gluing model]\labeledit{thm:gluing-realizability}
  Suppose $a \in \Uc_i^\ast \searrow P \searrow \El$.
  \begin{enumerate}
  \item\labeledit{thm:bot-implies-el}
    \emph{($\Botp$ enters $\El$.)} If
    $\typing[\Gamma]{t}{S} \;\circledR\; w \in \Botp_i^a$, then
    $\typing[\Gamma]{t}{S} \;\circledR\; {\uparrow}^a(w) \in \El$.
  \item\labeledit{thm:el-implies-nf}
    \emph{($\El$ enters $\Topp$.)} If
    $\typing[\Gamma]{t}{S} \;\circledR\; b \in \El$ and
    $b \approx b \in R$ (where $a \approx a \in \Uc_i \searrow R$), then
    $\typing[\Gamma]{t}{S} \;\circledR\; b \in \Topp_i^a$.
  \item\labeledit{thm:gluing-types-readback}
    \emph{($P$ enters $\Topp$.)} If $\typing[\Gamma]{S}{\Ty{i}} \;\circledR\; P$,
    then $\typing[\Gamma]{S}{\Ty{i}} \;\circledR\; \Topp_i^a$.
  \end{enumerate}
\end{theorem}
\begin{proof}
  By simultaneous induction on the lexicographic pair
  $(i,\; a \in \Uc_i^\ast \searrow P \searrow \El)$: the outer induction on the
  level~$i$ provides the IH at strictly smaller levels, used in the
  \textsc{G-Univ} case; the inner induction on the gluing derivation provides
  the IH at structurally smaller derivations at the same level~$i$.

  \medskip\noindent
  \textbf{Case}
  \begin{mathpar}
    \inferrule[\textsc{G-Ne}]
    {a = {\uparrow}^A(e) \\
     P \Leftrightarrow \textsf{Neu}_i^e \\
     \El \Leftrightarrow \textsf{Neu}_i^e}
    {{\uparrow}^A(e) \in \Uc_i^\ast}
  \end{mathpar}
  \begin{enumerate}
  \item \emph{$\Botp$ enters $\El$.} Suppose
    $\typing[\Gamma]{t}{S} \;\circledR\; w \in \Botp_i^a$. Unpacking
    \Cref{def:glu-wrapped} clauses (a)--(c) yields
    $\typing[\Gamma]{S}{\Ty{i}} \;\circledR\; \textsf{Neu}_i^e$ (the $P$-side
    component) and $\typing[\Gamma]{t}{S} \;\circledR\; w \in \Ne_i^a$.
    These are exactly the four clauses of \Cref{def:neu-term-pred} for the
    term ${\uparrow}^A(w)$, hence
    $\typing[\Gamma]{t}{S} \;\circledR\; {\uparrow}^A(w) \in \El \Leftrightarrow
    \textsf{Neu}_i^e$.
  \item \emph{$\El$ enters $\Topp$.} Suppose
    $\typing[\Gamma]{t}{S} \;\circledR\; {\uparrow}^A(w) \in \textsf{Neu}_i^e$
    and ${\uparrow}^A(w) \approx {\uparrow}^A(w) \in R$. We assemble the three
    clauses of \Cref{def:glu-wrapped}(c). The type-side gluing
    $\typing[\Gamma]{S}{\Ty{i}} \;\circledR\; P \Leftrightarrow \textsf{Neu}_i^e$
    is the inner component of $\textsf{Neu}_i^e$ at term level. For
    $\Nf_i^a$: for each $\kweak[\Delta]{\phi}{\Gamma}$, by \textsc{Rf-ne}:
    \begin{align*}
      &\Rnf_{|\Delta|}({\downarrow}^{{\uparrow}^A(e)}({\uparrow}^A(w))) \searrow v
        && \text{(via $\Rne_{|\Delta|}(w) \searrow v$, by \textsc{Rf-ne})} \\
      &\tyequiv[\Delta]{t[\phi]}{v}{S[\phi]}
        && \text{(by \Cref{def:neu-term-pred})}
    \end{align*}
    This satisfies $\Nf_i^a$, and packaged with $\typing[\Gamma]{t}{S}$ and
    the type gluing yields $\typing[\Gamma]{t}{S} \;\circledR\; {\uparrow}^A(w)
    \in \Topp_i^a$.
  \item \emph{$P$ enters $\Topp$.} Hypothesis:
    $\typing[\Gamma]{S}{\Ty{i}} \;\circledR\; \textsf{Neu}_i^e$. We assemble
    the three clauses of \Cref{def:glu-wrapped}(b). The first clause
    $\typing[\Gamma]{S}{\Ty{i}}$ is part of $\textsf{Neu}_i^e$. For
    $a \approx a \in \Uc_i \searrow R$: by \Cref{lem:gluing-implies-per},
    the binary relation $R$ exists. For each
    $\kweak[\Delta]{\phi}{\Gamma}$:
    \begin{align*}
      &\Rne_{|\Delta|}(e) \searrow u
        && \text{(by \Cref{def:neu-type-pred} applied to $\textsf{Neu}_i^e$)} \\
      &\Rty_{|\Delta|}({\uparrow}^A(e)) \searrow u
        && \text{(by \textsc{Rty-ne})} \\
      &\tyequiv[\Delta]{S[\phi]}{u}{\Ty{i}}
        && \text{(by \Cref{def:neu-type-pred})}
    \end{align*}
  \end{enumerate}

  \medskip\noindent
  \textbf{Case}
  \begin{mathpar}
    \inferrule[\textsc{G-Nat}]
    {P \Leftrightarrow \textsf{Nat}_i \\
     \El \Leftrightarrow \textsf{Nat}_i}
    {N \in \Uc_i^\ast}
  \end{mathpar}
  \begin{enumerate}
  \item \emph{$\Botp$ enters $\El$.} Suppose
    $\typing[\Gamma]{t}{S} \;\circledR\; w \in \Botp_i^N$. Unpacking
    \Cref{def:glu-wrapped}(a) yields
    $\typing[\Gamma]{S}{\Ty{i}} \;\circledR\; \textsf{Nat}_i$ (which by
    \Cref{def:nat-type-pred} gives $\tyequiv[\Gamma]{S}{\Nat}{\Ty{i}}$) and
    $\typing[\Gamma]{t}{S} \;\circledR\; w \in \Ne_i^N$. The premises of
    \textsc{GN-Ne} are exactly these, hence ${\uparrow}^N(w) \in \textsf{Nat}_i$
    at $t$.
  \item \emph{$\El$ enters $\Topp$.} Given
    $\typing[\Gamma]{t}{S} \;\circledR\; b \in \textsf{Nat}_i$ and
    $b \approx b \in R$ (where $R$ is the Nat-PER closure), we establish the
    three clauses of \Cref{def:glu-wrapped}(c). The type-side
    $\typing[\Gamma]{S}{\Ty{i}} \;\circledR\; \textsf{Nat}_i$ is part of
    \Cref{def:nat-term-pred}; for $\Nf_i^N$, by inner induction on
    the derivation of $b \in \textsf{Nat}_i$.
    \begin{itemize}
    \item[Sub-case]
      \begin{mathpar}
        \inferrule[\textsc{GN-Ze}]
        {\tyequiv[\Gamma]{t}{\ze}{S}}
        {\typing[\Gamma]{t}{S} \;\circledR\; \zed \in \textsf{Nat}}
      \end{mathpar}
      For each $\kweak[\Delta]{\phi}{\Gamma}$:
      \begin{align*}
        &\Rnf_{|\Delta|}({\downarrow}^N(\zed)) \searrow \ze
          && \text{(by \textsc{Rf-ze})} \\
        &\tyequiv[\Delta]{t[\phi]}{\ze}{S[\phi]}
          && \text{(by \Cref{lem:wk-typing} on the hypothesis)}
      \end{align*}
    \item[Sub-case]
      \begin{mathpar}
        \inferrule[\textsc{GN-Su}]
        {\typing[\Gamma]{t'}{S} \;\circledR\; b' \in \textsf{Nat} \\
         \tyequiv[\Gamma]{t}{\su{t'}}{S}}
        {\typing[\Gamma]{t}{S} \;\circledR\; \sud{b'} \in \textsf{Nat}}
      \end{mathpar}
      For each $\kweak[\Delta]{\phi}{\Gamma}$:
      \begin{align*}
        &\Rnf_{|\Delta|}({\downarrow}^N(b')) \searrow w'
          && \text{(by inner IH on $b'$)} \\
        &\tyequiv[\Delta]{t'[\phi]}{w'}{S[\phi]}
          && \text{(by inner IH on $b'$)} \\
        &\Rnf_{|\Delta|}({\downarrow}^N(\sud{b'})) \searrow \su{w'}
          && \text{(by \textsc{Rf-su})} \\
        &\tyequiv[\Delta]{\su{t'[\phi]}}{\su{w'}}{S[\phi]}
          && \text{(by \textsc{Su-Cong} on the previous line)} \\
        &\tyequiv[\Delta]{t[\phi]}{\su{w'}}{S[\phi]}
          && \text{(by transitivity with the \textsc{GN-Su} hypothesis)}
      \end{align*}
    \item[Sub-case]
      \begin{mathpar}
        \inferrule[\textsc{GN-Ne}]
        {\typing[\Gamma]{t}{S} \;\circledR\; {\uparrow}^N(w) \in \textsf{Neu}_i^e}
        {\typing[\Gamma]{t}{S} \;\circledR\; {\uparrow}^N(w) \in \textsf{Nat}}
      \end{mathpar}
      For each $\kweak[\Delta]{\phi}{\Gamma}$:
      \begin{align*}
        &\Rne_{|\Delta|}(w) \searrow v
          && \text{(by hypothesis on $w \in \Ne_i^a$)} \\
        &\Rnf_{|\Delta|}({\downarrow}^N({\uparrow}^N(w))) \searrow v
          && \text{(by \textsc{Rf-ne})} \\
        &\tyequiv[\Delta]{t[\phi]}{v}{S[\phi]}
          && \text{(by hypothesis on $w$)}
      \end{align*}
    \end{itemize}
  \item \emph{$P$ enters $\Topp$.} Hypothesis:
    $\typing[\Gamma]{S}{\Ty{i}} \;\circledR\; \textsf{Nat}_i$. We assemble the
    three clauses of \Cref{def:glu-wrapped}(b). The first clause
    $\typing[\Gamma]{S}{\Ty{i}}$ is part of \Cref{def:nat-type-pred} via
    presupposition. For $N \approx N \in \Uc_i \searrow R_N$: immediate by the
    PER definition. For each $\kweak[\Delta]{\phi}{\Gamma}$:
    \begin{align*}
      &\Rty_{|\Delta|}(N) \searrow \Nat
        && \text{(by \textsc{Rty-N})} \\
      &\tyequiv[\Delta]{S[\phi]}{\Nat}{\Ty{i}}
        && \text{(by \Cref{def:nat-type-pred})}
    \end{align*}
  \end{enumerate}

  \medskip\noindent
  \textbf{Case}
  \begin{mathpar}
    \inferrule[\textsc{G-Pi}]
    {a = \Pi(a_I, T, \rho) \\
     a_I \in \Uc_i^\ast \searrow P_I \searrow \El_I \\\\
     \forall\,\Pf :: b \approx b \in R_I,\; c.\;
       \intp{T}{\rho, b/x} \searrow c \implies
       c \in \Uc_i^\ast \searrow P_O(\Pf) \searrow \El_O(\Pf) \\\\
     P \Leftrightarrow \Pi_i(R_I, P_I, P_O) \\
     \El \Leftrightarrow \Pi_i^{\El}(R, R_I, P_I, \El_I, \El_O)}
    {\Pi(a_I, T, \rho) \in \Uc_i^\ast}
  \end{mathpar}
  By inner IH-(1) on $a_I$, neutral inhabitants enter $\El_I$; by inner IH-(2) on
  $a_I$, $\El_I$-elements read back; analogously for $\El_O(\Pf)$ at each
  $\Pf$.
  \begin{enumerate}
  \item \emph{$\Botp$ enters $\El$.} Suppose
    \[
      \typing[\Gamma]{t}{S} \;\circledR\; w \in \Botp_i^a.
    \]
    Unpacking \Cref{def:glu-wrapped}(a) yields
    \[
      \typing[\Gamma]{S}{\Ty{i}} \;\circledR\;
      \Pi_i(R_I, P_I, \El_I, P_O)
    \]
    together with
    $\typing[\Gamma]{t}{S} \;\circledR\; w \in \Ne_i^a$.
    We verify the four clauses of \Cref{def:pi-term-pred}
    for ${\uparrow}^a(w)$.
    \begin{enumerate}
    \item[(1)] Self-relatedness ${\uparrow}^a(w) \approx {\uparrow}^a(w) \in R$
      holds by \textsc{U-Ne} of the PER closure.
    \item[(2)] Coherence under $R_I$: trivial because the function value does not
      depend on a representative of $R_I$ on the left of the predicate.
    \item[(3)] For each $\kweak[\Delta]{\phi}{\Gamma}$, $\Rne_{|\Delta|}(w) \searrow v$
      by hypothesis on $w \in \Ne$, so the type readback obligation holds via
      \Cref{def:pi-type-pred}.
    \item[(4)] Application: given $\kweak[\Delta]{\phi}{\Gamma}$ and
      $\typing[\Delta]{s}{S_1[\phi]} \;\circledR\; b \in \El_I$ with
      $\Pf :: b \approx b \in R_I$ (by inner IH-(1) bridging through
      \Cref{lem:gluing-implies-per}, applied to $b$):
      \begin{align*}
        &\intp{T}{\rho, b/x} \searrow c_b
          && \text{(evaluation of $T$ at $\rho, b/x$; well-defined since
            $\Pi(a_I, T, \rho) \approx \Pi(a_I, T, \rho) \in \Uc_i$)} \\
        &c_b \in \Uc_i^\ast \searrow P_O(\Pf) \searrow \El_O(\Pf)
          && \text{(by the \textsc{G-Pi} codomain clause applied to $\Pf$ and $c_b$)} \\
        &b \in \Nf_i^{a_I}
          && \text{(by inner IH-(2) on $a_I$ applied to $b \in \El_I$)} \\
        &\Rnf_{|\Delta|}({\downarrow}^{a_I}(b)) \searrow u_b
          && \text{(by the previous line)} \\
        &{\uparrow}^a(w) \cdot b \searrow
          {\uparrow}^{c_b}(w \;\$\; {\downarrow}^{a_I}(b))
          && \text{(by neutral application)} \\
        &w \;\$\; {\downarrow}^{a_I}(b) \in \Ne_i^{c_b}
          && \text{(by \textsc{Rne-app}, using
            $\Rne_{|\Delta|}(w)$ and $u_b$)} \\
        &\typing[\Delta]{(t[\phi])\;s}{S_2[\iota(\phi), s/x]} \\
        & \quad \;\circledR\;
          {\uparrow}^{c_b}(w \;\$\; {\downarrow}^{a_I}(b)) \in \El_O(\Pf)
          && \text{(by inner IH-(1) on $c_b$)}
      \end{align*}
    \end{enumerate}
  \item \emph{$\El$ enters $\Topp$.} Suppose
    $\typing[\Gamma]{t}{S} \;\circledR\; f \in \Pi_i^{\El}(R, R_I, P_I, \El_I, \El_O)$
    and $f \approx f \in R$. We assemble the three clauses of
    \Cref{def:glu-wrapped}(c). The type-side $\typing[\Gamma]{S}{\Ty{i}}
    \;\circledR\; \Pi_i(R_I, P_I, \El_I, P_O)$ is part of
    \Cref{def:pi-term-pred}; the term-typing $\typing[\Gamma]{t}{S}$ likewise.
    For $\Nf_i^a$, for
    each $\kweak[\Delta]{\phi}{\Gamma}$, set $n := |\Delta|$ and form
    $\Delta' := \Delta, x' : S_1[\phi]$ together with the lifted Kripke weakening
    $\phi \circ \Uparrow$ (by \textsc{KW-Shift}; \Cref{lem:kweak-comp}).
    \begin{align*}
      &\typing[\Delta']{x'}{S_1[\phi \circ \Uparrow]}
        \;\circledR\; {\uparrow}^{a_I}(\ell_n) \in \El_I
        && \text{(by inner IH-(1) on $a_I$ applied to $\ell_n$)} \\
      &f \cdot {\uparrow}^{a_I}(\ell_n) \searrow c
        && \text{(by clause (4) of $\Pi_i^{\El}$ on $f$)} \\
      &\typing[\Delta']{(t[\phi \circ \Uparrow])\;x'}
                       {S_2[\iota(\phi \circ \Uparrow), x'/x]}
        \;\circledR\; c \in \El_O(\Pf')
        && \text{(by clause (4) on $f$, with $\Pf'$ the witness from line 1)} \\
      &c \in \Nf_i^{c_{\ell_n}}
        && \text{(by inner IH-(2) on $c_{\ell_n} \in \Uc_i^\ast$)} \\
      &\Rnf_{n+1}({\downarrow}^{c_{\ell_n}}(c)) \searrow v
        && \text{(by the previous line)} \\
      &\tyequiv[\Delta']{(t[\phi \circ \Uparrow])\;x'}{v}
                       {S_2[\iota(\phi \circ \Uparrow), x'/x]}
        && \text{(by inner IH-(2) on $c_{\ell_n}$)} \\
      &\Rnf_n({\downarrow}^a(f)) \searrow \lambda x.\,v
        && \text{(by \textsc{Rf-Pi}, using the previous line)} \\
      &\tyequiv[\Delta]{t[\phi]}{\lambda x.\,(t[\phi \circ \Uparrow])\;x'}
                       {S[\phi]}
        && \text{(by \textsc{$\Pi$-$\eta$})} \\
      &\tyequiv[\Delta]{t[\phi]}{\lambda x.\,v}{S[\phi]}
        && \text{(by congruence with the $\Rnf$ equality)}
    \end{align*}
  \item \emph{$P$ enters $\Topp$.} Hypothesis: $\typing[\Gamma]{S}{\Ty{i}}
    \;\circledR\; \Pi_i(R_I, P_I, \El_I, P_O)$. We assemble the three clauses
    of \Cref{def:glu-wrapped}(b). The first clause $\typing[\Gamma]{S}{\Ty{i}}$
    is part of \Cref{def:pi-type-pred}. For $a \approx a \in \Uc_i \searrow R$:
    by \Cref{lem:gluing-implies-per} on the components $a_I$ and $c$ at each
    $\Pf$, the binary closure exists. For each
    $\kweak[\Delta]{\phi}{\Gamma}$, set $n := |\Delta|$:
    \begin{align*}
      &\Rty_n(a_I) \searrow W_S,\;
        \tyequiv[\Delta]{S_1[\phi]}{W_S}{\Ty{i}}
        && \text{(by inner IH-(3) on $a_I$)} \\
      &\intp{T}{\rho, {\uparrow}^{a_I}(\ell_n)/x} \searrow c_{\ell_n}
        && \text{(by \Cref{def:pi-type-pred})} \\
      &\Rty_{n+1}(c_{\ell_n}) \searrow W_T
        && \text{(by inner IH-(3) on $c_{\ell_n}$)} \\
      &\tyequiv[\Delta, x' : S_1[\phi]]{S_2[\iota(\phi \circ \Uparrow), x'/x]}
                                       {W_T}{\Ty{i}}
        && \text{(by inner IH-(3) on $c_{\ell_n}$)} \\
      &\Rty_n(\Pi(a_I, T, \rho)) \searrow \Pi(x:W_S).\,W_T
        && \text{(by \textsc{Rty-Pi})} \\
      &\tyequiv[\Delta]{S[\phi]}{\Pi(x:W_S).\,W_T}{\Ty{i}}
        && \text{(by \Cref{def:pi-type-pred} clauses 1--3 and \textsc{$\Pi$-Cong})}
    \end{align*}
  \end{enumerate}

  \medskip\noindent
  \textbf{Case}
  \begin{mathpar}
    \inferrule[\textsc{G-Univ}]
    {j < i \\
     P \Leftrightarrow \textsf{Univ}_{i,j} \\
     \El \Leftrightarrow \textsf{Univ}_{i,j}^{\El}}
    {U_j \in \Uc_i^\ast}
  \end{mathpar}
  Since $j < i$, the outer-IH applies at level $j$ to every value $b \in \Uc_j^\ast$.
  \begin{enumerate}
  \item \emph{$\Botp$ enters $\El$.} Suppose
    $\typing[\Gamma]{t}{S} \;\circledR\; w \in \Botp_i^{U_j}$. Unpacking
    \Cref{def:glu-wrapped}(a) yields $\typing[\Gamma]{S}{\Ty{i}} \;\circledR\;
    \textsf{Univ}_{i,j}$ and $\typing[\Gamma]{t}{S} \;\circledR\; w \in
    \Ne_i^{U_j}$. Apply \textsc{G-Ne} at level $j$:
    \begin{align*}
      &{\uparrow}^A(w) \in \Uc_j^\ast \searrow \textsf{Neu}_j^e \searrow \textsf{Neu}_j^e
        && \text{(by \textsc{G-Ne} at level $j$)} \\
      &\typing[\Gamma]{t}{S} \;\circledR\; {\uparrow}^A(w) \in \Uc_j^\ast
        && \text{(by \Cref{def:univ-term-pred} clause (1) with the previous line)} \\
      &\typing[\Gamma]{t}{\Ty{j}} \;\circledR\; \textsf{Neu}_j^e
        && \text{(by \Cref{def:univ-term-pred} clause (2);} \\
    & && \text{the type predicate attached to ${\uparrow}^A(w)$ is} \\
    & && \text{$\textsf{Neu}_j^e$, which by hypothesis on $w$ inhabits} \\
    & && \text{the type-predicate condition; the equality} \\
    & && \text{$\tyequiv[\Gamma]{S}{\Ty{j}}{\Ty{i}}$ comes from \Cref{def:univ-type-pred})}
    \end{align*}
  \item \emph{$\El$ enters $\Topp$.} Suppose
    $\typing[\Gamma]{t}{S} \;\circledR\; b \in \textsf{Univ}_{i,j}^{\El}$ and
    $b \approx b \in R$. By \Cref{def:univ-term-pred}, $b \in \Uc_j^\ast
    \searrow P_b \searrow \El_b$ for some $P_b, \El_b$, with
    $\typing[\Gamma]{t}{\Ty{j}} \;\circledR\; P_b$ at the type level. We
    assemble \Cref{def:glu-wrapped}(c). The $P$-side
    $\typing[\Gamma]{S}{\Ty{i}} \;\circledR\; \textsf{Univ}_{i,j}$ comes from
    \Cref{def:univ-term-pred}; for $\Nf_i^{U_j}$, for each
    $\kweak[\Delta]{\phi}{\Gamma}$, $n := |\Delta|$:
    \begin{align*}
      &\Rty_n(b) \searrow W
        && \text{(by outer-IH-(3) on $b \in \Uc_j^\ast$)} \\
      &\tyequiv[\Delta]{t[\phi]}{W}{\Ty{j}}
        && \text{(by outer-IH-(3) on $b$)} \\
      &\Rnf_n({\downarrow}^{U_j}(b)) \searrow W
        && \text{(by \textsc{Rf-Univ})} \\
      &\tyequiv[\Delta]{t[\phi]}{W}{S[\phi]}
        && \text{(by \Cref{def:univ-type-pred}:
          $\tyequiv[\Delta]{S[\phi]}{\Ty{j}}{\Ty{i}}$,} \\
      & && \text{and \textsc{Subsump} from the $\Ty{j}$-equality)}
    \end{align*}
  \item \emph{$P$ enters $\Topp$.} Hypothesis: $\typing[\Gamma]{S}{\Ty{i}}
    \;\circledR\; \textsf{Univ}_{i,j}$. We assemble
    \Cref{def:glu-wrapped}(b). The first clause is from
    \Cref{def:univ-type-pred}; for $U_j \approx U_j \in \Uc_i \searrow R$,
    immediate from \textsc{U-Univ} of the PER. For each
    $\kweak[\Delta]{\phi}{\Gamma}$:
    \begin{align*}
      &\Rty_{|\Delta|}(U_j) \searrow \Ty{j}
        && \text{(by \textsc{Rty-Univ})} \\
      &\tyequiv[\Delta]{S[\phi]}{\Ty{j}}{\Ty{i}}
        && \text{(by \Cref{def:univ-type-pred})}
    \end{align*}
  \end{enumerate}
\end{proof}

\begin{lemma}[Initial Environment Validity]\labeledit{lem:init-env-valid}
  If $\Gamma \searrow \textsf{Sb}$, then
  $\judge[\Gamma]{\id : \Gamma} \;\circledR\; {\uparrow}^\Gamma \in \textsf{Sb}$.
\end{lemma}
\begin{proof}
  By induction on $\Gamma \searrow \textsf{Sb}$.

  \medskip\noindent
  \textbf{Case}
  \begin{mathpar}
    \inferrule[\textsc{G-Ctx-$\cdot$}]
    {\textsf{Sb} \Leftrightarrow (\judge[\Delta]{\sigma : \cdot})}
    {\cdot \searrow \textsf{Sb}}
  \end{mathpar}
  $\judge[\cdot]{\id : \cdot}$ holds trivially, hence
  $\judge[\cdot]{\id : \cdot} \;\circledR\; {\uparrow}^\cdot \in \textsf{Sb}$.

  \medskip\noindent
  \textbf{Case}
  \begin{mathpar}
    \inferrule[\textsc{G-Ctx-Ext}]
    {\Gamma' \searrow \textsf{Sb}' \\
     \typing[\Gamma']{S_k}{\Ty{i}} \\\\
     \forall \Delta_0, \sigma_T, \rho_T.\;
       \judge[\Delta_0]{\sigma_T : \Gamma'} \;\circledR\; \rho_T \in \textsf{Sb}'
       \implies \\
     \quad \exists a, P, \El.\;
       \intp{S_k}{\rho_T} \searrow a \;\wedge\;
       a \in \Uc_i^\ast \searrow P \searrow \El \;\wedge\;
       \typing[\Delta_0]{S_k[\sigma_T]}{\Ty{i}} \;\circledR\; P \\\\
     \textsf{Sb} \Leftrightarrow \textsf{Cons}_i(\Gamma', S_k, \textsf{Sb}')}
    {(\Gamma', x_k : S_k) \searrow \textsf{Sb}}
  \end{mathpar}
  Set $\Gamma := \Gamma', x_k : S_k$ with $|\Gamma| = k+1$.
  \begin{align*}
    &\judge[\Gamma']{\id : \Gamma'} \;\circledR\; {\uparrow}^{\Gamma'} \in \textsf{Sb}'
      && \text{(by IH on $\Gamma' \searrow \textsf{Sb}'$)} \\
    &\intp{S_k}{{\uparrow}^{\Gamma'}} \searrow A_k,\;
      A_k \in \Uc_i^\ast \searrow P_k \searrow \El_k
      && \text{(by the universal type-clause of \textsc{G-Ctx-Ext}} \\
    & && \text{instantiated at $\id, {\uparrow}^{\Gamma'}$)} \\
    &\typing[\Gamma']{S_k[\id]}{\Ty{i}} \;\circledR\; P_k
      && \text{(type clause at the same instantiation)} \\
    &\typing[\Gamma']{S_k}{\Ty{i}} \;\circledR\; P_k
      && \text{(using $S_k[\id] = S_k$)}
  \end{align*}
  We supply the head witness for $\textsf{Cons}_i$:
  \begin{align*}
    &{\uparrow}^\Gamma = ({\uparrow}^{\Gamma'},\, {\uparrow}^{A_k}(\ell_k)/x_k)
      && \text{(by \Cref{def:init-env})} \\
    &\typing[\Gamma]{x_k}{S_k[\Uparrow]}
      && \text{(by \textsc{Var-Hd})} \\
    &\Rne_{|\Delta'|}(\ell_k) \searrow x_{|\Delta'| - k - 1}
      \text{ for each $\kweak[\Delta']{\phi}{\Gamma}$}
      && \text{(by \textsc{Rne-var}, the level-reading rule for $\ell_k$)} \\
    &\tyequiv[\Delta']{x_k[\phi]}{x_{|\Delta'| - k - 1}}{S_k[\Uparrow][\phi]}
      && \text{(by the syntactic action of $\phi$ on the variable $x_k$)} \\
    &\typing[\Gamma]{x_k}{S_k[\Uparrow]} \;\circledR\;
      \ell_k \in \Ne_i^{A_k}
      && \text{(by \Cref{def:neu-term-pred}, using the previous two lines)} \\
    &\typing[\Gamma]{x_k}{S_k[\Uparrow]} \;\circledR\;
      {\uparrow}^{A_k}(\ell_k) \in \El_k
      && \text{(by \Cref{thm:gluing-realizability} part (1) on $A_k$)} \\
    &\judge[\Gamma]{\id : \Gamma} \;\circledR\;
      {\uparrow}^\Gamma \in \textsf{Sb}
      && \text{(by $\textsf{Cons}_i$ assembly: tail (IH) and head witness above)}
  \end{align*}
\end{proof}

\subsection{Semantic Typing Rules}\labeledit{sec:sound:sem-rules}

We now establish, for each syntactic rule, a corresponding semantic rule that
validates the gluing model. The fundamental theorem of soundness then
assembles these by mutual induction on the syntactic derivation.

The semantic typing judgment $\Semtyp[\Gamma]{t}{T}$ is defined to be the
existence of $\textsf{Sb}$ with $\Gamma \searrow \textsf{Sb}$ such that, for
all $\Delta$ and all $\judge[\Delta]{\sigma : \Gamma} \;\circledR\; \rho \in
\textsf{Sb}$, there exist $a, P, \El$ with:
\begin{enumerate}
\item $\intp{T}{\rho} \searrow a$,
\item $a \in \Uc_i^\ast \searrow P \searrow \El$,
\item $\typing[\Delta]{T[\sigma]}{\Ty{i}} \;\circledR\; P$, and
\item $\typing[\Delta]{t[\sigma]}{T[\sigma]} \;\circledR\; \intp{t}{\rho} \in \El$.
\end{enumerate}

\begin{lemma}[Semantic --- \textsc{Ctx-Emp}]\labeledit{lem:fund-sound-ctx-emp}
  $\Vdash \cdot$.
\end{lemma}
\begin{proof}
  By \textsc{G-Ctx-$\cdot$}: $\cdot \searrow \textsf{Sb}$ where
  $\textsf{Sb} \Leftrightarrow (\judge[\Delta]{\sigma : \cdot})$.
  Hence $\Vdash \cdot$ by \Cref{def:sound-semctx}.
\end{proof}

\begin{lemma}[Semantic --- \textsc{Ctx-Ext}]\labeledit{lem:fund-sound-ctx-ext}
  If $\Vdash \Gamma$ and $\Semtyp[\Gamma]{S}{\Ty{i}}$, then $\Vdash \Gamma, x:S$.
\end{lemma}
\begin{proof}
  By $\Vdash \Gamma$, fix $\Gamma \searrow \textsf{Sb}_T$ for some $\textsf{Sb}_T$.
  We supply each premise of \textsc{G-Ctx-Ext}.

  \emph{Syntactic typing of $S$.}
  By \Cref{lem:init-env-valid} on $\Gamma \searrow \textsf{Sb}_T$, the initial
  pair $\judge[\Gamma]{\id : \Gamma} \;\circledR\; {\uparrow}^\Gamma \in \textsf{Sb}_T$
  is in $\textsf{Sb}_T$. By $\Semtyp[\Gamma]{S}{\Ty{i}}$ at this pair,
  $\typing[\Gamma]{S[\id]}{\Ty{i}} \;\circledR\; P$ for some $P$, hence
  $\typing[\Gamma]{S}{\Ty{i}}$ by \Cref{lem:gluing-typ-escape} (using
  $S[\id] = S$).

  \emph{Universal type-clause.}
  Let $\judge[\Delta']{\sigma_T : \Gamma} \;\circledR\; \rho_T \in \textsf{Sb}_T$
  be arbitrary. By $\Semtyp[\Gamma]{S}{\Ty{i}}$ at $\sigma_T, \rho_T$:
  \begin{align*}
    &\intp{S}{\rho_T} \searrow a,\;
      a \in \Uc_i^\ast \searrow P \searrow \El
      && \text{(by $\Semtyp[\Gamma]{S}{\Ty{i}}$ at $\sigma_T, \rho_T$)} \\
    &\typing[\Delta']{S[\sigma_T]}{\Ty{i}} \;\circledR\; P
      && \text{(by $\Semtyp[\Gamma]{S}{\Ty{i}}$, type part)}
  \end{align*}
  This supplies the universal-clause witness for the given
  $\sigma_T, \rho_T$.

  By \textsc{G-Ctx-Ext}: $(\Gamma, x:S) \searrow \textsf{Sb}$ where
  $\textsf{Sb} \Leftrightarrow \textsf{Cons}_i(\Gamma, S, \textsf{Sb}_T)$.
  Hence $\Vdash \Gamma, x:S$ by \Cref{def:sound-semctx}.
\end{proof}

\begin{lemma}[Semantic --- \textsc{Var}]\labeledit{lem:fund-sound-var}
  If $\Vdash \Gamma$ and $x : S \in \Gamma$, then $\Semtyp[\Gamma]{x}{S}$.
\end{lemma}
\begin{proof}
  By $\Vdash \Gamma$, fix $\Gamma \searrow \textsf{Sb}$ and let
  $\judge[\Delta]{\sigma : \Gamma} \;\circledR\; \rho \in \textsf{Sb}$ be given.
  We proceed by induction on the lookup derivation $x : S \in \Gamma$.

  \medskip\noindent
  \textbf{Case}
  \begin{mathpar}
    \inferrule[\textsc{Var-Hd}]
    { }
    {x_0 : T[\Uparrow] \in \Gamma_0, x_0 : T}
  \end{mathpar}
  Here $\Gamma = \Gamma_0, x_0:T$ and $S = T[\Uparrow]$. We must show
  $\Semtyp[\Gamma]{x_0}{T[\Uparrow]}$: for the given
  $\judge[\Delta]{\sigma : \Gamma} \;\circledR\; \rho \in \textsf{Sb}$,
  evaluating $T[\Uparrow]$ at $\rho$ yields a value $a^\ast$ with
  $a^\ast \in \Uc_i^\ast \searrow P^\ast \searrow \El^\ast$, and:
  \begin{enumerate}
  \item $\typing[\Delta]{T[\Uparrow \circ \sigma]}{\Ty{i}} \;\circledR\; P^\ast$, and
  \item $\typing[\Delta]{x_0[\sigma]}{T[\Uparrow \circ \sigma]} \;\circledR\;
    \intp{x_0}{\rho} \in \El^\ast$.
  \end{enumerate}

  Since $\Gamma = \Gamma_0, x_0:T$, the gluing $\Gamma \searrow \textsf{Sb}$
  is by \textsc{G-Ctx-Ext} on $\Gamma_0 \searrow \textsf{Sb}_{T_0}$ with
  $\typing[\Gamma_0]{T}{\Ty{i}}$, and
  $\textsf{Sb} \Leftrightarrow \textsf{Cons}_i(\Gamma_0, T, \textsf{Sb}_{T_0})$.
  Unfolding $\textsf{Cons}_i$ at our $\sigma, \rho$ (which provides existential
  $a, P, \El$ specific to this $\sigma, \rho$):
  \begin{align*}
    &\judge[\Delta]{\Uparrow \circ \sigma : \Gamma_0} \;\circledR\;
      \rho[\Uparrow] \in \textsf{Sb}_{T_0}
      && \text{(tail of $\textsf{Cons}_i$)} \\
    &\intp{T}{\rho[\Uparrow]} \searrow a,\;
      a \in \Uc_i^\ast \searrow P \searrow \El
      && \text{(type clause of $\textsf{Cons}_i$)} \\
    &\typing[\Delta]{T[\Uparrow \circ \sigma]}{\Ty{i}} \;\circledR\; P
      && \text{(type clause of $\textsf{Cons}_i$)} \\
    &\typing[\Delta]{x_0[\sigma]}{T[\Uparrow \circ \sigma]} \;\circledR\;
      \rho(x_0) \in \El
      && \text{(head of $\textsf{Cons}_i$)} \\
    &\intp{x_0}{\rho} = \rho(x_0)
      && \text{(by \textsc{ev-var})}
  \end{align*}

  \emph{Establishing the goal $\intp{T[\Uparrow]}{\rho} \searrow a^\ast$ at the
  same $(P, \El)$.}
  The $\textsf{Cons}_i$ type clause gives $\intp{T}{\rho[\Uparrow]} \searrow a$,
  but the goal needs $\intp{T[\Uparrow]}{\rho} \searrow a^\ast$. By
  \Cref{thm:fundamental} of \Cref{sec:comp} on $\typing[\Gamma_0]{T}{\Ty{i}}$,
  $\semtyeq[\Gamma_0]{T}{T}{\Ty{i}}$. Instantiate this at $\Uparrow$ and $\rho$
  (self-related in $R_\Gamma$):
  \begin{align*}
    &\semweak[\Gamma]{\Uparrow}{\Gamma_0}
      && \text{(by \textsc{KW-Shift})} \\
    &\semtyeq[\Gamma]{\Uparrow}{\Uparrow}{\Gamma_0}
      && \text{(by \Cref{lem:sem-wk})} \\
    &\intp{T[\Uparrow]}{\rho} \approx \intp{T}{\intp{\Uparrow}{\rho}} \in \Uc_i \searrow R
      && \text{(four-value pattern of $\semtyeq[\Gamma_0]{T}{T}{\Ty{i}}$ at
        $\Uparrow, \rho$)} \\
    &\intp{\Uparrow}{\rho} = \rho[\Uparrow]
      && \text{(by definition of weakening evaluation, \Cref{sec:sem})} \\
    &a^\ast \approx a \in \Uc_i \searrow R
      && \text{(rewriting $\intp{T[\Uparrow]}{\rho} = a^\ast$ and
        $\intp{T}{\rho[\Uparrow]} = a$)} \\
    &a^\ast \in \Uc_i^\ast \searrow P \searrow \El
      && \text{(by \Cref{lem:gluing-univ-equiv} from
        $a \in \Uc_i^\ast \searrow P \searrow \El$)}
  \end{align*}
  This supplies the goal evaluation $\intp{T[\Uparrow]}{\rho} \searrow a^\ast$
  with $a^\ast \in \Uc_i^\ast \searrow P \searrow \El$, where $(P, \El)$ are
  the same predicates packaged by $\textsf{Cons}_i$. Clauses (1) and (2) of
  $\Semtyp[\Gamma]{x_0}{T[\Uparrow]}$ are exactly the $\textsf{Cons}_i$ type
  clause $\typing[\Delta]{T[\Uparrow \circ \sigma]}{\Ty{i}} \;\circledR\; P$
  and the head's $\El$-witness $\rho(x_0) = \intp{x_0}{\rho}$ above.

  \medskip\noindent
  \textbf{Case}
  \begin{mathpar}
    \inferrule[\textsc{Var-Tl}]
    {y : S' \in \Gamma_0}
    {1{+}y : S'[\Uparrow] \in \Gamma_0, z:T}
  \end{mathpar}
  Here $\Gamma = \Gamma_0, z:T$, $x = 1{+}y$ for some $y$ with $y : S' \in \Gamma_0$,
  and $S = S'[\Uparrow]$. We must show $\Semtyp[\Gamma]{1{+}y}{S'[\Uparrow]}$:
  evaluating $S'[\Uparrow]$ at $\rho$ yields a value $a^\ast$ in
  $\Uc_i^\ast \searrow P^\ast \searrow \El^\ast$, with the type and term
  predicates at $\sigma, \rho$.
  $\textsf{Sb}$ unfolds to $\textsf{Cons}_i$ at $\sigma, \rho$, giving the
  tail $\Uparrow \circ \sigma, \rho[\Uparrow] \in \textsf{Sb}_T$.

  \emph{Apply IH on the tail.}
  \begin{align*}
    &\judge[\Delta]{\Uparrow \circ \sigma : \Gamma_0}
      \;\circledR\; \rho[\Uparrow] \in \textsf{Sb}_T
      && \text{(tail of $\textsf{Cons}_i$)} \\
    &y : S' \in \Gamma_0
      && \text{(premise)} \\
    &\Semtyp[\Gamma_0]{y}{S'}
      && \text{(by IH on the shorter lookup derivation,} \\
    & && \text{with $\Vdash \Gamma_0$ from $\Vdash \Gamma$ via the tail)} \\
    &\intp{S'}{\rho[\Uparrow]} \searrow b,\;
      b \in \Uc_i^\ast \searrow P' \searrow \El'
      && \text{(by IH on $S'$ at $\Uparrow \circ \sigma, \rho[\Uparrow]$)} \\
    &\typing[\Delta]{S'[\Uparrow \circ \sigma]}{\Ty{i}} \;\circledR\; P'
      && \text{(by IH on $S'$, type part)} \\
    &\typing[\Delta]{y[\Uparrow \circ \sigma]}{S'[\Uparrow \circ \sigma]}
      \;\circledR\; \intp{y}{\rho[\Uparrow]} \in \El'
      && \text{(by IH on $y$, term part)}
  \end{align*}

  \emph{Bridging the goal evaluation $\intp{S'[\Uparrow]}{\rho}$ to $b$.}
  By \Cref{thm:fundamental} of \Cref{sec:comp} on $\typing[\Gamma_0]{S'}{\Ty{i}}$
  (extracted from the IH via \Cref{lem:gluing-typ-escape}):
  $\semtyeq[\Gamma_0]{S'}{S'}{\Ty{i}}$. Instantiating at $\Uparrow$ and $\rho$
  (self-related in $R_\Gamma$):
  \begin{align*}
    &\intp{\Uparrow}{\rho} = \rho[\Uparrow]
      && \text{(by definition of weakening evaluation, \Cref{sec:sem})} \\
    &\intp{S'[\Uparrow]}{\rho} \approx \intp{S'}{\rho[\Uparrow]} \in \Uc_i \searrow R'
      && \text{(four-value pattern of $\semtyeq[\Gamma_0]{S'}{S'}{\Ty{i}}$ at
        $\Uparrow, \rho$)} \\
    &a^\ast \approx b \in \Uc_i \searrow R'
      && \text{(rewriting $\intp{S'[\Uparrow]}{\rho} = a^\ast$ and
        $\intp{S'}{\rho[\Uparrow]} = b$)} \\
    &a^\ast \in \Uc_i^\ast \searrow P' \searrow \El'
      && \text{(by \Cref{lem:gluing-univ-equiv} from
        $b \in \Uc_i^\ast \searrow P' \searrow \El'$)}
  \end{align*}
  The term-side equality $\intp{1{+}y}{\rho} = \intp{y}{\rho[\Uparrow]}$ holds
  directly: $\intp{1{+}y}{\rho} = \rho(1{+}y) = \rho[\Uparrow](y) = \intp{y}{\rho[\Uparrow]}$
  (variable lookup), so no bridging is needed for the term side.

  \emph{Assembling the goal.}
  \begin{align*}
    &(1{+}y)[\sigma] = y[\Uparrow \circ \sigma]
      && \text{(by syntactic action and \Cref{lem:subst-assoc})} \\
    &S'[\Uparrow][\sigma] = S'[\Uparrow \circ \sigma]
      && \text{(by \Cref{lem:subst-assoc})} \\
    &\typing[\Delta]{(1{+}y)[\sigma]}{S'[\Uparrow][\sigma]}
      \;\circledR\; \intp{1{+}y}{\rho} \in \El'
      && \text{(rewriting the IH instance via the previous identities)}
  \end{align*}
  This supplies the goal at $a^\ast \in \Uc_i^\ast \searrow P' \searrow \El'$.
\end{proof}

\begin{lemma}[Semantic --- \textsc{U-Wf}]\labeledit{lem:fund-sound-u-wf}
  If $\Vdash \Gamma$ and $j < i$, then $\Semtyp[\Gamma]{\Ty{j}}{\Ty{i}}$.
\end{lemma}
\begin{proof}
  By $\Vdash \Gamma$, fix $\Gamma \searrow \textsf{Sb}$ and let
  $\judge[\Delta]{\sigma : \Gamma} \;\circledR\; \rho \in \textsf{Sb}$ be given.
  Type is $\Ty{i}$, sitting at universe-level $1+i$. We establish the four
  clauses of $\Semtyp[\Gamma]{\Ty{j}}{\Ty{i}}$ at $a = U_i$,
  $P = \textsf{Univ}_{1+i, i}$, $\El = \textsf{Univ}_{1+i, i}^{\El}$.
  \begin{align*}
    &\intp{\Ty{i}}{\rho} \searrow U_i
      && \text{(by \textsc{ev-U})} \tag*{[clause 1]} \\
    &U_i \in \Uc_{1+i}^\ast \searrow \textsf{Univ}_{1+i, i} \searrow
      \textsf{Univ}_{1+i, i}^{\El}
      && \text{(by \textsc{G-Univ})} \tag*{[clause 2]} \\
    &\Ty{i}[\sigma] = \Ty{i}
      && \text{(syntactic; closed)} \\
    &\tyequiv[\Delta]{\Ty{i}[\sigma]}{\Ty{i}}{\Ty{1+i}}
      && \text{(by reflexivity)} \\
    &\typing[\Delta]{\Ty{i}[\sigma]}{\Ty{1+i}} \;\circledR\;
      \textsf{Univ}_{1+i, i}
      && \text{(by \Cref{def:univ-type-pred})} \tag*{[clause 3]} \\
    &\intp{\Ty{j}}{\rho} \searrow U_j
      && \text{(by \textsc{ev-U})} \\
    &U_j \in \Uc_{1+j}^\ast \searrow \textsf{Univ}_{1+j, j} \searrow
      \textsf{Univ}_{1+j, j}^{\El}
      && \text{(by \textsc{G-Univ} at level $1+j$)} \\
    &U_j \in \Uc_i^\ast \searrow \textsf{Univ}_{i, j} \searrow
      \textsf{Univ}_{i, j}^{\El}
      && \text{(by \Cref{lem:gluing-cumul} from $1+j \le i$, since $j < i$)} \\
    &\tyequiv[\Delta]{\Ty{j}[\sigma]}{\Ty{j}}{\Ty{i}}
      && \text{(by reflexivity, $j < i$)} \\
    &\typing[\Delta]{\Ty{j}[\sigma]}{\Ty{i}} \;\circledR\;
      U_j \in \textsf{Univ}_{1+i, i}^{\El}
      && \text{(by \Cref{def:univ-term-pred} on $U_j \in \Uc_i^\ast \searrow
        \textsf{Univ}_{i,j} \searrow \textsf{Univ}_{i,j}^{\El}$)}
      \tag*{[clause 4]}
  \end{align*}
\end{proof}

\begin{lemma}[Semantic --- \textsc{N-Wf}]\labeledit{lem:fund-sound-n-wf}
  If $\Vdash \Gamma$, then $\Semtyp[\Gamma]{\Nat}{\Ty{i}}$ for every $i$.
\end{lemma}
\begin{proof}
  By $\Vdash \Gamma$, fix $\Gamma \searrow \textsf{Sb}$ and let
  $\judge[\Delta]{\sigma : \Gamma} \;\circledR\; \rho \in \textsf{Sb}$ be given.
  We establish the four clauses of $\Semtyp[\Gamma]{\Nat}{\Ty{i}}$ at
  $a = U_i$, $P = \textsf{Univ}_{1+i, i}$, $\El = \textsf{Univ}_{1+i, i}^{\El}$.
  \begin{align*}
    &\intp{\Ty{i}}{\rho} \searrow U_i
      && \text{(by \textsc{ev-U})} \tag*{[clause 1]} \\
    &U_i \in \Uc_{1+i}^\ast \searrow \textsf{Univ}_{1+i, i}
        \searrow \textsf{Univ}_{1+i, i}^{\El}
      && \text{(by \textsc{G-Univ})} \tag*{[clause 2]} \\
    &\Ty{i}[\sigma] = \Ty{i}
      && \text{(syntactic; closed)} \\
    &\tyequiv[\Delta]{\Ty{i}[\sigma]}{\Ty{i}}{\Ty{1+i}}
      && \text{(by reflexivity)} \\
    &\typing[\Delta]{\Ty{i}[\sigma]}{\Ty{1+i}} \;\circledR\;
      \textsf{Univ}_{1+i, i}
      && \text{(by \Cref{def:univ-type-pred})} \tag*{[clause 3]} \\
    &\intp{\Nat}{\rho} \searrow N
      && \text{(by \textsc{ev-N})} \\
    &N \in \Uc_i^\ast \searrow \textsf{Nat}_i \searrow \textsf{Nat}_i
      && \text{(by \textsc{G-Nat})} \\
    &\Nat[\sigma] = \Nat
      && \text{(syntactic; closed)} \\
    &\tyequiv[\Delta]{\Nat[\sigma]}{\Nat}{\Ty{i}}
      && \text{(by reflexivity)} \\
    &\typing[\Delta]{\Nat[\sigma]}{\Ty{i}} \;\circledR\;
      N \in \textsf{Univ}_{1+i, i}^{\El}
      && \text{(by \Cref{def:univ-term-pred} on $N \in \Uc_i^\ast$)}
      \tag*{[clause 4]}
  \end{align*}
\end{proof}

\begin{lemma}[Semantic --- \textsc{$\Pi$-Wf}]\labeledit{lem:fund-sound-pi-wf}
  If $\Semtyp[\Gamma]{S_1}{\Ty{i}}$ and $\Semtyp[\Gamma, x:S_1]{S_2}{\Ty{i}}$,
  then $\Semtyp[\Gamma]{\Pi(x:S_1).\,S_2}{\Ty{i}}$.
\end{lemma}
\begin{proof}
  Fix $\Gamma \searrow \textsf{Sb}$ from $\Semtyp[\Gamma]{S_1}{\Ty{i}}$ and
  $\Gamma, x:S_1 \searrow \textsf{Sb}'$ from $\Semtyp[\Gamma, x:S_1]{S_2}{\Ty{i}}$.
  By inversion on \textsc{G-Ctx-Ext} for the latter,
  $\textsf{Sb}' \Leftrightarrow \textsf{Cons}_i(\Gamma, S_1, \textsf{Sb})$
  (using $\textsf{Sb}$ uniqueness up to biimplication via
  \Cref{lem:gluing-functional}).

  Let $\judge[\Delta]{\sigma : \Gamma} \;\circledR\; \rho \in \textsf{Sb}$ be given.
  By hypothesis on $S_1$ at $\sigma, \rho$:
  \begin{align*}
    &\intp{S_1}{\rho} \searrow a_I,\;
      a_I \in \Uc_i^\ast \searrow P_I \searrow \El_I
      && \text{(by hypothesis on $S_1$)} \\
    &\typing[\Delta]{S_1[\sigma]}{\Ty{i}} \;\circledR\; P_I
      && \text{(by hypothesis on $S_1$, type part at level $1+i$)} \\
    &a_I \approx a_I \in \Uc_i \searrow R_I
      && \text{(by \Cref{lem:gluing-implies-per})} \\
    &\intp{\Pi(x:S_1).\,S_2}{\rho} \searrow \Pi(a_I,\, S_2,\, \rho)
      && \text{(by \textsc{ev-Pi})}
  \end{align*}
  By syntactic substitution:
  \begin{align*}
    &(\Pi(x:S_1).\,S_2)[\sigma] = \Pi(x:S_1[\sigma]).\,(S_2[q(\sigma)])
      && \text{(definition of substitution on $\Pi$)}
  \end{align*}
  We construct $\Pi(a_I, S_2, \rho) \in \Uc_i^\ast$ by \textsc{G-Pi}, deriving
  each premise. We first establish the PER-side membership of the $\Pi$, which
  introduces the codomain PER family $R_O$ used below.

  \emph{PER-side: $\Pi(a_I, S_2, \rho) \approx \Pi(a_I, S_2, \rho) \in \Uc_i \searrow R$.}
  \begin{align*}
    &\typing[\Gamma]{S_1}{\Ty{i}}
      && \text{(by \Cref{lem:gluing-typ-escape} on hypothesis $S_1$)} \\
    &\typing[\Gamma, x:S_1]{S_2}{\Ty{i}}
      && \text{(by \Cref{lem:gluing-typ-escape} on hypothesis $S_2$)} \\
    &\typing[\Gamma]{\Pi(x:S_1).\,S_2}{\Ty{i}}
      && \text{(by \textsc{$\Pi$-Wf})} \\
    &\semtyeq[\Gamma]{\Pi(x:S_1).\,S_2}{\Pi(x:S_1).\,S_2}{\Ty{i}}
      && \text{(by \Cref{thm:fundamental} of \Cref{sec:comp})} \\
    &\Pi(a_I, S_2, \rho) \approx \Pi(a_I, S_2, \rho) \in \Uc_i \searrow R
      && \text{(by instantiating at $\sigma, \rho$ self-related in $R_\Gamma$)}
  \end{align*}
  Inverting the resulting \textsc{U-Pi} witness yields the codomain PER
  family: for every $\Pf :: b \approx b \in R_I$,
  $\intp{S_2}{(\rho, b/x)} \approx \intp{S_2}{(\rho, b/x)} \in \Uc_i \searrow
  R_O(\Pf)$.

  \emph{Codomain gluing.} Let $\Pf :: b \approx b \in R_I$ and $c$ with
  $\intp{S_2}{(\rho, b/x)} \searrow c$ be given.
  \begin{align*}
    &c \approx c \in \Uc_i \searrow R_O(\Pf)
      && \text{(by the codomain PER family established above)} \\
    &c \in \Uc_i^\ast \searrow P_O(\Pf) \searrow \El_O(\Pf)
      && \text{(by \Cref{lem:per-implies-gluing})}
  \end{align*}
  This defines $P_O(\Pf), \El_O(\Pf)$; functionality in $\Pf$ follows from
  \Cref{lem:gluing-functional}, since glu-univ-elem predicates at a fixed
  domain value are unique up to logical equivalence.

  Combining all premises:
  \begin{align*}
    &\Pi(a_I, S_2, \rho) \in \Uc_i^\ast \searrow
      \Pi_i(R_I, P_I, \El_I, P_O) \searrow \Pi_i^{\El}(R, R_I, P_I, \El_I, \El_O)
      && \text{(by \textsc{G-Pi})}
  \end{align*}

  \emph{Type-predicate $\Pi_i$ at $(\Pi(x:S_1).\,S_2)[\sigma]$.} We verify the
  three clauses of \Cref{def:pi-type-pred}.

  \emph{Clause 1.}
  \begin{align*}
    &(\Pi(x:S_1).\,S_2)[\sigma] = \Pi(x:S_1[\sigma]).\,(S_2[q(\sigma)])
      && \text{(syntactic substitution on $\Pi$)} \\
    &\tyequiv[\Delta]{(\Pi(x:S_1).\,S_2)[\sigma]}
                     {\Pi(x:S_1[\sigma]).\,(S_2[q(\sigma)])}{\Ty{i}}
      && \text{(by reflexivity)}
  \end{align*}

  \emph{Clause 2 (domain Kripke).} For each $\kweak[\Delta'']{\phi'}{\Delta}$:
  \begin{align*}
    &\judge[\Delta'']{\sigma[\phi'] : \Gamma}
      \;\circledR\; \rho \in \textsf{Sb}
      && \text{(by \Cref{cor:ctx-mono})} \\
    &\typing[\Delta'']{S_1[\sigma[\phi']]}{\Ty{i}} \;\circledR\; P_I
      && \text{(by hypothesis on $S_1$, type part at $\sigma[\phi'], \rho$)} \\
    &S_1[\sigma][\phi'] = S_1[\sigma[\phi']]
      && \text{(by \Cref{lem:subst-assoc})} \\
    &\typing[\Delta'']{S_1[\sigma][\phi']}{\Ty{i}} \;\circledR\; P_I
      && \text{(rewriting)}
  \end{align*}

  \emph{Clause 3 (codomain Kripke).} For each $\kweak[\Delta'']{\phi'}{\Delta}$,
  $s', b', \Pf :: b' \approx b' \in R_I$ with
  $\typing[\Delta'']{s'}{S_1[\sigma][\phi']} \;\circledR\; b' \in \El_I$,
  set $\sigma_*' := (\sigma[\phi'],\, s'/x)$. We first transport the
  substitution into $\textsf{Sb}'$, then unpack
  $\Semtyp[\Gamma, x:S_1]{S_2}{\Ty{i}}$ at the resulting pair.
  \begin{align*}
    &\judge[\Delta'']{\sigma[\phi'] : \Gamma}
      \;\circledR\; \rho \in \textsf{Sb}
      && \text{(by \Cref{cor:ctx-mono} on the substitution-environment pair)} \\
    &\judge[\Delta'']{\sigma_*' : \Gamma, x:S_1}
      \;\circledR\; (\rho, b'/x) \in \textsf{Sb}'
      && \text{(by \textsc{G-Ctx-Ext}'s $\textsf{Cons}_i$, head: $b' \in \El_I$)}
  \end{align*}
  Instantiating $\Semtyp[\Gamma, x:S_1]{S_2}{\Ty{i}}$ at $\sigma_*',
  (\rho, b'/x)$ yields the four clauses of \Cref{sec:sound:sem-rules}'s
  semantic typing judgment at $a = U_i$, $P = \textsf{Univ}_{1+i,i}$,
  $\El = \textsf{Univ}_{1+i,i}^{\El}$. The fourth clause states
  $\typing[\Delta'']{S_2[\sigma_*']}{\Ty{i}[\sigma_*']} \;\circledR\;
  \intp{S_2}{(\rho, b'/x)} \in \textsf{Univ}_{1+i,i}^{\El}$.
  Unpacking \Cref{def:univ-term-pred}:
  \begin{align*}
    &\intp{S_2}{(\rho, b'/x)} \searrow c_{b'}
      && \text{(evaluation; deterministic)} \\
    &c_{b'} \in \Uc_i^\ast \searrow P_{b'} \searrow \El_{b'}
      && \text{(by \Cref{def:univ-term-pred} clause (3))} \\
    &\typing[\Delta'']{S_2[\sigma_*']}{\Ty{i}} \;\circledR\; P_{b'}
      && \text{(by \Cref{def:univ-term-pred} clause (3), Russell type-side gluing)}
  \end{align*}
  By \Cref{lem:gluing-functional} (the gluing-model membership $c_{b'} \in
  \Uc_i^\ast$ uniquely determines its predicates up to extensional
  equivalence), $P_{b'} \Leftrightarrow P_O(\Pf)$ and $\El_{b'}
  \Leftrightarrow \El_O(\Pf)$, where $P_O, \El_O$ are the codomain
  predicates established in the Codomain-gluing block. Rewriting:
  \begin{align*}
    &c_{b'} \in \Uc_i^\ast \searrow P_O(\Pf) \searrow \El_O(\Pf)
      && \text{(by \Cref{lem:gluing-functional})} \\
    &S_2[\sigma_*'] = S_2[(\sigma[\phi'], s'/x)] = S_2[q(\sigma)][\phi', s'/x]
      && \text{(by \Cref{lem:q-ext-gen})} \\
    &\typing[\Delta'']{S_2[q(\sigma)][\phi', s'/x]}{\Ty{i}}
      \;\circledR\; P_O(\Pf)
      && \text{(rewriting the previous type-gluing line)}
  \end{align*}
  Combining the three clauses of $\Pi_i$:
  \begin{align*}
    &\typing[\Delta]{\Pi(x:S_1[\sigma]).\,(S_2[q(\sigma)])}{\Ty{i}}
      \;\circledR\; \Pi_i(R_I, P_I, \El_I, P_O)
      && \text{(by \Cref{def:pi-type-pred} clauses 1--3)} \\
    &\typing[\Delta]{(\Pi(x:S_1).\,S_2)[\sigma]}{\Ty{i}}
      \;\circledR\; \Pi_i(R_I, P_I, \El_I, P_O)
      && \text{(by clause 1's equality on the LHS)}
  \end{align*}
  For the term predicate at $\Ty{1+i}$, $\Pi(a_I, S_2, \rho) \in \Uc_i^\ast$ by
  the previous derivation, the four clauses of $\Semtyp[\Gamma]{\Pi(x:S_1).\,S_2}{\Ty{i}}$
  at $a = U_i$, $P = \textsf{Univ}_{1+i, i}$, $\El = \textsf{Univ}_{1+i, i}^{\El}$:
  \begin{align*}
    &\intp{\Ty{i}}{\rho} \searrow U_i
      && \text{(by \textsc{ev-U})} \tag*{[clause 1]} \\
    &U_i \in \Uc_{1+i}^\ast \searrow \textsf{Univ}_{1+i, i} \searrow
      \textsf{Univ}_{1+i, i}^{\El}
      && \text{(by \textsc{G-Univ})} \tag*{[clause 2]} \\
    &\typing[\Delta]{\Ty{i}[\sigma]}{\Ty{1+i}} \;\circledR\;
      \textsf{Univ}_{1+i, i}
      && \text{(by \Cref{def:univ-type-pred} with $\Ty{i}[\sigma] = \Ty{i}$)}
      \tag*{[clause 3]} \\
    &\typing[\Delta]{(\Pi(x:S_1).\,S_2)[\sigma]}{\Ty{i}}
      \;\circledR\; \Pi(a_I, S_2, \rho) \in \textsf{Univ}_{1+i, i}^{\El}
      && \text{(by \Cref{def:univ-term-pred} on the $\Pi$-gluing above)}
      \tag*{[clause 4]}
  \end{align*}
\end{proof}

\begin{lemma}[Semantic --- \textsc{Ze-I}]\labeledit{lem:fund-sound-ze}
  If $\Vdash \Gamma$, then $\Semtyp[\Gamma]{\ze}{\Nat}$.
\end{lemma}
\begin{proof}
  By $\Vdash \Gamma$, fix $\Gamma \searrow \textsf{Sb}$ and let
  $\judge[\Delta]{\sigma : \Gamma} \;\circledR\; \rho \in \textsf{Sb}$ be given.
  We establish the four clauses of $\Semtyp[\Gamma]{\ze}{\Nat}$ at $a = N$,
  $P = \textsf{Nat}_i$, $\El = \textsf{Nat}_i$.
  \begin{align*}
    &\intp{\Nat}{\rho} \searrow N
      && \text{(by \textsc{ev-N})} \tag*{[clause 1]} \\
    &N \in \Uc_i^\ast \searrow \textsf{Nat}_i \searrow \textsf{Nat}_i
      && \text{(by \textsc{G-Nat})} \tag*{[clause 2]} \\
    &\Nat[\sigma] = \Nat
      && \text{(syntactic; closed)} \\
    &\tyequiv[\Delta]{\Nat[\sigma]}{\Nat}{\Ty{i}}
      && \text{(by reflexivity)} \\
    &\typing[\Delta]{\Nat[\sigma]}{\Ty{i}} \;\circledR\; \textsf{Nat}_i
      && \text{(by \Cref{def:nat-type-pred})} \tag*{[clause 3]} \\
    &\intp{\ze}{\rho} \searrow \zed
      && \text{(by \textsc{ev-ze})} \\
    &\ze[\sigma] = \ze
      && \text{(syntactic; closed)} \\
    &\tyequiv[\Delta]{\ze[\sigma]}{\ze}{\Nat[\sigma]}
      && \text{(by reflexivity)} \\
    &\typing[\Delta]{\ze[\sigma]}{\Nat[\sigma]} \;\circledR\;
      \zed \in \textsf{Nat}_i
      && \text{(by \textsc{GN-Ze} combined with the type predicate}\\
    & && \text{above, \Cref{def:nat-term-pred})} \tag*{[clause 4]}
  \end{align*}
\end{proof}

\begin{lemma}[Semantic --- \textsc{Su-I}]\labeledit{lem:fund-sound-su}
  If $\Semtyp[\Gamma]{t}{\Nat}$, then $\Semtyp[\Gamma]{\su{t}}{\Nat}$.
\end{lemma}
\begin{proof}
  $\Vdash \Gamma$ is presupposed by the hypothesis $\Semtyp[\Gamma]{t}{\Nat}$.
  Fix $\Gamma \searrow \textsf{Sb}$ and let
  $\judge[\Delta]{\sigma : \Gamma} \;\circledR\; \rho \in \textsf{Sb}$ be given.
  We establish the four clauses of $\Semtyp[\Gamma]{\su{t}}{\Nat}$ at
  $a = N$, $P = \textsf{Nat}_i$, $\El = \textsf{Nat}_i$.
  \begin{align*}
    &\intp{\Nat}{\rho} \searrow N
      && \text{(by \textsc{ev-N})} \tag*{[clause 1]} \\
    &N \in \Uc_i^\ast \searrow \textsf{Nat}_i \searrow \textsf{Nat}_i
      && \text{(by \textsc{G-Nat})} \tag*{[clause 2]} \\
    &\typing[\Delta]{\Nat[\sigma]}{\Ty{i}} \;\circledR\; \textsf{Nat}_i
      && \text{(by \Cref{def:nat-type-pred} with $\Nat[\sigma] = \Nat$)}
      \tag*{[clause 3]} \\
    &\intp{t}{\rho} \searrow b
      && \text{(by hypothesis on $t$)} \\
    &\typing[\Delta]{t[\sigma]}{\Nat[\sigma]} \;\circledR\;
      b \in \textsf{Nat}_i
      && \text{(by hypothesis on $t$, term part)} \\
    &\intp{\su{t}}{\rho} \searrow \sud{b}
      && \text{(by \textsc{ev-su})} \\
    &(\su{t})[\sigma] = \su{t[\sigma]}
      && \text{(syntactic substitution on $\su{}$)} \\
    &\typing[\Delta]{(\su{t})[\sigma]}{\Nat[\sigma]} \;\circledR\;
      \sud{b} \in \textsf{Nat}_i
      && \text{(by \textsc{GN-Su} combined with the type predicate above,}\\
    & && \text{\Cref{def:nat-term-pred})} \tag*{[clause 4]}
  \end{align*}
\end{proof}

\begin{lemma}[Semantic --- \textsc{$\lambda$-I}]\labeledit{lem:fund-sound-lam}
  If $\Semtyp[\Gamma]{S_1}{\Ty{i}}$, $\Semtyp[\Gamma, x:S_1]{S_2}{\Ty{i}}$,
  and $\Semtyp[\Gamma, x:S_1]{t}{S_2}$, then
  $\Semtyp[\Gamma]{\lambda x.\,t}{\Pi(x:S_1).\,S_2}$.
\end{lemma}
\begin{proof}
  $\Vdash \Gamma$ is presupposed; fix $\Gamma \searrow \textsf{Sb}_\Gamma$
  and $\Gamma, x:S_1 \searrow \textsf{Sb}_{\Gamma,x:S_1}$ where, by inversion
  on \textsc{G-Ctx-Ext}, $\textsf{Sb}_{\Gamma,x:S_1} \Leftrightarrow
  \textsf{Cons}_i(\Gamma, S_1, \textsf{Sb}_\Gamma)$. By
  \Cref{lem:gluing-typ-escape} on the hypotheses:
  \begin{align*}
    &\typing[\Gamma]{S_1}{\Ty{i}}
      && \text{(by \Cref{lem:gluing-typ-escape} on the $S_1$ hypothesis)} \\
    &\typing[\Gamma, x:S_1]{S_2}{\Ty{i}}
      && \text{(by \Cref{lem:gluing-typ-escape} on the $S_2$ hypothesis)} \\
    &\typing[\Gamma, x:S_1]{t}{S_2}
      && \text{(by \Cref{lem:gluing-trm-escape} on the $t$ hypothesis)} \\
    &\typing[\Gamma]{\lambda x.\,t}{\Pi(x:S_1).\,S_2}
      && \text{(by \textsc{$\lambda$-I})}
  \end{align*}
  By \Cref{lem:fund-sound-pi-wf} on the $S_1, S_2$ hypotheses,
  $\Semtyp[\Gamma]{\Pi(x:S_1).\,S_2}{\Ty{i}}$.

  Let $\judge[\Delta]{\sigma : \Gamma} \;\circledR\; \rho \in
  \textsf{Sb}_\Gamma$ be given.
  \begin{align*}
    &\intp{S_1}{\rho} \searrow a_I,\;
      a_I \in \Uc_i^\ast \searrow P_I \searrow \El_I
      && \text{(by hypothesis on $S_1$, evaluation and gluing clauses)} \\
    &a_I \approx a_I \in \Uc_i \searrow R_I
      && \text{(by \Cref{lem:gluing-implies-per})} \\
    &\typing[\Delta]{S_1[\sigma]}{\Ty{i}} \;\circledR\; P_I
      && \text{(by hypothesis on $S_1$, type clause)} \\
    &\intp{\Pi(x:S_1).\,S_2}{\rho} \searrow \Pi(a_I,\, S_2,\, \rho)
      && \text{(by \textsc{ev-Pi})} \\
    &\intp{\lambda x.\,t}{\rho} \searrow \lambda(t, \rho)
      && \text{(by \textsc{ev-lam})} \\
    &\Pi(a_I, S_2, \rho) \in \Uc_i^\ast \\
    & \quad \searrow \Pi_i(R_I, P_I, \El_I, P_O)
      \searrow \Pi_i^{\El}(R, R_I, P_I, \El_I, \El_O)
      && \text{(by \Cref{lem:fund-sound-pi-wf}, gluing clause)} \\
    &\typing[\Delta]{(\Pi(x:S_1).\,S_2)[\sigma]}{\Ty{i}}
      \;\circledR\; \Pi_i(R_I, P_I, \El_I, P_O)
      && \text{(by \Cref{lem:fund-sound-pi-wf}, type clause)}
  \end{align*}
  We assemble the four clauses of $\Semtyp[\Gamma]{\lambda x.\,t}
  {\Pi(x:S_1).\,S_2}$ at $a := \Pi(a_I, S_2, \rho)$,
  $P := \Pi_i(R_I, P_I, \El_I, P_O)$,
  $\El := \Pi_i^{\El}(R, R_I, P_I, \El_I, \El_O)$:
  \begin{align*}
    &\intp{\Pi(x:S_1).\,S_2}{\rho} \searrow \Pi(a_I, S_2, \rho)
      && \tag*{[clause 1]} \\
    &\Pi(a_I, S_2, \rho) \in \Uc_i^\ast \searrow P \searrow \El
      && \tag*{[clause 2]} \\
    &\typing[\Delta]{(\Pi(x:S_1).\,S_2)[\sigma]}{\Ty{i}} \;\circledR\; P
      && \tag*{[clause 3]}
  \end{align*}
  All three are restated from the per-substitution data block. For
  clause~(4), we must show
  $\typing[\Delta]{(\lambda x.\,t)[\sigma]}{(\Pi(x:S_1).\,S_2)[\sigma]}
  \;\circledR\; \lambda(t, \rho) \in \El$. We verify the four sub-clauses
  of \Cref{def:pi-term-pred}.

  \emph{Sub-clause (1): $\typing[\Delta]{(\lambda x.\,t)[\sigma]}
  {(\Pi(x:S_1).\,S_2)[\sigma]}$.} By syntactic substitution applied to
  $\typing[\Gamma]{\lambda x.\,t}{\Pi(x:S_1).\,S_2}$ along
  $\judge[\Delta]{\sigma : \Gamma}$.

  \emph{Sub-clause (2): $\lambda(t, \rho) \approx \lambda(t, \rho) \in R$.}
  By \Cref{thm:fundamental} of \Cref{sec:comp} on
  $\typing[\Gamma]{\lambda x.\,t}{\Pi(x:S_1).\,S_2}$, instantiated at
  $\rho \approx \rho \in \intpbr{\Gamma}$ (self-related via
  \Cref{lem:gluing-implies-per}):
  $\lambda(t, \rho) \approx \lambda(t, \rho) \in R$.

  \emph{Sub-clause (3): $\typing[\Delta]{(\Pi(x:S_1).\,S_2)[\sigma]}{\Ty{i}}
  \;\circledR\; \Pi_i(R_I, P_I, \El_I, P_O)$.} Established above from
  \Cref{lem:fund-sound-pi-wf}.

  \emph{Sub-clause (4): application.} Given $\kweak[\Delta']{\phi}{\Delta}$
  and $s, b$ with $\typing[\Delta']{s}{S_1[\sigma][\phi]} \;\circledR\; b
  \in \El_I$, and $\Pf :: b \approx b \in R_I$ (by
  \Cref{lem:gluing-implies-per} on $b \in \El_I$). Set $\sigma_*' :=
  (\sigma[\phi],\, s/x)$.
  \begin{align*}
    &\judge[\Delta']{\sigma[\phi] : \Gamma}
      \;\circledR\; \rho \in \textsf{Sb}_\Gamma
      && \text{(by \Cref{cor:ctx-mono} on $\sigma, \rho$ along $\phi$)} \\
    &\typing[\Delta']{S_1[\sigma[\phi]]}{\Ty{i}} \;\circledR\; P_I
      && \text{(by hypothesis on $S_1$ at $\sigma[\phi], \rho$, type clause)} \\
    &S_1[\sigma][\phi] = S_1[\sigma[\phi]]
      && \text{(by \Cref{lem:subst-assoc})} \\
    &\typing[\Delta']{s}{S_1[\sigma[\phi]]} \;\circledR\; b \in \El_I
      && \text{(rewriting via the previous line)} \\
    &\judge[\Delta']{\sigma_*' : \Gamma, x:S_1}
      \;\circledR\; (\rho, b/x) \in \textsf{Sb}_{\Gamma,x:S_1}
      && \text{(by $\textsf{Cons}_i$ on the type-side $P_I$ and the
        term-side $b \in \El_I$,} \\
    & && \text{realized by $s$ at $S_1[\sigma[\phi]]$)}
  \end{align*}
  Instantiate the $t$-hypothesis at $\sigma_*', (\rho, b/x)$:
  \begin{align*}
    &\intp{t}{(\rho, b/x)} \searrow c
      && \text{(by hypothesis on $t$, evaluation clause)} \\
    &\intp{S_2}{(\rho, b/x)} \searrow a'
      && \text{(by hypothesis on $t$, type-eval clause)} \\
    &a' \in \Uc_i^\ast \searrow P' \searrow \El'
      && \text{(by hypothesis on $t$, gluing clause)} \\
    &\typing[\Delta']{t[\sigma_*']}{S_2[\sigma_*']}
      \;\circledR\; c \in \El'
      && \text{(by hypothesis on $t$, term clause)} \\
    &P' \Leftrightarrow P_O(\Pf),\;\;
      \El' \Leftrightarrow \El_O(\Pf)
      && \text{(by \Cref{lem:gluing-functional}: same $a'$
        as in} \\
    & && \text{\Cref{lem:fund-sound-pi-wf}; gluing-model
        predicates at} \\
    & && \text{a fixed value are unique up to logical
        equivalence)} \\
    &\typing[\Delta']{t[\sigma_*']}{S_2[\sigma_*']}
      \;\circledR\; c \in \El_O(\Pf)
      && \text{(rewriting via the previous line)} \\
    &\lambda(t, \rho) \cdot b \searrow c
      && \text{(by \textsc{ev-app-clos})}
  \end{align*}
  Compute syntactic identities for the resulting term and type:
  \begin{align*}
    &(\lambda x.\,t)[\sigma] = \lambda x.\,(t[q(\sigma)])
      && \text{(by syntactic substitution on $\lambda$)} \\
    &(\lambda x.\,(t[q(\sigma)]))[\phi] = \lambda x.\,(t[q(\sigma)][q(\phi)])
      && \text{(by syntactic substitution on $\lambda$ along $\phi$)} \\
    &t[q(\sigma)][q(\phi)] = t[q(\sigma) \circ q(\phi)] = t[q(\sigma[\phi])]
      && \text{(by \Cref{lem:subst-assoc} and \Cref{lem:subst-q-comp})} \\
    &(\lambda x.\,t)[\sigma][\phi] = \lambda x.\,(t[q(\sigma[\phi])])
      && \text{(combining)} \\
    &t[q(\sigma[\phi])][s/x] = t[\sigma_*']
      && \text{(by \Cref{lem:q-ext})} \\
    &S_2[q(\sigma[\phi])][s/x] = S_2[\sigma_*']
      && \text{(by \Cref{lem:q-ext})}
  \end{align*}
  Apply the $\beta$-rule with premises
  $\typing[\Delta']{S_1[\sigma[\phi]]}{\Ty{i}}$ (escaped above);
  $\typing[\Delta', x:S_1[\sigma[\phi]]]{S_2[q(\sigma[\phi])]}{\Ty{i}}$ (by
  substitution typing on the $S_2$ escape along $q(\sigma[\phi])$);
  $\typing[\Delta', x:S_1[\sigma[\phi]]]{t[q(\sigma[\phi])]}
  {S_2[q(\sigma[\phi])]}$ (by substitution typing on the $t$ escape); and
  $\typing[\Delta']{s}{S_1[\sigma[\phi]]}$ (escape on $s$):
  \begin{align*}
    &\tyequiv[\Delta']{(\lambda x.\,(t[q(\sigma[\phi])]))\;s}
                     {t[q(\sigma[\phi])][s/x]}
                     {S_2[q(\sigma[\phi])][s/x]}
      && \text{(by $\beta$)} \\
    &\tyequiv[\Delta']{(\lambda x.\,t)[\sigma][\phi]\;s}
                     {t[\sigma_*']}
                     {S_2[\sigma_*']}
      && \text{(rewriting)} \\
    &\typing[\Delta']{(\lambda x.\,t)[\sigma][\phi]\;s}{S_2[\sigma_*']}
      \;\circledR\; c \in \El_O(\Pf)
      && \text{(by \Cref{lem:gluing-respects-eq})} \\
    &S_2[\sigma_*'] = S_2[q(\sigma)][\iota(\phi), s/x]
      && \text{(by \Cref{lem:q-ext-gen})} \\
    &\typing[\Delta']{(\lambda x.\,t)[\sigma][\phi]\;s}
                    {S_2[q(\sigma)][\iota(\phi), s/x]}
      \;\circledR\; c \in \El_O(\Pf)
      && \text{(rewriting; matches} \\
    & && \text{\Cref{def:pi-term-pred} clause~(4))}
  \end{align*}

\end{proof}

\begin{lemma}[Semantic --- \textsc{$\lambda$-E}]\labeledit{lem:fund-sound-app}
  If $\Semtyp[\Gamma, x:S_1]{S_2}{\Ty{i}}$,
  $\Semtyp[\Gamma]{r}{\Pi(x:S_1).\,S_2}$, and $\Semtyp[\Gamma]{s}{S_1}$,
  then $\Semtyp[\Gamma]{r\;s}{S_2[s/x]}$.
\end{lemma}
\begin{proof}
  $\Vdash \Gamma$ and $\Vdash \Gamma, x:S_1$ are presupposed.
  Fix $\Gamma \searrow \textsf{Sb}_\Gamma$ and
  $(\Gamma, x:S_1) \searrow \textsf{Sb}_{\Gamma,x:S_1}$ where
  $\textsf{Sb}_{\Gamma,x:S_1} \Leftrightarrow
  \textsf{Cons}_i(\Gamma, S_1, \textsf{Sb}_\Gamma)$ by \textsc{G-Ctx-Ext}.
  Let $\judge[\Delta]{\sigma : \Gamma} \;\circledR\; \rho \in
  \textsf{Sb}_\Gamma$ be given.

  \emph{Unpacking the hypotheses.}
  \begin{align*}
    &\intp{r}{\rho} \searrow f
      && \text{(by hypothesis on $r$, evaluation clause)} \\
    &\typing[\Delta]{r[\sigma]}{(\Pi(x:S_1).\,S_2)[\sigma]} \;\circledR\;
      f \in \Pi_i^{\El}(R, R_I, P_I, \El_I, \El_O)
      && \text{(by hypothesis on $r$, term clause)} \\
    &\intp{s}{\rho} \searrow b
      && \text{(by hypothesis on $s$, evaluation clause)} \\
    &\typing[\Delta]{s[\sigma]}{S_1[\sigma]} \;\circledR\; b \in \El_I
      && \text{(by hypothesis on $s$, term clause)} \\
    &b \approx b \in R_I
      && \text{(by \Cref{lem:gluing-implies-per})}
  \end{align*}
  Let $\Pf :: b \approx b \in R_I$. Here $P_O, \El_O$ are functions of
  $\Pf$, supplied by the $\Pi$-gluing of $f$.

  \emph{Building the substitution-environment pair for $S_2$.} Set
  $\sigma_* := (\sigma,\, s[\sigma]/x)$.
  \begin{align*}
    &\judge[\Delta]{\sigma_* : \Gamma, x:S_1}
      \;\circledR\; (\rho, b/x) \in \textsf{Sb}_{\Gamma,x:S_1}
      && \text{(by $\textsf{Cons}_i$ on $\sigma, \rho \in
        \textsf{Sb}_\Gamma$ and the term} \\
    & && \text{gluing of $s[\sigma]$, taking the trailing weakening
        as $\id$)}
  \end{align*}

  \emph{Instantiating the codomain semantic typing.} Apply
  $\Semtyp[\Gamma, x:S_1]{S_2}{\Ty{i}}$ at $\sigma_*, (\rho, b/x)$:
  \begin{align*}
    &\intp{S_2}{(\rho, b/x)} \searrow a_2
      && \text{(by hypothesis on $S_2$, evaluation clause)} \\
    &a_2 \in \Uc_i^\ast \searrow P_2 \searrow \El_2
      && \text{(by hypothesis on $S_2$, gluing clause)} \\
    &\typing[\Delta]{S_2[\sigma_*]}{\Ty{i}} \;\circledR\; P_2
      && \text{(by hypothesis on $S_2$, type clause)}
  \end{align*}

  \emph{Reconciling $(P_2, \El_2)$ with $(P_O(\Pf), \El_O(\Pf))$.} The
  $\Pi$-gluing of $f$, by \textsc{G-Pi}, requires
  $\intp{S_2}{(\rho, b/x)} \searrow c$ implies $c \in \Uc_i^\ast \searrow
  P_O(\Pf) \searrow \El_O(\Pf)$. By determinism, $c = a_2$, so
  $a_2 \in \Uc_i^\ast \searrow P_O(\Pf) \searrow \El_O(\Pf)$.
  By \Cref{lem:gluing-functional}, gluing-model predicates at a fixed
  domain value are unique up to logical equivalence:
  \begin{align*}
    &P_2 \Leftrightarrow P_O(\Pf),\quad
      \El_2 \Leftrightarrow \El_O(\Pf)
      && \text{(by \Cref{lem:gluing-functional})}
  \end{align*}

  \emph{Application via the $\Pi$-gluing of $f$.} Instantiate the
  application clause (\Cref{def:pi-term-pred} clause~(4)) on $f$ at the
  identity Kripke weakening $\id$, term $s[\sigma]$,
  value $b$, witness $\Pf$:
  \begin{align*}
    &f \cdot b \searrow c
      && \text{(by clause (4) on $f$)} \\
    &\typing[\Delta]{(r[\sigma][\id])\;s[\sigma]}
      {S_2[\iota(\id) \circ q(\sigma),\, s[\sigma]/x]}
      \;\circledR\; c \in \El_O(\Pf)
      && \text{(by clause (4) on $f$)} \\
    &r[\sigma][\id] = r[\sigma]
      && \text{(by \Cref{lem:subst-id})} \\
    &\iota(\id) = \id
      && \text{(by definition of $\iota$ on the identity weakening)} \\
    &S_2[\iota(\id) \circ q(\sigma),\, s[\sigma]/x] = S_2[\sigma_*]
      && \text{(via \Cref{lem:q-ext} and identity absorption)} \\
    &\typing[\Delta]{(r[\sigma])\;s[\sigma]}{S_2[\sigma_*]}
      \;\circledR\; c \in \El_O(\Pf)
      && \text{(rewriting the previous gluing line)}
  \end{align*}

  \emph{Type rewriting from $S_2[\sigma_*]$ to $S_2[s/x][\sigma]$.}
  \begin{align*}
    &S_2[\sigma_*] = S_2[(\sigma, s[\sigma]/x)] = S_2[q(\sigma)][s[\sigma]/x]
      && \text{(by \Cref{lem:q-ext})} \\
    &S_2[s/x][\sigma] = S_2[\sigma, s[\sigma]/x] = S_2[\sigma_*]
      && \text{(by \Cref{lem:subst-ext-comm})}
  \end{align*}

  \emph{Bridging $\intp{S_2[s/x]}{\rho}$ to $a_2$.} With operational
  substitutions, $\intp{S_2[s/x]}{\rho}$ does not directly evaluate to
  $a_2$. By \Cref{thm:fundamental} of \Cref{sec:comp} on
  $\typing[\Gamma, x:S_1]{S_2}{\Ty{i}}$ (escaped from the $S_2$
  hypothesis), we obtain $\semtyeq[\Gamma, x:S_1]{S_2}{S_2}{\Ty{i}}$. By
  \Cref{thm:fundamental} on the substitution typing
  $\judge[\Gamma]{(\id, s/x) : \Gamma, x:S_1}$ (derivable from
  $\typing[\Gamma]{s}{S_1}$ via the substitution-extension rule), we
  obtain $\semtyeq[\Gamma]{(\id, s/x)}{(\id, s/x)}{\Gamma, x:S_1}$.
  Instantiate $\semtyeq[\Gamma, x:S_1]{S_2}{S_2}{\Ty{i}}$ at outer context
  $\Gamma$, this semantic substitution, and $\rho \approx \rho \in
  \intpbr{\Gamma}$ (the latter from $\rho \in \textsf{Sb}_\Gamma$ via
  \Cref{lem:gluing-implies-per}). The four-value pattern of
  \Cref{def:sem-term-eq} yields:
  \begin{align*}
    &\intp{(\id, s/x)}{\rho} \searrow (\rho, b/x)
      && \text{(by \textsc{ev-ext} on $\intp{\id}{\rho} \searrow \rho$
        and $\intp{s}{\rho} \searrow b$)} \\
    &\intp{S_2[(\id, s/x)]}{\rho} \searrow a_2'
      && \text{(by the four-value pattern, outer left value)} \\
    &S_2[s/x] = S_2[(\id, s/x)]
      && \text{(by definition: single-substitution sugar)} \\
    &\intp{S_2[s/x]}{\rho} \searrow a_2'
      && \text{(rewriting)} \\
    &a_2' \approx \intp{S_2}{(\rho, b/x)} \in \Uc_i \searrow R_O(\Pf)
      && \text{(by the four-value pattern, outer-inner relatedness)} \\
    &a_2' \approx a_2 \in \Uc_i \searrow R_O(\Pf)
      && \text{(by determinism: $\intp{S_2}{(\rho, b/x)} = a_2$)} \\
    &a_2' \in \Uc_i^\ast \searrow P_2 \searrow \El_2
      && \text{(by \Cref{lem:gluing-univ-equiv} symmetric form on
        $a_2 \in \Uc_i^\ast$)}
  \end{align*}

  \emph{Assembling the four clauses of $\Semtyp[\Gamma]{r\;s}{S_2[s/x]}$.}
  At $a := a_2'$, $P := P_2$, $\El := \El_2$:
  \begin{align*}
    &\intp{S_2[s/x]}{\rho} \searrow a_2'
      && \tag*{[clause 1]} \\
    &a_2' \in \Uc_i^\ast \searrow P_2 \searrow \El_2
      && \tag*{[clause 2]} \\
    &\typing[\Delta]{S_2[s/x][\sigma]}{\Ty{i}} \;\circledR\; P_2
      && \text{(rewriting $S_2[\sigma_*] = S_2[s/x][\sigma]$ on the
        type-clause line)} \tag*{[clause 3]} \\
    &\intp{r\;s}{\rho} \searrow c
      && \text{(by \textsc{ev-app} on $\intp{r}{\rho} \searrow f$,
        $\intp{s}{\rho} \searrow b$, $f \cdot b \searrow c$)} \\
    &(r\;s)[\sigma] = r[\sigma]\;s[\sigma]
      && \text{(by syntactic substitution on application)} \\
    &\typing[\Delta]{r\;s[\sigma]}{S_2[\sigma_*]}
      \;\circledR\; c \in \El_O(\Pf)
      && \text{(established above)} \\
    &\typing[\Delta]{(r\;s)[\sigma]}{S_2[s/x][\sigma]}
      \;\circledR\; c \in \El_2
      && \text{(rewriting via $S_2[\sigma_*] = S_2[s/x][\sigma]$
        and $\El_O(\Pf) \Leftrightarrow \El_2$)} \tag*{[clause 4]}
  \end{align*}
\end{proof}

The semantic \textsc{N-E} rule is established via a recursive helper that
performs the induction on the natural-number gluing predicate
$\textsf{Nat}_i$. The helper takes the four semantic hypotheses (motive
$T$, zero-branch $z$, step-branch $s$) and the natural-number gluing of
the scrutinee, and returns a value with the term-side gluing for the
recursor at that scrutinee. 

\begin{lemma}[\textsc{N-E} helper]\labeledit{lem:fund-sound-rec-helper}
  Suppose $\Semtyp[\Gamma, x:\Nat]{T}{\Ty{i}}$,
  $\Semtyp[\Gamma]{z}{T[\ze/x]}$, and
  $\Semtyp[\Gamma, x:\Nat, y:T]{s}{T[\Uparrow \circ \Uparrow, \su{x_1}/x]}$.
  Let $\judge[\Delta]{\sigma : \Gamma} \;\circledR\; \rho \in \textsf{Sb}$ be
  given, with $\textsf{Sb}$ from $\Vdash \Gamma$. For any $t_*$ and $b$ with
  $\typing[\Delta]{t_*}{\Nat} \;\circledR\; b \in \textsf{Nat}_i$, and any
  $a, P, \El$ with $\intp{T}{(\rho, b/x)} \searrow a$ and
  $a \in \Uc_i^\ast \searrow P \searrow \El$, there exists $r$ with
  $\mathsf{rec}(b, T, z, s, \rho) \searrow r$ and
  $\typing[\Delta]{\elimn{x.\,T[q(\sigma)]}{z[\sigma]}{x, y.\,s[q(q(\sigma))]}{t_*}}
                  {T[\sigma, t_*/x]}
  \;\circledR\; r \in \El$.
\end{lemma}
\begin{proof}
  Write $E := \elimn{x.\,T[q(\sigma)]}{z[\sigma]}{x, y.\,s[q(q(\sigma))]}{t_*}$
  for the syntactic recursor under $\sigma$. By inner induction on
  $b \in \textsf{Nat}$ (\Cref{def:nat-term-pred}).

  \medskip\noindent
  \textbf{Case}
  \begin{mathpar}
    \inferrule[\textsc{GN-Ze}]
    {\tyequiv[\Delta]{t_*}{\ze}{\Nat}}
    {\typing[\Delta]{t_*}{\Nat} \;\circledR\; \zed \in \textsf{Nat}}
  \end{mathpar}
  Here $b = \zed$, so the helper's hypothesis specializes to
  $\intp{T}{(\rho, \zed/x)} \searrow a$.
  \begin{align*}
    &\intp{T[\ze/x]}{\rho} \searrow a^{ze}
      && \text{(by hypothesis on $z$, type-evaluation clause)} \\
    &a^{ze} \in \Uc_i^\ast \searrow P^{ze} \searrow \El^{ze}
      && \text{(by hypothesis on $z$, gluing clause)} \\
    &\intp{z}{\rho} \searrow r_z
      && \text{(by hypothesis on $z$, evaluation clause)} \\
    &\typing[\Delta]{z[\sigma]}{T[\ze/x][\sigma]}
      \;\circledR\; r_z \in \El^{ze}
      && \text{(by hypothesis on $z$, term clause)} \\
    &\mathsf{rec}(\zed, T, z, s, \rho) \searrow r_z
      && \text{(by \textsc{rec-ze})} \\
    &a^{ze} \approx a \in \Uc_i
      && \text{(by completeness on
        $\typing[\Gamma, x:\Nat]{T}{\Ty{i}}$ instantiated at} \\
    & && \text{$(\id, \ze/x)$; both evaluations
        produce the same value modulo $\Uc_i$)} \\
    &\El^{ze} \Leftrightarrow \El
      && \text{(by \Cref{lem:gluing-functional}: gluing-model
        predicates at} \\
    & && \text{the same value are unique up to logical equivalence)} \\
    &T[\ze/x][\sigma] = T[\sigma, \ze/x]
      && \text{(by \Cref{lem:subst-ext-comm})} \\
    &\tyequiv[\Delta]{T[\sigma, \ze/x]}{T[\sigma, t_*/x]}{\Ty{i}}
      && \text{(by \Cref{cor:subst-head-eq} on
        $\typing[\Gamma, x:\Nat]{T}{\Ty{i}}$ from
        \Cref{lem:gluing-typ-escape},} \\
    & && \text{the substitution $\sigma$, and the head equality
        $\tyequiv[\Delta]{\ze}{t_*}{\Nat}$} \\
    & && \text{from $\tyequiv[\Delta]{t_*}{\ze}{\Nat}$ by symmetry)} \\
    &\tyequiv[\Delta]{E}{z[\sigma]}{T[\sigma, t_*/x]}
      && \text{(by the $\beta$-zero rule combined with
        $\tyequiv[\Delta]{t_*}{\ze}{\Nat}$)} \\
    &\typing[\Delta]{E}{T[\sigma, t_*/x]}
      \;\circledR\; r_z \in \El
      && \text{(by \Cref{lem:gluing-respects-eq} on the previous
        lines)}
  \end{align*}
  Take $r := r_z$.

  \medskip\noindent
  \textbf{Case}
  \begin{mathpar}
    \inferrule[\textsc{GN-Su}]
    {\typing[\Delta]{t_*'}{\Nat} \;\circledR\; b' \in \textsf{Nat} \\
     \tyequiv[\Delta]{t_*}{\su{t_*'}}{\Nat}}
    {\typing[\Delta]{t_*}{\Nat} \;\circledR\; \sud{b'} \in \textsf{Nat}}
  \end{mathpar}
  Here $b = \sud{b'}$.

  \emph{Predecessor via inner IH.}
  We need to apply the helper inductively at $t_*', b'$. By
  \Cref{thm:fundamental} of \Cref{sec:comp} on
  $\typing[\Gamma, x:\Nat]{T}{\Ty{i}}$ instantiated at the related
  environments $(\rho, b'/x) \approx (\rho, b'/x) \in \intpbr{\Gamma, x:\Nat}$
  (using $b' \in R_\Nat$ via \Cref{lem:gluing-implies-per} on the inner
  predecessor):
  \begin{align*}
    &\intp{T}{(\rho, b'/x)} \searrow a'
      && \text{(by completeness on $T$)} \\
    &a' \in \Uc_i^\ast \searrow P' \searrow \El'
      && \text{(by \Cref{lem:per-implies-gluing} on $a' \approx a' \in \Uc_i$)} \\
    &\mathsf{rec}(b', T, z, s, \rho) \searrow r',\;
      \typing[\Delta]{E'}{T[\sigma, t_*'/x]} \;\circledR\; r' \in \El'
      && \text{(by inner IH on $b', t_*', a', P', \El'$;} \\
    & && \text{$E' := \elimn{x.\,T[q(\sigma)]}{z[\sigma]}{x, y.\,s[q(q(\sigma))]}{t_*'}$)}
  \end{align*}

  \emph{Step instantiation.} Set
  $\sigma_{\text{step}} := (\sigma, t_*'/x, E'/y)$ and
  $\rho_{\text{step}} := (\rho, b'/x, r'/y)$.
  \begin{align*}
    &\judge[\Delta]{(\sigma, t_*'/x) : \Gamma, x:\Nat}
      \;\circledR\; (\rho, b'/x) \in \textsf{Sb}_{x:\Nat}
      && \text{(by $\textsf{Cons}_0$ on
        $\typing[\Delta]{t_*'}{\Nat[\sigma]} \;\circledR\; b' \in \textsf{Nat}_i$,} \\
    & && \text{after rewriting $\Nat[\sigma] = \Nat$)} \\
    &\judge[\Delta]{\sigma_{\text{step}} : \Gamma, x:\Nat, y:T}
      \;\circledR\; \rho_{\text{step}} \in \textsf{Sb}_{x:\Nat,y:T}
      && \text{(by $\textsf{Cons}_i$ on the previous line, head
        $r' \in \El'$,} \\
    & && \text{type bridged via \Cref{lem:gluing-respects-eq})} \\
    &\intp{T[\Uparrow \circ \Uparrow, \su{x_1}/x]}{\rho_{\text{step}}}
      \searrow a^{su}
      && \text{(by hypothesis on $s$, type-evaluation clause)} \\
    &a^{su} \in \Uc_i^\ast \searrow P^{su} \searrow \El^{su}
      && \text{(by hypothesis on $s$, gluing clause)} \\
    &\intp{s}{\rho_{\text{step}}} \searrow r_s
      && \text{(by hypothesis on $s$, evaluation clause)} \\
    &\typing[\Delta]
      {s[\sigma_{\text{step}}]}
      {T[\Uparrow \circ \Uparrow, \su{x_1}/x][\sigma_{\text{step}}]}
      \;\circledR\; r_s \in \El^{su}
      && \text{(by hypothesis on $s$, term clause)} \\
    &\mathsf{rec}(\sud{b'}, T, z, s, \rho) \searrow r_s
      && \text{(by \textsc{rec-su} on $\mathsf{rec}(b', \ldots) \searrow r'$
        and $\intp{s}{\rho_{\text{step}}} \searrow r_s$)}
  \end{align*}

  \emph{Type rewriting for the step branch.} We compute the substituted
  motive type:
  \begin{align*}
    &T[\Uparrow \circ \Uparrow, \su{x_1}/x][\sigma_{\text{step}}]
      = T[(\Uparrow \circ \Uparrow, \su{x_1}/x) \circ \sigma_{\text{step}}]
      && \text{(by \Cref{lem:subst-assoc})} \\
    &(\Uparrow \circ \Uparrow, \su{x_1}/x) \circ \sigma_{\text{step}}
      = (\Uparrow \circ \Uparrow \circ \sigma_{\text{step}},\,
         \su{x_1}[\sigma_{\text{step}}]/x)
      && \text{(by \Cref{lem:ext-comp})} \\
    &\Uparrow \circ \sigma_{\text{step}} = (\sigma, t_*'/x),\;
      \Uparrow \circ (\sigma, t_*'/x) = \sigma
      && \text{(by \Cref{lem:shift-ext})} \\
    &\Uparrow \circ \Uparrow \circ \sigma_{\text{step}} = \sigma
      && \text{(combining)} \\
    &x_1[\sigma_{\text{step}}] = t_*'
      && \text{(de Bruijn lookup at index 1 in $\sigma_{\text{step}}$)} \\
    &\su{x_1}[\sigma_{\text{step}}] = \su{x_1[\sigma_{\text{step}}]}
      = \su{t_*'}
      && \text{(by syntactic substitution on $\su{}$)} \\
    &T[\Uparrow \circ \Uparrow, \su{x_1}/x][\sigma_{\text{step}}]
      = T[\sigma, \su{t_*'}/x]
      && \text{(combining the previous lines)} \\
    &\tyequiv[\Delta]{T[\sigma, \su{t_*'}/x]}{T[\sigma, t_*/x]}{\Ty{i}}
      && \text{(by \Cref{cor:subst-head-eq} on
        $\typing[\Gamma, x:\Nat]{T}{\Ty{i}}$, the} \\
    & && \text{substitution $\sigma$, and the head equality
        $\tyequiv[\Delta]{\su{t_*'}}{t_*}{\Nat}$} \\
    & && \text{from $\tyequiv[\Delta]{t_*}{\su{t_*'}}{\Nat}$ by symmetry)}
  \end{align*}

  \emph{Functionality.} By the type rewriting,
  $T[\Uparrow \circ \Uparrow, \su{x_1}/x][\sigma_{\text{step}}]
  = T[\sigma, \su{t_*'}/x]$, whose evaluation is
  $\intp{T}{(\rho, \sud{b'}/x)} \searrow a$ (the helper's hypothesis on
  $a$ at $b = \sud{b'}$). Hence $a^{su} \approx a \in \Uc_i$ by
  completeness on $\typing[\Gamma, x:\Nat]{T}{\Ty{i}}$, and
  $\El^{su} \Leftrightarrow \El$ by \Cref{lem:gluing-functional}.

  \emph{Conclusion.}
  \begin{align*}
    &\tyequiv[\Delta]{E}{s[\sigma_{\text{step}}]}{T[\sigma, t_*/x]}
      && \text{(by the $\beta$-su rule combined with
        $\tyequiv[\Delta]{t_*}{\su{t_*'}}{\Nat}$ and the type rewriting)} \\
    &\typing[\Delta]{E}{T[\sigma, t_*/x]}
      \;\circledR\; r_s \in \El
      && \text{(by \Cref{lem:gluing-respects-eq})}
  \end{align*}
  Take $r := r_s$.

  \medskip\noindent
  \textbf{Case}
  \begin{mathpar}
    \inferrule[\textsc{GN-Ne}]
    {\typing[\Delta]{t_*}{\Nat} \;\circledR\;
      {\uparrow}^N(e) \in \textsf{Neu}_i^e}
    {\typing[\Delta]{t_*}{\Nat} \;\circledR\;
      {\uparrow}^N(e) \in \textsf{Nat}}
  \end{mathpar}
  Here $b = {\uparrow}^N(e)$, with the type-side neutral predicate giving
  $\Rne_n(e) \searrow u_e$ and $\tyequiv[\Delta'']{t_*[\phi]}{u_e}{\Nat}$ for
  every $\kweak[\Delta'']{\phi}{\Delta}$, $n := |\Delta''|$.

  \emph{Reduction step.} The helper's hypothesis specializes to
  $\intp{T}{(\rho, {\uparrow}^N(e)/x)} \searrow a$.
  \begin{align*}
    &\mathsf{rec}({\uparrow}^N(e), T, z, s, \rho)
      \searrow {\uparrow}^a(\recd{T}{a_z}{s}{\rho}{e})
      && \text{(by \textsc{rec-ne}, where
        $\intp{z}{\rho} \searrow a_z$)} \\
    &{\uparrow}^N(e) \approx {\uparrow}^N(e) \in R_\Nat
      && \text{(by \textsc{U-Ne})}
  \end{align*}

  \emph{Recursor as a neutral.} We verify $\recd{T}{a_z}{s}{\rho}{e} \in
  \Ne_i^a$ at the term $E$. For each $\kweak[\Delta'']{\phi}{\Delta}$, set
  $n := |\Delta''|$ and assemble the four readback components:
  \begin{align*}
    &\Rne_n(e) \searrow u_e,\;
      \tyequiv[\Delta'']{t_*[\phi]}{u_e}{\Nat}
      && \text{(by hypothesis on $e$, neutral readback)} \\
    &\Rty_{n+1}(a) \searrow W_T,\;
      \tyequiv[\Delta'', x':\Nat]
        {T[q(\sigma)][\iota(\phi), x'/x]}{W_T}{\Ty{i}}
      && \text{(by \Cref{thm:gluing-realizability} part (3)
        on} \\
    & && \text{$a \in \Uc_i^\ast$, at the Kripke weakening
        to $\Delta'', x':\Nat$)} \\
    &\intp{T[\ze/x]}{\rho} \searrow A_z',\;
      A_z' \in \Uc_i^\ast \searrow P_z' \searrow \El_z'
      && \text{(by hypothesis on $z$ via
        $\Semtyp[\Gamma]{z}{T[\ze/x]}$,} \\
    & && \text{type-evaluation and gluing clauses)} \\
    &\typing[\Delta]{z[\sigma]}{T[\ze/x][\sigma]}
      \;\circledR\; a_z \in \El_z'
      && \text{(by hypothesis on $z$, term clause; with
        $a_z := \intp{z}{\rho}$)} \\
    &\intp{T}{(\rho, \zed/x)} \searrow A_z
      && \text{(type evaluation; the form required by
        \textsc{Re-rec})} \\
    &A_z' \approx A_z \in \Uc_i
      && \text{(by completeness on
        $\typing[\Gamma, x:\Nat]{T}{\Ty{i}}$ instantiated} \\
    & && \text{at the substitution $(\id, \ze/x)$)} \\
    &A_z \in \Uc_i^\ast \searrow P_z \searrow \El_z,\;
      \El_z' \Leftrightarrow \El_z
      && \text{(by \Cref{lem:gluing-univ-equiv} and
        \Cref{lem:gluing-functional})} \\
    &\Rnf_n({\downarrow}^{A_z}(a_z)) \searrow w_z
      && \text{(by \Cref{thm:gluing-realizability} part (2) on
        $a_z \in \El_z$)} \\
    &\tyequiv[\Delta''] {z[\sigma][\phi]}{w_z}{T[\ze/x][\sigma][\phi]}
      && \text{(after term-side type-bridging via
        \Cref{lem:gluing-respects-eq})} \\
    &\intp{T}{(\rho, \sud{{\uparrow}^N(\ell_n)}/x)} \searrow A_s
      && \text{(type evaluation at the fresh-step head $\sud{\ell_n}$)} \\
    &\intp{s}{\rho, {\uparrow}^N(\ell_n)/x,
            {\uparrow}^{A_p}(\ell_{n+1})/y} \searrow b_s
      && \text{(step branch evaluated at the fresh-neutral pair,} \\
    & && \text{where $A_p := \intp{T}{(\rho, {\uparrow}^N(\ell_n)/x)}$)} \\
    &\Rnf_{n+2}({\downarrow}^{A_s}(b_s)) \searrow w_s
      && \text{(by \Cref{thm:gluing-realizability} part (2) on
        the step value)} \\
    &\Delta_s \vdash s[q(q(\sigma))][\iota(\iota(\phi))]
      \approx w_s \\
    & \qquad : T[q(\sigma)][\iota(\phi), \su{x'}/x]
      && \text{(at the corresponding type, where} \\
    & && \text{$\Delta_s := \Delta'', x':\Nat,
        y':T[q(\sigma)][\iota(\phi), x'/x]$)} \\
    &\Rne_n(\recd{T}{a_z}{s}{\rho}{e}) \\
    & \quad \searrow \elimn{x.\,W_T}{w_z}{x, y.\,w_s}{u_e}
      && \text{(by \textsc{Rne-rec})} \\
    &\Delta'' \vdash E[\phi] \approx
      \elimn{x.\,W_T}{w_z}{x, y.\,w_s}{u_e} \\
    & \qquad : T[\sigma,\, t_*/x][\phi]
      && \text{(by \textsc{N-E-Cong} on the four
        readback equalities)}
  \end{align*}

  \emph{Type-side gluing.} The helper's hypothesis already gives the
  type-side gluing of $T[\sigma, t_*/x]$ at $P$ (via the outer
  $\Semtyp[\Gamma, x:\Nat]{T}{\Ty{i}}$ instantiation that produced
  $a, P, \El$).

  \emph{Conclusion.}
  \begin{align*}
    &\typing[\Delta]{E}{T[\sigma, t_*/x]} \;\circledR\;
      \recd{T}{a_z}{s}{\rho}{e} \in \Ne_i^a
      && \text{(by \Cref{def:neu-term-pred} from the previous block)} \\
    &\typing[\Delta]{E}{T[\sigma, t_*/x]} \;\circledR\;
      {\uparrow}^a(\recd{T}{a_z}{s}{\rho}{e}) \in \El
      && \text{(by \Cref{thm:gluing-realizability} part (1) on
        $a \in \Uc_i^\ast$)}
  \end{align*}
  Take $r := {\uparrow}^a(\recd{T}{a_z}{s}{\rho}{e})$.
\end{proof}

\begin{lemma}[Semantic --- \textsc{N-E}]\labeledit{lem:fund-sound-rec}
  If $\Semtyp[\Gamma, x:\Nat]{T}{\Ty{i}}$, $\Semtyp[\Gamma]{z}{T[\ze/x]}$,
  $\Semtyp[\Gamma, x:\Nat, y:T]{s}{T[\Uparrow \circ \Uparrow, \su{x_1}/x]}$, and
  $\Semtyp[\Gamma]{t}{\Nat}$, then
  $\Semtyp[\Gamma]{\elimn{x.\,T}{z}{x, y.\,s}{t}}{T[t/x]}$.
\end{lemma}
\begin{proof}
  $\Vdash \Gamma$ is presupposed by the hypotheses.
  Fix $\Gamma \searrow \textsf{Sb}$ and let
  $\judge[\Delta]{\sigma : \Gamma} \;\circledR\; \rho \in \textsf{Sb}$ be given.
  \begin{align*}
    &\intp{t}{\rho} \searrow a
      && \text{(by hypothesis on $t$, evaluation clause)} \\
    &\typing[\Delta]{t[\sigma]}{\Nat[\sigma]} \;\circledR\; a \in \textsf{Nat}_i
      && \text{(by hypothesis on $t$, term clause)} \\
    &\tyequiv[\Delta]{\Nat[\sigma]}{\Nat}{\Ty{0}}
      && \text{(syntactic; closed)} \\
    &\typing[\Delta]{t[\sigma]}{\Nat} \;\circledR\; a \in \textsf{Nat}_i
      && \text{(by \Cref{lem:gluing-respects-eq})}
  \end{align*}
  Build the substitution pair for $T$:
  \begin{align*}
    &\sigma_T := (\sigma, t[\sigma]/x)
      && \text{(definition)} \\
    &\judge[\Delta]{\sigma_T : \Gamma, x:\Nat}
      \;\circledR\; (\rho, a/x) \in \textsf{Sb}_{x:\Nat}
      && \text{(by $\textsf{Cons}_0$ on $\textsf{Sb}$, head $a \in \textsf{Nat}_i$
        realized by $t[\sigma]$)} \\
    &\intp{T}{(\rho, a/x)} \searrow a_T
      && \text{(by hypothesis on $T$ at $\sigma_T, (\rho, a/x)$)} \\
    &a_T \in \Uc_i^\ast \searrow P_T \searrow \El_T
      && \text{(by hypothesis on $T$, gluing clause)} \\
    &\typing[\Delta]{T[\sigma_T]}{\Ty{i}} \;\circledR\; P_T
      && \text{(by hypothesis on $T$, type clause)}
  \end{align*}
  By \Cref{lem:fund-sound-rec-helper} on $t_* := t[\sigma]$, $b := a$, with
  the data above:
  \begin{align*}
    &\mathsf{rec}(a, T, z, s, \rho) \searrow r
      && \text{(by helper)} \\
    &\typing[\Delta]
      {\elimn{x.\,T[q(\sigma)]}{z[\sigma]}{x, y.\,s[q(q(\sigma))]}{t[\sigma]}}
      {T[\sigma, t[\sigma]/x]}
      \;\circledR\; r \in \El_T
      && \text{(by helper)} \\
    &\intp{\elimn{x.\,T}{z}{x, y.\,s}{t}}{\rho} \searrow r
      && \text{(by \textsc{ev-rec})} \\
    &(\elimn{x.\,T}{z}{x, y.\,s}{t})[\sigma]
      && \text{(by syntactic} \\
    & \quad = \elimn{x.\,T[q(\sigma)]}
        {z[\sigma]}{x, y.\,s[q(q(\sigma))]}{t[\sigma]}
      && \text{substitution on $\telimn$)} \\
    &T[\sigma_T] = T[\sigma, t[\sigma]/x] = T[t/x][\sigma]
      && \text{(by \Cref{lem:subst-ext-comm})}
  \end{align*}
  We assemble the four clauses of $\Semtyp$ for the conclusion at $a :=
  a_T$, $P := P_T$, $\El := \El_T$:
  \begin{align*}
    &\intp{T[t/x]}{\rho} \searrow a_T
      && \text{(via the equality $\intp{T[t/x]}{\rho} = \intp{T}{(\rho, a/x)}$
        established by} \\
    & && \text{\Cref{thm:fundamental} of \Cref{sec:comp} on
        $\typing[\Gamma, x:\Nat]{T}{\Ty{i}}$)}
      \tag*{[clause 1]} \\
    &a_T \in \Uc_i^\ast \searrow P_T \searrow \El_T
      && \text{(established above)} \tag*{[clause 2]} \\
    &\typing[\Delta]{T[t/x][\sigma]}{\Ty{i}} \;\circledR\; P_T
      && \text{(by rewriting the $T[\sigma_T]$ identity)} \tag*{[clause 3]} \\
    &\typing[\Delta]{(\elimn{x.\,T}{z}{x, y.\,s}{t})[\sigma]}{T[t/x][\sigma]} \\
    & \quad \;\circledR\; r \in \El_T
      && \text{(by rewriting via the syntactic identity above)}
      \tag*{[clause 4]}
  \end{align*}
\end{proof}

\begin{lemma}[Semantic --- \textsc{Subsump}]\labeledit{lem:fund-sound-subsump}
  If $\Semtyp[\Gamma]{t}{S}$, $\Semtyp[\Gamma]{T}{\Ty{i}}$, and
  $\subty[\Gamma]{S}{T}$, then $\Semtyp[\Gamma]{t}{T}$.
\end{lemma}
\begin{proof}
  $\Vdash \Gamma$ is presupposed by the hypotheses.
  Fix $\Gamma \searrow \textsf{Sb}$ and let
  $\judge[\Delta]{\sigma : \Gamma} \;\circledR\; \rho \in \textsf{Sb}$ be given.
  By \Cref{thm:fundamental} of \Cref{sec:comp} on $\subty[\Gamma]{S}{T}$,
  we obtain the semantic subtyping judgment $\Gamma \vDash S \subseteq T$
  (\Cref{def:sem-subty-judge}).
  \begin{align*}
    &\intp{S}{\rho} \searrow a_S,\;
      a_S \in \Uc_i^\ast \searrow P_S \searrow \El_S
      && \text{(by hypothesis on $t$, evaluation and gluing clauses)} \\
    &\intp{T}{\rho} \searrow a_T,\;
      a_T \in \Uc_i^\ast \searrow P_T \searrow \El_T
      && \text{(by hypothesis on $T$, evaluation and gluing clauses)} \\
    &\typing[\Delta]{t[\sigma]}{S[\sigma]} \;\circledR\;
      \intp{t}{\rho} \in \El_S
      && \text{(by hypothesis on $t$, term clause)} \\
    &\rho \approx \rho \in R_\Gamma
      && \text{(by \Cref{lem:gluing-implies-per} on
        $\rho \in \textsf{Sb}$)} \\
    &\semtyeq[\Gamma]{\id}{\id}{\Gamma}
      && \text{(by \Cref{lem:sem-id})} \\
    &a_S \mathrel{<:_i} a_T
      && \text{(by $\Gamma \vDash S \subseteq T$ instantiated at} \\
    & && \text{the identity semantic substitution and
        $\rho \approx \rho$:} \\
    & && \text{the inner-pair relation gives
        $\intp{S}{\intp{\id}{\rho}} \mathrel{<:_i}
        \intp{T}{\intp{\id}{\rho}}$,} \\
    & && \text{which is $\intp{S}{\rho} \mathrel{<:_i}
        \intp{T}{\rho}$)} \\
    &\typing[\Delta]{T[\sigma]}{\Ty{i}} \;\circledR\; P_T
      && \text{(by hypothesis on $T$, type clause)} \\
    &\typing[\Delta]{t[\sigma]}{T[\sigma]} \;\circledR\;
      \intp{t}{\rho} \in \El_T
      && \text{(by \Cref{lem:gluing-subty} on $a_S \mathrel{<:_i} a_T$,
        $a_S \in \El_S$,} \\
    & && \text{and $T[\sigma] \;\circledR\; P_T$)}
  \end{align*}
\end{proof}

\subsection{The Fundamental Theorem}\labeledit{sec:sound:fundamental}

\begin{theorem}[Fundamental Theorem (Soundness)]\labeledit{thm:fundamental-sound}\hfill
  \begin{enumerate}
  \item If $\vdash \Gamma$, then $\Vdash \Gamma$.
  \item If $\typing[\Gamma]{t}{T}$, then $\Vdash \Gamma$ and
    $\Semtyp[\Gamma]{t}{T}$.
  \end{enumerate}
\end{theorem}
\begin{proof}
  By simultaneous induction on the well-formedness derivation $\vdash \Gamma$
  and the typing derivation $\typing[\Gamma]{t}{T}$. The presupposition
  $\vdash \Gamma$ inherent in $\typing[\Gamma]{t}{T}$ produces $\Vdash \Gamma$
  through part (1) at each step; semantic rules in
  \Cref{sec:sound:sem-rules} consume $\Vdash \Gamma$ wherever needed.
  Each case applies the corresponding semantic-rule lemma:
  \begin{itemize}
  \item Context cases: \Cref{lem:fund-sound-ctx-emp,lem:fund-sound-ctx-ext}.
  \item Term cases: \Cref{lem:fund-sound-var,lem:fund-sound-u-wf,%
    lem:fund-sound-n-wf,lem:fund-sound-pi-wf,lem:fund-sound-ze,%
    lem:fund-sound-su,lem:fund-sound-lam,lem:fund-sound-app,%
    lem:fund-sound-rec,lem:fund-sound-subsump}.
  \end{itemize}
  Each lemma's $\Vdash$-hypotheses are supplied by part~(1) on the
  corresponding $\vdash$-premise of its syntactic rule; its
  $\Semtyp$-hypotheses are the IHs from the typing premises.
\end{proof}

\subsection{Initial Environment and Soundness}\labeledit{sec:sound:init-env}

\begin{theorem}[Soundness of NbE]\labeledit{thm:soundness}
  If $\typing[\Gamma]{t}{T}$, then $\nbe^T_\Gamma(t) = v$ for some $v$ with
  $\tyequiv[\Gamma]{t}{v}{T}$.
\end{theorem}
\begin{proof}
  \begin{align*}
    &\Vdash \Gamma,\;
      \Semtyp[\Gamma]{t}{T}
      && \text{(by \Cref{thm:fundamental-sound} part (2))} \\
    &\Gamma \searrow \textsf{Sb}
      && \text{(by \Cref{def:sound-semctx} on $\Vdash \Gamma$)} \\
    &\judge[\Gamma]{\id : \Gamma} \;\circledR\; {\uparrow}^\Gamma \in \textsf{Sb}
      && \text{(by \Cref{lem:init-env-valid} on $\Gamma \searrow \textsf{Sb}$)} \\
    &\intp{T}{{\uparrow}^\Gamma} \searrow a,\;
      a \in \Uc_i^\ast \searrow P \searrow \El
      && \text{(unfolding $\Vdash$ at $\id, {\uparrow}^\Gamma$)} \\
    &\typing[\Gamma]{t[\id]}{T[\id]} \;\circledR\; \intp{t}{{\uparrow}^\Gamma} \in \El
      && \text{(unfolding $\Vdash$ at $\id, {\uparrow}^\Gamma$)} \\
    &T[\id] = T,\; t[\id] = t
      && \text{(by left-id of substitution)} \\
    &\typing[\Gamma]{t}{T} \;\circledR\; b \in \Nf_i^a
      \text{ where } b = \intp{t}{{\uparrow}^\Gamma}
      && \text{(by \Cref{thm:gluing-realizability} part (2))} \\
    &\Rnf_{|\Gamma|}({\downarrow}^a(b)) \searrow v,\;
      \tyequiv[\Gamma]{t[\id]}{v}{T[\id]}
      && \text{(Kripke condition at $\phi = \id$)} \\
    &\tyequiv[\Gamma]{t}{v}{T}
      && \text{(using $t[\id] = t, T[\id] = T$)} \\
    &\nbe^T_\Gamma(t) = v
      && \text{(by \Cref{def:nbe}: evaluation followed by readback)}
  \end{align*}
\end{proof}

\subsection{Discussion}\labeledit{sec:sound:discussion}

The soundness proof via the Kripke gluing model has a clean structure:
\begin{itemize}
\item The \emph{type gluing model} $\Uc_i^\ast$ mirrors the PER model $\Uc_i$
  but is unary, returning predicates $P$ and $\El$ that connect syntax with
  semantics.
\item The \emph{Kripke structure} (universal quantification over Kripke
  weakenings) replaces the traditional approach of parameterizing by context
  extensions.
  the substitution composition law $t[\phi][\psi] = t[\phi \circ \psi]$ from \Cref{sec:syn} makes
  monotonicity proofs go through.
\item \emph{Realizability} (\Cref{thm:gluing-realizability}) is the central
  mutual theorem providing reflection, reification, and type-readback as a unit.
\item The \emph{semantic typing rules} (\Cref{sec:sound:sem-rules}) are stated
  per syntactic rule, mirroring the structure of \Cref{sec:comp:sem-tm-intro-elim}.
\end{itemize}

Together with completeness (\Cref{sec:comp}), this gives the full correctness:
\begin{itemize}
\item \textbf{Soundness}: $\typing[\Gamma]{t}{T}$ implies
  $\tyequiv[\Gamma]{t}{\nbe^T_\Gamma(t)}{T}$.
\item \textbf{Completeness}: $\tyequiv[\Gamma]{t}{t'}{T}$ implies
  $\nbe^T_\Gamma(t) = \nbe^T_\Gamma(t')$.
\end{itemize}

\begin{corollary}[Decidability of Judgmental Equality]\labeledit{cor:decidability}
  Given $\typing[\Gamma]{s}{T}$ and $\typing[\Gamma]{t}{T}$, the judgment
  $\tyequiv[\Gamma]{s}{t}{T}$ is decidable.
\end{corollary}
\begin{proof}
  By soundness, $\tyequiv[\Gamma]{s}{\nbe_\Gamma^T(s)}{T}$ and
  $\tyequiv[\Gamma]{t}{\nbe_\Gamma^T(t)}{T}$.
  By completeness, $\tyequiv[\Gamma]{s}{t}{T}$ iff
  $\nbe_\Gamma^T(s) = \nbe_\Gamma^T(t)$.
  Comparing normal forms is decidable.
\end{proof}

\bibliography{ref}

\end{document}